\documentclass[12pt,a4paper,oneside]{book}

\usepackage[english]{babel}
\usepackage{layaureo}
\usepackage{cite}
\usepackage{amsmath}
\usepackage{amssymb}
\usepackage{graphicx}
\usepackage{tikz}
\usepackage[T2C]{fontenc}

\usepackage{imakeidx}
\makeindex[columns=2, title=Subject Index, intoc]

\usepackage{amsthm}
\theoremstyle{definition} 
\newtheorem{ex}{Exercise}[chapter] 

\usepackage{tikz-3dplot}

\usepackage{hyperref}

\begin{document}
\pagestyle{empty}
\begin{center}
\begin{LARGE}                 
{\textit{Lecture Notes in Cosmology-v2}\footnote{v1 is published by Springer and can be found at \url{https://link.springer.com/book/10.1007/978-3-319-95570-4}}}
\end{LARGE}
\end{center}

\begin{center}
\vspace{1cm}
Oliver F. Piattella\footnote{E-mail: \url{of.piattella@unisubria.it} --- Webpage: \url{https://www.oliverfpiattella.eu}}\\
\vspace{1cm}
Dipartimento di Scienza e Alta Tecnologia,\\
Universit\`a degli Studi dell'Insubria,\\
Como, Italy\\
\vspace{0.1cm}
-----------------------\\
\vspace{0.1cm} 
N\'ucleo Cosmo-UFES,\\
Universidade Federal do Esp\'{\i}rito Santo\\
Vit\'oria, ES, Brazil\\
\vspace{0.1cm}
-----------------------\\
\vspace{0.1cm} 
Como Lake center for AstroPhysics (CLAP)\\
Universit\`a degli Studi dell'Insubria,\\
Como, Italy\\
\end{center}
\vspace{\stretch{1.25}}       


%

\vphantom{mark} 
\vskip1.8truein
\noindent{\large \it To Giulio,\\The energy content of my Universe}
\vskip0.25truein


%
%
\tableofcontents
 
%
%
%
%
\vphantom{mark}
\vskip 0.truein
\noindent{\Huge\bf Notation}
\addcontentsline{toc}{chapter}{Notation}
\vskip 0.3truein

{\rightskip = 3truepc\leftskip = 3truepc\noindent
{\it Humans are good, she knew, at discerning subtle patterns that are really there, but equally so at imagining them when they are altogether absent}
\vskip 0.10 in
\centerline{\it ---Carl Sagan, Contact}
\vskip 0.20 in}

\noindent Latin indices ($i,j,k,\dots$) run over the three spatial coordinates and take values $1$, $2$, and $3$. In a Cartesian coordinate system, we use $x^1 \equiv x$, $x^2 \equiv y$, and $x^3 \equiv z$.\\
\vspace{0.1cm}

\noindent Greek indices ($\mu,\nu,\rho,\dots$) run over the four spacetime coordinates and take values $0$, $1$, $2$, $3$, with $x^0$ being the time coordinate.\\
\vspace{0.1cm}

\noindent Repeated high and low Greek indices are summed over their values, unless otherwise stated: $x^\mu y_\mu \equiv \sum_{\mu = 0}^3x^\mu y_\mu$. Repeated Latin indices are always summed; for example, $x_i y_i \equiv \sum_{i = 1}^3x_i y_i$.\\
\vspace{0.1cm}

\noindent The signature used for the metric is $(-,+,+,+)$.\\
\vspace{0.1cm}

\noindent 3-vectors are indicated in boldface, for example $\textbf{v}$.\\
\vspace{0.1cm}

\noindent The unit vector corresponding to any 3-vector $\textbf{v}$ is denoted with a hat; that is, $\hat v \equiv \textbf{v}/|\textbf{v}|$, where $|\textbf{v}|$ is the modulus of the 3-vector $\textbf{v}$, defined as $|\textbf{v}|^2 = \delta_{ij}v^iv^j$, with $\delta_{ij}$ being the usual Kronecker delta.\\
\vspace{0.1cm}

\noindent The cosmic time is denoted by $t$, and the conformal time is denoted by $\eta$.\\
\vspace{0.1cm}

\noindent A dot over any quantity denotes the derivative of that quantity with respect to cosmic time. The prime of any quantity denotes the derivative of that quantity with respect to the conformal time.\\
\vspace{0.1cm}

\noindent The operator $\nabla^2$ is the Laplacian operator in Euclidean space: $\nabla^2 \equiv \delta^{ij}\partial_i\partial_j$, where $\partial_\mu$ is used as a shorthand for the partial derivative with respect to the coordinate $x^\mu$.\\
\vspace{0.1cm}

\noindent Except for vectors and tensors, a $0$ subscript indicates that a time-dependent quantity is evaluated today, i.e., at $t = t_0$ or $\eta = \eta_0$, where $t_0$ and $\eta_0$ represent the age of the universe in cosmic time or in conformal time.\\
\vspace{0.1cm}

\noindent The subscripts $b$, $c$, $\gamma$, and $\nu$ attached to matter quantities such as density and pressure refer to baryons, cold dark matter, photons, and neutrinos, respectively. The subscripts $m$ and $r$ refer to matter and radiation, respectively.\\
\vspace{0.1cm}

\noindent The comoving wavenumber is denoted as $k$.\\
\vspace{0.1cm}

\noindent Starting from Chapter~\ref{Chap:CosmoPertTheory}, natural units $\hbar = c = 1$ are employed.\\
\vspace{0.1cm}

\noindent The azimuthal coordinate is denoted by $\phi$. A scalar field is denoted by $\varphi$.\\

\clearpage
\vphantom{mark}
\vskip 0.truein
\noindent{\Huge\bf Acknowledgments}
\addcontentsline{toc}{chapter}{Acknowledgments}
\vskip 0.2truein

{\rightskip=3truepc\leftskip=3truepc\noindent
{\it Omnia mea mecum sunt\\(All that is mine, I carry with me)}
\vskip 0.10 in
\centerline{\it ---Seneca, Epistulae Morales}
\vskip 0.20 in
}

I thank all the students who contacted me for pointing out errors, typos, or simply for asking for some clarifications about the material that I covered in the first edition of these notes. I am particularly indebted to Ali Rida Khalife, whose review has been particularly thorough.

\clearpage
\pagenumbering{arabic}
\pagestyle{plain}
\chapter{Cosmology}\label{Chap:Cosmology}

{\rightskip=3truepc\leftskip=3truepc\noindent
{\it And that inverted Bowl we call the Sky,\\ Whereunder crawling coop't we live and die,\\ \indent Lift not thy hands to It for help---for It,\\ Rolls impotently on as Thou or I.}
\vskip 0.10 in
\centerline{\it ---Omar Khayy\'am, Rub\'aiy\'at}
\vskip 0.20 in
}

This chapter intends to offer an overview of cosmology, covering both the observational evidence and the unresolved issues.   

\section{The expanding universe}

The starting point of our study of cosmology is the evidence that most of the galaxies we observe are receding from us; in other words, \textbf{our universe is expanding}. 

This was a major breakthrough in Western philosophy (but not in Eastern philosophy), which considered the universe static and immutable.\footnote{Although many pieces of evidence, such as supernova explosions and comets, showed long ago to the open-minded natural philosophers that immutability is \textit{not} an attribute of the heavens.} More than that, faraway galaxies recede from us in the same way (obeying the so-called \textbf{Hubble-Lemaître law}) independently of their directions. Are we then the center of the universe? 

Although some people think they are, there is no need to advocate for a center of the universe. The fact that every sufficiently distant galaxy recedes from us, no matter the direction of observation, has another explanation: space itself expands, carrying the galaxies within it. An effective similitude is to imagine the expanding universe as the surface of an inflating balloon on which many dots (representing the galaxies) and waves (representing electromagnetic radiation) have been drawn.\footnote{See, for example, \url{https://astro.ucla.edu/~wright/balloon0.html}.}


This explanation suggests that we do not occupy any privileged position in the universe from which every galaxy is moving away. This statement is elevated to the status of a principle, the \textbf{Copernican principle}:\index{Copernican Principle} the universe looks (on average) the same from any vantage point.  

\subsection{A bit of history}

The recession of galaxies is a landmark discovery made in the 1920s and is usually attributed to E. Hubble \cite{Hubble:1929ig}. However, evidence that \textit{nebulae} (this is how galaxies were called until the 1930s) are mostly receding from us rather than approaching us was already present in the work of V. M. Slipher \cite{Slipher:1917zz}. In Table I of \cite{Slipher:1917zz}, in fact, the predominance of the $+$ sign (meaning the recession of the corresponding galaxy) is striking: 21 of 25. On the other hand, the non-uniform distribution of the observed galaxies prevented Slipher from drawing the conclusion that \textit{most} galaxies are receding from us \textit{independently} of their angular positions in the sky. Indeed, as he writes:

\textit{``As noted before the majority of the nebulae here discussed have positive velocities, and they are located in the region of sky near right ascension twelve hours which is rich in spiral nebulae. In the opposite point of the sky some of the spiral nebulae have negative velocities, i.e., are approaching us; and it is expected that when more are observed there, still others will be found to have approaching motion.''}

The last sentence, and what Slipher writes in the following, suggests that the astronomer had in mind a drift motion of our Solar System through space rather than a cosmic expansion, since some of the nebulae appeared to be approaching us, whereas others appeared to be receding from us. He computed such a drift velocity to be about $700$ km/s in the direction $22$ hours right-ascension and $-22$ degrees declination. We are able today to determine such a drift motion of our Solar System cosmologically by measuring the dipole signal of the CMB radiation.  

The last conclusion in Slipher's paper is very interesting:

\textit{``While the number of nebulae is small and their distribution poor this result may still be considered as indicating that we have some such drift through space. For us to have such motion and the stars not show it means that our whole stellar system moves and carries us with it. It has for a long time been suggested that the spiral nebulae are stellar system seen at great distances. This is the so-called ``island-universe'' theory, which regards our stellar system and the Milky Way as a great spiral nebula which we see from within. This theory, it seems to me, gains favor in the present observations.''} 

At that time, it was not yet clear that nebulae were other galaxies, similar to our own Milky Way, and Slipher's results, although not interpreted as an expanding universe, were an important step forward in that understanding. 

There actually is a peculiar motion of our Solar System relative to other stars in our galaxy and relative to other galaxies (and also relative to the CMB). In this regard, in the 1920s, a problem arose which, in Hubble's words from the beginning of \cite{Hubble:1929ig}, is the following:

\textit{``Determinations of the motion of the sun with respect to the extragalactic nebulae have involved a $K$ term of several hundred kilometers which appears to be variable. Explanations of this paradox have been sought in a correlation between apparent radial velocities and distances, but so far the results have not been convincing. The present paper is a re-examination of the question, based on only those nebular distances which are believed to be fairly reliable.''}  

The paradox was that the relative motion between the Sun and the nebulae seemed to be too large and to present some ``not convincing'' correlation with the distance of the nebulae. This issue had already been taken into account before \cite{Hubble:1929ig} by other astronomers; for example, by K. Lundmark \cite{1925MNRAS..85..865L}, who is cited by Hubble in \cite{Hubble:1929ig} (perhaps, the ``has not been convincing'' refers to him). 

In his 1929 paper, Hubble reexamined the problem with more precise data on the distance of the nebulae. He used the relation:
\begin{equation}
	Kr + X\cos\alpha\cos\delta + Y\sin\alpha\cos\delta + Z\sin\delta = v\;,
\end{equation}   
where $K$ is the sought correction, which we call today Hubble constant\index{Hubble constant} and denote with $H_0$ (pronounced ``H-naught'');\footnote{It would be interesting to know when the symbol $H_0$ was used the first time. In A. Sandage's 1958 paper \cite{Sandage1958ApJ} $H$ was already used, with no subscript $0$, and, in Sandage's words, \textit{``$H$ is often called the Hubble's constant.''}.} $r$ is the distance to the galaxy; $X,Y,Z$ are the components of the relative velocity between the Sun and the observed galaxy; $\alpha$ and $\delta$ are the right ascension and declination of the observed galaxy; finally, $v$ is the radial velocity of the observed galaxy and is thus given by two contributions: the ``new'' one, $Kr$, and the ``expected'' peculiar relative velocity projected along the line of sight.

For sufficiently large distances, we can neglect the peculiar relative motion and, in modern notation, we have the celebrated \textbf{Hubble-Lemaître law}\index{Hubble-Lemaître law}:
\begin{equation}\label{Hubblelaw}
	\boxed{v = H_0r}
\end{equation}
The value of $H_0$ determined by Hubble himself from the estimated distances of 24 objects is:
\begin{equation}\label{H0byHubble}
	H_0 = (465 \pm 50)\;\textrm{km s}^{-1}\;\textrm{Mpc}^{-1}\;.
\end{equation}
The error in Hubble's determination \eqref{H0byHubble} is about 10\%. What might be surprising is instead the use of a linear regression for data as scattered as those in Fig. 1 of Hubble's paper, where a linear trend is far from evident. 


Perhaps such a choice was based on simplicity. Indeed, Hubble writes on page 171 of \cite{Hubble:1929ig}:

\textit{``A constant term, introduced into the equations, was found to be small and negative. This seems to dispose of the necessity for the old constant $K$ term. Solutions of this sort have been published by Lundmark, who replaced the old $K$ by $k+ lr+ mr^2$. His favored solution gave $k= 513$, as against the former value of the order of $700$, and hence offered little advantage.''}   

Indeed, Lundmark proposed in \cite{1925MNRAS..85..865L} the relation: 
\begin{equation}
	X\cos\alpha\cos\delta + Y\sin\alpha\cos\delta + Z\sin\delta + k + lr + mr^2 - v = 0\;,
\end{equation}  
where one can already see the Hubble-Lemaître law, along with a quadratic correction. The value found by Lundmark for $l$, which represents the Hubble constant in this case, is greater than $10000$ km/s/Mpc. Lundmark mentions that the observed redshifts could also have some explanation within General Relativity (Hubble also comments on this at the end of his paper).\index{Lundmark}

Given this, it is not clear why Lundmark is almost completely unknown when he seems to have had the brilliant idea before Hubble, and the latter seems to have just simplified Lundmark's ansatz and used more reliable distance determinations. In this respect, see \cite{steer2012history}.

A more accurate (but not precise) estimate of the Hubble constant was made by Sandage \cite{Sandage1958ApJ} in 1958: 
\begin{equation}
	H_0 = 50-100\;\textrm{km s}^{-1}\;\textrm{Mpc}^{-1}\;,
\end{equation}
with huge uncertainty. 

Since then, the precision in the measurement of $H_0$ has greatly improved. Let us cite two figures. The first is the final 2018 result of the Planck mission \cite{Planck:2018vyg}:
\begin{equation}
	H_0 = (67.4 \pm 0.5)\;\textrm{km s}^{-1}\;\textrm{Mpc}^{-1}\;.
\end{equation}
Here, the Hubble constant is derived as a by-product from the analysis of CMB data. Another figure comes from the Hubble Space Telescope (HST) observations of 70 long-period Cepheids in the Large Magellanic Cloud \cite{Riess:2021jrx}:
\begin{equation}
	H_0 = (73.04 \pm 1.04)\;\textrm{km s}^{-1}\;\textrm{Mpc}^{-1}\;.
\end{equation}
This measure is then done like Hubble's, using Cepheids. The two determinations above are very precise but also quite distinct. They are said to be in \textbf{tension}. Understanding the cause of this tension is one of the hottest tasks in cosmology today.\index{Hubble tension}

From the theoretical side, Georges Lema\^{\i}tre \cite{Lemaitre:1927zz} obtained a linear relation between the recessional velocity and the distance of a galaxy in 1927. He did so by finding a cosmological model that mediates between the static universe of Albert Einstein \cite{einstein1917kosmologische} and the homogeneous spacetime of Willem de Sitter \cite{dS1917A}. He even applied the law to the then available observational data, finding a value of 625 $\textrm{km s}^{-1}\;\textrm{Mpc}^{-1}$. It should be noted that the model, or class of models, proposed by Lema\^{\i}tre had already been used by Alexander Friedmann in 1922 \cite{F1922}. 

In the English translation of Lemaître's paper \cite{Lemaitre:1931zz}, the formula corresponding to the Hubble-Lemaître law is present, but the estimate of the value of the Hubble constant is not included. This raised some suspicion of conspiracy, but it seems that Lemaître himself translated his paper and omitted that part \cite{livio2011mystery}.

In October 2018, the International Astronomical Union (IAU) voted in favor of changing the name ``Hubble's law'' to ``Hubble-Lemaître law,'' acknowledging the contribution of Lemaître. However, a renaming such as ``Lundmark-Lema\^\i tre-Hubble law'' would have been even fairer. However, the fairest approach would probably have been to avoid naming people altogether and simply refer to ``the law of the expansion of the universe''. 

\subsection{Olbers's paradox}

It is often ruminated upon that the expansion of the universe could have been predicted a century before Hubble by solving Olbers's paradox, also known as the \textbf{dark night sky paradox}.\footnote{Dennis Sciama gives a wonderful account of the paradox in Chapter VI of his book \textit{The Unity of the Universe} \cite{Sciama1959book}.}\index{Olbers's paradox}\index{Dark night sky paradox} 

In 1823, H. W. M. Olbers (1758–1840) published in the \textit{Astronomisches Jahrbuch f\"ur das Jahr 1826} \cite{1826astronomisches} an essay titled \textit{Ueber die Durchsichtigkeit des Weltraums}, that is, \textbf{On the transparency of the universe}. In the first part of the article, Olbers mentions the previous works of E. Halley (1656-1742) and I. Kant (1724-1804) and discusses the possibility of an infinite space filled with infinite celestial bodies (unfortunately, there are no precise citations). Then Olbers states:
\\\\
\textit{Sind wirklich im ganzen unendlichen Raum Sonnen vorhanden, sie mögen nun in ungefähr gleichen Abständen von einander, oder in Milchstrassen-Systeme vertheilt sein, so wird ihre Menge unendlich, und da müsste der ganze Himmel eben so hell sein, wie die Sonne. Denn jede Linie, die ich mir von unserem Auge gezogen denken kann, wird nothwendig auf irgend einen Fixstern treffen, und also müsste uns jeder Punkt am Himmel Fixsternlicht, also Sonnenlicht zusenden.}
\\\\
A translation into twenty-first-century English (provided by the author of these notes, who is an Italian native speaker) is as follows:
\\\\
\emph{If there are indeed suns in the entirety of infinite space, whether they are at roughly equal distances from each other or distributed in Milky Way-like systems, then their number would be infinite, and the entire sky would have to be just as bright as one sun. Because every line that I can imagine drawn by our eyes will necessarily hit some fixed star, every point in the sky must send us the light of a fixed star; that is, sunlight.}


A few lines after the above-reported passage in \cite{1826astronomisches}, Olbers writes:
\\\\
\textit{Halley läugnet die Folgerung, dass bei einer unendlichen Menge von Fixsternen der ganze Himmel so hell aussehen müsse, wie die Sonne, aber aus ganzen irrigen Gründen.}
\\\\
Again, a translation into twenty-first-century English (again provided by the author of these notes) is as follows:
\\\\
\textit{Halley denies the conclusion that, with an infinite number of fixed stars, the whole sky must look as bright as the Sun, but for entirely erroneous reasons.}
\\\\
Therefore, the possibility that an infinite number of stars in an infinitely large universe might make the night sky bright had been considered by Halley well before Olbers, and the conclusion was that the night sky should \textbf{not} have been bright; however, according to Olbers, Halley made the mistake of confounding the apparent size of a star with its actual size. Thus, while correctly considering that the number of fixed stars grows as the square of the distance, he mistakenly assumed that their separation grows as the fourth power of the distance. 
\\\\
Let us be more quantitative. Olbers's strategy is to compute the fraction of area occupied by stars pinned on spheres centered on Earth and of increasing radius, the first being that containing our Sun. We shall do something similar; however, since we know today that stars are packed in galaxies, we will rephrase Olbers's paradox using galaxies instead of stars.

First of all, given the infinite vastness of the universe and the infinite number of galaxies uniformly distributed within it, each line of sight must end on the surface of a galaxy or on a portion of it (since it might be eclipsed by other galaxies in front of it). Let $N$ be the number (possibly huge, but finite) of such portions of galaxies' surfaces covering the sky, and let $a_i$ ($i = 1,\dots,N$) be their areas. If the $i$-th portion lies at a distance $r_i$, we receive from it the flux $F_i = a_i\mathcal{L}/(4\pi r_i^2)$, where $\mathcal{L}$ is the luminosity per unit area of the galaxy. But $a_i/r_i^2 = \Omega_i$ is the solid angle subtended by the $i$-th portion of the sky. Therefore:
\begin{equation}\label{Olbersparadoxformula}
	F_i = \frac{\mathcal{L}}{4\pi}\Omega_i\;.
\end{equation}
Summing over all $i$, we have a total flux equal to $\mathcal{L}$, meaning that the whole sky should be as bright as a galaxy. The key point is that it is true that the farther a galaxy is, the fainter it appears; but we can pack more of them into the same patch of sky, thereby compensating for the lack of flux with more sources.
\\\\
Let us look for ways to solve Olbers's paradox. Clearly, we must modify or eliminate one or more of the initial assumptions. For example:
\begin{itemize}
	\item The universe is not eternal, so the light of some galaxies has not yet arrived at us. This is plausible, but eventually, this light will arrive at us, making the night sky bright. Worse than that, this would be a gradual process, i.e., the sky would be getting brighter and brighter with time. This is something for which we have no observational evidence.

\item Maybe there is no infinite number of galaxies. In this case, not every line of sight would end on a galaxy, and we could indeed expect a dark night sky. However, why in an infinitely large universe would we have a limited number of galaxies? 

\item What if the galaxies are not distributed uniformly? Perhaps in this case, the night sky would still be bright, even though not uniformly.

\item Maybe the light of the galaxies somehow gets lost during its journey to us, absorbed by some ``fog''. This was the explanation adopted by Olbers. However, this ``fog'', the intergalactic gas, would be heated up by the absorption of light, eventually becoming bright itself.
\end{itemize}
In order to solve the paradox, we must find a way for the light of some galaxies to not reach us. Which galaxies contribute the most to lighting up Olbers' sky? The farthest ones. Let us see why. 

The probability of a line of sight intercepting a galaxy follows a Poisson distribution, whose derivation is provided in the Appendix \ref{App:Poisson}. Here we just need to consider that the probability of a line of sight intersecting a galaxy between $r$ and $r + dr$ can be expressed as:
\begin{equation}
	dP = dr/\lambda\,,
\end{equation}  
where $\lambda$ is the mean free path, and it is calculated as follows (for large $N$):
\begin{equation}
	\frac{dr}{\lambda} = \frac{N\sigma}{A} = n\sigma dr\,,
\end{equation}
where $N$ is the number of galaxies, all assumed to have the same section $\sigma$, contained in a small volume of sky with a (flat) section $A$. In the above equation, the probability $dP = dr/\lambda$ is reformulated as the area covered by the $N$ galaxies, that is, $N\sigma$, divided by the total area $A$ considered. Then, one uses $N = nAdr$. So, we have the known result:
\begin{equation}
	\lambda = \frac{1}{n\sigma}\,.
\end{equation}
The probability of a line of sight not intercepting any galaxy up to a distance $r$ is:
\begin{equation}
	P(r) = \frac{1}{\lambda}e^{-r/\lambda}\,.
\end{equation}
The expectation value for $r$ is then precisely $\lambda$:
\begin{equation}
	\langle r\rangle = \lambda = \frac{1}{n\sigma}\,.
\end{equation}
Let us estimate $\lambda$ by taking, as a representative volume, a sphere of radius $1$ Mpc, roughly the distance between the Milky Way and Andromeda, and as a representative section, the Milky Way's, $\sigma = (15 \mbox{ kpc})^2$. Then:
\begin{equation}
	\lambda \approx \frac{1\mbox{ Mpc}^3}{225\mbox{ kpc}^2} \approx 10^4 \mbox{ Mpc}\,.
\end{equation}
This is a huge number, indeed of the order of the size of the observable universe (which is about $10^4$ Mpc).

Therefore, a way to solve the paradox is to eliminate the farthest spherical shells of galaxies by dropping the assumption of staticity. 

Suppose that the spherical shells move away from us and do so faster and faster the more distant they are (Hubble-Lemaître law). From a certain distance onward, the recession occurs at a speed greater than that of light. This distance is called \textbf{Hubble radius} $R_H$\index{Hubble radius}. Its present value is:
\begin{align}
    R_H = \frac{c}{H_0} \approx 10^4\mbox{ Mpc}\,,
\end{align}
and established \textbf{the size of the visible universe}. \index{Size of the visible universe}

There is no need to get upset when reading \textit{a speed larger than that of light} because space can expand in this way.\footnote{An easy example is the following: imagine two spaceships traveling in opposite directions with speeds $2c/3$. In the reference frame of one spaceship, the other is moving at speed $12c/13$, but the distance between the two spaceships is increasing at a rate of $4c/3$ times the time elapsed.} Thus, beyond the Hubble radius, light from stars cannot reach us, and the paradox is solved. So, \textit{it is dark at night because the universe is expanding}.

\subsection{An exercise}\label{Subsec:Mukhanovexercise}

This is a good point to solve Exercise 1.1 of \cite{Mukhanov:2005sc}. This exercise asks us to derive the Hubble-Lemaître law from the requirement that a universal expansion law exists, embodying the requirements of isotropy and homogeneity. This exercise is interesting because it does not require any dynamical equations; it is based on pure kinematics.

In Euclidean space, consider a generic expansion law $\mathbf v(\mathbf r, t)$, where we use Eulerian coordinates (these correspond to the comoving coordinates that we will encounter later in Chapter \ref{Chap:ExpandingUniverse}), so $\mathbf r$ does not depend on time. Since $\mathbf v(\mathbf r, t)$ is a universal expansion law, the relative velocity between any pair of points is given by $\mathbf v(\mathbf r, t)$, with $\mathbf{r}$ being the relative position between the two points. In this way, homogeneity is taken into account.

Again, since $\mathbf v(\mathbf r, t)$ is a universal expansion law, we have that:
\begin{align}\label{isotropiccond}
    \mathbf{v}(R\mathbf{r},t) = R\mathbf{v}(\mathbf{r},t)\,,
\end{align}
that is, the velocity field at the rotated vector $R\mathbf{r}$ is the velocity field $\mathbf{v}(\mathbf{r},t)$ rotated by the same rotation $R$. In this way, isotropy is taken into account.

Now, let us consider an infinitesimal rotation. In components:
\begin{align}
	R_j{}^i = \delta_j{}^i + \varepsilon \omega_j{}^i\,,
\end{align} 
where $\varepsilon \ll 1$. Expanding the above relation \eqref{isotropiccond} up to the first order in $\varepsilon$, one obtains:
\begin{align}
	v_j(r_k,t) + \frac{\partial v_j(r_k,t)}{\partial r_i}\varepsilon\omega_i{}^kr_k =  v_j(r_k,t) + \varepsilon\omega_j{}^kv_k(r_k,t)\,.
\end{align}
Equating the contributions $\propto\varepsilon$, we have:
\begin{align}
	\frac{\partial v_j(r_k,t)}{\partial r_i}\omega_i{}^kr_k = \omega_j{}^kv_k(r_k,t)\,,
\end{align}
that is:
\begin{align}
	\frac{\partial v_j(r_k,t)}{\partial r_i}r_k = \delta_j{}^iv_k(r_k,t)\,.
\end{align}
Taking $i \neq j$, we can conclude that the $i$-th component of the velocity depends only on the $i$-th component of the relative position vector. Dropping the indices for simplicity, we are left with an ordinary differential equation:
\begin{align}
	v'r = v\,,
\end{align}
where the prime represents the derivative with respect to $r$, and its solution is:
\begin{align}
	v = \dot a(t)r\,,
\end{align}
where a generic function of time, called $\dot a(t)$ for reasons that will become clear later, has appeared upon integration. The velocity field should be a derivative of some notion of distance (the physical or proper distance that we will encounter later in Chapter \ref{Chap:ExpandingUniverse}), so:
\begin{align}
    \mathbf{v} = \dot{\mathbf{D}}\,,
\end{align}
and then we have:
\begin{align}
    \mathbf v = \dot{\mathbf{D}} = \dot a(t)\mathbf r\,,
\end{align}
so that:
\begin{align}
    \mathbf{D} = a(t)\mathbf r\,,
\end{align}
where a possible integration constant has been set to zero, without loss of generality, because such a choice can be considered part of the definition of $\mathbf{D}$ (that is, if $\mathbf{r} = 0$, then $\mathbf{D} = 0$). So, we have:
\begin{align}
    \mathbf{v} = H(t)\mathbf{D}\,,
\end{align}
with $H(t) = \dot a/a$. The function of time $a(t)$ is called \textbf{scale factor} and $H(t)$ is the \textbf{Hubble parameter}.

\section{The content of the universe}

A key point of GR and its application to cosmology is that the details of the expansion of the universe depend on its content in a way not dissimilar to Newton's dynamics and gravitation. It turns out that the content of our universe is not as expected.

\subsection{Photons: the cosmic microwave background radiation}

The CMB is a thermal radiation relic from a hot, dense phase in the early evolution of our universe, which has now been cooled by cosmic expansion to just three degrees above absolute zero. Its existence had been predicted in the 1940s by Alpher and Gamow \cite{Alpher:1948ve}, and its discovery by Penzias and Wilson at Bell Labs in New Jersey, announced in 1965 \cite{Penzias:1965wn}, was convincing evidence that the cosmos we see today emerged from a \textbf{ hot big bang} more than 10 billion years ago.

Since its discovery, many experiments have been performed to observe the CMB radiation at different frequencies, directions, and polarizations, mostly with ground-based and balloon-based detectors. These have established the remarkable uniformity of the CMB radiation at a temperature of 2.7 Kelvin in all directions, with a small $\pm 3.3$~ mK dipole resulting from the Doppler shift due to our local motion (at 1 million kilometers per hour) with respect to this cosmic background. 

However, the study of the CMB has been transformed over the last 30 years by three pivotal satellite experiments. The first of these was the \textit{Cosmic Background Explorer} (CoBE), launched by NASA in 1990. In 1992, CoBE reported the detection of statistically significant temperature anisotropies in the CMB, at the level of $\pm 30$ $\mu$K on 10 degree scales~\cite{Smoot:1992td}, and confirmed the blackbody spectrum with astonishing precision, with deviations of less than 50 parts per million \cite{Smoot:1992td}. 


CoBE was followed by the \textit{Wilkinson Microwave Anisotropy Probe} (WMAP) satellite, launched by NASA in 2001. This satellite produced full-sky maps in five frequencies (from 23 to 94 GHz), mapping the temperature anisotropies to sub-degree scales and determining the CMB polarization on large angular scales for the first time.

The \textit{Planck} satellite, launched by the ESA in 2009, set the state-of-the-art with nine separate frequency channels, measuring temperature fluctuations to a millionth of a degree at an angular resolution of up to 5 arc-minutes. Planck's mission ended in 2013, and the full-mission data can be found in the Planck Legacy Archive. 


A great effort is being devoted to the detection of the B-mode of CMB polarization because it contains the signature of the background of primordial gravitational waves, as we shall see in Chapter \ref{Chap:CMBEvo}.

Among the history-making CMB experiments, we must mention the \textit{Balloon Observations Of Millimetric Extragalactic Radiation ANd Geophysics} (BOOMERanG), which was a balloon-based mission that flew in 1998 and 2003 and measured CMB anisotropies with great precision (higher than CoBE). From these data, the Boomerang collaboration first determined that the spatial curvature of the universe is small \cite{deBernardis:2000sbo}.

\subsection{Neutrinos}

Neutrinos also form a cosmic background, dubbed C$\nu$B. This should also contain a wealth of information; however, it is unobservable with our present technological level, given the difficulty in detecting neutrinos.\index{Neutrinos}

The C$\nu$B should also present a black-body spectrum, but with a temperature of about $1.9$ K, slightly colder than the CMB, for reasons that we will discuss in chapter \ref{Chap:KinTh}.

Neutrinos are also important in the formation of structures, and from their impact on the cosmological setting, it is possible to provide a constraint on their total mass, which is approximately $\leq 0.12$ eV \cite{Planck:2018vyg}. \index{Neutrinos!Mass}

\subsection{The accelerated expansion of the universe and dark energy}

The possibility of using type Ia supernovae as standard candles \cite{Phillips1993ApJ} has allowed astronomers to extend the Hubble diagram (distance \textit{vs.} redshift) to the farthest distances, where they made a revolutionary discovery. The analysis of the emission of this type of supernova indicates that the expansion of our universe is \textbf{accelerated} \cite{Perlmutter:1998np, Riess:1998cb}. 

This is baffling. Gravity, \textit{as we know it}, is an attractive force; therefore, a universe filled with matter, \textit{as we understand it}, should cause the expansion to decelerate. So, what causes the acceleration of the expansion? There is the possibility that neither matter in the universe nor gravity is \textit{as we know it} in our cosmic neighborhood. Maybe gravity is a repulsive force at very large scales, those on which we do cosmology. If this is the case, GR might need to be modified or extended. Einstein already took care of this task, albeit accidentally, by introducing the cosmological constant $\Lambda$, which, if of the correct sign, provides the sought-after effect of anti-gravity. On the other hand, maybe GR is just fine without $\Lambda$, and it is a new form of matter, or rather energy, that acts as antigravity. Whatever it might be, the unknown cause of the expansion of the universe is generically called \textbf{dark energy}\index{Dark Energy} (DE). 

\subsection{Dark matter}
 
Different kinds of observations indicate that there seems to be some extra gravitational pull here and there in the universe; a pull that cannot be explained by the stuff we see and that we associate with another dark component called \textbf{dark matter}\index{Cold Dark Matter} (DM).
\\\\
Let us examine some of the most important observational evidence for DM.

\paragraph{Rotation curves of disc galaxies.} By measuring the spectral displacement in the emission of a galaxy (from stars and interstellar gas), it is possible to build a velocity curve, i.e., a plot of velocity \textit{vs.} distance from the center of the galaxy.\footnote{The spectral displacement is due to the Doppler effect.}

Let us assume spherical symmetry (so that we can use Gauss' theorem) and consider a small mass moving on a circular orbit of radius $R$ about the center of the galaxy. Using Newtonian gravity (which seems to be good enough for galaxies) and setting the centrifugal force equal to the gravitational attraction of the galaxy on the small source, we obtain the velocity $V$ of the small portion:
\begin{equation}\label{galvelcurv}
	\frac{V^2}{R} = \frac{GM}{R^2}\;,
\end{equation}
where $G$ is Newton's constant, and $M$ is the mass of the galaxy. We then expect, at the outskirts of the galaxy, that $V \propto 1/\sqrt{R}$. What is observed, instead, is that the velocity profile is relatively flat. 


To explain this, we need either more matter or a different theory of gravity, a dichotomy similar to the one concerning dark energy. In the first case, we advocate for DM, which should be distributed as a spheroidal \textbf{halo}\index{Cold Dark Matter!Halo} around the galaxy. In the second case, a popular candidate is the so-called \textit{MOdified Newtonian Dynamics} (MOND), i.e., a modification of Newton's second law such that, basically, the left-hand side of Eq.~\eqref{galvelcurv} is $V^2/R^2$, and therefore $V$ turns out to be constant. \index{Modified Newtonian Dynamics}

It must be stressed that the inner velocity profile of disk galaxies can be satisfactorily explained by Newtonian gravity, assuming an appropriate model for the mass distribution and \textit{without} DM \cite{Kalnajs1999}. The necessity of the latter arises when one is able to extend the observation very far from the center of the galaxy. Here, one cannot see stars; rather, one observes the intergalactic medium HI 21 cm radio emission.\footnote{This is the transition between the two hyperfine states of neutral hydrogen. The hyperfine states take into account the spin-spin interaction between the electron and the proton.} Here, a flat velocity curve cannot be sustained by the bulge, and one must hypothesize a dark matter halo.  

\paragraph{The dynamics of galaxies in clusters.} The pioneering applications of the virial theorem to the Coma cluster by Fritz Zwicky \cite{Zwicky:1933} resulted in a mass-to-luminosity ratio $M/L$ much larger than that established for the individual galaxies. Let us report here the first paragraph of page 125 of \cite{Zwicky:1933} (translated from German by the author of these notes):

\textit{``In order to obtain, as observed, a mean Doppler effect of 1000 km/s or more, the average density in the Coma system should be at least 400 times higher than that derived from the observation of luminous matter. If this were the case, then we would be presented with the surprising result that dark matter is present with a density much larger than that of luminous matter.''}

To the best of our knowledge, this is the first instance in which the terminology ``dark matter'' (``dunkle Materie'', in German) is used.

\paragraph{The formation of structures in the universe.} The existence of structures such as galaxies and clusters of galaxies reveals that, in some regions of the universe, the density of matter is much higher than the cosmological average. More specifically, one calls the relative variation of the density of matter about the cosmological average the \textbf{density contrast} and denotes it as $\delta_{\rm b}$, where the subscript means \textit{baryonic}, an adjective that in cosmology refers to standard matter. At present, we have $\delta_{\rm b} \gg 1$ in many places in the universe, such as where galaxies and galaxy clusters are located. As we shall see in chapter \ref{Chap:Evopert}, relativistic cosmology predicts that $\delta_{\rm b}$ \textit{alone} would grow by a factor $10^3$ between recombination and the present time. Moreover, from the analysis of the CMB spectrum, we compute that the value of $\delta_{\rm b}$ at recombination is $\delta_{\rm b} \approx 10^{-5}$. Combining these two bits of information, we infer that today $\delta_{\rm b} \approx 10^{-2}$, which means that no structure could have formed. In this case, DM is required to enhance $\delta_{\rm b}$ growth after recombination.


\paragraph{The CMB.} The anisotropies in the CMB are quite well understood, as we will see in Chapter \ref{Chap:CMBEvo}, but this understanding requires DM. In particular, DM is required to explain the relative heights of the first and third peaks of the CMB TT spectrum.

\paragraph{Gravitational Lensing and the Bullet Cluster.} The bending of light is a method for measuring mass. When a background field of galaxies is distorted by a foreground lens (for example, a DM halo), the lensing phenomenon is referred to as \textit{weak}. Analyzing the distortion in the shapes of the background galaxies allows one to map the gravitational potential across the lens (more precisely, the \textbf{shear field}) and, consequently, its matter distribution. Weak gravitational lensing is a powerful tool for the study of the geometry of the universe, and its observation is one of the primary targets of surveys such as the ESA satellite \textit{Euclid}. 

A remarkable combination of X-ray and weak lensing observations made the \textbf{Bullet Cluster}\index{Bullet Cluster} famous \cite{Clowe:2006eq}. Indeed, X-ray maps show the result of a merger between the hot gasses of two galaxy clusters, which gravitational lensing maps reveal to be lagging behind their respective centers of mass. Therefore, most parts of the clusters simply passed through one another, leaving behind a smaller fraction of hot gas. This is considered direct empirical proof of the existence of dark matter, forming a massive halo and a gravitational potential well in which gas and galaxies lie.  


The observational evidence listed above points not only to the existence of dark matter but also to its necessity of being \textbf{cold}, meaning that DM particles should have small velocities. Hence, the acronym CDM (cold dark matter).\index{Cold Dark Matter!Coldness}

The combined observational successes of $\Lambda$ and CDM form the so-called $\Lambda$CDM model, which is the standard model of cosmology.\index{$\Lambda$CDM model} 

\subsection{The periodic table of the elements}

Observation and data analysis tell us that, within the $\Lambda$CDM model, $\Lambda$ accounts for more than 70\% of the energy content of the universe, whereas CDM accounts for about 25\%, thus leaving only 5\% of matter ``that we know about'': the elements of the periodic table or, in other words, protons and neutrons. 

\section{Open problems in cosmology}

The fundamental issue in cosmology is to understand what DM and DE are. The effort to answer this question causes cosmology, particle physics, and quantum field theory (QFT) to merge. The methods adopted to tackle these problems are essentially the search for particles beyond the Standard Model and the investigation of new theories of gravity, which, in most cases, are extensions of general relativity. For a review dedicated to the challenges to the $\Lambda$CDM model, see \cite{Perivolaropoulos:2021jda}.

\subsection{The cosmological constant problem and dark energy}

The pure geometrical $\Lambda$ and the quantum vacuum energy have the same dynamical behavior in general relativity. Estimating the latter via QFT calculations and comparing the result with the observed value leads to the famous \textbf{fine-tuning} problem of the cosmological constant. See \cite{Weinberg:1989, Martin:2012bt, Buchbinder:2021wzv}.\index{Cosmological constant!Problem} This roughly goes as follows: the observed value of $\rho_\Lambda$ is approximately $10^{-47}$ GeV$^4$ \cite{Ade:2015xua}. The natural scale for the vacuum energy density is the Planck scale, that is, $10^{76}$ GeV$^4$. \textit{Ergo}, there are 123 orders of magnitude of difference! This is a sensationalistic way of presenting the cosmological constant problem, but it can be stated more precisely within the framework of semiclassical gravity and attributed entirely to the electroweak phase transition, in which the Higgs classical vacuum jumps a gap of $10^8$ GeV$^4$ to reach the observed value $10^{-47}$ GeV$^4$. The precision in the jump is 55 orders of magnitude smaller than the gap. This is the typical \textit{fine tuning} problem that we would like to explain otherwise.\footnote{Note that, framed in this way, the cosmological constant problem affects not only the $\Lambda$CDM model but also any cosmological model attempting to substitute $\Lambda$ with a dynamical DE component.} 

Forgetting quantum fields and thinking about $\Lambda$ in a classical manner, a problem arises due to its constancy: the so-called \textbf{cosmic coincidence}\index{Cosmic coincidence} \cite{Zlatev:1998tr}. This problem stems from the fact that the density of matter increases as $(1 + z)^3$, whereas the energy density of the cosmological constant is, as its name indicates, constant. However, observation tells us that these two densities are approximately equal at the present time. This coincidence becomes all the more intriguing when we consider that if the cosmological constant had dominated the energy content of the universe earlier, galaxies would not have had time to form; on the other hand, had the cosmological constant dominated later, then the universe would still be in a decelerated phase of expansion or younger than some of its oldest structures, such as clusters of stars \cite{Velten:2014nra}.

The cosmic coincidence problem can be rephrased as a fine-tuning problem in the initial conditions of our universe. Indeed, consider the ratio $\rho_{\rm m}/\rho_\Lambda$ of the matter content to the cosmological constant. This ratio goes as $z^3$ for large redshifts. Suppose that we could extrapolate our classical theory (GR) up to the Planck scale, for which $z \approx 10^{32}$. Then, on the Planck scale, we have $\rho_{\rm m}/\rho_\Lambda \approx 10^{96}$. This means that, at trans-Planckian energies, possibly in the quantum universe, there must be a mechanism that establishes the ratio $\rho_{\rm m}/\rho_\Lambda$ with a precision of 96 significant digits! Not a digit can be missed, or we would have today 10 times more cosmological constant than matter, or 10 times less, with consequences that would strongly disagree with observation.

So, we find ourselves in an \textit{impasse}. On the one hand, $\Lambda$ is the simplest and most successful DE candidate. On the other hand, it suffers from the aforementioned issues. What do we do? Much of today research in cosmology addresses this question.\index{Alternatives to $\Lambda$} Answers are sought in new theories of gravity, including extensions or modifications of GR, of which DE would be a manifestation. There are so many papers addressing extended theories of gravity that it is quite difficult to choose representatives. Probably, the best option is to start with a textbook \cite{amendola2010dark}. 

A different approach is to accept that $\Lambda$ has the value it does by chance, and it turns out to be just the right value for structures to form and for us to be here doing cosmology. This is also known as \textbf{the Anthropic Principle}\index{Anthropic principle} and exists in many forms, some stronger than others. See, for example, \cite{Weinberg:1987dv}, where the value of $\Lambda$ is estimated based on the anthropic principle.

It is also possible that ours is one universe out of an infinite number of realizations, called \textbf{Multiverse}\index{Multiverse}, with different values of the fundamental constants. Life, as we know it, develops only in those universes where the conditions are favorable (e.g., not too large or too small $\Lambda$). Again, it is difficult to cite papers on these topics, but given the author's admiration for Steven Weinberg, see \cite{Weinberg:1992nd}.  

\subsection{What is dark matter?}

Most of the scientific community considers DM to be fundamental particles that do not belong to the Standard Model. It is pointless to provide a summary of the candidate theories, their motivations, and the experimental efforts to prove them, as it would take up too much space. Instead, see \cite{Bertone:2016nfn} for a historical account of dark matter and \cite{profumo2017introduction} for a textbook on particle dark matter. It must be stressed that, despite the enormous effort put into the search for DM particles, no detection, either direct or indirect, has been reported thus far.

\subsection{The cosmological tensions}

\textbf{Tensions}\index{Tensions in cosmology} occur when observations of different phenomena provide constraints on the same parameter that differ significantly, exceeding a confidence level of 99\%. We have seen the tension between the determination of $H_0$ using low-redshift probes and the CMB. Another important tension, or discrepancy—since it is less severe—regards the parameter $\sigma_8$. Roughly, this parameter quantifies the amount of structures in the universe, and it is found from local probes (mainly through weak-lensing observations) to be significantly smaller than the prediction derived from the analysis of the CMB. For a general overview of tensions in cosmology, see \cite{CosmoVerseNetwork:2025alb}.

\subsection{The cosmological principle and the Copernican principle}. 

These principles are the pillars on which the theoretical structure of cosmology is built. They state that our universe is isotropic and homogeneous on large distance scales and that it looks, on average, the same irrespective of the point from which it is observed. This allows us, mathematically, to single out a very simple metric, the Friedmann-Lemaître-Robertson-Walker metric, for describing the geometry of our universe on very large scales. Employing this metric, as we will see throughout these notes, we can provide many successfully confirmed predictions. But do these principles really hold true? How can we test them? See \cite{Euclid:2022ucc}.\index{Copernican Principle} See also \cite{Secrest:2025wyu} for a discussion of the \textbf{Cosmic Dipole Anomaly}. \index{Cosmic Dipole Anomaly} This represents an excess of power in the CMB dipole signal compared to that of the matter signal (although there is agreement in direction). Other anomalies in the CMB sky, in the sense of unexpected, statistically relevant features \cite{Schwarz:2015cma}, might have some relation to our fundamental assumptions of homogeneity and isotropy.

\subsection{The lithium problem}  

The abundance of lithium predicted by Big Bang Nucleosynthesis (BBN) is larger by a factor of 3 to 4 than the observed value, \cite{Fields:2011zzb, Fields:2019pfx}.\index{Lithium problem} Indeed, the primordial value of the lithium-to-hydrogen abundance is found to be:
\begin{align}
    \left(\frac{\rm Li}{\rm H}\right)_p = (1.6 \pm 0.3)\times 10^{-6}\,,
\end{align}
from the analysis of the spectra of low-metallicity stars. From Big Bang Nucleosynthesis, the prediction is:
\begin{align}
    \left(\frac{\rm Li}{\rm H}\right)_p = (4.72 \pm 0.72)\times 10^{-6}\,,
\end{align}
\clearpage
\chapter{The universe in expansion}\label{Chap:ExpandingUniverse}

{\rightskip = 3truepc\leftskip = 3truepc\noindent
{\it oras ubicumque locaris extremas, quaeram: quid telo denique fiet?\\ (wherever you shall set the boundaries, I will ask: what will then happen to the arrow?)}
\vskip 0.10 in
\centerline{\it ---Lucretius, De Rerum Natura}
\vskip 0.20 in}

We introduce in this chapter the geometric basis of cosmology and the expansion of the universe. Apart from the technical treatment, historical, theological, and mythological introductions to cosmology can be found in \cite{Ryden:2003yy} and \cite{bonometto2008cosmologia}. A basic knowledge of General Relativity is welcome.

\section{Newtonian cosmology}

In order to study cosmology, we need a theory of gravity because gravity is a long-range interaction, and the universe is quite vast. One might protest that electromagnetism is also a long-range interaction, but considering the lack of evidence that the universe is charged or made up of charges here and there, it seems reasonable to conclude that gravity is the only force we need to describe the universe on large scales.

Which theory of gravity do we use to describe the universe? Of course, GR.\footnote{Although extensions of GR are investigated today in relation to DE, as we have seen in Chapter \ref{Chap:Cosmology}.} However, it turns out that Newtonian physics works surprisingly well. It is also surprising that attempts to do cosmology with Newtonian gravity are well postdated to relativistic cosmology itself.

The first work on Newtonian cosmology can be dated back to McCrea and Milne in the 1930s \cite{1934QJMat...5...73M}, \cite{1934QJMat...5...64M}. These were models of pure dust, while pressure was introduced later by McCrea \cite{1951RSPSA.206..562M} and Harrison \cite{1965AnPhy..35..437H}. More recently, the issue of pressure corrections in Newtonian cosmology has been tackled again in \cite{Lima:1996at}, \cite{Fabris:2012xt}, \cite{Hwang:2013sia}, and \cite{Baqui:2015dqp}. 

Newtonian cosmology works as follows.\index{Newtonian cosmology} Imagine a sphere of \textbf{dust}\footnote{Dust is, in the jargon of cosmology, a pressureless fluid. In GR, the geodesic equations of test particles are identical to the Euler equation for a pressureless fluid.} of radius $r$. This radius must be time-dependent because the configuration is not stable, as there is no pressure. Thus, let $r = r(t)$. Assume also the homogeneity of the sphere during the evolution, i.e., its density depends only on time: 
\begin{equation}\label{dustdensity}
	\rho(t) = \frac{3M}{4\pi r(t)^3}\;,
\end{equation}
where $M$ is the mass of the sphere of dust, and it is constant. Now, imagine a small test particle of mass $m$ on the surface of the sphere. By Newton's law of gravitation and Gauss's theorem, one has:
\begin{equation}\label{newtequationcosmo}
	\ddot r = - \frac{GM}{r^2} \quad \Rightarrow \quad \ddot r = - \frac{4\pi G}{3}\rho r\,,
\end{equation}
where we have used Eq.~\eqref{dustdensity}, and the dot denotes derivation with respect to time $t$. The second form of the equation above is the same acceleration equation that we shall find later using GR, cf. Eq.~\eqref{accEq}. 

\hrulefill

\begin{ex} Integrate Eq.~\eqref{newtequationcosmo} and show that:
\begin{equation}\label{friedeqnewtcosmo}
	\frac{\dot r^2}{r^2} = \frac{2GM}{r^3} - \frac{K}{r^2} \quad \Rightarrow \quad \frac{\dot r^2}{r^2} = \frac{8\pi G}{3}\rho - \frac{K}{r^2}\;,
\end{equation}
where $K$ is an integration constant.
\end{ex}

\hrulefill

We shall also see that the second expression of Eq.~\eqref{friedeqnewtcosmo} is the same as the Friedmann equation in GR, cf. \eqref{FriedEq2}. The integration constant $K$ can be interpreted as the total energy of the particle. Indeed, we can rewrite Eq.~\eqref{friedeqnewtcosmo} as follows:
\begin{equation}
	E \equiv -\frac{mK}{2} = \frac{m}{2}\dot r^2 - \frac{GMm}{r}\;.
\end{equation}
This is the expression for the total energy of a particle of mass $m$ in the gravitational field of the mass $M$.

\hrulefill

\begin{ex} Solve Eq.~\eqref{friedeqnewtcosmo} for $r(t)$. Consider the different cases of $K > 0$, $K < 0$ and $K = 0$. In the first two cases look for a parametric solution: assume $t = t(p)$ and $r = r(p)$, where $p$ is some parameter. Then, Eq.~\eqref{friedeqnewtcosmo} becomes:
\begin{align}
    t'^2 = \frac{rr'^2}{2GM - Kr}\,,
\end{align}
\end{ex}
where the prime denotes derivation with respect to $p$. The strategy is now to provide a simple ansatz for $t'$ so that the right hand side becomes easily integrable (use $t' = \alpha r$, where $\alpha$ is a coefficient to be determined).

\hrulefill

Equation \eqref{friedeqnewtcosmo} (with $K > 0$) is used to study the nonlinear evolution of spherically symmetric dust perturbations (the so-called Top-Hat model).\index{Top-Hat Model} \index{Spherical Collapse}

\section{The cosmological principle and the Friedmann-Lema\^{\i}tre-Robertson-Walker metric}

In GR, we have geometry and matter related by the Einstein equations:
\begin{equation}
	G_{\mu\nu} \equiv R_{\mu\nu} - \frac{1}{2}g_{\mu\nu}R = \frac{8\pi G}{c^4} T_{\mu\nu}\;,
\end{equation}
where $G_{\mu\nu}$ is the Einstein tensor; $R_{\mu\nu}$ is the Ricci tensor:
\begin{equation}
	R_{\mu\nu} = \partial_\rho\Gamma_{\mu\nu}^\rho - \partial_\mu\Gamma_{\rho\nu}^\mu + \Gamma_{\mu\nu}^\rho\Gamma_{\rho\sigma}^\sigma - \Gamma_{\mu\rho}^\sigma\Gamma_{\nu\sigma}^\rho\,,
\end{equation}
where the Christoffel symbols of the Levi-Civita connection are:
\begin{equation}\label{Christoffelsymbols}
	\Gamma_{\mu\nu}^\rho = \frac{1}{2}g^{\rho\sigma}\left(\partial_\nu g_{\mu\sigma} + \partial_\mu g_{\nu\sigma} - \partial_\rho g_{\mu\nu}\right)\,,
\end{equation}
and $g_{\mu\nu}$ is the metric. Finally, $R$ is the Ricci scalar, or scalar curvature, which is the trace of the Ricci tensor $R = g^{\mu\nu}R_{\mu\nu}$. In the right-hand side of the Einstein equations, $T_{\mu\nu}$ is the energy-momentum or stress-energy tensor, and it describes the matter content.

In cosmology, what is the metric that describes the universe, and what is its matter content? It turns out that both questions are very difficult to answer; indeed, there are no final answers, as we stressed in Chapter~\ref{Chap:Cosmology}.

The metric used to describe the universe on large scales is the Friedmann-Lema\^{\i}tre-Robertson-Walker (FLRW) metric. This is based on a requirement of very high symmetry for the universe, called the \textbf{cosmological principle}\index{Cosmological principle}, which is minimally stated as follows: the universe is isotropic and homogeneous; i.e., there is no way of identifying a special direction or a special position.\footnote{The terminology ``cosmological principle'' was used for the first time by E. A. Milne in \cite{1935rgws.book.....M}. H. P. Robertson writes in the introduction of his paper \cite{R1935}: \textit{``The uniformity postulate, which Milne fittingly calls ``the cosmological principle,'' asserts that the description of the whole system, ...''}.}  

A more formal definition can be found in \cite[pag. 412]{Weinberg:1972} and is based on the following two requirements:
\begin{enumerate}
 \item The hypersurfaces with constant cosmic standard time are \textbf{maximally symmetric subspaces} of the entirety of spacetime;
 \item The global metric and all the cosmic tensors, such as the stress-energy one $T_{\mu\nu}$, are form-invariant with respect to the isometries of those subspaces.
\end{enumerate}
We shall return momentarily to maximally symmetric spaces. The above requirements together demand the existence of coordinate systems that are equivalent (that is, related by isometries) and in which the spatial sections (constant time) are maximally symmetric (therefore, isotropic and homogeneous). Equivalent coordinate systems all use the same time coordinate, called \textbf{cosmic time}. Moreover, due to maximal symmetry, in the above coordinate systems, physical quantities can depend only on the cosmic time.\footnote{The Copernican principle states that we are not a privileged observer in the universe and that any other observer would see, on average, the same universe from their vantage point as we do from ours. The cosmological principle can be formulated as the requirement of isotropy plus the Copernican principle since a space that is isotropic at any point is also homogeneous.}\index{Copernican Principle}

The cosmological principle seems to be compatible with observations at very large scales. According to \cite{Wu:1998ad}: \textit{on a scale of about 100 $h^{-1}$ Mpc, the rms density fluctuations are at the level of $\sim$10\% and on scales larger than 300 $h^{-1}$ Mpc, the distribution of both mass and luminous sources safely satisfies the cosmological principle of isotropy and homogeneity}.\footnote{Here $h$ is the Hubble constant in units of 100 km s$^{-1}$ Mpc$^{-1}$.} 

In a recent work, \cite{Sarkar:2016fir} it was found that the quasar distribution is homogeneous on scales larger than 250 $h^{-1}$ Mpc. 

According to the cosmological principle, the constant-time spatial hypersurfaces are maximally symmetric.\footnote{This means that they possess 6 Killing vectors, i.e., there are six transformations (isometries) that leave the spatial metric invariant \cite{Weinberg:1972}.} A maximally symmetric space\index{Maximally symmetric space} is completely characterized by one number only, i.e., its scalar curvature, which is also a constant. See \cite[Chapter 13]{Weinberg:1972}. 

Let $R$ be this constant scalar curvature. The Riemann tensor of a maximally symmetric $D$-dimensional space is written as:
\begin{equation}
	R_{\mu\nu\rho\sigma} = \frac{R}{D(D - 1)}(g_{\mu\rho}g_{\nu\sigma} - g_{\mu\sigma}g_{\nu\rho})\;.
\end{equation}
Contracting with $g^{\mu\rho}$, we obtain the Ricci tensor:
\begin{equation}
	R_{\nu\sigma} = \frac{R}{D}g_{\nu\sigma}\;,
\end{equation}
and then $R$ is the scalar curvature, as we stated, since $g^{\nu\sigma}g_{\nu\sigma} = D$. Since any given number can be negative, positive, or zero, we have three possible maximally symmetric spaces. 

In the 3-dimensional spatial case, the three possibilities of $R$ being zero, positive, or negative can be realized as follows:
\begin{enumerate}
	\item $ds_3^2 = |d\textbf{x}|^2 \equiv \delta_{ij}dx^idx^j$, i.e., the three-dimensional Euclidean space. The scalar curvature is zero, i.e., the space is flat. This metric is invariant under 3-translations and 3-rotations. 

\item $ds_3^2 = |d\textbf{x}|^2 + dz^2$, with the constraint $z^2 + |\textbf{x}|^2 = a^2$. This is a 3-sphere of radius $a$ embedded in a 4-dimensional Euclidean space. It is invariant under the six 4-dimensional rotations.

\item $ds_3^2 = |d\textbf{x}|^2 - dz^2$, with the constraint $z^2 - |\textbf{x}|^2 = a^2$.\footnote{This is the Gauss-Bolyai-Lobachevsky geometry, or hyperbolic geometry. Developed in the first half of the 19th century, it is the first example of geometry (of a non-compact space, since the sphere was already well-known) in which the fifth postulate of Euclid does not hold true.} This is a 3-hypersphere, or a hyperboloid, in a 4-dimensional pseudo-Euclidean space. It is invariant under the six 4-dimensional pseudo-rotations (i.e., the Lorentz transformations).
\end{enumerate}

\hrulefill

\begin{ex} Why are there six independent 4-dimensional rotations in the 4-dimensional Euclidean space? How many are there in a $D$-dimensional Euclidean space?
\end{ex}

\hrulefill

Let us write the above 3-dimensional line elements in a compact form as follows:
\begin{equation}\label{rescaling}
	ds_3^2 = |d\textbf{x}|^2 \pm dz^2\;, \qquad z^2 \pm |\textbf{x}|^2 = a^2\;.
\end{equation}
Differentiating $z^2 \pm |\textbf{x}|^2 = a^2$, one obtains:
\begin{equation}
	zdz = \mp \textbf{x}\cdot d\textbf{x}\;.
\end{equation}
Now put this back into $ds_3^2$:
\begin{equation}
	ds_3^2 = |d\textbf{x}|^2 \pm \frac{(\textbf{x}\cdot d\textbf{x})^2}{a^2 \mp |\textbf{x}|^2}\;.
\end{equation}
In a more compact form:
\begin{equation}\label{compactform}
	ds_3^2 = |d\textbf{x}|^2 + K\frac{(\textbf{x}\cdot d\textbf{x})^2}{a^2 - K|\textbf{x}|^2}\;,
\end{equation}
with $K = 0$ for the Euclidean case, $K = 1$ for the spherical case, and $K = -1$ for the hyperbolic case. The components of the spatial metric in Eq.~\eqref{compactform} can be immediately read off and are:
\begin{equation}\label{componentsspatial0}
	ds_3^2 = {}^{(3)}g_{ij}dx^idx^j\,, \qquad {}^{(3)}g_{ij} = \delta_{ij} + K\frac{x_ix_j}{a^2 - K|\textbf{x}|^2}\;. 
\end{equation}
Recall that $a^2$ is the curvature radius of the maximally symmetric space. If we normalize the coordinates as $\bar x_i \equiv x_i/a$, we can write:
\begin{equation}\label{componentsspatial}
	ds_3^2 = a^2\gamma_{ij}d\bar x^id\bar x^j\,, \qquad \gamma_{ij} = \delta_{ij} + K\frac{\bar x_i\bar x_j}{1 - K|\bar{\mathbf x}|^2}\;. 
\end{equation}
However, another possibility is to redefine $K$ as $\bar K \equiv K/a^2$. In this case, $\bar K$ is no longer restricted to having values $\pm 1$ or zero; any real value is possible. We will prefer this choice throughout these notes.

\hrulefill

\begin{ex}
	Calculate the Riemann tensor, the Ricci tensor and the scalar curvature ${}^{(3)}R$ for $\gamma_{ij}$. Show that ${}^{(3)}R_{ij} = 2K\gamma_{ij}$ and thus ${}^{(3)}R = 6K$, as it is expected for a maximally symmetric space.
\end{ex}

\hrulefill

\begin{ex} Write metric \eqref{compactform} in spherical coordinates. Use the fact that $|d\textbf{x}|^2 = dr^2 + r^2d\Omega^2$, where 
\begin{equation}
  d\Omega^2 \equiv d\theta^2 + \sin^2\theta d\phi^2\;,
\end{equation}
and use:
\begin{equation}
	\textbf{x}\cdot d\textbf{x} = \frac{1}{2}d|\textbf{x}|^2 = \frac{1}{2}d(r^2) = rdr\;.
\end{equation}
Show that the result is:
\begin{equation}\label{spatialFLRWmetricsphericalcoords}
  \boxed{ds_3^2 = \frac{a^2dr^2}{a^2 - Kr^2} + r^2d\Omega^2}
\end{equation}

\end{ex}

\hrulefill

If we normalize $\bar r \equiv r/a$ in metric \eqref{spatialFLRWmetricsphericalcoords}, we can write:
\begin{equation}\label{spatialFLRWmetricsphericalcoords2}
  ds_3^2 = a^2\left(\frac{d\bar r^2}{1 - K\bar r^2} + \bar r^2d\Omega^2\right)\;.
\end{equation} 
Let us now include the time interval $-c^2dt^2$, let $a$ be a function of time, and drop the bar over $r$. We have finally arrived at the FLRW metric:
\begin{equation}\label{FLRWmet}
  \boxed{ds^2 = -c^2dt^2 + a^2(t)\left(\frac{dr^2}{1 - Kr^2} + r^2d\Omega^2\right)}
\end{equation}
The time coordinate used here is called \textbf{cosmic time}\index{Cosmic time}, whereas the spatial coordinates are referred to as \textbf{comoving coordinates}\index{Comoving coordinates}. For each $t$, the spatial slices are maximally symmetric; $a(t)$ is called \textbf{scale factor}\index{Scale factor}, since it tells us how the distance between two points on a spatial hypersurface scales with time. 

From Eq.~\eqref{FLRWmet} we see that:
\begin{equation}\label{gammaijsphercoords}
	\gamma_{ij}dx^idx^j = \left(\frac{dr^2}{1 - Kr^2} + r^2d\Omega^2\right)\;.
\end{equation}
It is useful to write $\gamma$ in isotropic coordinates, i.e., in diagonal form:
\begin{equation}
	\gamma_{ij} = B(x)\delta_{ij}\;,
\end{equation}
with $B$ as a function of the spatial coordinates. 

\hrulefill

\begin{ex}
	In order to find $B$, let us use Eq.~\eqref{gammaijsphercoords} and define a new radius $R$ such that:
	\begin{equation}
		B(R)R^2 = r^2\;, \qquad \frac{dr^2}{1 - Kr^2} = B(R)dR^2\;.
	\end{equation}
	Combining the above definitions, show that (there is a non-trivial integration involved):
	\begin{equation}
		B(R) = \frac{4}{K(1 + R^2)^2}\;.
	\end{equation}
Redefining:
\begin{equation}
	\bar R^2 = \frac{4R^2}{K}\;,
\end{equation}
and forgetting about the bar, we can finally write:
\begin{equation}\label{gammaijB}
	\gamma_{ij} = \frac{1}{(1 + \frac{K}{4}R^2)^2}\delta_{ij}\;.
\end{equation}

\end{ex}

\hrulefill

\begin{ex}
	Using metric \eqref{gammaijB}, with $R^2 = \delta_{ij}x^ix^j$, show that:
	\begin{equation}\label{3Gammaijk}
{}^{(3)}\Gamma^i_{jk} = -\frac{K/2}{1 + KR^2/4}(x_k\delta^i{}_{j} + x_j\delta^i{}_{k} - x^i\delta_{jk})\;.	
\end{equation}

\end{ex}

\hrulefill

The FLRW metric\index{FLRW metric} (with positive curvature of its spatial hypersurfaces) was first used as an ansatz to go beyond the cosmological solutions of Einstein and de Sitter by Friedmann in \cite{F1922} and Lema\^{\i}tre \cite{Lemaitre:1927zz}. Indeed, Friedmann's ansatz in \cite{F1922} is (using his notation):
\begin{equation}
	ds^2 = -\frac{R^2}{c^2}(dx^2_1 + \sin^2 x_1 dx_2^2 + \sin^2 x_1 \sin^2 x_2 dx^2_3) + M^2 dx^2_4\;,
\end{equation}
where $x_4$ is the time coordinate, $R$ is the scale factor, which depends only on $x_4$, and $M$ is another function that depends on all the coordinates. Indeed, the spatial part of the above metric is the metric of a 3-sphere. Lema\^{\i}tre's ansatz is the same as Friedmann's, but with $M =1$. On the other hand, Lema\^{\i}tre includes pressure, whereas Friedmann does not. Already in 1929, Robertson established in \cite{Robertson822} the foundations of relativistic cosmology, where, in particular, he presented the above metric with three possible choices for spatial curvature. 

The above metric is also named after Robertson and Walker. They followed Milne's pioneering work \cite{1935rgws.book.....M}, in which the expansion of the universe is discussed solely on the basis of kinematics, without relying on a theory of gravitation, and the idea of the ``cosmological principle'' is stated for the first time. Then, they showed how to find the above metric starting from the cosmological principle alone. Robertson used a technique \cite{R1935} based on the Helmholtz-Lie theorem, whereas Walker based his derivation \cite{walker1937milne} on continuous groups of transformations.

\subsection{The conformal time}

A useful form of rewriting the metric \eqref{FLRWmet} is via the \textbf{conformal time}\index{Conformal time} $\eta$, defined as follows:
\begin{equation}\label{conftimedefinition}
	ad\eta = dt \qquad \Rightarrow \qquad \eta - \eta_i = \int_{t_i}^t\frac{dt'}{a(t')}\;.
\end{equation}
As we can guess from the above integration, $c(\eta - \eta_i)$ represents the comoving distance traveled by a photon between the times $\eta_i$ and $\eta$, or $t_i$ and $t$. 

When we change coordinates, say from $x$ to $ \bar x$, the metric transforms as follows:
\begin{equation}
	 ds^2 = g_{\mu\nu}(x)dx^\mu dx^\nu = g_{\mu\nu}(x(\bar x))\frac{\partial x^\mu}{\partial \bar x^\rho}d\bar x^\rho\frac{\partial x^\nu}{\partial \bar x^\sigma}d\bar x^\sigma\,,
\end{equation}
because the line element $ds^2$ is an invariant. Therefore, the components of the metric in the new coordinates are:
\begin{equation}
	g_{\mu\nu}(x(\bar x))\frac{\partial x^\mu}{\partial \bar x^\rho}\frac{\partial x^\nu}{\partial \bar x^\sigma} = \bar g_{\rho\sigma}(\bar x)\,.
\end{equation}
In general, one cannot simply replace functional dependencies in the components of a tensor because these are mixed up by the coordinate transformation.

Using the above transformation (it is simpler if we work with $ds^2$ instead of the components), the FLRW metric \eqref{FLRWmet} with the conformal time is as follows:
\begin{equation}\label{FRWmetconf}
  \boxed{ds^2 = a(\eta)^2\left(-c^2d\eta^2 + \frac{dr^2}{1 - Kr^2} + r^2d\Omega^2\right)}
\end{equation}
Note that the scale factor has become a conformal factor. Recalling the earlier discussion about dimensionality, if $a$ has dimensions of length, then $c\eta$ is dimensionless. On the other hand, if $a$ is dimensionless, then $\eta$ indeed has dimensions of time. In these notes, we prefer to keep $a$ dimensionless. So, $K$ is a real number and not simply $0,\pm 1$.

Note also that the metric \eqref{FRWmetconf} for $K = 0$ is the Minkowski metric multiplied by a conformal factor. More generally, since its Weyl tensor is vanishing, the FLRW metric is \textbf{conformally flat}.

\subsection{The comoving distance}

The FLRW metric \eqref{FLRWmet} can be written as follows:
\begin{align}
    ds^2 = -c^2dt^2 + a^2(t)d\chi^2\,,
\end{align}
defining:
\begin{equation}\label{d2comdist}
	d\chi^2 \equiv \gamma_{ij}dx^idx^j = \frac{dr^2}{1 - Kr^2} + r^2d\Omega^2\;.
\end{equation}
The quantity $\chi$ is called \textbf{comoving distance}\index{Comoving distance}. The comoving distance is a notion of distance that does not account for the expansion of the universe and, thus, does not depend on time. For a photon ($ds^2 = 0$), one can express $\chi$ as:
\begin{align}
    \chi = c\int \frac{dt}{a(t)}\,.
\end{align}
Here, the time enters, so we would have a $\chi(t)$, an apparent contradiction with what has just been said about $\chi$. That is because we are measuring $\chi$ through the time of flight of the photon, and this is affected by the expansion of the universe. In order to compensate, we divide by $a(t)$. Compare the above formula with Eq. \eqref{conftimedefinition}: one can regard the comoving distance as the conformal time of flight for a photon.

The comoving distance $\chi$ introduced above is defined between any two spatial points. Observationally, it is convenient to place the observer at the origin of the comoving reference frame and to introduce a radial comoving distance by integrating Eq.~\eqref{d2comdist} with $d\Omega = 0$:
\begin{equation}
	\chi_r \equiv \int_0^r\frac{dr'}{\sqrt{1 - Kr^{'2}}} = 
	\begin{cases}
		\frac{1}{\sqrt{K}}\arcsin(\sqrt{K}r)\;, &\mbox{ for } K > 0\;,\\
		r\;, &\mbox{ for } K = 0\;,\\
		\frac{1}{\sqrt{|K|}}\mbox{arcsinh}(\sqrt{|K|}r)\;, &\mbox{ for } K < 0\;.
	\end{cases}
\end{equation}
In the literature \cite{Hogg:1999ad}, $\chi_r$ is called \textbf{line-of-sight comoving distance}.\index{Line-of-sight comoving distance}

The infinitesimal tangential comoving distance would then be:
\begin{align}
    d\chi_t = r(\chi_r)d\Omega\,,
\end{align}
where:
\begin{equation}\label{transversecomdist}
	r(\chi_r) = 
	\begin{cases}
		\frac{1}{\sqrt{K}}\sin(\sqrt{K}\chi_r)\;, &\mbox{ for } K > 0\;,\\
		\chi_r\;, &\mbox{ for } K = 0\;,\\
		\frac{1}{\sqrt{|K|}}\sinh(\sqrt{|K|}\chi_r)\;, &\mbox{ for } K < 0\;.
	\end{cases}
\end{equation}
In the literature \cite{Hogg:1999ad}, $r(\chi_r)$ is called \textbf{transverse comoving distance}.\index{Transverse comoving distance}

\subsection{The proper distance}

The \textbf{proper distance}\index{Proper distance} between two points at a certain time $t$ is:\footnote{It is customary to denote the proper distance with a $d$, but since we are going to consider its differential, it would be awkward to write $dd$.} 
\begin{align}
    D(t,\chi) = a(t)\chi\,.
\end{align}
The proper distance is the distance that would be measured instantaneously by rulers. For example, imagine extending a ruler between any distant galaxy and us. The reading at time $t$ would be the proper distance. 

Deriving $D$ with respect to cosmic time, one obtains:
\begin{equation}
	\dot{{D}} = \dot a \chi = \frac{\dot a}{a} D = H(t)D\;,
\end{equation}
where:
\begin{equation}
	\boxed{H \equiv \frac{\dot a}{a}}
\end{equation}
is the \textbf{Hubble parameter}.\index{Hubble parameter} The dot denotes differentiation with respect to cosmic time. As we have seen already in subsection \ref{Subsec:Mukhanovexercise}, the above formula generalizes the Hubble-Lemaître law at any time.\index{Hubble-Lemaître law!Derivation}

Another useful way to write the FLRW metric \eqref{FLRWmet} is by using the proper distance.

\hrulefill

\begin{ex} Using the proper distance $D$, show that the FLRW metric \eqref{FLRWmet} becomes:
\begin{eqnarray}\label{FRWmetpropdist}
  ds^2 = -c^2dt^2\left(1 - \frac{H^2{D}^2}{c^2}\right) - 2HDdtdD + dD^2\;.
\end{eqnarray}

\end{ex}

\hrulefill

Note that the quantity $c/H$ has appeared. This is $R_H$, the \textbf{Hubble radius}.

\section{Light-cone structure in the FLRW geometry}

Let us consider the propagation of a photon and build the light-cones for the FLRW metric.\index{FLRW metric!Light-cone structure}

\subsection{Cosmic time-comoving distance}

From the FLRW metric \eqref{FLRWmet}, the condition $ds^2 = 0$ gives us:
\begin{equation}\label{lightconecomdist}
	\frac{cdt}{d\chi} = \pm a(t)\;.
\end{equation}
We position ourselves, the observers, at $\chi = 0$ and $t = t_0$. The plus sign in the above equation describes an outgoing photon, i.e., the future light cone, whereas the negative sign describes an incoming photon, i.e., the past-light cone, which is much more interesting to us. So, let us keep the negative sign and discuss the shape of the light-cone.

At $t = t_0$, the slope of the past light-cone is $-a(t_0)$. We will see that we can normalize $a(t_0)$ to $-1$. This means that at $t = t_0$, the past light-cone is identical to the one in Minkowski space. Let us assume that $a(0) = 0$: a cosmological model with the Big-Bang. Then, at the crossing of the $\chi$-axis, the light-cone becomes flat, encompassing more radii than it would in Minkowski space. We can show this analytically by taking the second derivative of Eq.~\eqref{lightconecomdist} with the minus sign:
\begin{equation}
	\frac{c^2d^2t}{d\chi^2} = -\dot a\frac{cdt}{d\chi} = a\dot a\;.
\end{equation}
Being $a > 0$ and $\dot a > 0$ (we consider only the case of an expanding universe), the function $t(\chi)$ is convex (i.e., it ``bends upwards'').

Let us make a concrete example. Consider $a(t) = (t/t_0)^{2/3}$. This, as we shall see, is the evolution of the scale factor with cosmic time for a spatially flat universe filled with pressureless matter.

\begin{figure}[ht]
\centering
	\includegraphics[width=0.75\columnwidth]{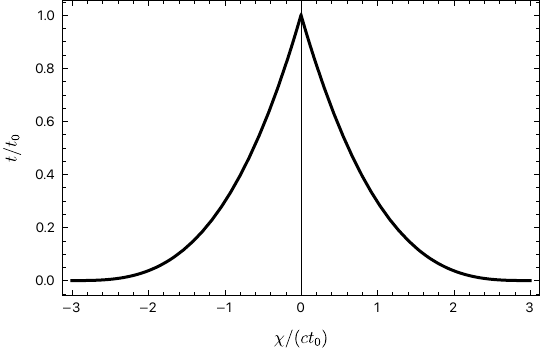}
	\caption{Light-cone for the case $a(t) = (t/t_0)^{2/3}$, using cosmic time and comoving distance.}
	\label{Fig:figlct}
\end{figure}

Then, Eq. \eqref{lightconecomdist}, with the minus sign, can be easily integrated to give:
\begin{equation}\label{lightconeat23examplechi}
	\frac{t}{t_0} = \left(-\frac{\chi}{3ct_0} + 1\right)^3\,. 
\end{equation}
The shape of this light-cone is given in Fig. \ref{Fig:figlct}. The comoving distance $\chi = 3ct_0$ is the maximum distance from which we can receive photons, and it is called \textbf{the comoving particle horizon}.\index{Particle horizon} As we shall see later, it is, in general, given by the formula:
\begin{equation}
	\chi_p = c\int_0^{t_0}\frac{dt}{a(t)} = c\eta_0\,,
\end{equation} 
where, in the last equality, we have used the definition of conformal time and $\eta_0 \equiv \eta(t_0)$.
 
\hrulefill

\subsection{Conformal time-comoving distance}

For the FLRW metric \eqref{FRWmetconf}, the condition $ds^2$ gives:
\begin{equation}
	\frac{cd\eta}{d\chi} = \pm 1\;.
\end{equation}
Of course, this equation can also be obtained from Eq. \eqref{lightconecomdist}, recalling that $dt = ad\eta$.

The light-cone structure is the same as that of Minkowski space. Indeed, the Friedmann metric, written in conformal time and comoving distance, is formally equivalent to the Minkowski metric multiplied by a conformal factor $a(\eta)$.

\subsection{Cosmic time-proper distance}

From Eq. \eqref{FRWmetpropdist} with $ds^2 = 0$, we need to solve the following equation:
\begin{equation}\label{lightconeproperdist}
	-\frac{c^2dt^2}{d{D}^2}\left(1 - \frac{H^2{D}^2}{c^2}\right) - \frac{2H{D}}{c}\frac{cdt}{d{D}} + 1 = 0\;.
\end{equation}

\hrulefill

\begin{ex} Solve Eq.~\eqref{lightconeproperdist} algebraically for $cdt/d{D}$ and show that:
\begin{equation}\label{lightconeproperdistsol}
	\frac{cdt}{d{D}} = \left(\frac{H{D}}{c} \pm 1\right)^{-1}\;.
\end{equation}
\end{ex}

\hrulefill

Alternatively, we can start from Eq. \eqref{lightconecomdist} and use $\chi = D/a$, so that:
\begin{equation}
	d\chi = \frac{d D}{a} - \frac{ D}{a^2}\dot a dt = \frac{d D}{a} - \frac{ D}{a}H dt\,.
\end{equation}
Therefore:
\begin{equation}
	\frac{d D}{a} - \frac{ D}{a}H dt = \pm \frac{1}{a}cdt\,,
\end{equation}
from which:
\begin{equation}
	d D = \left(\frac{H 
	D}{c} \pm 1\right)cdt\,.
\end{equation}
For $t = t_0$, we have $H(t_0) > 0$ and ${D} = 0$. Therefore, from Eq.~\eqref{lightconeproperdistsol}, we have that $(cdt/d{D})(t_0) = \pm 1$, and thus we must choose the minus sign to describe the past light-cone. 

Sufficiently close to $D = 0$, $HD/c$ is less than one, and thus the slope $cdt/dD$ remains negative, as in the previous instances of light-cones. Therefore, as we increase $D$, we go backward in time, and $H(t)$ grows (we again assume a model with a cosmological singularity). So, it happens for a certain time $t_H$ that the proper distance is equal to the Hubble radius $R_H(t_H)$:
\begin{equation}\label{Hubbleradiusgeneraltime}
	\frac{H(t_H)D_H}{c} = 1\;,
\end{equation}
and that $cdt/d{D}$ diverges at that event. This means that no signal can reach us (at $t = t_0$ and $D = 0$) from beyond $c/H(t_H)$. 

Going further back in time, $HD/c$ grows larger than unity, and so the slope of the light cone changes sign. Since $H \to \infty$ for $a \to 0$, then $cdt/d{D} \to 0$. Indeed, the light-cone flattens and closes at $t = 0$ (the Big-Bang singularity).

\begin{figure}[ht]
\centering
	\includegraphics[width=0.75\columnwidth]{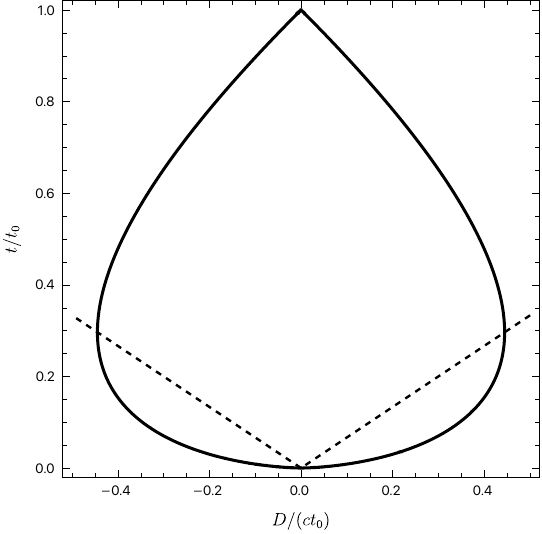}
	\caption{Light-cone for the case $a(t) = (t/t_0)^{2/3}$, using cosmic time and proper distance. The dashed lines represent the Hubble radius $c/H = 3ct/2$. The change of the slope takes place when $D = c/H$.}
	\label{Fig:figlctD}
\end{figure}

Coming back to our example $a(t) = (t/t_0)^{2/3}$, using the solution \eqref{lightconeat23examplechi} with $D$ as the independent variable, we can write:
\begin{equation}
	D = 3(ct_0)^{1/3}(ct)^{2/3} - 3ct\,.
\end{equation}
Through an implicit plot, the light-cone is shown in Fig. \ref{Fig:figlctD}.

The maximum proper distance is $D_H = (2/3)^2ct_0$ at $t_H = (2/3)^3t_0$. At this time, $a(t_H) = (2/3)^{2}$ and $H(t_H) = (3/2)^2/t_0$. Thence, $H(t_H)D_H/c = 1$, as expected.

Such a light cone is quite strange because it seems that the photon changes direction at a certain distance. But this is just an effect due to the reference frame that we used. Indeed, close to the Big-Bang, the proper distance traveled by a photon grows much faster than the Hubble radius (compare $\dot D = 2c(t_0/t)^{2/3} - 3c$ with $\dot R_H = 3c/2$). So, new proper distances can be probed. However, $D$ loses ground, and at a certain point, the photon is ``left behind'' by the expansion of space, appearing as if it has moved back to the origin (the upper part of the light-cone). 

In our example $a(t) = (t/t_0)^{2/3}$, the comoving distance to $t = 0$ is, in fact, $3ct_0$, but here we also have $a = 0$; henceforth $D = 0$. On the other hand, at $t = t_0$, the comoving distance is zero, and so is $D$. The same reasoning explains, in general, why, at any given $D$ that is less than its maximum, there are two different cosmic times.

Note also that the maximum physical distance from which we can receive a signal is not $c/H_0$, but rather less than that. In our example, $c/H_0 = 3ct_0/2$, whereas $D_H = 4ct_0/9$.

\section{Connection and geodesics in the FLRW geometry}

Let us perform our first step towards the construction of the Einstein equations for the FLRW metric: let us compute the Christoffel symbols.\footnote{The Christoffel symbols are the coefficients, with respect to a given basis, of the Levi-Civita connection.} Using the definition given in Eq.~\eqref{Christoffelsymbols}, it is time for a few exercises.

\hrulefill

\begin{ex}
	Using the cosmic time, show that:
	\begin{equation}\label{FLRWmetricGammatgen}
	\Gamma^0_{00} = 0\;, \quad \Gamma_{0i}^0 = 0\;, \quad \Gamma_{ij}^0 = \frac{1}{c}a\dot a\gamma_{ij}\;, \quad \Gamma^i_{00} = 0\;, \quad \Gamma^i_{0j} = \frac{1}{c}\frac{\dot a}{a}\delta^i{}_j\;, \quad \Gamma^i_{jk} = {}^{(3)}\Gamma^i_{jk}\;,
\end{equation}
where ${}^{(3)}\Gamma^i_{jk}$ are the Christoffel symbols of $\gamma_{ij}$.
\end{ex}

\hrulefill

\begin{ex}
	Using the conformal time, show that:
	\begin{equation}\label{FLRWmetricGammatgenconftime}
	\Gamma^0_{00} = \frac{a'}{a}\;, \quad \Gamma_{0i}^0 = 0\;, \quad \Gamma_{ij}^0 = \frac{1}{c}\frac{a'}{a}\gamma_{ij}\;, \quad \Gamma^i_{00} = 0\;, \quad \Gamma^i_{0j} = \frac{1}{c}\frac{a'}{a}\delta^i{}_j\;, \quad \Gamma^i_{jk} = {}^{(3)}\Gamma^i_{jk}\;.
\end{equation}
One can make the calculation directly from the FLRW metric written in the conformal time or use the results already found in Eq.~\eqref{FLRWmetricGammatgen} for the FLRW metric written in the cosmic time and use the transformation relation for the Christoffel symbol:
\begin{equation}
	\bar{\Gamma}'^\mu_{\nu\rho} = \bar{\Gamma}^\alpha_{\beta\gamma}\frac{\partial x^\beta}{\partial x'^{\nu}}\frac{\partial x^\gamma}{\partial x'^{\rho}}\frac{\partial x'^\mu}{\partial x^{\alpha}} + \frac{\partial x'^\mu}{\partial x^{\sigma}}\frac{\partial^2 x^\sigma}{\partial x'^{\nu}\partial x'^\rho}\;, 
\end{equation}
where the primed coordinates are in the conformal time and hence:
\begin{equation}
	\frac{\partial x'^0}{\partial x^0} = \frac{1}{a}\;, \qquad \frac{\partial x'^0}{\partial x^i} = 0\;, \qquad \frac{\partial x'^i}{\partial x^0} = 0\;, \qquad \frac{\partial x'^l}{\partial x^m} = \delta^l{}_m\;.
\end{equation}
It is a good and reassuring exercise to do in both ways and check that the result is the same.
\end{ex}

\hrulefill

We wonder now how a test particle moves in the expanding universe. We might also ask ourselves: what can be considered a test particle in the expanding universe? Given the vastness of the latter, probably any sufficiently isolated gravitationally bound system could be considered a test particle. So, we also have galaxies or clusters of galaxies.

From General Relativity, we know that a test particle moves along a \textbf{geodesic}. That is, its trajectory $x^\mu(\lambda)$, where $\lambda$ is an affine parameter, satisfies the \textbf{geodesic equation}:\index{Geodesic equation}
\begin{equation}
	\boxed{\frac{dP^\mu}{d\lambda} + \Gamma^\mu_{\nu\rho}P^\nu P^\rho = 0}
\end{equation}
Here, $P^\mu \equiv dx^\mu/d\lambda$ is the four-momentum. For a particle of mass $m$, one can choose $\lambda = \tau/m$, where $\tau$ is the proper time. 

For an observer moving along a world-line of 4-velocity $u^\mu$, the energy of the photon is:
\begin{equation}
	E = -g_{\mu\nu}u^\mu P^\nu\,.
\end{equation}
In particular, in a reference frame in which the observer is at rest, we have $u^0 = c$, and thus:
\begin{equation}
	E = -g_{0\nu}P^\nu\,,
\end{equation}
which, in the FLRW metric written in cosmic time and comoving coordinates, becomes:
\begin{equation}
	E = cP^0\,.
\end{equation}
It is useful for future applications to write the norm of the four-momentum as follows, again for our FLRW metric expressed in cosmic time and comoving coordinates:
\begin{equation}\label{fourmomentumnorm}
	g_{\mu\nu}P^\mu P^\nu = -(P^0)^2 + p^2 = -m^2c^2\;. 
\end{equation}
Here we have defined the \textbf{physical momentum} modulus (or \textbf{proper momentum}) as follows:\index{Proper momentum}
\begin{equation}\label{propermomentum}
	 \boxed{p^2 \equiv g_{ij}P^iP^j = a^2\gamma_{ij}P^iP^j}
\end{equation}
The last equality of Eq.~\eqref{fourmomentumnorm}, which applies only to massive particles, comes from:
\begin{equation}
	\frac{ds^2}{d\lambda^2} = \frac{m^2ds^2}{d\tau^2} = -m^2c^2\;,
\end{equation}
since, by definition, $ds^2 = -c^2d\tau^2$. As expected from the equivalence principle, we recover the usual mass-shell relationship of special relativity.

\subsection{Photons}

Now consider a photon: $m = 0$ and so $P^0 = p$. The time-component of the geodesic equation is as follows:
\begin{equation}
	\frac{dP^0}{d\lambda} + \frac{a\dot a}{c}\gamma_{ij}P^i P^j = 0 = \frac{dP^0}{d\lambda} + \frac{H}{c}p^2\;.
\end{equation}
Since $P^0 = p$:
\begin{equation}\label{photongeod}
	c\frac{dp}{d\lambda} + Hp^2 = 0\;.
\end{equation}

\hrulefill

\begin{ex} Solve Eq.~\eqref{photongeod} and show that $p = E/c \propto 1/a$, i.e. the energy of the photon is proportional to the inverse scale factor. Note that this conclusion is independent of the value of the spatial curvature $K$.
\end{ex}

\hrulefill

Since $p \propto 1/a$, let us define $q = ap$. In this way:
\begin{equation}
    \frac{dq}{d\lambda} = 0\,,
\end{equation}
and, using $p^2 = g_{ij}P^iP^j = a^2\gamma_{ij}P^iP^j$, we can establish that:
\begin{equation}
    P^i = \frac{q}{a^2}\hat P^i\,, \qquad \gamma_{ij}\hat P^i\hat P^j = 1\,.
\end{equation}
So, the contravariant momentum $P^i \propto 1/a^2$. On the other hand:
\begin{equation}
    P_i = g_{ij}P^j = a^2\gamma_{ij}\frac{q}{a^2}\hat P^j = q\gamma_{ij}\hat{P}^j = q\hat{P}_i\,.
\end{equation}
So, the covariant momentum $P_i$ is independent of time. These relations will become useful in the following chapters.

\hrulefill

\begin{ex} Show that the $i$-component of the geodesic equation for a photon can be written as:
\begin{equation}
	\frac{d\hat P^i}{d\lambda} + a^2q\Gamma^i_{jk}\hat P^j\hat P^k = 0\;.
\end{equation}

\end{ex}

\hrulefill

Since $E \propto 1/a$, we can then write:
\begin{equation}
	\frac{E_{\rm obs}}{E_{\rm em}} = \frac{a_{\rm em}}{a_{\rm obs}}\;.
\end{equation}
On the other hand, the photon energy is $E = hf$, with $f$ as its frequency. Therefore:
\begin{equation}
	\frac{a_{\rm em}}{a_{\rm obs}} = \frac{E_{\rm obs}}{E_{\rm em}} = \frac{f_{\rm obs}}{f_{\rm em}} = \frac{\lambda_{\rm em}}{\lambda_{\rm obs}} = \frac{1}{1 + z}\;.
\end{equation}
This is the relationship between the redshift\index{Redshift} and the scale factor.\index{Scale factor} We have connected observation (the redshift) with theory (the scale factor). Usually, the time of observation is called $t_0$, the present time, and the corresponding scale factor is denoted as $a_{\rm obs} = a(t_0) = a_0$. Calling the scale factor at the time of emission simply $a$, the above relationship is written as: 
\begin{equation}\label{redshiftscaleactorrel}
	\boxed{1 + z = \frac{a_0}{a}}
\end{equation}
Knowing the evolution over time of $a(t)$, we also know from the above formula $z(t)$. This means that by measuring a certain value of the redshift from a specific source, we would be able to determine the cosmic time $t$ when that signal was emitted (this is also called lookback time).

Note that we have treated the photon as a test particle and found that $E \propto 1/a$. We can also think of treating it as a wave; then, since all lengths are stretched by $a$, we would infer that $\lambda \propto a$. Combining the two treatments, we would find $E \propto 1/\lambda$, i.e., the Planck relation. Quite strangely, a fundamental quantum-mechanical relation pops out.\footnote{This fact was commented on to the author by Luciano Casarini.}

\subsection{Massive particles}

The time-geodesic equation for massive particles is as follows: 
\begin{equation}
	\frac{dP^0}{d\lambda} + \frac{H}{c}p^2 = 0\;,
\end{equation}
but now $P^0$ is not equal to $p$.

\hrulefill

\begin{ex} Using $(P^0)^2 + p^2 = m^2c^2$, solve the above equation for $p$. 

\end{ex}

\hrulefill

The result is again $p \propto 1/a$, identical to that for photons. However, the physical implications are dramatically different. In fact, a photon always moves with speed $c$, irrespective of the fact that $p \propto 1/a$. For a massive particle, instead, $p \propto 1/a$ means that, eventually, it will be stopped by the expansion of the universe (as $a \to \infty$).

\section{The Friedmann equations}\label{Sec:Friedeq}

The Friedmann equations are the Einstein equations:
\begin{equation}\label{Einseq}
 G_{\mu\nu} + \Lambda g_{\mu\nu} = R_{\mu\nu} - \frac{1}{2}g_{\mu\nu}R + \Lambda g_{\mu\nu} = \frac{8\pi G}{c^4} T_{\mu\nu}\;,
\end{equation}
for the case of the FLRW metric. Here, $\Lambda$ is the cosmological constant. 

\hrulefill

\begin{ex} Calculate the components of the Riemann tensor for the FLRW metric, using the cosmic time. Show that:\index{FLRW metric!Riemann tensor}
\begin{align}
    R_{0i0j} &= -\frac{a\ddot a}{c^2}\gamma_{ij} = -\frac{1}{c^2}\frac{\ddot a}{a}g_{ij}\,, \\ 
    R_{ikjl} &= \frac{a^2(Kc^2 + \dot a^2)}{c^2}(\gamma_{ij}\gamma_{kl} - \gamma_{il}\gamma_{kj}) = \left(\frac{K}{a^2} + \frac{1}{c^2}H^2\right)(g_{ij}g_{kl} - g_{il}g_{kj})\,.
\end{align}
Now show that the Ricci tensor has components:\index{FLRW metric!Ricci tensor}
\begin{align}
	R_{00} &= -\frac{3}{c^2}\frac{\ddot a}{a}\;, \qquad R_{0i} = 0\;,\\
	R_{ij} &= \frac{1}{c^2}a^2\gamma_{ij}\left(2H^2 + \frac{\ddot a}{a} + 2\frac{Kc^2}{a^2}\right) = \frac{1}{c^2}a^2\gamma_{ij}\left(2H^2 + \frac{\ddot a}{a}\right) + {}^{(3)}R_{ij}\;.
\end{align}
Show that the scalar curvature is:
\begin{equation}\label{Ricciscalcosmo}
	R = \frac{6}{c^2}\left(\frac{\ddot a}{a} + H^2 + \frac{Kc^2}{a^2}\right) = \frac{6}{c^2}\left(\frac{\ddot a}{a} + H^2\right) + {}^{(3)}R\;.
\end{equation}
Finally, put together these results in the Einstein equations:\index{Friedmann equations}
\begin{eqnarray}
\label{Friedeq1}	\boxed{H^2 + \frac{Kc^2}{a^2} = \frac{8\pi G}{3c^2} T_{00} + \frac{\Lambda c^2}{3}}\\ 
\label{acceq1} \boxed{g_{ij}\left(H^2 + 2\frac{\ddot a}{a} + \frac{Kc^2}{a^2} - \Lambda c^2\right) = -\frac{8\pi G}{c^2}T_{ij}}
\end{eqnarray} 
These are called \textbf{Friedmann equations} or \textbf{Friedmann equation} and \textbf{acceleration equation} or \textbf{Friedmann equation} and \textbf{Raychaudhuri equation}.\index{Acceleration equation}
\end{ex}

\hrulefill

The Weyl tensor is defined as:
\begin{equation}
    C_{\mu\nu\rho\sigma} = R_{\mu\nu\rho\sigma} + \frac{1}{2}(R_{\mu\sigma}g_{\nu\rho} - R_{\mu\rho}g_{\nu\sigma} + R_{\nu\rho}g_{\mu\sigma} - R_{\nu\sigma}g_{\mu\rho}) + \frac{R}{6}(g_{\mu\rho}g_{\nu\sigma} - g_{\mu\sigma}g_{\nu\rho})\,.
\end{equation}

\hrulefill

\begin{ex}
Show that the Weyl tensor vanishes identically for the FLRW metric. For this reason, the FLRW metric is \textbf{conformally flat}.
\end{ex}

\hrulefill

\subsection{The stress-energy tensor}

Which stress-energy tensor $T_{\mu\nu}$ do we use in Eqs.~\eqref{Friedeq1} and \eqref{acceq1}? The cosmological principle requires the use of the FLRW metric. In the same way, we also have some strong constraints for the choice of $T_{\mu\nu}$:
\begin{itemize}
	\item First of all, $G_{0i} = 0$ implies that $T_{0i} = 0$; i.e., there cannot be a flux of energy in any direction because it would violate isotropy;
	\item Second, since $G_{ij} \propto g_{ij}$, then $T_{ij} \propto g_{ij}$.
	\item Finally, since $G_{\mu\nu}$ depends only on $t$, it must also be the case for $T_{\mu\nu}$. 
\end{itemize}
Therefore, let us stipulate that:
\begin{equation}\label{perfectfluidtcomps}
 T_{00} = \rho(t)c^2 = \varepsilon(t)\;, \qquad T_{0i} = 0\;, \qquad T_{ij} = g_{ij}P(t)\;.
\end{equation}
It turns out that $\rho(t)$ is the rest mass density, $\varepsilon(t)$ is the energy density, and $P(t)$ is the pressure. For the components of $T^{\mu}{}_\nu$, we have:
\begin{equation}\label{perfectfluidtcomps2}
 T^0{}_{0} = -\rho(t)c^2 = -\varepsilon(t)\;, \qquad T^0{}_{i} = 0\;, \qquad T^i{}_{j} = \delta^i{}_{j}P(t)\;.
\end{equation}
Matter described by a stress-energy tensor for which there exists a reference frame in which $T^\mu{}_\nu$ is diagonal and has all equal spatial entries is called a \textbf{perfect fluid}.\index{Perfect fluid}

Introducing the 4-velocity of the fluid element, which must have coordinates $u_\mu = (-c,0,0,0)$ in our cosmic time-comoving coordinates, we can write the perfect fluid stress-energy tensor as follows:
\begin{equation}\label{perfectfluidtensor}
 \boxed{T_{\mu\nu} = \left(\rho + \frac{P}{c^2}\right)u_{\mu}u_\nu + Pg_{\mu\nu}}
\end{equation}
A word is in order about the $c^2$ appearing above. The stress-energy tensor has the dimension of energy density, as well as $\rho c^2$ and $P$. The four-velocity has dimensions of velocity; in particular, it is normalized as $u_\mu u^\mu = -c^2$, whereas $g_{\mu\nu}$ is dimensionless. For this reason, a $P/c^2$ appears above.\footnote{Sometimes, one finds $-Pg_{\mu\nu}$ instead of $Pg_{\mu\nu}$ as the last term of the perfect fluid energy-momentum tensor. This depends on the choice of the signature of the metric.}

In this form of Eq.~\eqref{perfectfluidtensor}, the stress-energy tensor does not contain either viscosity or energy transport terms.\index{Perfect fluid} For more details about perfect fluids, see \cite{Schutz:1985jx}. For more details about viscosity, heat fluxes, and imperfect fluids, see, e.g., \cite{Weinberg:1972} and \cite{Maartens:1996vi}.

If we introduce the projector onto the spatial hyper-surfaces orthogonal to the fluid worldline, i.e.:
\begin{equation}
	h_{\mu\nu} = g_{\mu\nu} + u_\mu u_\nu/c^2\,,
\end{equation}
one can write:
\begin{equation}\label{perfectfluidtensor2}
 \boxed{T_{\mu\nu} = \rho u_{\mu}u_\nu + Ph_{\mu\nu}}
\end{equation}
Combine Eqs.~\eqref{Friedeq1}, \eqref{acceq1}, and \eqref{perfectfluidtcomps}. The Friedmann equation becomes:
\begin{equation}\label{FriedEq2}
 \boxed{H^2 = \frac{8\pi G}{3} \rho + \frac{\Lambda c^2}{3} - \frac{Kc^2}{a^2}}
\end{equation}
The acceleration equation is as follows:
\begin{equation}\label{accEq}
 \boxed{\frac{\ddot{a}}{a} = -\frac{4\pi G}{3}\left(\rho + \frac{3P}{c^2}\right) + \frac{\Lambda c^2}{3}}
\end{equation}
The scale factor $a$ is, by definition, positive; however, its derivative can be negative. This would describe a contracting universe. Note that the left hand side of the Friedmann equation \eqref{FriedEq2} is non-negative. Therefore, $\dot{a}$ can vanish only if $K > 0$, i.e., for a spatially closed universe. This implies that if $K \leqslant 0$ and if there exists an instant for which $\dot{a} > 0$, then the universe will expand forever.

Moreover, note from \eqref{accEq} that standard matter, with $\rho$ and $P$ positive, contributes to a decelerated expansion of the universe ($\ddot a < 0$). On the other hand, the cosmological constant $\Lambda > 0$ naturally provides a source of acceleration and, thus, it is a rightful candidate for dark energy.

\hrulefill

\begin{ex} Use now the conformal time introduced in Eq.~\eqref{conftimedefinition}. Show that the Friedmann equation becomes:
\begin{equation}\label{FriedEq2conft}
 \boxed{\mathcal H^2 = \frac{8\pi G}{3} \rho a^2 + \frac{\Lambda c^2 a^2}{3} - Kc^2}
\end{equation}
and that the acceleration equation becomes:
\begin{equation}\label{accEqconft}
 \boxed{\frac{a''}{a} = \frac{4\pi G}{3}\left(\rho - \frac{3P}{c^2}\right)a^2 + \frac{2\Lambda c^2a^2}{3} - Kc^2}
\end{equation}
where the prime denotes derivation with respect to the conformal time $\eta$ and 
\begin{equation}
	\boxed{\mathcal H \equiv \frac{a'}{a}}
\end{equation}
is the \textbf{conformal Hubble factor}.\index{Conformal Hubble factor} 
\end{ex}

\hrulefill

The evolution of the scale factor is given by the acceleration equation; this is a second-order ordinary differential equation, and it is typically non-linear (it is linear only if $P = \rho c^2/3$ and $\Lambda = 0$). Then, a solution is singled out if we fix two initial conditions: one on $a$ itself and one on $\dot a$. However, these two initial conditions are not independent; rather, they are constrained by the Friedmann equation. So, we actually have the freedom to choose only one initial condition. As we will see in Sec. \ref{Sec:solFriedeq}, one typically chooses $a(t = 0) = 0$ for those cosmological models that feature a Big-Bang. 

When this initial condition is set, the present time $t_0$, which is the time at which we observe the universe, is also the \textbf{age of the universe}. To the time $t_0$ corresponds the scale factor $a(t_0) = a_0$, which we have already encountered in Eq. \eqref{redshiftscaleactorrel}. At $t = t_0$, the redshift is zero, meaning that the emission time is precisely $t_0$, and so the source is at a distance of zero from us.\footnote{Clearly, zero is a fictitious value for the redshift, an extrapolation. The lowest values for $z$ are of the order of $10^{-3}$, since below these, peculiar velocities start to dominate, and the Hubble flow is negligible.} 

It is customary to normalize the scale factor $a(t)$ to $a_0$, independently of the choice of an initial condition. This amounts to fixing $a_0 = 1$. This can be done because $a$ enters as $\ddot a/a$ in the acceleration equation \eqref{accEq} and as $\dot a/a$ in the Friedmann equation \eqref{FriedEq2}. However, in the latter, a new, rescaled spatial curvature parameter, $K \to K/a_0^2$, must be defined.\index{Age of the universe} One can see the possibility of this rescaling directly in the metric \eqref{FLRWmet}.

How is $t_0$ calculated? It is established observationally from $H_0$. As we will see, for any given cosmological model, we can express its Hubble parameter as a function of $a$ (or the redshift) as:
\begin{equation}
    H(a) = H_0E(a)\,,
\end{equation}
where $E(a)$ is such that $E(a_0) = 1$. Then, one computes:
\begin{equation}\label{ageofu}
	t_0 = \int_{0}^{t_0}dt = \int_{0}^{1}\frac{da}{H(a)a} = \frac{1}{H_0}\int_{0}^{1}\frac{da}{E(a)a} = \frac{1}{H_0}\int_{0}^{\infty}\frac{dz}{E(z)(1 + z)}\;,
\end{equation}
where $a(t = 0) = 0$ and $a_0 = 1$ have been used, and relation \eqref{redshiftscaleactorrel} has been employed for the last equality. 

\hrulefill

\begin{ex} Prove the last equality of Eq.~\eqref{ageofu}. \end{ex}

\hrulefill

Hence, $t_0 \propto 1/H_0$, where the proportionality factor is determined by the cosmological model. Then, the value of $t_0$ is inferred by measuring the Hubble constant.

For some cosmological models, such as the de Sitter universe, for which $a$ vanishes only when $t \to -\infty$, the age of the universe is infinite.

In cosmology, when a quantity has a subscript $0$, it usually means that it is evaluated at $t = t_0$.

\section{The continuity equation}

The conservation equation 
\begin{equation}\label{nablaTmunuzero}
	\boxed{\nabla_\nu T^{\mu\nu}{} = 0}
\end{equation}
is encapsulated in GR through the invariance under diffeomorphisms, and it is compatible with the \textbf{Bianchi identities} $\nabla_\nu G^{\mu\nu}{} = 0$. Therefore, it is not independent of the Friedmann equations \eqref{FriedEq2} and \eqref{accEq}. For the FLRW metric and a perfect fluid, it has a particularly simple form:
\begin{equation}\label{Encons}
 \boxed{\dot{\rho} + 3H\left(\rho + \frac{P}{c^2}\right) = 0}
\end{equation}
This is the $\mu = 0$ component of $\nabla_\nu T^{\mu\nu}{} = 0$, and it is also known in fluid dynamics as \textbf{the continuity equation}.\index{Continuity equation}

\hrulefill

\begin{ex} Derive the continuity equation \eqref{Encons} by combining Friedmann and acceleration equations \eqref{FriedEq2} and \eqref{accEq}. Derive it in a second way by explicitly calculating the four-divergence of the energy-momentum tensor. Show that the $\mu = i$ components of the conservation equation are identically vanishing.
\end{ex}

\hrulefill

\noindent So, the two Friedmann equations and the continuity equations are not independent. We have two independent differential equations for three unknown functions: the scale factor $a(t)$, the density $\rho(t)$, and the pressure $P(t)$. In order to find an explicit solution, we then need a third equation. Typically, this is an \textbf{equation of state} relating the pressure to the density:\index{Equation of state}
\begin{align}
    P = P(\rho)\,.
\end{align}
This is also known as a \textbf{barotropic equation of state}.

Typically, the continuity equation is handled by trading $t$ for $a$. Indeed, in so doing, Eq. \eqref{Encons} becomes:
\begin{equation}\label{Enconsa}
 \frac{d\rho}{da} + \frac{3}{a}\left(\rho + \frac{P}{c^2}\right) = 0\,.
\end{equation}
If we assume an equation of state of the form $P = w\rho c^2$, with $w$ constant, the general solution is:
\begin{equation}\label{rhowgen}
 \boxed{\rho = \rho_0a^{-3(1 + w)}} \qquad (w = \mbox{ constant})\;,
\end{equation}
where $\rho_0 \equiv \rho(a_0 = 1)$.

\hrulefill

\begin{ex} Prove the above result of Eq.~\eqref{rhowgen}. \end{ex}

\hrulefill

\noindent There are three particular values of $w$ that play a major role in cosmology:

\paragraph{Cold matter:} $w = 0$, i.e. $P = 0$, for which $\rho = \rho_0a^{-3}$. The adjective cold refers to the fact that the particles making up this kind of matter have a kinetic energy much smaller than the mass energy; i.e., they are non-relativistic. If they are thermally produced, i.e., if they were in thermal equilibrium with the primordial plasma, they have a mass much larger than the temperature of the thermal bath. We shall see this characteristic in more detail in Chapter~\ref{Chap:KinTh}. 

Cold matter is also called \textbf{dust}, and it encompasses all the non-relativistic known elementary particles, which are generically dubbed \textbf{baryons} in the jargon of cosmology.\footnote{These are, essentially, protons and neutrons (the latter bound in nuclei). Electrons also count, of course, but they are much lighter and thus contribute much less to the energy density.} If they exist, yet undetected non-relativistic particles might contribute to \textbf{cold dark matter} (CDM).\index{Dust}\index{Cold Dark Matter}\index{Baryons}

\paragraph{Hot matter:} $w = 1/3$, i.e., $P = \rho c^2/3$, for which $\rho = \rho_0a^{-4}$. The adjective hot refers to the fact that the particles making up this kind of matter are relativistic. For this reason, they are known, in the jargon of cosmology, as \textbf{radiation}\index{Radiation}, and they encompass not only the relativistically known elementary particles but also possibly unknown ones (i.e., \textbf{hot dark matter}). Primordial neutrinos would belong to this class if they were massless. However, neutrinos do have a mass of approximately 0.1 eV, so they must now be considered cold matter. We shall see this in more detail in Chapter~\ref{Chap:KinTh}.\index{Hot Dark Matter}

\paragraph{Vacuum energy:} $w = -1$, i.e., $P = -\rho c^2$, and $\rho$ are constants. 

\hrulefill

\begin{ex}
    From the Friedmann equations \eqref{FriedEq2} and \eqref{accEq} deduce that the cosmological constant can be considered as a matter-energy, non-interacting component with the following density and pressure:
\begin{equation}
	\rho_\Lambda \equiv \frac{\Lambda c^2}{8\pi G}\;, \qquad P_\Lambda \equiv -\rho_\Lambda c^2\;.
\end{equation}
\end{ex}

\hrulefill 

Cold matter, radiation, and the cosmological constant make up the standard model of cosmology, the $\Lambda$CDM, which we discuss in more detail in Sec. \ref{Sec:LCDMmodel}.

\subsection{Combining different forms of matter}

In the Friedmann and acceleration equations, $\rho$ and $P$ are to be considered as the total density and pressure. They can be written as sums of the contributions of individual components:
\begin{equation}
	\rho \equiv \sum_x\rho_x\;, \qquad P \equiv \sum_{x}P_x\;.
\end{equation}
This is useful if we know how $\rho_x$ and $P_x$ evolve with time or with the scale factor. More generally, one can write the total stress-energy tensor as a sum:
\begin{equation}
	T_{\mu\nu} = \sum_x T^{(x)}_{\mu\nu}\;,
\end{equation}
over matter components $x$. As we saw, the total stress-energy tensor satisfies the conservation equation \eqref{nablaTmunuzero}, but this does not imply that each individual $T^{(x)}_{\mu\nu}$ has a vanishing divergence. In fact, different matter components might interact. For example, we might have two components, 1 and 2, for which the total energy-momentum tensor is conserved:
\begin{align}
    \nabla^\mu\left(T^{(1)}_{\mu\nu} + T^{(2)}_{\mu\nu}\right) = 0\,,
\end{align}
but, individually, do not:
\begin{align}
    \nabla^\mu T^{(1)}_{\mu\nu} = Q\,, \qquad \nabla^\mu T^{(2)}_{\mu\nu} = -Q\,,
\end{align}
where $Q$ is a function of time (and only of time, as long as the cosmological principle is enforced) that encapsulates the physics of the interaction.

\section{The Hubble constant, the deceleration parameter and the density parameters}

When the Hubble parameter $H$ is evaluated at the present time $t_0$ (the age of the universe), it becomes a number: the Hubble constant $H_0$\index{Hubble constant}, which we have already encountered in Chapter~\ref{Chap:Cosmology} in the Hubble-Lemaître law \eqref{Hubblelaw}. Usually $H_0$ is conveniently written as 
\begin{equation}\label{heq}
 H_0 = 100\;h\;{\rm km}\;{\rm s}^{-1}\;{\rm Mpc}^{-1}\;.
\end{equation}
The unit of measure of the Hubble constant\index{Hubble constant} is an inverse time:
\begin{equation}\label{H0valuetime}
 H_0 = 3.24\;h\times 10^{-18}\;{\rm s}^{-1}\;,
\end{equation}
whose inverse gives the order of magnitude of the age of the universe:\index{Age of the universe}
\begin{equation}\label{H0valuetime2}
 \frac{1}{H_0} = 3.09\;h^{-1}\times 10^{17}\;{\rm s} = 9.78\;h^{-1}\;{\rm Gyr}\;,
\end{equation}
and multiplied by $c$ gives the order of magnitude of the size of the visible universe, known as the \textbf{Hubble radius}:\index{Size of the visible universe}
\begin{equation}\label{Hubbleradius}
 \frac{c}{H_0} = 9.27\;h^{-1}\times 10^{25}\;{\rm m} = 3.00\;h^{-1}\;{\rm Gpc}\;.
\end{equation}
The Hubble radius is the distance at which, according to the Hubble-Lemaître law, the receding velocity equals the speed of light. 

On the other hand, time flows, so $t_0$ is not really a constant. That is true, but if we compare a time span of 100 years (the span of some human lives) to the age of the universe (about 14 billion years), we see that the ratio is about $10^{-8}$. Since this is pretty small, we can consider $t_0$ to be a constant.\footnote{Pretty much the same happens with the redshift. A certain source has a redshift $z$, which, actually, is not a constant but varies slowly because the source recedes faster and faster, according to the Hubble-Lemaître law. This is called \textbf{redshift drift}, and it was first considered by Sandage and McVittie in \cite{1962ApJ...136..319S} and \cite{1962ApJ...136..334M}. Applications of the redshift drift phenomenon to gravitational lensing are proposed in \cite{Piattella:2017uat}.\index{Redshift drift}} 

\subsection{The deceleration parameter} 

The \textbf{deceleration parameter} is defined as follows:\index{Deceleration parameter}
\begin{equation}
 \boxed{q \equiv -\frac{\ddot{a}a}{\dot{a}^2}}
\end{equation}
In \cite{Riess:1998cb} and \cite{Perlmutter:1998np}, analysis based on type Ia supernova observations has shown that $q_0 < 0$; i.e., the deceleration parameter today is negative; therefore, \textit{the universe is, at present, in a state of accelerated expansion}. We perform a similar but simplified analysis in Sec.~\ref{Sec:BayesiananalysisSNIa} in order to provide an example of Bayesian analysis.

\subsection{Critical density and density parameters}

Let us now rewrite Eq.~\eqref{FriedEq2} incorporating $\Lambda$ in the total density $\rho$:
\begin{equation}\label{FriedEqH}
 H^2 = \frac{8\pi G \rho}{3} - \frac{Kc^2}{a^2}\;.
\end{equation}
The value of the total $\rho$ such that $K = 0$ is called \textbf{critical energy density}\index{Critical density} and has the following form:
\begin{equation}
 \boxed{\rho_{\rm cr} \equiv \frac{3H^2}{8\pi G}}
\end{equation}
Its present value can be written as:
\begin{equation}
 \boxed{\rho_{\rm cr,0} = 1.878\;h^2\times 10^{-29}\;{\rm g}\;{\rm cm}^{-3}}
\end{equation}
and depends only on the value of $H_0$, which must be determined observationally.

Instead of densities, it is very common and useful to employ the density parameter $\Omega$, which is defined as 
\begin{eqnarray}\index{Density parameter}
	\boxed{\Omega \equiv \frac{\rho}{\rho_{\rm cr}} = \frac{8\pi G\rho}{3H^2}}
\end{eqnarray}
i.e., the energy density normalized to the critical density. We can then rewrite the Friedmann equation \eqref{FriedEq2} as follows:
\begin{equation}\label{FriedEqHOmega}
 1 = \Omega - \frac{Kc^2}{H^2a^2}\;.
\end{equation}
Defining 
\begin{equation}\label{OmegaK}
 \Omega_{K} \equiv -\frac{Kc^2}{H^2a^2}\;,
\end{equation}
i.e., associating a fictitious energy density 
\begin{equation}\label{rhoK}
 \rho_{K} \equiv -\frac{3Kc^2}{8\pi Ga^2}\;,
\end{equation}
to the spatial curvature, we can recast Eq.~\eqref{FriedEqHOmega} in the following simple form: 
\begin{equation}\label{FriedEqHOmegaOmegaK}
 1 = \Omega + \Omega_{K}\;.
\end{equation}
Therefore, the sum of all the density parameters, {\it the curvature one included}, is equal to unity at all times.\footnote{This is a manifestation of the fact that the Friedmann equation is a \textit{constraint}.} 

On the other hand, in the literature, it is more common to normalize $\rho$ to the \emph{present-time} critical density: 
\begin{equation}
	\boxed{\Omega \equiv \frac{\rho}{\rho_{\rm cr,0}} = \frac{8\pi G\rho}{3H_0^2}}
\end{equation}
In this way, the time or $a$-dependence of each component's $\Omega$ is the same as that component's density. If:
\begin{equation}
    \rho_x(a) = \rho_{x0}f_x(a)\,,
\end{equation}
where $f_x(a)$ is a function that gives the $a$-dependence of the material components $x$ and $f_{x}(a_0 = 1) = 1$, then:
\begin{equation}
	\Omega_{x0} = \frac{8\pi G\rho_{x0}}{3H_0^2}\,,
\end{equation}
and the Friedmann equation \eqref{FriedEq2} is written as:
\begin{equation}\label{FriedEqOmega2def}
	\frac{H^2}{H_0^2} = \sum_{x}\Omega_{x0}f_x(a) + \frac{\Omega_{K0}}{a^2} = E(a)^2\;.
\end{equation}
Consistently:
\begin{equation}\label{closurerelation}
	\boxed{\Omega_{0} + \Omega_{K0} = 1} \qquad \Omega_0 \equiv \sum_{x}\Omega_{x0}\,,
\end{equation}
also known as \textbf{the closure relation}.\index{Closure relation} We shall use the definition $\Omega_x \equiv \rho_x / \rho_{\rm cr,0}$ throughout these notes.

The standard strategy to solve the Friedmann equations is thus:
\begin{itemize}
    \item Establish a cosmological model: the matter components, along with their equations of state;
    \item Solve the continuity equation for each component and find $\rho_x(a)$;
    \item Solve the Friedmann equation or the acceleration equation.
    \item If solving the acceleration equation, one of the two integration constants is determined by the Friedmann equation.
\end{itemize}

\section{The \texorpdfstring{$\Lambda$CDM}{LambdaCDM} model}\label{Sec:LCDMmodel}

The most successful cosmological model is called $\Lambda$CDM and is composed of $\Lambda$, CDM, baryons, and radiation (photons and massless neutrinos). Therefore, the Friedmann equation for the $\Lambda$CDM model\index{$\Lambda$CDM model} is the following:
\begin{equation}\label{LCDMFriedeq}
	E(a)^2 = \frac{H^2}{H_0^2} = \Omega_\Lambda + \frac{\Omega_{\rm c0}}{a^3} + \frac{\Omega_{\rm b0}}{a^3} + \frac{\Omega_{\rm r0}}{a^4} + \frac{\Omega_{K0}}{a^2}\;.
\end{equation}
From the latest Planck data \cite{Planck:2018vyg}, we have:
\begin{equation}\label{omegacurv}
	\boxed{\Omega_{K0} = 0.0007 \pm 0.0019}
\end{equation}
at a 68\% confidence level for the joint analysis of lensed CMB temperature and polarization spectra and BAO. The universe is close to being spatially flat.\index{Spatial curvature density} The values for the other density parameters are (these values are derived for $H_0 = 67.66 \pm 0.42$ km s$^{-1}$ Mpc$^{-1}$):
\begin{equation}
	\boxed{\Omega_\Lambda = 0.6889 \pm 0.0056\;, \qquad \Omega_{\rm m0} = 0.3111\pm 0.0056}
\end{equation}
where $\Omega_{\rm m0} = \Omega_{\rm c0} + \Omega_{\rm b0}$, i.e., it includes the contributions from both CDM and baryons, since they have the same dynamics (i.e., they are both cold).\index{Cosmological constant!Density parameter} It is possible to disentangle them, and one obtains:\footnote{CMB physics is sensitive to the energy densities, and these are proportional to $H_0^2$. So, a coefficient $h^2$ appears.}
\begin{equation}
	\boxed{\Omega_{\rm b0}h^2 = 0.02242 \pm 0.00014\;, \qquad \Omega_{\rm c0}h^2 = 0.11933 \pm 0.00091}
\end{equation}
\index{Cold Dark Matter!Density parameter}\index{Baryons!Density parameter}The radiation content, i.e., photons and neutrinos, can be easily calculated from the temperature of the CMB, as we shall see in Chapter~\ref{Chap:ThermalHistory}. It turns out that:
\begin{equation}
	\boxed{\Omega_{\gamma0}h^2 \approx 2.47\times 10^{-5}\;, \qquad \Omega_{\nu0}h^2 \approx 1.68\times 10^{-5}}
\end{equation}
Recalling the closure relation of Eq.~\eqref{closurerelation}, we can conclude that today 69\% of our universe is composed of the cosmological constant, 26\% of cold dark matter, and 5\% of baryons. Radiation and spatial curvature are negligible.\index{Photons!Density parameter}\index{Neutrinos!Density parameter}

Due to their different scaling, there are special epochs in the history of the $\Lambda$CDM universe when the energy densities of different species become equal. For example, the radiation-matter equality epoch occurs at a scale factor $a_{\rm eq}$ such that:\index{Radiation-matter equality}
\begin{equation}\label{radmattequalityscalefactor}
    \rho_{\rm r} = \rho_{\rm m}\,, \quad \Rightarrow \quad \Omega_{\rm r0} = \Omega_{\rm m0}a_{\rm eq}\,, \quad \Rightarrow \quad a_{\rm eq} = \frac{\Omega_{\rm r0}}{\Omega_{\rm m0}} \approx 2.93\times 10^{-4}\,.
\end{equation}
Similarly, we can define a matter-cosmological constant epoch of equality at a scale factor $a_\Lambda$ as follows:\index{Matter-Cosmological constant equality}
\begin{equation}\label{mattLambdaequalityscalefactor}
    \rho_{\rm m} = \rho_{\Lambda}\,, \quad \Rightarrow \quad \Omega_{\rm m0} = \Omega_{\Lambda}a_{\Lambda}^3\,, \quad \Rightarrow \quad a_{\Lambda} = \left(\frac{\Omega_{\rm m0}}{\Omega_{\Lambda}}\right)^{1/3} \approx 0.77\,.
\end{equation}
This corresponds to a redshift $z_\Lambda \approx 0.3$.

\subsection{Age of the universe in the \texorpdfstring{$\Lambda$CDM}{LambdaCDM} model}

Let us now calculate the age of the universe for the $\Lambda$CDM model. Using Eq.~\eqref{ageofu}, we obtain:
\begin{equation}\label{ageofuLCDM}
	t_0 = \frac{1}{H_0}\int_0^1da\frac{a}{\sqrt{\Omega_\Lambda a^4 + \Omega_{\rm m0}a + \Omega_{\rm r0} + \Omega_{K0}a^2}}\;.
\end{equation}\index{$\Lambda$CDM model}
Using the numbers shown thus far with $h = 0.68$ and Eq. \eqref{H0valuetime2}, we obtain, upon numerical integration:
\begin{equation}\label{obsvalue}
	\boxed{t_0 = \frac{0.95}{H_0} = 13.73\;{\rm Gyr}}
\end{equation}
The value reported by \cite{Planck:2018vyg} is $13.787 \pm 0.020$ at a 68\% confidence level. 

\begin{figure}[ht]
\centering
	\includegraphics[width=0.5\columnwidth]{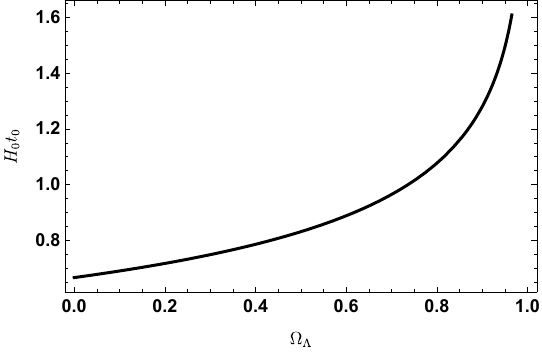}\includegraphics[width=0.5\columnwidth]{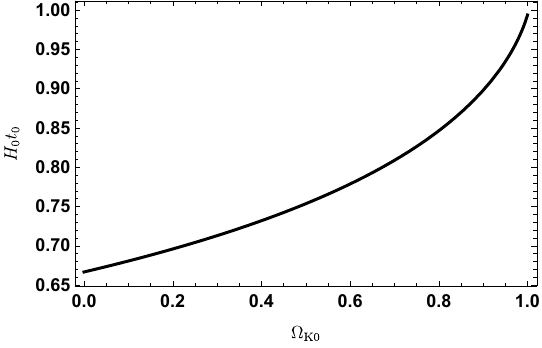}
	\caption{Dimensionless age of the universe $H_0t_0$ as function of $\Omega_\Lambda$, keeping fixed the curvature and radiation content. The value of $\Omega_{\rm m0}$ is then obtained by the closure relation as $\Omega_{\rm m0} = 1 - \Omega_{\Lambda} - \Omega_{\rm r0} - \Omega_{K0}$.}
	\label{Fig:figage}
\end{figure}

In Fig.~\ref{Fig:figage}, we plot the dimensionless age of the universe $H_0t_0$ for varying $\Omega_\Lambda$ and $\Omega_{\rm m0}$, while keeping the other density parameters constant (so the closure relation implies that $\Omega_{\rm m0} = 1 - \Omega_{\Lambda} - \Omega_{\rm r0} - \Omega_{K0}$). 

As one can see, in the presence of $\Lambda$, the dimensionless age of the universe reaches values larger than unity. This, mathematically, is due to the $a^4$ factor multiplying $\Omega_\Lambda$ in Eq.~\eqref{ageofuLCDM}. 

In Fig.~\ref{Fig:figage}, we also plot the dimensionless age of the universe $H_0t_0$ in a cosmological model with no $\Omega_\Lambda$ and for varying $\Omega_{K0}$ and $\Omega_{\rm m0}$, keeping the density parameter of radiation constant (so the closure relation implies that $\Omega_{\rm m0} = 1 - \Omega_{\rm r0} - \Omega_{K0}$). 

Also, in this case, we can obtain the observed value $H_0t_0 \approx 0.95$, despite the absence of a cosmological constant. The amount of curvature required is, however, enormous, i.e., $\Omega_{K0} \approx 0.97$. This implies $\Omega_{\rm m0} \approx 0.03$, which is incompatible with observational results of different kinds.

\subsection{The flatness problem}\label{subsec:flatnessproblem}

It turns out from observation that our universe has a very small spatial curvature, cf. Eq.~\eqref{omegacurv}. This requires an enormous fine-tuning in $\Omega_K$ at early times, known as the \textbf{flatness problem}.\index{Flatness problem} Consider:
\begin{equation}
 |\Omega_{K}| = \left|-\frac{Kc^2}{H^2a^2}\right| = \left|-\frac{Kc^2}{H^2a^2}\frac{H_0^2a_0^2}{H_0^2a_0^2}\right| = \left|\Omega_{K0}\right|\frac{H_0^2}{H^2a^2}\;,
\end{equation}
where we have used $a_0 = 1$ in the last equality. Now, how does the function on the right hand side scale? From \eqref{LCDMFriedeq}:
\begin{equation}
    \frac{H_0^2}{H^2a^2} = \frac{a^2}{\Omega_\Lambda a^4 + \Omega_{\rm m0}a + \Omega_{\rm r0} + \Omega_{K0}a^2}\,.
\end{equation}
Unlike matter and radiation, $|\Omega_{K}|$ does not diverge for $a \to 0$ but rather vanishes. In particular:
\begin{equation}
	\frac{H_0^2}{H^2a^2} \propto a^2\;, \qquad \mbox{radiation domination.}
\end{equation}
This is the point at the heart of the flatness problem.

\hrulefill

\begin{ex}
	Show that at $z = 10^4$ one has $\Omega_{K}/\Omega_{K0} = 10^{-4}$. Show that at $z = 10^{10}$ one has $\Omega_{K}/\Omega_{K0} = 10^{-16}$. Finally, show that the Planck time corresponds roughly to $z = 10^{32}$, and there we have $\Omega_{K}/\Omega_{K0} = 10^{-60}$.
\end{ex}

\hrulefill

The Planck time is chosen since it is the farthest point to which we can extrapolate our classical theory. Note that $10^{-60}$ is something \emph{very} close to zero. 

During matter domination, we have:
\begin{equation}
	\frac{H_0^2}{H^2a^2} \propto a\;, \qquad \mbox{matter domination.}
\end{equation}
Instead, during $\Lambda$ domination, we have:
\begin{equation}
	\frac{H_0^2}{H^2a^2} \propto \frac{1}{a^2}\;, \qquad \Lambda\mbox{ domination.}
\end{equation}
Looking the other way, $|\Omega_K|$ has grown for almost the entire history of the universe, \textit{proportional to a power of the scale factor}. Yet, today its value is close to zero rather than large.

Let us make another calculation in the opposite direction. Suppose that the earliest time at which our theory is reliable is the Planck scale. So, here $a_P = 10^{-32}$ and also suppose that $\Omega_{K,P}$ has some unknown value, which we consider to be the \textit{initial value} for the curvature density parameter. Up to matter-radiation equality, where $a_{\rm eq} = 10^{-4}$, the scale factor has grown by a factor of $10^{28}$, and thus $\Omega_{K,\rm eq} = 10^{56}\Omega_{K,P}$. From matter-radiation equality up to the present time, the scale factor has grown by over 4 orders of magnitude; thus (neglecting $\Lambda$) $\Omega_{K0} = 10^{60}\Omega_{K,P}$ 

If, for some reason, $\Omega_{K,P} \simeq 10^{-59}$, then today $\Omega_{K0} \simeq 10$ is in complete disagreement with the observation. 

We conclude that, in order to match the observation, $\Omega_{K,P}$ must be determined by some physical mechanism to be zero with a precision of at least 60 significant digits! This is an example of \textbf{fine-tuning}\index{Fine tuning}. 

Fine tuning might not be a severe problem. After all, we live in a universe where every single fundamental constant seems to be of the ``right'' magnitude for us to be here.\footnote{About this ``rightness'', see \cite{bonaventura2021universo}.} A fine-tuned theory is not \textit{wrong}; it is not falsified by observation. On the other hand, fine-tuning conveys a sense of \textit{being ad hoc} to the theory, something unnatural or not fully understood that we would like to explain better.

Such an explanation is possibly provided by inflationary theory, which we shall see in detail in Chapter~\ref{Chap:Inflation}. How it works for the flatness problem can already be seen from the above equation $H_0^2/(H^2a^2) \sim 1/a^2$: if $H$ is almost constant, the curvature density parameter \textit{decreases}. So, if before radiation-domination an evolutionary phase exists in which $H$ is almost constant for a sufficiently long time, then we might be able to explain why the curvature density parameter was so small to begin with.

\subsection{The fine-tuned cosmological constant and the cosmic coincidence problem}

The essence of the flatness problem is that $\Omega_K$ does not diverge for $a \to 0$, but rather vanishes. There is another component in the $\Lambda$CDM model whose density does not diverge for $a \to 0$: the cosmological constant. Should we expect fine-tuning for $\Lambda$ too?

Let us consider the ratio between matter and $\rho_\Lambda$. It grows as:
\begin{equation}
	\frac{\rho_{\rm m}}{\rho_\Lambda} = \frac{\Omega_{\rm m0}}{\Omega_\Lambda}(1 + z)^3\,.
\end{equation}
This ratio grows incessantly as $z \to \infty$. In particular, at the Planck time $z = 10^{32}$, we would have the matter density being $10^{96}$ times larger than the density of the cosmological constant. We face, then, another fine-tuning. In fact, it would be problematic if, at the Planck time, the above ratio were $10^{97}$ or $10^{95}$, because, in the first case, we would not observe the present acceleration of the expansion, and in the second case, structures possibly would not form (or their distribution would be much sparser than the observed one). The ratio $\rho_{\rm m}/\rho_\Lambda$ should be set at the Planck scale with a precision of 96 decimal places.\footnote{The fine-tuning is even harsher if one chooses the radiation density as a reference.} 

Seen from the other perspective, as a \textbf{coincidence problem},\index{Cosmic coincidence} one might ask why the densities of matter and the cosmological constant are of the same order at the present time \cite{Zlatev:1998tr}.

The way out of this problem is typically to dismiss the cosmological constant in favor of \textbf{Dynamical Dark Energy}.\index{Dark Energy} That is, one assumes that what causes the accelerated expansion of the universe is also a dynamical component, subdominant at early-times, but not constant and, rather, with density diverging for $a \to 0$. Although this seems to work, it is a pity to withdraw from the simplicity and beauty of $\Lambda$.

\section{Solutions to the Friedmann equations}\label{Sec:solFriedeq}

The Friedmann equations can be solved exactly for many cases of interest.

\subsection{The Einstein Static Universe}

As the first application of his theory to cosmology, Einstein was concerned with boundary conditions at infinity \cite{einstein1917kosmologische}. It is in this context that the cosmological constant was born, denoted by Einstein as $\lambda$. Within Newton's theory of gravity, he comes to the conclusion that in order to have a viable density distribution of matter (implicitly assumed to be static), one should modify the Poisson equation as follows:\footnote{Einstein's notation of \cite{einstein1917kosmologische} is used here.}
\begin{equation}
	\Delta\phi - \lambda\phi = 4\pi K\varrho\;.
\end{equation}
In this way, it is possible to have a ``background'' constant density of matter:\index{Cosmological constant}
\begin{equation}
	\phi = -\frac{4\pi K}{\lambda}\varrho_0\;,
\end{equation} 
extending to infinity, with a constant potential there, meaning that no forces act at infinity. In GR, Einstein supposes an isotropic line element:
\begin{equation}
	ds^2 = -A(dx_1^2 + dx_2^2 + dx_3^2) + Bdx_4^2\;,
\end{equation}
with the strange requirement (best known today as \textbf{the unimodular constraint})\index{Unimodular constraint} that $\sqrt{-g} = \sqrt{A^3B} = 1$, and furthermore with $A$ and $B$ dependent only on $x_1^2 + x_2^2 + x_3^2$, but not on the time $x_4$. Staticity is not explicitly assumed by Einstein; probably because it was a natural hypothesis to make at that time. Einstein makes another very interesting hypothesis: in accordance with Mach's principle, inertia must vanish at infinity. This requires that $A \to 0$ and $B \to \infty$ (so that $\sqrt{A^3B}$ stays 1), but this requirement is incompatible with the gravitational field of a distribution of matter in which the velocities are much smaller than the speed of light. Indeed, $B$ is the gravitational potential, and small orbital velocities imply a small gravitational potential. The only way out that Einstein could find was to postulate a ``closed world'' of radius $R$, i.e., a world described by the metric
\begin{equation}
	g_{\mu\nu} = -\left(\delta_{\mu\nu} + \frac{x_\mu x_\nu}{R^2 - (x_1^2 + x_2^2 + x_3^2)}\right)\;,
\end{equation}   
where $\mu$ and $\nu$ can assume only spatial values (the use of Latin indices referring only to spatial components was not yet in force). We recognize metric \eqref{componentsspatial} with $K = +1$. Yet, this metric does not provide a solution to the Einstein equations unless $\lambda$ is introduced, as in Eq.~\eqref{Einseq}. With this extra ingredient, Einstein found:\footnote{The $2\pi^2$ factor is the surface of the unit hypersphere in 3 dimensions.}
\begin{equation}
	\lambda = \frac{\kappa\varrho}{2} = \frac{1}{R^2}\;, \qquad M = \varrho 2\pi^2R^3 = \frac{4\pi^2R}{\kappa} = \frac{\sqrt{32}\pi^2}{\sqrt{\kappa^3\varrho}}\;.
\end{equation}
Returning to our notation and formalism, we can reproduce Einstein's results by using the FLRW metric and requiring staticity. To this purpose, we must set $\dot a = \ddot a = 0$, and since $\rho$ is positive, we must choose $K > 0$. Therefore, the Einstein Static Universe (ESU)\index{Einstein static universe} is a closed universe. Its radius is:
\begin{equation}
	\frac{8\pi G}{3}\rho = Kc^2 \qquad \Rightarrow \qquad R \equiv \frac{1}{\sqrt{K}} = c\sqrt{\frac{3}{8\pi G\rho}}\;,
\end{equation}
where we have normalized $a$, which is a constant, to unity (indeed, $a = a_0 = 1$).

From the acceleration equation, we obtain that:
\begin{equation}
	\rho + 3P/c^2 = 0\;.
\end{equation}
Therefore, there cannot be just ordinary matter because we need negative pressure. Here enters the cosmological constant $\Lambda$. We assume that $\rho = \rho_{\rm m} + \rho_\Lambda$, so that
\begin{equation}
	\rho + 3P/c^2 = 0\;, \qquad \Rightarrow \qquad \rho_{\rm m} + \rho_\Lambda - 3\rho_\Lambda = 0\;,
\end{equation}
and therefore $\rho_{\rm m} = 2\rho_\Lambda$. The radius can thus be written as:
\begin{equation}
	R = \frac{c}{\sqrt{4\pi G\rho_{\rm m}}} = \frac{1}{\sqrt{\Lambda}}\;.
\end{equation}
Until here, everything seems to be fine. But it is not. The problem is indeed the condition $\rho_{\rm m} = 2\rho_\Lambda$, which makes the ESU unstable. In fact, if this condition is broken, then the universe either expands or collapses.

\hrulefill

\begin{ex} Prove that the ESU is unstable. 
\end{ex}

A way to solve this exercise is to write the acceleration equation as:
\begin{equation}
	\ddot a = -\frac{d}{da}U\,,
\end{equation}
introducing a potential $U$ and then using the known results of classical mechanics. In particular, writing $\rho_{\rm m} = \rho_{\rm m0}/a^3$:
\begin{equation}
	U = -\frac{4\pi G}{3a}(\rho_{\rm m0} + \rho_\Lambda a^3)\,.
\end{equation}
Then one sees that the extremal point: 
\begin{equation}
	\bar a = \left(\frac{\rho_{\rm m0}}{2\rho_\Lambda}\right)^{1/3}\,,
\end{equation}
is a maximum and thus a point of unstable equilibrium.

Perturbing the acceleration equation about $\bar a$, i.e., writing $a = \bar a + \delta a$, one obtains:
\begin{equation}
	\ddot{\delta a} = 8\pi G\rho_\Lambda\delta a\,.
\end{equation}
If $\delta a > 0$, so is $\ddot{\delta a}$, and thus the scale factor keeps growing. Conversely, if $\delta a < 0$, so is $\ddot{\delta a}$, and thus the scale factor keeps decreasing.

Since:
\begin{equation}
	\rho_{\rm m} - 2\rho_\Lambda = -6\rho_\Lambda\frac{\delta a}{\bar a}\,,
\end{equation}
one sees that $\delta a > 0$ represents an excess of the cosmological constant, whereas $\delta a < 0$ indicates a deficit of the same.

\hrulefill

The ESU is also known as the ``cylindrical universe'' because its topology is $R\times S^3$, that of a four-dimensional cylinder.

The introduction of the cosmological constant by Einstein is frequently reported to be regarded by him as ``the biggest blunder of my life''. This seems to be a personal comment made by Einstein to George Gamow and reported by the latter in a 1956 article in \textit{Scientific American} \cite{1956SciAm.195c.136G}. The dissatisfaction of Einstein regarding $\lambda$ can be read in a 1931 paper \cite{einstein1931kosmologische} by Einstein himself, in which he writes:

\textit{``Unter diesen Umst\"anden mu\ss\; man sich die Frage vorlegen, ob man den Tatsachen ohne die Einf\"uhrung des theoretisch ohnedies unbefriedigenden $\lambda$-Gliedes gerecht werden kann.''}  

That is:

\textit{``Under these circumstances, one should ask whether the observational facts can be accounted for without the inclusion of the theoretically, in all respects unsatisfactory, $\lambda$-term.''} 

\subsection{The de Sitter universe}

The Dutch mathematician Willem de Sitter had the idea in \cite{dS1917A} of extending Einstein hypothesis of a closed world to the whole spacetime.\footnote{In a footnote of \cite{dS1917A}, de Sitter reveals that such an idea was, in fact, suggested by Ehrenfest.} In de Sitter's notation:
\begin{equation}
	g_{\mu\nu} = -\delta_{\mu\nu} - \frac{x_\mu x_\nu}{R^2 - \sum x_\mu^2}\;,
\end{equation}
where the sum $\sum$ goes from 1 to 4.\footnote{de Sitter uses in his paper the convention of using Latin indices for the spatial components.} From this metric, he finds:
\begin{equation}
	G_{\mu\nu} = 12\sigma g_{\mu\nu}\;,  
\end{equation}
with
\begin{equation}
	\sigma = \frac{1}{4R^2}\;, \qquad \lambda = 12\sigma\;, \qquad \varrho = 0\;.
\end{equation}
This solution describes a universe with curvature but no matter. W. de Sitter also assesses the impact of $\lambda$ on the perihelion precession and estimates that $2\sigma < 10^{-50}$ cm$^{-2}$ (which gives $\lambda < 10^{-49}$ cm$^{-2}$, remarkably close to the measured value today $\lambda \approx 10^{-48}$ cm$^{-2}$). The debate that gave rise to the ESU and the de Sitter universe was based on the relativity of inertia, i.e., on the postulate that at infinity the metric should be invariant. The impossibility of doing this compelled Einstein to consider a closed world and to introduce the cosmological constant. In \cite{dS1917A}, W. de Sitter seems to be very dissatisfied with the above postulate and with $\lambda$. In fact, he concludes it as follows:

\textit{``Finally we can also reject both systems A and B, and stick to the original field-equations (3) and the values (1) of the $g_{\mu\nu}$, which are not invariant at infinity. Then, of course, inertia is not explained: we must then prefer to leave it unexplained rather than explain it by the undetermined and undeterminable constant $\lambda$. It cannot be denied that the introduction of this constant detracts from the symmetry and elegance of Einstein's original theory, one of whose chief attractions was that it explained so much without introducing any new hypothesis or empirical constants.''}  
Coming back to our formalism, let us see which kind of cosmology the de Sitter universe gives rise to. It must be noted that the de Sitter space is maximally symmetric and, as such, it possesses a time-like Killing vector. This means that it is \textit{static}; thus, no evolution actually takes place. The solutions that we are going to find will nonetheless display an evolution; however, but this is due to a particular choice of the coordinates. In fact, note that, since the de Sitter space is maximally symmetric, any spatial hypersurface is also maximally symmetric; thus, there are different possible choices for the cosmic time. Moreover, there are also choices of coordinates in which the staticity of de Sitter space is manifest.  

Suppose $\rho = 0$ and a cosmological constant $\Lambda$. the Friedmann equation \eqref{FriedEq2} becomes:\index{de Sitter universe}
\begin{equation}\label{FriedEqdeSitter}
 H^2 = \frac{\Lambda c^2}{3} - \frac{Kc^2}{a^2}\;.
\end{equation}
When spatial curvature is taken into account, it is more convenient to solve the acceleration equation \eqref{accEq} rather than the Friedmann equation. Indeed:
\begin{equation}
	\ddot a = \frac{\Lambda c^2}{3}a\;,
\end{equation}
is straightforwardly integrated:
\begin{equation}\label{gensol}
	a(t) = C_+\exp\left(\sqrt{\frac{\Lambda}{3}}ct\right) + C_-\exp\left(-\sqrt{\frac{\Lambda}{3}}ct\right)\;,
\end{equation}
where $C_+$ and $C_-$ are two integration constants. One of these is constrained by the Friedmann equation \eqref{FriedEqdeSitter}. Calculating $\dot a^2$ and $a^2$ from Eq.~\eqref{gensol}, we get:
\begin{eqnarray}
	\dot a^2 = \frac{\Lambda c^2}{3}\left[C_+^2\exp\left(2\sqrt{\frac{\Lambda}{3}}ct\right) + C_-^2\exp\left(-2\sqrt{\frac{\Lambda}{3}}ct\right) - 2C_+C_-\right]\;,\\
	a^2 = C_+^2\exp\left(2\sqrt{\frac{\Lambda}{3}}ct\right) + C_-^2\exp\left(-2\sqrt{\frac{\Lambda}{3}}ct\right) + 2C_+C_-\;.
\end{eqnarray}
Combining them, one finds:
\begin{equation}
	\dot a^2 = \frac{\Lambda c^2}{3}(a^2 - 4C_+C_-)\;.
\end{equation}
Using Eq.~\eqref{FriedEqdeSitter}, we can impose a constraint on the product of the two integration constants:
\begin{equation}
	\frac{4\Lambda}{3}C_+C_- = K\;,
\end{equation}
so that we have the freedom to fix just one of them. Let us do that by imposing an initial condition at $t = 0$:
\begin{equation}
	\bar a = C_+ + C_-\,,
\end{equation} 
where $\bar a \ge 0$, i.e., we avoid negative values for the scale factor since they are unphysical.

If $K = 0$, apart from the trivial identically zero solution, there are two solutions:
\begin{equation}
	a(t) = \bar a \exp\left(\pm\sqrt{\frac{\Lambda}{3}}ct\right)\;,
\end{equation} 
of which one represents an exponential expansion and the other an exponential contraction of the universe.

If $K > 0$, we need $C_{\pm} > 0$ to avoid negative scale factors. In this case, it is useful to choose $\bar a$ such that $C_+ = C_-$ (e.g., $\bar a = 2C_+$). Then:
\begin{equation}
	C_+ = C_- = \sqrt{\frac{3K}{4\Lambda}}\;,
\end{equation}
and the solution can be written as:
\begin{equation}
	a(t) = \sqrt{\frac{3K}{\Lambda}}\cosh\left(\sqrt{\frac{\Lambda}{3}}ct\right)\,.
\end{equation}
If $K < 0$, the integration constants must have opposite signs. We choose $C_+ > 0$ so that the scale factor remains positive for large $t$. In this case, it is convenient to choose $\bar a = 0$ so that $C_+ = -C_-$ and the solution can be written as:
\begin{equation}
	a(t) = \sqrt{\frac{3|K|}{\Lambda}}\sinh\left(\sqrt{\frac{\Lambda}{3}}ct\right)\,.
\end{equation}
Summarizing:
\begin{equation}\label{solcoshsinh}
	a(t) =
	\begin{cases}
		\sqrt{\frac{3|K|}{\Lambda}}\sinh\left(\sqrt{\Lambda/3}ct\right)\;, \qquad &\mbox{ for } K < 0\;,\\
		\exp\left(\pm \sqrt{\Lambda/3}ct\right)\;, \qquad &\mbox{ for } K = 0\;,\\
		\sqrt{\frac{3K}{\Lambda}}\cosh\left(\sqrt{\Lambda/3}ct\right)\;, \qquad &\mbox{ for } K > 0\;.\\
	\end{cases}
\end{equation}
Here, we have chosen $\bar a = 1$ for the case $K = 0$. Note that there is no free $\bar a$ for the solutions with non-vanishing spatial curvature because we fixed it in order to have the hyperbolic sine and cosine functions.

\hrulefill

\begin{ex} From the Einstein equations \eqref{Einseq} show that $R = 4\Lambda$ for the de Sitter universe. Verify that the above solutions \eqref{solcoshsinh} satisfy this relation by substituting them into the expression in Eq.~\eqref{Ricciscalcosmo} for the Ricci scalar.
\end{ex}

\hrulefill 

For $K > 0$, the de Sitter universe is eternal, with no Big-Bang ($a = 0$), but with a bounce at the minimum value of the scale factor. For $K = 0$, there is a Big-Bang at $t = -\infty$. For $K < 0$, we might have negative scale factors, which we, however, neglect, and consider the evolution as starting only at $t = 0$, for which there is another Big-Bang. Note that the Big-Bang's we are mentioning here are not actual singularities, but rather coordinate singularities (the coordinates chosen do not cover the whole space). There is no geometrical singularity in de Sitter space since it is maximally symmetric.

As one might expect, the deceleration parameter is always negative:
\begin{equation}
	q = -\frac{\ddot a a}{\dot a^2} = -\frac{\Lambda c^2}{3}\frac{a^2}{\dot a^2} = -\left(1 - \frac{3K}{\Lambda a^2}\right)^{-1}\;.
\end{equation}\index{de Sitter universe!Deceleration parameter} 

\subsection{Radiation-dominated universe}

For the simple case $\rho = \rho_0 a^{-4}$, $K = 0$, and $\Lambda = 0$, the solution of Eq.~\eqref{FriedEq2} can be expressed in the form:
\begin{equation}\label{atraddom}
 a = \sqrt{\frac{t}{t_0}}\;,
\end{equation}
where the integration constant is chosen such that $a(t_0) = 1$. The deceleration parameter is $q_0 = 1$, and the age of the universe is:
\begin{equation}\label{radage}
 t_0 = \dfrac{1}{2H_0}\;.
\end{equation}
Note the coefficient $1/2$ instead of the $0.95$ of Eq. \eqref{obsvalue} of the $\Lambda$CDM model.

\hrulefill

\begin{ex} Prove the results of Eqs.~\eqref{atraddom} and \eqref{radage}. \end{ex}

\hrulefill

It is quite complicated to analytically solve the Friedmann equation \eqref{FriedEq2} for a radiation-dominated universe when $K \neq 0$. On the other hand, solving the acceleration equation using conformal time, cf. Eq. \eqref{accEqconft}, is much easier. Indeed, for $\rho c^2 = 3P$, Eq.~\eqref{accEqconft} becomes:
\begin{equation}\label{acceqconftraddom}
	a'' + Kc^2a = 0\;,
\end{equation}
whose general solution is:\index{Radiation-dominated universe!Scale factor solution}
\begin{equation}\label{gensolraddom}
	a(\eta) = 
	\begin{cases}
		C_1\sinh(\sqrt{|K|}c\eta) + C_2\cosh(\sqrt{|K|}c\eta)\;, \qquad &\mbox{ for } K < 0\;,\\
		C_3 + C_4\eta\;, \qquad &\mbox{ for } K = 0\;,\\
		C_5\sin(\sqrt{K}c\eta) + C_6\cos(\sqrt{K}c\eta)\;, \qquad &\mbox{ for } K > 0\;.\\
	\end{cases}
\end{equation}
Since $\rho = \rho_0/a^4$, there is a singularity when $a = 0$. So, let us choose the zero of the conformal time such that $a(\eta = 0) = 0$. Thus, the general solution \eqref{gensolraddom} becomes:
\begin{equation}\label{gensolraddom1}
	a(\eta) = 
	\begin{cases}
		C_1\sinh(\sqrt{|K|}c\eta)\;, \qquad &\mbox{ for } K < 0\;,\\
		C_4\eta\;, \qquad &\mbox{ for } K = 0\;,\\
		C_5\sin(\sqrt{K}c\eta)\;, \qquad &\mbox{ for } K > 0\;.\\
	\end{cases}
\end{equation}
The remaining constant is determined from the constraint of the Friedmann equation \eqref{FriedEq2conft}:
\begin{equation}
	a'^2 = \frac{8\pi G}{3}\rho a^4 - Kc^2a^2 = \frac{8\pi G\rho_0}{3} - Kc^2a^2\;,
\end{equation}
where it should be noted that $\rho a^4 = \rho_0$, i.e., a constant. When $a = 0$, i.e. $\eta = 0$, then:
\begin{equation}
	a'^2(\eta = 0) = \frac{8\pi G\rho_0}{3}\;,
\end{equation}
and the solutions \eqref{gensolraddom1} become:
\begin{equation}\label{gensolraddom2}
	a(\eta) = \sqrt{\frac{8\pi G\rho_0}{3c^2}}
	\begin{cases}
		\frac{1}{\sqrt{|K|}}\sinh(\sqrt{|K|}c\eta)\;, \qquad &\mbox{ for } K < 0\;,\\
		c\eta\;, \qquad &\mbox{ for } K = 0\;,\\
		\frac{1}{\sqrt{K}}\sin(\sqrt{K}c\eta)\;, \qquad &\mbox{ for } K > 0\;.\\
	\end{cases}
\end{equation}
Corresponding to $t_0$, we also have $\eta_0$, for which $a(\eta_0) = a_0 = 1$. Therefore, from the above solutions, we can establish the relation between $\eta_0$ and the present time density and spatial curvature.

If we want to recover cosmic time from the above solutions, we need to perform the following integration:
\begin{equation}
	\int_0^\eta a(\eta')d\eta' = t\;.
\end{equation}
Using Eq.~\eqref{gensolraddom2}, one obtains:
\begin{equation}\label{gensolraddom2tetarel}
	ct = \sqrt{\frac{8\pi G\rho_0}{3c^2}}
	\begin{cases}
		\frac{1}{|K|}[\cosh(c\sqrt{|K|}\eta) - 1]\;, \qquad &\mbox{ for } K < 0\;,\\
		(c\eta)^2/2\;, \qquad &\mbox{ for } K = 0\;,\\
		\frac{1}{K}[1 - \cos(c\sqrt{K}\eta)]\;, \qquad &\mbox{ for } K > 0\;.\\
	\end{cases}
\end{equation}
Inverting these relations allows one to find $\eta = \eta(t)$, which, once substituted in Eq.~\eqref{gensolraddom2}, returns $a = a(t)$.

For example, for the $K = 0$ case, we have:
\begin{equation}
	a^2 = \frac{8\pi G\rho_0}{3c^2}2ct\sqrt{\frac{3c^2}{8\pi G\rho_0}} = 2H_0t = \frac{t}{t_0}\,, 
\end{equation}
thereby obtaining the result of Eq. \eqref{atraddom}.

\subsection{Dust-dominated universe}

Assuming $\rho = \rho_0 a^{-3}$, $K = 0$, and $\Lambda = 0$, the solution of the Friedmann equation \eqref{FriedEq2} is straightforwardly obtained:
\begin{equation}\label{adustflat}
 a = \left(\frac{t}{t_0}\right)^{2/3}\;.
\end{equation}
The deceleration parameter is $q_0 = 1/2$, and the age of the universe is:
\begin{equation}\label{EdSage}
 t_0 = \dfrac{2}{3H_0} = 6.52~h^{-1}~{\rm Gyr}\;.
\end{equation}
This model of the universe is also known as the \textbf{Einstein-de Sitter universe} \cite{einstein1932relation}.\index{Einstein-de Sitter universe} It does not predict accelerated expansion, and there is also a significant problem with its age: the universe would be younger than some globular clusters \cite{Velten:2014nra}.

\hrulefill

\begin{ex} Prove the results of Eqs.~\eqref{adustflat} and \eqref{EdSage}. \end{ex}

\hrulefill

\noindent As for the radiation-dominated case, it is impossible to analytically solve the Friedmann equation in cosmic time \eqref{FriedEq2} for a dust-dominated universe when $K \neq 0$. However, as in the radiation-dominated case, it is possible to find an exact solution for $a(\eta)$ using the acceleration equation and the conformal time. Equation~\eqref{accEqconft} for the dust-dominated case reads:
\begin{equation}\label{acceqdustdometa}
	a'' + Kc^2 a = \frac{4\pi G}{3}\rho a^3 = \frac{4\pi G}{3}\rho_0\;.
\end{equation}
Note that $\rho a^3 = \rho_0 =$ is constant. We have already encountered this equation, cf. \eqref{newtequationcosmo}. The parameter $p$ introduced to solve the latter is the conformal time.

The general solution of Eq.~\eqref{acceqdustdometa} is the general solution of its homogeneous part, cf. Eq.~\eqref{acceqconftraddom}, plus a particular solution. That is:\index{Dust-dominated universe!Scale factor solution}
\begin{equation}\label{gensoldustdom}
	a(\eta) = 
	\begin{cases}
		C_1\sinh(c\sqrt{|K|}\eta) + C_2\cosh(c\sqrt{|K|}\eta) - \frac{4\pi G}{3c^2|K|}\rho_0\;, \qquad &\mbox{ for } K < 0\;,\\
		C_3 + C_4\eta + \frac{2\pi G}{3}\rho_0\eta^2\;, \qquad &\mbox{ for } K = 0\;,\\
		C_5\sin(c\sqrt{K}\eta) + C_6\cos(c\sqrt{K}\eta) + \frac{4\pi G}{3c^2K}\rho_0\;, \qquad &\mbox{ for } K > 0\;.\\
	\end{cases}
\end{equation}

\hrulefill

\begin{ex} Using the condition $a(0) = 0$ and employing the Friedmann equation for a dust-dominated universe, i.e.
\begin{equation}
	a'^2 + Kc^2a^2 = \frac{8\pi G}{3}\rho a^4 = \frac{8\pi G}{3}\rho_0 a\;,
\end{equation}
as constraint, show that Eq.~\eqref{gensoldustdom} can be cast as:
\begin{equation}\label{gensoldustdom1}
	a(\eta) = \frac{4\pi G\rho_0}{3c^2}
	\begin{cases}
		\frac{1}{|K|}[\cosh(c\sqrt{|K|}\eta) - 1]\;, \qquad &\mbox{ for } K < 0\;,\\
		(c\eta)^2/2\;, \qquad &\mbox{ for } K = 0\;,\\
		\frac{1}{K}[1 - \cos(c\sqrt{K}\eta)]\;, \qquad &\mbox{ for } K > 0\;.\\
	\end{cases}
\end{equation}
\end{ex}

\hrulefill

Recovering the cosmic time from Eq.~\eqref{gensoldustdom1}, one has:
\begin{equation}\label{gensoldustdom1teta}
	ct = \frac{4\pi G\rho_0}{3c^2}
	\begin{cases}
		\frac{1}{|K|^{3/2}}\sinh(c\sqrt{|K|}\eta) - \frac{1}{|K|}c\eta\;, \qquad &\mbox{ for } K < 0\;,\\
		(c\eta)^3/6\;, \qquad &\mbox{ for } K = 0\;,\\
		\frac{1}{K}c\eta - \frac{1}{K^{3/2}}\sin(c\sqrt{K}\eta)\;, \qquad &\mbox{ for } K > 0\;.\\
	\end{cases}
\end{equation}
Unfortunately, the above relations for $K \neq 0$ cannot be explicitly inverted to yield $\eta(t)$ and then $a(t)$.

\subsection{Radiation plus dust universe}\label{Subsec:radplusdustuni}

The mixture of radiation and dust is a realistic model of the universe over a very large timespan. Consider the total density:
\begin{equation}
	\rho = \rho_{m} + \rho_{r} = \frac{\rho_{m0}}{a^3} + \frac{\rho_{r0}}{a^4} = \frac{\rho_{\rm eq}}{2}\frac{a_{\rm eq}^3}{a^3} + \frac{\rho_{\rm eq}}{2}\frac{a_{\rm eq}^4}{a^4}\;,
\end{equation}
where $a_{\rm eq}$ is the equivalence scale factor introduced in Eq. \eqref{radmattequalityscalefactor}. At this time, we dub the total density as $\rho_{\rm eq}$. The acceleration equation \eqref{accEqconft} for the radiation plus dust model:
\begin{equation}\label{acceqcontdustrad}
	a'' = \frac{4\pi G}{3}\rho_m a^3 - Kc^2a = \frac{4\pi G}{3}\rho_{m0} - Kc^2a\;,
\end{equation}
is identical to that for the dust-dominated case, cf. Eq.~\eqref{acceqdustdometa}; thus, the general solutions are the same: 
\begin{equation}
	a(\eta) = 
	\begin{cases}
		C_1\sinh(c\sqrt{|K|}\eta) + C_2\cosh(c\sqrt{|K|}\eta) - \frac{4\pi G\rho_{m0}}{3c^2|K|}\;, \qquad &\mbox{ for } K < 0\;,\\
		C_3 + C_4\eta + \frac{2\pi G\rho_{m0}}{3}\eta^2\;, \qquad &\mbox{ for } K = 0\;,\\
		C_5\sin(c\sqrt{K}\eta) + C_6\cos(c\sqrt{K}\eta) + \frac{4\pi G\rho_{m0}}{3c^2K}\;, \qquad &\mbox{ for } K > 0\;.\\
	\end{cases}
\end{equation}
Imposing the condition $a(0) = 0$ leads to the following solutions:\index{Radiation plus dust universe!Scale factor solution}
\begin{equation}
	a(\eta) = 
	\begin{cases}
		C_1\sinh(c\sqrt{|K|}\eta) + \frac{4\pi G\rho_{m0}}{3c^2|K|}[\cosh(c\sqrt{|K|}\eta) - 1]\;, \qquad &\mbox{ for } K < 0\;,\\
		C_4\eta + \frac{2\pi G\rho_{m0}}{3}\eta^2\;, \qquad &\mbox{ for } K = 0\;,\\
		C_5\sin(c\sqrt{K}\eta) + \frac{4\pi G\rho_{m0}}{3c^2K}[1 - \cos(c\sqrt{K}\eta)]\;, \qquad &\mbox{ for } K > 0\;.\\
	\end{cases}
\end{equation}
The presence of the radiation component enters the Friedmann equation:
\begin{equation}
	a'^2 + Kc^2a^2 = \frac{8\pi G}{3}\left(\rho_{m0}a + \rho_{r0}\right)\;,
\end{equation}
which we use to determine the integration constants introduced above. Since:
\begin{equation}
	a'^2(\eta = 0) = \frac{8\pi G\rho_{r0}}{3}\;,
\end{equation}
then we have:
\begin{equation}
	a(\eta) = 
	\begin{cases}
		\sqrt{\frac{8\pi G\rho_{r0}}{3}}\frac{1}{c\sqrt{|K|}}\sinh(c\sqrt{|K|}\eta) + \frac{4\pi G\rho_{m0}}{3c^2|K|}[\cosh(c\sqrt{|K|}\eta) - 1]\;, &\mbox{ for } K < 0\;,\\
		\sqrt{\frac{8\pi G\rho_{r0}}{3}}\eta + \frac{2\pi G\rho_{m0}}{3}\eta^2\;, &\mbox{ for } K = 0\;,\\
		\sqrt{\frac{8\pi G\rho_{r0}}{3}}\frac{1}{c\sqrt{K}}\sin(c\sqrt{K}\eta) + \frac{4\pi G\rho_{m0}}{3c^2K}[1 - \cos(c\sqrt{K}\eta)]\;, &\mbox{ for } K > 0\;.\\
	\end{cases}
\end{equation}

\hrulefill

\begin{ex}
Show that the solution for $K = 0$ can be written as:
\begin{equation}
	a(\eta) = \sqrt{\Omega_{\rm r0}}H_0\eta + \frac{\Omega_{\rm m0}H_0^2}{4}\eta^2\;.
\end{equation}
From here, show that the conformal time at equivalence $\eta_{\rm eq}$ can be written as:
\begin{equation}
	\eta_{\rm eq} = (\sqrt{2} - 1)\frac{2\sqrt{\Omega_{\rm r0}}}{\Omega_{\rm m0}H_0} \qquad (K = 0)\,.
\end{equation}

\end{ex}

\hrulefill

\begin{ex} Solve Friedmann equation for the $\Lambda$CDM model, neglecting radiation and spatial curvature:
\begin{equation}
\frac{H^2}{H_0^2} = \frac{\Omega_{\rm m0}}{a^3} + \Omega_\Lambda\;.
\end{equation}
Show that:
\begin{equation}
a(t) = \left[\frac{\Omega_{\rm m0}}{\Omega_\Lambda}\sinh^2\left(\frac{3}{2}\sqrt{\Omega_\Lambda}H_0t\right)\right]^{1/3}\;.
\end{equation}\index{$\Lambda$CDM model}The limit $\Omega_\Lambda \to 0$ gives the correct result $(t/t_0)^{2/3}$. Why, instead, the limit $\Omega_{\rm m0} \to 0$ does not reproduce the known result $\exp(H_0\sqrt{\Omega_\Lambda}t)$?
\end{ex}

\hrulefill

\begin{ex} Solve the Friedmann equation for the curvature-dominated universe:
\begin{equation}
 H^2 = -\frac{Kc^2}{a^2}\;.	
\end{equation} 
This is the Milne model \cite{1935rgws.book.....M}. Clearly, only $K < 0$ is allowed.\index{Milne universe} The equivalence of Milne's kinematic model to a world with FLRW metric with $K < 0$ was shown by Walker in \cite{walker1935formal}.

Show then that $a = ct$. Substitute this solution into the expression for the Ricci scalar \eqref{Ricciscalcosmo}. Show that $R = 0$. 

Write down explicitly the FLRW metric with $a = ct$ and show that it is Minkowski metric written in a coordinates system different from the usual.
\end{ex}

\hrulefill

This result is not surprising since the Milne model has no matter (an empty universe) and no cosmological constant. The spatial hypersurfaces are already maximally symmetric due to the cosmological principle, and the absence of matter adds even more symmetry to the spacetime.

\section{Distance measures in cosmology}\label{Sec:distances}

We present and discuss in this section the various notions of distance that are employed in cosmology. We have already encountered the comoving distance and the proper distance. The line-of-sight comoving distance used in this section has no subscript $r$ in order not to burden the notation.

\subsection{Relation of the comoving and proper distances with the redshift}

It is useful, when considering photon propagation in the observation of sources, to relate distances to the redshift, as this is an observable quantity. This task is not difficult to complete, given a model of the universe, i.e., by establishing its matter content. On the other hand, establishing the redshift dependence of distance in a \textit{model-independent} way can be very valuable.

From Eq. \eqref{d2comdist}, we have that: 
\begin{equation}
	d\chi = \frac{cdt}{a(t)} = cd\eta\;.
\end{equation}
By integrating $cdt/a(t)$ from $t_{\rm em}$ to $t_0$, we obtain the line-of-sight comoving distance from the source to us, or the conformal time spent by the photon traveling from the source to us:
\begin{equation}\label{comdist}
	\chi = \int_{t_{\rm em}}^{t_0}\frac{cdt'}{a(t')} = \int_a^1\frac{cda'}{H(a')a^{'2}} = \int_0^z\frac{cdz'}{H(z')}\;.
\end{equation}
For the EdS case, one has $H = H_0/a^{3/2}$, and the line-of-sight comoving distance as a function of the scale factor and the redshift is:
\begin{equation}\label{comdistdust}
	\chi(a) = \frac{c}{H_0}\int_a^1\frac{da'}{\sqrt{a'}} = \frac{2c}{H_0}\left(1 - \sqrt{a}\right)\;, \quad \chi(z) = \frac{2c}{H_0}\left(1 - \frac{1}{\sqrt{1 + z}}\right)\;.
\end{equation}
In order to obtain the line-of-sight proper distance, we simply multiply by the scale factor:
\begin{equation}\label{propdistdust}
	D(a) = \frac{2c}{H_0}a\left(1 - \sqrt{a}\right)\;, \quad D(z) = \frac{2c}{H_0}\frac{1}{1 + z}\left(1 - \frac{1}{\sqrt{1 + z}}\right)\;.
\end{equation}
When $z \to 0$, both distances have the same dependence on the redshift: $\chi \sim D \sim cz/H_0$. Note that if we fix $a_0 = 1$, the line-of-sight proper distance at $t = t_0$ is equal to the line-of-sight comoving distance.

\hrulefill

\begin{ex} Calculate the line-of-sight comoving distance and the line-of-sight proper distance as functions of the scale factor and of the redshift for a radiation-dominated universe and for the de Sitter universe.
\end{ex}

\hrulefill

Without specifying a cosmological model, we can expand $\chi$ in Eq. \eqref{comdist} in powers of the redshift:
\begin{align}
    \chi = \frac{c}{H_0}\left[\frac{1}{E(0)}z + \frac{1}{2}\left(-\frac{1}{E^2}\frac{dE}{dz}\right)_{z = 0}z^2 + O(z^3)\right]\,.
\end{align}
By definition, we have $E(0) = 1$; moreover:
\begin{align}
    \frac{dE}{dz} = \frac{1}{H_0}\frac{dH}{dz} = \frac{1}{H_0}\frac{\dot H}{\dot z} = \frac{1}{H_0}\left(aH - \frac{\ddot a}{H}\right)\,.
\end{align}
Therefore, the line-of-sight comoving distance can be written as:
\begin{align}\label{chiexpredshift}
    \chi = \frac{c}{H_0}\left[z - \frac{1}{2}\left(1 + q_0\right)z^2 + O(z^3)\right]\,.
\end{align}    
Of course, one can proceed in any desired order in the redshift, and the $n$-th power of $z$ will be multiplied by the $n$-th derivative of the scale factor, evaluated at $t_0$. The third derivative of $a$ evaluated at $t_0$ is denoted $j_0$ (\textit{jerk}), and the fourth derivative of $a$ evaluated at $t_0$ is denoted $s_0$ (\textit{snap}).\index{Jerk parameter}\index{Snap parameter}

Despite Eq. \eqref{chiexpredshift} being very interesting, it is not particularly useful because the only measurable quantity within it is the redshift.

\subsection{Horizons}

If the cosmological model in which we are interested has a singularity in $a = 0$ (the Big Bang), it makes sense to set it as the lower integration limit in Eq.~\eqref{comdist}. In this way, one defines the \textbf{comoving horizon} $\chi_{\rm p}$ (also known as \textbf{the particle horizon}\index{Particle horizon} or \textbf{cosmological horizon}). This is the conformal time spent from the Big Bang until the cosmic time $t$ or scale factor $a$. It is also the maximum comoving distance that a photon can travel (hence the name particle horizon), and so it is the comoving size of the visible universe.

In the EdS case, using Eq.~\eqref{comdistdust}, with $a=0$ or $z=\infty$, one obtains:
\begin{equation}
	\chi_{\rm p} = c\eta_0 = \frac{2c}{H_0}\;.
\end{equation}
Note that this is not the age of the universe given in Eq.~\eqref{EdSage}, but three times its value. The reason is that it is the \textit{conformal} age of the universe; since the conformal time is defined as an integral in the cosmic time of $1/a(t)$ and $a(t) < 1$, it is expected that $\eta_0 > t_0$.

When the upper integration limit of Eq.~\eqref{comdist} is infinite, one defines the \textbf{event horizon}:\index{Event horizon}
\begin{equation}
	\chi_{\rm e}(t) \equiv c\int_t^\infty\frac{dt'}{a(t')} = c\int_a^\infty\frac{da'}{H(a')a^{'2}}\;,
\end{equation}
which, of course, makes sense only if the universe does not collapse. This represents the maximum distance traveled by a photon from a time $t$. If it diverges, then no event horizon exists; therefore, eventually, all the events in the universe will be causally connected. This happens, for example, in the dust-dominated case:
\begin{equation}
	\chi_{\rm e} = \frac{c}{H_0}\int_a^\infty\frac{da'}{\sqrt{a'}} = \infty\;.
\end{equation}
But in the de Sitter universe, we have
\begin{equation}
	\chi_{\rm e} = \frac{c}{H_0}\int_a^\infty\frac{da'}{a^{'2}} = \frac{c}{H_0a}\;.
\end{equation}
The proper event horizon for the de Sitter universe is constant:
\begin{equation}
	a\chi_{\rm e} = \frac{c}{H_0}\;.
\end{equation}

\subsection{The lookback time} 

Imagine a photon emitted by a galaxy at a time $t_{\rm em}$ and detected at $t_0$ on Earth. A very basic notion of distance is $c(t_0 - t_{\rm em})$, i.e., it is the light-travel distance, based on the fact that light always travels at speed $c$.\footnote{We do not enter the discussion of the measure of the \textit{one-way} speed of light and simply adopt Einstein's synchronization convention.} The quantity $t_0 - t_{\rm em}$ is called \textbf{lookback time}\index{Lookback time} and suggestively reminds us of the fact that when we observe some source in the sky, we are actually looking into the past because of the finiteness of $c$.

From the FLRW metric, by inserting $ds^2 = 0$, we can relate the lookback time to the comoving distance as follows: 
\begin{equation}
	cdt = a(t)d\chi\;.
\end{equation}
This seems quite similar to the proper distance, but be careful: the proper distance is defined as $a\chi$ and evidently $ad\chi \neq d(a\chi)$. The lookback time is the photon time of flight and, thus, it cumulatively includes the expansion of the universe. On the other hand, the proper distance is the distance considered between two simultaneous events; therefore, the expansion of the universe is not taken into account cumulatively.

Is there a way to calculate the lookback time from $z$? In principle, yes: one solves the Friedmann equation, finds $a(t)$, inverts this function to find $t = t(a)$, uses $1 + z = 1/a$, and finally obtains a relation $t = t(z)$. 

For example, for the EdS universe, using Eqs.~\eqref{adustflat} and \eqref{EdSage}, one obtains:
\begin{equation}
	1 + z = \left(\frac{2}{3H_0t}\right)^{2/3} \qquad \Rightarrow \qquad t = \frac{2}{3H_0(1 + z)^{3/2}}\;.
\end{equation}
This approach is \textit{model-dependent} because, in order to solve the Friedmann equation, we must know it, and this is possible only if we know or \textit{model} the energy content of the universe.

As we did for the comoving distance in Eq. \eqref{chiexpredshift}, we can expand the lookback time in powers of the redshift. Starting from:
\begin{align}
    t_0 - t = \int_t^{t_0}dt = \int_a^1\frac{da'}{H(a')a'} = \frac{1}{H_0}\int_0^z\frac{dz'}{E(z')(1 + z')}\,,
\end{align}
one obtains:
\begin{align}
    t_0 - t = \frac{1}{H_0}\left[z - \frac{1}{2}(2 + q_0)z^2 + \dots\right]\,.
\end{align}
Here is a direct, model-independent relation between the redshift and the lookback time $(t_0 - t)$.

\subsection{The luminosity distance}

The luminosity distance is a very important notion of distance because it can be measured for standard candles. These are sources whose intrinsic luminosity $L$ is known; therefore, by measuring their flux, a distance can be obtained. See \ref{Sec:cosmicdistladder}. 

Now we characterize the notion of luminosity distance in the cosmological setting, relating it to the energy content of the universe and, thus, to its dynamical properties.

Imagine a source at a certain redshift $z$ with intrinsic luminosity $L = dE/dt$. The observed flux is given by the following formula:
\begin{equation}
	F_0 = \frac{dE_0}{dt_0A_0}\;.
\end{equation}
Here, $A_0$ is the area of the surface on which the radiation is spread:
\begin{equation}
	A_0 = 4\pi r(\chi)^2\;,
\end{equation}
i.e., over a sphere with the proper distance as the radius.\footnote{Recall that $a_0 = 1$, so the comoving distance and the proper distance are equal at $t_0$.} We must use the proper distance because this is the instantaneous distance between the source and the observer at the time of detection. Note that $r(\chi)$ is the transverse comoving distance introduced in Eq. \eqref{transversecomdist}. It is the correct distance to be used here since the comoving line element is written as $d\chi^2 + r(\chi)^2d\Omega^2$.

We do not observe the same photon energy as that which is emitted because photons suffer from cosmological redshift, cf. Eq. \eqref{photongeod}; thus:
\begin{equation}
	\frac{dE}{dE_0} = \frac{1}{a}\;.
\end{equation}
Finally, the rate of detection is different. Indeed, consider a certain number of photons emitted by the source during a specific interval of time $dt$. After being emitted, they cover an infinitesimal comoving distance $d\chi$ related to $dt$ by $dt = a(t)d\chi$. However, the same comoving distance right before detection is related to the infinitesimal time interval $dt_0$ as $dt_0 = d\chi$. Therefore:
\begin{equation}
	\frac{dt}{dt_0} = a\;.
\end{equation}
Putting all the contributions together, we obtain
\begin{equation}
	F = \frac{dE_0}{dt_0A_0} = \frac{a^2dE}{dt4\pi r(\chi)^2} = \frac{dE}{dt4\pi r(\chi)^2(1 + z)^2}\;.
\end{equation} 
Hence, the luminosity distance\index{Luminosity distance} is defined as:
\begin{equation}\label{lumdistdef}
	\boxed{d_{\rm L} \equiv (1 + z)r(\chi)}
\end{equation}
We can now apply what we know about the comoving distance $\chi$ to the luminosity distance as well. In particular, we can employ the expansion \eqref{chiexpredshift}, obtaining:
\begin{equation}\label{lumdistfunzredshift}
	d_{\rm L} = \frac{c}{H_0}\left[z + \frac{1}{2}(1 - q_0)z^2 + \cdots\right]\,.
\end{equation}
Note that in expanding $r(\chi)$, the spatial curvature $K$ enters starting from the third order in $z$. See \cite[page 33]{Weinberg:2008zzc}. 

For standard candles, $d_{\rm L}$ is measured independently via the flux. Therefore, knowing $z$, one can fit the data to this quadratic function for many sources and determine $H_0$ and $q_0$, thereby establishing whether the expansion of the universe is accelerated or not. This procedure is presented in Sec. \ref{Sec:BayesiananalysisSNIa}. Note that $H_0$ is an overall multiplicative factor; thus, it does not determine the shape of the function $d_L(z)$.

\subsection{Angular diameter distance}

\begin{figure}[ht]
\centering
	\begin{tikzpicture}
	\draw (0,0) -- (7,0);
	\draw [line width=2pt] (7,0) -- (7,1);
	\fill (0,0) circle (2pt) node[above left] {O};
	\fill (3.5,0) node[below] {$\chi$};
	\fill (7,0.5) node[right] {S};
	\draw (0,0) -- (7, 1);
	\draw (1,0) arc (0:8:1);
    \draw (1,0) node[below] {$d\phi$};
    \draw (7,0.5) node[left] {$ds$};
\end{tikzpicture}
\caption{Defining the angular diameter distance.}
\label{Fig:angdiamdistfig}
\end{figure}
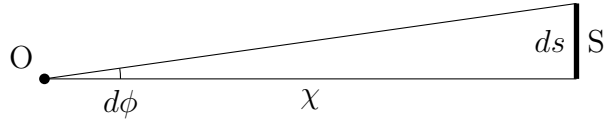

The angular diameter distance is based on the knowledge of proper sizes. Objects with a known proper size are called \textbf{standard rulers}.\index{Standard rulers} Suppose a standard ruler of transverse proper size $ds$ (small) is located at a redshift $z$ and a \textit{radial} comoving distance $\chi$. Moreover, this object has an angular dimension $d\phi$, which is also small. See Fig.~\ref{Fig:angdiamdistfig} for reference.

At a fixed time $t$, we can write the FLRW metric as:
\begin{equation}
	ds^2 = a(t)^2d\chi^2\;.
\end{equation}
The transversal dimension of the object is given by the transverse comoving distance introduced in Eq. \eqref{transversecomdist}, so the transversal proper distance is:
\begin{equation}
	ds = a(t)r(\chi)d\phi\;.
\end{equation}
Dividing the proper dimension of the object by its angular size provides us with the angular diameter distance:\index{Angular diameter distance}
\begin{equation}
	\boxed{d_{\rm A} = \frac{1}{1 + z}r(\chi)}
\end{equation}
Note the relation:
\begin{equation}
	\boxed{d_{\rm L} = (1 + z)^2d_{\rm A}}
\end{equation}
known as \textbf{Etherington's distance duality} \cite{1933PMag...15..761E}.\index{Etherington's distance duality}

Again, we can easily apply the results obtained for the comoving distance to the angular-diameter distance. For the case of a dust-dominated universe, one has:
\begin{equation}
	d_{\rm A} = \frac{2c}{H_0}\left[\frac{1}{1 + z} - \frac{1}{(1 + z)^{3/2}}\right]\;,
\end{equation}
whereas, using the expansion \eqref{chiexpredshift}, one obtains:
\begin{align}\label{dAexpredshift}
    d_{\rm A} = \frac{c}{H_0}\left[z - \frac{1}{2}\left(3 + q_0\right)z^2 + O(z^3)\right]\,.
\end{align}
In the limit of small $z$, we find $d_{\rm A} \sim cz/H_0$. All the distances that we have defined thus far coincide in the first order expansion in $z$.

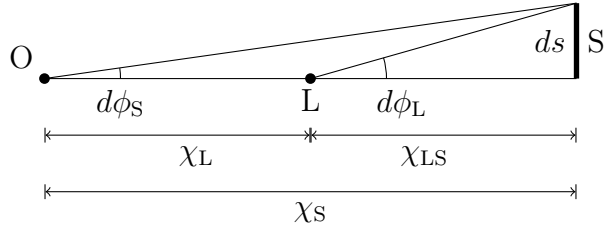
\begin{figure}[ht]
\centering
	\begin{tikzpicture}
	\draw (0,0) -- (7,0);
	\draw[|<->|] (0,-1.5) -- (7,-1.5);
	\draw[|<->|] (0,-0.75) -- (3.5,-0.75);
	\draw[|<->|] (3.5,-0.75) -- (7,-0.75);
	\draw [line width=2pt] (7,0) -- (7,1);
	\draw (2,-0.75) node[below] {$\chi_{\rm L}$};
	\draw (5,-0.75) node[below] {$\chi_{\rm LS}$};
	\draw (3.5,-1.5) node[below] {$\chi_{\rm S}$};
	\fill (0,0) circle (2pt) node[above left] {O};
	\fill (3.5,0) circle (2pt) node[below] {L};
	\fill (7,0.5) node[right] {S};
	\draw (3.5,0) -- (7, 1);
	\draw (0,0) -- (7, 1);
	\draw (4.5,0) arc (0:16:1);
	\draw (4.7,0) node[below] {$d\phi_{\rm L}$};
	\draw (1,0) arc (0:8:1);
    \draw (1,0) node[below] {$d\phi_{\rm S}$};
    \draw (7,0.5) node[left] {$ds$};
\end{tikzpicture}
\caption{The angular diameter distance between two different redshifts.}
\label{Fig:angdiamdistfigtworeds}
\end{figure}

In gravitational lensing applications, it is often necessary to know the angular-diameter distance between two sources at different redshifts (i.e., the angular-diameter distance between the lens and the background source). In order to compute this, let us refer to Fig.~\ref{Fig:angdiamdistfigtworeds}.

The problem is to determine the angular-diameter distance between L and S, say $d_{\rm A}(\rm LS)$. Is this the difference between the angular-diameter distances $d_{\rm A}(\rm S) - d_{\rm A}(\rm L)$? We now show that this is not the case. Simple trigonometry is sufficient to establish that:
\begin{equation}
	ds = a(t_{\rm S})r(\chi_{\rm S})d\phi_{\rm S} = a(t_{\rm S})r(\chi_{\rm LS})d\phi_{\rm L}\;,
\end{equation} 
And for the line-of-sight comoving distances, we do have that $\chi_{\rm LS} = \chi_{\rm S} - \chi_{\rm L}$. Therefore, we have:
\begin{equation}
	\boxed{d_{\rm A}({\rm LS}) = a(t_{\rm S})r(\chi_{\rm LS}) = a(t_{\rm S})r(\chi_{\rm S} - \chi_{\rm L}) = \frac{1}{1 + z_{\rm S}}r(\chi_{\rm S} - \chi_{\rm L})}
\end{equation}
For $K < 0$, according to our definition \eqref{transversecomdist}, we have, using the sum formula for the hyperbolic sine:
\begin{align}
	r(\chi_{\rm S} - \chi_{\rm L}) = \frac{1}{\sqrt{|K|}}\nonumber\\
    \left[\sinh\left(\sqrt{|K|}\chi_{\rm S}\right)\cosh\left(\sqrt{|K|}\chi_{\rm L}\right) - \cosh\left(\sqrt{|K|}\chi_{\rm S}\right)\sinh\left(\sqrt{|K|}\chi_{\rm L}\right)\right]\,.
\end{align}
Using the fundamental relation between the hyperbolic functions, one obtains:
\begin{align}
	r(\chi_{\rm S} - \chi_{\rm L}) = \left[r(\chi_{\rm S})\sqrt{1 + |K|r(\chi_{\rm L})^2} - r(\chi_{\rm L})\sqrt{1 + |K|r(\chi_{\rm S})^2}\right]\,.
\end{align}
This formula is obtained in \cite{Peebles:1994xt} and reported in \cite{Hogg:1999ad}.\footnote{In \cite{Hogg:1999ad}, a comment states that, according to private communication, the formula cannot be extended to $K > 0$. However, in the thread \url{https://github.com/astropy/astropy/issues/4661}, such a comment is retracted.} 

For $K > 0$, according to our definition \eqref{transversecomdist}, we have, using the sum formula for the sine, the following:
\begin{align}
	r(\chi_{\rm S} - \chi_{\rm L}) = \frac{1}{\sqrt{K}}\nonumber\\
    \left[\sin\left(\sqrt{K}\chi_{\rm S}\right)\cos\left(\sqrt{K}\chi_{\rm L}\right) - \cos\left(\sqrt{K}\chi_{\rm S}\right)\sin\left(\sqrt{K}\chi_{\rm L}\right)\right]\,.
\end{align}
Using the fundamental relation between the trigonometric functions, one obtains:
\begin{align}
	r(\chi_{\rm S} - \chi_{\rm L}) = \left[r(\chi_{\rm S})\sqrt{1 - K r(\chi_{\rm L})^2} - r(\chi_{\rm L})\sqrt{1 - K r(\chi_{\rm S})^2}\right]\,.
\end{align}
For any $K$, we can thus conclude that:
\begin{equation}
	\boxed{d_{\rm A}({\rm LS}) = \frac{1}{1 + z_{\rm S}}\left[r(\chi_{\rm S})\sqrt{1 - K r(\chi_{\rm L})^2} - r(\chi_{\rm L})\sqrt{1 - K r(\chi_{\rm S})^2}\right]}
\end{equation}
Apart from gravitational lensing on cosmological scales, the angular diameter distance is also important for the analysis of the Baryon Acoustic Oscillations (BAO), as they provide a standard ruler (albeit statistically).

\clearpage
\chapter{Kinetic theory in the expanding universe}\label{Chap:KinTh}

{\rightskip=3truepc\leftskip=3truepc\noindent
{\it L'umanit\`a non sopporta il pensiero che il mondo sia nato per caso, per sbaglio. Solo perch\'e quattro atomi scriteriati si sono tamponati sull'autostrada bagnata\\
(Humanity cannot bear the thought that the world was born by accident, by mistake. Just because four mindless atoms crashed on the wet highway)}
\vskip 0.10 in
\centerline{\it ---Umberto Eco, Il Pendolo di Foucault}
\vskip 0.20 in
}

We have seen in the previous chapters the evidence for an expanding universe and how to address this within the framework of General Relativity. Now we want to accompany this expansion from primordial times, a few instants after the Big Bang, up to the present, in order to understand how the abundance of light elements is fixed and from where the cosmic microwave background originates.

In order to treat these problems, we need to equip ourselves with kinetic theory and the Boltzmann equation as applied to cosmology. The classical references for these topics are the book by Kolb and Turner \cite{Kolb:1990vq} and the monograph by Bernstein \cite{1988kteu.book.....B}.

\section{Thermal equilibrium in cosmology}

Let us try to apply thermodynamical concepts in cosmology by first recalling the continuity equation \eqref{Encons} for a perfect fluid in a space described by the Friedmann metric:
\begin{equation}\label{enconsepsilon}
	\dot\varepsilon + 3H(\varepsilon + P) = 0\;.
\end{equation}
As usual, the dot represents derivation with respect to cosmic time. 

It is possible to derive this equation from thermodynamics. For an isolated system, as the universe should be, the first law of thermodynamics states that:
\begin{equation}\label{thermorel}
	0 = PdV + dU\;.
\end{equation}
By taking $V$ as the physical volume (which includes the cube of the scale factor) and $U$ as the internal energy of the cosmic fluid, and writing $U = \varepsilon V$, one obtains:
\begin{equation}\label{dVeq}
	PdV + d(\varepsilon V) = 0 \qquad \Rightarrow \qquad Vd\varepsilon + (\varepsilon + P)dV = 0\;.
\end{equation}

\hrulefill

\begin{ex} Since $V \propto a^3$, show that Eq.~\eqref{dVeq} leads to the the continuity equation \eqref{enconsepsilon}.\index{Continuity equation!Using thermodynamics} Such thermodynamic reformulation of the continuity equation was noted by Lema\^{\i}tre \cite{Lemaitre:1927zz}.
\end{ex}

\hrulefill 

The above calculation works for the total density and pressure; however, in general, it does not apply to the individual components if they interact, as is expected for particles such as photons, electrons, and protons in the early universe. On the other hand, we also expect that, in the past, our universe was sufficiently dense for the various interacting species to attain \textbf{thermal equilibrium} and share a common temperature. At this stage, their particle distributions can be taken as the Maxwell-Boltzmann, Fermi-Dirac, or Bose-Einstein distributions.

Indeed, consider a scattering between two particle species:
\begin{align}
    1 + 2 \longrightarrow 1' + 2'\,,
\end{align}
where the primes denote different energies and momenta.

The \textbf{interaction rate}\index{Interaction rate} $\Gamma$ is defined as:
\begin{equation}\label{intratedef}
	\boxed{\Gamma \equiv n\sigma v_{\rm rel}}
\end{equation}
where $n$ is the particle number density of the projectiles (of either of the two particles at the position of the other), $v_{\rm rel}$ is the relative velocity between projectiles and targets (hence $nv_{\rm rel}$ is the flux), and $\sigma$ is the cross section. The latter is established by the theory of the interaction among the particles under examination. 

In an expanding universe $n \propto a^{-3}$, whereas $\sigma$ and $v_{\rm rel}$ are related to the energies and momenta of the particles. So, for very small scale factors, one expects $\Gamma$ to be very large. If it so happens that: 
\begin{equation}\label{TEcond}
	\boxed{\Gamma \gg H}
\end{equation}
then we have \textbf{thermal equilibrium}. Roughly speaking, Eq.~\eqref{TEcond} can be rephrased as the fact that the mean-free path (which is $\propto 1/\Gamma$) is much smaller than the Hubble radius (which is $\propto 1/H$). In this situation, particles interact so frequently that they do not ``feel'' the cosmological expansion, and any fluctuation in their energy density is rapidly smoothed out, thus recovering thermal equilibrium. On the other hand, the universe expands, and the equilibrium can thus only be approximate or local. Hence, we will deal with \textbf{local thermal equilibrium} (LTE), where local is intended with respect to time. 

It is important to make a distinction between \textbf{kinetic equilibrium}\index{Kinetic equilibrium} and \textbf{chemical equilibrium}.\index{Chemical equilibrium} When $\Gamma \gg H$ refers to a process of the type:
\begin{equation}\label{1234reaction}
	1 + 2 \longleftrightarrow 3 + 4\;,
\end{equation}
i.e., we have four different particle species that transform into each other in a balanced way; thus, we have \textbf{chemical equilibrium}. This can also be reformulated as:
\begin{equation}\label{chemicalequilibrium}
	\boxed{\mu_1 + \mu_2 = \mu_3 + \mu_4}
\end{equation}
where the $\mu$'s are the chemical potentials. 

On the other hand, when $\Gamma \gg H$ refers to a reaction such as scattering:
\begin{equation}
	1 + 2 \longleftrightarrow 1 + 2\;,
\end{equation}
then we have \textbf{kinetic equilibrium}. Kinetic or chemical equilibrium (or both) implies thermal equilibrium. In general, it is possible for a species to break chemical equilibrium and still remain in kinetic equilibrium with the rest of the cosmic plasma through scattering processes.\footnote{This occurs in some DM particle models. For example, a $m = 100$ GeV WIMP chemically decouples at 5 GeV and kinetically decouples at 25 MeV. See, e.g., \cite{Profumo:2006bv}.}

Note that $\Gamma$ is different for different interactions and different particle species. Therefore, the above condition \eqref{TEcond} is generally valid for all known (and perhaps unknown) particles in the early universe, but it is broken at different times for different species. This is the essence of the \textbf{thermal history of the universe}.\index{Hot Big Bang!Thermal history}

So, at the very beginning (we are talking about tiny fractions of a second after the Big Bang), all the particles were in thermal equilibrium in a ``primordial soup'', also known as the \textbf{primordial plasma}.\index{Primordial plasma} A brutal dimensional analysis shows that since $n \propto a^{-3}$, then also:
\begin{align}
    \Gamma \propto a^{-3}\,,
\end{align}
whereas:
\begin{align}
    H \propto a^{-2}\,,
\end{align}
during the radiation-dominated epoch. Therefore, the condition:
\begin{align}
    \boxed{\Gamma \sim H}
\end{align}
is eventually reached. When this occurs, the LTE approximation fails, and we need to follow the evolution of the particle species via kinetic theory and the Boltzmann equation. From the latter, we will see that Eq.~\eqref{enconsepsilon} can be obtained by assuming a very high rate of interactions. Interestingly, Eq.~\eqref{enconsepsilon} can also be obtained if $\Gamma = 0$, i.e., if there are no interactions at all. When, for a species, the condition $\Gamma \sim H$ is reached, it \textbf{decouples} from the primordial plasma. If it does this by breaking the chemical equilibrium, then it is said to \textbf{freeze-out}\index{Freeze out} and attains some fixed abundance.\footnote{It attains a fixed abundance if it is a stable particle, of course. If not, it disappears.}

When we want to explicitly calculate the residual abundance of some species, we have to track its evolution until $\Gamma\sim H$. In this instance, equilibrium thermodynamics fails, and we must use the Boltzmann equation. For example, we shall use the Boltzmann equation when analyzing: $i)$ the formation of light elements during the BBN; $ii)$ the recombination of protons and electrons in neutral hydrogen atoms; $iii)$ the relic abundance of CDM. 

The fundamental interactions that characterize the above-mentioned processes compel some particles to react and transform into others and vice-versa, as shown in Eq.~\eqref{1234reaction}. When $\Gamma \gg H$ these reactions take place with equal probability in both directions, hence the $\longleftrightarrow$ symbol; however, when $\Gamma\sim H$ eventually one direction is preferred over the other. This is the characteristic of irreversibility that demands the use of the Boltzmann equation.

Finally, we will see that the primordial plasma will remain composed of photons only: the cosmic microwave background. Since, as we will see, photons maintain a thermal distribution throughout cosmic history, it is their temperature to which one usually refers as the temperature of the universe up to the present.

\subsection{The entropy density}

When the LTE approximation is valid, we can introduce the entropy as a function of the state and describe the evolution of the species in thermal equilibrium as an adiabatic process, that is, with $dS = 0$. 

In order to describe the variation in temperature when some species decouple, it is useful to introduce the entropy density $s \equiv S/V$. Since $V \propto a^3$, we can write $dS = 0$ as:
\begin{equation}\label{entropydensityconservation}
	\boxed{\frac{d}{dt}(sa^3) = 0}
\end{equation}
Now we find a more convenient way to express $s$, which will be useful in the forthcoming applications. 

Assuming thermal equilibrium, we have $\varepsilon = \varepsilon(T)$, $P = P(T)$, and we expect $s = s(T)$. Therefore:
\begin{equation}
	U = \varepsilon(T)V\,,
\end{equation} 
and:
\begin{align}
    TdS = Td(sV) = TV\frac{ds}{dT}dT + TsdV\,,
\end{align}
and:
\begin{align}
    Vd\varepsilon + (\varepsilon + P)dV = V\frac{d\varepsilon}{dT}dT + (\varepsilon + P)dV\,.
\end{align}
Putting these together:
\begin{equation}\label{thermorel2}
	TV\frac{ds}{dT}dT + TsdV = V\frac{d\varepsilon}{dT}dT + (\varepsilon + P)dV\;.
\end{equation}
Hence, equating the coefficients of $dV$ yields:
\begin{equation}\label{entropydensity}
	\boxed{s = \frac{\varepsilon + P}{T}}
\end{equation}
Using this result, equating the coefficients of $VdT$ in Eq. \eqref{thermorel2}, we obtain a useful reformulation of the continuity equation:
\begin{equation}
    \boxed{T\frac{dP}{dT} = \varepsilon + P}
\end{equation}
now involving the derivative with respect to the temperature. For example, for radiation ($P = \varepsilon/3$) one has:
\begin{align}
    T\frac{d\varepsilon}{dT} = 4\varepsilon\,,
\end{align}
from which one obtains:
\begin{align}
    \varepsilon \propto T^4\,,
\end{align}
the well-known result of the Stefan-Boltzmann law.

Finally, note that, taking into account the chemical potential $\mu$, the entropy density is defined as
\begin{equation}
	\boxed{s = \frac{\varepsilon + P - \mu n}{T}}
\end{equation}

\section{Short summary of the thermal history of the universe}

We present a brief list of the main events characterizing the primordial universe.

\paragraph{Planck energy, inflation and Grand Unified Theory.} The Planck energy is usually considered an upper threshold in the energy beyond which quantum gravity becomes mandatory. The Planck energy can be obtained by suitably combining the fundamental constants $c$, $G$, and $\hbar$:
\begin{equation}
	E_{\rm Pl} = M_{\rm Pl}c^2 = \sqrt{\frac{\hbar c}{G}}c^2 = 10^{19} \mbox{ GeV}\;,
\end{equation} 
where the Planck mass $M_{\rm Pl}$ is implicitly defined. Another combination of the fundamental constants returns the Planck time:
\begin{equation}
	t_{\rm Pl} = \sqrt{\frac{\hbar G}{c^5}} = 10^{-43} \mbox{ s}\;,
\end{equation}
and, of course, multiplying it by $c$ gives us the Planck length $l_{\rm Pl} = 10^{-35}$ m. 

Inflation is a very important aspect of the current description of the primordial universe, and we shall dedicate Chapter~\ref{Chap:Inflation} to it. For now, it is enough to say that it is expected to occur at an energy scale of the order $10^{16}$ GeV, which is the same as that of the Grand Unified Theory (GUT), i.e., a model in which electromagnetic, weak, and strong interactions are unified.\index{Grand unified theory (GUT)}

\paragraph{Baryogenesis and Leptogenesis.} We know that antimatter exists, but we also know that matter is the most abundant substance in the universe. If they were produced in exactly the same quantity, they should have annihilated almost completely when in equilibrium in the primordial plasma, leaving only photons. See Sec. \ref{Sec:relicbaryonselectrons}. Instead, we have a small but non-vanishing baryon-to-photon ratio $\eta_{\rm b} = 5.5\times 10^{-10}$.\index{Baryon-to-photon ratio} 

Baryogenesis\index{Baryogenesis} is the creation, via a still unclear mechanism, of a positive baryon number. In other words, the creation of a primordial quark-antiquark asymmetry, by virtue of which protons and neutrons are much more common than anti-protons and anti-neutrons, is suggested. The original idea and necessary conditions for an asymmetry in the creation of matter and anti-matter in the early universe were put forward in 1967 by A. D. Sakharov \cite{Sakharov:1967dj}. See also \cite{Dolgov:1997qr}.  

Leptogenesis is a class of scenarios in which the cosmic baryon asymmetry originates from an initial
lepton asymmetry generated in the decays of heavy sterile neutrinos in the early Universe. See \cite{Fukugita:1986hr} for the original proposal and \cite{Davidson:2008bu} for a review of the various proposals of Leptogenesis.\index{Leptogenesis}

In order to maintain the neutrality of the universe, Baryogenesis and Leptogenesis must include a mechanism that produces a non-vanishing lepton number in the form of an excess of electrons over positrons. Such an excess would presumably also be present for neutrinos over anti-neutrinos. On the other hand, both Baryogenesis and Leptogenesis are processes concerning massive particles. In these notes, we assume neutrinos and anti-neutrinos to be massless and present in the same abundance. Since they do not annihilate, we must take both into account in the cosmic energy budget while neglecting anti-protons, anti-neutrons, and positrons.

\paragraph{Electroweak phase transition.}\index{Electroweak phase transition} At thermal energies of about 200 GeV, which correspond to $10^{-12}$ seconds after the Big Bang, the electromagnetic and weak forces start to behave distinctly. This occurs because the vector bosons $W^\pm$ and $Z^0$ gain their masses, which are roughly 80 and 90 GeV, respectively, through the Higgs mechanism, and the weak interaction ``weakens'' since it is now governed by the Fermi constant, whose value is as follows \cite{Mohr:2024kco}:
\begin{equation}\label{GFermivalue}
	\frac{G_{\rm F}}{(\hbar c)^3} = 1.166 3787(6) \times 10^{-5}\; \mbox{GeV}^{-2}\;. 
\end{equation}
Fermions might also acquire their masses through the Higgs mechanism, assuming a coupling of them with the Higgs field (the Yukawa coupling). If this is so, then before the electroweak phase transition, the universe was populated only by effectively massless species.

\paragraph{Decoupling of the Top quark.} The Top quark is the most massive of all observed fundamental particles, $m_{\rm Top}c^2 = 173$ GeV, so it is probably the first to decouple from the primordial plasma. For this reason, it is included on this list.\index{Top quark}

\paragraph{QCD phase transition.} Below 150 MeV, quarks transition from their asymptotic freedom to bound states composed of two (mesons) or three (baryons) particles. The above energy corresponds roughly to 20 $\mu$s after the Big Bang.

\paragraph{DM freeze-out.}\index{Dark Matter!Decoupling} With freeze-out, one usually refers to an epoch when particle-antiparticle annihilation becomes inefficient due to the expansion of the universe, and thus the particle abundance becomes fixed. For the neutralino case, the freeze-out occurs at approximately 25 MeV.

\paragraph{Neutrino decoupling.}\index{Neutrino decoupling} Neutrinos maintain thermal equilibrium with the primordial plasma through interactions such as:
\begin{equation}\label{neutrinosinteractions}
	p + e^- \leftrightarrow n + \nu\;, \qquad p + \bar\nu \leftrightarrow n + e^+\;, \qquad n \leftrightarrow p + e^- + \bar{\nu}\;,
\end{equation}
down to a thermal energy of 1 MeV, which corresponds roughly to 1 second after the Big Bang. Below this energy threshold, they decouple. We can roughly calculate the 1 MeV scale of decoupling in the following way. Since neutrinos are relativistic, we can use Eq.~\eqref{intratedef} with $v_{\rm rel} = c$, obtaining:
\begin{equation}
	\frac{\Gamma}{H} = \frac{n_\nu\sigma c}{H} \approx \frac{G_{\rm F}^2}{(\hbar c)^6}M_{\rm Pl}c^2(k_{\rm B}T)^3 \approx \left(\frac{k_{\rm B}T}{1\mbox{ MeV}}\right)^3\;,
\end{equation} 
where we have used some results that we shall prove later, such as that the particle number density $n_\nu$ goes as $T^3$ and $H \propto T^2/M_{\rm Pl}$. When $k_{\rm B}T \approx 1\mbox{ MeV}$, decoupling occurs.

We have just used the \textbf{effective field theory} of the weak interaction in assuming $\sigma \propto G_{\rm F}^2T^2$. An effective field theory is a low energy approximation of the full theory. Indeed, the cross section $\sigma \propto G_{\rm F}^2T^2$ diverges at high energies, and this is unphysical. In the case of the weak interaction, we have chosen to work at an energy scale much smaller than 80 GeV, which is the mass of the vector boson $W^\pm$ that mediates the weak interaction. In this approximation, the boson vectors $W^\pm$ and $Z^0$ have infinite masses; therefore, the range of the weak interaction is zero. In other words, we have considered the interactions of Eq.~\eqref{neutrinosinteractions} as if they occurred in a point. The result 1 MeV $\ll$ 80 GeV is consistent with the effective field theory approximation and, thus, is reliable.

Another approximation that we have used is the masslessness of neutrinos. Even considering a small mass $m_\nu c^2 \lesssim 0.1$ eV, the above calculation is solid since $m_\nu c^2 \ll k_{\rm B}T \sim 1$ MeV. The condition $m c^2 \ll k_{\rm B}T$ for a generic particle of mass $m$ in thermal equilibrium guaranties that such a particle is relativistic.

Other kinds of interactions involving neutrinos are those of annihilation, such as:
\begin{equation}\label{nuantinuannihil}
	e^- + e^+ \leftrightarrow \nu_e + \bar\nu_e\;.
\end{equation}
These are also no more efficient on energy scales below 1 MeV. 

\paragraph{Electron-positron annihilation.}\index{Electron-positron annihilation} The interaction 
\begin{equation}
	e^- + e^+ \leftrightarrow \gamma + \gamma\;,
\end{equation}
is balanced for energies higher than the mass of electrons and positrons $m_ec^2 = 511$ keV. When the temperature of the thermal bath drops below this value, pair production is no longer possible, and annihilation takes over. The cross section for pair production is \cite{Berestetsky:1982aq}:
\begin{equation}\label{pairprodsigma}
	\sigma_{\gamma\gamma} = \frac{3}{16}\sigma_{\rm T}(1 - v^2)\left[(3 - v^4)\ln\left(\frac{1+v}{1-v}\right) - 2v(2-v^2)\right]\;,
\end{equation}
where the parameter $v$ is defined as follows:
\begin{eqnarray}\label{pairprodcond}
	v \equiv \sqrt{1 - (m_ec^2)^2/(\hbar^2\omega_1\omega_2)}\;;
\end{eqnarray}
the \textbf{Thomson cross section}\index{Thomson scattering!Cross section} is
\begin{equation}
	\sigma_{\rm T} = \frac{8\pi}{3}\left(\frac{\alpha\hbar c}{m_ec^2}\right)^2 \approx 66.52 \mbox{ fm}^2\;,
\end{equation}
with $\alpha \approx 1/137$ being the \textbf{fine structure constant}.\index{Fine structure constant} Finally, $\omega_1$ and $\omega_2$ are the frequencies of the two photons. Being in a thermal bath, we have that $\hbar\omega_1 \approx \hbar\omega_2 \approx k_{\rm B}T$. From Eq.~\eqref{pairprodcond}, it is then clear that the condition
\begin{equation}
	\hbar^2\omega_1\omega_2 \approx (k_{\rm B}T)^2 > (m_ec^2)^2\;,
\end{equation}
must be satisfied in order to produce pairs. Therefore, when the temperature of the thermal bath drops below the value of the electron mass, annihilation becomes the only relevant process. Thanks to Leptogenesis, positrons disappear, and only electrons remain. 

Here, we can comment on the following. If chemical equilibrium holds, then:
\begin{equation}
	\mu_{e^-} + \mu_{e^+} = 2\mu_\gamma\;.
\end{equation}
If the lepton number is conserved, then $\mu_{e^-} =  -\mu_{e^+}$, and thus the photon chemical potential is zero.

\paragraph{Big Bang Nucleosynthesis (BBN).} We shall discuss BBN in great detail in Sec.~\ref{Sec:BBN}. It occurs at about 0.1 MeV (some three minutes after the Big Bang) as deuterium and Helium form.

\paragraph{Recombination and photon decoupling.} Protons and electrons form neutral hydrogen at about 0.3 eV. Having no more free electrons with which to scatter, photons decouple and can be seen today as the CMB. We shall study this process in detail in Sec.~\ref{Sec:Recombination}.

\section{The phase space and the distribution function}

Consider a test particle whose trajectory is given by $x^\mu(\lambda)$, where $\lambda$ is an affine parameter. The covariant momentum is obtained as:
\begin{equation}
	P_\mu = \frac{dx_\mu}{d\lambda}\;,
\end{equation}
and it is the conjugate momentum for the position $x^\mu$, in the sense that:
\begin{equation}
	P_\mu = \frac{\partial L}{\partial (dx^\mu/d\lambda)}\;,
\end{equation}
where $L$ is the Lagrangian for a test particle, which is:
\begin{equation}
	L = \frac{1}{2}g_{\mu\nu}\frac{dx^\mu}{d\lambda}\frac{dx^\nu}{d\lambda}\;.
\end{equation}
The canonical quantities $(x^i,P_j)$ describe the \textbf{phase space}. Note that $P_0$ is related to $P_i$ via the relation $g^{\mu\nu}P_\mu P_\nu = -m^2c^2$, so it is not an independent degree of freedom.

The evolution of an ensemble of particles is statistically described by a \textbf{distribution function} $f(x^0, x^i, P_j)$ of the phase space variables and time. In particular, the number of particles in a small phase space volume element at a given time coordinate $x^0$ is:
\begin{equation}
	dN = dx^1dx^2dx^3dP_1dP_2dP_3f(x^0, x^i,P_j)\;.
\end{equation}
This quantity must not depend on the coordinates chosen to describe the ensemble. Indeed, $f$ is a scalar, and the volume elements can be expressed in the following form:
\begin{equation}
	dN = dx^1dx^2dx^3P^0\sqrt{-g}\frac{dP_1dP_2dP_3}{P^0\sqrt{-g}}f(x^i,P_j,x^0)\;.
\end{equation}
Each of the two volume elements $dx^1dx^2dx^3P^0\sqrt{-g}$ and $\frac{dP_1dP_2dP_3}{P^0\sqrt{-g}}$ is invariant.

\hrulefill

\begin{ex}
	Prove the latter claim. 
\end{ex}

\hrulefill

Because of the Heisenberg uncertainty principle, no particle can be localized in a point $(x^i,P_j)$ in phase space (the very notion of a ``point'' particle loses its meaning in quantum physics); rather, it can be localized only in a small volume $\mathcal{V} = h^3$ around that point, where $h$ is the Planck constant. Therefore, the actual volume element that we use is the following:
\begin{equation}
	dN = \frac{dx^1dx^2dx^3dP_1dP_2dP_3}{h^3}f(x^i,P_j,x^0)\;.
\end{equation}
Note the dimensions: $h$ has dimensions of energy $\times$ time, or momentum $\times$ space. Hence, $f$ is dimensionless. In the following, when possible, we will incorporate $h^3$ within $f$ in order to avoid unnecessarily burdening the notation.

For a given configuration in the phase space, we might still have more particle states because, for example, of the spin of the particle species. We will indicate this degeneracy with $g_{\rm s}$, and, when possible, we will incorporate it into $f$ as well.

In a cosmological setting, enforcing the cosmological principle, the distribution function cannot depend on $x^i$, nor on the direction $\hat{P}_j$ of the momentum, but only on its modulus. Recalling the definition~\eqref{propermomentum} of the square modulus of the proper momentum:
\begin{equation}
	p^2 = g_{ij}P^iP^j = g^{ij}P_iP_j = \frac{1}{a^2}\gamma^{ij}P_iP_j = \frac{1}{a^2}P^2\;,
\end{equation}
where $\gamma^{ij}P_iP_j \equiv P^2$, we have:
\begin{align}
    f = f(x^0,P)\,, \qquad \mbox{or} \qquad f = f(x^0,p)\,.
\end{align}
We can also choose $P^0$ instead of $P$ or $p$ if this provides some convenience. 

Given such a restricted functional form, it is then useful to express the volume element of momentum space in spherical coordinates, that is, using the modulus of the momentum:
\begin{equation}
	P_i = P\sqrt{B}\hat P_i\,, \qquad \delta^{ij}\hat P_i\hat P_j = 1\,, \qquad \gamma^{ij} = \frac{1}{B}\delta^{ij}\,, \qquad \sqrt{\gamma} = B^{3/2}\,,
\end{equation}
where we have chosen isotropic spatial comoving coordinates. For the proper momentum, we have:
\begin{equation}
	p_i = p\hat p_i\,, \qquad \delta^{ij}\hat p_i\hat p_j = 1\,, \qquad p = P/a\,.
\end{equation}
Note that $\hat P_i = \hat p_i$. The volume element is then:
\begin{equation}
	dP_1dP_2dP_3 = B^{3/2}P^2dPd^2\hat{P}\;,
\end{equation}
where $d^2\hat{P}$ is the solid angle element in momentum space.

\subsection{Volume of the fundamental cell}

Why is the volume of the fundamental cell equal to $h^3$? We will see that this value is very important for correctly calculating the abundances of the energy components of the universe. For example, if that volume were $2h^3$ instead of $h^3$, we would estimate half of the present abundance of photons; thus, we must get it right. Let us prove that $\mathcal V = h^3$.

Consider a quantum particle confined in a cube of side $L$. The eigenfunctions $u_\textbf{p}(\textbf{x})$ of the momentum operator $\hat{\textbf{p}}$ are determined by the equation:
\begin{equation}\label{momeigen}
	\hat{\textbf{p}}u_\textbf{p}(\textbf{x}) = \textbf{p}u_\textbf{p}(\textbf{x}) \qquad \Rightarrow \qquad -i\hbar\nabla u_\textbf{p}(\textbf{x}) = \textbf{p}u_\textbf{p}(\textbf{x})\;.
\end{equation}
Assuming variable separation,\index{Phase-space!Volume of the fundamental cell} 
\begin{equation}
	u_\textbf{p}(\textbf{x}) = u_x(x)u_y(y)u_z(z)\;,
\end{equation}
Eq.~\eqref{momeigen} is easily solved:
\begin{equation}\label{solmomeigen}
	u_\textbf{p}(\textbf{x}) = \frac{1}{L^{3/2}}\exp\left(\frac{i\textbf{p}\cdot\textbf{x}}{\hbar}\right)\;,
\end{equation}
where the factor $1/L^{3/2}$ comes from the normalization of the eigenfunction, which has an integrated square modulus equal to 1 in the box.

\hrulefill

\begin{ex} 
Prove the result \eqref{solmomeigen}.	
\end{ex}

\hrulefill

Since we have used variable separation, from now on, let us focus only on the $x$ dimension for simplicity. The particle is constrained to be in the box, so we must impose periodic boundary conditions in Eq.~\eqref{solmomeigen}:\index{Periodic boundary conditions}
\begin{equation}
	u_{x}(0) = u_{x}(L)\;.
\end{equation}
These conditions imply that the momentum is quantized, i.e.,
\begin{equation}\label{quantmom}
	\frac{p_xL}{\hbar} = 2\pi n_x\;,
\end{equation}
where $n_x \in \mathbb{Z}$.

\hrulefill

\begin{ex} 
Prove the result \eqref{quantmom}.	
\end{ex}

\hrulefill

The phase space occupied by a single state is thus:
\begin{equation}
	\Delta (p_x L) = 2\pi\hbar = h\;,
\end{equation}
By recovering the three dimensions, we obtain the expected result
\begin{equation}
	\mathcal{V} = \Delta (p_x L)\Delta (p_y L)\Delta (p_z L) = h^3\;.
\end{equation}

\subsection{Momentum integrals of the distribution function}

Integrating physical quantities weighted with the distribution function over momenta yields the average properties of the ensemble of particles, thus simplifying the dynamical treatment of a large number of degrees of freedom. 

For example, consider \textbf{the particle number current}:
\begin{equation}\label{particlenumbercurrent}
	\boxed{N^\mu(x^0,x^i) = \int d_3\mathbf P\frac{P^\mu}{\sqrt{-g}P^0}f(x^0,x^i,P_j)}
\end{equation}
where $d_3\mathbf P \equiv dP_1dP_2dP_3$. It is essentially the averaged four-momentum of the particle distribution. Averaging two momenta instead, we obtain \textbf{the energy-momentum tensor}:
\begin{equation}\label{enmomtensdistrfun}
	\boxed{T^\mu{}_\nu(x^0,x^i) = \int d_3\mathbf P\frac{cP^\mu P_\nu}{\sqrt{-g}P^0}f(x^0,x^i,P_j)}
\end{equation}
The volume elements, as we have seen earlier, are invariant, and so $N^\mu$ and $T^\mu{}_\nu$ are tensors. Note that $P^0$ is always determined by the relation $g_{\mu\nu}P^\mu P^\nu = -m^2c^2$. Using the FLRW metric \eqref{FLRWmet}, we have:
\begin{equation}
	\frac{dP_1dP_2dP_3}{\sqrt{-g}P^0} = \frac{dP_1dP_2dP_3}{a^3\sqrt{\gamma}P^0} = \frac{P^2B^{3/2}dPd^2\hat P}{a^3B^{3/2}P^0} = \frac{P^2dPd^2\hat P}{a^3P^0}\,,
\end{equation}
so that:
\begin{equation}
	N^\mu(t) = \frac{1}{a(t)^3}\int P^2dPd^2\hat{P} \frac{P^\mu}{P^0}f(t,P)\,.
\end{equation}
The particle number density is then:\index{Distribution function!Particle number density}
\begin{equation}\label{partnumdens}
	n(t) = N^0 = \frac{1}{a(t)^3}\int P^2dPd^2\hat Pf(t,P) = \frac{4\pi}{a(t)^3}\int P^2dPf(t,P)\;,
\end{equation}
The standard $n(t) \propto 1/a(t)^3$ result is then recovered. On the other hand, $N^i = 0$ because:

\hrulefill

\begin{ex} Show that:
\begin{equation}
	\int d^2\hat P \hat P_i = 0\,.
\end{equation}
\end{ex}

\hrulefill

This is also expected because, due to the cosmological principle, we cannot have any flux of particles. Since $P = ap$, the particle number density can also be expressed using the proper momentum as:
\begin{equation}\label{numdenspropmom}
	\boxed{n(t) = \int d^3\textbf{p}f(t,p)}
\end{equation}
Let us now write the energy-momentum tensor Eq.~\eqref{enmomtensdistrfun} for an FLRW background. With $P^0 = E/c$, as in Eq. \eqref{EecP0}, Eq.~\eqref{enmomtensdistrfun} becomes:
\begin{equation}\label{enmomtensdistrfun2}
	T^\mu{}_\nu(t) = \frac{1}{a(t)^3}\int P^2dPd^2\hat P\frac{c^2P^\mu P_\nu}{E}f(t,P)\;.
\end{equation}
For $\mu = \nu = 0$:
\begin{equation}
	T^0{}_0(t) = \frac{1}{a(t)^3}\int P^2dPd^2\hat PcP_0f(t,P) = -\frac{4\pi}{a(t)^3}\int P^2dPEf(t,P)\;.
\end{equation}
The energy density in the rest frame of an observer with 4-velocity $u^\mu$ is given by $T^\mu{}_\mu u_\mu u^\nu/c^2 = T^0{}_0u_0u^0/c^2 = -T^0{}_0$. So, there are no worries about the minus sign. Note that we recover the $1/a^3$ scaling for non relativistic particles $m^2c^2a^2 \gg P^2$ and the $1/a^4$ scaling for relativistic particles ($m = 0$). Using the proper momentum, we can write:
\begin{equation}\label{endensf}
	\boxed{\varepsilon(t) = \int d^3\textbf{p}E(p)f(t,p)}
\end{equation}
For $\mu = 0$ and $\nu = i$:
\begin{equation}
	T^0{}_i(t) = \frac{1}{a(t)^3}\int P^2dPd^2\hat P cP_if(t,P)\;.
\end{equation}
As for $N^i$, this integral also vanishes.  

Finally, for $\mu = i$ and $\nu = j$:
\begin{equation}
	T^i{}_j(t) = \frac{1}{a(t)^3}\int P^2dPd^2\hat P\frac{c^2P^i P_j}{E}f(t,P)\;.
\end{equation}
Taking the spatial trace and dividing it by 3, we obtain a formula for the pressure:
\begin{equation}\label{pressf}
	\boxed{P(t) = \int d^3\textbf{p}\frac{p^2c^2}{3E(p)}f(t,p)}
\end{equation}
For photons, one has $E(p) = pc$, and therefore, combining Eqs.~\eqref{endensf} and \eqref{pressf}, one obtains the familiar result $P = \varepsilon/3$.

\section{Thermal distribution functions}

In a Euclidean setting, bosons and fermions in thermal equilibrium are described by the \textbf{Bose-Einstein and Fermi-Dirac distributions}:\index{Bose-Einstein distribution}\index{Fermi-Dirac distribution}
\begin{equation}\label{FDBEdistributions}
	\boxed{f_{\rm eq}^{-1} = \exp\left[\frac{E - \mu(T)}{k_{\rm B}T}\right] \pm  1}
\end{equation}
Here $E$ is the energy, $\mu(T)$ the chemical potential (which, in thermal equilibrium, depends only on the temperature), and $T$ the temperature of the thermal bath (well-defined since there is thermal equilibrium). The sign is $-$ for bosons (Bose-Einstein distribution) and $+$ for fermions (Fermi-Dirac distribution).

For classical particles, we have the \textbf{Maxwell-Boltzmann distribution}:\index{Maxwell-Boltzmann distribution}
\begin{equation}
	\boxed{f_{\rm eq} = \exp\left[-\frac{E - \mu(T)}{k_{\rm B}T}\right]} 
\end{equation}
For cosmological applications, the MB turns out to be a valid approximation since the temperatures of interest (from 0.1 eV up to a few MeV) are far larger than those required for the $\pm 1$ in the statistics to be of some relevance (for example, in order to form a Bose-Einstein condensate).

We derive these distributions in Appendix~\ref{App:Thermaldistr}, using statistical methods. Indeed, thermodynamics alone is unable to provide us with the above two functional dependencies, whereas the Boltzmann equation tells us how $f$ evolves, but not its initial functional form.

In an FLRW background, the energy $E$ is: 
\begin{equation}\label{EecP0}
	E = cP^0 = \sqrt{p^2c^2 + m^2c^4} = \sqrt{P^2c^2/a^2 + m^2c^4}\;.
\end{equation}
Note that the modulus of the conjugate momentum $P$ is time-independent in an FLRW background.

Equilibrium distributions are, by definition, time-independent (we will check this explicitly later by solving the collisionless Boltzmann equation). Hence, we expect $f_{\rm eq} = f_{\rm eq}(P)$. On the other hand, $E = \sqrt{P^2c^2/a^2 + m^2c^4}$ does present an explicit time-dependence through the scale factor $a(t)$. Indeed, as shown in \cite{1988kteu.book.....B}, the equilibrium distribution functions \eqref{FDBEdistributions} \emph{are incompatible} with an FLRW background unless, somehow, the quantity:
\begin{align}
    \frac{\sqrt{P^2c^2/a^2 + m^2c^4} - \mu(T)}{k_{\rm B}T}\,,
\end{align}
is time-independent. This can be the case if the temperature is a suitable function of the scale factor; however, since $\mu$ cannot depend on $P$, this can happen in only two instances:
\begin{itemize}
	\item The particle is massless and $\mu = 0$, so that:
	\begin{equation}\label{FDBEdistributionsmassless}
	\boxed{f_{\rm eq}^{-1} = \exp\left(\frac{Pc}{k_{\rm B}aT}\right) \pm  1}
\end{equation}
If $T \propto 1/a$, then $f_{\rm eq}(P)$ has no explicit time dependence on the scale factor, and so its shape is preserved throughout cosmic evolution.
\item The non-relativistic case. In this case:
\begin{equation}
	E = \sqrt{P^2c^2/a^2 + m^2c^4} \approx mc^2 + \frac{P^2}{2ma^2}\;.
\end{equation}
We also need to assume $\mu = mc^2$. Then, we are left with: 
\begin{equation}\label{FDBEdistributionsmass}
	\boxed{f_{\rm eq}^{-1} = \exp\left(\frac{P^2}{2mk_{\rm B}a^2T}\right) \pm  1} 
\end{equation}
In this case, $T \propto 1/a^2$.
\end{itemize}
Therefore, strictly speaking, we are entitled to use the known thermal distributions only in the two cases mentioned above. Fortunately, the one with $m = 0$ and $\mu = 0$ is satisfactorily suitable for the photons of the CMB. On the other hand, what can we do for more general instances?

Following \cite{1988kteu.book.....B}, consider:
\begin{equation}
    f(P) = f_0(t,P)[1 + \phi(t,P)]\,,
\end{equation}
where $f_0$ is one of the aforementioned thermal distributions, and $\phi$ is a correction meant to make $f$ an equilibrium distribution, at least at first order. 

One can then show that, due to a sort of gauge freedom in the determination of $\phi$, the number density and the energy density can be computed as in Eqs. \eqref{numdenspropmom} and \eqref{endensf} using the thermal distribution $f_0$. That is:
\begin{align}
    n(t) = \int d^3\textbf{p} f_0(t,p)\;, \quad \varepsilon(t) = \int d^3\textbf{p}E(p)f_0(t,p)\,.
\end{align}
The pressure, on the other hand, does acquire a correction due to $\phi$:
\begin{align}
    P_0(t) = \int d^3\textbf{p}\frac{p^2c^2}{3E(p)}f_0(t,p)\;, \\ 
    P_1(t) = \int d^3\textbf{p}\frac{p^2c^2}{3E(p)}f_0(t,p)\phi(t,p)\;.
\end{align}
This $P_1$ is known as \textit{bulk viscosity}\index{Bulk viscosity}, and it will be discussed somewhat more in Chapter \ref{Chap:CosmoPertTheory}. Furthermore, it can be shown that the production of entropy is of order $\phi^2$, and thus the entropy density is:
\begin{align}
    s = \frac{\varepsilon + P_0 - \mu n}{T} + \mathcal{O}(\phi^2)\,.
\end{align}
Therefore, even if slightly out of equilibrium, we can use the standard formulas that we have seen for LTE. Finally, it can also be shown that the requirement of $|\phi| \ll 1$, which is necessary for the above approximations, corresponds to the condition $\Gamma \gg H$.   

Finally, if we have a thermal bath with both massless particles with $\mu = 0$, for which $T \propto 1/a$, and massive, non-relativistic particles with $\mu = mc^2$, for which $T \propto 1/a^2$, how would the temperature actually scale with $a$? In general, the scaling would be neither $T \propto 1/a$ nor $T \propto 1/a^2$, so none of the distributions will be preserved in shape. In reality, as we will see, in our universe, there are so many more photons than protons or electrons (about a billion more) that the scaling is driven to be $T \propto 1/a$.

\subsection{The equilibrium number density}

Using the equilibrium distributions introduced above, the number density per degree of freedom (normalized to $g_{\rm s}$) is expressed as:
\begin{align}
     n = \frac{1}{2\pi^2\hbar^3}\int_0^\infty dp\,p^2\frac{1}{e^{(E(p) - \mu)/(k_{\rm B}T)} \pm 1}\,,
\end{align}
with, as usual:
\begin{equation}
	E(p) = \sqrt{p^2c^2 + m^2c^4}\,.
\end{equation}
Let us assume $\mu = 0$ and rewrite the above equation as follows:
\begin{equation}
	n = \frac{1}{2\pi^2\hbar^3}\int_0^\infty dp\;p^2\frac{1}{\exp\left[A\sqrt{p^2/(m^2c^2) + 1}\right] \pm 1}\;, \qquad A \equiv \frac{mc^2}{k_{\rm B}T}\;.
\end{equation}

\hrulefill

\begin{ex} Calling $x \equiv A\sqrt{p^2/(m^2c^2) + 1}$, show that the number density can be written as:
\begin{equation}\label{numdensmassivenu}
	n = \frac{(k_{\rm B}T)^3}{2\pi^2(\hbar c)^3}\int_A^\infty dx\;\frac{x\sqrt{x^2 - A^2}}{e^{x} \pm 1}\;.
\end{equation}
\end{ex}

\hrulefill

For $A = 0$, the above integral yields:
\begin{eqnarray}
\int_0^\infty dx\;\frac{x^2}{e^{x} - 1} = 2\zeta(3) \;,\\
	\int_0^\infty dx\;\frac{x^2}{e^{x} + 1} = \frac{3\zeta(3)}{2}\;, \\ \int_0^\infty dx\;x^2e^{-x} = 2\,,
\end{eqnarray}
where $\zeta(3) = 1.20206\dots$. So, for massless particles in LTE, we have:
\begin{equation}\label{partnumdensthermrel}
	n_{\rm boson} = \frac{\zeta(3)}{\pi^2}\frac{(k_{\rm B}T)^3}{(\hbar c)^3}\;, \quad 
	n_{\rm fermion} = \frac{3}{4}\frac{\zeta(3)}{\pi^2}\frac{(k_{\rm B}T)^3}{(\hbar c)^3}\;, \quad n_{\rm cl} = \frac{1}{\pi^2}\frac{(k_{\rm B}T)^3}{(\hbar c)^3}
\end{equation}
On the other hand, consider $A \gg 1$ ($mc^2 \gg k_{\rm B}T$). In this case, for the temperatures of interest in cosmology, we can ignore the $\pm 1$ in the denominator and use the Maxwell-Boltzmann equation:
\begin{equation}
	n = \frac{1}{2\pi^2\hbar^3}e^{-mc^2/(k_{\rm B}T)}\int_0^{\infty} dp\;p^2 e^{-p^2/(2mk_{\rm B}T)}\;.
\end{equation}
Changing the variable to $x^2 \equiv p^2/(2mk_{\rm B}T)$, we get:
\begin{equation}
	n = \frac{1}{2\pi^2\hbar^3}e^{-mc^2/(k_{\rm B}T)}(2mk_{\rm B}T)^{3/2}\int_0^{\infty} dx\;x^2 e^{-x^2}\;,
\end{equation}
and the result of the integration is:
\begin{equation}\label{partnumdensthermnonrel}
	n = \left(\frac{mk_{\rm B}T}{2\pi\hbar^2}\right)^{3/2}e^{-mc^2/(k_{\rm B}T)}\;.
\end{equation}
In this case, it is easy to recover the chemical potential:
\begin{equation}\label{partnumdensthermnonrelwithmu}
	n = \left(\frac{mk_{\rm B}T}{2\pi\hbar^2}\right)^{3/2}e^{-(mc^2 - \mu)/(k_{\rm B}T)}\;.
\end{equation}
When $\mu = mc^2$ from $n \propto a^{-3}$, one recovers the behavior $T \propto a^{-2}$ for massive, non-relativistic particles.

\subsection{The equilibrium energy density and pressure}

The energy density per degree of freedom is obtained by solving:
\begin{align}
\varepsilon = \frac{1}{2\pi^2\hbar^3}\int_0^\infty dp\,p^2\frac{E(p)}{e^{(E(p) - \mu)/(k_{\rm B}T)} \pm 1}\,,
\end{align}
with:
\begin{equation}
	E(p) = \sqrt{p^2c^2 + m^2c^4}\,.
\end{equation}
Let us assume again $\mu = 0$ and rewrite the above equation as follows:
\begin{equation}\label{endensmassivenu2}
	\varepsilon = \frac{mc^2}{2\pi^2\hbar^3}\int_0^\infty dp\;p^2\frac{\sqrt{p^2/(m^2c^2) + 1}}{\exp\left[A\sqrt{p^2/(m^2c^2) + 1}\right] \pm 1}\;, \qquad A \equiv \frac{mc^2}{k_{\rm B}T}\;.
\end{equation}

\hrulefill

\begin{ex} Calling $x \equiv A\sqrt{p^2/(m^2c^2) + 1}$, show that the above integration \eqref{endensmassivenu2} becomes:
\begin{equation}\label{endensmassivenu3}
	\varepsilon = \frac{m^4 c^5}{2\pi^2\hbar^3A^4}\int_A^\infty dx\;\frac{x^2\sqrt{x^2 - A^2}}{e^{x} \pm 1} = \frac{(k_{\rm B}T)^4}{2\pi^2(\hbar c)^3}\int_A^\infty dx\;\frac{x^2\sqrt{x^2 - A^2}}{e^{x} \pm 1}\;.
\end{equation}
Note the lower integration limit.
\end{ex}

\hrulefill

\begin{figure}[ht!]
\centering
	\includegraphics[width=\columnwidth]{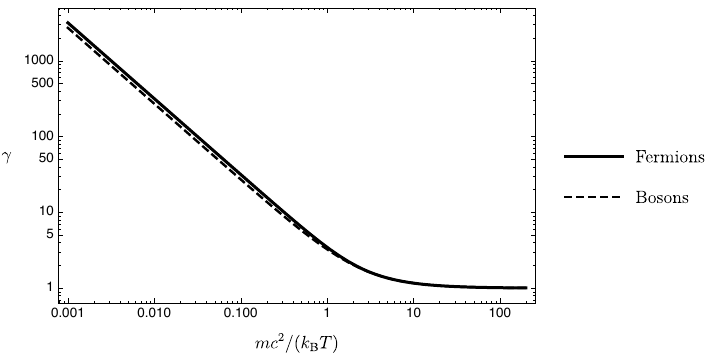}
	\caption{Plot of $\gamma$ as function of $A$. The solid line is for fermions whereas the dashed one for bosons.}
	\label{Fig:gammaPlot}
\end{figure}

When $A \ll 1$, the integral in Eq.~\eqref{endensmassivenu3} can be expanded as follows:
\begin{eqnarray}
\int_A^\infty dx\;\frac{x^2\sqrt{x^2 - A^2}}{e^{x} - 1} = \frac{\pi^4}{15} - \frac{\pi^2A^2}{12} + \mathcal{O}(A^3)\;,\\
	\int_A^\infty dx\;\frac{x^2\sqrt{x^2 - A^2}}{e^{x} + 1} = \frac{7}{8}\frac{\pi^4}{15} - \frac{\pi^2A^2}{24} + \mathcal{O}(A^3)\;.
\end{eqnarray}
For massless particles in LTE, we have:
\begin{equation}
	\varepsilon_{\rm boson} = \frac{\pi^2}{30}\frac{(k_{\rm B}T)^4}{(\hbar c)^3}\;, \qquad \varepsilon_{\rm fermion} = \frac{7}{8}\frac{\pi^2}{30}\frac{(k_{\rm B}T)^4}{(\hbar c)^3}\;, \qquad \varepsilon_{\rm cl} = \frac{3}{\pi^2}\frac{(k_{\rm B}T)^4}{(\hbar c)^3}\;.
\end{equation}
The opposite limit $mc^2 \gg k_{\rm B}T$ is much trickier to investigate analytically; therefore, we shall address it numerically. The ratio $\varepsilon/(nmc^2)$ can be written as:
\begin{equation}
	\gamma \equiv \frac{\varepsilon}{(nmc^2)} = \frac{1}{A}\frac{\int_A^\infty dx\;\frac{x^2\sqrt{x^2 - A^2}}{e^{x} \pm 1}}{\int_A^\infty dx\;\frac{x\sqrt{x^2 - A^2}}{e^{x} \pm 1}}\;,
\end{equation}
where $\gamma$ is indeed some sort of averaged Lorentz factor, as it is the ratio of the energy density to the mass energy density of the ensemble of particles. The behavior of $\gamma$ as a function of $A$ is shown in Fig.~\ref{Fig:gammaPlot}. From this figure, we can infer that $\varepsilon/(nmc^2) \approx 1$ when $mc^2 \gg k_{\rm B}T$, and therefore, the particle becomes non-relativistic. Hence, for $mc^2 \gg k_{\rm B}T$ one has:
\begin{equation}
	\varepsilon = nmc^2\,.
\end{equation}
One can easily show that the pressure is equal to $\varepsilon/3$ in the relativistic case (regardless of the statistics) and: 
\begin{equation}
	P = \frac{1}{2\pi^2\hbar^3}e^{-mc^2/(k_{\rm B}T)}\int_0^{mc} dp\;p^2 e^{-p^2/(2mk_{\rm B}T)}\frac{p^2}{3m} = nk_{\rm B}T \ll \varepsilon\;,
\end{equation}
in the non-relativistic case.

In general, we can state that \textbf{particles with mass $m$ in thermal equilibrium at temperature $T$ behave as relativistic particles when $mc^2 \ll k_{\rm B}T$}. The transition from relativistic to non-relativistic occurs for $k_{\rm B}T \approx 10 mc^2$.

\subsection{The equilibrium entropy density}

We compile here the results obtained thus far and compute the entropy density (with $\mu = 0$) $s = (\varepsilon + P)/T$. For the relativistic contributions, suppose that we have various species of particles, each in LTE with its own temperature $T_i$ and non-interacting with the others. In this case, the total energy density of the relativistic species is:
\begin{align}
	\varepsilon = \sum_{i = \textrm{bosons}}\frac{\pi^2(k_{\rm B}T_i)^4}{30(\hbar c)^3}g_i + \frac{7}{8}\sum_{i = \textrm{fermions}}\frac{\pi^2(k_{\rm B}T_i)^4}{30(\hbar c)^3}g_i
\end{align}
It is customary to use the temperature of the photons, say $T$, as a reference. That is:
\begin{equation}\label{endensradgstar1}
	\boxed{\varepsilon = \frac{\pi^2(k_{\rm B}T)^4}{30(\hbar c)^3}g_*}
\end{equation}
thereby defining:
\begin{equation}\label{gstar}
	g_* \equiv \sum_{i = \textrm{bosons}}g_i\frac{T_i^4}{T^4} + \frac{7}{8}\sum_{i = \textrm{fermions}}g_i\frac{T_i^4}{T^4}\;,
\end{equation}
the \textbf{effective number of relativistic degrees of freedom}. In the standard cosmological model, only neutrinos and photons form thermal baths with different temperatures, following the decoupling of neutrinos from the primordial soup. All the other species, when in thermal equilibrium, share the same temperature as the photons.

For the entropy density $s$, we have the contribution from relativistic species with $\mu = 0$ as follows:
\begin{align}
	s_{\rm rel} = \frac{4\varepsilon}{3T}\,.
\end{align}
Therefore:
\begin{equation}
	\boxed{s_{\rm rel} = \frac{2\pi^2k_{\rm B}^4T^3}{45(\hbar c)^3}g_{*S}}
\end{equation}
where $g_{*S}$ is the \textbf{effective number of degrees of freedom for the entropy}:\index{Effective number of relativistic degrees of freedom}
\begin{equation}
	g_{*S} \equiv \sum_{i = \textrm{bosons}}g_i\frac{T_i^3}{T^3} + \frac{7}{8}\sum_{i = \textrm{fermions}}g_i\frac{T_i^3}{T^3}\;.
\end{equation}
In general, $g_{*} \neq g_{*S}$, unless all the relativistic species share the same temperature. This is indeed the case for most of the history of the universe. As already mentioned, massless neutrinos have a different, smaller temperature than that of photons below $k_{\rm B}T \approx 16$ keV.

The effective number of relativistic degrees of freedom $g_{*}$, or $g_{*S}$, is actually a function of the temperature, as it decreases when a certain species becomes non-relativistic, i.e., when $k_{\rm B}T$ drops below the $mc^2$ of such species. Thus, $g_*$ varies close to the thresholds of the mass energies. The particles in the Standard Model, along with their properties useful for computing $g_{*}$ and $g_{*S}$, are gathered in Table \ref{Tab:Standmodelpart}. 

\begin{table}[ht]
\begin{center}
	\begin{tabular}{|l|l|l|l|l|}
\hline
Particle &  & Mass & Spin & $g_s$\\
\hline	
Quarks   & $t,\bar{t}$ & 173 GeV & $\frac{1}{2}$ & $2\cdot 2 \cdot 3 = 12$\\
         & $b,\bar{b}$ &   4 GeV &               &  \\
              & $c,\bar{c}$ &   1 GeV &               &  \\
     & $s,\bar{s}$ &   100 MeV &               &  \\
     & $d,\bar{d}$ &   5 MeV &               &  \\
     & $u,\bar{u}$ &   2 MeV &               &  \\
     \hline
Leptons   & $\tau^\pm$ & 1777 MeV & $\frac{1}{2}$ & $2\cdot 2 = 4$\\
         & $\mu^\pm$ &   106 MeV &               &  \\
              & $e^\pm$ &   511 keV &               &  \\
     & $\nu_\tau,\bar{\nu}_\tau$ &   $< 0.6$ eV & $\frac{1}{2}$ & $2\cdot 1 = 2$ \\
     & $\nu_\mu,\bar{\nu_\mu}$ & $< 0.6$ eV &               &  \\
     & $\nu_e,\bar{\nu}_e$ & $< 0.6$ eV &               &  \\
     \hline     
Gauge Bosons  & $W^+$ & 80 GeV & 1 & 3\\
         & $W^-$ & 80 GeV &               &  \\
              & $Z^0$ & 91 GeV &               &  \\
     & $\gamma$ &   0 &  & 2 \\
     & g & 0 & & $8\cdot 2 = 16$ \\
\hline     
Higgs Bosons  & $H^0$ & 125 GeV & 0 & 1\\
\hline
\end{tabular}
\caption{The standard model particles, with their mass, spin and degeneracies $g_s$.}
\label{Tab:Standmodelpart}
\end{center}	
\end{table}

\begin{figure}[htbp]
\centering
	\includegraphics[width=\columnwidth]{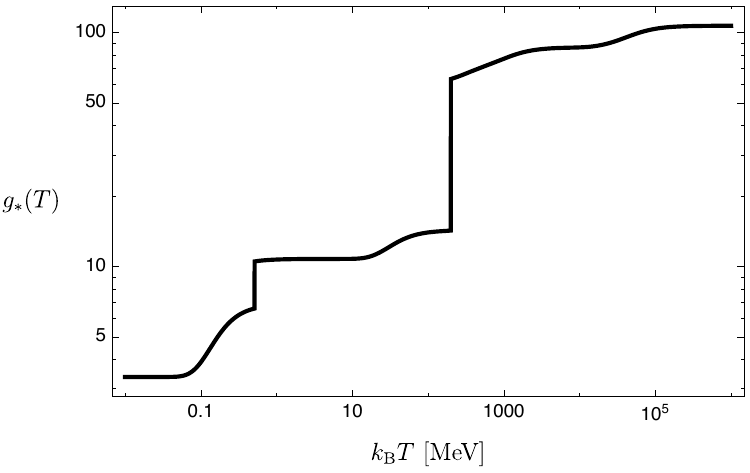}
	\caption{Evolution of $g_*$ calculated from Eq.~\eqref{endensmassivenu2} and Eq.~\eqref{endensradgstar1} for the species of Table~\ref{Tab:Standmodelpart}. Note the falls in $g_*$ due to the QCD phase transition at 200 MeV and to the electron-positron annihilation at 0.5 MeV. Note that this plot is merely an illustrative example because it assumes LTE at all times. Moreover, the QCD phase transition at 200 MeV and the difference between photon and neutrino temperatures after electron-positron annihilation at 0.5 MeV have been put ``by hand''. The correct calculation should employ the energy density obtained from the solutions of the Boltzmann equations of the various species, which correctly track the evolutions when $\Gamma \sim H$, i.e. out of equilibrium.}
	\label{Fig:gstarPlot}
\end{figure}

In Fig.~\ref{Fig:gstarPlot}, we plot $g_*$ calculated from Eq.~\eqref{endensmassivenu2} and Eq.~\eqref{endensradgstar1} for the species listed in Table~\ref{Tab:Standmodelpart}. The residual non-vanishing value of $g_*(T)$ for low temperatures is due to the massless species: photons and neutrinos. 

To the $s_{\rm rel}$ calculated for relativistic species, we should, in principle, add the one for non-relativistic ones, i.e., for baryons. For the temperature of interest in cosmology, and for the applications of the next chapter, baryons are essentially protons and neutrons. We have then:
\begin{align}
    s_{\rm b} = \frac{n_{\rm b}m_{\rm b}c^2 + n_{\rm b}k_{\rm B}T - \mu_{\rm b}n_{\rm b}}{T}\,.
\end{align}
If we take $\mu_{\rm b} = m_{\rm b}c^2$, then:
\begin{align}
    s_{\rm b} = n_{\rm b}k_{\rm B}\,.
\end{align}
Again, using Eq. \eqref{partnumdensthermnonrelwithmu} for the dependence of the number density on temperature and employing the result $s \propto a^{-3}$, one obtains $T \propto a^{-2}$. One can actually conclude this already from the fact that $n \propto a^{-3}$ if no interactions are present that destroy or create particles. The problem that we anticipated earlier is the following: if we have a thermal bath with both massless particles with $\mu = 0$, for which $T \propto 1/a$, and massive, non-relativistic particles with $\mu = mc^2$, for which $T \propto 1/a^2$, how would the temperature actually scale with $a$? Now we can answer this question thanks to the entropy density, since $s$ is calculated for \textit{all} the species of the thermal bath:
\begin{align}
    s = s_{\rm rel} + s_{\rm b} = \frac{2\pi^2k_{\rm B}^4T^3}{45(\hbar c)^3}g_{*S} + n_{\rm b}k_{\rm B}\,.
\end{align}
Introducing the number density of photons: 
\begin{align}
    s = \frac{2\pi^4g_{*S}}{45\zeta(3)}n_\gamma k_{\rm B} + n_{\rm b}k_{\rm B}\,.
\end{align}
So that:
\begin{align}
     s = n_\gamma k_{\rm B}(\alpha  + \eta_{\rm b})\,, \qquad \alpha \equiv \frac{2\pi^4g_{*S}}{45\zeta(3)}\,,
\end{align}
where $\alpha$ is a numerical factor of order $10-10^2$ (see Tab. \ref{Tab:Standmodelpart}), and $\eta_{\rm b} \equiv n_{\rm b}/n_\gamma$ is the \textbf{baryon-to-photon ratio}, i.e., the number of baryons per photon. We will see that this number is very small (of order $10^{-10}$), so the entropy density is dominated by the relativistic component, and the scaling of the temperature is essentially given by:
\begin{equation}\label{gstarSTscaling}
	g_{*S}(T)T^3 \propto 1/a^3\,.
\end{equation}
Therefore, during the epochs in which $g_{*S}(T)$ is constant, we have:
\begin{equation}
	T \propto 1/a\,.
\end{equation}  
On the other hand, when species become non-relativistic, $g_{*S}$ decreases, and the temperature of the thermal bath decreases \textit{more slowly} than $1/a$.

\section{Application: photons}

In these notes, we use ``photons'' as a synonym for CMB, though there exist photons whose origin is not cosmological, such as those produced in stars or emitted by hot interstellar gas. The energy density of these non-cosmological photons is approximately one order of magnitude less than that of CMB photons \cite{Torres:2016fmj}.

The assumption $\mu_\gamma = 0$ for CMB photons is compatible with observation. Indeed, $\mu/(k_{\rm B}T_0) < 9 \times 10^{-5}$, with $T_0 \approx 3$ K, as reported by \cite{Fixsen:1996nj}. Hence, the Bose-Einstein distribution for photons is compatible with the cosmological expansion. 

Exploiting the results of the previous section and recalling the two states of polarization, the photon energy density is as follows:
\begin{equation}\label{photonenergydensity}
	\boxed{\varepsilon_\gamma = \frac{\pi^2}{15(\hbar c)^3}(k_{\rm B}T)^4}\;.
\end{equation}
Knowing the value of the CMB temperature today, $T_0 = 2.725$ K, we can estimate the photon energy content today (and thus at any time):
\begin{equation}
	\Omega_{\gamma 0} = \frac{\varepsilon_{\gamma 0}}{\varepsilon_{\rm cr0}} = \frac{8\pi^3G}{45(\hbar c)^3H_0^2c^2}(k_{\rm B}T_0)^4\;.
\end{equation}

\hrulefill

\begin{ex} Using $H_0 = 100$ $h$ km s$^{-1}$ Mpc$^{-1}$ and the known constants of nature show that:
\begin{equation}\label{Omegagamma}\index{Photons!Density parameter}
	\boxed{\Omega_{\gamma 0} h^2 = 2.47\times 10^{-5}}\;.
\end{equation}
\end{ex}

\hrulefill

The photon number density is:\index{Photons!Number density}
\begin{align}\label{photonnumdens}
    \boxed{n_\gamma = \frac{2\zeta(3)}{\pi^2(\hbar c)^3}(k_{\rm B}T)^3}
\end{align}

\hrulefill

\begin{ex} Calculate the photon number density today. Show that it is $n_{\gamma 0} = 411$ cm$^{-3}$.	
\end{ex}

\hrulefill

Note that if we had used the MB distribution instead of the BE one, we would have obtained $\Omega_{\gamma 0} h^2 = 2.28\times 10^{-5}$ and $n_{\gamma 0} = 342$ cm$^{-3}$.

\subsection{The baryon-to-photon ratio}

A very important number for the forthcoming applications is the \textbf{baryon-to-photon ratio}:
\begin{equation}\label{baryontophotonratio}
	\boxed{\eta_{\rm b} \equiv \frac{n_{\rm b}}{n_\gamma} = 5.5 \times 10^{-10}\left(\frac{\Omega_{\rm b0}h^2}{0.020}\right)}
\end{equation}\index{Baryon-to-photon ratio}
It is the ratio between the number of baryons (protons and neutrons, the latter combined in light nuclei) and the number of photons. We have defined it via number densities, which are individually time-dependent quantities, but whose ratio is fixed since both scale as $1/a^3$. The numbers in Eq.~\eqref{baryontophotonratio} can be found as follows:
\begin{equation}
	\eta_{\rm b} = \frac{n_{\rm b}}{n_\gamma} = \frac{\varepsilon_{\rm b0}}{m_{\rm b}c^2n_{\gamma 0}}\;,
\end{equation}
where we have assumed baryons to be non-relativistic, which is indeed the case at temperatures $k_{\rm B}T \lesssim $ 1 GeV, the mass of the proton.

Using Eq.~\eqref{photonenergydensity} and Eq.~\eqref{photonnumdens}, we can write
\begin{equation}
	\frac{n_{\gamma 0}}{\varepsilon_{\gamma 0}} = \frac{30\zeta(3)}{\pi^4k_{\rm B}T_0}\;.
\end{equation}
Therefore,
\begin{equation}\label{etabproof}
	\eta_{\rm b} = \frac{\pi^4k_{\rm B}T_0\varepsilon_{\rm b0}}{30\zeta(3)m_{\rm b}c^2\varepsilon_{\gamma 0}} = \frac{\pi^4k_{\rm B}T_0\Omega_{\rm b0}}{30\zeta(3)m_{\rm b}c^2\Omega_{\gamma 0}}\;,
\end{equation}
where in the last equality, we have multiplied and divided by the present critical energy density in order for the density parameters to appear.

\hrulefill

\begin{ex} Show that Eq.~\eqref{etabproof} leads to Eq.~\eqref{baryontophotonratio}. Use $m_{\rm b}c^2 = m_{\rm p}c^2 = 1$ GeV (roughly, the mass energy density of the proton and of the neutron).
\end{ex}

\hrulefill

The fact that there are a billion photons for each proton (and electron) is very important for the following reason: \textbf{bound states among particles in the primordial plasma do not form when the thermal energy equals the binding energy, but when it is much smaller}.\footnote{Such particles must interact electromagnetically, of course.}  

That happens because photons are thermally distributed. Let $B$ represent a certain binding energy among baryons. The fraction of photons with energy larger than $B$ is:
\begin{align}
	\frac{n_\gamma(pc/B > 1)}{n_\gamma} = \frac{\int_{\frac{B}{k_{\rm B}T}}^\infty dx\frac{x^2}{e^x - 1}}{\int_0^\infty dx\frac{x^2}{e^x - 1}}\,.
\end{align} 
Suppose that all the baryons are bound in such a state, with binding energy $B$. Then, when the ratio:
\begin{align}
	\frac{n_\gamma(pc/B > 1)}{n_{\rm b}} = \frac{n_\gamma(pc/B > 1)}{n_\gamma\eta_{\rm b}}\,,
\end{align} 
is equal to $1$, we expect photons to progressively become unable to overcome the bounds. Numerically, one can check that this occurs for:
\begin{align}
    \frac{B}{k_{\rm B}T} \approx 27\,.
\end{align}
So, a given bound with binding energy $B$ starts to be stable against the ``attack'' of photons for a thermal energy of approximately $k_{\rm B}T \approx B/27$.

\section{Application: neutrinos}

The same comment made at the beginning of the previous section also applies here: with ``neutrinos'' we mean the cosmological, or primordial, neutrinos and not those produced, for example, in supernova explosions.

We assume neutrinos to be massless, so their energy density also scales as $\varepsilon_\nu = \varepsilon_{\nu 0}/a^4$. Moreover, we also assume their chemical potential to be vanishing so that the Fermi-Dirac distribution preserves its shape along with the cosmological expansion. 

Neutrinos are spin $1/2$ fermions, so we would expect $g_\nu = 2$. On the other hand, neutrinos are particles that interact only via weak interaction, and this violates parity. In other words, only left-handed neutrinos exist, and so $g_\nu = 1$. 

The neutrino energy density, using the results of the previous section, is as follows: 
\begin{equation}\label{neutrinoendens2}
	\boxed{\varepsilon_\nu = \frac{7}{8}N_\nu g_\nu\frac{\pi^2}{30\hbar^3c^3}\left(k_{\rm B}T_\nu\right)^4}
\end{equation}
where $N_\nu$ is the number of neutrino families (3 in the Standard Model). Of course, an equivalent expression holds true for antineutrinos.

Note that we have used a temperature $T_\nu$, hinting at the possibility that $T_\nu \neq T_\gamma$. As long as neutrinos and photons are in thermal equilibrium, we do have $T_\nu = T_\gamma$. This equilibrium is attained through reactions such as:
\begin{align}
    e^- + e^+ \longleftrightarrow \nu_e + \bar\nu_e\,,
\end{align}
and, as we have seen, it is maintained down to $k_{\rm B}T \approx 1$ MeV.

After neutrinos decouple from the primordial plasma, one expects both temperatures to decrease as $1/a$, since neutrinos are relativistic, and then to remain equal throughout cosmic history. This is true until electrons and positrons start to annihilate ($k_{\rm B}T \approx 16$ keV).

\subsection{Temperature of the massless neutrino thermal bath}

When the temperature of the photon thermal bath was sufficiently high ($k_{\rm B}T \gtrsim 16$ keV, and not twice the rest mass of the electron due to the huge photon-to-baryon ratio), positron-electron annihilation and positron-electron pair production were balanced reactions:
\begin{equation}\label{e+e-2gamma}
	e^+ + e^- \longleftrightarrow \gamma + \gamma\;.
\end{equation} 
Recalling the condition of chemical equilibrium:
\begin{equation}
	\mu_{e^+} + \mu_{e^-} = 2\mu_\gamma\,.
\end{equation}
Since $\mu_{e^+} = -\mu_{e^-}$ (in the thermodynamic limit, removing or adding a positron is the same as adding or removing an electron), we get, as expected, $\mu_\gamma = 0$.

\hrulefill

\begin{ex} Knowing that today ($z = 0$) the photon thermal bath has temperature $T_0 = 2.725$ K, estimate the redshift at which $k_{\rm B} T = 1$ MeV (assuming thermal equilibrium).	
\end{ex}

\hrulefill

When the temperature of the thermal bath drops below $k_{\rm B} T \approx 16$ keV (about 511/27 keV, note the key factor 27 computed earlier), the above reaction becomes unbalanced: pair production is no longer possible. Only annihilation is. 

Annihilation injects high-energy photons into the primordial plasma; however, this remains in thermal equilibrium due to the high rates of other interactions, which also include photons. The extra energy coming from the annihilation is then ``evenly distributed'' in the plasma. For this reason, we expect the photon temperature to drop more slowly than $1/a$ and then to be different from the neutrino temperature. 

We now quantify this difference using the conservation of $sa^3$. Using the results of the previous section again, at a certain scale factor $a_1$ earlier than neutrino decoupling, the entropy density of the primordial plasma is:
\begin{equation}
	s(a_1) = \frac{2\pi^2k_{\rm B}^4}{45\hbar^3c^3}T_1^3\left[2 + \frac{7}{8}\left(2 + 2\right) + \frac{7}{8}N_\nu\left(g_\nu + g_{\bar{\nu}}\right)\right]\;,
\end{equation}
where we have left explicit all the degrees of freedom: 2 for the photons, 2 for the electrons, 2 for the positrons, $g_\nu$ for the neutrinos, and $g_{\bar{\nu}}$ for the antineutrinos. Being still coupled, $T_1$ is the same temperature for both photons and neutrinos.

On the other hand, at a certain scale factor $a_2$, after neutrino decoupling and after electron-positron annihilation, the entropy density of the neutrinos and the remaining plasma (composed of photons, residual electrons, protons, and Helium nuclei, all of which are non-relativistic) is:
\begin{equation}
	s(a_2) = \frac{2\pi^2k_{\rm B}^4}{45\hbar^3c^3}\left(2T_\gamma^3 + \frac{7}{4}N_\nu g_\nu T_\nu^3\right)\;,
\end{equation}
where we have now made a distinction between the two temperatures and set $g_\nu = g_{\bar{\nu}}$. At this epoch, photons are still in LTE with the other non-relativistic species, whereas neutrinos have decoupled but have maintained their thermal distribution since we are assuming them to be massless and to have a vanishing chemical potential. 

Equating:
\begin{equation}
	s(a_1)a_1^3 = s(a_2)a_2^3\;,
\end{equation}
we obtain:
\begin{equation}
	(a_1T_1)^3\left[2 + \frac{7}{4}\left(N_\nu g_\nu + 2\right)\right] = (a_2T_\nu)^3\left[2\left(\frac{T_\gamma}{T_\nu}\right)^3 + \frac{7}{4}N_\nu g_\nu\right]\;.
\end{equation}
Since $(a_1T_1)^3 = (a_2T_\nu)^3$ and because the neutrino temperature did not change its $\propto 1/a$ behavior, we are left with:
\begin{equation}\label{TgammaTnu}
	\boxed{\frac{T_\nu}{T_\gamma} = \left(\frac{4}{11}\right)^{1/3} \approx 0.714}
\end{equation}
Therefore, since the CMB temperature today is $T_{\gamma 0} = 2.725$ K, we expect a thermal neutrino background of temperature $T_{\nu 0} \approx 1.945$ K. Remarkably, the result of Eq.~\eqref{TgammaTnu} does not depend on the neutrino $g_\nu$.\index{Neutrinos!Temperature}

Note that electrons and positrons are not the only particles that annihilate (for example, protons and antiprotons also annihilate at much larger temperatures), but they are the only particles that annihilate \textit{after} neutrinos decouple; thus, we took only them into account in order to determine the temperature difference.



For a more detailed description of the positron-electron annihilation epoch in the early universe, see Ref. \cite{Thomas:2019ran}.

Using Eq.~\eqref{TgammaTnu}, the neutrino and antineutrino energy densities can be related to the photon energy density as follows:\footnote{Recall that we include antineutrinos in the calculations because we are assuming that, due to their masslessness, there is no excess production of $\nu$ over $\bar\nu$, and $\varepsilon_\nu = \varepsilon_{\bar{\nu}} \neq 0$ since they do not annihilate.}
\begin{equation}\label{neutrinoendens2phot}
	\boxed{\varepsilon_\nu = \varepsilon_{\bar{\nu}} = \frac{7}{8}\frac{N_\nu g_\nu}{2}\left(\frac{4}{11}\right)^{4/3}\varepsilon_\gamma}
\end{equation}
where we have left unspecified the values of the neutrino degeneracy $g_\nu$ and the number of neutrino families $N_\nu$.

The total radiation energy content can thus be expressed as
\begin{equation}\label{epsilonR}
	\varepsilon_{\rm r} \equiv \varepsilon_\gamma + \varepsilon_\nu + \varepsilon_{\bar{\nu}} = \varepsilon_\gamma\left[1 + \frac{7}{8}N_\nu g_\nu\left(\frac{4}{11}\right)^{4/3}\right]\;.
\end{equation}
The Planck collaboration \cite{Planck:2018vyg} has put forth the constraint: 
\begin{equation}
	\boxed{N_{\rm eff} \equiv N_\nu g_\nu = 2.99 \pm 0.17}
\end{equation}
Therefore, three neutrino families ($N_\nu = 3$) and one spin state for each neutrino ($g_\nu = 1$) are values that work well.\index{Neutrinos!Effective number of families $N_{\rm eff}$}

\hrulefill

\begin{ex} Taking into account \eqref{TgammaTnu}, show that
\begin{equation}\label{neutrinonumberdensity}
	\boxed{n_\nu = \frac{3}{11}N_\nu g_\nu n_\gamma}
\end{equation}
\end{ex}

\hrulefill

\subsection{Massive neutrinos}

The 2015 Nobel Prize in Physics was awarded to Takaaki Kajita and Arthur B. McDonald \textit{for the discovery of neutrino oscillations, which shows that neutrinos have mass} (quoting from the Nobel Prize website). Indeed, neutrino flavors oscillate among the leptonic families (electron, muon, and tau); e.g., an electron neutrino can turn into a muon one or a tau one, depending on its energy and on how far it travels. 

The most stringent constraints on neutrino mass do not come from particle accelerators but from cosmology: $\sum m_\nu c^2 < 0.12$ eV from the Planck collaboration \cite{Planck:2018vyg}. The neutrino mass is thus very small, and this does not significantly change the early history of the universe; however, it has some impact at late-times for structure formation \cite{Lesgourgues:2006nd}.\index{Neutrinos!Mass constraints}


Suppose that a single family of neutrinos and antineutrinos has become non-relativistic only recently. What would their energy density and density parameter be today? Being non-relativistic, we can express their energy density as follows:
\begin{equation}
	\varepsilon_{m_\nu} = \rho_\nu c^2 = n_\nu m_\nu c^2\;.
\end{equation}
Thus, the density parameter today is:
\begin{equation}\label{massivenuomega}
	\Omega_{m_\nu 0} = \frac{8\pi Gn_{\nu 0} m_\nu c^2}{3H_0^2}\;.
\end{equation}

\hrulefill

\begin{ex} Using Eq.~\eqref{neutrinonumberdensity} and the result for $n_{\gamma 0}$, prove that Eq.~\eqref{massivenuomega} can be written as:
\begin{equation}\label{massivenuomegaeV}
	\boxed{\Omega_{m_\nu 0}h^2 = \frac{1}{94}g_\nu\frac{m_\nu c^2}{\mbox{eV}}}
\end{equation}\index{Neutrinos!Massive neutrinos density parameter}
\end{ex}

\hrulefill



\subsection{Matter-Radiation equality}

The epoch, or instant, at which the energy density of matter (i.e., baryons plus CDM) equals the energy density of radiation (i.e., photons plus massless neutrinos) is particularly important from the point of view of the evolution of perturbations, as we shall see in Chapter~\ref{Chap:Evopert}. From Eq.~\eqref{epsilonR} we have that:
\begin{equation}
	\boxed{\Omega_{\rm r0} = \Omega_{\gamma 0}\left[1 + \frac{7}{8}N_{\rm eff}\left(\frac{4}{11}\right)^{4/3}\right]}
\end{equation}
In order to calculate the scale factor $a_{\rm eq}$ of the equivalence, we only need to solve the following equation:\index{Matter-radiation equality}
\begin{equation}
	\frac{\Omega_{\rm r0}}{a_{\rm eq}^4} = \frac{\Omega_{\rm m0}}{a_{\rm eq}^3}\;,
\end{equation}
which gives
\begin{equation}
	a_{\rm eq} = \frac{\Omega_{\rm r0}}{\Omega_{\rm m0}} = \frac{\Omega_{\gamma 0}}{\Omega_{\rm m0}}\left[1 + \frac{7}{8}N_{\rm eff}\left(\frac{4}{11}\right)^{4/3}\right]\;.
\end{equation}
Using Eq.~\eqref{Omegagamma} and $N_{\rm eff} = 3$, one obtains:
\begin{equation}
	\boxed{a_{\rm eq} = \frac{4.15\times 10^{-5}}{\Omega_{\rm m0} h^2}} \qquad \Rightarrow \qquad \boxed{1 + z_{\rm eq} = 2.4 \times 10^4\;\Omega_{\rm m0} h^2}
\end{equation}

\subsubsection{Massive neutrinos}

What happens to the equivalence redshift $z_{\rm eq}$ if one of the neutrino species has mass $m_\nu > 0$? 

If that neutrino species becomes non-relativistic \textit{after} the equivalence epoch, then the above calculation still holds true. Therefore, let us assume that the neutrino species becomes non-relativistic, thereby counting as matter, before the equivalence. Since there is more matter and less radiation, we expect the equivalence to occur earlier, i.e.,
\begin{equation}
	a_{\rm eq} = \frac{3.59\times 10^{-5}}{(\Omega_{\rm m0} + \Omega_{m_\nu 0})h^2} \quad 1 + z_{\rm eq} = 2.79 \times 10^4\;(\Omega_{\rm m0} + \Omega_{m_\nu 0})h^2\;.
\end{equation}
Since $T = T_0(1 + z)$, the photon temperature at equivalence is:
\begin{equation}
	T_{\gamma,\rm eq} = 2.79 \times T_0 \cdot 10^4\;(\Omega_{\rm m0} + \Omega_{m_\nu 0}) h^2 = 7.60 \times 10^4\;(\Omega_{\rm m0} + \Omega_{m_\nu 0}) h^2\;\mbox{K}\;.
\end{equation}
Using Eq.~\eqref{TgammaTnu}, the temperature of neutrinos at equivalence is:
\begin{equation}
	T_{\nu,\rm eq} = 5.43 \times 10^4\;(\Omega_{\rm m0} + \Omega_{m_\nu 0}) h^2\;\mbox{K}\;.
\end{equation}
The neutrino mass energy $m_\nu c^2$ has to be larger than $k_{\rm B} T_{\nu,\rm eq}$ in order for the above calculation to be consistent. This yields:
\begin{equation}
	m_\nu c^2 > 5.43\;k_{\rm B} \times 10^4\;(\Omega_{\rm m0} + \Omega_{m_\nu 0}) h^2\;\mbox{K} = 4.68\;(\Omega_{\rm m0} + \Omega_{m_\nu 0}) h^2\;\mbox{eV}\;.
\end{equation}
Using $\Omega_{\rm m0} h^2 = 0.14$ and Eq.~\eqref{massivenuomegaeV}, we get $m_\nu c^2 > 0.69$ eV, which is incompatible with the constraint $\sum m_\nu c^2< 0.194$ eV found by the Planck collaboration. 

Therefore, we conclude that massive neutrino species do not alter the equivalence epoch.

\section{The Boltzmann equation}

In a Euclidean and classical setting, following \cite{1987stme.book.....H}, the Boltzmann equation is:
\begin{equation}
	\frac{df}{dt} = C[f]\;,
\end{equation}
where $f$ is the one-particle distribution function and $C[f]$ is the collisional term: a functional of $f$ describing the interactions among the particles that constitute the system under investigation. 

For the forthcoming applications, we shall need to investigate two-particle interactions; that is:
\begin{equation}\label{12into34}
	1 + 2 \longleftrightarrow 3 + 4\;,
\end{equation}
with generic species (possibly even the same) that we designate as 1, 2, 3, and 4. This reaction can describe scattering or annihilation among particles and will be suitable for discussing the primordial formation of Helium, the recombination of the primordial plasma into neutral atoms, and for calculating the relic abundance of hypothetical CDM particles. 

For the process \eqref{12into34}, the collisional term for the species $1$ is as follows:
\begin{equation}\label{collisionaltermgen}
	C[f_1] = \int d^3p_2 d\Omega |\mathbf v_2 - \mathbf v_1|(d\sigma/d\Omega)(f_3f_4 - f_1f_2)\;,
\end{equation}
where the differential cross section is defined as:
\begin{eqnarray}
	d\sigma = \frac{1}{2E_12E_2|\mathbf v_2 - \mathbf v_1|}\int\frac{d^3\textbf{p}_3}{2E_3}\int\frac{d^3\textbf{p}_4}{2E_4}\nonumber\\
	(2\pi)^4\delta^{(3)}(\textbf{p}_1 + \textbf{p}_2 - \textbf{p}_3 - \textbf{p}_4)\delta(E_1 + E_2 - E_3 - E_4)|\mathcal{M}|^2\;.
\end{eqnarray} 
Here, the interaction amplitude $|\mathcal{M}|$ is assumed to be invariant under the exchange of ingoing and outgoing momenta. Note that the $\delta$ enforces the conservation of energy and momentum.

At equilibrium, one expects $df/dt = 0$, and so the collisional term must vanish when evaluated on the equilibrium distribution function.

Let us focus on the total time derivative of $f$. Since $f$ is a function of time $t$, the particle position $\textbf{x}(t)$ and the particle momentum $\textbf{p}(t)$, and $\textbf{x}(t)$ and $\textbf{p}(t)$ are functions of time due to particle motion, one has:
\begin{equation}
	\frac{df}{dt} = \frac{\partial f}{\partial t} + \frac{d\textbf{x}}{dt}\cdot\nabla_\textbf{x}f + \frac{d\textbf{p}}{dt}\cdot\nabla_\textbf{p}f = \frac{\partial f}{\partial t} + \textbf{v}\cdot\nabla_\textbf{x}f + \textbf{F}\cdot\nabla_\textbf{p}f\;,
\end{equation}
where $\textbf{v}$ is the particle velocity and $\textbf{F} = d\textbf{p}/dt$ (according to Newton's second law) is the force acting on the particle. 

If interactions are absent, then 
\begin{equation}\label{colllessboltz}
	\frac{df}{dt} = 0\;,
\end{equation}\index{Boltzmann equation!Non-relativistic case}is the \textbf{collisionless Boltzmann equation}, or \textbf{Vlasov equation} \cite{vlasov1945theory}. It mathematically represents the fact that the number of particles in a phase space volume element does not change with time. 
 
The collisionless Boltzmann equation $df/dt = 0$ is a direct consequence of the Liouville theorem, cf. Appendix \ref{App:LiouvilleTh}:
\begin{equation}\label{Liouvilletheorem}
	\frac{d\rho(t,\textbf{x}_i,\textbf{p}_i)}{dt} = 0\;, \qquad i = 1,\cdots,N
\end{equation}
where $\rho(\textbf{x}_i,\textbf{p}_i,t)$ is the $N$-particle distribution function, the meaning of which is provided by
\begin{equation}
	\rho(t,\textbf{x}_i,\textbf{p}_i)d^N\textbf{x}d^N\textbf{p}\;,
\end{equation}
as the probability of finding our system of $N$ particles in a small volume $d^N\textbf{x}d^N\textbf{p}$ of the phase space, centered on $(\textbf{x}_i,\textbf{p}_i)$.\index{Liouville theorem}

If the particles are not interacting, then the probability of finding $N$ particles in some configuration is the product of the individual probabilities. That is, the positions in the phase space of the individual particles are independent stochastic variables. Therefore:
\begin{equation}
	\rho \propto f^N\;, \qquad \frac{d\rho}{dt} = Nf^{N-1}\frac{df}{dt}\;,
\end{equation} 
Using Liouville's theorem \eqref{Liouvilletheorem}, one obtains the collisionless Boltzmann equation $df/dt = 0$. 

In general, the $i$-particle distribution functions for $i = 1,\dots,N$ follow evolutions that are correlated with one another. For example, the evolution of the 1-particle distribution function depends on the 2-particle distribution function. Therefore, we have a hierarchy of evolution equations called the \textbf{Bogoliubov-Born-Green-Kirkwood-Yvon hierarchy}.\index{Bogoliubov-Born-Green-Kirkwood-Yvon hierarchy} A full description of a system of particles based only on the 1-particle distribution function and its evolution equation, namely the Boltzmann equation, is possible only if we make the assumption of \textbf{molecular chaos}.\index{Molecular chaos} This assumption states that the positions and velocities of the particles are uncorrelated. For our forthcoming purposes, this assumption is acceptable since the temperatures of the cosmic thermal bath that we consider are sufficiently low.

\subsection{Boltzmann equation in Cosmology}

In order to comply with general covariance, the Boltzmann equation must be slightly reformulated. First of all, $f$ is now to be considered a function $f(x^\mu,P_\nu)$ of the particle four-position $x^\mu = x^\mu(\lambda)$ and conjugate momentum $P_\nu = P_\nu(\lambda)$, with $\lambda$ being an affine parameter. Note that, since $g^{\mu\nu}P_\mu P_\nu = -m^2c^2$, the distribution function actually depends on 7 variables, not 8.

Secondly, the total derivative with respect to time must be substituted with the derivative with respect to the affine parameter $\lambda$. So, we have:\index{Boltzmann equation!Relativistic case}
\begin{equation}\label{boltzeqrel}
	\frac{df}{d\lambda} = \frac{\partial f}{\partial x^\mu}\frac{dx^\mu}{d\lambda} + \frac{\partial f}{\partial P_\nu}\frac{dP_\nu}{d\lambda} = \frac{\partial f}{\partial x^\mu}P^\mu + \frac{\partial f}{\partial P_\nu}\frac{dP_\nu}{d\lambda}\;.
\end{equation}
Instead of Newton's second law, we use the geodesic equation here, in the absence of other interactions besides gravity:
\begin{equation}
	\frac{dP_\nu}{d\lambda} - \Gamma^\rho_{\nu\sigma}P_\rho P^\sigma = 0\;,
\end{equation}
In Cosmology, in order to comply with the cosmological principle, we must assume that $f = f(t,P)$: the distribution function is allowed to depend only on cosmic time $t$ and on $P^2 \equiv \gamma^{jl}P_jP_l$. However, since $P$ is time-independent, the Boltzmann equation in Cosmology reduces to: 
\begin{equation}
	\boxed{\frac{\partial f(t,P)}{\partial t} = 0}
\end{equation}
which is simply the statement that:
\begin{equation}
	\boxed{f = f(P)}
\end{equation}
the distribution function depends only on the spatial modulus of the conjugate momentum. This is the result that must also hold true in the case of equilibrium, as we have seen in the previous sections.

Note that the above derivation is valid for any $K$. The spatial curvature is absorbed into the definition of $P$.

\hrulefill

\begin{ex} Recall the definition of proper momentum:
\begin{equation}
	p^2 = P^iP_i = \frac{1}{a^2}\gamma^{ij}P_iP_j = \frac{1}{a^2}P^2\,.
\end{equation}
	Show that the Boltzmann equation written using the proper momentum has the following form:
\begin{equation}\label{boltzeqp}
	\boxed{\frac{\partial f}{\partial t} - Hp\frac{\partial f}{\partial p} = 0}
\end{equation}\index{Boltzmann equation!Coupled to FLRW metric} where $f = f(t,p)$. 

\end{ex}

\hrulefill


\subsection{The particle number conservation and the continuity equation}

Rather than directly solving the Boltzmann equation for $f$, it will be more useful to consider its integral in momentum space, weighted with suitable functions of the momentum.


In particular, consider:
\begin{equation}\label{zeromomboltzeq}
	\int d^3\textbf{p}\left(\frac{\partial f}{\partial t} - Hp\frac{\partial f}{\partial p}\right) = 0\;.
\end{equation}

\hrulefill

\begin{ex} Using the definition of the particle number density \eqref{numdenspropmom}, show that Eq.~\eqref{zeromomboltzeq} becomes:
\begin{equation}\label{partnumcons}
	\boxed{\dot n + 3Hn = 0 \quad \Rightarrow \quad \frac{1}{a^3}\frac{d(na^3)}{dt} = 0}
\end{equation}
\end{ex}

\hrulefill

The particle number $na^3$ is conserved. This is an expected result since we have considered $df/d\lambda = 0$. Hence, the particle number is conserved either because there are no interactions and thus no source of creation or destruction of particles, or because we are in equilibrium.

Weighing Eq.~\eqref{boltzeqp} with the energy and integrating in momentum space, we get:
\begin{equation}
	\dot\varepsilon - H\int d^3\textbf{p}\, pE(p)\frac{\partial f}{\partial p} = 0\;.\index{Particle number conservation}
\end{equation}

\hrulefill

\begin{ex} Integrate by parts the above equation and show that it becomes:
\begin{equation}
	\dot \varepsilon + 3H\left(\varepsilon + P\right) = 0\;,
\end{equation}
the continuity equation.\index{Continuity equation}
\end{ex}

\hrulefill

\subsection{The collisional term}\label{Sec:BEcollisional}

Focusing on the interaction \eqref{12into34}, let us take particle 1 as a reference and focus on its Boltzmann equation:
\begin{equation}\label{boltzeqpart1}
	\frac{df_1}{dt} = \frac{1}{P_1^0}C[f]\;.
\end{equation}\index{Boltzmann equation!Collisional term} The four-momentum $P_1^0$ appears because we switch from the affine parameter to the cosmic time.

Taking into account quantum statistics, the collisional term that we have seen earlier can be expressed in the following form:
\begin{eqnarray}\label{collterm}
	\frac{1}{P_1^0}C[f] = \frac{1}{2E_1}\int\frac{d^3\textbf{p}_2}{2E_2}\int\frac{d^3\textbf{p}_3}{(2\pi\hbar)^32E_3}\int\frac{d^3\textbf{p}_4}{(2\pi\hbar)^32E_4}\nonumber\\
	(2\pi)^4\delta^{(3)}(\textbf{p}_1 + \textbf{p}_2 - \textbf{p}_3 - \textbf{p}_4)\delta(E_1 + E_2 - E_3 - E_4)|\mathcal{M}|^2\nonumber\\
	\left[f_3f_4(1 \pm f_1)(1 \pm f_2) - f_1f_2(1 \pm f_3)(1 \pm f_4)\right]\;.
\end{eqnarray}
Note that $E_i = \sqrt{p_i^2 + m_i^2}$. We have to provide several comments. 

$\bullet$ We have incorporated the factors $1/h^3$ in the distribution functions, but not the particle degeneracies $g_{\rm s}$, which are absent in this equation because, in general, $|\mathcal{M}|^2$ depends on the spins and helicities of the particles. Therefore, we could not factor out $g_{\rm s}$. In computing $|\mathcal{M}|^2$, one sums or averages them.

$\bullet$ The integrals are over the particles phase spaces, and the volume element $d^3\textbf{p}/2E$ is covariant. But why does a factor of 2 now appear in the denominator? This actually comes from the normalization of the in and out states that define the scattering matrix and hence $\mathcal M$. For example, in \cite{Weinberg:1995mt}, the cross section is defined without those factors of 2.

$\bullet$ The total four-momentum must be conserved; hence, Dirac deltas appear in the second line. 

$\bullet$ The fundamental physics of the interaction is represented by the amplitude $|\mathcal{M}|^2$. We have assumed symmetric interaction ($T$ invariance): for fixed particles four-momenta, the amplitude probability for $1 + 2 \rightarrow 3 + 4$ is the same as for $1 + 2 \leftarrow 3 + 4$.

$\bullet$ In the last line, we have a balance: the more particles 3 and 4 we have, the more they react and produce particles 1 and 2. And vice-versa. Since our reference particle is 1, we have the combination $f_3f_4 - f_1f_2$. 

$\bullet$ Finally, the contributions of the types $1 + f$ and $1 - f$ are referred to as Bose enhancement and Pauli blocking, respectively. They represent the fact that it is easier to produce a boson rather than a fermion because, due to the Pauli exclusion principle, there are more states available to the former than to the latter.

For the calculations that we are going to perform in the next chapter, we do not need to compute $f$, but rather its integral in momentum space; that is, the number density. Doing this for Eq.~\eqref{boltzeqpart1}, we get:
\begin{eqnarray}\label{boltzeqcollterm}
	\frac{1}{a^3}\frac{d(n_1a^3)}{dt} = \int\frac{d^3\textbf{p}_1}{2E_1}\int\frac{d^3\textbf{p}_2}{2E_2}\int\frac{d^3\textbf{p}_3}{(2\pi\hbar)^32E_3}\int\frac{d^3\textbf{p}_4}{(2\pi\hbar)^32E_4}\nonumber\\
	(2\pi)^4\delta^{(3)}(\textbf{p}_1 + \textbf{p}_2 - \textbf{p}_3 - \textbf{p}_4)\delta(E_1 + E_2 - E_3 - E_4)|\mathcal{M}|^2\left[f_3f_4 - f_1f_2\right]\;,
\end{eqnarray}
where we have neglected the Pauli blocking and Bose enhancement terms since the temperatures of our interest are large enough (we do not expect Bose condensation, for example). Note that the left hand side is the time derivative of the number of particles of species 1. See also Eq.~\eqref{partnumcons}. 

From here, one can see how the known thermal distributions are not equilibrium ones in the expanding universe. In fact, taking the MB distribution for all the species (as they are supposed to be in equilibrium), one has in the collisional term:
\begin{eqnarray}
	f_3 f_4 - f_1 f_2 = \frac{1}{h^6}e^{-(E_1 + E_2)/(k_{\rm B}T)}\left[e^{(\mu_3 + \mu_4)/(k_{\rm B}T)} - e^{(\mu_1 + \mu_2)/(k_{\rm B}T)}\right]\;,
\end{eqnarray}
where we have used $E_1 + E_2 = E_3 + E_4$. Now, this is zero if:
\begin{equation}
	\mu_1 + \mu_2 = \mu_3 + \mu_4\,,
\end{equation}
that is condition \eqref{chemicalequilibrium}. On the other hand: 
\begin{align}
	n = \frac{g}{2\pi^2\hbar^3}e^{\mu/(k_{\rm B}T)}\int_0^{\infty}dp\,p^2e^{-E(p)/(k_{\rm B}T)}\,,
\end{align}
is time-independent only for the two special cases seen earlier: $\mu = m = 0$ or $\mu = mc^2 \gg p^2c^2$. However, we have also seen that if the overall interaction rate is large $\Gamma \gg H$, then it is still possible to employ the thermal distribution for computing the number density and the energy density, even if they are slightly out of equilibrium.

For the applications we address in the next chapter, we will consider some interactions that go out of equilibrium (thereby leading, for example, to the production of Helium nuclei); however, other interactions maintain the particles in $1 + 2 \longleftrightarrow 3 + 4$ in \textit{kinetic} equilibrium, allowing us to use thermal distributions, and the Boltzmann equation effectively becomes an evolution equation for the chemical potential $\mu$. 

To this end, let $n^{(0)}$ be the number density for $\mu = 0$:
\begin{align}
	n = \frac{g}{2\pi^2\hbar^3}e^{\mu/(k_{\rm B}T)}\int_0^{\infty}dp\,p^2e^{-E(p)/(k_{\rm B}T)} \equiv e^{\mu/(k_{\rm B}T)}n^{(0)}\,.
\end{align}
We can write:
\begin{eqnarray}
	f_3f_4 - f_1f_2 = \frac{1}{h^6}e^{-(E_1 + E_2)/(k_{\rm B}T)}\left(\frac{n_3n_4}{n_3^{(0)}n_4^{(0)}} - \frac{n_1n_2}{n_1^{(0)}n_2^{(0)}}\right)\;,
\end{eqnarray}
and then:
\begin{align}
	\frac{1}{a^3}\frac{d(n_1a^3)}{dt} = \left(\frac{n_3 n_4}{n_3^{(0)}n_4^{(0)}} - \frac{n_1n_2}{n_1^{(0)}n_2^{(0)}}\right)\times\nonumber\\
    \int\frac{d^3\textbf{p}_1}{2h^3E_1}\int\frac{d^3\textbf{p}_2}{2h^3E_2}\int\frac{d^3\textbf{p}_3}{(2\pi\hbar)^32E_3}\int\frac{d^3\textbf{p}_4}{(2\pi\hbar)^32E_4}\nonumber\\
	(2\pi)^4\delta^{(3)}(\textbf{p}_1 + \textbf{p}_2 - \textbf{p}_3 - \textbf{p}_4)\delta(E_1 + E_2 - E_3 - E_4)|\mathcal{M}|^2e^{-(E_1 + E_2)/(k_{\rm B}T)}\;.
\end{align}
The interaction amplitude integrated in the phase space with the Boltzmann weight:
\begin{eqnarray}
	\langle\sigma v\rangle \equiv \frac{1}{n_1^{(0)}n_2^{(0)}}\int\frac{d^3\textbf{p}_1}{h^32E_1}\int\frac{d^3\textbf{p}_2}{h^32E_2}\int\frac{d^3\textbf{p}_3}{h^32E_3}\int\frac{d^3\textbf{p}_4}{h^32E_4}\nonumber\\
	(2\pi)^4\delta^{(3)}(\textbf{p}_1 + \textbf{p}_2 - \textbf{p}_3 - \textbf{p}_4)|\mathcal{M}|^2e^{-(E_1 + E_2)/(k_{\rm B}T)}\;,
\end{eqnarray}
is called \textbf{thermally averaged cross section}\index{Cross section!Thermally averaged}:

We can thus recast the integrated collisional Boltzmann's equation \eqref{boltzeqcollterm} as follows:
\begin{equation}\label{Boltzmannequationfinal}
	\boxed{\frac{1}{a^3}\frac{d(n_1a^3)}{dt} = n_1^{(0)} n_2^{(0)}\langle\sigma v\rangle\left(\frac{n_3n_4}{n_3^{(0)}n_4^{(0)}} - \frac{n_1n_2}{n_1^{(0)}n_2^{(0)}}\right)}
\end{equation}
This will be our fundamental equation to explore in the next chapter.

When interactions are negligible, then $\langle\sigma v\rangle = 0$ and the number of particles per comoving volume, $n_1a^3$, is constant, and so $n_1 \propto 1/a^3$.

This also happens when the interaction rate is very high (much larger than $H$). The LHS is $\propto H$, so if $\langle\sigma v\rangle \gg H$ the equation can hold true only if the rhs of \eqref{Boltzmannequationfinal} vanishes, and again $n_1a^3$ is constant. This is the condition for LTE. 

This similarity between a high interaction rate and no interaction at all is intriguing and is responsible for the very low degree of CMB polarization, as we shall see in Sec.~\ref{Chap:CMBEvo}.
 
\clearpage
\chapter{Applications of thermal equilibrium and kinetic theory in the expanding universe}\label{Chap:ThermalHistory}

{\rightskip=3truepc\leftskip=3truepc\noindent
{\it It was a pleasure to burn.}
\vskip 0.10 in
\centerline{\it ---Ray Bradbury, Fahrenheit 451}
\vskip 0.20 in
}

In this chapter, we address the primordial synthesis of Helium, or Big Bang Nucleosynthesis (BBN), the recombination of protons and electrons in neutral hydrogen atoms, and the relic abundance of Cold Dark Matter (CDM) particles.

\section{Big-Bang Nucleosynthesis}\label{Sec:BBN}

Big-Bang Nucleosynthesis (BBN) is the formation of light elements, mainly Helium, in the primordial plasma. It took place at a temperature of about $0.1$ MeV, which corresponds to a redshift $z \approx 10^{9}$.\index{Big Bang Nucleosynthesis} Since $T \propto 1/a$ and in the radiation-dominated epoch $a \propto \sqrt{t}$, we can relate time and temperature as follows, using Eq. \eqref{endensradgstar1}:
\begin{equation}\label{t2T4relation1}
	\frac{1}{4t^2} = H^2 = \frac{4\pi^3G(k_{\rm B}T)^4}{45c^2(\hbar c)^3}g_*\;.
\end{equation} 

\hrulefill

\begin{ex} Show from Eq.~\eqref{t2T4relation1} that:
\begin{equation}
	t \approx 157\left(\frac{0.07\mbox{ MeV}}{k_{\rm B}T}\right)^2\left(\frac{10}{g_*}\right)^{1/2}\;\textrm{s}\;.
\end{equation}
\end{ex}

\hrulefill

The value $g_* = 10$ is introduced as an order-of-magnitude estimate, and we will determine it more precisely later. Hence, primordial nucleosynthesis occurs only a few minutes after the Big Bang.

The characters of the story are: photons, electrons, positrons, neutrinos, antineutrinos, protons, and neutrons.\footnote{Antiprotons have already disappeared after annihilation with protons.} In particular, we focus on neutrons. Their interactions relevant to BBN are:
\begin{align}
\label{betadecay}	
    n &\longleftrightarrow p + e^- + \bar{\nu}_e \qquad &(\mbox{Beta decay})\;,\\
	p + e^- &\longleftrightarrow \nu_e + n \qquad &(\mbox{Electron capture})\;,\\
	p + \bar{\nu}_e &\longleftrightarrow e^+ + n \qquad &(\mbox{Inverse beta decay})\;,\\
	p + n &\longleftrightarrow D + \gamma \qquad &(\mbox{Deuteron formation})\,.
\end{align}
As we have seen, at a temperature of about 1 MeV, neutrinos decouple. Therefore, the $\beta$-decay reaction in Eq.~\eqref{betadecay} (from left to right) takes over, and the number of neutrons starts to diminish. On the other hand, neutrons can also be captured by protons to form deuterium nuclei. The BBN is essentially a competition in capturing neutrons before they decay. Things are also a little bit complicated by the fact that, in the meanwhile, electrons and positrons start to annihilate (from about 0.3 MeV down to 16 keV \cite{Thomas:2019ran}) and become non-relativistic (at about 0.05 MeV).

As mentioned earlier, when we introduced the baryon-to-photon ratio, the fact that there is a billion photons for each proton implies that even if the temperature of the thermal bath is lower than the binding energy of the deuteron (2.2 MeV), there are still many photons with energy higher than 2.2 MeV that can break newly formed deuterium nuclei. This is also known as \textbf{the deuteron bottleneck}.\index{Deuteron bottleneck} According to our rough estimate from the previous chapter, we may estimate BBN to start at a temperature $\approx 2.2/27 \approx 0.08$ MeV.  

It is not the deuteron's fault if BBN takes place at a temperature much smaller than 2.2 MeV. Rather, the smallness of $\eta_{\rm b}$ is the culprit.

\subsection{The deuteron bottleneck}

Let us start by considering the reaction:
\begin{equation}
	p + n \longleftrightarrow D + \gamma\;,
\end{equation}
for sufficiently high temperatures, such that chemical equilibrium is established:
\begin{equation}
	\mu_p + \mu_n = \mu_D + \mu_\gamma\;.
\end{equation}
From Eq. \eqref{Boltzmannequationfinal}, at (approximately) equilibrium, we have:
\begin{equation}\label{SahaequationDeuterium}
	\frac{n_D n_\gamma}{n^{(0)}_Dn^{(0)}_\gamma} = \frac{ n_p n_n}{n^{(0)}_pn^{(0)}_n}\;.
\end{equation}
For large enough temperatures, we can neglect the photon chemical potential. This is generally true for perfect thermal radiation since the number of photons is not conserved.\footnote{The concept of chemical potential, although rather fundamental, is often not easy to grasp fully. See the nice discussion about it in \cite{baierlein2001elusive}. For thermal radiation, one can conclude that the chemical potential for photons is zero by combining Planck's law with Bose-Einstein statistics. In Appendix \ref{App:Thermaldistr}, we derive the Bose-Einstein statistics, and one can appreciate there that the chemical potential arises as a Lagrange multiplier enforcing particle number conservation. Since, for thermal radiation, the number of photons is not conserved, their chemical potential is zero. At present, the constraint on $\mu$ is still the COBE-FIRAS one.} So, $n_\gamma = n^{(0)}_\gamma$. We then obtain:
\begin{equation}
	\frac{n_D}{ n_p n_n} = \frac{n^{(0)}_D}{n^{(0)}_pn^{(0)}_n}\;.
\end{equation}
For $k_{\rm B}T \ll m_pc^2 \approx 1$ GeV, but much larger than 1 MeV, all the particles in play are in equilibrium and non-relativistic. So, using Eq.~\eqref{partnumdensthermnonrel}, we obtain:
\begin{equation}\label{SahaequationDeuterium2}
	\frac{ n_D}{ n_p n_n} = \frac{n^{(0)}_D}{n^{(0)}_pn^{(0)}_n} = \frac{g_D}{g_pg_n}\left(\frac{2\pi\hbar^2m_D}{m_pm_nk_{\rm B}T}\right)^{3/2}e^{-(m_D - m_p - m_n)c^2/(k_{\rm B}T)}\;.
\end{equation}
The deuteron has spin $1$, whereas protons and neutrons have spin $1/2$. Therefore:
\begin{equation}\label{SahaequationDeuterium3}
	\frac{ n_D}{ n_p n_n} = \frac{3}{4}\left(\frac{2\pi\hbar^2m_D}{m_pm_nk_{\rm B}T}\right)^{3/2}e^{B_D/(k_{\rm B}T)}\;,
\end{equation}
where $B_D = 2.22$ MeV is the modulus of the deuteron binding energy. We can write the ratio of the masses as follows:
\begin{equation}
	\frac{m_D}{m_pm_n} = \frac{m_p + m_n - B_D/c^2}{m_pm_n} = \frac{1}{m_n} + \frac{1}{m_p} - \frac{B_D}{m_pm_nc^2}\;.
\end{equation}
Introducing the neutron-proton mass difference:
\begin{equation}\label{Qenergy}
	Q \equiv (m_n - m_p)c^2 = 1.239 \mbox{ MeV}\;,
\end{equation}
we can write
\begin{equation}
	\frac{m_D}{m_pm_n} = \frac{1}{m_p(1 + Q/m_pc^2)} + \frac{1}{m_p} - \frac{B_D}{m_p^2(1 + Q/m_pc^2)c^2}\;.
\end{equation}
Since $Q/m_pc^2 \approx B_D/m_pc^2 \approx 10^{-3}$, we approximate:
\begin{equation}
	\frac{m_D}{m_pm_n} \approx \frac{2}{m_p}\;.
\end{equation} 
Introducing the abundances:
\begin{align}
    X_D \equiv \frac{n_D}{n_{\rm b}}\,, \quad X_p \equiv \frac{n_p}{n_{\rm b}}\,, \quad X_n \equiv \frac{n_n}{n_{\rm b}}\,, 
\end{align}
we have:
\begin{equation}
	X_D = \frac{3}{4}X_pX_nn_{\rm b}\left(\frac{4\pi\hbar^2}{m_pk_{\rm B}T}\right)^{3/2}e^{B_D/(k_{\rm B}T)}\;.
\end{equation}
Introducing the baryon-to-photon number $n_{\rm b} = \eta_{\rm b}  n_\gamma = \eta_{\rm b} n_\gamma^{(0)}$, we can write:
\begin{equation}
	X_D = \frac{12\zeta(3)}{\sqrt{\pi}}X_pX_n\eta_{\rm b}\left(\frac{k_{\rm B}T}{m_pc^2}\right)^{3/2}e^{B_D/(k_{\rm B}T)}\;.
\end{equation}
Neglecting $O(1)$ coefficients, we have:
\begin{equation}\label{SahaequationDeuterium5}
	X_D \approx \eta_{\rm b}\left(\frac{k_{\rm B}T}{m_pc^2}\right)^{3/2}e^{B_D/(k_{\rm B}T)}\;.
\end{equation}
Even when $k_{\rm B}T \approx B_D$ the abundance of deuterium is very small, due to $\eta_{\rm b}$. This is the deuteron bottleneck. 

One can define the temperature $T_{\rm BBN}$ at which BBN takes place as the one at which the bottleneck is overcome. This is because, as numerical calculations show, newly formed deuterons rapidly combine into Helium nuclei.

So, let us define $T_{\rm BBN}$ as the temperature at which $X_D$ becomes $O(1)$:
\begin{equation}\label{SahaequationDeuterium6}
	\log\left(\eta_{\rm b}\right) + \frac{3}{2}\log\left(\frac{k_{\rm B}T_{\rm BBN}}{m_pc^2}\right) \approx -\frac{B_D}{k_{\rm B}T_{\rm BBN}}\;.
\end{equation}
Numerically solving this equation, one finds:
\begin{equation}
	\boxed{k_{\rm B}T_{\rm BBN} \approx 0.07 \mbox{ MeV}}
\end{equation}
Note that the above derivation presupposes chemical equilibrium, but this does not last forever. When the reaction $p + n \longleftrightarrow D + \gamma$ unbalances and a deuteron is formed, we must use the full Boltzmann equation \eqref{Boltzmannequationfinal}. However, it can be shown numerically that the two equations provide compatible results up to the moment when equilibrium is broken.

Finally, note that only two-body reactions are efficient during the epoch considered here, so Helium-4 is formed only after a chain of two-body processes that is shown in the next subsection. It is mainly for this reason that the formation of the deuteron provides a bottleneck.

\subsection{The neutron abundance}

After the deuteron bottleneck is overcome, Helium-4 is rapidly formed according to the following two-body reactions:
\begin{eqnarray}
	p + n &\longrightarrow & D + \gamma\;,\\
	D + D &\longrightarrow & {}^3\textrm{He} + n\;,\\
	D + {}^3{\rm He} &\longrightarrow & {}^4{\rm He} + p\;.
\end{eqnarray}
In order to simplify calculations, we assume that all the deuterons and neutrons available are instantaneously converted into Helium-4 nuclei. This is not what actually occurred, of course, but it is something that we can easily compute, and it turns out to be a good approximation.

Note that Lithium ${}^7{\rm Li}$ and Beryllium ${}^7{\rm Be}$ are also produced: 
\begin{eqnarray}
	{}^3\textrm{H} + {}^4\textrm{He} &\longrightarrow & {}^7\textrm{Li} + \gamma\;,\\
	{}^3\textrm{H} + {}^4\textrm{He} &\longrightarrow & {}^7\textrm{Be} + \gamma\;,\\
	{}^7\textrm{Be} + n &\longrightarrow & {}^7\textrm{Li} + p\;,\\
    {}^7\textrm{Li} + p &\longrightarrow & {}^4\textrm{He} + {}^4\textrm{He}\;,
\end{eqnarray}
but in a tiny fraction (one billionth of the hydrogen abundance for Lithium). However, measuring the Lithium abundance in the universe is a very important independent measure of $\Omega_{\rm b0}$ \cite{Fields:2011zzb}. 

The main prediction of BBN is on the abundance of ${}^4{\rm He}$ because this is the element that is mainly formed, due to both its high binding energy per nucleon, which is about 7 MeV, and the fact that there is not much time for forming heavier nuclei since the thermal bath is rapidly cooling and $\eta_{\rm b}$ is so small.

Our objective is thus to determine the neutron abundance at $T_{\rm BBN}$. This is done by considering three reactions. The electron and positron capture:
\begin{equation}
	p + e^- \longleftrightarrow n + \nu_e\;,
\end{equation}
\begin{equation}
	n + e^+ \longleftrightarrow p + \bar\nu_e\;,
\end{equation}
and the $\beta$-decay:
\begin{equation}
	n \longleftrightarrow p + e^- + \bar\nu_e\;.
\end{equation}
The $\beta$-decay can be easily accounted for by considering an exponential suppression of the neutron abundance once this has been calculated. Note that the three reactions are governed by the weak interaction.

For temperatures $k_{\rm B}T \gg 1$ MeV, but much smaller than 1 GeV (the proton mass), protons and neutrons are non-relativistic, and all the particles are in kinetic and chemical equilibrium. So, we have:
\begin{align}
	\mu_p + \mu_{e^-} = \mu_n + \mu_\nu\;,\\
    \mu_n + \mu_{e^+} = \mu_p + \mu_{\bar\nu}\,,\\
    \mu_n = \mu_p + \mu_{e^-} + \mu_{\bar\nu}\,.
\end{align}
Taking into account that, due to reactions such as $\bar\nu + \nu = e^+ + e^-$ and $e^+ + e^- = \gamma + \gamma$, one has that $\mu_\nu = -\mu_{\bar\nu}$ and $\mu_{e^-} = -\mu_{e^+}$, and thus the three equations above are essentially one. Nonetheless, even when the aforementioned reactions are no longer efficient, it turns out to be a good approximation to neglect the chemical potentials of electrons and neutrinos. Hence:
\begin{align}
    \mu_e = \mu_\nu = 0\,, \qquad \mu_p = \mu_n\,.
\end{align}
The unknown nucleon chemical potential can be bypassed by computing the relative abundance:
\begin{equation}\label{Sahaeqprotonneutron}
	\frac{ n_p}{ n_n} = \frac{n^{(0)}_p}{n^{(0)}_n} = \left(\frac{m_p}{m_n}\right)^{3/2}e^{Q/(k_{\rm B}T)}\;,
\end{equation}
where $Q = 1.239$ MeV is the neutron-proton mass difference. When 1 GeV $\gg k_{\rm B}T \gg Q$, there are as many protons as neutrons. This will be our initial condition for solving the Boltzmann equation for neutrons. The latter has to be done When $k_{\rm B}T$ drops and becomes comparable with $Q$. 

As we mentioned earlier, in order to reliably predict the residual abundance of neutrons, we cannot resort to kinetic and chemical equilibrium forever. We need to solve the full Boltzmann equation. On the other hand, we use Eq.~\eqref{Boltzmannequationfinal} and assume that protons, electrons, and neutrinos are in kinetic equilibrium, so we can use their thermal distributions. This assumption is acceptable for protons and electrons since they interact via Coulomb scattering at a high rate, and for neutrinos since they do not interact at all.\footnote{Neutrinos are considered massless and have a vanishing chemical potential here, so when they decouple from the primordial plasma, they maintain their thermal distribution.}  

Continuing to assume that electrons and neutrinos have vanishing chemical potentials, we then have:
\begin{equation}\label{Boltzmannequationprotonneutron2}
	\frac{1}{a^3}\frac{d( n_na^3)}{dt} = n^{(0)}_\nu\langle\sigma v\rangle\left(\frac{ n_pn^{(0)}_n}{n^{(0)}_p} -  n_n\right)\;,
\end{equation}
for $p + e^- \longleftrightarrow n + \nu_e$ and:
\begin{equation}
	\frac{1}{a^3}\frac{d( n_na^3)}{dt} = n^{(0)}_{e^+}\langle\sigma v\rangle\left(\frac{ n_pn^{(0)}_n}{n^{(0)}_p} -  n_n\right)\;,
\end{equation}
for $n + e^+ \longleftrightarrow p + \bar\nu_e$. The two equations look very similar. The cross-section of the two processes is comparable (it is a weak interaction process), and the only difference is in the factors $n^{(0)}_\nu$ and $n^{(0)}_{e^+}$. On the other hand, when $k_{\rm B}T \simeq Q$, positrons are non-relativistic, and so their $n^{(0)}_{e^+}$ is exponentially suppressed with respect to $n^{(0)}_\nu$.\footnote{Throughout BBN, positrons have not yet been annihilated away, and their abundance is the same as that of electrons down to about 20 keV \cite{Thomas:2019ran}} Therefore, only Eq.~\eqref{Boltzmannequationprotonneutron2} is relevant for our calculation, and we will focus only on it from now on. Let:
\begin{equation}\label{Xnlambdanpdefs}
	X_n \equiv \frac{ n_n}{n_n +  n_p}\;, \qquad n^{(0)}_\nu\langle\sigma v\rangle \equiv \lambda_{np}\;.
\end{equation}\index{Neutron abundance}

\hrulefill

\begin{ex} Show that Eq.~\eqref{Boltzmannequationprotonneutron2} can be cast, using Eq.~\eqref{Xnlambdanpdefs}, as follows:
\begin{equation}\label{Boltzmannequationprotonneutron3}
	\frac{dX_n}{dt} = \lambda_{np}\left[(1 - X_n)e^{-Q/(k_{\rm B}T)} - X_n\right]\;.
\end{equation}
\end{ex}

\hrulefill 

In order to solve (numerically) Eq.~\eqref{Boltzmannequationprotonneutron3}, we introduce the variable:
\begin{equation}
	x \equiv \frac{Q}{k_{\rm B}T}\;.
\end{equation}

\hrulefill

\begin{ex} Show that the time derivative of $x \equiv Q/(k_{\rm B}T)$ can be cast as follows:
\begin{equation}\label{xdot}
	\frac{dx}{dt} = Hx = x\sqrt{\frac{8\pi G\varepsilon}{3c^2}}\;.
\end{equation}
\end{ex}

\hrulefill 

Since we are deep in the radiation-dominated era, the energy density in Eq.~\eqref{xdot} can be written as follows:
\begin{equation}\label{endensradgstar}
	\varepsilon = \frac{\pi^2(k_{\rm B}T)^4}{30(\hbar c)^3}g_*\,,
\end{equation}
where the effective number of relativistic degrees of freedom was introduced in the last Chapter. For temperatures larger than 1 MeV (up to 200 MeV), $g_*$ is roughly a constant, and its value is:
\begin{equation}\label{gstarcalculation}
	g_* = 2 + \frac{7}{8}(3 + 3 + 2 + 2) = 10.75\;,
\end{equation}
because there are two degrees of freedom coming from the photons, 3 + 3 coming from neutrinos and anti-neutrinos, and 2 + 2 coming from electrons and positrons. We have considered, as usual, just a single spin state for each neutrino and anti-neutrino and we have not taken into account yet the temperature difference between photons and neutrinos, as electrons and positrons have not yet annihilated appreciably.\index{Effective number of relativistic degrees of freedom}

\hrulefill

\begin{ex} Show that the Hubble parameter can thus be cast in the following form:
\begin{equation}
	H^2 = \frac{8\pi G}{3c^2}\frac{\pi^2(k_{\rm B}T)^4}{30(\hbar c)^3}g_* = \frac{4\pi^3Gg_*Q^4}{45c^2(\hbar c)^3}\frac{1}{x^4}\;.
\end{equation}
Calculate:
\begin{equation}
	H(x = 1) = 1.13\; s^{-1}\;.
\end{equation}
Show that Eq.~\eqref{Boltzmannequationprotonneutron3} can be cast as:
\begin{equation}\label{Boltzmannequationprotonneutron4}
	\frac{dX_n}{dx} = \frac{x\lambda_{np}}{H(x = 1)}\left[e^{-x} - X_n(1 + e^{-x})\right]\;.
\end{equation}
\end{ex}

\hrulefill

We only need a last piece of information, the interaction rate $\lambda_{np}$:
\begin{equation}\label{lambdanpscatteringrate}
	\lambda_{np} = \frac{255}{\tau_n x^5}(12 + 6x + x^2)\;,
\end{equation}
where: 
\begin{equation}
	\tau_n = 886.7\; \mbox{s}\;,
\end{equation}
is the neutron lifetime. The above scattering rate can be found in \cite{1988kteu.book.....B}.

\begin{figure}[htbp]
	\centering
	\includegraphics[width=\columnwidth]{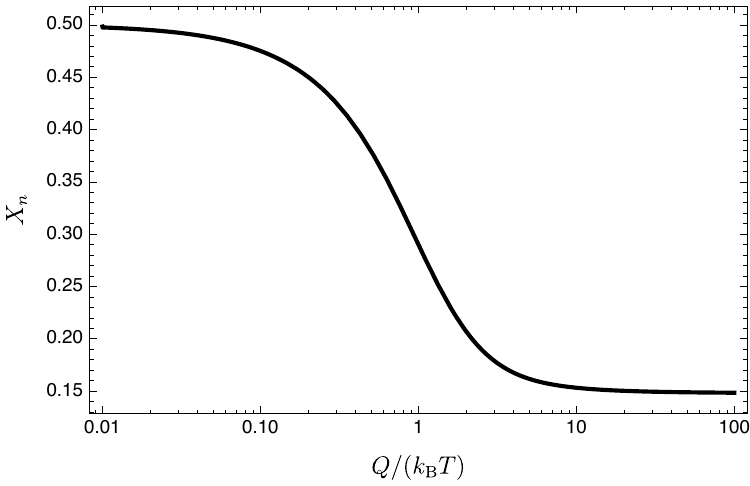}
	\caption{Evolution of $X_n$ from Eq.~\eqref{Boltzmannequationprotonneutron4}.}
	\label{Fig:Xnevo}
\end{figure}

We can now solve numerically Eq.~\eqref{Boltzmannequationprotonneutron4} together with Eq.~\eqref{lambdanpscatteringrate}. The initial condition on $X_n$ is, of course, $X_n(x \to 0) = 1/2$, as we discussed after Eq.~\eqref{Sahaeqprotonneutron}. We plot the evolution of $X_n$ in Fig.~\ref{Fig:Xnevo}.

As we can appreciate from Fig.~\ref{Fig:Xnevo}, $X_n$ settles on a value of $\approx 0.15$ for $x \approx 10$, corresponding to $k_{\rm B}T \approx 0.12$ MeV. This is below the mass-energy of electrons and positrons, which we have considered to be relativistic when computing $g_*$ in Eq. \eqref{gstarcalculation}. On the other hand, we have seen in Fig. \ref{Fig:gammaPlot} that a species in thermal equilibrium becomes effectively non-relativistic when $k_{\rm B}T \approx mc^2/10$; this is $\approx 0.05$ MeV for electrons and positrons, so, from this side, our computation of $g_*$ is consistent. Moreover, at $k_{\rm B}T \approx 0.12$ MeV the electron-positron annihilation has just started \cite{Thomas:2019ran}, so there is still no appreciable difference in the neutrinos and photons temperatures. So, also from this side, our computation of $g_*$ is consistent.

Nonetheless, $X_n \approx 0.15$ is not a good approximation of the residual abundance of neutrons at $T_{\rm BBN}$ because there are other relevant processes that contribute to depleting or enhancing the number of neutrons. Namely:
\begin{eqnarray}
	n \rightarrow p + e^- + \bar{\nu}\;,\\
	n + p \rightarrow D + \gamma\;,
\end{eqnarray}
i.e., the $\beta$-decay and the neutron-proton capture. These processes lower the number of free neutrons. The Helium-3 formation:
\begin{equation}
	D + D \rightarrow {}^3{\rm He} + n\;,
\end{equation}
puts another neutron back into play and the formation of Helium-4:
\begin{equation}
	{}^3{\rm He} + D \rightarrow {}^4{\rm He} + p\;,
\end{equation}
reinserts into play another proton, which helps in capturing neutrons, thus lowering their number.

The correct way to calculate the abundances of the light elements produced during BBN is to consider all the coupled Boltzmann equations for all relevant reactions taking place. This is, of course, done numerically \cite{Wagoner:1972jh}.

We now show that correcting $X_n = 0.15$ by taking into account only the $\beta$-decay gives a result that is surprisingly in agreement with the more reliable result that takes into account all the reactions. 

What we have to do is weigh $X_n = 0.15$ with $\exp(-t_{\rm BBN}/\tau_n)$,\footnote{This exponential weight comes from the Poisson distribution, which governs stochastic processes such as the $\beta$-decay. We derive it in Appendix \ref{App:Poisson}} where $t_{\rm BBN}$ is the time corresponding to $k_{\rm B}T_{\rm BBN} = 0.07$ MeV, i.e., the time at which BBN starts. Moreover, we suppose that at this time all the free neutrons are immediately captured and produce Helium-4. Thereby, estimating $X_n$ gives a direct estimation of $X_{{}^4\rm He}$.

At $k_{\rm B}T_{\rm BBN} = 0.07$ MeV, electrons and positrons are quite non-relativistic and their annihilation has not yet started. Therefore, the effective relativistic degrees of freedom are:
\begin{equation}
	g_* = 2 + \frac{7}{8}6 = 7.25\;.
\end{equation}

\hrulefill

\begin{ex} Show from Eq.~\eqref{t2T4relation1} that:
\begin{equation}
	t = 184\left(\frac{0.07\mbox{ MeV}}{k_{\rm B}T}\right)^2\;\textrm{s}\;.
\end{equation}
\end{ex}

\hrulefill

Therefore, the expected abundance of neutrons at $k_{\rm B}T_{\rm BBN} = 0.07$ MeV is:
\begin{equation}
	X_n(T_{\rm BBN}) = 0.15 \cdot e^{-184/886.7} \approx 0.12\;.
\end{equation}
Assuming that all the neutrons end up in Helium-4 nuclei, we have the prediction:
\begin{equation}
	Y_P \equiv 4X_{{}^4\rm He} \equiv \frac{4n_{{}^4\rm He}}{n_{\rm b}} = 2X_n(T_{\rm BBN}) = 0.24\;. 
\end{equation}
In order to understand the numerical coefficients, consider that after BBN, we have:
\begin{equation}
	n_{\rm b} = n_p + 4n_{{}^4\rm He}\;,
\end{equation}
i.e., baryons (protons and neutrons) are present only within hydrogen nuclei (1 proton) or Helium nuclei (2 protons and 2 neutrons, hence the coefficient 4 above). So, dividing by $n_{\rm b}$:
\begin{equation}
	1 = \frac{n_p}{n_{\rm b}} + 4\frac{n_{{}^4\rm He}}{n_{\rm b}} = X_{\rm H} + 4X_{{}^4\rm He} \equiv X_P + Y_P\;.
\end{equation}
The factor 2 in front of $X_n$ comes from: 
\begin{equation}
	X_n = \frac{n_n}{n_{\rm b}} = \frac{2n_{{}^4\rm He}}{n_{\rm b}} = 2X_{{}^4\rm He}\,,
\end{equation}
since a Helium-4 nucleus contains two neutrons.\index{Primordial Helium}

At a 68\% confidence level, the observed value is \cite{Peimbert:2016bdg}:
\begin{equation}
	Y_P = 0.2446 \pm 0.0029\;,
\end{equation}
which is in very good agreement with our ``back-of-the-envelope'' calculation. 

\section{Recombination and decoupling}\label{Sec:Recombination}

\textbf{Recombination} is the process by which neutral hydrogen is formed via the combination of protons and electrons. \textbf{Decoupling} is the epoch when photons stop interacting with free electrons, their mean free path becomes larger than the Hubble radius, and we are able to detect them as CMB coming from the \textbf{last scattering surface}. For the two events, the relevant interactions are:\index{Recombination}\index{Photons!Decoupling from electrons}\index{Last scattering surface}
\begin{eqnarray}
	p + e^- &\longleftrightarrow& H + \gamma\;,\\
	e^- + \gamma &\longleftrightarrow& e^- + \gamma \qquad (\mbox{Compton/Thomson scattering})\;.
\end{eqnarray}
Recombination and decoupling temporally occur close to each other for the following reason. At sufficiently low temperatures, which we will calculate, photons are no longer able to break hydrogen atoms, and so these start to form in larger numbers (recombination). The number of free electrons dramatically drops because they are captured into hydrogen atoms, and the Thomson scattering rate goes to zero (decoupling). The seminal paper on recombination is \cite{Peebles:1968ja}.

In order to estimate the epoch of recombination, let us use our simple calculation: $B = 27k_{\rm B}T$, with 13.6 eV, the ionization energy of the hydrogen atom, as binding energy. One gets $k_{\rm B}T \approx 0.5$ eV. 

Now, using chemical equilibrium, along with the vanishing chemical potential for the photon:
\begin{equation}
	\frac{n_en_p}{n_H} = \frac{n_e^{(0)}n_p^{(0)}}{n_H^{(0)}}\;.
\end{equation}
Let us also assume the neutrality of the universe, i.e. $n_e = n_p$, and define the free electron fraction as:
\begin{equation}\label{Xedef}
	X_e \equiv \frac{n_e}{n_e + n_H} = \frac{n_p}{n_p + n_H}\;.
\end{equation}\index{Free electron fraction}
This ratio represents the fraction of free electrons ($n_e$) to the total, which includes those bound in atoms (of which we consider only hydrogen, neglecting Helium).

\hrulefill

\begin{ex} Considering that the degeneracy of the hydrogen atom, in the state $1s$, is $g_{1s} = 4$ (it has two hyperfine states, one of spin 0 and the other of spin 1), show that Saha equation can be written as follows:
\begin{equation}
	\frac{X_e^2}{1 - X_e} = \frac{1}{n_e + n_H}\left(\frac{m_em_pk_{\rm B}T}{2m_H\pi\hbar^2}\right)^{3/2}e^{-(m_e + m_p - m_H)c^2/(k_{\rm B}T)}\;.
\end{equation}
\end{ex}

\hrulefill

Consider the contribution in the denominator of the right hand side as: 
\begin{equation}
	n_e + n_H = n_p + n_H = n_{\rm b} = \eta_{\rm b} n_\gamma = \eta_{\rm b}\frac{2\zeta(3)}{\pi^2\hbar^3c^3}(k_{\rm B}T)^3 \approx 10^{-9}\frac{(k_{\rm B}T)^3}{\hbar^3c^3}\;.
\end{equation}
We have again used that $n_e = n_p$ and that all the baryons are either free protons or protons in a hydrogen atom (recall that we are neglecting the abundance of Helium nuclei). 

We again neglect the mass difference outside of the exponential and write:
\begin{equation}\label{Sahaeqrecombination}
	\frac{X_e^2}{1 - X_e} \approx 10^9 \left(\frac{m_ec^2}{2\pi k_{\rm B}T}\right)^{3/2}\exp\left(-\frac{13.6 \mbox{ eV}}{k_{\rm B}T}\right)\;,
\end{equation}
where we used 
\begin{equation}
	\varepsilon_0 \equiv (m_e + m_p - m_H)c^2 = 13.6 \mbox{ eV}\;,
\end{equation}
i.e. the ionization energy of the hydrogen atom.

The high photon-to-baryon number delays recombination as well as it delayed BBN. Indeed, when $k_{\rm B}T = 13.6$ eV, we get from Eq.~\eqref{Sahaeqrecombination}:
\begin{equation}
	\frac{X_e^2}{1 - X_e} \approx 10^{15}\;.
\end{equation}
From this, one gets that $X_e \approx 1$. This means that even when the energy of the thermal bath drops below the ionization energy of the hydrogen atom, no hydrogen is formed and the electrons remain free. This, again, happens because there are still many photons with energy much higher than 13.6 eV. In Tab.~\ref{Tab:Recombtable} we show numerical calculations of $X_e$ from Eq.~\eqref{Sahaeqrecombination} in order to provide insight into the time of recombination.

\begin{table}[ht]
	\begin{center}
    \begin{tabular}{ | l | l |}
    \hline
    $k_{\rm B}T\mbox{ [eV]}$ & $X_e$ \\ \hline
    0.5 & 1\\ \hline
    0.38 & 0.995\\ \hline
    0.36 & 0.970\\ \hline
    0.34 & 0.819\\ \hline
    0.32 & 0.434\\ \hline
    0.30 & 0.137\\ \hline
    0.29 & 0.067\\ \hline
    0.25 & 0.001\\
    \hline
    \end{tabular}
\end{center}
\caption{Free electron fraction at different photon temperatures.}
\label{Tab:Recombtable}
\end{table} 

From Tab.~\ref{Tab:Recombtable} we see that the free electron fraction falls abruptly at about $k_{\rm B}T \approx 0.30$ eV. Our previous estimate $k_{\rm B}T \approx 0.5$ eV must be considered a superior limit for the thermal energy.

\hrulefill

\begin{ex} Calculate at which redshift corresponds the energy $k_{\rm B}T = 0.3$ eV of the photon thermal bath.
\end{ex}

\hrulefill

\begin{figure}[ht]
\centering
 \includegraphics[width=\columnwidth]{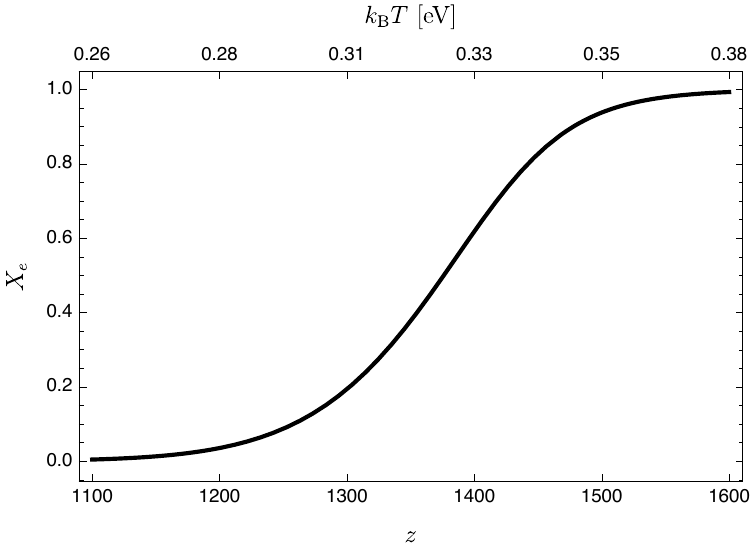}
	\caption{Numerical solution of Eq.~\eqref{Sahaeqrecombination}.}
	\label{Fig:Sahaeqrec}
\end{figure}

In Fig.~\ref{Fig:Sahaeqrec}, we numerically solve Eq.~ \eqref{Sahaeqrecombination} and use both $k_{\rm B}T$ and the redshift as variables.

In order to accurately calculate $X_e$, i.e., the residual ionization, we need to use the Boltzmann equation. Using Eq.~\eqref{Boltzmannequationfinal}, we get:  
\begin{equation}\label{BoltzmannequationRec2}
	\frac{1}{a^3}\frac{d(n_ea^3)}{dt} = \langle\sigma v\rangle\left[n_H\left(\frac{m_ek_{\rm B}T}{2\pi\hbar^2}\right)^{3/2}e^{-\varepsilon_0/(k_{\rm B}T)} - n_e^2\right]\;.
\end{equation}
Introducing now the free electron fraction $X_e$ defined in Eq.~\eqref{Xedef}, we write $n_H = n_{\rm b} (1-X_e)$ and $n_e = n_{\rm b} X_e$, so that we get:
\begin{equation}\label{BoltzmannequationRec3}
	\frac{dX_e}{dt} = n_{\rm b}\langle\sigma v\rangle\left[(1 - X_e)\left(\frac{m_ek_{\rm B}T}{2\pi\hbar^2}\right)^{3/2}e^{-\varepsilon_0/(k_{\rm B}T)} - X_e^2n_{\rm b}\right]\;.
\end{equation}
As we did for BBN, we can replace $n_{\rm b}$ with:
\begin{equation}
	n_{\rm b} = 2\eta_{\rm b}\frac{(k_{\rm B}T)^3}{\pi^2\hbar^3c^3}\;.
\end{equation}
Now we need the fundamental physics of the capture process. It is given by:
\begin{equation}\label{capturecrosssection}
	\langle\sigma v\rangle \equiv \alpha^{(2)} = 9.78\;\alpha^2\frac{\hbar^2}{m_e^2c}\left(\frac{\varepsilon_0}{k_{\rm B}T}\right)^{1/2}\log\left(\frac{\varepsilon_0}{k_{\rm B}T}\right)\;,
\end{equation}
where $\alpha = 1/137$ is the fine structure constant. The superscript $(2)$ serves to indicate that the most efficient way to form hydrogen is not via the capture of an electron in the $1s$ state because this generates a 13.6 eV photon that ionizes another newly formed hydrogen.

The most efficient way to form hydrogen is to create it in an excited state. When it relaxes to the ground state, the photons emitted do not have enough energy to ionize other hydrogen atoms.

For example, an electron captured in the $n = 2$ state generates a 3.4 eV photon. Subsequently, when the electron falls to the ground state, hydrogen releases another 10.2 eV photon. Neither of the two photons has sufficient energy to ionize another hydrogen atom in the ground state. We refer the reader to \cite{Weinberg:2008zzc} for a comprehensive treatment of the recombination process.

We now solve numerically Eq.~\eqref{BoltzmannequationRec3} together with Eq.~\eqref{capturecrosssection}. We use the redshift as the independent variable:
\begin{equation}
	\frac{dX_e}{dt} = \frac{dX_e}{dz}\frac{dz}{dt} = -\frac{dX_e}{dz}H(1 + z)\;,
\end{equation}
and the following Hubble parameter (note that $\Lambda$ is quite irrelevant at high redshifts):
\begin{equation}
	\frac{H^2}{H_0^2} = \Omega_{\rm m0}(1 + z)^3 + \Omega_{\rm r0}(1 + z)^4 + \Omega_\Lambda\;,
\end{equation}
where for $z \approx 1100$ the matter contribution is the dominant one.

Therefore, we rewrite Eq.~\eqref{BoltzmannequationRec3} as follows:
\begin{equation}\label{BoltzmannequationRec4}
	\frac{dX_e}{dz} = -\frac{\langle\sigma v\rangle}{H(1 + z)}\left[(1 - X_e)\left(\frac{m_ek_{\rm B}T}{2\pi\hbar^2}\right)^{3/2}e^{-\varepsilon_0/(k_{\rm B}T)} - X_e^2n_{\rm b}\right]\;,
\end{equation}
also taking into account that the photon temperature $T$ scales as:
\begin{equation}
	T = T_0(1 + z)\;,
\end{equation}
with $T_0 = 2.725$ K.

\begin{figure}[ht]
\centering
\includegraphics[width=\columnwidth]{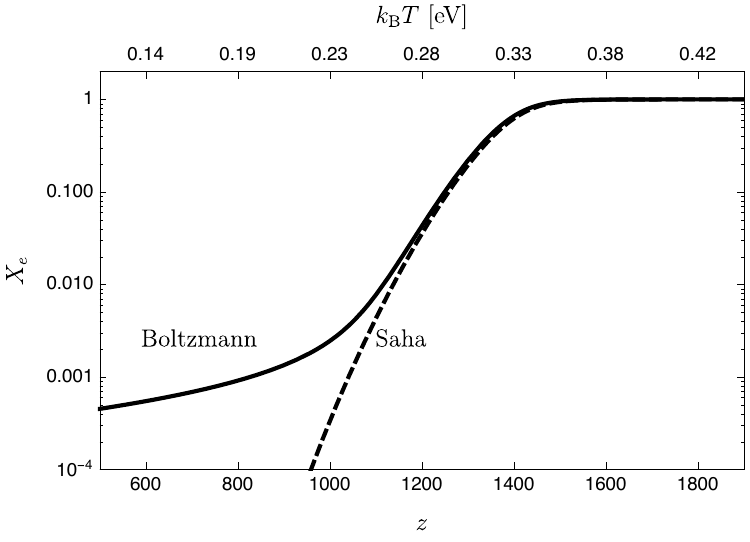}
	\caption{Solid line: numerical solution of the Boltzmann equation \eqref{BoltzmannequationRec4}. Dashed line: solution of the Saha equation \eqref{Sahaeqrecombination}.}
	\label{Fig:BoltzandSahaeqrec}
\end{figure}

In Fig.~\ref{Fig:BoltzandSahaeqrec}, we show the numerical solution of the Boltzmann equation \eqref{BoltzmannequationRec4}, compared with the solution of the Saha equation \eqref{Sahaeqrecombination}. Note how the two solutions are compatible for high redshifts, but the Boltzmann equation predicts a residual free electron fraction of about $X_e \approx 10^{-3}$.

At the same time as recombination, the decoupling of photons from electrons takes place. As we anticipated, this happens because almost no electrons remain after hydrogen formation; therefore, photons are free to propagate undisturbed (and are seen by us as the CMB).

\subsection{Decoupling}

As we have already mentioned many times by now, roughly speaking, in the expanding universe, any kind of reaction stops being efficient when its interaction rate $\Gamma$ becomes on the order of $H$. In the case of photons and electrons, the relevant process is Thomson scattering, for which:\index{Thomson scattering!Rate}
\begin{equation}
	\Gamma_{\rm T} = n_e\sigma_{\rm T}c = X_en_{\rm b}\sigma_{\rm T}c\;,
\end{equation}
where $n_e$ is the free-electron number density, which we have written as $X_en_{\rm b}$ because we have neglected the Helium abundance. The baryon number density can be expressed as
\begin{equation}
	n_{\rm b} = \frac{\rho_{\rm b}}{m_{\rm b}} = \frac{3H_0^2\Omega_{\rm b0}}{8\pi G m_p a^3}\;,
\end{equation}
where we have identified $m_{\rm b} = m_p$ since it is the proton mass that dominates the baryon energy density.

\hrulefill

\begin{ex} Show that:
\begin{equation}
	\frac{\Gamma_{\rm T}}{H} = \frac{n_e\sigma_{\rm T}c}{H} = 0.0692\;h\frac{X_e\Omega_{\rm b0}H_0}{Ha^3}\;.
\end{equation}
\end{ex}

\hrulefill

As for $H$, we consider a matter plus radiation universe, for which:
\begin{equation}
	H^2 = H_0^2\frac{\Omega_{\rm m0}}{a^3}\left(1 + \frac{a_{\rm eq}}{a}\right)\;.
\end{equation}

\hrulefill

\begin{ex} Using the above Hubble parameter show that:
\begin{equation}\label{GammaoH}
	\frac{\Gamma_{\rm T}}{H} = 113X_e\left(\frac{\Omega_{\rm b0}h^2}{0.02}\right)\left(\frac{0.15}{\Omega_{\rm m0}h^2}\right)^{1/2}\left(\frac{1 + z}{1000}\right)^{3/2}\left(1 + \frac{1 + z}{3600}\frac{0.15}{\Omega_{\rm m0}h^2}\right)^{-1/2}\;.
\end{equation}
\end{ex}

\hrulefill

The decoupling redshift $z_{\rm dec}$ is defined as the redshift for which $\Gamma_{\rm T} = H$. Note that in the above equation $X_e$ is a function of the redshift, which we have plotted in Fig.~\ref{Fig:BoltzandSahaeqrec}. On the other hand, $X_e$ drops abruptly during recombination, so the factor of 113 is easily overcome. For this reason, recombination and decoupling take place at roughly the same time.

\begin{figure}[ht]
\centering
\includegraphics[width=\columnwidth]{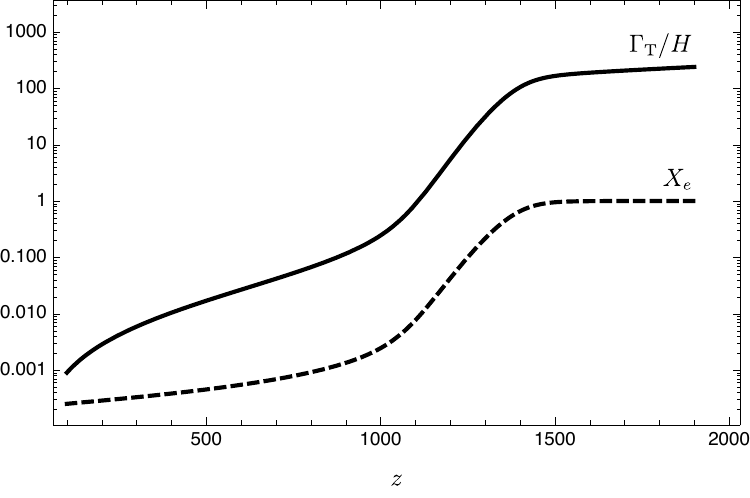}
\caption{Solid line: numerical solution of the ratio $\Gamma_{\rm T}/H$ of Eq.~\eqref{GammaoH}. Dashed line: numerical solution of Eq.~\eqref{BoltzmannequationRec4} for $X_e$.}
	\label{Fig:Decouplingplot}
\end{figure}

Now imagine that no recombination takes place; this means that $X_e = 1$. The above equation then gives the decoupling redshift:
\begin{equation}
	1 + z_{\rm dec} = 43\left(\frac{0.02}{\Omega_{\rm b0}h^2}\right)^{2/3}\left(\frac{\Omega_{\rm m0}h^2}{0.15}\right)^{1/3}\;.
\end{equation}
This is the freeze-out redshift of the electrons, i.e., eventually photons and electrons do not interact anymore because they are too diluted by the cosmological expansion. This would happen for a redshift of 42. This number is important because of the following. Well after decoupling, ultraviolet light emitted by stars and interstellar gas is able to ionize hydrogen atoms again. This phase is called \textbf{reionization}. If the latter occurred for redshifts smaller than 42, the newly freed electrons would not interact with photons because they are too diluted by the cosmological expansion. Indeed, reionization takes place for $z_{\rm reion} \approx 10$, so that the CMB spectrum is poorly affected, as we shall see in Chapter~\ref{Chap:CMBEvo}.\index{Reionization}

\section{Thermal relics}

In this section, we investigate in detail the freeze-out and relic abundance of CDM. In general, with \textbf{thermal relics}, one refers to the abundance of a certain species left over from the annihilation that occurred in the thermal bath and, after its decoupling, from the dilution caused by the expansion of the universe. To this purpose, consider the following process:\index{Thermal relics}
\begin{equation}\label{XXll}
	X + \bar{X} \longleftrightarrow S + \bar{S}\;,
\end{equation}
where $X$ represents the DM particle and $S$ a Standard Model particle.\footnote{The general form of this problem is referred to in \cite{1988kteu.book.....B} as ``the generalized Lee-Weinberg problem''. In the original work by Lee and Weinberg \cite{Lee:1977ua}, the purpose was to compute the residual abundance of massive neutrinos, thereby inferring constraints on their mass.}\index{The generalized Lee-Weinberg problem} Due to the expansion of the universe, at a certain point, the annihilation of DM is no longer efficient, and thus its abundance is fixed. This is precisely what we want to calculate. We could then put constraints on the cross-section of the above process and on the mass of the DM particle by measuring the abundance of DM necessary today to be in agreement with observation.

We are assuming here that DM is made up of massive particles that were in thermal equilibrium with the rest of the standard model particles in the primordial universe; hence the name \textbf{thermal relics}. Moreover, since we are considering CDM, or more generally, particles that decouple from the primordial plasma when nonrelativistic, ours is a calculation of the abundance of \textbf{cold relics}.

The Boltzmann equation \eqref{Boltzmannequationfinal} for the process \eqref{XXll} has the following form:
\begin{equation}
	\frac{1}{a^3}\frac{d(n_Xa^3)}{dt} = \langle\sigma v\rangle\left(n_X^{(0)2} - n_X^2\right)\;,
\end{equation}
where we have assumed the $S$ particles to have a vanishing chemical potential (as we are implicitly referring to sufficiently high temperatures).

Now we take advantage of the scaling $T \propto 1/a$ and define the following dimensionless quantities:
\begin{equation}
	Y \equiv \frac{n_X(\hbar c)^3}{(k_{\rm B}T)^3}\;, \qquad Y_{\rm eq} \equiv \frac{n^{(0)}_X(\hbar c)^3}{(k_{\rm B}T)^3}\;, \qquad x \equiv \frac{m_Xc^2}{k_{\rm B}T}\;, \qquad \lambda \equiv \frac{(m_Xc^2)^3\langle\sigma v\rangle}{H(m_X)(\hbar c)^3}\;,
\end{equation}
where $H(m_X) = H(x = 1)$ is the Hubble parameter corresponding to a thermal energy $k_{\rm B}T = m_Xc^2$, with $m_X$ being the mass of the DM particle. Note that, being deep in the radiation-dominated era, the Hubble parameter scales as follows:
\begin{equation}
	H = \frac{H(x=1)}{x^2}\;.
\end{equation}
The Boltzmann equation thus becomes:
\begin{equation}
	\frac{(k_{\rm B}T)^3}{(\hbar c)^3}\frac{dY}{dx}\frac{H(x = 1)}{x} = \langle\sigma v\rangle \frac{(k_{\rm B}T)^6}{(\hbar c)^6}\left(Y_{\rm eq}^{2} - Y^2\right)\;,
\end{equation}
and finally:
\begin{equation}\label{CDMboltzeq}
	\boxed{\frac{dY}{dx} = - \frac{\lambda}{x^2}\left(Y^{2} - Y_{\rm eq}^2\right)}
\end{equation}
In this case, the Saha equation is simply $Y = Y_{\rm eq}$, and from Eq.~\eqref{numdensmassivenu}, we know that $Y_{\rm eq} \to 0$ for $x \to \infty$ because of the dilution due to cosmological expansion. On the other hand, we also expect, as we saw for recombination, that $Y$ attains an asymptotic value, which we call $Y_\infty$, and with which we shall calculate the present abundance of DM. The departure between the solution of the Saha equation and that of the Boltzmann equation defines the \textbf{freeze-out}, and, as we know, it takes place approximately when $\Gamma \sim H$.\index{Cold relics!Freeze-out}

\begin{figure}[htbp]
\centering
	\includegraphics[width=\columnwidth]{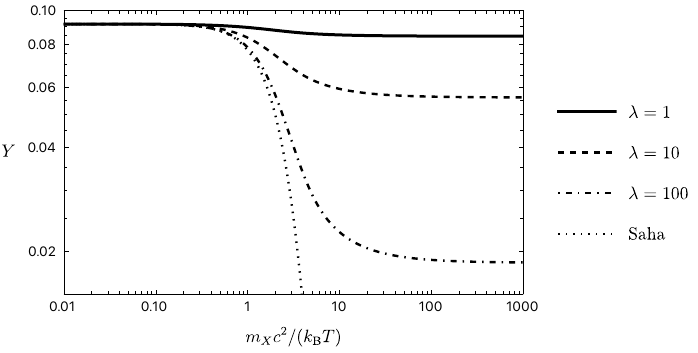}
	\caption{Numerical solution of Eq.~\eqref{CDMboltzeq} for the case of fermionic DM.}
	\label{Fig:FermDMplot}
\end{figure}

\begin{figure}[htbp]
\centering
	\includegraphics[width=\columnwidth]{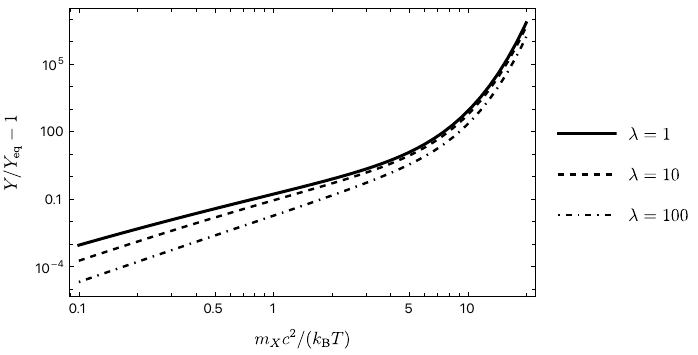}
	\caption{Relative difference $Y/Y_{\rm eq} -1$.}
	\label{Fig:xfplot}
\end{figure}

Using Eq.~\eqref{numdensmassivenu} to express $Y_{\rm eq}$, we can numerically solve Eq.~\eqref{CDMboltzeq}. To do this, we set a small initial value $x = x_i$ such that $Y(x_i) = Y_{\rm eq}(x_i)$ and fix $\lambda$ to be a constant. In Figs.~\ref{Fig:FermDMplot} and \ref{Fig:xfplot}, we choose $x_i = 0.01$ and $\lambda=1, 10, 100$ for a fermionic DM species.

From Fig.~\ref{Fig:FermDMplot}, one can appreciate that $Y$ attains a residual abundance, and that, as expected, the latter is smaller the larger $\lambda$ is. This happens because larger values of $\lambda$ result in a more efficient annihilation. 

In Fig.~\ref{Fig:xfplot}, we show the behavior of the relative difference $Y/Y_{\rm eq} -1$ to understand when the freeze-out approximately takes place. Of course, the freeze-out is not a specific instant but depends on the criterion that we choose. For example, from the inspection of Fig.~\ref{Fig:xfplot}, we see that $Y/Y_{\rm eq} -1$ at $x \approx 10$ starts to increase with greater steepness; therefore, we might establish that $x_{\rm f} \approx 10$. Note how this value is very weakly dependent on $\lambda$.

\begin{figure}[htbp]
\centering
	\includegraphics[width=\columnwidth]{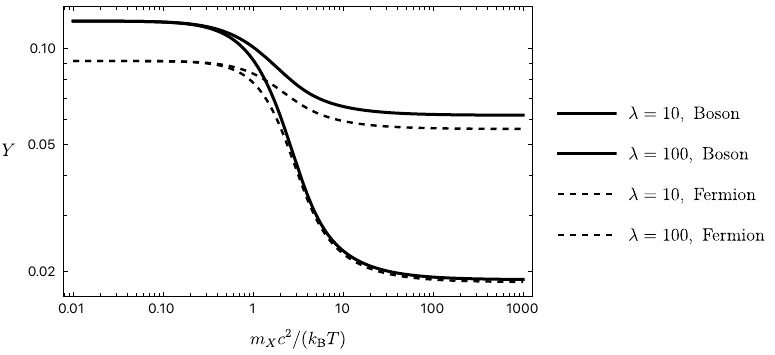}
	\caption{Comparison of the numerical solutions of Eq.~\eqref{CDMboltzeq} for the cases of bosonic DM (solid lines) and fermionic DM (dashed lines) and $\lambda = 10$ (top two lines) and $\lambda = 100$ (bottom two lines, nearly superposed).}
	\label{Fig:BosFermDMplot}
\end{figure}

Now, what is the difference between fermionic and bosonic DM? In Fig.~\ref{Fig:BosFermDMplot} we plot $Y$ for bosonic DM (solid lines) and fermionic DM (dashed lines) for $\lambda = 10, 100$. As one can see, there are two differences: $i)$ When $x$ is small, the bosonic abundance is larger than the fermionic one (due to the $\pm 1$ in the denominator of the distribution function) by a factor of 4/3; $ii)$ for large $x$, the larger $\lambda$ is, the smaller the difference between the fermionic and bosonic relic abundances becomes. 

We now relate the relic abundance to the DM annihilation cross-section. For $x \gtrsim x_{\rm f}$, $Y_{\rm eq}$ is exponentially vanishing, and the Boltzmann equation can be written as:
\begin{equation}
	\frac{dY}{dx} \approx - \frac{\lambda}{x^2}Y^2\;,
\end{equation}
whose solution for $\lambda$ constant is:
\begin{equation}
	\frac{1}{Y_\infty} - \frac{1}{Y_f} \approx \frac{\lambda}{x_{\rm f}}\;.
\end{equation}
As we can see from Fig.~\ref{Fig:FermDMplot}, $Y_{\rm f} - Y_\infty > 0$ increases as $\lambda$ becomes larger. Let us simplify:
\begin{equation}
	Y_\infty \approx \frac{x_{\rm f}}{\lambda}\;.
\end{equation}
When $Y$ has attained $Y_\infty$, the abundance of particles is fixed, and their number density starts to dilute as $n_X \propto a^{-3}$. Suppose that $Y_\infty$ is attained at a scale factor $a_1$; we can then write the present-time mass density as follows:
\begin{equation}\label{DMrelicabundance}
	\rho_{X0} = n_1m_X\frac{a_1^3}{a_0^3} = m_XY_\infty\frac{(k_{\rm B}T_1)^3}{(\hbar c)^3}\frac{a_1^3}{a_0^3} = m_X\frac{x_{\rm f}}{\lambda} \frac{(k_{\rm B}T_0)^3}{(\hbar c)^3}\left(\frac{a_1T_1}{a_0T_0}\right)^3\;.
\end{equation}
We have introduced the photon temperature here (the only one we can measure). The ratio in parentheses is not just equal to 1. We have seen an example of this discrepancy when we calculated the ratio of the photon to neutrino temperatures. The reason, as we have seen, is that not throughout the whole cosmological evolution $T$ decays as the inverse scale factor. When there are processes such as electron-positron annihilation, more photons are injected into the thermal bath, and the temperature scales in a milder way than $1/a$. In particular, from Eq. \eqref{gstarSTscaling}, we have $g_{*S}(T)T^3 \propto 1/a^3$, where $g_{*S}(T)$ is the number of effective degrees of freedom in entropy.

We can therefore compute the density parameter for the $X$ particle abundance from Eq.~\eqref{DMrelicabundance} as follows:
\begin{equation}\label{DMrelicabundance2}
	\Omega_{X0} = \frac{\rho_{X0}}{\rho_{\rm cr,0}} = \frac{m_X}{\rho_{\rm cr,0}}\frac{x_{\rm f}}{\lambda}\frac{(k_{\rm B}T_0)^3}{(\hbar c)^3}\frac{g_{*S}(T_0)}{g_{*S}(T_1)}\;.
\end{equation}
Now, we have: 
\begin{equation}
	g_{*S}(T_0) = 2 + \frac{7}{8}\cdot 6 \cdot \left(\frac{T_\nu}{T_0}\right)^3 = 2 + \frac{7}{8}\cdot 6 \cdot \frac{4}{11} = 3.91\;,
\end{equation}
Whereas for $g_{*S}(T_1)$, we can suppose that $T_1$ is so large that it exceeds the masses of all the particles in Tab. \ref{Tab:Standmodelpart}. Therefore, we have contributions from bosons and fermions:
\begin{equation}
	g_{\rm bosons} = 28\;, \qquad g_{\rm fermions} = 90\;,
\end{equation}
so that
\begin{equation}
	g_{*S}(T_1) = 28 + \frac{7}{8}\cdot 90 = 106.75\;.
\end{equation}
Recall that before neutrino decoupling $g_* = g_{*S}$, we can then remove the subscript $S$. Moreover, we suppose that the DM particle is so massive that $g_{*S}(T_1) = g_{*S}(m_X)$ (approximately, from the previous plots, one has $k_{\rm B}T_1 = m_Xc^2/100$).

Equation \eqref{DMrelicabundance2} is then rewritten as follows:
\begin{equation}\label{DMrelicabundance3}
	\Omega_{X0} = \frac{8\pi G H(m_X) x_{\rm f} (k_{\rm B}T_0)^3}{3H_0^2m_X^2c^6\langle\sigma v\rangle}\frac{g_{*S}(T_0)}{g_{*}(m_X)}\;,
\end{equation}
where we have also made $\lambda$ explicit. The Hubble parameter $H(m_X)$ can be written as:
\begin{equation}
	H^2(m_X) = \frac{8\pi G}{3c^2}g_*(m_X)\frac{\pi^2}{30}\frac{(m_Xc^2)^4}{(\hbar c)^3}\;,
\end{equation}
so that we have finally:
\begin{equation}\label{DMrelicabundance4}
	\Omega_{X0}h^2 \approx 0.3\frac{x_{\rm f}}{\sqrt{g_*(m_X)}}\frac{10^{-37}\mbox{ cm}^2}{\langle\sigma v/c\rangle}\;,
\end{equation}
or, passing from cm$^2$ to GeV$^{-2}$, we have equivalently:
\begin{equation}\label{DMrelicabundance4bis}
	\Omega_{X0}h^2 \approx \frac{x_{\rm f}}{\sqrt{g_*(m_X)}}\frac{10^{-10}\mbox{ GeV}^{-2} (\hbar c)^2}{\langle\sigma v/c\rangle}\;.
\end{equation}
Introducing the value \eqref{GFermivalue} of the Fermi constant, one obtains:
\begin{equation}\label{DMrelicabundance5}
	\Omega_{X0}h^2 \approx \frac{x_{\rm f}}{\sqrt{g_*(m_X)}}\frac{G_{\rm F}^2\mbox{ GeV}^{2}}{\langle\sigma v/c\rangle (\hbar c)^4}\;.
\end{equation}
For sufficiently large masses $m_X$, $g_*(m_X) \approx 100$, and thus $x_{\rm f}/\sqrt{g_*(m_X)} \approx 1$. If the thermally averaged cross section is of the order of $G_{\rm F}^2$, meaning that the weak interaction governs the behavior of the DM particle, then we have the correct order of magnitude for the abundance of the $X$ particles (as DM) today. This means that weakly interacting massive particles (WIMP) are interesting candidates for the role of particle DM. The ``coincidence'' of having $\Omega_{X0}h^2$ compatible with the observed value for massive particles is sometimes referred to as \textbf{WIMP miracle}.\index{WIMP miracle}

Now, let us make a rough estimate of the mass of the WIMP DM particle. For $k_{\rm B}T \ll m_Xc^2$, the weak thermally averaged cross section is:
\begin{align}
    \langle\sigma v/c\rangle (\hbar c)^4 = G_{\rm F}^2(m_Xc^2)^2\mathcal{F}^2\,,
\end{align}
where $\mathcal{F}^2$ is a ``fudge factor'' accounting for the details of the annihilation process (for example, in how many channels the $X$ particles can annihilate). So, we have:
\begin{equation}\label{DMrelicabundance6}
	\Omega_{X0}h^2 \approx \frac{x_{\rm f}}{\sqrt{g_*(m_X)}}\left(\frac{\mbox{GeV}}{m_Xc^2\mathcal{F}}\right)^2\;.
\end{equation}
For a mass larger than the weak interaction scale $m_Wc^2 \approx 100$ GeV, we have $g_* \approx 100$, and the fudge factor should assume the following form:\footnote{This is because when $m_X > m_W$, we can no longer use an effective contact theory where the mass of the vector boson is infinite. Rather, in the scattering amplitudes, at the tree level, the vector boson propagator gives a contribution $1/m_X^2$. Therefore, it becomes $1/m_X^4$ in the cross section.}
\begin{align}
    \mathcal{F}^2 \approx \frac{m_W^4}{m_X^4}N_A\,,
\end{align}
where $N_A$ are the annihilation channels. Therefore:
\begin{equation}\label{DMrelicabundance8}
	\Omega_{X0}h^2 \approx \frac{1}{N_A}10^{-4}\left(\frac{m_X}{m_W}\right)^2\;.
\end{equation}
Taking $N_A \approx 50$ (all the possible particle-antiparticle couples of Tab. \ref{Tab:Standmodelpart}), we see that in order to have $\Omega_{X0}h^2 \approx 0.15$, we need a huge DM particle mass: $m_Xc^2 \approx 10$ TeV.

\subsection{Relic abundance of baryons}\label{Sec:relicbaryonselectrons}

The very same result of Eq.~\eqref{DMrelicabundance4} can be used for the annihilation of baryons:
\begin{equation}
	\textrm{b} + \bar{\textrm{b}} \longleftrightarrow \gamma + \gamma\;.
\end{equation}
Let us consider only nucleons, i.e., protons and neutrons, since their mass is the dominant one for the baryonic energy density. The annihilation cross-section is of the order of:
\begin{equation}
	\langle\sigma v/c\rangle \approx \frac{(\hbar c)^2}{(m_{\pi}c^2)^2}\;,
\end{equation}
where $m_\pi c^2 \approx 140$ MeV is the mass of the meson $\pi$, which can be thought of as the mediator of the strong interaction among nucleons (in an effective quantum field theory, of course, for sufficiently low energies). Substituting into Eq.~\eqref{DMrelicabundance4} we get:
\begin{equation}\label{baryonrelicabundance}
	\Omega_{\rm b0}h^2 \approx 10^{-11}\;,
\end{equation}
i.e., a value many orders of magnitude below the observed one, which makes baryogenesis a necessity.

Focusing on electrons and positrons, the annihilation cross section is of the order
\begin{equation}
	\langle\sigma v/c\rangle \approx \frac{\alpha^2 (\hbar c)^2}{(m_ec^2)^2} \approx 204 \mbox{ GeV}^{-2}\;,
\end{equation}
and from Eq.~\eqref{DMrelicabundance4} we get:
\begin{equation}\label{baryonrelicabundance2}
	\Omega_{e0}h^2 \approx 10^{-12}\;.
\end{equation}
These simple calculations show how baryogenesis and leptogenesis are fundamental mechanisms that must have taken place in some form in the very early universe.

\clearpage
\chapter{Cosmological perturbations}\label{Chap:CosmoPertTheory}

{\rightskip=3truepc\leftskip=3truepc\noindent
\CYRB\cyre\cyrz~\cyrt\cyrr\cyru\cyrd\cyra~\cyrn\cyre~\cyrv\cyrery\cyrn\cyre\cyrsh\cyrsftsn~\cyrr\cyrery\cyrb\cyrk\cyru~\cyri\cyrz~\cyrp\cyrr\cyru\cyrd\cyra\\ (Without effort, you can't pull a fish out of the pond)
\vskip 0.10 in
\centerline{\it ---Russian proverb}
\vskip 0.20 in
}

The cosmological principle remarkably simplifies the relativistic treatment of cosmology, but does it find correspondence in the real world? 

The observation of the CMB reveals that the early universe ($z$ larger than 1100) was indeed close to homogeneity and isotropy, with relative deviations on the order of $10^{-5}$. Since these are so small, it is reasonable to treat them as \textbf{linear} \textbf{perturbations} around a perfectly homogeneous and isotropic FLRW background. Setting up the formalism for this task is the purpose of this Chapter and the following ones. 

As for late times, on scales smaller than about 200 Mpc, the universe is clearly not homogeneous and isotropic, as there are galaxies and their clusters that represent significant deviations from the cosmological principle. Therefore, a perturbative approach would allow us to understand in more detail the evolution of the universe on very large scales, but it would not fully describe how structures form. This ultimately requires powerful machines and numerical simulations. 

The material for this Chapter is mainly drawn from the standard textbooks cited in the preface, but also from some papers and reviews such as \cite{Bardeen:1980kt}, \cite{Kodama:1985bj}, \cite{Mukhanov:1990me}, and \cite{Ma:1995ey}. 

We assume hereafter flat spatial hypersurfaces for the background FLRW metric; that is, we set $K = 0$. This choice is made not only because the mathematical framework is simpler and more transparent, but it is also supported by observation. Indeed, when one derives the equations for the evolution of cosmological perturbations, one can see that the spatial curvature always enters in the combination $H^2 - K/a^2$ multiplying a perturbative quantity. See, for example, \cite{Mukhanov:1990me}. From the latest Planck data \cite{Planck:2018vyg}, we have learned that $\Omega_{K0} = 0.0007 \pm 0.0019$, and that spatial curvature is subdominant throughout the entire history of the universe. Therefore, it is a fair assumption to set $H^2 - K/a^2 \approx H^2$, which is equivalent to setting $K = 0$.

From this chapter onward, we adopt natural $\hbar = c = 1$ units.

\section{From the perturbations of the FLRW metric to the linearized Einstein tensor}

Let $\bar{g}_{\mu\nu}$ be the FLRW metric \eqref{FLRWmet}, with $K = 0$, and let us write it using the conformal time and comoving spatial coordinates (which we refer to generically as $\bar x^\mu$):
\begin{equation}\label{FRWmeteta}
  \bar{ds}^2 = \bar{g}_{\mu\nu}d\bar x^\mu d\bar x^\nu = a^{2}(\eta)(-d\eta^2 + \delta_{ij}dx^idx^j)\;.
\end{equation}
In order to define a perturbation of this metric, we expect that a more general (less symmetric) metric $g_{\mu\nu}$ exists such that, in some sense, we can define:
\begin{equation}\label{metricdec}
	\delta g_{\mu\nu}(x(\bar x)) := g_{\mu\nu}(x(\bar x)) - \bar{g}_{\mu\nu}(\bar x)\;,
\end{equation}
and this is, in some sense to be specified in what follows, small.

In the above equation, we express the components of the full metric $g_{\mu\nu}$ as functions of the background coordinates $\bar x$, and the \textbf{background metric} $\bar{g}_{\mu\nu}(\bar x)$ is \eqref{FRWmeteta} and is \textit{fixed}. The choice of $x(\bar x)$ is arbitrary and establishes a \textbf{gauge}\index{Cosmological perturbations!Gauge}: a certain dependence of the perturbations on the background coordinates. This leads to the so-called \textbf{problem of the gauge}, which we discuss later.

In an arbitrary gauge, we write down $g_{\mu\nu}(x(\bar x))$ in the following form:
\begin{equation}\label{genmet}
  g_{\mu\nu} = a^{2}(\eta)\left\{
\begin{array}{cc}
  -[1 + 2\psi(\eta,\mathbf x)] & w_{i}(\eta,\mathbf x)\\ \\
 w_{i}(\eta,\mathbf x) & \delta_{ij}[1 + 2\phi(\eta,\mathbf x)] + \chi_{ij}(\eta,\mathbf x)
\end{array}
\right\}\;, \quad \delta^{ij}\chi_{ij} = 0\;,
\end{equation}
where $\psi,~\phi,~w_{i},~\chi_{ij}$ ($i,j = 1,2,3$) are \textbf{perturbations}. This means that they \textbf{are small}, or $\ll 1$, so we can disregard powers with exponents larger than one in the quantities themselves and in their derivatives. Also, combinations among different perturbations are neglected. For example, $\psi^2$, $\phi w_i$, $w^i\chi_{ij}$, $\phi'\phi$, and so on, are all second order perturbations and therefore negligible.\footnote{The reason why $\psi$ and $\phi$ are multiplied by 2 in the perturbed metric is simply for future convenience in calculation.} From now on, we omit their explicit functional dependence wherever possible to keep notation light.\index{FLRW metric!Perturbation} 

The perturbations in Eq.~\eqref{genmet} can be regarded as usual 3-tensors, defined from the rotation group SO(3). Indeed, consider the coordinate transformation:
\begin{equation}
	\frac{\partial x^\mu}{\partial x^{\nu'}} = \left(
\begin{array}{cc}
  1 & 0\\ \\
 0 & R^i{}_{j}
\end{array}
\right)\;,
\end{equation}
where $R^i{}_{j}$ is a rotation. By definition, a rotation is characterized by the orthogonality condition $R^TR = RR^T = I$, or, in components, $\delta_{kl}R^{k}{}_iR^{l}{}_j = \delta_{ij}$. Applying this transformation to the metric \eqref{genmet}, we get:
\begin{eqnarray}
	g'_{00} = g_{00}\;, \quad g'_{0i} = R^k{}_{i}g_{0k}\;, \quad g'_{ij} = R^k{}_{i}R^l{}_{j}g_{kl}\;,
\end{eqnarray}
and hence, recalling that $R^k{}_{i}R^l{}_{j}\delta_{kl} = \delta_{ij}$, we finally find:
\begin{eqnarray}
	\psi' = \psi\;, \quad w'_{i} = R^k{}_{i}w_{k}\;, \quad \phi' = \phi\;, \quad \chi'_{ij} = R^k{}_{i}R^l{}_{j}\chi_{kl}\;.
\end{eqnarray}
Therefore, $\psi$ and $\phi$ are two 3-scalars; $w_{i}$ ($i = 1,2,3$) is a 3-vector, and $\chi_{ij}$ is a 3-tensor ($i,j = 1,2,3$). Note that $\chi_{ij}$ is traceless, as we have already highlighted the spatial trace of the metric through $\phi$.

An important comment: $w_i$ and $\chi_{ij}$ are \textit{defined} to have only lower indices.

\subsection{The perturbed Christoffel symbols}

Let us adopt a decomposition similar to the one in Eq. \eqref{metricdec} for the contravariant metric $g$ as well:
\begin{align}
    g^{\mu\nu} = \bar g^{\mu\nu} + \delta g^{\mu\nu}\,.
\end{align}
Therefore, for the Christoffel symbols, we have:\index{Christoffel symbols!Perturbation}
\begin{eqnarray}\label{pertChristsymb}
	\Gamma^\mu_{\nu\rho} = \frac{1}{2}g^{\mu\sigma}\left(g_{\sigma\nu,\rho} + g_{\sigma\rho,\nu} - g_{\nu\rho,\sigma}\right) = \frac{1}{2}\bar{g}^{\mu\sigma}\left(\bar{g}_{\sigma\nu,\rho} + \bar{g}_{\sigma\rho,\nu} - \bar{g}_{\nu\rho,\sigma}\right) \nonumber\\ + \frac{1}{2}\bar{g}^{\mu\sigma}\left(\delta g_{\sigma\nu,\rho} + \delta g_{\sigma\rho,\nu} - \delta g_{\nu\rho,\sigma}\right) + \frac{1}{2}\delta g^{\mu\sigma}\left(\bar{g}_{\sigma\nu,\rho} + \bar{g}_{\sigma\rho,\nu} - \bar{g}_{\nu\rho,\sigma}\right)\;,
\end{eqnarray}
where the comma denotes the usual partial derivative. Note that, as prescribed by our notion of smallness, we have neglected terms such as $\delta g^{\mu\sigma}\delta g_{\sigma\nu,\rho}$. 

It is important to realize that $\delta g^{\mu\sigma}$ is not simply $\delta g_{\mu\sigma}$ with indices raised by $\bar{g}^{\mu\sigma}$.

\hrulefill

\begin{ex} 
Since
\begin{equation}
	g^{\mu\rho}g_{\rho\nu} = \delta^\mu{}_\nu\;, \qquad \bar{g}^{\mu\rho}\bar{g}_{\rho\nu} = \delta^\mu{}_\nu\;,
\end{equation}
because both are metrics, using Eq.~\eqref{metricdec} show that
\begin{equation}\label{hmunuinversion}
	\boxed{\delta g^{\mu\nu} = - \bar{g}^{\mu\rho}\delta g_{\rho\sigma}\bar{g}^{\nu\sigma}}
\end{equation}
In particular, in our scenario of cosmological perturbations, we shall write the full metric, using the conformal time, as follows:
\begin{equation}
	g_{\mu\nu} = a^2(\eta_{\mu\nu} + h_{\mu\nu})\;,
\end{equation}
where $h_{\mu\nu}$ is defined to have only lower indices. Hence, the perturbed contravariant metric is the following:
\begin{equation}
	\delta g^{00} = -\frac{1}{a^2}h_{00}\; \quad \delta g^{0i} = \frac{1}{a^2}\delta^{il}h_{0l} = \frac{1}{a^2}h_{0i}\;, \quad \delta g^{ij} = -\frac{1}{a^2}\delta^{il}h_{lm}\delta^{mj} = -\frac{1}{a^2}h_{ij}\;,
\end{equation}
where we have used our definition $h^{ij} = h_{ij}$. Operatively, one raises the indices of the contravariant perturbed metric with the background one, but a minus sign must be taken into account. This fact is not dissimilar to considering the Taylor expansion
\begin{equation}
	\frac{1}{1 + x} = 1 - x + \mathcal O(x^2)\;.
\end{equation}
\end{ex}

\hrulefill

It is clear from Eq.~\eqref{pertChristsymb} that we can decompose the Christoffel symbols as follows:
\begin{equation}\label{Christdec}
	\Gamma^\mu_{\nu\rho} = \bar{\Gamma}^\mu_{\nu\rho} + \delta \Gamma^\mu_{\nu\rho}\;,
\end{equation}
where the barred one is computed from the background metric only.

\hrulefill

\begin{ex} Show that:
\begin{equation}
	\boxed{\delta \Gamma^\mu_{\nu\rho} = \frac{1}{2}\bar{g}^{\mu\sigma}\left(\delta g_{\sigma\nu,\rho} + \delta g_{\sigma\rho,\nu} - \delta g_{\nu\rho,\sigma} - 2\delta g_{\sigma\alpha}\bar{\Gamma}^\alpha_{\nu\rho}\right)}
\end{equation}
\end{ex}

\hrulefill

The background Christoffel symbols were already calculated in Chapter \ref{Chap:ExpandingUniverse}. For $K = 0$, they are:
\begin{equation}
	\bar\Gamma^0_{00} = \frac{a'}{a}\;, \quad \bar\Gamma^0_{ij} = \frac{a'}{a}\delta_{ij}\;, \quad \bar\Gamma^i_{0j} = \frac{a'}{a}\delta^i{}_j\;,
\end{equation}\index{FLRW metric!Christoffel symbols}
where the prime denotes derivation with respect to the conformal time. 

\hrulefill

\begin{ex} We are now in the position of writing the perturbed Christoffel symbols. We do so without making $h_{\mu\nu}$ explicit via Eq.~\eqref{genmet}. Find that: 
\begin{eqnarray}
	\delta \Gamma^0_{00} = -\frac{1}{2}h_{00}'\;, \quad \delta \Gamma^0_{i0} =  -\frac{1}{2}\left(h_{00,i} - 2\mathcal Hh_{0i}\right)\;,\\
	\delta \Gamma^i_{00} = h_{i0}' + \mathcal H h_{i0} - \frac{1}{2}h_{00,i}\;,\\ \delta\Gamma^0_{ij} = -\frac{1}{2}\left(h_{0i,j} + h_{0j,i} - h'_{ij} - 2\mathcal Hh_{ij} - 2\mathcal H\delta_{ij}h_{00}\right)\;,\\
	\delta\Gamma^i_{j0} = \frac{1}{2}h_{ij}' + \frac{1}{2}\left(h_{i0,j} - h_{0j,i}\right)\;,\\
	\delta\Gamma^i_{jk} = \frac{1}{2}\left(h_{ij,k} + h_{ik,j} - h_{jk,i} - 2\mathcal H\delta_{jk}h_{i0}\right)\;.
\end{eqnarray}\index{FLRW metric!Perturbed Christoffel symbols}The prime denotes derivative with respect to the conformal time. The indices might seem unbalanced, but recall that $h_{\mu\nu}$ is defined with lower indices only. Recall that:
\begin{equation}
	\mathcal H \equiv \frac{a'}{a}\;,
\end{equation}
\end{ex}

\hrulefill

We shall mostly use the conformal time throughout these notes since, as we saw earlier, it represents the comoving particle horizon and allows us to clearly distinguish the evolution of super-horizon (hence causally disconnected) scales from sub-horizon ones.

On the other hand, the most economical way to compute the linearized Einstein equations is by using cosmic time and calculating the Ricci tensor only. In fact, one can write the Einstein equations in the following form:
\begin{equation}
	R_{\mu\nu} = 8\pi G\left(T_{\mu\nu} - \frac{1}{2}g_{\mu\nu}T\right)\;,
\end{equation}
as done, e.g., in \cite{Weinberg:2008zzc}. The ``economy'' comes from the fact that, using cosmic time, we have only 2 non-vanishing Christoffel symbols, whereas in conformal time we have three. Moreover, we are spared from computing the perturbed Ricci scalar if we need only the Ricci tensor and not the Einstein tensor. 

Compare the results found here with those in \cite[Chapter 5]{Weinberg:2008zzc} using the tensorial properties:
\begin{equation}
	R_{\mu\nu} = \tilde{R}_{\rho\sigma}\frac{\partial\tilde{x}^\rho}{\partial x^\mu}\frac{\partial\tilde{x}^\sigma}{\partial x^\nu}\;, \quad h_{\mu\nu} = \tilde{h}_{\rho\sigma}\frac{\partial\tilde{x}^\rho}{\partial x^\mu}\frac{\partial\tilde{x}^\sigma}{\partial x^\nu}\;,
\end{equation} 
where the quantities with a tilde are in cosmic time. Since the change from cosmic to conformal time does not affect the spatial coordinates, we have that:
\begin{eqnarray}
	R_{00} = \tilde{R}_{00}a^2\;, \quad R_{0i} = \tilde{R}_{0i}a\;, \quad R_{ij} = \tilde{R}_{ij}\;,
\end{eqnarray}
and similarly for $h_{\mu\nu}$. With this map, we can compare the results that we are going to find here with those in \cite{Weinberg:2008zzc}. Note that in \cite{Weinberg:2008zzc}, the Ricci tensor is defined with the opposite sign compared to ours.

\subsection{The perturbed Ricci tensor and the Einstein tensor}

Substituting in:
\begin{equation}
	R_{\mu\nu} = \Gamma^\rho_{\mu\nu,\rho} - \Gamma^\rho_{\mu\rho,\nu} + \Gamma^\rho_{\mu\nu}\Gamma^\sigma_{\rho\sigma} - \Gamma^\rho_{\mu\sigma}\Gamma^\sigma_{\nu\rho}\;,
\end{equation}
the expansion in Eq.~\eqref{Christdec} yields:
\begin{eqnarray}
	R_{\mu\nu} = \bar\Gamma^\rho_{\mu\nu,\rho} - \bar\Gamma^\rho_{\mu\rho,\nu}  + \bar\Gamma^\rho_{\mu\nu}\bar\Gamma^\sigma_{\rho\sigma} - \bar\Gamma^\rho_{\mu\sigma}\bar\Gamma^\sigma_{\nu\rho}\nonumber\\ + \delta\Gamma^\rho_{\mu\nu,\rho} - \delta\Gamma^\rho_{\mu\rho,\nu} + \bar\Gamma^\rho_{\mu\nu}\delta\Gamma^\sigma_{\rho\sigma} + \delta\Gamma^\rho_{\mu\nu}\bar\Gamma^\sigma_{\rho\sigma} - \bar\Gamma^\rho_{\mu\sigma}\delta\Gamma^\sigma_{\nu\rho} - \delta\Gamma^\rho_{\mu\sigma}\bar\Gamma^\sigma_{\nu\rho}\;,
\end{eqnarray}\index{Ricci tensor!Perturbation}by neglecting second order terms in the connection. It is clear that we can also expand the Ricci tensor as:
\begin{equation}\label{Rdec}
	R_{\mu\nu} = \bar{R}_{\mu\nu} + \delta R_{\mu\nu}\;,
\end{equation}
Now, we compute its perturbed components.

\hrulefill

\begin{ex} Show that:
\begin{equation}
	\delta R_{00} = \delta\Gamma^l_{00,l} - \delta\Gamma^l_{0l,0} - \mathcal H\delta\Gamma^l_{0l} + 3\mathcal H\delta\Gamma^0_{00}\;, 
\end{equation}
and, substituting the previous results, one gets:
\begin{eqnarray}
	\delta R_{00} = -\frac{1}{2}\nabla^2h_{00} - \frac{3}{2}\mathcal Hh_{00}' + h_{k0,k}' + \mathcal Hh_{k0,k} - \frac{1}{2}\left(h_{kk}'' + \mathcal Hh_{kk}'\right)\;.
\end{eqnarray}
Here we have defined $\delta^{ij}\partial_i\partial_j \equiv \nabla^2$ as the Laplacian in comoving coordinates. Note that repeated spatial indices $k$ are summed over their values.\index{FLRW metric!Perturbed Ricci tensor}
\end{ex}

\hrulefill

\begin{ex} Find that:
\begin{eqnarray}
	\delta R_{0i} = -\mathcal Hh_{00,i} - \frac{1}{2}\left(\nabla^2h_{0i} - h_{k0,ik}\right) + \left(\frac{a''}{a} + \mathcal H^2\right)h_{0i} - \frac{1}{2}\left(h_{kk,i}' - h_{ki,k}'\right)\;,
\end{eqnarray}
and
\begin{eqnarray}
\label{deltaRijgeneralhmunu}	\delta R_{ij} = \frac{1}{2}h_{00,ij} + \frac{\mathcal H}{2}h_{00}'\delta_{ij} + \left(\mathcal H^2 + \frac{a''}{a}\right)h_{00}\delta_{ij}\nonumber\\ - \frac{1}{2}\left(\nabla^2h_{ij} - h_{ki,kj} - h_{kj,ki} + h_{kk,ij}\right) + \frac{1}{2}h_{ij}'' + \mathcal Hh_{ij}' + \left(\mathcal H^2 + \frac{a''}{a}\right)h_{ij}\nonumber\\ + \frac{\mathcal H}{2}h_{kk}'\delta_{ij} - \mathcal Hh_{k0,k}\delta_{ij} - \frac{1}{2}(h_{0i,j}' + h_{0j,i}') - \mathcal H(h_{0i,j} + h_{0j,i})\;.
\end{eqnarray}
\end{ex}

\hrulefill

In the same way that we decomposed the metric \eqref{metricdec}, we can also decompose the Einstein tensor. We shall work with mixed indices:
\begin{equation}\label{Gdec}
	G^{\mu}{}_{\nu} = g^{\mu\rho}R_{\rho\nu} - \frac{1}{2}\delta^{\mu}{}_{\nu}R = \bar{g}^{\mu\rho}\bar{R}_{\rho\nu} - \frac{1}{2}\delta^{\mu}{}_{\nu}\bar{R} + \bar{g}^{\mu\rho}\delta R_{\rho\nu} + \delta g^{\mu\rho}\bar{R}_{\rho\nu} - \frac{1}{2}\delta^{\mu}{}_{\nu}\delta{R}\;,
\end{equation}
where $\bar{G}^{\mu}{}_{\nu} = \bar{R}^{\mu}{}_{\nu} - \frac{1}{2}\delta^{\mu}{}_{\nu}\bar{R}$ is the background Einstein tensor and depends solely on the background metric $\bar{g}_{\mu\nu}$, whereas 
\begin{equation}
	\delta G^{\mu}{}_{\nu} = \bar{g}^{\mu\rho}\delta R_{\rho\nu} + \delta g^{\mu\rho}\bar{R}_{\rho\nu} - \frac{1}{2}\delta^{\mu}{}_{\nu}\delta{R}\;,
\end{equation}\index{Einstein tensor!Perturbation}
is the linearly perturbed Einstein tensor, which depends on both $\bar{g}_{\mu\nu}$ and $h_{\mu\nu}$.

\hrulefill

\begin{ex} Compute the perturbed Ricci scalar:
\begin{equation}
	\delta R = \delta(g^{\mu\nu}R_{\mu\nu}) = \bar{g}^{\mu\nu}\delta R_{\mu\nu} + \delta g^{\mu\nu}\bar{R}_{\mu\nu}\;.
\end{equation}\index{Ricci scalar!Perturbation}
Expand the above expression and use formula \eqref{hmunuinversion} in order to find:
\begin{equation}
	\delta R = -\frac{1}{a^2}\delta R_{00} + \frac{1}{a^2}\delta^{ij}\delta R_{ij} - a^2h_{\rho\sigma}\bar{g}^{\rho\mu}\bar{g}^{\sigma\nu}\bar{R}_{\mu\nu}\;,
\end{equation}
and then, recalling that the background Ricci tensor is:
\begin{equation}
	\bar{R}_{00} = 3\left(\mathcal H^2 - \frac{a''}{a}\right)\;, \quad \bar{R}_{ij} = \delta_{ij}\left(\mathcal H^2 + \frac{a''}{a}\right)\;,
\end{equation}
one can write:
\begin{eqnarray}
	\delta R = -\frac{1}{a^2}\delta R_{00} - \frac{3}{a^2}h_{00}\left(\mathcal H^2 -\frac{a''}{a}\right) + \frac{1}{a^2}\delta^{ij}\delta R_{ij} - \frac{1}{a^2}h_{kk}\left(\mathcal H^2 + \frac{a''}{a}\right)\;,
\end{eqnarray}
Substituting the formulae for the perturbed Ricci tensor, one finds:
\begin{eqnarray}
	a^2\delta R = \nabla^2h_{00} + 3\mathcal Hh_{00}' + 6\frac{a''}{a}h_{00} - 2h_{k0,k}' - 6\mathcal Hh_{k0,k}\nonumber\\ + \underbrace{h_{kk}'' + 3\mathcal Hh_{kk}' - \nabla^2h_{kk} + h_{kl,kl}}_{a^2\delta R^{(3)}}\;.
\end{eqnarray}\index{FLRW metric!Perturbed Ricci scalar}The second line represents $a^2\delta R^{(3)}$, the intrinsic spatial perturbed curvature scalar.
\end{ex}

\hrulefill

Now, let us calculate the mixed components of the perturbed Einstein tensor.

\hrulefill

\begin{ex} Show that:\index{FLRW metric!Perturbed Einstein tensor}
\begin{equation}
\label{deltaG00gen}	\boxed{2a^2\delta G^0{}_0 = -6\mathcal H^2h_{00} + 4\mathcal Hh_{k0,k} - 2\mathcal Hh_{kk}' + \nabla^2h_{kk} - h_{kl,kl}}
\end{equation}
and
\begin{equation}
\label{deltaG0igen}	\boxed{2a^2\delta G^0{}_i = 2\mathcal Hh_{00,i} + \nabla^2h_{0i} - h_{k0,ki} + h_{kk,i}' - h_{ki,k}'}
\end{equation}
and
\begin{eqnarray}
\label{deltaGijgen}	2a^2\delta G^i{}_j = \left[-4\frac{a''}{a}h_{00} - 2\mathcal Hh_{00}' - \nabla^2h_{00} + 2\mathcal H^2 h_{00} - 2\mathcal H h_{kk}' \right.\nonumber\\
	\left. + \nabla^2h_{kk} - h_{kl,kl} + 2h_{k0,k}' + 4\mathcal Hh_{k0,k} - h_{kk}''\right]\delta^i{}_j + h_{00,ij} - \nabla^2h_{ij} \nonumber\\ + h_{ki,kj} + h_{kj,ki}
	- h_{kk,ij} + h_{ij}'' + 2\mathcal Hh_{ij}' - (h_{0i,j}' + h_{0j,i}') - 2\mathcal H(h_{0i,j} + h_{0j,i})\;.
\end{eqnarray}
\end{ex}

\hrulefill

Now, after a bit of recovery from this \textit{tour de force}, we turn to the right hand side of the Einstein equations: the energy-momentum tensor.

\section{Perturbation of the energy-momentum tensor}

In the following, we shall mostly use the energy-momentum tensor defined through the distribution function and its perturbation. However, let us see how to perturb the background perfect fluid energy-momentum tensor. This was introduced as:
\begin{equation}
	\bar{T}_{\mu\nu} = (\bar{\rho} + \bar{P})\bar{u}_\mu \bar{u}_\nu + \bar{P}\bar{g}_{\mu\nu}\;.
\end{equation}
This tensor describes a fluid in which no dissipative processes take place. 

\hrulefill

\begin{ex}
	Compute $\bar{u}^\nu\nabla_\nu \bar{u}^\mu$, demand its vanishing, and check via the second law of thermodynamics that this corresponds to a constant entropy. 
\end{ex}

\hrulefill

Let us rewrite $\bar{T}_{\mu\nu}$ as follows:
\begin{equation}\label{barTmunutheta}
	\bar{T}_{\mu\nu} = \bar{\rho}\bar{u}_\mu \bar{u}_\nu + \bar{P}\bar\theta_{\mu\nu}\;, \qquad \bar\theta_{\mu\nu} \equiv \bar{g}_{\mu\nu} + \bar{u}_\mu \bar{u}_\nu\;.
\end{equation}
The tensor $\bar\theta_{\mu\nu}$ acts as a projector on the hypersurfaces orthogonal to the four-velocity. The plus sign in its definition is due to our choice of signature, which is mostly positive; due to this, one has $\bar u_\mu \bar u^\mu = -1$. It is immediate to check that $\bar\theta_{\mu\nu}\bar u^\mu = 0$. Equation \eqref{barTmunutheta} is an example of a 3+1 decomposition, which is particularly useful when we study fluid flow. 

\hrulefill

\begin{ex}
	Show that:
\begin{equation}
	\bar{\rho} = \bar{T}_{\mu\nu}\bar{u}^\mu \bar{u}^\nu\;, \quad 3\bar{P} = \bar{T}_{\mu\nu}\bar\theta^{\mu\nu}\;, \quad T \equiv \bar{g}^{\mu\nu}\bar{T}_{\mu\nu} = -\rho + 3P\;.
\end{equation}
The fluid density is the projection of the energy-momentum tensor along the 4-velocity of the fluid element and the pressure is the projection of the energy-momentum tensor on the 3-hypersurface orthogonal to the four-velocity.

When we project $\bar{T}_{\mu\nu}$ only once along the 4-velocity, we have:
\begin{align}
    \bar{T}_{\mu\nu}\bar{u}^\mu = -\rho \bar u_\nu\,.
\end{align}
This equation tells us that energy flows only in the direction of the 4-velocity. Therefore, there is no dissipation occurring.
\end{ex}

\hrulefill

The most general energy-momentum tensor, which also includes the possibility of dissipation, can be written by generalizing Eq.~\eqref{barTmunutheta}, i.e.
\begin{equation}\label{genTmunu3p1dec}\index{Energy momentum tensor!Imperfect fluid}
	\boxed{T_{\mu\nu} = \rho u_\mu u_\nu + Q_\mu u_\nu + Q_\nu u_\mu + (P + \Pi)\theta_{\mu\nu} + \Pi_{\mu\nu}}
\end{equation} 
where $Q_\mu$ is the \textbf{heat transfer}\index{Heat transfer} contribution, satisfying $Q_\mu u^\mu = 0$ and thus contributing 3 independent components; $\Pi_{\mu\nu}$ is the \textbf{anisotropic stress},\index{Anisotropic stress} it is traceless and satisfies $\Pi_{\mu\nu}u^\mu = 0$, hence providing 5 independent components; $\Pi$ is called \textbf{bulk viscosity}\index{Bulk viscosity} and provides a correction to the pressure, although its origin is dissipative. Bulk viscosity is compatible with the cosmological principle and can also be contemplated at the background level. It plays a central role in the so-called \textbf{bulk viscous cosmology}; see \cite{Zimdahl:1996ka}. For more details about dissipative processes in cosmology and the above decomposition of the energy-momentum tensor, see the review \cite{Maartens:1996vi}. 

The anisotropic stress $\Pi_{\mu\nu}$ is not necessarily related to viscosity, but can exist for relativistic species such as photons and neutrinos because of the quadrupole moments of their distributions, as we shall see later. On the other hand, heat fluxes and bulk viscosity are related to dissipative processes, and we neglect them in these notes starting from the next section.

Now, consider a small perturbation about the perfect fluid energy-momentum tensor:
\begin{equation}
	\rho = \bar{\rho} + \delta\rho(\eta,\mathbf x)\;, \quad P = \bar{P} + \delta P(\eta,\mathbf x)\;, \quad u^\mu = \bar{u}^\mu + \delta u^\mu(\eta,\mathbf x)\;.
\end{equation} 
The barred quantities depend only on $\eta$, since they are defined in the FLRW background. On the other hand, heat fluxes, bulk viscosity, and anisotropic stresses are purely perturbed quantities. Therefore, we have:
\begin{equation}\label{genTmunu}
	T_{\mu\nu} = \bar T_{\mu\nu} + \delta\rho\bar{u}_\mu \bar{u}_\nu + \bar{\rho}\delta u_\mu \bar{u}_\nu + \bar{\rho}\bar u_\mu \delta{u}_\nu + Q_\mu \bar u_\nu + Q_\nu \bar u_\mu + \bar\theta_{\mu\nu}(\delta P + \Pi) + \bar P\delta\theta_{\mu\nu} + \Pi_{\mu\nu}\;.
\end{equation}
The perturbed energy-momentum tensor can be straightforwardly identified as follows:
\begin{equation}\label{Tmunupert}
	\delta T_{\mu\nu} = \delta\rho\bar{u}_\mu \bar{u}_\nu + \bar{\rho}\delta u_\mu \bar{u}_\nu + \bar{\rho}\bar u_\mu \delta{u}_\nu + Q_\mu \bar u_\nu + Q_\nu \bar u_\mu + \bar\theta_{\mu\nu}(\delta P + \Pi) + \bar P\delta\theta_{\mu\nu} + \Pi_{\mu\nu}\;.
\end{equation}
Moreover, the perturbed projector has the following form:
\begin{equation}
	\delta\theta_{\mu\nu} = \delta g_{\mu\nu} + \delta u_\mu \bar{u}_\nu + \bar u_\mu \delta{u}_\nu\;.
\end{equation}
The background four-velocity normalization:
\begin{equation}\label{bgnormvel}
	\bar{g}_{\mu\nu}\bar{u}^\mu \bar{u}^\nu = -1\;,
\end{equation}
for our choice of conformal time and comoving coordinates, for which one has $\bar{u}^i = 0$, simplifies as follows:
\begin{equation}\label{bgnormvel2}
	a^2(\bar{u}^0)^2 = 1\;.
\end{equation}
We choose the positive solution $\bar{u}^0 = 1/a$, which means that the conformal time and the fluid element proper time flow in the same direction. Therefore, $u_0 = -a$.

The relations $Q_\mu u^\mu = \Pi_{\mu\nu}u^\mu = 0$ imply that $Q_0 = \Pi_{\mu 0} = 0$. Thus, heat fluxes and anisotropic stress have only spatial components. For future convenience, we redefine these as:
\begin{equation}
	Q_i \equiv aq_i\;, \qquad \Pi_{ij} \equiv a^2\pi_{ij}\;,
\end{equation}
where $q_i$ and $\pi_{ij}$ are tensors under rotations in the same way that $w_i$ and $\chi_{ij}$ are.

\hrulefill

\begin{ex} Calculate the components of the energy-momentum tensor \eqref{genTmunu}. Show that:\index{Energy momentum tensor!Perturbations}
\begin{align}
	T_{00} &= \bar\rho(1 + \delta)a^2 - 2a(\bar{\rho} + \bar P)\delta u_0 + \bar P\delta g_{00}\;,\\
	T_{0i} &= -a(\bar{\rho} + \bar P)\delta u_i - a^2q_i + \bar P\delta g_{0i}\;,\\
	T_{ij} &= (\bar P + \delta P + \pi)a^2\delta_{ij} + \bar P\delta g_{ij} + a^2\pi_{ij}\;.
\end{align}
We have introduced one of the main characters of these notes, the \textbf{density contrast}:\index{Density contrast!Definition}
\begin{equation}
 \boxed{\delta \equiv \frac{\delta\rho}{\bar\rho}}
\end{equation}
The density contrast quantifies small deviations from homogeneity and isotropy, and the study of its evolution is the study of how structure formation begins. 

Moreover, note how the presence of perturbations in the four-velocity gives rise to mixed time-space components in the energy-momentum tensor. The breaking of homogeneity and isotropy allows for extra fluxes beyond the Hubble one (the so-called peculiar velocities).
\end{ex}

\hrulefill

The total four-velocity also satisfies a normalization relation:
\begin{equation}
	g_{\mu\nu}u^\mu u^\nu = -1\;.
\end{equation}
Expanding this relation:
\begin{equation}
	\bar{g}_{\mu\nu}\bar{u}^\mu \bar{u}^\nu + \delta g_{\mu\nu}\bar{u}^\mu \bar{u}^\nu  + 2\bar{g}_{\mu\nu}\delta u^\mu \bar{u}^\nu = -1\;,
\end{equation}
and using Eq.~\eqref{bgnormvel} and $\bar{u}^i = 0$, we find that:
\begin{equation}
	\delta g_{00} + 2\bar{g}_{00}a\delta u^0 = 0\;.
\end{equation} 
Therefore, we can relate the metric perturbation $h_{00}$ to $\delta u^0$ as follows:
\begin{equation}\label{u0psirel}
	\boxed{\delta u^0 = \frac{\delta g_{00}}{2a^3} = \frac{h_{00}}{2a}}
\end{equation}
Care is needed when we want to compute the covariant components of the perturbed four-velocity $\delta u_\mu$. These are not simply $\delta u_\mu = \bar{g}_{\mu\nu}\delta u^\nu$. It is the same care we had to apply when considering the relation between $\delta g^{\mu\nu}$ and $h_{\mu\nu}$. So, let us define:
\begin{equation}
	\boxed{\delta u_i \equiv av_i}
\end{equation}
with $v_i$ components of a 3-vector.\footnote{Note that $v_i$, as $w_i$, is \textit{defined} with lower indices only.} Let us now compute the components $\delta u^i$. We must start from the covariant expression for the total four velocity:
\begin{equation}
	\bar{u}_\mu + \delta u_\mu = u_\mu = g_{\mu\nu}u^\nu = g_{\mu\nu}(\bar{u}^\nu + \delta u^\nu)\;,
\end{equation}
and, expanding up to first order, we get
\begin{equation}
	\bar{u}_\mu + \delta u_\mu = \bar{g}_{\mu\nu}\bar{u}^\nu + \bar{g}_{\mu\nu}\delta u^\nu + \delta g_{\mu\nu}\bar{u}^\nu\;.
\end{equation}
Equating order by order, we obtain $\bar{u}_\mu = \bar{g}_{\mu\nu}\bar{u}^\nu$, as expected, and:
\begin{equation}\label{covdeltaumu}
	\boxed{\delta u_\mu = \bar{g}_{\mu\nu}\delta u^\nu + \delta g_{\mu\nu}\bar{u}^\nu}
\end{equation}
so that:
\begin{equation}\label{contravdeltaumu}
	\boxed{\delta u^\mu = \bar{g}^{\mu\nu}\delta u_\nu - \bar{g}^{\mu\rho}\delta g_{\rho\nu}\bar{u}^\nu}
\end{equation}
The term $\delta_{\mu\nu}\bar{u}^\nu$ comes from the fact that $\bar{g}_{\mu\nu}$ raises or lowers indices for the background quantities only, whereas $g_{\mu\nu}$ raises or lowers indices for the full quantities only, and $\delta u^\mu$ is neither. From Eq.~\eqref{covdeltaumu} and \eqref{contravdeltaumu} we have that:
\begin{equation}\label{covdeltau0deltaui}
	\boxed{\delta u_0 = \frac{\delta g_{00}}{2a} = \frac{a h_{00}}{2}\;, \qquad a\delta u^i = v_i - \frac{1}{a^2}\delta g_{i0} = v_i - h_{0i}}
\end{equation}\index{Four-velocity!Perturbations}

\hrulefill

\begin{ex} Rewrite the components of the energy-momentum tensor as:
\begin{align}
\label{genT00}	T_{00} &= \bar\rho(1 + \delta)a^2 - \bar{\rho}a^2h_{00}\;,\\
\label{genT0i}	T_{0i} &= -a^2(\bar{\rho} + \bar P)v_i - a^2q_i + \bar Pa^2h_{0i}\;,\\
\label{genTij}	T_{ij} &= (\bar P + \delta P + \pi)a^2\delta_{ij} + \bar Pa^2h_{ij} + a^2\pi_{ij}\;,
\end{align}
\end{ex}

\hrulefill

In order to calculate the mixed components, we use the standard relation:
\begin{equation}
	T^{\mu}{}_{\nu} = g^{\mu\rho}T_{\rho\nu} = \bar g^{\mu\rho}\bar T_{\rho\nu} + \bar g^{\mu\rho}\delta T_{\rho\nu} + \delta g^{\mu\rho}\bar T_{\rho\nu}\;.
\end{equation}

\hrulefill

\begin{ex} Compute the mixed components of the energy-momentum tensor. Show that:
\begin{align}
\label{genT00mixed}	T^0{}_{0} &= -\bar\rho(1 + \delta)\;,\\
\label{genT0imixed}	T^0{}_{i} &= \left(\bar\rho + \bar{P}\right)v_{i} + q_{i}\;,\\
\label{genTi0mixed}	T^i{}_{0} &= -\left(\bar\rho + \bar{P}\right)(v_{i} - h_{0i}) - q_{i}\;,\\
\label{genTijmixed}	T^i{}_{j} &= \delta^{i}{}_{j}(\bar{P} + \delta P + \pi) + \pi_{ij}\;.
\end{align}
\end{ex}

\hrulefill

So much for the perturbed energy-momentum tensor of a fluid. However, in Chapter \ref{Chap:PertubedBoltzmannEquations}, we are going to need the energy-momentum tensor computed from kinetic theory (cf. Eq.~\eqref{enmomtensdistrfun}):
\begin{equation}\label{emtpertgen}
	T^\mu{}_\nu(x^i,\eta) = \int d_3\mathbf{P}\frac{cP^\mu P_\nu}{\sqrt{-g}P^0}f(x^i,P_j,\eta)\;,
\end{equation}
where $d_3\mathbf{P} := dP_1dP_2dP_3$. How are perturbations implemented here? 

The most obvious place where they come into play is $\sqrt{-g}$, in the integral of Eq. \eqref{emtpertgen}. Computing the determinant of \eqref{genmet} will be done later, in some specific cases.

The conjugate momentum and the distribution function also change. The spacetime coordinates, through the choice of a gauge, are still the conformal time and the comoving spatial coordinates. Therefore, a particle worldline is still described by such coordinates, say $x^\mu(\lambda)$, where $\lambda$ is an affine parameter. The conjugate momentum is still defined as $P^\mu = dx^\mu(\lambda)/d\lambda$, but the geodesic equation:
\begin{align}
    \frac{dP^\mu}{d\lambda} + \Gamma^\mu_{\nu\rho}P^\nu P^\rho = 0\,,
\end{align}
now includes a perturbed connection $\Gamma^\mu_{\nu\rho} = \bar\Gamma^\mu_{\nu\rho} + \delta \Gamma^\mu_{\nu\rho}$. Therefore, the conjugate momentum will also be characterized by the background contribution plus a perturbation $P^\mu = \bar P^\mu + \delta P^\mu$. The proper momentum is similarly split, since:
\begin{align}
    p^2 = g_{ij}P^iP^j = (\bar g_{ij} + \delta g_{ij})(\bar P^i + \delta P^i)(\bar P^j + \delta P^j)\,,
\end{align}
and, keeping only the first order contributions:
\begin{align}
    p^2 = \bar g_{ij}\bar P^i\bar P^j + 2\bar g_{ij}\bar P^i\delta P^j + \delta g_{ij}\bar P^i\bar P^j\,.
\end{align}
Since $\bar g_{ij}\bar P^i\bar P^j = \bar p^2$, we can write, at first order:
\begin{align}
    p = \bar p\left(1 + \frac{\bar g_{ij}\bar P^i\delta P^j}{\bar p^2} + \frac{1}{2}\frac{\delta g_{ij}\bar P^i\bar P^j}{\bar p^2}\right)\,.
\end{align}
We will see how to deal with these perturbations later in this chapter and in Chapter \ref{Chap:PertubedBoltzmannEquations}. Let us anticipate that such decompositions of the momenta in background plus perturbations are unnecessary when integrating over the phase-space. Note that the proper momentum $p^2 = g^{ij}P_iP_j$ is computed with the spatial metric, which, in the general case, is not simply $g_{ij}$, but rather $g_{ij} - g_{0i}g_{0j}/g_{00}$. See \cite{Landau:1982dva} for a nice explanation of this fact. On the other hand, when considering perturbations, $g_{0i}g_{0j}$ is a second-order contribution and is thus negligible.

Finally, it will prove convenient to express the distribution function as follows:
\begin{equation}\label{distributionfunctionperturbation}
	\boxed{f(x^i,P_j,\eta) = \bar f(ap)[1 + \mathcal F(x^i,P_j,\eta)]}\index{Distribution function!Perturbation}
\end{equation}
that is, the background part $\bar f(ap)$, which depends on the total proper momentum and the perturbation $\mathcal F(x^i,P_j,\eta)$. Our $\mathcal F$ corresponds to the $\Psi$ of \cite{Ma:1995ey}. We have reserved $\Psi$ to denote one of the Bardeen potentials. 

From the definition \eqref{emtpertgen}, we then have:
\begin{equation}\label{deltaT00kinetictheory}
	T^0{}_0(x^i,\eta) = \int d_3\mathbf{P}\frac{P_0}{\sqrt{-g}}f(x^i,P_j,\eta) = -\bar\rho - \delta\rho\;,
\end{equation}
with a minus sign because we define $P^0$ as positive; thus, $P_0$, using our signature, is negative. 

The mixed time-space components are:
\begin{equation}\label{deltaT0ikinetictheory}
	T^0{}_i(x^i,\eta) = \int d_3\mathbf{P}\frac{P_i}{\sqrt{-g}}f(x^i,P_j,\eta) = (\bar\rho + \bar P)v_i\;.
\end{equation}
Note the factor $\bar\rho + \bar P$ multiplying the velocity flow $v_i$. Physically, the integral gives the perturbed spatial momentum density, which can be decomposed into the velocity flow times the inertial mass density, which is $\bar\rho + \bar P$ in GR.

Finally:\footnote{Note that the pressure always enters as $\bar P$ in the background or as $\delta P$ for the perturbation. Therefore, there should be no confusion with the conjugate momentum.}
\begin{equation}\label{deltaTijkinetictheory}
	\delta T^i{}_j(x^i,\eta) = \int d_3\mathbf{P}\frac{P^i P_j}{\sqrt{-g}P^0}f(x^i,P_j,\eta) = (\bar P + \delta P)\delta^i{}_j + \pi^i{}_{j}\;.
\end{equation}
We will elaborate further on these definitions later in this chapter.

The evolution equations for the perturbations are given by the \textbf{linearized Einstein equations}:\index{Linearized Einstein equations}
\begin{equation}\label{pertEE}
 \delta G^{\mu}{}_{\nu} = 8\pi G\delta T^{\mu}{}_{\nu}\;.
\end{equation}
Unfortunately, Eq.~\eqref{pertEE} is not sufficient to completely describe the behavior of both matter and metric quantities if the fluid components are more than one. See \cite{Gorini:2007ta} for a discussion of this issue. We shall make use of the perturbed Boltzmann equations for each component of our cosmological model, and we shall derive them in chapter \ref{Chap:PertubedBoltzmannEquations}.

\section{Gauge transformations and the problem of the gauge}

We defined the perturbation of the metric in Eq. \eqref{metricdec}, stating that the background metric $\bar{g}_{\mu\nu}(\bar x)$ is \textit{fixed} and that $x(\bar x)$ represents a choice of gauge. 

We now discuss in some detail what a \textbf{gauge transformation} is and what the \textbf{problem of the gauge} entails. We follow the references \cite{Stewart:1974uz, Stewart:1990fm, Mukhanov:1990me, Malik:2008yp, Malik:2012dr}.

Let us consider a tensor field $Q$ on a manifold, which we qualify as \textbf{physical}. Being a geometric quantity, $Q$ is defined by its transformation properties under a change of coordinates. In a system of coordinates $x$, we define a \textit{perturbation} of $Q$ as:
\begin{align}
	\delta Q(x) = Q(x) - \bar Q(x)\,,
\end{align}
where $\bar Q(x)$ is the \textit{background counterpart} of $Q$. The crucial point is that $\bar Q$ \textit{is not a geometric quantity but a fixed function of the coordinates}. This makes the above splitting \textit{not covariant}; thus, $\delta Q$ \textit{is not a geometric quantity}. This is the origin of the gauge transformation. Indeed, upon a change of coordinates $x \to \hat x$, $Q(x)$ changes to $\hat Q(\hat x)$ according to its tensorial properties, but $\bar Q(x)$ simply turns into $\bar Q(\hat x)$, i.e., it remains the same function (or functions) as before. So, the perturbation changes as:
\begin{align}
	\hat \delta Q(\hat{x}) = \hat Q(\hat x) - \bar Q(\hat x)\,.
\end{align}  
The \textbf{gauge transformation} is the change in the functional form of $\delta Q$, i.e., $\delta Q \to \hat\delta Q$, which is induced by a change of coordinates on the physical manifold and stems from the non-geometric character of $\delta Q$ implicit in its definition.

The change in the functional form of $\delta Q$ can be made explicit if we consider a coordinate transformation:
\begin{align}
	x \to \hat x = x + \xi(x)\,,
\end{align}
where $\xi$ is considered as small as $\delta Q$, in order to preserve the linear order of the perturbations. Then:
\begin{align}
	\hat \delta Q(x + \xi) - \delta Q(x) = \hat Q(x + \xi) - Q(x)\,.
\end{align} 
Since $\delta Q$ and $\xi$ are small, $\hat \delta Q(x + \xi) = \hat \delta Q(x)$. Therefore:
\begin{align}
	\hat \delta Q(x) - \delta Q(x) = \mathcal L_{\xi} Q(x)\,,
\end{align} 
where $\mathcal L_{\xi}$ is the Lie derivative along $\xi$.

Let us make an explicit example, using a scalar field $\varphi$. By the definition of a scalar, upon a coordinate transformation $x \to \hat x(x)$ we have:
\begin{equation}
	\varphi(x) = \hat{\varphi}(\hat{x})\;.
\end{equation}
It is important to understand that the above equation is not a result; rather, it is \textit{the definition} of a scalar quantity. The scalar function associates a number (real or complex) with any point on the manifold, and this value does not depend on the chart employed.\footnote{If we change the chart, the function $\varphi$ must change to $\hat{\varphi}$ in order to provide the same value as before. This is the passive view, where the point of the manifold is kept fixed.}

Now we introduce the perturbation as defined earlier:
\begin{equation}
	\delta\varphi(x) = \varphi(x) - \bar\varphi(x)\;,
\end{equation}
with $\bar\varphi(x)$ the fixed background part. Considering an infinitesimal coordinate transformation $x \to \hat x = x + \xi(x)$, then:
\begin{equation}
	\hat\delta\varphi(\hat x) = \hat\varphi(\hat x) - \bar\varphi(\hat x)\;.
\end{equation}
From this, up to first order in $\xi$, we have:
\begin{equation}
	\hat\delta\varphi(x) = \varphi(x) - \bar\varphi(x) - \partial_\mu\bar\varphi(x)\xi^\mu\;, 
\end{equation}
or:
\begin{equation}
	\hat\delta\varphi(x) = \delta\varphi(x) - \partial_\mu\bar\varphi(x)\xi^\mu(x)\;.
\end{equation}
From the above equation, one can see that there is \textbf{gauge invariance}:
\begin{equation}
	\delta\hat\varphi(x) = \delta\varphi(x)\;,
\end{equation}
if and only if:
\begin{equation}
	\partial_\mu\bar\varphi(x) = 0\;.
\end{equation}
That is, if and only if the scalar field in the background is constant. This is a special case of the Stewart-Walker lemma \cite{Stewart:1974uz} and refers to a restrictive property known as Stewart and Walker \textit{identification gauge invariance}. We do not need to be so restrictive in cosmological perturbation theory because, as we will see, we can combine different kinds of perturbations to make them gauge-invariant.

Another approach to gauge transformations is the one proposed by Stewart \cite{Stewart:1990fm}. Along the physical manifold $\mathcal M$, one considers a \textbf{background} manifold $\mathcal M_0$ on which $\bar Q$ is defined as a geometrical quantity. On the background manifold, the coordinates are fixed, say $\bar x$, and these are to be used also for the physical manifold. In order to do this, we need an extra ingredient: a map that identifies points of the background manifold with those of the physical manifold. This map is a diffeomorphism that establishes a gauge. We have expressed its action as $x(\bar x)$ in Eq. \eqref{metricdec}.

Let us call this map $\mathcal D$. Fix a point $p \in \mathcal M_0$, then $\mathcal D(p) = P \in \mathcal M$. If we change the map to $\hat{\mathcal D}$, then $\hat{\mathcal D}(p)$ will, in general, be a different point on the physical manifold. Alternatively, fix a point $P \in \mathcal M$, then $\mathcal D^{-1}(P) = p \in \mathcal M_0$. If we change the map to $\hat{\mathcal D}$, then $\hat{\mathcal D}^{-1}(P)$ will, in general, be a different point on the background manifold. The two views are, of course, equivalent and are called \textit{passive} and \textit{active}.\footnote{This is when referring to the background manifold. It is the opposite if we refer to the physical manifold. This difference can be seen in \cite{Mukhanov:1990me} and \cite{Malik:2008yp}.}  

Let us consider the active view and see how to define a gauge transformation. We need to compare a geometric quantity $Q(P)$ with its background counterpart $\bar Q(\mathcal D^{-1}(P))$. In order to do this, we need the \textit{pullback} of $Q(P)$ by $\mathcal D$ to compare two quantities in the same space; however, let us avoid complications and write the comparison directly in coordinates:
\begin{align}
	\delta Q((\mathcal D \circ\bar\phi^{-1})(\bar x)) = Q((\mathcal D \circ\bar\phi^{-1})(\bar x)) - \bar Q(\bar\phi^{-1}(\bar x))\,,
\end{align} 
where $\bar\phi$ is the background chart. We can make this definition simpler: 
\begin{align}
	\delta Q(x(\bar x)) = Q(x(\bar x)) - \bar Q(\bar x)\,.
\end{align}
When we change gauge, say $\mathcal D \to \hat{\mathcal D}$, this induces a change of coordinates $x \to \hat x$. So:
\begin{align}
	\hat \delta Q(\hat x(\bar x)) = \hat Q(\hat x(\bar x)) - \bar Q(\bar x)\,.
\end{align}
Note that the background quantity $\bar Q$ has remained the same because we have not changed the background coordinates. Letting $\hat x = x + \xi(x)$, we obtain the same result as before.
 
A change of gauge corresponds to a change of coordinates on the physical manifold. For example, upon a change of gauge, metric~\eqref{genmet} can be rewritten as:\index{Gauge!Problem}
\begin{equation}\label{genmethatgauge}
  g_{\mu\nu} = a^{2}(\eta)\left\{
\begin{array}{cc}
  -[1 + 2\hat\psi(\eta,\mathbf x)] & \hat{w}_{i}(\eta,\mathbf x)\\ \\
 \hat{w}_{i}(\eta,\mathbf x) & \delta_{ij}[1 + 2\hat\phi(\eta,\mathbf x)] + \hat{\chi}_{ij}(\eta,\mathbf x)
\end{array}
\right\}\;, \quad \delta^{ij}\hat{\chi}_{ij} = 0\;.
\end{equation}
The structure of the perturbed metric is the same as before, but the perturbative quantities are now different (hatted) functions of the background coordinates, which we have fixed. 

Since the choice of the gauge is arbitrary, we might find one for which:
\begin{align}
	g_{\mu\nu}(x(\bar x)) = \bar{g}_{\mu\nu}(\bar x)\,,
\end{align}
and then conclude that there are no perturbations, even if $g$ is a different metric.\footnote{This is not possible for perturbations of the FLRW metric, since the gauge freedom is not ample enough to set all the perturbative quantities to zero.} Conversely, we might have $g = \bar g$ and choose a gauge such that:
\begin{align}
	\bar g_{\mu\nu}(x(\bar x)) \neq \bar{g}_{\mu\nu}(\bar x)\,,
\end{align}
concluding that there are perturbations, even if there are none. Apart from these limiting cases, the problem of the gauge is the very dependence of perturbations on the gauge, which does not allow for their unambiguous definition. This issue is overcome by using \textbf{gauge-invariant variables} \cite{Bardeen:1980kt}.

\subsection{Gauge transformations for the components of the metric and of the energy-momentum tensor}

Let us now specify for the metric tensor the procedure previously sketched for a generic geometric quantity $Q$. As we have seen, a gauge transformation induces the following infinitesimal coordinate transformation:\index{Gauge transformation}
\begin{equation}\label{gaugetrans}
 x^{\mu} \rightarrow \hat{x}^{\mu} = x^{\mu} + \xi^{\mu}(x)\;,
\end{equation}
where $x^\mu$ are the background coordinates and $\xi^{\mu}$ is a generic vector field, \textbf{the gauge generator}, which must be $|\xi^\mu| \ll 1$ in order to preserve the smallness of the perturbation. 

Under a coordinate transformation, the metric tensor $g_{\mu\nu}$, as well as any other tensor of the same rank, transforms in the following way:
\begin{equation}\label{mettrans}
  g_{\mu\nu}(x) = \frac{\partial\hat{x}^{\rho}}{\partial x^{\mu}}\frac{\partial\hat{x}^{\sigma}}{\partial x^{\nu}}\hat{g}_{\rho\sigma}(\hat{x})\;, 
\end{equation}  
which, using Eq.~\eqref{gaugetrans}, can be expressed as follows:  
\begin{equation}\label{mettrans2}  
   g_{\mu\nu}(x) = \left(\delta^{\rho}{}_{\mu} + \partial_\mu\xi^{\rho}\right)\left(\delta^{\sigma}{}_{\nu} + \partial_\nu\xi^{\sigma}\right)\hat{g}_{\rho\sigma}(\hat{x})\;.
\end{equation}
Since:
\begin{equation}
	\hat{g}_{\rho\sigma}(\hat{x}) = \hat{g}_{\rho\sigma}(x + \xi) = \hat{g}_{\mu\nu}(x) + \partial_\alpha\hat{g}_{\mu\nu}(x)\xi^{\alpha} + \dots\;,
\end{equation}
writing down Eq.~\eqref{mettrans2} up to first order, one obtains:
\begin{equation}\label{mettransfirsto}
  g_{\mu\nu}(x) = \hat{g}_{\mu\nu}(x) + \partial_\alpha\hat{g}_{\mu\nu}(x)\xi^{\alpha} + \partial_\mu\xi^{\rho}{}\hat{g}_{\rho\nu}(x) + \partial_\nu\xi^{\rho}\hat{g}_{\rho\mu}(x)\;.
\end{equation}

\hrulefill

\begin{ex} Show that Eq.~\eqref{mettransfirsto} can be cast in the following form:
\begin{equation}\label{mettransfirsto2}
  g_{\mu\nu}(x) = \hat{g}_{\mu\nu}(x) + \partial_\alpha g_{\mu\nu}(x)\xi^{\alpha} + \partial_\mu\xi^{\rho}{}g_{\rho\nu}(x) + \partial_\nu\xi^{\rho}g_{\rho\mu}(x)\;.
\end{equation}
That is, prove that we can remove the hat from the metric when it is multiplied by $\xi$. Then, cast the above equation as follows: 
\begin{equation}
  g_{\mu\nu}(x) = \hat{g}_{\mu\nu}(x) + \nabla_\nu\xi_{\mu} + \nabla_\mu\xi_{\nu}\;.
\end{equation}
This equation shows how the functional form of the metric components, the gauge, changes upon a coordinate transformation.\index{Gauge transformation!Metric}
\end{ex}

\hrulefill

In general, if $g_{\mu\nu}(x) = \hat{g}_{\mu\nu}(x)$, then Eq.~\eqref{gaugetrans} is called \textbf{isometry}, and the vector field $\xi$ satisfies the \textbf{Killing equations}:
\begin{equation}
	\nabla_\nu\xi_{\mu} + \nabla_\mu\xi_{\nu} = 0\;.
\end{equation}
Now we employ the perturbed FLRW metric \eqref{genmet} in Eq.~\eqref{mettransfirsto2}, considering the conformal time and the comoving coordinates. Let us subtract $\bar g_{\mu\nu}(\bar x)$ in order to define the perturbations: 
\begin{equation}\label{mettransfirsto2b}
  \delta g_{\mu\nu} = \hat\delta{g}_{\mu\nu} + (\partial_\alpha \bar g_{\mu\nu})\xi^{\alpha} + (\partial_\mu\xi^{\rho}{})\bar g_{\rho\nu} + (\partial_\nu\xi^{\rho})\bar g_{\rho\mu}\;.
\end{equation}  

\hrulefill

\begin{ex} Show that at first-order the following relations hold:
\begin{equation}\label{psiwtrans}
 \boxed{\hat{\psi} = \psi - \mathcal H\xi^{0} - \xi^{0'}} \qquad \boxed{\hat{w}_{i} = w_{i} - \zeta_i' + \partial_i\xi^{0}}
\end{equation}	  

\begin{equation}\label{phichitrans}	  
\boxed{\hat{\phi} = \phi - \mathcal H\xi^{0} - \frac{1}{3}\partial_l\xi^l} \qquad \boxed{\hat{\chi}_{ij} = \chi_{ij} - \partial_j\zeta_i - \partial_i\zeta_j + \frac{2}{3}\delta_{ij}\partial_l\xi^l}
\end{equation}
where $\zeta_i \equiv \delta_{il}\xi^l$. We have introduced $\zeta_i$ in order not to make confusion with the spatial part of $\xi_\mu$, which is $\xi_i = a^2\delta_{il}\xi^l = a^2\zeta_i$.
\end{ex}

\hrulefill

In the very same fashion that we adopted for the metric, we can also find the transformation rules for the components of the energy-momentum tensor. That is, through the same steps that we have just used for the metric, we can write:
\begin{equation}
  \hat\delta{T}_{\mu\nu} = \delta T_{\mu\nu} - (\partial_\alpha \bar T_{\mu\nu})\xi^{\alpha} - (\partial_\mu\xi^{\rho}{})\bar T_{\rho\nu} - (\partial_\nu\xi^{\rho})\bar T_{\rho\mu}\;.
\end{equation}\index{Gauge transformation!Energy-momentum tensor}

\hrulefill

\begin{ex} Use the above transformation with $T_{\mu\nu}$ given by Eq.~\eqref{genT00}-\eqref{genTij}. Find that:
\begin{equation}\label{deltarhoviqitrans}
	 \boxed{\hat{\delta\rho} = \delta\rho - \bar{\rho}'\xi^{0}} \qquad \boxed{\hat v_{i} = v_{i} + \partial_i\xi^{0}} \qquad \boxed{\hat q_i = q_i}
\end{equation}

\begin{equation}\label{deltaPpipiijtrans}
	\boxed{\hat{\pi} = \pi} \qquad \boxed{\hat{\delta P} = \delta P - \bar{P}'\xi^{0}} \qquad \boxed{\hat{\pi}_{ij} = \pi_{ij}}
\end{equation}	 
\textit{Hint.} In order to find these relations you have to use those for the metric quantities. Moreover, one obtains $\hat q_i = q_i$ noticing that $\bar\rho + \bar P$ is arbitrary. One the other hand, we do not have a mathematical way to separate the transformations for $\delta P$ and $\pi$. We do that by giving the physical argument by which $\pi$ is related to dissipative processes whereas $\delta P$ is not.
\end{ex}

\hrulefill

As we anticipated, perturbations of quantities that are vanishing or constant in the background are automatically gauge-invariant. One can see this explicitly for the heat flux, the bulk viscosity, and the anisotropic stress. This property suggests the use of the Weyl tensor instead of the Einstein tensor in order to treat perturbations of the FLRW metric (since the Weyl tensor vanishes in the FLRW metric). From the perturbed Weyl tensor, one can derive perturbative equations that are known as ``quasi-Maxwellian'' \cite{Hawking:1966qi}, \cite{jordan2009republication}.

\section{The Scalar-Vector-Tensor decomposition}\index{Scalar-Vector-Tensor decomposition}

The scalar-vector-tensor (SVT) decomposition was introduced in 1946 by Evgeny Lifshitz \cite{Lifshitz:1945du}, who was the first to address cosmological perturbations. See also \cite{Lifshitz:1963ps} and \cite{Ma:1995ey}. It consists of the following procedure. We have already seen that the perturbed metric can be written in terms of two scalars $\psi$ and $\phi$, a 3-vector $w_i$, and a 3-tensor $\chi_{ij}$. However, we can ``squeeze out'' two more scalars from $w_i$ and $\chi_{ij}$ and one more vector from $\chi_{ij}$.

The Helmholtz theorem (see Sec.~\ref{App:Helmholtztheorem} for a brief reminder) states that, under certain conditions of regularity,\footnote{The divergence and the curl of the vector must vanish at infinity faster than $1/r^2$, which seems reasonable for perturbations.} any vector $\mathbf w$ can be uniquely decomposed into a ``longitudinal'' part plus an ``orthogonal'' contribution:
\begin{equation}
	w_i = w_{i}^{\parallel} + w_{i}^{\perp}\;,
\end{equation}
which are, respectively, irrotational and solenoidal (divergenceless), namely:
\begin{equation}
	\nabla\times\mathbf w^{\parallel} = 0\;, \qquad \nabla\cdot\mathbf w^{\perp} = 0\;,
\end{equation}
or, in components:
\begin{equation}\label{vdec}
 \epsilon^{ijk}\partial_j w_k^{\parallel} = 0\;, \qquad \partial^kw_{k}^{\perp} = 0\;,
\end{equation}
where $\epsilon^{ijk}$ is the Levi-Civita symbol.\footnote{Recall that $\epsilon^{123} = 1$ and the symbol changes sign upon any odd permutation of its indices. It follows that $\epsilon^{ijk} = 0$ if two or three indices are equal.} Thanks to irrotationality, we write:
\begin{equation}
	w_{i}^{\parallel} = \partial_i w\;,
\end{equation} 
where $w$ is a scalar. Therefore, we can write $w_{i}$ as follows:
\begin{equation}\label{SVTdecwi}
  \boxed{w_{i} = \partial_i w + S_{i}}
\end{equation}
where $S_i \equiv w_{i}^{\perp}$; $w$ is the \textbf{scalar} part of $w_i$, and $S_i$ is the \textbf{vector} part of $w_i$. Usually, when discussing a \textbf{vector perturbation} in cosmology, one refers to a vector that cannot be written as the gradient of a scalar.\index{Vector perturbations}

Similarly, $\chi_{ij}$ can be decomposed into its longitudinal part $\chi^{\parallel}_{ij}$, its orthogonal part $\chi^{\perp}_{ij}$, and the transverse contribution $\chi^{T}_{ij}$:
\begin{equation}
	\chi_{ij} = \chi^{\parallel}_{ij} + \chi^{\perp}_{ij} + \chi^{T}_{ij}\;,
\end{equation}
defined as follows:
\begin{equation}
 \epsilon^{ijk}\partial^l\partial_j\chi^{\parallel}_{lk} = 0\;, \qquad \partial^i\partial^j\chi^{\perp}_{ij} = 0\;, \qquad \partial^j\chi^{T}_{ij} = 0\;. 
\end{equation}
The strategy is as follows: one builds a vector by taking the divergence of $\chi_{ij}$ and then applies the Helmholtz theorem to it. This implies that the longitudinal and the orthogonal parts can be further decomposed in the same spirit as Eq.~\eqref{vdec} in the following way:
\begin{equation}
 \chi^{\parallel}_{ij} = \left(\partial_i\partial_j - \frac{1}{3}\delta_{ij}\nabla^2\right)2\mu\;, \qquad \chi^{\perp}_{ij} = \partial_jA_{i} + \partial_iA_{j}\;, \qquad \partial^iA_i = 0\;,
\end{equation}
where $\mu$ is a scalar, $A_{i}$ is a divergenceless vector. Recall that $\nabla^2 \equiv \delta^{lm}\partial_l\partial_m$ is the Laplacian in comoving coordinates. We can thus write $\chi_{ij}$ in the following form: 
\begin{equation}\label{tenssplit}
  \boxed{\chi_{ij} = \left(\partial_i\partial_j - \frac{1}{3}\delta_{ij}\nabla^2\right)2\mu + \partial_jA_{i} + \partial_iA_{j} + \chi^{T}_{ij}}
\end{equation}
The transverse part $\chi^{T}_{ij}$ cannot be decomposed into any scalar or divergenceless vector. Therefore, it constitutes a \textbf{tensor perturbation}.\index{Tensor perturbations}

The SVT decomposition is a fundamental tool for the investigation of first order perturbations because the three classes do not mix; therefore, they can be independently analyzed. The absence of mixing is due to the fact that any kind of interaction term among the three categories would be of second order and therefore negligible.

Let us see how each class of perturbations transforms. Let us apply the Helmholtz theorem also to the spatial part of $\xi^{\mu}$ as follows:
\begin{equation}\label{xisplit}
 \xi^0 \equiv \alpha\;, \qquad \zeta_i = \partial_i\beta + \epsilon_{i}\;, \qquad (\partial^l\epsilon_l = 0)\;,
\end{equation}
where $\alpha$ and $\beta$ are scalars and $\epsilon^{i}$ is a divergenceless vector. Now let us write the transformations found in Eqs.~\eqref{psiwtrans} and \eqref{phichitrans} using the SVT decomposition:
\begin{eqnarray}
	 \hat{\psi} = \psi - \mathcal H\alpha - \alpha'\;,\\ 
\label{wtransSVT}	 \partial_i\hat{w} + \hat{S}_i = \partial_i w + S_i - \partial_i\beta' - \epsilon_{i}' + \partial_i\alpha\;,\\ 
\hat{\phi} = \phi - \mathcal H\alpha - \frac{1}{3}\nabla^2\beta\;,\\
\label{chitransSVT} \left(\partial_i\partial_j - \frac{1}{3}\delta_{ij}\nabla^2\right)2\hat{\mu} + \partial_j\hat{A}_{i} + \partial_i\hat{A}_{j} + \hat{\chi}^{T}_{ij} = \left(\partial_i\partial_j - \frac{1}{3}\delta_{ij}\nabla^2\right)2\mu\nonumber\\ + \partial_jA_{i} + \partial_iA_{j} + \chi^{T}_{ij} - 2\partial_j\partial_i\beta - \partial_j\epsilon_i - \partial_i\epsilon_j + \frac{2}{3}\delta_{ij}\nabla^2\beta\;.
\end{eqnarray}
We are now in a position to explicitly write the transformation rules for each class of perturbation.

\subsubsection{Scalar perturbations and their gauge-invariant combinations}

By taking the divergence $\partial^i$ of Eq.~\eqref{wtransSVT} and twice the divergence $\partial^i\partial^j$ of Eq.~\eqref{chitransSVT}, we eliminate all the vector and tensor contributions and are left with the transformation equations for the scalar perturbations only:
\begin{eqnarray}
	 \hat{\psi} &=& \psi - \mathcal H\alpha - \alpha'\;,\\ 
	 \nabla^2\hat{w} &=& \nabla^2\left(w - \beta' + \alpha\right)\;,\\ 
\hat{\phi} &=& \phi - \mathcal H\alpha - \frac{1}{3}\nabla^2\beta\;,\\
\nabla^2\nabla^2\hat{\mu} &=& \nabla^2\nabla^2(\mu - \beta)\;.
\end{eqnarray}
From the above transformations, we obtain:
\begin{eqnarray}
\label{psitrans}	 \boxed{\hat{\psi} = \psi - \mathcal H\alpha - \alpha'}\\ 
	 \boxed{\hat{w} = w - \beta' + \alpha}\\
\boxed{\hat{\phi} = \phi - \mathcal H\alpha - \frac{1}{3}\nabla^2\beta}\\
\label{mutrans} \boxed{\hat{\mu} = \mu - \beta}
\end{eqnarray}
How did we get rid of the Laplacian operators? Since $\nabla^2\left(\hat{w} - w + \beta' - \alpha\right) = 0$, $\hat{w} - w + \beta' - \alpha$ is a harmonic function. On the other hand, $\hat{w} - w + \beta' - \alpha$ is a perturbation; thus, it is reasonable to suppose that it is defined over all $\mathbb R^3$ and that it is limited. Therefore, by Liouville's theorem, it must be a constant. This constant is zero because $\hat{w} - w + \beta' - \alpha$ vanishes at infinity. The same reasoning applies to $\nabla^2\nabla^2(\hat{\mu}- \mu + \beta)$.

The following combinations of scalar perturbations are gauge-invariant:
\begin{eqnarray}\label{Bardeenpot}
  \boxed{\Psi = \psi + \frac{1}{a}\left[\left(w - \mu'\right)a\right]'} \qquad
  \boxed{\Phi = \phi + \mathcal H\left(w - \mu'\right) - \frac{1}{3}\nabla^2\mu}
\end{eqnarray}
They are the famous \textbf{Bardeen's potentials} \cite{Bardeen:1980kt}.\index{Bardeen's potentials} 

\hrulefill

\begin{ex} Prove that the Bardeen potentials are gauge-invariant. Why are there only two of them?
\end{ex} 
 
\hrulefill

The same technique that we have just used for the metric perturbations can be applied to the matter quantities in Eqs.~\eqref{deltarhoviqitrans} and \eqref{deltaPpipiijtrans}. 

\hrulefill

\begin{ex} Applying the SVT decomposition to $v_i$:
\begin{equation}\label{viSVTdec}
	v_i = \partial_i v + U_i\;, \qquad (\partial^lU_l = 0)\;,
\end{equation}
show that, for scalar perturbations, one gets:
\begin{eqnarray}
\label{deltarhotrans}	 \boxed{\hat{\delta\rho} = \delta\rho - \bar{\rho}'\alpha} \quad \Rightarrow \quad \boxed{\hat{\delta} = \delta + 3\mathcal H(1 + \bar P/\bar\rho)\alpha}\\  
	 \boxed{\hat{v} = v + \alpha}\\
\label{deltaPtrans}     \boxed{\hat{\delta P} = \delta P - \bar{P}'\alpha}
\end{eqnarray}
Does a fluid with equation of state $w = -1$ have gauge-invariant perturbations?
\end{ex}

\hrulefill

As we did earlier for the geometric quantities, we can combine the above transformations for matter in order to obtain gauge-invariant variables. The strategy is, in general, to combine the transformations in order to eliminate $\alpha$ and $\beta$. The result is then manifestly gauge-invariant. Using matter quantities only, we can eliminate $\alpha$ from Eqs.~\eqref{deltarhotrans} and \eqref{deltaPtrans}, thus obtaining the following gauge-invariant perturbation:
\begin{equation}\label{entropypert}
	\boxed{\Gamma \equiv \delta P - \frac{\bar P'}{\bar\rho'}\delta\rho}
\end{equation}
This is the \textbf{entropy perturbation}.\index{Entropy perturbation} The ratio $\delta P/\delta\rho$ is called \textbf{the effective speed of sound},\index{Effective speed of sound} whereas the ratio $\bar P'/\bar\rho'$ is called \textbf{the adiabatic speed of sound}.\index{Adiabatic speed of sound} 

\hrulefill

\begin{ex} Show that when the effective speed of sound and the adiabatic one are equal, i.e., when $\Gamma = 0$, one has $\delta S = 0$, i.e., there is adiabaticity (hence the name ``entropy perturbation''). 
	
\end{ex}

\hrulefill

The other gauge-invariant combinations are:
\begin{equation}
	\boxed{\delta\rho_m^{(\rm gi)} \equiv \bar\rho\Delta \equiv \delta\rho + \bar{\rho}'v} \qquad \boxed{\delta P_m^{(\rm gi)} \equiv \delta P + \bar{P}'v}
\end{equation}
i.e., the gauge-invariant density, also called \textbf{the comoving-gauge density perturbation},\index{Comoving-gauge density perturbation} and pressure perturbations. The subscript $m$ refers to ``matter'', since it is also possible to build gauge-invariant perturbations of the density and pressure using metric quantities. We are borrowing this notation from \cite{Bardeen:1980kt}. 

We can now think of combining the geometric and matter transformations, a total of 7 relations, trying to eliminate $\alpha$ and $\beta$ in order to create new gauge-invariant variables. 

Indeed, we can form the so-called \textbf{comoving curvature perturbation}
\begin{equation}\label{Rperturb}
	\boxed{\mathcal{R} \equiv \phi + \mathcal Hv - \frac{1}{3}\nabla^2\mu}
\end{equation}\index{Comoving curvature perturbation}
also known as \textbf{Lukash variable} \cite{Lukash:1980iv}, and we can form the quantity:\index{Lukash variable}
\begin{equation}\label{zetaperturb}
	\boxed{\zeta \equiv \phi + \frac{\delta\rho}{3(\bar\rho + \bar P)} - \frac{1}{3}\nabla^2\mu}
\end{equation}
which was introduced first in \cite{Bardeen:1983qw} but started to be ``seriously'' employed in \cite{Wands:2000dp}. These $\mathcal{R}$ and $\zeta$ are especially important in the framework of inflation because they are conserved on large scales and for adiabatic perturbations, as already noted in \cite{Bardeen:1980kt} (at least for $\mathcal R$), and as we shall prove in Sec.~\ref{App:Rconslargescales}.

Again, we can form gauge-invariant density, velocity, and pressure perturbations as follows:
\begin{equation}
	\boxed{\delta\rho_g^{(\rm gi)} \equiv \delta\rho + \bar{\rho}'\left(w - \mu'\right)} \quad \boxed{v^{\rm (gi)} \equiv v - (w - \mu')} \quad \boxed{\delta P_g^{(\rm gi)} \equiv \delta P + \bar{P}'\left(w - \mu'\right)}
\end{equation}
In general, having 2 gauge variables $\alpha$ and $\beta$ and 7 transformations, we can build 5 independent scalar gauge-invariant perturbations.

\subsubsection{Vector perturbations and their gauge-invariant combinations}

We can now eliminate the scalar contribution from Eq.~\eqref{wtransSVT} and consider the divergence $\partial^j$ of Eq.~\eqref{chitransSVT}. In this way, we shall find the transformations for vector perturbations:
\begin{eqnarray}\label{Sitransf}
	 \boxed{\hat{S}_i = S_i - \epsilon'_{i}}\\ 
\nabla^2\hat{A}_i = \nabla^2(A_i - \epsilon_{i})\;.
\end{eqnarray}
From the second equation, we can find the following transformation:
\begin{equation}\label{metfuntransvector3}
 \boxed{\hat{A}_{i} = A_{i} - \epsilon_{i}}
\end{equation}
It is possible to define a new gauge-invariant vector potential that has the following form:
\begin{equation}
  \boxed{W_{i} \equiv S_{i} - A'_{i}}
\end{equation}

\hrulefill

\begin{ex} Prove that $W_i$ is gauge-invariant. Why is there only one gauge-invariant vector perturbation?

Using the SVT decomposition of Eq.~\eqref{viSVTdec}, show that from the matter sector we just have:
\begin{equation}\label{vectveltrans}
	\hat{U}_l = U_l\;,
\end{equation}
i.e. the vector contribution of $v_i$ is already gauge-invariant.
\end{ex}

\hrulefill

\subsubsection{Tensor perturbations} 

Since an infinitesimal gauge transformation, cf. Eq.~\eqref{gaugetrans}, cannot be realized by any rank-2 tensor, the following result is not unexpected:
\begin{equation}\label{metfuntranstensor}
\boxed{\hat{\chi}^T_{ij} = \chi^T_{ij}}
\end{equation}
i.e., that the transverse part of $\chi_{ij}$ is already gauge-invariant.

\subsubsection{Summary} 

We have thus seen that a generic perturbation of the metric can be split into:
\begin{itemize}
	\item 4 scalar functions;
	\item 2 divergenceless 3-vectors, for a total of 4 independent components (2 each);
	\item A transverse, traceless spatial tensor of rank 2, $\chi^T_{ij}$. It has 2 independent components.
\end{itemize}
The total number of independent components sums up to 10, as it should.

\hrulefill

\begin{ex} Why does $\chi^T_{ij}$ have two independent components only? \end{ex}

\hrulefill

The above decomposition holds true not only for the metric but also for any rank-2 symmetric tensor.

\subsection{The synchronous gauge and the Newtonian gauge}

Thanks to gauge freedom, we can set any 4 perturbed components of the metric to zero. There are two particularly useful choices: the synchronous gauge and the Newtonian gauge.

\paragraph{Synchronous gauge.} This is realized by the choice:\index{Synchronous gauge}
\begin{equation}
	\hat\psi = 0\;, \qquad \hat{w}_i = 0\;.
\end{equation} 
Note that $\hat{w}_i = 0$ means that both the scalar and vector parts of $w_i$ are set to zero. Using the transformations found in the previous subsection, we find:
\begin{equation}\label{synchgaugefix}
\left\{ 
\begin{array}{l}
 \psi - \mathcal H\alpha - \alpha' = 0\\ \\
 w - \beta' + \alpha = 0\\ \\
 S_i - \epsilon'_{i} = 0\\
\end{array}
\right.\;.
\end{equation}
This must be interpreted as a system of equations for the unknowns $\alpha$, $\beta$, and $\epsilon_i$. That is, from a generic gauge, we want to know which transformations we have to perform in order to go to the synchronous gauge. We have 4 equations for 4 unknowns, so we expect to determine a single solution. However, the above equations are differential, and this implies the following:
\begin{equation}\label{synchgaugefix2}
\left\{ 
\begin{array}{l}
 \alpha = \frac{1}{a}\int d\eta\; a\psi + f(\textbf{x})\\ \\
 \beta = \int d\eta\; (w + \alpha) + g(\textbf{x})\\ \\
 \epsilon_{i} = \int d\eta\; S_i + h_i(\textbf{x})\\
\end{array}
\right.\;.
\end{equation}
Since we have only time derivatives, the integrations give rise to purely space-dependent functions, which we have called $f$, $g$, and $h_i$ here, and which are \textbf{spurious gauge modes}.\index{Spurious gauge modes}

\paragraph{Newtonian gauge.} This is realized by the choice:\index{Newtonian gauge}
\begin{equation}
	\hat{w} = 0\;, \qquad \hat{\mu} = 0\;, \qquad \hat{\chi}^\perp_{ij} = 0\;.
\end{equation} 
Using the transformations found in the previous subsection, we find:
\begin{equation}\label{confgaugefix}
\left\{ 
\begin{array}{l}
 w - \beta' + \alpha = 0\\ \\
 \mu - \beta = 0\\ \\
 A_{i} - \epsilon_{i} = 0\\
\end{array}
\right.\;.
\end{equation}
The second and third equations are algebraic and determine $\beta$ and $\epsilon_i$. Substituting the solution for $\beta$ into the first equation, we then find $\alpha$. There is no integration to perform; therefore, no spurious gauge mode appears. 

The literature is mostly concerned with scalar cosmological perturbations, and the Newtonian gauge is simply defined as $\hat{w} = 0$ and $\hat{\mu} = 0$. We have added the condition $\hat{\chi}^\perp_{ij} = 0$ here to make the vector perturbations of the spatial part of the metric vanish. With this choice, as shown above, there is no integration to perform for finding $\alpha$ and $\beta$, and thus, no spurious gauge modes arise.

On the other hand, another possible definition of the Newtonian gauge is as follows:
\begin{equation}\label{2newtoniangaugecondition}
	\hat{w}_i = 0\;, \qquad \hat{\mu} = 0\;.
\end{equation} 
That is, instead of using the vector variable in the mixed space-time components of the perturbed metric, we use the one in the spatial part, as for the synchronous gauge. However, as it happens with the synchronous gauge, the choice $\hat{w}_i = 0$ gives rise to the condition $S_i - \epsilon_i' = 0$ and the consequent emergence of a vector spurious gauge mode.

We choose to characterize the Newtonian gauge as the one free from spurious gauge modes, so, where the gauge is completely fixed. Another feature that makes the conformal Newtonian gauge somewhat appealing is that $\hat\psi = \Psi$ and $\hat\phi = \Phi$, i.e., the metric perturbations become identical to the Bardeen potentials. In these lecture notes, we employ the Newtonian gauge.

\paragraph{Transformations between the two gauges}

It is useful to provide the transformation rules among the metric perturbations in the two gauges for those readers who might want to translate the results of these notes into the synchronous gauge and compare them with the extensive literature in which this gauge is employed. For the scalar case, using Eqs.~\eqref{psitrans}-\eqref{mutrans} and assuming that the hatted quantities are the synchronous ones, whereas the non-hatted perturbations are the conformal Newtonian ones, we get:
\begin{eqnarray}
	 0 &=& \psi_N - \mathcal H\alpha - \alpha'\;,\\ 
	 0 &=& 0 - \beta' + \alpha\;,\\ 
\phi_S &=& \phi_N - \mathcal H\alpha - \frac{1}{3}\nabla^2\beta\;,\\
 \mu_S &=& - \beta\;.
\end{eqnarray}
The second and fourth equations completely specify the transformation in terms of $\mu_S$; i.e., we have:
\begin{equation}
	\boxed{\alpha = \beta' = - \mu_S'}
\end{equation}
and the metric potentials are related by:
\begin{equation}\label{SynchronoustoNewtoniangauge}
	\boxed{\psi_N = -\mathcal H\mu_S' - \mu_S''} \qquad \boxed{\phi_N = \phi_S - \mathcal H\mu_S' - \frac{1}{3}\nabla^2\mu_S}
\end{equation}
In the literature, see e.g. \cite{Ma:1995ey}, the Fourier transforms of $\phi_S$ and $\mu_S$ are usually named $h$ and $6\eta$, respectively.

For vector perturbations, we have:
\begin{eqnarray}
	0 = S^i_N - \epsilon^{i'}\;,\\
	A^i_S = 0 - \epsilon^i\;,
\end{eqnarray}
from which it is straightforward to obtain:
\begin{equation}
	S^i_N = - A^{i'}_S\;.
\end{equation}
Of course, tensor perturbations are naturally gauge-invariant and thus have the same functional form in the two gauges. This is also true for any tensor of rank equal to or higher than 2.

\section{Normal mode decomposition}\index{Normal mode decomposition}

We are now in the position of writing down explicitly the Einstein equations, fixing a gauge of our choice (which will be the Newtonian one). We expect these equations to be linear second order partial differential equations, given the perturbation scheme employed. Because of this, it is very convenient to express the perturbations as superpositions of normal modes, i.e., the eigenmodes $Q(\mathbf k, \mathbf x)$ of the Laplacian operator, defined via the Helmholtz equation:
\begin{equation}
	\nabla^2Q(\mathbf k, \mathbf x) = -k^2Q(\mathbf k, \mathbf x)\;.
\end{equation}
For flat spatial slicing, this normal mode decomposition, of course, amounts to a Fourier transform:
\begin{equation}
	Q(\mathbf k, \mathbf x) = e^{i\mathbf k\cdot \mathbf x}\;,
\end{equation}\index{Fourier transform}
and a given quantity $X(\eta,\textbf{x})$ is expressed as:
\begin{equation}
	\tilde{X}(\eta,\textbf{k}) = \int d^3\textbf{x}\;X(\eta,\textbf{x})e^{-i\textbf{k}\cdot\textbf{x}}\;, \quad X(\eta,\textbf{x}) = \int \frac{d^3\mathbf k}{(2\pi)^3}\;\tilde{X}(\eta,\textbf{k})e^{i\textbf{k}\cdot\textbf{x}}\;.
\end{equation}
Consider, for example, the metric perturbations $\psi$, $\phi$, $w_i$, and $\chi_{ij}$. For $\psi$ and $\phi$, we simply have that:
\begin{equation}
	\psi(\eta,\textbf{x}) = \int \frac{d^3\mathbf k}{(2\pi)^3}\;\tilde{\psi}(\eta,\textbf{k})e^{i\textbf{k}\cdot\textbf{x}}\;, \quad \phi(\eta,\textbf{x}) = \int \frac{d^3\mathbf k}{(2\pi)^3}\;\tilde{\phi}(\eta,\textbf{k})e^{i\textbf{k}\cdot\textbf{x}}\;,
\end{equation}
and similar expressions also apply for $\delta\rho$ and $\delta P$.

Using its SVT decomposition in Eq.~\eqref{SVTdecwi}, we have for $w_i$ the following Fourier transformation:
\begin{equation}
	w_i(\eta,\textbf{x}) = \int \frac{d^3\mathbf k}{(2\pi)^3}\;\tilde{w}_i(\eta,\textbf{k})e^{i\textbf{k}\cdot\textbf{x}} = \int \frac{d^3\mathbf k}{(2\pi)^3}\;[\tilde{\partial_i w}(\eta,\textbf{k}) + \tilde{S}_i(\eta,\textbf{k})]e^{i\textbf{k}\cdot\textbf{x}}\;.
\end{equation}
The Fourier transform of a partial spatial derivative is 
\begin{equation}
	\tilde{\partial_i w}(\eta,\textbf{k}) = ik_i\tilde{w}(\eta,\textbf{k})\;,
\end{equation}
and therefore, we have:
\begin{equation}
	w_i(\eta,\textbf{x}) = i\int \frac{d^3\mathbf k}{(2\pi)^3}\;k_i\tilde{w}(\eta,\textbf{k})e^{i\textbf{k}\cdot\textbf{x}} + \int \frac{d^3\mathbf k}{(2\pi)^3}\;\tilde{S}_i(\eta,\textbf{k})e^{i\textbf{k}\cdot\textbf{x}}\;.
\end{equation}
So, we see that the Fourier transforms of $\psi$ or $\phi$ and $w$ are not treated on an equal footing, because $\tilde{w}$ is multiplied by a $k_i$. A similar argument also applies to $v_i$. 

\begin{figure}[h!]
\center
\includegraphics[width=\columnwidth]{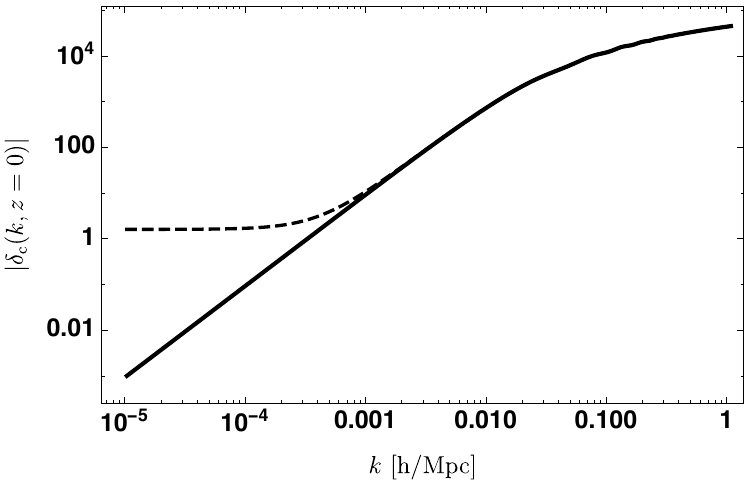}
\caption{Plot of the modulus of the CDM density contrast at $z = 0$ (today) as function of $k$, using CLASS. The solid line is the result obtained using the synchronous gauge whereas the dashed one is obtained using the Newtonian one. The initial conditions are adiabatic and normalised in order to have $\mathcal R = 1$. All the cosmological parameters have been set, as default, corresponding to the Planck best fit of the $\Lambda$CDM model \cite{Planck:2018vyg}.}
\label{Fig:deltaCDMSynchrNewt}
\end{figure}

Finally, doing the same for $\chi_{ij}$ in Eq.~\eqref{tenssplit}, one gets:
\begin{eqnarray}
	  \chi_{ij}(\eta,\textbf{x}) = \int \frac{d^3\mathbf k}{(2\pi)^3}\;\left(-k_ik_j + \frac{1}{3}\delta_{ij}k^2\right)2\tilde\mu(\eta,\textbf{k}) e^{i\textbf{k}\cdot\textbf{x}}\nonumber\\ + i\int \frac{d^3\mathbf k}{(2\pi)^3}\;[k_j\tilde A_{i}(\eta,\textbf{k}) + k_i\tilde A_{j}(\eta,\textbf{k})]e^{i\textbf{k}\cdot\textbf{x}} + \int \frac{d^3\mathbf k}{(2\pi)^3}\;\tilde\chi^{T}_{ij}(\eta,\textbf{k})e^{i\textbf{k}\cdot\textbf{x}}\;,
\end{eqnarray}
with a similar expansion holding true for $\pi^i{}_j$. We see that $\tilde\mu$ is multiplied by a factor $k^2$ and $\tilde A_i$ is multiplied by a factor $k$. Therefore, in order to properly compare perturbations, \textit{par condicio} is restored by ``correcting'' as follows:
\begin{equation}\label{correctionintheFTquantities}
	\tilde w \equiv -\frac{1}{k}\tilde B\;, \quad \tilde v \equiv -\frac{1}{k}\tilde V\;, \quad \tilde\mu \equiv \frac{1}{k^2}\tilde E\;, \quad \tilde A_i \equiv -\frac{1}{2k}\tilde F_i\;.
\end{equation}
In this way, all scalar quantities are treated on the same footing. Let us see what happens to the comoving gauge density perturbation:
\begin{eqnarray}
	\boxed{\tilde\Delta = \tilde\delta + 3(1 + \bar P/\bar\rho)\frac{\mathcal H}{k}\tilde V} 
\end{eqnarray}
Since $\mathcal H \propto 1/\eta$, on sub-horizon scales, i.e., for $k\eta \gg 1$, the density contrast becomes gauge-invariant. We present this fact in Fig.~\ref{Fig:deltaCDMSynchrNewt}, where we plot the evolution of the modulus of the CDM density contrast $\delta_{\rm c}$ as a function of $k$ and for $z = 0$, computed with CLASS \cite{Lesgourgues:2011re} in the synchronous (solid line) and Newtonian (dashed line) gauges.

The plots in Fig.~\ref{Fig:deltaCDMSynchrNewt} are drawn for $z = 0$; hence, the value of the Hubble parameter is the Hubble constant, $H_0 = 3\times 10^{-4}$ $h$ Mpc$^{-1}$. Indeed, when $k > H_0$, the two evolutions coincide.

Now we have to understand how to extract the scalar and vector parts from a full 3-vector quantity. Let us reformulate the FT of $w_i$ as follows:
\begin{equation}
	\tilde{w}_i(\eta,\textbf{k}) = -i\hat{k}_i\tilde{B}(\eta,\textbf{k}) + \tilde{S}_i(\eta,\textbf{k})\;.
\end{equation} 
We see that we can isolate the scalar part of the FT transform of a 3-vector perturbation by contracting it with $i\hat{k}^i$. Indeed:
\begin{equation}\label{extractingscalarpartfromvector}
	i\hat{k}^i\tilde{w}_i(\eta,\textbf{k}) = \tilde{B}(\eta,\textbf{k})\;,
\end{equation}
because $\hat{k}^i\tilde{S}_i = 0$, since $\partial^iS_i = 0$. The vector part is therefore obtained by subtracting the scalar part:
\begin{equation}\label{extractingvectorpartfromvector}
	\tilde S_i = \left(\delta^j{}_i - \hat{k}^j\hat{k}_i\right)\tilde{w}_j\;.
\end{equation}
We can easily check that this formula satisfies $\hat{k}^i\tilde{S}_i = 0$.

What about a tensorial quantity? For the traceless spatial metric perturbation in Eq.~\eqref{tenssplit}, we can write:
\begin{equation}\label{tenssplitFT}
  \tilde\chi_{ij} = -\left(\hat{k}_i\hat{k}_j - \frac{1}{3}\delta_{ij}\right)2\tilde{E} - \frac{i}{2}\hat{k}_j\tilde{F}_{i} - \frac{i}{2}\hat{k}_i\tilde{F}_{j} + \tilde{\chi}^{T}_{ij}\;.
\end{equation}
Here, the scalar contribution can be isolated by contracting with $-3\hat{k}^i\hat{k}^j/2$. In fact:
\begin{equation}\label{extractingscalarfromtensor}
	-\frac{3}{2}\hat{k}^i\hat{k}^j\tilde\chi_{ij} = 2\tilde E\;. 
\end{equation} 

\hrulefill

\begin{ex}
Show that the vector contribution is obtained contracting once with $2i\hat{k}^i$ and using Eq.~\eqref{extractingscalarfromtensor}, i.e.
\begin{equation}\label{extractingvectorpartfromtensor}
	\tilde F_i = 2i\left(\delta^j{}_i - \hat{k}^j\hat{k}_i\right)\hat{k}^l\tilde{\chi}_{jl}\;.
\end{equation} 
For the tensor part, show that:
\begin{equation}\label{extractingtensorperturbationfromtensor}
	\tilde{\chi}^{T}_{ij} = \left(\delta^l{}_i - \hat{k}^l\hat{k}_i\right)\left(\delta^m{}_j - \hat{k}^m\hat{k}_j\right)\tilde{\chi}_{lm} + \frac{1}{2}(\delta_{ij} - \hat k_i\hat k_j)\hat{k}^l\hat{k}^m\tilde{\chi}_{lm}\;.
\end{equation}
Verify that $\hat k^i\tilde{\chi}^{T}_{ij} = \hat k^j\tilde{\chi}^{T}_{ij} = 0$ and $\delta^{ij}\tilde{\chi}^{T}_{ij} = 0$. Recall that $\tilde{\chi}_{ij}$ is already traceless.
\end{ex}

\hrulefill

\section{Einstein equations for scalar perturbations}\index{Einstein equations!Scalar perturbations}

In this section, we focus on scalar perturbations and employ the Newtonian gauge. Our perturbed metric is then:
\begin{equation}\label{confnewtmetric}
	g_{00} = -a(\eta)^2[1 + 2\Psi(\eta,\textbf{x})]\;, \quad g_{0i} = 0\;, \quad g_{ij} = a(\eta)^2\delta_{ij}\left[1 + 2\Phi(\eta,\textbf{x})\right]\;,
\end{equation}
where we are employing the Bardeen potentials, exploiting the fact that $w = \mu = 0$. In metric \eqref{confnewtmetric} $\Psi$ plays the role of the Newtonian potential and $\Phi$ is the spatial curvature perturbation. 

\hrulefill

\begin{ex} Calculate from scratch the perturbed Einstein tensor $\delta G^\mu{}_\nu$ from metric \eqref{confnewtmetric}, AND do it again using also Eqs.~\eqref{deltaG00gen}-\eqref{deltaGijgen}. Show that:
\begin{eqnarray}
a^2\delta G^0{}_0 = -6\mathcal H\Phi' + 6\mathcal H^2\Psi + 2\nabla^2\Phi\;,\\
\label{deltaG0i} a^2\delta G^0{}_i = 2\partial_i(\Phi' - \mathcal H\Psi)\;,\\
\label{deltaGij} a^2\delta G^i{}_j = \left[-2\Phi'' - 4\mathcal H\Phi' + 2\mathcal H\Psi' + 4\frac{a''}{a}\Psi - 2\mathcal H^2\Psi + \nabla^2(\Phi + \Psi)\right]\delta^i{}_j\nonumber\\ -\partial^i\partial_j(\Phi + \Psi)\;.
\end{eqnarray}
Note again that $\partial^i = \partial_i$ since it is the partial derivative with respect to comoving coordinates and $\nabla^2 = \delta^{lm}\partial_l\partial_m$ is the comoving Laplacian. Being comoving, it is always accompanied by a factor $1/a^2$, since together they form the physical Laplacian.
\end{ex}

\hrulefill 

\subsection{The relativistic Poisson equation}\index{Relativistic Poisson equation} 

The $0-0$ Einstein equation is as follows:
\begin{equation}
	-3\mathcal H\Phi' + 3\mathcal H^2\Psi + \nabla^2\Phi = 4\pi Ga^2\delta T^0{}_0\;.
\end{equation}
Using the perturbed energy-momentum tensor component $\delta T^0{}_0$, as we can read from Eq.~\eqref{genT00mixed}, we have:
\begin{equation}
	\boxed{3\mathcal H\Phi' - 3\mathcal H^2\Psi - \nabla^2\Phi = 4\pi Ga^2\delta\rho_{\rm tot}}
\end{equation}
where the total perturbed density is:
\begin{equation}
	\delta\rho_{\rm tot} = \sum_{i}\delta\rho_i = \sum_i\rho_i\delta_i\;,
\end{equation} 
i.e., it is the sum of the perturbed densities of all the material components that constitute our cosmological model, in the same way that we did for the background. We start here to eliminate the bar over the background quantities. We shall consider the $\Lambda$CDM as our standard cosmological model. Thus, we have to deal with 4 contributions:
\begin{enumerate}
	\item Photons; 
	\item Neutrinos; 
	\item CDM; 
	\item Baryons.
\end{enumerate} 
Each of these has its own energy-momentum tensor, and the total one, which enters the right hand side of the Einstein equations, is their sum. The cosmological constant contributes only at the background level.

We have found the \textbf{relativistic Poisson equation} above. Indeed, if we consider $a = 1$, and hence $\mathcal H = 0$, we recover the usual Newtonian Poisson equation (but not quite, since we have two potentials in GR). Written in terms of the density contrast, using Eq.~\eqref{genT00mixed}, the relativistic Poisson equation is as follows:
\begin{equation}\label{relPoisseqdeltas}
	\boxed{3\mathcal H\Phi' - 3\mathcal H^2\Psi - \nabla^2\Phi = 4\pi Ga^2\left(\rho_{\rm c}\delta_{\rm c} + \rho_{\rm b}\delta_{\rm b} + \rho_\gamma\delta_\gamma + \rho_\nu\delta_\nu\right)}
\end{equation}
As one can see, Eq.~\eqref{relPoisseqdeltas} is a second order partial differential equation (PDE) that is \textit{linear} because we are doing first-order perturbation theory; thus, all the perturbative variables appear with power 1. Because of this, as we anticipated, it is very convenient to introduce the Fourier transform of the latter, thus transforming Eq.~\eqref{relPoisseqdeltas} into a linear ordinary differential equation (ODE). 

Hereafter, we shall constantly employ the Fourier transform but drop the tilde above the transformed quantities, as it is customary in cosmology because almost always one deals directly with the Fourier modes rather than with the configuration space. Therefore, Eq.~\eqref{relPoisseqdeltas}, Fourier-transformed and written in conformal time, is as follows:
\begin{equation}\label{relativisticPoissonequation}
	\boxed{3\mathcal H\Phi' - 3\mathcal H^2\Psi + k^2\Phi = 4\pi Ga^2\left(\rho_{\rm c}\delta_{\rm c} + \rho_{\rm b}\delta_{\rm b} + \rho_\gamma\delta_\gamma + \rho_\nu\delta_\nu\right)}
\end{equation}
Now that we have specified a gauge, and a time slicing, we can establish that:
\begin{equation}\label{sqrtgNewtoniangauge}
	\sqrt{-g} = \left[a^8(1 + 2\Psi)(1 + 2\Phi)^3\right]^{1/2} \stackrel{\text{1st}}{=} a^4\left(1 + \Psi + 3\Phi\right)\;.
\end{equation}
Moreover, using the mass-shell relation:
\begin{equation}
	g^{\mu\nu}P_\mu P_\nu = -m^2 = -E^2 + p^2\;,
\end{equation}
which defines our energy $E$, we can write $P_0$ as follows:
\begin{equation}
	\frac{1}{a^2(1 + 2\Psi)}(P_0)^2 = E^2\;.
\end{equation}
By choosing the minus sign for $P_0$ (because we want the plus sign for $P^0$), we get:
\begin{equation}\label{P0Erel}
	P_0 \stackrel{\text{1st}}{=} -a(1 + \Psi)E(p)\;.
\end{equation}
Hence, we can express the total energy density from Eq.~\eqref{deltaT00kinetictheory} as follows:
\begin{equation}
	\bar\rho + \delta\rho = \int d_3\mathbf{P}\frac{(1 + \Psi)E(p)}{a^3\left(1 + \Psi + 3\Phi\right)}f(x^i,P_j,\eta)\;,
\end{equation}
or
\begin{equation}
	\bar\rho + \delta\rho \stackrel{\text{1st}}{=} \frac{1}{a^3}\int d_3\mathbf{P}(1 - 3\Phi)E(p)f(x^i,P_j,\eta)\;.
\end{equation}

\hrulefill

\begin{ex}
From the definition of the proper momentum:
\begin{equation}
	p^2 = g^{ij}P_iP_j = \frac{\delta^{ij}}{a^2(1 + 2\Phi)}P_iP_j \stackrel{\text{1st}}{=} \frac{\delta^{ij}}{a^2}(1 - 2\Phi)P_iP_j = \frac{1}{a^2}(1 - 2\Phi)P^2\;,
\end{equation}
obtain the following relation between the proper momentum and the conjugate one:
\begin{equation}\label{conjPproperpprel}
	P_i = a(1 + \Phi)p_i\;.
\end{equation}
\end{ex}

\hrulefill

Using the result of the exercise, it is convenient to change the variable from the conjugate momentum to the proper one:
\begin{equation}
	d_3\mathbf{P} \stackrel{\text{1st}}{=} a^3(1 + 3\Phi)d^3\mathbf p\;,
\end{equation}
because we can write: 
\begin{equation}
	\bar\rho + \delta\rho \stackrel{\text{1st}}{=} \frac{1}{a^3}\int d^3\mathbf pE(p)f(x^i,p_j,\eta)\;.
\end{equation}
Now, no metric perturbation explicitly enters the integration, so we can conclude that:
\begin{equation}\label{deltarhod3p}
	\boxed{\delta\rho = \int d^3\mathbf pE(p)\bar f(ap)\mathcal F(x^i,p_j,\eta)}
\end{equation}
i.e., the perturbed energy density is obtained by integrating the proper momentum of the perturbed distribution function only.

It will be useful, when building the Boltzmann equations in Chapter \ref{Chap:PertubedBoltzmannEquations}, to introduce a sort of comoving momentum:
\begin{equation}
	q_i = ap_i\;,
\end{equation} 
in order to remove the background redshift of the temperature from the perturbations. Using $q_i$, the perturbed energy density becomes:
\begin{equation}\label{deltarhod3q}
	\boxed{\delta\rho = \frac{1}{a^4}\int d^3\mathbf q\sqrt{q^2 + m^2a^2}\bar f(q)\mathcal F(x^i,q_j,\eta)}
\end{equation}

\subsection{The equation for the anisotropic stress}

The next Einstein equation that we present is the traceless part of $\delta G^i{}_j$, which we know from Eq.~\eqref{genTijmixed} to be related to the anisotropic stress $\pi_{ij}$.\index{Anisotropic stress} 

\hrulefill

\begin{ex} From Eq.~\eqref{deltaGij}, calculate the trace and the traceless part of $\delta G^i{}_j$. Show that:
\begin{eqnarray}
\label{tracedeltaGij} a^2\delta G^l{}_l = -6\Phi'' - 12\mathcal H\Phi' + 6\mathcal H\Psi' + 12\frac{a''}{a}\Psi - 6\mathcal H^2\Psi + 2\nabla^2(\Phi + \Psi)\;,	
\end{eqnarray}
and then:
\begin{equation}
	a^2\delta G^i{}_j - \frac{1}{3}\delta^i{}_{j}a^2\delta G^l{}_l = -\left(\partial^i\partial_j - \frac{1}{3}\delta^i{}_{j}\nabla^2\right)(\Phi + \Psi)\;.
\end{equation}
\end{ex}

\hrulefill

The Fourier transform of the latter can be written in the following form:
\begin{equation}
	a^2\delta G^i{}_j - \frac{1}{3}\delta^i{}_{j}a^2\delta G^l{}_l = k^2\left(\hat{k}^i\hat{k}_j - \frac{1}{3}\delta^i{}_{j}\right)(\Phi + \Psi)\;,
\end{equation}
where we have used $k^i = k\hat{k}^i$ and $\hat{k}^i$, the unit vector denotes the direction of $\textbf{k}$. The spatial traceless Einstein equation can thus be written as:
\begin{equation}
	\boxed{k^2\left(\hat{k}^i\hat{k}_j - \frac{1}{3}\delta^i{}_{j}\right)(\Phi + \Psi) = 8\pi G a^2\pi^i{}_{j}}
\end{equation}
since $\pi^i{}_j$ is the spatial traceless part of the energy-momentum tensor, as we know from Eq.~\eqref{genTijmixed}. On the left hand side, note the same operator multiplying the scalar contribution of $\chi_{ij}$ in Eq.~\eqref{tenssplitFT}. Hence, only the scalar contribution of $\pi^i{}_{j}$ would contribute to the right hand side. Indeed, contracting the above equation with $\hat{k}_i\hat{k}^j$, as in Eq.~\eqref{extractingscalarfromtensor}, we obtain:
\begin{equation}\label{anisotropicstressscalarequation}
	\boxed{k^2(\Phi + \Psi) = 12\pi G a^2\hat{k}_i\hat{k}^j\pi^i{}_j}
\end{equation}
Our $\hat{k}_i\hat{k}^j\pi^i{}_j$ corresponds to the $-(\rho + P)\sigma$ used in \cite{Ma:1995ey}. We leave this equation as it is for the moment. We shall see that $\hat{k}_i\hat{k}^j\pi^i{}_j$ is sourced by the quadrupole moment of the photon and neutrino distributions. 

This equation tells us that $\Phi = -\Psi$, i.e., there exists only one gravitational potential unless a quadrupole moment of the energy content distribution is present. For example, when CDM dominated the universe, $\Phi = -\Psi$; however, this is not the case in the early universe due to neutrinos. Even when CDM or DE dominates, if the underlying theory of gravity is not GR, one might have $\Phi \neq -\Psi$.\footnote{One can probe the value of $\Phi + \Psi$ via weak lensing, which we do not address in these notes. See e.g., \cite{2017grle.book.....D} for a treatise on gravitational lensing.}

Using Eq.~\eqref{deltaTijkinetictheory} here, we get:
\begin{equation}
	(\bar P + \delta P)\delta^i{}_j + \pi^i{}_{j} = \int d_3\mathbf{P}\frac{P^i P_j}{a^3\left(1 + 3\Phi\right)E(p)}f(x^i,P_j,\eta)\;,
\end{equation}
where we have already substituted the expressions \eqref{sqrtgNewtoniangauge} and \eqref{P0Erel} for $\sqrt{-g}$ and $P^0$. Using Eq. \eqref{conjPproperpprel} to change the integration variable from the conjugate momentum to the proper one, we obtain:
\begin{equation}
	(\bar P + \delta P)\delta^i{}_j + \pi^i{}_{j} = \int d^3\mathbf p\frac{p^i p_j}{E(p)}f(x^i,p_j,\eta)\;.
\end{equation}
Again, no metric perturbations are explicitly present in the integrand. The pressure perturbation is:
\begin{equation}\label{deltaPd3p}
	\delta P = \int d^3\mathbf p\frac{p^2}{3E(p)}\bar f(ap)\mathcal F(x^i,p_j,\eta)\;,
\end{equation}
whereas the anisotropic stress is:
\begin{equation}\label{piijd3p}
	\pi^i{}_{j} = \int d^3\mathbf p\frac{(p^i p_j - p^2\delta^i{}_j/3)}{E(p)}\bar f(ap)\mathcal F(x^i,p_j,\eta)\;.
\end{equation}
The scalar contribution of the latter, as we have seen, is obtained via contraction with $\hat k_i\hat k^j$.

Using $q_i$, the pressure perturbation becomes:
\begin{equation}\label{deltaPd3q}
	\delta P = \frac{1}{a^4}\int d^3\mathbf q\frac{q^2}{3\sqrt{q^2 + m^2a^2}}\bar f(q)\mathcal F(x^i,q_j,\eta)\;,
\end{equation}
and the anisotropic stress:
\begin{equation}\label{piijd3q}
	\pi^i{}_{j} = \frac{1}{a^4}\int d^3\mathbf q\frac{(q^i q_j - q^2\delta^i{}_j/3)}{\sqrt{q^2 + m^2a^2}}\bar f(q)\mathcal F(x^i,q_j,\eta)\;.
\end{equation}

\subsection{The equation for the velocity}

The $0-i$ Einstein equation can be written using Eqs.~\eqref{deltaG0i} and \eqref{genT0imixed} as follows:
\begin{equation}\label{0iEinsteineq}
 \partial_i(\Phi' - \mathcal H\Psi) = 4\pi Ga^2\left(\rho + P\right)v_i\;.
\end{equation}
Upon FT, we obtain:
\begin{equation}
 ik_i(\Phi' - \mathcal H\Psi) = 4\pi Ga^2\left(\rho + P\right)v_i\;,
\end{equation}
and now we must obtain the scalar part of this vectorial equation by contracting with $i\hat{k}^i$, as we showed in Eq.~\eqref{extractingscalarpartfromvector}. Hence, we get:
\begin{equation}
 \boxed{k(-\Phi' + \mathcal H\Psi) = 4\pi Ga^2\left(\rho + P\right)V}
\end{equation}
where:
\begin{align}\label{kVdefinition}
    kV := i\hat{k}^iv_i\,.
\end{align}
Compared with the notation employed in \cite{Ma:1995ey}, one has $\theta = kV$. Note that the $\left(\rho + P\right)V$ in the above equation is the total; hence, to make explicit the various contributions, one has:
\begin{equation}\label{EEqV}
 \boxed{k(-\Phi' + \mathcal H\Psi) = 4\pi Ga^2\left(\rho_{\rm c}V_{\rm c} + \rho_{\rm b}V_{\rm b} + \frac{4}{3}\rho_\gamma V_\gamma + \frac{4}{3}\rho_\nu V_\nu\right)}
\end{equation}
where we have considered the usual equations of state for the various components, i.e., $P_{\rm c} = P_{\rm b} = 0$, $P_\gamma = \rho_\gamma/3$, and $P_\nu = \rho_\nu/3$.

Using Eq.~\eqref{deltaT0ikinetictheory}
\begin{equation}
	(\bar\rho + \bar P)v_i = \int d_3\mathbf{P}\frac{P_i}{\sqrt{-g}}f(x^i,P_j,\eta)\;,
\end{equation}
and again, exploiting the previous formulae, we get:
\begin{equation}
	(\bar\rho + \bar P)v_i = \int d^3\mathbf p(1 + \Phi - \Psi)p_if(x^i,p_j,\eta)\;.
\end{equation}
This time, the metric perturbations remain; however, note the following. There is no velocity flow in the background, so in the above integral we need to use $\bar f\mathcal F$ in order to obtain a non-vanishing result. But then $(\Phi - \Psi)\bar f\mathcal F$ is a second order quantity, and we are entitled to discard it. Hence:
\begin{equation}\label{rhoPvid3p}
	\boxed{(\bar\rho + \bar P)v_i = \int d^3\mathbf pp_i\bar f(ap)\mathcal F(x^i,p_j,\eta)}
\end{equation}
Contracting with $i\hat k^i$, we obtain the scalar part $V$ of the velocity flow perturbation.

Using $q_i$, the flow velocity becomes:
\begin{equation}\label{rhoPvid3q}
	\boxed{(\bar\rho + \bar P)v_i = \frac{1}{a^4}\int d^3\mathbf qq_i\bar f(q)\mathcal F(x^i,q_j,\eta)}
\end{equation}

\subsection{The equation for the pressure perturbation}

Using Eq.~\eqref{tracedeltaGij}, we can immediately write down the last Einstein equation for scalar perturbations:
\begin{equation}\label{GiideltaPeq2}
 \boxed{\Phi'' + 2\mathcal{H}\Phi' - \mathcal{H}\Psi' - (2\mathcal{H}' + \mathcal{H}^2)\Psi + \frac{k^2}{3}(\Phi + \Psi) = - 4\pi Ga^2 \delta P}
\end{equation}

\hrulefill

\begin{ex}
	Use the previously derived transformations \eqref{SynchronoustoNewtoniangauge} from the Newtonian to the synchronous gauge and write there the Einstein equations.
\end{ex}

\hrulefill

The expression of $\delta P$ as an integral in the proper momentum space of $\bar f\mathcal F$ has been presented earlier.

Finally, note an important property of the set of equations \eqref{relativisticPoissonequation}, \eqref{anisotropicstressscalarequation}, \eqref{EEqV}, and \eqref{GiideltaPeq2}. They depend only on the modulus of the wavenumber $k$. Hence, any dependence on the direction of $\mathbf{k}$, which is necessary to describe the formation of structure, must be contained in the initial conditions. We will discuss the latter in Chapter \ref{Chap:IC}.

\section{Einstein equations for tensor perturbations}\label{Sec:EEtenspert}
\index{Einstein equations!Tensor perturbations}

Our perturbed FLRW metric, with tensor perturbations only, can be expressed as follows:
\begin{equation}\label{tenspertmetric}
	g_{00} = -a^2\;, \qquad g_{0i} = 0\;, \qquad g_{ij} = a^2(\delta_{ij} + h^T_{ij})\;,
\end{equation}
where $h^T_{ij}$ is divergence-free and traceless.

\hrulefill

\begin{ex} Start from metric \eqref{tenspertmetric} and calculate the perturbed Einstein tensor $\delta G^\mu{}_\nu$. Verify the calculations also using Eqs.~\eqref{deltaG00gen}-\eqref{deltaGijgen}. Show that the only non-vanishing components are:
\begin{equation}\label{Gijtens}
	\boxed{2a^2\delta G^i{}_j = h^{T''}_{ij} + 2\mathcal Hh^{T'}_{ij}  - \nabla^2 h^T_{ij}}
\end{equation}
Note that the wave operator has appeared.
\end{ex}

\hrulefill  

The calculation of the above exercise is straightforward because the tensor nature of the perturbation $h^T_{ij}$ already suggests that it cannot contribute to $R_{00}$, $R_{0i}$, and the Ricci scalar $R$ (at first-order). 

The tensor part of the Einstein equations is as follows:
\begin{equation}\label{GijTensEinsteineqs}
	\boxed{h^{T''}_{ij} + 2\mathcal Hh^{T'}_{ij}  + k^2 h^T_{ij} = 16\pi Ga^2\pi_{ij}^T}
\end{equation}
where $\pi^T_{ij}$ is the tensorial part of the anisotropic stress, which can be computed from the total as shown in Eq.~\eqref{extractingtensorperturbationfromtensor}:
\begin{equation}
	\pi^{T}_{ij} = \left(\delta^l{}_i - \hat{k}^l\hat{k}_i\right)\left(\delta^m{}_j - \hat{k}^m\hat{k}_j\right)\pi_{lm} + \frac{1}{2}\hat{k}^l\hat{k}^m\pi_{lm}\left(\delta_{ij} - \hat{k}_i\hat{k}_j\right)\;.
\end{equation}
In Fourier space, the divergenceless condition can be written as:
\begin{equation}\label{divergencelessconditionhT}
	\hat{k}^ih^T_{ij} = 0\;.
\end{equation}
Therefore, $h^T_{ij}$ can be expanded with respect to a 2-dimensional basis $\{\hat e_1, \hat e_2\}$ defined on the 2-dimensional subspace orthogonal to $\hat k$. The basis thus satisfies the condition:
\begin{equation}
	\gamma^{ij}e_{a,i}\hat k_j = 0\;, \qquad \gamma^{ij}e_{a,i}e_{b,j} = \delta_{ab}\;,
\end{equation}
where $\gamma_{ij}$ is the metric on the 2-dimensional subspace and $a, b \in \{1,2\}$. We can then write $h^T_{ij}$ as follows:
\begin{equation}
	h^T_{ij}(\mathbf k) = (e_{1,i}e_{1,j} - e_{2,i}e_{2,j})(\hat k)h_+(\mathbf k) + (e_{1,i}e_{2,j} + e_{2,i}e_{1,j})(\hat k)h_\times(\mathbf k)\;.
\end{equation}
Note the dependence on $\hat k$ of the combinations of the 2-dimensional basis vectors. In fact, these depend on the orientation of $\hat k$.

Substituting the above expansion into Eq.~\eqref{GijTensEinsteineqs} and, in the absence of quadrupole moments, $h_{+,\times}$ satisfies the equation:
\begin{equation}\label{GWeqconftime}
	\boxed{h_{+,\times}'' + 2\mathcal Hh_{+,\times}' + k^2h_{+,\times} = 0}
\end{equation}
which we will employ in order to investigate the GW production during inflation in Sec.~\ref{Sec:ProductionofGWduringinfl}.\index{Gravitational waves!Equation}

If we choose a Cartesian reference frame and $\hat{k} = \hat{z}$, i.e., a propagation direction of a gravitational wave along $\hat{z}$, then a natural choice is $\hat e_1 = \hat x$ and $\hat e_2 = \hat y$, and the perturbed metric can be written as:
\begin{equation}\label{tensorperthphx}
	h^T_{ij}(k\hat z) = \left(
	\begin{array}{lll}
		h_+ & h_\times & 0\\
		h_\times & -h_+ & 0\\
		0 & 0 & 0
	\end{array}
	\right)\;.
\end{equation}
Though this is a convenient way of expressing $h_{ij}^T(k\hat z)$, one usually prefers to use the combinations:
\begin{equation}
	h_+ \mp ih_\times\;,
\end{equation}
since these have helicity $\pm 2$, see e.g. \cite{Weinberg:1972}. In order to see this, we apply a rotation about $\hat{z}$ and calculate how $h_{ij}^T(k\hat z)$ transforms.\index{Gravitational waves!Helicity}

\hrulefill

\begin{ex}
	Apply the rotation:
	\begin{equation}
		R^i{}_j(\theta) = \left(\begin{array}{ccc}
			\cos\theta & -\sin\theta & 0\\
			\sin\theta & \cos\theta & 0\\
			0 & 0 & 1
		\end{array}
		\right)\;,
	\end{equation}
about the $\hat z$ axis	to $h_{ij}^T$, i.e. compute the components:
	\begin{equation}
		\bar h_{lm}^T = R^i{}_{l}R^j{}_{m}h_{ij}^T\;,
	\end{equation}
	and show that:
	\begin{eqnarray}
		\bar h_+ = \cos^2\theta h_+ + 2\sin\theta\cos\theta h_\times - \sin^2\theta h_+ = \cos2\theta h_+ + \sin2\theta h_\times\;,\\
		\bar h_\times = \cos^2\theta h_\times - 2\sin\theta\cos\theta h_+ - \sin^2\theta h_\times = \cos2\theta h_\times - \sin2\theta h_+\;.
	\end{eqnarray}
\end{ex}

\hrulefill

Hence, the aforementioned combinations $h_+ \pm ih_\times$ transform as follows:
\begin{eqnarray}
	\boxed{\bar h_+ \pm i\bar h_\times = e^{\mp 2i\theta}(h_+ \pm ih_\times)}
\end{eqnarray}
and have thus helicity $\mp 2$. Sometimes, the sign could be a bit confusing, depending on which direction the rotation is performed. By convention, $\theta > 0$ denotes an anti-clockwise rotation, so a rotation of $\theta$ about the $\hat z$ axis corresponds to a $-\theta$ rotation about the $-\hat z$ axis, which is the line of sight and therefore the relevant direction for the observer. Thus, the observed helicities have opposite signs with respect to the propagating ones. 

So, we write the total tensor perturbation as a sum over the helicities:
\begin{equation}\label{hTijhelicitiesexpansionkparz}
	h_{ij}^T(\eta, k\hat z) = \sum_{\lambda = \pm 2}e_{ij}(\hat{z},\lambda)h(\eta, k\hat z, \lambda)\;,
\end{equation}
where: 
\begin{equation}\label{normalizationpolarizationGW}
	e_{11}(\hat z,\pm 2) = -e_{22}(\hat z,\pm 2) = \mp ie_{12}(\hat z,\pm 2) = \mp ie_{21}(\hat z,\pm 2) = \frac{1}{\sqrt{2}}\;,
\end{equation}
and of course $e_{3i} = e_{i3} = 0$. Therefore, from Eq.~\eqref{hTijhelicitiesexpansionkparz} we get:
\begin{eqnarray}
	h_+(\eta, k\hat z) = \frac{1}{\sqrt{2}}h(\eta, k\hat z, +2) + \frac{1}{\sqrt{2}}h(\eta, k\hat z, -2)\;,\\
	h_\times(\eta, k\hat z) = \frac{i}{\sqrt{2}}h(\eta, k\hat z, +2) - \frac{i}{\sqrt{2}}h(\eta, k\hat z, -2)\;,
\end{eqnarray}
and inverting:
\begin{eqnarray}
	\sqrt{2}h(\eta, k\hat z, +2) = h_+(\eta, k\hat z) - ih_\times(\eta, k\hat z)\;,\\
	\sqrt{2}h(\eta, k\hat z, -2) = h_+(\eta, k\hat z) + ih_\times(\eta, k\hat z)\;.
\end{eqnarray}
For $\mathbf k$ in a generic direction, we have that:
\begin{equation}\label{hTijhelicitiesexpansion}
	h_{ij}^T(\eta, \mathbf k) = \sum_{\lambda = \pm 2}e_{ij}(\hat{k},\lambda)h(\eta, \mathbf k, \lambda)\;,
\end{equation}
where the polarization tensor is defined as:
\begin{equation}
	e_{ij}(\hat{k},\pm 2) = \sqrt{2}e_{\pm,i}e_{\pm,j}\;,
\end{equation}
where the \textbf{polarization vectors} are:\index{Polarization vectors}
\begin{equation}\label{polarizationvectors}
	e_{\pm,i}(\hat k) \equiv \frac{(e_{1} \pm ie_2)_i}{\sqrt 2}\;.
\end{equation}
Of course $e_{ij}(\hat{k},\lambda)$ has the same symmetry as $h_{ij}^T(\eta, \mathbf k)$ and thus is traceless and transverse, i.e.
\begin{equation}
	\hat k^le_{lm}(\hat{k},\lambda) = 0\;,
\end{equation}
Again, the two $h(\eta, \mathbf k, \pm 2)$ satisfy the same Eq.~\eqref{GWeqconftime} as for $h_{+,\times}$.

When we discuss the effect of GW on photon propagation and the CMB, we shall have to deal with two directions: one is the GW direction $\hat k$ and the other is the direction of photon propagation $\hat p$. The strategy to simplify the calculations is to choose a frame in which $\hat k = \hat z$. Afterwards, before taking the anti-Fourier transform and obtaining the physical quantities in real space, one must perform a rotation that brings $\hat k$ back in a generic direction. We shall see this in detail in Chapter \ref{Chap:CMBEvo}.

\section{Einstein equations for vector perturbations}
\index{Einstein equations!Vector perturbations}

Finally, we address vector perturbations. The vector-perturbed FLRW metric, in the Newtonian gauge, has the following form:
\begin{equation}\label{vectpertmetric}
	g_{00} = -a^2\;, \qquad g_{0i} = 0\;, \qquad g_{ij} = a^2(\delta_{ij} + h^V_{ij})\;,
\end{equation}
where
\begin{equation}\label{hvectexp}
	h^V_{ij} = \partial_iA_j + \partial_jA_i\;,
\end{equation}
and $A_i$ is a divergenceless vector, i.e. $\partial_iA^i = 0$.

\hrulefill

\begin{ex} Repeat the very same calculations performed in the tensor case but now for $h^V_{ij}$. Note a very important difference: $h^V_{ij}$ is traceless but NOT divergenceless. Compare the results with those found using Eqs.~\eqref{deltaG00gen}-\eqref{deltaGijgen}. Show that the non-vanishing components of the perturbed Einstein tensor are:
\begin{eqnarray}
\label{deltaG0ivect}	2a^2\delta G^0{}_{i} &=& -\partial_l h^{V'}_{li} = - \nabla^2A_i'\;,\\
\label{deltaGijvect}	2a^2\delta G^i{}_{j} &=& h^{V''}_{ij} + 2\mathcal H h^{V'}_{ij} = (\partial_iA_j + \partial_jA_i)'' + 2\mathcal H(\partial_iA_j + \partial_jA_i)'\;.
\end{eqnarray}
With the Laplacian missing, the last equation is no more a wave equation, such as the one for tensor perturbations. With no vector sources, show then that in the early, radiation-dominated universe, for which $\mathcal H = 1/\eta$, one has: 
\begin{equation}
	h^V_{ij} \propto 1/\eta^2\;,
\end{equation}
and hence vector perturbations vanish, if not sourced.
\end{ex}

\hrulefill

Einstein equations are thus:
\begin{eqnarray}
	kF_i' = -32\pi Ga^2(\rho + P)U_i\;,\\
	(i\hat{k}_iF_j + i\hat k_jF_i)'' + 2\mathcal H(i\hat k_iF_j + i\hat k_jF_i)' = -32\pi Ga^2\pi_{ij}^V\;,
\end{eqnarray}
where $U_i$ is the vector part of $v_i$, defined in Eq.~\eqref{viSVTdec}; $F_i$ is defined in Eq.~\eqref{correctionintheFTquantities}; and $\pi_{ij}^V$ is the vector part of the anisotropic stress, defined as from Eq.~\eqref{extractingvectorpartfromtensor}:
\begin{equation}
	\pi_{ij}^V = 2i\left(\delta^m{}_i - \hat{k}^m\hat{k}_i\right)\hat{k}^l\pi_{lm}\hat{k}_j + 2i\left(\delta^m{}_j - \hat{k}^m\hat{k}_j\right)\hat{k}^l\pi_{lm}\hat{k}_i\;,
\end{equation}
Of course, we have that:
\begin{equation}
	\hat k^iF_i = 0\;, \qquad \hat k^iU_i = 0\;, \qquad \hat k^i\hat k^j\pi_{ij}^V = 0\;.
\end{equation}
Contracting the second Einstein equation with $i\hat k^j$ leaves us with:
\begin{equation}
	\boxed{F_i'' + 2\mathcal HF_i' = 32\pi Ga^2i\hat k^j\pi_{ij}^V}
\end{equation}
We choose, as in the tensor case, to align $\hat{k}$ to $\hat{z}$. Therefore, the divergencelessness of $A_i$ implies that
\begin{equation}
	ik^iA_i = 0 \quad \Rightarrow \quad A_3 = 0\;,
\end{equation}
and from Eq.~\eqref{hvectexp} we have:
\begin{equation}
	h^V_{ij} = ik_iA_j + ik_jA_i = -\frac{i}{2}\hat{k}_iF_j - \frac{i}{2}\hat{k}_jF_i\;.
\end{equation}
So, $h^V_{ij}$ can be written as follows:
\begin{equation}\label{vectoreperturbedmetricFi}
	h^V_{ij} = \left(
	\begin{array}{lll}
	0 & 0 & -iF_{1}/2\\
		0 & 0 & -iF_{2}/2\\
		-iF_{1}/2 & -iF_{2}/2 & 0
	\end{array}
	\right)\;.
\end{equation}

\hrulefill

\begin{ex}
	Applying the same rotation about $\hat z$ as we did for tensor perturbations. Show that:
	\begin{eqnarray}
		\bar F_1 = F_1\cos\theta - F_2\sin\theta\;,\\
		\bar F_2 = F_1\sin\theta + F_2\cos\theta\;.
	\end{eqnarray}
\end{ex}

\hrulefill

Hence, we have:
\begin{eqnarray}
	\bar F_1 + i\bar F_2 = (F_1 + iF_2)e^{i\theta}\;,\\
	\bar F_1 - i\bar F_2 = (F_1 - iF_2)e^{-i\theta}\;,
\end{eqnarray}
and therefore, the quantities:
\begin{equation}
	F_\pm \equiv F_1 \pm iF_2\;,
\end{equation}
are fields with helicity $\pm 1$.

\clearpage 
\chapter{Perturbed Boltzmann equations}\label{Chap:PertubedBoltzmannEquations}

{\rightskip=3truepc\leftskip=3truepc\noindent
{\it the Boltzmann equation plays a similar role for physicists and astronomers: no one ever talks about it, but everyone is always thinking about it}
\vskip 0.10 in
\centerline{\it ---Scott Dodelson, Modern Cosmology}
\vskip 0.20 in
}

In this chapter, we work out the perturbed Boltzmann equations for cold dark matter, massless neutrinos, photons, and baryons. We shall use these equations to track the evolution of small fluctuations in these components and couple them to the Einstein equations. For the derivation of the hierarchy of temperature and polarization equations for photons, we primarily follow \cite{Ma:1995ey}, \cite{Hu:1997hp}, and \cite{Tram:2013ima}. We shall dedicate the largest part of the chapter to the photon perturbed Boltzmann equation because it is much more laborious than the others for two reasons: photons are massless, and they interact. Therefore, we cannot truncate the hierarchy of moment equations, and a collisional term must be taken into account. The latter comes from Thomson scattering, whose cross-section also depends on the polarization, thus further complicating the treatment of photon fluctuations, which we nonetheless will bravely face.

\section{General form of the perturbed Boltzmann equation}

We wrote in Eq.~\eqref{distributionfunctionperturbation} the distribution function $f(x^i,P_j,\eta)$ as the sum of a background contribution plus a perturbation:
\begin{equation}
	f(x^i,P_j,\eta) = \bar f(ap)[1 + \mathcal F(x^i,P_j,\eta)]\;.
\end{equation}
The functional dependence explicitly shows that homogeneity and isotropy are broken. As we commented for Eq.~\eqref{distributionfunctionperturbation}, in principle, the conjugate and proper momenta can also be split into a background contribution plus a perturbation. However, we do not make this separation explicit because there is no point in doing so. In fact, at the end, we will perform an integration in the phase space, and the evolution equations will not contain the particles momenta anymore. Therefore, $P$ and $p$ are to be treated as integration variables.

It turns out to be convenient to choose, instead of $P_j$, the proper momentum modulus $p$ and its direction $\hat{p}_i$, defined to satisfy:
\begin{equation}
	p^i \equiv p\hat{p}^i\,, \qquad \delta_{ij}\hat{p}^i\hat{p}^j = 1\;, \qquad \delta_{ij}p^ip^j = p^2\,.
\end{equation} 
With such a choice, the Liouville operator is as follows:
\begin{equation}\label{Liouvilloppert}
	\frac{df}{d\lambda} = P^0\frac{\partial f}{\partial x^0} + \underbrace{\frac{\partial f}{\partial x^i}}_\text{At least 1st order}P^i + \frac{\partial f}{\partial p}\frac{dp}{d\lambda} + \underbrace{\frac{\partial f}{\partial \hat{p}_i}\frac{d\hat{p}_i}{d\lambda}}_\text{At least 2nd order}\;.
\end{equation}
We have stressed that $\partial f/\partial x^i$, $\partial f/\partial \hat{p}_i$, and $d\hat{p}_i/d\lambda$ have no background contributions and are thus at least first-order quantities. For this reason, the last term of Eq.~\eqref{Liouvilloppert} is at least of second order and thus negligible. Let us examine in more detail why.\index{Boltzmann equation!Perturbation}

The term $\partial f/\partial x^i$ is at least first order because, in order for it to be different from zero, $f$ must depend on $x^i$, thereby breaking homogeneity. The same reasoning applies to $\partial f/\partial \hat{p}_i$, which, if non-vanishing, is related to the breaking of isotropy. Finally, $d\hat{p}_i/d\lambda$ represents the deviation of the direction of a photon along its path, and it is a quantity very useful to analyze when studying gravitational lensing. On the other hand, in a homogeneous and isotropic space, there are no clumps of matter that can act as lenses, so $d\hat{p}_i/d\lambda$ vanishes in the background: it is at least a first-order quantity.\footnote{The effect of the expansion of the universe on the usual gravitational deflections by a matter lens is treated, for example, in \cite{Piattella:2017uat}.}

Dividing by $P^0$, we can write the first-order perturbed Boltzmann equation as follows:
\begin{equation}\label{PertBoltzeqetagen}
	\boxed{\frac{df}{dx^0} = \frac{\partial f}{\partial x^0} + \frac{\partial f}{\partial x^i}\frac{P^i}{P^0} + \frac{\partial f}{\partial p}\frac{dp}{dx^0} = \frac{1}{P^0}C[f]}
\end{equation}
Let us work with the conformal time from now on, so $x^0 = \eta$.

The above equation contains a background and a perturbed part. The background part is:
\begin{equation}\label{PertBoltzeqetagenothorder}
	\frac{\partial \bar f}{\partial \eta} + \frac{\partial \bar f}{\partial p}\left.\frac{dp}{d\eta}\right|_\text{0th} = 0\;,
\end{equation} 
where we already know that:
\begin{equation}
	\left.\frac{dp}{d\eta}\right|_\text{0th} = -\mathcal H p\;.
\end{equation}
We take the background distribution to be the quasi equilibrium one. In this way, the background part of the collisional term is vanishing, and it must be considered a purely perturbative quantity. This assumption means that we cannot apply the formalism that we are describing here to those epochs studied in Chapter \ref{Chap:ThermalHistory}. This is fine for dark matter freeze-out and for Big-Bang nucleosynthesis, as we are interested in much later epochs; however, regarding recombination, we will need to make some extra assumptions.

So, the linearly perturbed Boltzmann equation is:
\begin{equation}
	\frac{\partial (\bar f\mathcal F)}{\partial\eta} + \frac{\partial (\bar f\mathcal F)}{\partial x^i}\left.\frac{P^i}{P^0}\right|_\text{0th} + \frac{\partial \bar f}{\partial p}\left.\frac{dp}{d\eta}\right|_\text{1st} - \mathcal Hp\frac{\partial (\bar f\mathcal F)}{\partial p} = \frac{1}{P^0}C[\mathcal F]\;.
\end{equation}
The first and fourth terms on the left hand side can be worked out using the Leibniz rule and the zeroth order Boltzmann equation, leaving us with:
\begin{equation}\label{PertBoltzeqetagen1st}
	\frac{\partial \mathcal F}{\partial \eta} + \frac{\partial \mathcal F}{\partial x^i}\left.\frac{P^i}{P^0}\right|_\text{0th} + \frac{\partial \ln\bar f}{\partial p}\left.\frac{dp}{d\eta}\right|_\text{1st} - \mathcal Hp\frac{\partial \mathcal F}{\partial p} = \frac{1}{P^0\bar f}C[\mathcal F]\;.
\end{equation}
The structure of the first and fourth terms on the left hand side suggests a change of variable. As we anticipated in chapter \ref{Chap:CosmoPertTheory}, if we use:
\begin{equation}
	\boxed{q_i = ap_i}
\end{equation}
instead of the proper momentum, we get:
\begin{equation}\label{PertBoltzeqetagen1stq}
	\frac{\partial \mathcal F}{\partial \eta} + \frac{\partial \mathcal F}{\partial x^i}\left.\frac{P^i}{P^0}\right|_\text{0th} + \frac{\partial \ln\bar f}{\partial q}\left.\frac{dq}{d\eta}\right|_\text{1st}  = \frac{1}{P^0\bar f}C[\mathcal F]\;.
\end{equation}
Indeed:
\begin{equation}
	\left.\frac{dq}{d\eta}\right|_\text{0th} = \frac{da}{d\eta}\bar p + a\left.\frac{dp}{d\eta}\right|_\text{0th} = a'\bar p - a\mathcal H\bar p = 0\;.
\end{equation}
Physically, as we will see, the transformation $q = ap$ allows us to normalize the fluctuations of the temperature (for photons and neutrinos) to their background behavior $1/a$. Thereby, what remains is purely a perturbative effect.

Finally, we have to deal with 3 terms:
\begin{enumerate}
	\item The velocity term $P^i/P^0$.
	\item The force term $dq/d\eta$ or $dp/d\eta$.
	\item The collisional term $C[\mathcal F]$.
\end{enumerate}
The velocity term is the easiest to compute because it is of order zero, as $\partial \mathcal F/\partial x^i$ is a first-order quantity. Let us explicitly write the mass-shell relation:
\begin{equation}
	g_{\mu\nu}P^\mu P^\nu = -a^2(1 - h_{00})(P^0)^2 + a^2(\delta_{ij} + h_{ij})P^iP^j = -m^2\;,
\end{equation}
and define the energy and proper momentum modulus as:
\begin{equation}
	E = a\left(1 - \frac{1}{2}h_{00}\right)P^0\;, \qquad p^2 = a^2(\delta_{ij} + h_{ij})P^iP^j\;.
\end{equation}
Since we have to keep order zero, we can neglect the perturbed metric contribution and write immediately:
\begin{equation}
	\frac{P^i}{P^0} \stackrel{\text{0th}}{=} \frac{p}{E}\hat p^i\;.
\end{equation}
The force term is made explicit through the geodesic equation and hence depends on the metric perturbations. The only collisional term that we shall consider explicitly is the one related to Thomson scattering, which will be of interest only for photons and free electrons.

\subsection{Force term}\index{Boltzmann equation!Force term}

The force term originates from the metric perturbations. We first calculate it for a perturbed metric $a^2h_{\mu\nu}$ with $h_{0i} = 0$, and then specify the result obtained for the various types of cosmological perturbations: scalar, tensor, and vector.

\hrulefill

\begin{ex}
Determine $dE/d\eta$ using the geodesic equation:
\begin{equation}
	\frac{dP^0}{d\lambda} = -\Gamma^0_{\alpha\beta}P^\alpha P^\beta\;.
\end{equation}
First show that:
\begin{equation}
	\frac{dE}{d\eta} = \mathcal H E - E\frac{d(h_{00}/2)}{d\eta} - \Gamma^0_{\alpha\beta}\frac{P^\alpha P^\beta}{E}a^2(1 - h_{00})\;.
\end{equation}
Working out the Christoffel symbol term:
\begin{eqnarray}
	-2\Gamma^0_{\alpha\beta}P^\alpha P^\beta a^2(1 - h_{00}) = -2aa'(1 - h_{00})(P^0)^2 + a^2h_{00}'(P^0)^2 + 2a^2\partial_lh_{00}P^0P^l\nonumber\\
	-2aa'(\delta_{ij} + h_{ij})P^iP^j - a^2h_{ij}'P^iP^j\;. \qquad
\end{eqnarray}
Hence, using again the definitions of $P^0$ and of the proper momentum, get:
\begin{equation}
	- \Gamma^0_{\alpha\beta}\frac{P^\alpha P^\beta}{E}a^2(1 - h_{00}) = -\mathcal HE + E\frac{h_{00}'}{2} + p\hat{p}^l\partial_lh_{00} - \mathcal H\frac{p^2}{E} - \frac{1}{2E}h_{ij}'p^ip^j\;.
\end{equation}
\end{ex}

\hrulefill

Collecting all the contributions found in the above exercise, we have: 
\begin{equation}
	\frac{dE}{d\eta} = - \mathcal H\frac{p^2}{E} + p\hat{p}^l\partial_l\frac{h_{00}}{2} - \frac{1}{2E}h_{ij}'p^ip^j\;,
\end{equation}
and using the differentiated mass-shell relation ($EdE =pdp$), we can write:
\begin{equation}\label{generalforceterm}
	\boxed{\frac{dp}{d\eta} = - \mathcal Hp + E\hat{p}^l\partial_l\frac{h_{00}}{2} - \frac{p}{2}h_{ij}'\hat p^i\hat p^j}
\end{equation}
Using the variable $q$, one obtains:
\begin{equation}\label{generalforcetermq}
	\boxed{\frac{dq}{d\eta} = aE\hat{p}^l\partial_l\frac{h_{00}}{2} - \frac{q}{2}h_{ij}'\hat p^i\hat p^j}
\end{equation}
Note that if we choose the synchronous gauge, then $h_{00} = 0$ and all the perturbation types are elegantly taken into account in the last term.\footnote{Note that since we have already chosen $h_{0i} = 0$, the Newtonian gauge can be realized as in the second definition, Eq. \eqref{2newtoniangaugecondition}, given in Chapter \ref{Chap:CosmoPertTheory}.} 

The perturbed Boltzmann equation thus becomes:
\begin{equation}
	\frac{\partial \mathcal F}{\partial\eta} + \frac{q\hat{p}^i}{aE}\frac{\partial \mathcal F}{\partial x^i} + \left(\frac{aE}{q}\hat{p}^l\partial_l\frac{h_{00}}{2} - \frac{1}{2}h_{ij}'\hat p^i\hat p^j\right)\frac{\partial \ln \bar f}{\partial \ln q} = \frac{a}{\bar fE}C[\mathcal F]\;,
\end{equation}
where we have taken into account that the collisional term is at least a first-order quantity and thus neglected its multiplication by $h_{00}$. 

Introducing the normal mode decomposition, the perturbed Boltzmann equation can be written as:
\begin{equation}\label{pertBoltzeqgeneralFT}
	\boxed{\frac{\partial \mathcal F}{\partial\eta} + \frac{ik\mu q}{aE}\mathcal F + \left(ik\mu \frac{aE}{q}\frac{h_{00}}{2} - \frac{1}{2}h_{ij}'\hat{p}^i\hat{p}^j\right)\frac{\partial \ln \bar f}{\partial \ln q} = \frac{a}{\bar fE}C[\mathcal F]}
\end{equation}
where we have defined 
\begin{equation}\label{definitionofmu}
	\boxed{\mu \equiv \hat{k}\cdot\hat{p}}
\end{equation}
This quantity is the cosine of the angle between the wave vector of the perturbation and the direction of the particle, which are the two relevant directions in this context.

Unlike the perturbed Einstein equations that we derived in Chapter \ref{Chap:CosmoPertTheory}, the perturbed Boltzmann equation induces a dependence on the direction of the wave vector, due to the presence of $\mu$. Beyond that, further angular dependencies might also arise due to the term $h_{ij}'\hat{p}^i\hat{p}^j$.

\subsection{Perturbed Boltzmann equation sourced by scalar perturbations}

Using the Newtonian gauge, cf. Eq.~\eqref{confnewtmetric}, we can identify $h_{00} = -2\Psi$ and $h_{ij} = 2\Phi\delta_{ij}$. As we saw in Chapter \ref{Chap:CosmoPertTheory}, the proper momentum modulus is defined as:
\begin{equation}
	p^2 = \frac{1}{a^2}(1 - 2\Phi)\delta^{ij}P_iP_j\;, \qquad P_i = a(1 + \Phi)p_i\;,
\end{equation}
and
\begin{equation}\label{P0CDM}
	P^0 = \frac{E}{a}(1 - \Psi)\;.
\end{equation}
From Eq.~\eqref{generalforceterm}, we have:
\begin{equation}
	\frac{dE}{d\eta} = -\mathcal H\frac{p^2}{E} - \frac{p^2}{E}\Phi' - p\hat{p}^i\partial_i\Psi\;,
\end{equation}
and then:
\begin{equation}
	\boxed{\frac{dq}{d\eta} = - q\Phi' - aE\hat{p}^i\partial_i\Psi}
\end{equation}
The first term on the right hand side takes into account the time variation of the spatial curvature, and the second is the gradient of the gravitational potential along the direction of the particle. The perturbed Boltzmann equation with scalar perturbations has the following general form:
\begin{equation}\label{pertBoltzeqgeneralscalarpertFT}
	\boxed{\frac{\partial \mathcal F}{\partial\eta} + ik\mu\frac{q}{aE}\mathcal F - \left(ik\mu \frac{aE}{q}\Psi + \Phi'\right)\frac{\partial \ln \bar f}{\partial \ln q} = \frac{a}{\bar fE}C[\mathcal F]}
\end{equation}

\subsection{Perturbed Boltzmann equation sourced by tensor perturbations}\index{Boltzmann equation!Force term!Tensor perturbations}

From Eq.~\eqref{tenspertmetric} we have that $h_{00} = 0$ and $h_{ij} = h_{ij}^T$, with $h_{ij}^T$ transverse and traceless. From Eq.~\eqref{generalforceterm}, we have:
\begin{equation}
	\frac{dE}{d\eta} = -\mathcal H\frac{p^2}{E} - \frac{p^2}{2E}h_{ij}^{T'}\hat{p}^i\hat{p}^j\;, \qquad \frac{dq}{d\eta} = - \frac{q}{2}h_{ij}^{T'}\hat{p}^i\hat{p}^j\;.
\end{equation}
The Boltzmann equation can be expressed as follows:
\begin{equation}\label{pertBoltzeqgeneraltensorpertFT}
	\boxed{\frac{\partial \mathcal F}{\partial\eta} + \frac{ik\mu q}{aE}\mathcal F - \frac{1}{2}h_{ij}^{T'}\hat{p}^i\hat{p}^j\frac{\partial \ln \bar f}{\partial \ln q} = \frac{a}{\bar fE}C[\mathcal F]}
\end{equation}
Let us choose $\hat k = \hat z$. From Eq.~\eqref{tensorperthphx}, we then have that $h_{11}^T = -h_{22}^T = h_+$ and $h_{12}^T = h_{21}^T = h_\times$; hence, the contributions $\hat{p}^1$ and $\hat{p}^2$ shall be selected in the term $h_{ij}^{T'}\hat{p}^i\hat{p}^j$. The particle direction unit vector can be written in spherical coordinates as follows:
\begin{equation}\label{hatpiinsphericalcoordinates}
	\hat{p} = (\sqrt{1 - \mu^2}\cos\phi,\sqrt{1 - \mu^2}\sin\phi,\mu)\;.
\end{equation}

\hrulefill

\begin{ex}
	Show that:
	\begin{eqnarray}\label{hijpipjexpansionthetaphi}
	h_{ij}^{T'}\hat{p}^i\hat{p}^j = h'_+(1 - \mu^2)\cos 2\phi + h'_\times(1 - \mu^2)\sin 2\phi =\nonumber\\ 2\sqrt{\frac{2\pi}{15}}\left[Y_2^{2}(\mu,\phi)(h_+' - ih_\times') + Y_2^{-2}(\mu,\phi)(h_+' + ih_\times')\right] = \nonumber\\
	4\sqrt{\frac{\pi}{15}}\left[Y_2^{2}(\mu,\phi)h'(\lambda = +2) + Y_2^{-2}(\mu,\phi)h'(\lambda = -2)\right]\;.
\end{eqnarray}
The spherical harmonic functions with $l = 2$ and $m = \pm 2$ have appeared.
\end{ex}

\hrulefill

So, the metric contribution $h_{ij}'\hat{p}^i\hat{p}^j$ to the perturbed Boltzmann equation carries, in the case of tensor perturbations, a $Y_2^{\pm 2}(\mu,\phi)$ proportionality, each coupled to the respective helicity of the gravitational wave. The appearance of this simple azimuthal dependence is due to the choice $\hat k = \hat z$. Rotational symmetry allows us to align one of $\hat{k}$ and $\hat{p}$ along $\hat{z}$, but not both. We will see how to treat the general case in due time.

\subsection{Perturbed Boltzmann equation sourced by vector perturbations}\index{Boltzmann equation!Force term!Vector perturbations} 

From Eq.~\eqref{vectpertmetric}, we have that $h_{00} = 0$ and $h_{ij} = h_{ij}^V$ are traceless and have a vanishing double divergence. From Eq.~\eqref{generalforceterm}, we have:
\begin{equation}
	\frac{dE}{d\eta} = -\mathcal H\frac{p^2}{E} - \frac{p^2}{2E}h_{ij}^{V'}\hat{p}^i\hat{p}^j\;, \qquad \frac{dq}{d\eta} = - \frac{q}{2}h_{ij}^{V'}\hat{p}^i\hat{p}^j\;.
\end{equation}
The Boltzmann equation can be expressed in a form similar to that of the tensor case:
\begin{equation}\label{pertBoltzeqgeneralvectorpertFT}
	\boxed{\frac{\partial \mathcal F}{\partial\eta} + \frac{ik\mu q}{aE}\mathcal F - \frac{1}{2}h_{ij}^{V'}\hat{p}^i\hat{p}^j\frac{\partial \ln \bar f}{\partial \ln q} = \frac{a}{\bar fE}C[\mathcal F]}
\end{equation}
Let us assume again $\hat k = \hat z$. In Eq.~\eqref{vectoreperturbedmetricFi}, we have defined $h_{13}^V = -iF_1/2$ and $h_{23}^V = -iF_2/2$; hence, again, an azimuthal dependence appears. In particular, using Eq.~\eqref{hatpiinsphericalcoordinates}:
\begin{eqnarray}\label{hijVpertthetaphiBoltzeq}
	h_{ij}^{V'}\hat{p}^i\hat{p}^j = -iF'_1\sqrt{1 - \mu^2}\mu\cos\phi - iF'_2\sqrt{1 - \mu^2}\mu\sin\phi = \nonumber\\
	-\sqrt{\frac{2\pi}{15}}\left[Y_2^{1}(\mu,\phi)(iF_1' + F_2') + Y_2^{-1}(\mu,\phi)(iF_1' - F_2')\right] =\nonumber\\
	-i\sqrt{\frac{2\pi}{15}}\left[Y_2^{1}(\mu,\phi)F_-' + Y_2^{-1}(\mu,\phi)F_+'\right]\;.
\end{eqnarray}
Thus, the metric contribution $h_{ij}'\hat{p}^i\hat{p}^j$ carries, in the case of vector perturbations, a $Y_2^{\pm 1}(\mu,\phi)$ proportionality.

One might ask about a possible $Y_1^{\pm 1}(\mu,\phi)$ contribution. This is absent because of our choice $h_{0i} = 0$. In general, the monopole ($\ell = 0$) and dipole ($\ell = 1$) contributions can be set to zero upon a suitable choice of the gauge.

We now treat the perturbed Boltzmann equation for the relevant species that constitute our universe. Since each of the following sections is devoted to a species, we do not use any subscripts specifying the same. This is intended to make the notation somewhat less cluttered.

\section{Moments of the perturbed Boltzmann equation. The continuity and Euler equations for cold dark matter.}

We have come to understand that solving the Boltzmann equation analytically is a challenging task, and we would gladly accept the support of special techniques and approximations. For example, would it be viable to treat a matter component solely by considering its density and velocity flow? This is the fluid approximation, and it is possible when the mass of the particle in the thermal bath is very large, much larger than the temperature of the thermal bath. This is the case for the hypothetical CDM particle in the epochs of interest, which are those well after the Big Bang nucleosynthesis.

In the fluid approximation, we integrate the Boltzmann equation in momentum space, weighing it with increasing powers of momentum, thus creating tensors of increasing rank. With each integration, a new variable appears, thus requiring us to take the next order moment in order to describe its evolution. If the particle mass is sufficiently large, we can stop this procedure at some point. It is clear, then, that the treatment of the perturbed Boltzmann equation for massless neutrinos and photons must be completely different.   

We can limit ourselves here to studying only scalar perturbations of the metric that source the evolution of perturbations because we restrict the hierarchy of moments to just two equations: one for the density contrast and the other for the fluid velocity. The former has only a scalar contribution, whereas the latter does have a vector contribution, albeit negligible, as vector perturbations are expected to be, for the reasons illustrated in Chapter \ref{Chap:CosmoPertTheory}.

Let us consider the perturbed Boltzmann equation Eq.~\eqref{pertBoltzeqgeneralscalarpertFT} with a vanishing collisional term. We have seen that the latter might be important for CDM, at least for WIMPs, at energies on the order of tens of MeV, before kinetic decoupling. However, as already mentioned, we are not interested in epochs that are so primordial. The reason is that the theory of cosmological perturbations has the primary purpose of describing the fluctuations in the temperature and polarization of the CMB, as well as characterizing the onset of the process of structure formation.

\subsection{Moment zero}

Consider Eq.~\eqref{pertBoltzeqgeneralscalarpertFT} with no $C[\mathcal F]$. Multiply it by $\bar f\sqrt{q^2 + m^2a^2}$ and integrate in the $\mathbf q$ space:
\begin{eqnarray}
	\int d^3\mathbf q\sqrt{q^2 + m^2a^2}\bar f\frac{\partial \mathcal F}{\partial\eta} + ik_i\int d^3\mathbf q\bar fq\hat{p}^i\mathcal F\nonumber\\ 
	- \int d^3\mathbf q\bar f\left(\frac{q^2 + m^2a^2}{q}\hat{p}^iik_i\Psi + \sqrt{q^2 + m^2a^2}\Phi'\right)\frac{\partial \ln \bar f}{\partial \ln q} = 0\;.
\end{eqnarray}
The integral containing $\Psi$ vanishes because $\bar f$ has no angular dependence; thus, when integrated with $\hat p^i$, the result is zero. Since $\partial\bar f/\partial\eta = 0$ and it is, of course, independent from $x^i$, we have:
\begin{eqnarray}
	\frac{\partial}{\partial\eta}\int d^3\mathbf q\sqrt{q^2 + m^2a^2}\bar f\mathcal F - \int d^3\mathbf q\frac{m^2aa'}{\sqrt{q^2 + m^2a^2}}\bar f\mathcal F\nonumber\\ + ik_i\int d^3\mathbf qq\hat{p}^i\bar f\mathcal F
	- \Phi'\int d^3\mathbf q q\sqrt{q^2 + m^2a^2}\frac{\partial \bar f}{\partial q} = 0\;.
\end{eqnarray}
Using the definitions of $\delta\rho$ and $v_i$ given in Eqs.~\eqref{deltarhod3q} and \eqref{rhoPvid3q}, and $a' = \mathcal H a$, we can write:
\begin{eqnarray}
	\frac{\partial}{\partial\eta}(a^4\delta\rho)' - \mathcal H\int d^3\mathbf q\frac{m^2a^2}{\sqrt{q^2 + m^2a^2}}\bar f\mathcal F + a^4(\bar\rho + \bar P)ik_i v^i\nonumber\\
	- \Phi'\int d^3\mathbf q q\sqrt{q^2 + m^2a^2}\frac{\partial \bar f}{\partial q} = 0\;.
\end{eqnarray}

\hrulefill

\begin{ex}
	Show, using also the definition of $\delta P$ of Eq.~\eqref{deltaPd3q}, that:
	\begin{equation}
		\int d^3\mathbf q\frac{m^2a^2}{\sqrt{q^2 + m^2a^2}}\bar f\mathcal F = a^4(\delta\rho - 3\delta P)\;. 
	\end{equation}
	Then, show that:
	\begin{equation}
		\int d^3\mathbf q q\sqrt{q^2 + m^2a^2}\frac{\partial \bar f}{\partial q} = 3a^4(\bar\rho + \bar P)\;.
	\end{equation}
	There is partial integration to be performed here, so be careful of taking into account the $q^2$ contribution in $d^3\mathbf q$.
\end{ex}

\hrulefill

Thus, the integrated perturbed Boltzmann equation, weighted with the energy, becomes:
\begin{equation}\label{deltarhoeqCDM}
	\delta \rho' + (\bar\rho + \bar P)kV + 3\mathcal H(\delta \rho + \delta P) + 3(\bar\rho + \bar P)\Phi' = 0\;,
\end{equation}
where $kV = ik_iv^i$, cf. Eq. \eqref{kVdefinition}.

\hrulefill

\begin{ex}
Show, using $\delta\rho = \rho\delta$, that:
\begin{equation}\label{deltaeqgeneric}
	\boxed{\delta' + (1 + w)kV + 3\mathcal H\left(\frac{\delta P}{\delta\rho} - w\right)\delta + 3(1 + w)\Phi' = 0}
\end{equation}
where we have introduced the background equation of state
\begin{equation}
	w \equiv \frac{P}{\rho}\;.
\end{equation} 
It is necessary here to use the background conservation equation $\rho' = -3\mathcal H(\rho + P)$ at some point during the calculation.
\end{ex}\index{Continuity equation!Perturbation}

\hrulefill

The procedure leading to Eq.~\eqref{deltaeqgeneric} is valid for non-interacting particles of any species. In the specific case of CDM, one simply takes $w = 0$ and $\delta P = 0$. 

Equation \eqref{deltaeqgeneric} alone is not enough to describe fluctuations in a non-interacting matter component. We have derived an evolution equation for $\delta\rho$ (or $\delta$) from the Boltzmann equation, but we do not know how $V$ and $\delta P$ evolve. 

\subsection{Moment one}

In order to find an equation describing the evolution of $V$, we take the first moment of Eq.~\eqref{pertBoltzeqgeneralscalarpertFT}. That is, we multiply it by $q\hat{p}^i\bar f$ and then integrate in the $\mathbf q$ space:
\begin{equation}
	\int d^3\mathbf q q\hat p^i\bar f\left[\frac{\partial \mathcal F}{\partial\eta} + ik\mu\frac{q}{aE}\mathcal F - \left(ik\mu \frac{aE}{q}\Psi + \Phi'\right)\frac{\partial \ln \bar f}{\partial \ln q}\right] = 0\;.
\end{equation} 
Let us take care of the four contributions above separately.

\paragraph{1.} The first term is straightforward to compute:
\begin{equation}
	\int d^3\mathbf q q\hat p^i\bar f\frac{\partial \mathcal F}{\partial\eta} = \frac{\partial}{\partial\eta}\int d^3\mathbf q q\hat p^i\bar f\mathcal F = \frac{\partial}{\partial\eta}\left[a^4(\bar\rho + \bar P)v^i\right]\;.
\end{equation} 

\paragraph{2.} The second term can be written as follows:
\begin{equation}
	ik^l\int d^3\mathbf q\frac{q^2\hat p^i\hat p_l}{\sqrt{q^2 + m^2a^2}}\bar f\mathcal F = ik^la^4\delta T^{i}{}_l = ik^la^4\left(\delta P\delta^{i}{}_l + \pi^{i}{}_l\right)\;.
\end{equation}

\paragraph{3.} The third integration is as follows:
\begin{equation}
	-ik^l\Psi\int d^3\mathbf q q\hat p^i\hat p_l \sqrt{q^2 + m^2a^2}\frac{\partial \bar f}{\partial q}\;.
\end{equation} 

\hrulefill

\begin{ex}
	Integrating by parts show that:
	\begin{equation}
		\int d^3\mathbf q q\hat p^i\hat p_l \sqrt{q^2 + m^2a^2}\frac{\partial \bar f}{\partial q} = -3\int d^3\mathbf q \hat p^i\hat p_l \sqrt{q^2 + m^2a^2}\bar f - \int d^3\mathbf q \frac{q^2\hat p^i\hat p_l}{\sqrt{q^2 + m^2a^2}}\bar f\;.
	\end{equation}
\end{ex}

\hrulefill

Since $\bar f$ does not depend on $\hat p^i$ and since:
\begin{equation}
	\int d^2\hat p \hat p^i\hat p_l = \delta^i{}_l/3\;,
\end{equation}
we can conclude that:
\begin{equation}
	-ik^l\Psi\int d^3\mathbf q q\hat p^i\hat p_l \sqrt{q^2 + m^2a^2}\frac{\partial \bar f}{\partial q} = ik^i\Psi a^4(\bar\rho + \bar P)\;.
\end{equation} 

\paragraph{4.} The fourth and last contribution is vanishing because $\bar f$ does not depend on $\hat p^i$.

\hrulefill

\begin{ex}
	Putting together all the contributions show that we get:
\begin{equation}
	v^{i'} + \mathcal H(1 - 3w)v^i + \frac{w'}{1 + w}v^i + \frac{ik^i\delta P}{\bar\rho(1 + w)} + \frac{ik^l\pi^{i}{}_l}{\bar\rho(1 + w)} + ik^i\Psi = 0
\end{equation}
\end{ex}

\hrulefill

Contracting with $i\hat k_i$, and thereby extracting the scalar contribution to the above equation, we get:
\begin{equation}\label{Vequationgeneral}
	\boxed{V' + \mathcal H(1 - 3w)V + \frac{w'}{1 + w}V - \frac{\delta P/\delta\rho}{1 + w}k\delta - \frac{k\hat{k}_i\hat{k}^l\pi^{i}{}_l}{\rho(1 + w)} - k\Psi = 0}
\end{equation}
This equation corresponds to the Euler equation.\index{Euler equation!Perturbation} If one extracts the vector part, it turns out that it is not sourced by $\Psi$, as expected.

\hrulefill

\begin{ex}
	Find Eqs.~\eqref{deltaeqgeneric} and \eqref{Vequationgeneral} starting now from:
\begin{equation}
	\nabla_\mu \delta T^{\mu\nu} = 0\;,
\end{equation}
i.e. using the fluid approximation directly, without passing through kinetic theory.
\end{ex}

\hrulefill

Now, after having taken the first moment of the perturbed Boltzmann equation, we do have an equation describing the evolution of $V$, but we still do not know how $\delta P$ evolves Moreover, a new variable has appeared: $\pi^{i}{}_l$.

Both new variables enter $\delta T^i{}_j$, as we have seen, so in order to describe their evolution, we need to compute the second moment of the perturbed Boltzmann equation, i.e., we should multiply Eq.~\eqref{pertBoltzeqgeneralscalarpertFT} by $q^2\hat p^i\hat p_l/\sqrt{q^2 + m^2a^2}$ and then integrate over the $\mathbf q$ space. Then, new variables appear, and we reiterate the procedure. We do not do that, though.\footnote{In \cite{Piattella:2013cma} and  \cite{Piattella:2015nda}, the CDM velocity dispersion is taken into account, and the second moment of the Boltzmann equation is computed.} We assume that for the particle under consideration, its mass is so large that we can neglect any term in which $m$ is found in the denominator.

\subsection{The equations for CDM}

This is the case for CDM. If no $m$ can stay in the denominator of any term, then $w = 0$, $\delta P = 0$, and $\pi^{i}{}_l = 0$, along with all the higher order quantities. Hence, the only equations describing fluctuations in CDM are:
\begin{equation}\label{deltaequationCDM}
	\boxed{\delta_{\rm c}' + kV_{\rm c} + 3\Phi' = 0}
\end{equation}
and
\begin{equation}\label{VequationCDM}	
	\boxed{V_{\rm c}' + \mathcal HV_{\rm c} - k\Psi = 0}
\end{equation}
Similar equations also describe the evolution of baryonic fluctuations within the same approximation made here. Eq.~\eqref{deltaequationCDM} is identical; just change the subscript ``c'' to ``b''. The other equation is different because we need to take into account the collisional term arising from Thomson scattering in the Boltzmann equation, which couples photons with free electrons. For this reason, we postpone the discussion of the equations describing the evolution of baryonic fluctuations to the end of this chapter, after the treatment of photons.

\hrulefill

\begin{ex}
	Transform Eqs.~\eqref{deltaequationCDM} and \eqref{VequationCDM} to the synchronous gauge. Show that one can exploit the residual gauge freedom in order to choose $V^{\rm syn}_{\rm c} = 0$. Therefore, in the synchronous gauge CDM is described by a single equation only.
\end{ex}

\hrulefill

The treatment of the Boltzmann equation for CDM is the simplest among the species of the standard model of cosmology because CDM is non-relativistic and non-interacting. In the next section, we start to see how things become complicated by considering a massless species: neutrinos.

\section{The perturbed Boltzmann equation for massless neutrinos}\label{Sec:masslessnuBoltzeq}

We assume neutrinos to be massless, although we know that at least one of their families does have mass. For a treatment of the Boltzmann equation for massive neutrinos, see \cite{Ma:1995ey}. 

For massless particles $p/E = 1$. Therefore, it is not convenient to tackle Eq.~\eqref{pertBoltzeqgeneralFT} by taking its moments because we are unable to truncate the procedure at any point. Instead, it is more convenient to build a hierarchy of differential equations via a partial wave expansion. 

The Eq.~\eqref{pertBoltzeqgeneralFT} for massless neutrinos is obtained by setting $m = 0$ and neglecting the collisional term:
\begin{equation}\label{pertBoltzeqgeneralFTnu}
	\frac{\partial \mathcal F}{\partial\eta} + ik\mu\mathcal F + \left(ik\mu \frac{h_{00}}{2} - \frac{1}{2}h_{ij}'\hat{p}^i\hat{p}^j\right)\frac{\partial \ln \bar f}{\partial \ln q} = 0\;.
\end{equation}
Recall that $\mathcal F$ is a function of $(\eta, k^i, q, \hat p^i)$. On the other hand, no term apart from the distribution functions contains $q$ explicitly in Eq. \eqref{pertBoltzeqgeneralFTnu}. Therefore, it is convenient to define:
\begin{equation}\label{mathcalNdefinition}
	\boxed{\mathcal N(\eta, \mathbf k, \hat{p}) \equiv \frac{\int dq\;q^2 q\bar f\mathcal F}{4\int dq\;q^2 q\bar f}}
\end{equation}
We have kept the factor $4$ in the denominator, unlike \cite{Ma:1995ey} in their definition of a similar quantity, which is referred to as $F_\nu$. Hence $4\mathcal N = F_\nu$ (for scalar perturbations only). Moreover, $4\mathcal{N}$ corresponds to the Fourier transform of Weinberg's $J$ \cite[Chapter 6]{Weinberg:2008zzc}.

\hrulefill

\begin{ex}
	Multiply Eq.~\eqref{pertBoltzeqgeneralFTnu} by $q^3\bar f$ and perform the $dq$ integration. Using the definition of Eq.~\eqref{mathcalNdefinition} show that one obtains:
	\begin{equation}\label{nuBoltzeqNgen}
	\boxed{\mathcal N' + ik\mu \mathcal N - \left(ik\mu\frac{h_{00}}{2} - \frac{1}{2}h_{ij}'\hat{p}^i\hat{p}^j\right) = 0}
\end{equation}
Recall that $\bar f$ does not depend explicitly on $\eta$.
\end{ex}

\hrulefill

From Eq. \eqref{deltarhod3q}, it is clear that the angular integral of $\mathcal{N}$ is proportional to the density contrast of neutrinos. Since $\varepsilon_\nu \propto T_\nu^4$, we then have $\delta_\nu = 4\delta T_\nu/T_\nu$. Therefore, although formally $\mathcal{N}$ is the angular differential relative density perturbation in the neutrino component, we can think of it informally as $\delta T_\nu/T_\nu$.\footnote{The reason our definition is different by a factor of four from those of \cite{Ma:1995ey} and \cite{Weinberg:2008zzc} is precisely to avoid a factor of 4 in the temperature fluctuation.}

\subsection{Scalar perturbations}

For scalar perturbations, Eq.~\eqref{nuBoltzeqNgen} becomes:
\begin{equation}\label{nuBoltzeqNscal}
	\boxed{\mathcal N^{(S)'} + ik\mu \mathcal N^{(S)} + \Phi' + ik\mu \Psi = 0}
\end{equation}
We note that this equation contains $\hat p$ only within $\mu$, so the differential operator here has axial symmetry in the sense that the physics depends only on the angle between $\hat k$ and $\hat p$. As we will demonstrate in Chapter \ref{Chap:IC}, the initial conditions for all the perturbative quantities do not depend at all on the particle direction $\hat{p}$. Therefore, we have $\mathcal N^{(S)}(\eta, \mathbf k, \mu)$. Moreover, as is the case with the Einstein equations, Eq. \ref{nuBoltzeqNscal} does not contain the direction of the wave vector $\mathbf{k}$ (aside from the scalar product in $\mu$, which we have already considered as an independent variable associated with the particle momentum). Therefore, again, an explicit dependence on $\hat{k}$ can be present only in the initial conditions.

Having a scalar function that depends on an angular variable, it is natural to perform an expansion in Legendre polynomials or partial waves. Following the convention of \cite{Ma:1995ey}:\index{Partial wave expansion}\index{Legendre polynomials!Expansion}
\begin{eqnarray}
	\mathcal N^{(S)}(\eta, \mathbf k, \mu) = \sum_{\ell = 0}^\infty(-i)^\ell(2\ell + 1)\mathcal N^{(S)}_\ell(\eta, \mathbf k)\mathcal P_\ell(\mu)\;,\\ 
\label{Nl}	\mathcal N^{(S)}_{\ell}(\eta, \mathbf k) = \frac{1}{(-i)^\ell}\int_{-1}^{1}\frac{d\mu}{2}\mathcal{P}_\ell(\mu)\mathcal N^{(S)}(\eta, \mathbf k, \mu)\;.
\end{eqnarray}
So, let us apply the integral operator on the right hand side of Eq.~\eqref{Nl} to the Boltzmann equation \eqref{nuBoltzeqNscal}. We obtain the following equation:
\begin{eqnarray}
	\mathcal N^{(S)'}_\ell + \frac{ik}{(-i)^{\ell}}\int_{-1}^{1}\frac{d\mu}{2}\mu\mathcal{P}_{\ell}\mathcal N^{(S)} + \frac{\Phi'}{(-i)^\ell}\int_{-1}^{1}\frac{d\mu}{2}\mathcal{P}_{\ell} + \frac{ik\Psi}{(-i)^{\ell}}\int_{-1}^{1}\frac{d\mu}{2}\mu\mathcal{P}_{\ell} = 0\;.
\end{eqnarray}
Employing the orthogonality of the Legendre polynomials: 
\begin{equation}\label{orthrelLegPol}
	\int_{-1}^{1} \frac{dx}{2} \mathcal{P}_\ell(x)\mathcal{P}_{\ell'}(x) = \frac{\delta_{\ell\ell'}}{2\ell + 1}\;,
\end{equation}\index{Legendre polynomials!Orthogonality relation}
the integrals of the above equation are easily computed as follows:
\begin{eqnarray}
	\int_{-1}^{1}\frac{d\mu}{2}\mathcal{P}_{\ell} = \frac{\delta_{\ell0}}{2\ell + 1}\;, \quad \int_{-1}^{1}\frac{d\mu}{2}\mu\mathcal{P}_{\ell} = \frac{\delta_{\ell1}}{2\ell + 1}\;.
\end{eqnarray}
Therefore, we must distinguish 3 cases: when only one of the above integrals contributes, or none of them does (for $\ell \ge 2$). Moreover, we employ the following recurrence relation:
\begin{equation}\label{recurrencerelationPl}
	(\ell + 1)\mathcal{P}_{\ell + 1}(\mu) = (2\ell + 1)\mu\mathcal{P}_{\ell}(\mu) - \ell\mathcal{P}_{\ell - 1}(\mu)\;,
\end{equation}\index{Legendre polynomials!Recurrence relation}
which allows us to write:
\begin{eqnarray}
	\frac{ik}{(-i)^{\ell}}\int_{-1}^{1}\frac{d\mu}{2}\mu\mathcal{P}_{\ell}\mathcal N^{(S)} = \frac{k(\ell + 1)}{2\ell + 1}\mathcal N^{(S)}_{\ell + 1} - \frac{k\ell}{2\ell + 1}\mathcal N^{(S)}_{\ell - 1}\;.
\end{eqnarray}

\hrulefill

\begin{ex}
Show that the hierarchy of Boltzmann equations for neutrinos is the following:
\begin{eqnarray}
\label{lge2equationnu} \boxed{(2\ell + 1)\mathcal N^{(S)'}_\ell + k\left[(\ell + 1)\mathcal N^{(S)}_{\ell + 1} - \ell\mathcal N^{(S)}_{\ell - 1}\right] = 0 \qquad (\ell \ge 2)\;,}\\
\label{le1equationnu} \boxed{3\mathcal N^{(S)'}_1 + 2k\mathcal N^{(S)}_{2} - k\mathcal N^{(S)}_{0} = k\Psi \qquad (\ell = 1)\;,}\\
\label{le0equationnu} \boxed{\mathcal N^{(S)'}_0 + k\mathcal N^{(S)}_1 = -\Phi' \qquad (\ell = 0)\;.}
\end{eqnarray}
\end{ex}\index{Neutrinos!Temperature equations hierarchy}

\hrulefill

This infinite set of equations is called a hierarchy because the $\ell$ equation is sourced by the $\ell + 1$ and $\ell - 1$ multipoles. Of course, it is not possible to numerically solve an infinite number of equations, so some truncation is necessary at a certain $\ell_{\rm max}$. This would not be done, however, on the basis of some expansion in $p/E$ (this being always one) but rather on the basis of some observational angular resolution, since $\ell \simeq 1/\theta$.

Using the definitions \eqref{mathcalNdefinition} and \eqref{Nl}, the \textbf{monopole} $\mathcal N^{(S)}_0$ can be expressed as:
\begin{equation}\label{monopoledeltanu}
	4\mathcal N^{(S)}_0 = \frac{4\pi}{a^4\bar \rho_\nu}\int\frac{d\mu}{2}\int dq\;q^2 q\bar f\mathcal F = \frac{1}{a^4\bar \rho_\nu}\int d^3\mathbf q q\bar f\mathcal F = \frac{\delta\rho_\nu}{\rho_\nu} = \delta_\nu\;.
\end{equation}
Hence $4\mathcal N^{(S)}_0 = \delta_\nu$, i.e., the monopole is proportional to the density contrast.\footnote{It is redundant to write $\mathcal N^{(S)}_0$ because the monopole can be sourced only by scalar perturbations.} In Chapter~\ref{Chap:IC}, we shall see that indeed the primordial mode excited is just the monopole. For the \textbf{dipole} $\mathcal N^{(S)}_1$:
\begin{equation}\label{dipoleVnu}
	4\mathcal N^{(S)}_1 = \frac{i4\pi}{a^4\bar\rho_\nu}\int\frac{d\mu}{2}\mu\int dq\;q^2q\bar f\mathcal F = \frac{i\hat k^l}{a^4\bar\rho_\nu}\int d^3\mathbf qq\hat p_l\bar f\mathcal F = \frac{(\bar\rho_\nu + \bar P_\nu)}{\bar \rho_\nu}V_\nu = \frac{4}{3}V_\nu\;.
\end{equation}
Hence, $3\mathcal N^{(S)}_1 = V_\nu$. Finally, for the \textbf{quadrupole}:
\begin{eqnarray}\label{quadrupoleanisotropicstressnu}
	4\mathcal N^{(S)}_2 = -\frac{4\pi}{a^4\bar\rho_\nu}\int\frac{d\mu}{2}\mathcal P_2(\mu)\int dq\;q^2q\bar f\mathcal F = -\frac{3}{2a^4\bar \rho_\nu}\int d^3\mathbf q q(\hat{k}_l\hat{p}^l\hat{k}_m\hat{p}^m - 1/3)\bar f\mathcal F\nonumber\\
	 = -\frac{3}{2\bar\rho_\nu}\hat{k}_l\hat{k}_m\left(\delta T_\nu^{lm} - \frac{1}{3}\delta^{lm} \delta T_\nu^i{}_i\right) = -\frac{3}{2\rho_\nu}\hat{k}_l\hat{k}_m\pi_\nu^{lm}\;. \qquad
\end{eqnarray}
Therefore, $\hat{k}_l\hat{k}_m\pi_\nu^{lm} = -4\rho_\nu\mathcal N^{(S)}_2/3$. Similar expressions hold true for photons.

The above equations can be compared with those in \cite{Ma:1995ey} by making the identifications $4\mathcal N^{(S)}_0 = \delta_\nu$, $3k\mathcal N^{(S)}_1 = \theta_\nu$, and $2\mathcal N^{(S)}_2 = \sigma_\nu$. 

\subsection{Line-of-sight integral}\index{Line-of-sight integration}

\hrulefill

\begin{ex}
	Show that Eq.~\eqref{nuBoltzeqNscal} can be formally integrated as follows:
	\begin{equation}\label{nuBoltzeqNscal2}
	\mathcal N^{(S)}(\eta_0,\mathbf k,\mu) = \mathcal N^{(S)}(\eta_i,\mathbf k,\mu)e^{ik\mu (\eta_i - \eta_0)} - \int_{\eta_i}^{\eta_0}d\eta(\Phi' + ik\mu\Psi)e^{ik\mu(\eta - \eta_0)}\;,
\end{equation}
where $\eta_i$ is some initial conformal time.
\end{ex}

\hrulefill

This approach is called \textbf{the line-of-sight integral} \cite{Seljak:1996is}. It is a simple formal integration, but it is very effective for the numerical calculation of the evolution of CMB anisotropies. We shall see this in Chapter~\ref{Chap:CMBEvo}. Note that the gravitational potentials $\Phi$ and $\Psi$ couple only with $\mathcal N^{(S)}_{0,1,2}$ via the Einstein equations. Hence, a way to avoid truncating the hierarchy of Boltzmann equations is to solve Eq.~\eqref{nuBoltzeqNscal2} together with $\mathcal N^{(S)}_{0,1,2}$. This strategy was proposed in \cite{Weinberg:2006hh} (for CMB photons).

Consider now the expansion of a plane wave into spherical harmonics:
\begin{equation}
	e^{i\mathbf k\cdot\mathbf r} = 4\pi\sum_{\ell = 0}^\infty\sum_{m = -\ell}^{\ell}i^{\ell}Y_{\ell}^{m*}(\hat k)Y_{\ell}^{m}(\hat r)j_{\ell}(kr)\;,
\end{equation}\index{Spherical harmonics!Plane wave expansion}where $j_\ell$ is a spherical Bessel function. Spherical harmonics are omnipresent in cosmology, especially in CMB physics, because they form the natural basis over which to expand quantities defined on the celestial sphere. Since they are extremely important for our work to come, Sec.~\ref{App:SphericalHarmonics} is dedicated to them.

Recalling that $\mu = \hat k\cdot\hat p$, we introduce in Eq.~\eqref{nuBoltzeqNscal2} the following plane wave expansion:
\begin{equation}
	\boxed{e^{-ik\hat k\cdot\hat p(\eta_0 - \eta)} = 4\pi\sum_{\ell' = 0}^\infty\sum_{m' = -\ell'}^{\ell'}(-i)^{\ell'}Y_{\ell'}^{m'*}(\hat k)Y_{\ell'}^{m'}(\hat p)j_{\ell'}(kr)}
\end{equation}
where we have defined:
\begin{eqnarray}
	\boxed{r(\eta) \equiv \eta_0 - \eta}
\end{eqnarray}
Using the addition theorem for spherical harmonics, the above result can be written as:
\begin{equation}\label{Partialwaveexpansionplanewave}
	e^{ik\mu(\eta - \eta_0)} = \sum_\ell(-i)^\ell(2\ell + 1)\mathcal P_\ell(\mu)j_\ell(kr)\;.
\end{equation}
Now, note that:
\begin{equation}
	- ik\mu\Psi e^{ik\mu(\eta - \eta_0)} = \Psi\frac{d}{d\eta_0}e^{ik\mu(\eta - \eta_0)}\;.
\end{equation}
Therefore, Eq.~\eqref{nuBoltzeqNscal2} can be written as follows:
\begin{eqnarray}\label{nuBoltzeqNscal3}
	\mathcal N^{(S)}(\eta_0,\mathbf k,\mu) = \sum_\ell(-i)^\ell(2\ell + 1)\mathcal P_\ell(\mu)\nonumber\\
	\left[\mathcal N^{(S)}(\eta_i,\mathbf k,\mu)j_\ell(kr_i) - \int_{\eta_i}^{\eta_0}d\eta\left(\Phi' - \Psi\frac{d}{d\eta_0}\right)j_\ell(kr)\right]\;.
\end{eqnarray}
We shall see in Chapter \ref{Chap:IC} that, at early times ($\eta_i \to 0$), the monopole contribution is dominant; hence, we may approximate $\mathcal N^{(S)}(\eta_i,\mathbf k,\mu) \approx \mathcal N_0^{(S)}(\eta_i,\mathbf k)$ and thus write:
\begin{eqnarray}\label{nuBoltzeqNscal4}
	\mathcal N_\ell^{(S)}(\eta_0,\mathbf k) = \mathcal N_0^{(S)}(\eta_i,\mathbf k)j_\ell(k\eta_0) - \int_{0}^{\eta_0}d\eta\left[\Phi'j_\ell(kr) - \Psi j_\ell'(kr)\right]\;.
\end{eqnarray}
Of course, we cannot really set $\eta_i = 0$ since this is the cosmological singularity. Such an equation should be understood as $\eta_i \to 0$. Knowing the gravitational potentials allows us to determine $\mathcal N^{(S)}_\ell$ without solving the hierarchy of Boltzmann equations. The advantage is that $\Phi$ and $\Psi$ are determined from the Einstein equations, which couple only with $\mathcal N^{(S)}_{0,1,2}$. 

\hrulefill

\begin{ex}
	Write the Boltzmann equation for neutrinos in the cases of tensor and vector perturbations. It might be useful to first study those for photons in the next section.
\end{ex}

\hrulefill

\section{The perturbed Boltzmann equation for photons}

The Boltzmann equation for photons requires a collisional term due to the electromagnetic interaction among photons and free electrons (Compton scattering):
\begin{equation}
	e^- + \gamma \longleftrightarrow e^- + \gamma\;.
\end{equation}
From 4-momentum conservation, it is a standard exercise in Special Relativity to prove that:
\begin{equation}\label{kinrelcomptonsc}
	p' = \frac{p}{1 + \frac{p}{m_e}(1 - \cos\theta)}\;,
\end{equation}
where $p = E$ and $p' = E'$ are the initial and final energies of the photon, respectively; $m_e$ is the mass of the electron, and $\theta$ is the deviation angle of the photon. 

The collisional term is the scattering rate; thus, we need the differential cross-section for Compton scattering. This is obtained from the scattering amplitude, whose calculation (at the tree level) is a standard exercise in Quantum Electrodynamics. See, for example, \cite{Weinberg:1995mt}. 

If the interaction takes place with an electron in any spin state, one averages over the initial electron spin states. Furthermore, if one is interested only in the properties of the scattered photon (as we are in the case of the CMB), then one sums over the final electron spin states. As for the photon, if it is in an unpolarized state before scattering, which means that it can have either of its two polarizations, then one also averages over the initial photon polarizations. Therefore, the differential cross-section is calculated as follows:
\begin{equation}
	\frac{1}{4}\sum_{s,s',\epsilon}\frac{d\sigma}{d\Omega} = \frac{e^4}{64\pi^2m_e^2}\frac{p'^2}{p^2}\left[\frac{p'}{p} + \frac{p}{p'} - 2(\hat p\cdot \epsilon')^2\right]\;,
\end{equation}
where we denote $s$ and $s'$ as the initial and final electron spin states, respectively, and $\epsilon$ and $\epsilon'$ as the initial and final photon polarizations, respectively. We have adopted the laboratory frame here, where the electron is initially at rest. 

Observation of the CMB is also able to discern its polarization, providing a valuable source of information about the primordial universe. On the other hand, let us focus only on the momentum of the scattered photon. Therefore, summing over the final polarization states gives us:
\begin{equation}
	\frac{1}{4}\sum_{s,s',\epsilon,\epsilon'}\frac{d\sigma}{d\Omega} = \frac{e^4}{32\pi^2m_e^2}\frac{p'^2}{p^2}\left(\frac{p'}{p} + \frac{p}{p'} - \sin^2\theta\right)\;.
\end{equation}
In the non-relativistic case $p \ll m_e$, one can see from Eq.~\eqref{kinrelcomptonsc} that $p = p'$. Therefore:
\begin{equation}\label{Thomsonscattamplesigma}
	 \frac{1}{4}\sum_{s,s',\epsilon,\epsilon'}\frac{d\sigma}{d\Omega} = \frac{e^4}{32\pi^2m_e^2}\left(1 + \cos^2\theta\right)\;.
\end{equation}
This can be readily integrated for a total cross-section:
\begin{equation}
	\sigma_T = \frac{e^4}{6\pi m_e^2}\;,
\end{equation}
which is \textbf{the Thomson cross-section}.

Note that, in order to use the above formulae in our cosmological setting, the momenta that appear must be interpreted as proper momenta (or equivalently as $q$-momenta, since only ratios of $p$-momenta appear).

Moreover, in the following, we assume we are dealing with non-relativistic electrons, so we will use Eq.~\eqref{Thomsonscattamplesigma}. This is a reasonable assumption for redshifts $\ll 10^{10}$.

\subsection{The Boltzmann equation for photons, neglecting polarization}

As a first, simple calculation, let us neglect the polarization of the photons. In this case, the treatment of the left-hand side of the Boltzmann equation is very similar to the one employed for neutrinos. In particular, we use the scalar distribution function $f_\gamma = f^{ij}\epsilon_i\epsilon_j^*$ since we do not have to specify the polarization state. Therefore, as we did in the massless neutrino case, we similarly define:\index{Photons!Relative temperature fluctuations}
\begin{equation}\label{Thetadefinition}
	\boxed{\Theta(\eta, \mathbf k, \hat{p}) \equiv \frac{\int dq\;q^2 q\bar f\mathcal F}{4\int dq\;q^2 q\bar f}}
\end{equation}
This $\Theta$ corresponds to $F_\gamma/4$ of \cite{Ma:1995ey} and to the trace of the dimensionless intensity matrix $J_{ij}/4$ of \cite{Weinberg:2008zzc}. Then, we can write Eq.~\eqref{pertBoltzeqgeneralFT} as follows:
\begin{equation}\label{Thetaeqgen}
	\Theta' + ik\mu \Theta - \left(ik\mu\frac{h_{00}}{2} - \frac{1}{2}h_{ij}'\hat{p}^i\hat{p}^j\right) = \frac{1}{4\int dq\;q^2 q\bar f}\int dq\;a^2q^2\;C[\mathcal F]\;.
\end{equation}
The left hand side is identical to the one in Eq.~\eqref{nuBoltzeqNgen}, so we are left to work out the collisional term.

\subsection{The collisional term, neglecting polarization}

The collisional term for the total photon distribution function is, cf. Eq. \eqref{collisionaltermgen}, with unity relative velocity, the following:
\begin{equation}
	\left(\frac{\partial f_\gamma}{\partial t}\right)_{\rm coll} = \int d^3\mathbf P d^2\hat p'\frac{d\sigma}{d\Omega}(\hat p')\left[f_e(\mathbf p + \mathbf P - \mathbf p')f_\gamma(\mathbf p') - f_e(\mathbf P)f_\gamma(\mathbf p)\right]\;.
\end{equation}
Note that $\mathbf P$ is the proper momentum of the incoming electron, not a conjugate momentum; $\mathbf p$ is the proper momentum of the incoming photon, and $\mathbf p'$ is the proper momentum of the outgoing photon. The differential cross section is the one in Eq.~\eqref{Thomsonscattamplesigma}. The partial derivative on the left hand side is with respect to cosmic time. We will need the one with respect to conformal time, so at the end, we will multiply by $a$. Note that a Dirac delta has already been used, eliminating the integration in the momentum of the outgoing electron and in the energy of the outgoing photon. 

We are interested in epochs in which the electrons are non relativistic. Therefore, in the subsequent calculations, we only keep terms that are linear in the velocity of the electrons. This is quite convenient because, in the first place, we do not have to compute the differential cross section again. The one computed in the lab frame, cf. Eq. \eqref{Thomsonscattamplesigma}, is adequate because corrections due to the initial electron velocity are of second order. 

Introducing the perturbed distribution functions in the collisional term, we have:
\begin{align}
	\left(\frac{\partial f_\gamma}{\partial t}\right)_{\rm coll} &= \int d^3\mathbf P d^2\hat p'\frac{d\sigma}{d\Omega}(\hat p')[\bar f_e(|\mathbf p + \mathbf P - \mathbf p'|)\delta f_\gamma(\mathbf p')\nonumber\\ &+ \delta f_e(\mathbf p + \mathbf P - \mathbf p')\bar f_\gamma(p') - \bar f_e(P)\delta f_\gamma(\mathbf p) - \delta f_e(\mathbf P)\bar f_\gamma(p)]\;.
\end{align}
Now let us discuss $\bar f_e$. It is a background distribution, so it depends only on the energy of the electron. For the recoiling electron, it is as follows:
\begin{equation}
	\frac{(\mathbf p + \mathbf P - \mathbf p')^2}{2m_e} = \frac{P^2}{2m_e} + \frac{\textbf{P}\cdot(\textbf{p} - \textbf{p}')}{m_e} + \frac{(\textbf{p} - \textbf{p}')^2}{2m_e}\;.
\end{equation}
The energy available to the particles in a thermal bath is, of course, of the order of the temperature $T$. This means that, for the photons, $|\textbf{p} - \textbf{p}'| \simeq T$. On the other hand, since the electrons are non-relativistic, for them one has $P^2/m_e \simeq T \ll m_e$. The above relation can thus be approximated as follows:
\begin{equation}
	\frac{(\mathbf p + \mathbf P - \mathbf p')^2}{2m_e} \simeq \frac{P^2}{2m_e} + \frac{P^3}{m_e^2} + \frac{P^4}{m_e^3}\;.
\end{equation}
Therefore, the first term is the dominant one, and since $f_e(\mathbf p + \mathbf P - \mathbf p')$ is multiplied by $\delta f_\gamma(\mathbf p')$, we simply approximate $\bar f_e(|\mathbf p + \mathbf P - \mathbf p'|) = \bar f_e(P)$. The same reasoning applies to $\delta f_e(\mathbf p + \mathbf P - \mathbf p')$. Therefore, we have:
\begin{eqnarray}
	\left(\frac{\partial f_\gamma}{\partial t}\right)_{\rm coll} = \int d^3\mathbf P d^2\hat p'\frac{d\sigma}{d\Omega}(\hat p')\bar f_e(P)\left[\delta f_\gamma(\mathbf p') - \delta f_\gamma(\mathbf p)\right] +\nonumber\\ 
	\int d^3\mathbf P d^2\hat p'\frac{d\sigma}{d\Omega}(\hat p')\delta f_e(\mathbf P)\left[\bar f_\gamma(p') - \bar f_\gamma(p)\right]\;. \qquad
\end{eqnarray}
Finally, it is not difficult to see that, at the lowest order, one has:
\begin{equation}
	p' = p - \frac{\textbf{P}\cdot(\textbf{p} - \textbf{p}')}{m_e}\;.
\end{equation} 
So, the energy of the scattered photon does change (it is the Doppler effect due to the motion of the electron). We have then:
\begin{equation}
	\bar f_\gamma(p') - \bar f_\gamma(p) = -\frac{d\bar f_\gamma}{dp}\frac{\textbf{P}\cdot(\textbf{p} - \textbf{p}')}{m_e} = -p\frac{d\bar f_\gamma}{dp}\frac{\textbf{P}\cdot(\hat{p} - \hat{p}')}{m_e}\;.
\end{equation}
In the last equation, we can take $p' = p$, since the expression already contains $\mathbf{P}/m_e$, which is a perturbed quantity.

Now, we use the definitions given in Chapter \ref{Chap:CosmoPertTheory}:
\begin{equation}
	n_e \equiv \int d^3\mathbf P\bar f_e(P)\;, \qquad n_em_e\textbf{v}_{\rm b} \equiv \int d^3\mathbf P \delta f_e(\mathbf P)\mathbf{P}\;, 
\end{equation}
where we have assumed that $\textbf{v}_e = \textbf{v}_p = \textbf{v}_{\rm b}$. This is justified by the fact that Coulomb scattering tightly couples electrons and protons. So, we can almost finally write the collision term as follows:
\begin{eqnarray}
	\left(\frac{\partial f_\gamma}{\partial t}\right)_{\rm coll} = n_e\int d^2\hat p'\frac{d\sigma}{d\Omega}(\hat p')\left[\delta f_\gamma(\mathbf p') - \delta f_\gamma(\mathbf p)\right] -\nonumber\\ 
	n_e\textbf{v}_{\rm b}\cdot\int d^2\hat p'\frac{d\sigma}{d\Omega}(\hat p')p\frac{d\bar f_\gamma}{dp}(\hat{p} - \hat{p}')\;. \qquad
\end{eqnarray}
Now we have that:
\begin{equation}
	\int d^2\hat p'\frac{d\sigma}{d\Omega}(\hat p') = \sigma_{\rm T}\;, \qquad \int d^2\hat p'\frac{d\sigma}{d\Omega}(\hat p')\hat p' = 0\;,
\end{equation}
where the latter equation simply establishes the fact that Thomson scattering does not have \textit{a priori} a preferred direction. 

Finally, the collisional term becomes:
\begin{eqnarray}\label{fullcolltermnopol}
	\left(\frac{\partial f_\gamma}{\partial t}\right)_{\rm coll} = -n_e\sigma_{\rm T}\delta f_\gamma(\mathbf p) + n_e\int d^2\hat p'\frac{d\sigma}{d\Omega}(\hat p')\delta f_\gamma(\mathbf p') - 
	n_e\sigma_{\rm T}\textbf{v}_{\rm b}\cdot\mathbf p\frac{d\bar f_\gamma}{dp}\;. \qquad
\end{eqnarray}
Multiplying the collisional term by $a$ to be consistent with the use of conformal time and introducing \textbf{the optical depth} $\tau$:
\begin{equation}\label{opticaldepthdefinition}
	\boxed{\tau(\eta) \equiv \int_\eta^{\eta_0}d\eta' n_e\sigma_Ta\;, \qquad \tau' = -n_e\sigma_{\rm T}a}
\end{equation}\index{Optical depth}
the Boltzmann equation~\eqref{Thetaeqgen} is rewritten as follows:
\begin{equation}
	\Theta' + ik\mu \Theta - \left(ik\mu\frac{h_{00}}{2} - \frac{1}{2}h_{ij}'\hat{p}^i\hat{p}^j\right) = \tau'\Theta + an_e\int d^2\hat p'\frac{d\sigma}{d\Omega}(\hat p')\Theta(\hat p') - \tau'\hat p\cdot\mathbf v_{\rm b}\;.
\end{equation}
The second term on the right hand side describes the anisotropic character of Thomson scattering. We can split it as follows, using Eq.~\eqref{Thomsonscattamplesigma}:
\begin{equation}\label{decompositionThomsonscatteringC}
	an_e\int d^2\hat p'\frac{d\sigma}{d\Omega}(\hat p')\Theta(\hat p') = -\tau'\int \frac{d^2\hat p'}{4\pi}\Theta(\hat{p}') - \frac{3\tau'}{16\pi}\int d^2\hat p'\left(\cos^2\theta - \frac{1}{3}\right)\Theta(\hat{p}')\;.
\end{equation}
On the right-hand side, the first term contains the so-called monopole, which represents the temperature fluctuation averaged over the sphere. The second term, instead, contains Legendre polynomials of the second degree:
\begin{equation}
	\frac{3}{16\pi}\int d^2\hat p'\left(\cos^2\theta - \frac{1}{3}\right)\Theta(\hat{p}') = \frac{1}{8\pi}\int d^2\hat p'\mathcal P_2(\hat p\cdot\hat p')\Theta(\hat{p}')
\end{equation}
It is useful to ``free'' the $\hat p$ dependence in the above term by using the addition theorem:
\begin{equation}
	\mathcal P_2(\hat p\cdot\hat p') = \frac{4\pi}{5}\sum_{m = -2}^{2}Y_2^m(\hat p)Y_2^{m*}(\hat{p}')\;,
\end{equation}
so the equation for $\Theta$ can be written as:
\begin{eqnarray}
	\Theta' + ik\mu \Theta - \left(ik\mu\frac{h_{00}}{2} - \frac{1}{2}h_{ij}'\hat{p}^i\hat{p}^j\right) = \nonumber\\ 
	\tau'\left[\Theta - \Theta_0 - \hat p\cdot\mathbf v_{\rm b} - \frac{1}{10}\sum_{m = -2}^2Y_2^m(\hat p)\int d^2\hat p'Y_2^{m*}(\hat{p}')\Theta(\hat{p}')\right]\;.
\end{eqnarray}
We now introduce the contribution from polarization. 

\subsection{The Boltzmann equation for photons, including polarization}

The differential cross-section in Eq.~\eqref{Thomsonscattamplesigma} is averaged over the initial polarization states and summed over the final ones. On the other hand, if we want to keep track of how Thomson scattering shapes the distribution of the polarization of the photons in the CMB sky, we need to use the full polarization-dependent cross section. In this case, it is useful to decompose the full Boltzmann equation for polarized photons into the Boltzmann equations for the Stokes parameters $\Theta$, $Q$, and $U$. If the reader is not familiar with Stokes parameters, Appendix \ref{App:polarization} offers a brief reminder. 

The original derivation of the Thomson scattering matrix can be found in \cite{1960ratr.book.....C}. The theoretical framework for CMB polarization can be found, for example, in \cite{Kosowsky:1994cy}, but also in \cite{Weinberg:2008zzc}. We derive the collisional term in Appendix \ref{App:Thomsonscattering} and write the final equation that we are going to analyze here, following \cite{Tram:2013ima}:
\begin{eqnarray}\label{photonBoltzmannequationfull}
	\left(\frac{\partial}{\partial\eta} + ik\mu\right)\left(\begin{array}{c}
		\Theta \\ Q\\ iU
	\end{array}\right) 
	 + \left(\begin{array}{c}
		- ik\mu\frac{h_{00}}{2} + \frac{1}{2}h_{ij}'\hat{p}^i\hat{p}^j \\ 0\\ 0
	\end{array}\right) =  \tau'\left(\begin{array}{c}
		\Theta - \Theta_0 - \hat p\cdot \mathbf v_{\rm b}\\ Q\\ iU
	\end{array}\right)\nonumber\\ 
	 -\frac{\tau'}{10}\sum_{m = -2}^2\left(\begin{array}{c}
		Y_2^m \\ \frac{1}{2}\mathcal E^m\\ \frac{1}{2}\mathcal B^m
	\end{array}\right)_{\hat p}\int d^2\hat p'\left(\begin{array}{c}
		Y_2^{m*}\Theta - \sqrt{\frac{3}{2}}\mathcal E^{m*} Q - \sqrt{\frac{3}{2}}\mathcal B^{m*} iU\\ -\sqrt{6}Y_2^{m*}\Theta + 3\mathcal E^{m*}Q + 3\mathcal B^{m*}iU\\ -\sqrt{6}Y_2^{m*}\Theta + 3\mathcal E^{m*}Q + 3\mathcal B^{m*}iU
	\end{array}\right)_{\hat p'}\;, \qquad
\end{eqnarray}\index{Stokes parameters!Boltzmann equation}where the subscripts $\hat p$ and $\hat p'$ denote the dependence on the photon direction, and we have used the definition:
\begin{equation}\label{EandBdefinitions}
	\mathcal E^m \equiv {}_2Y_2^m + {}_{-2}Y_2^m\;, \qquad \mathcal B^m \equiv {}_2Y_2^m - {}_{-2}Y_2^m\;,
\end{equation}
where ${}_{\pm 2}Y_2^m$ are the spin-2 weighted spherical harmonics. Their explicit form is given in Tabs.~\ref{Tab:SphericalHarmonics} and \ref{Tab:EandB}.

Inspecting the above trio of Boltzmann equations in Eq.~\eqref{photonBoltzmannequationfull}, one can see that $iU$ and $\mathcal B^m Q/\mathcal E^m$ satisfy the same Boltzmann equation. Hence, since they have the same initial condition (which is zero, as we shall see in Chapter~\ref{Chap:IC}), we just need a single polarization hierarchy. This is the principal result of \cite{Tram:2013ima}, which helps to make the CLASS code faster. 

Working with $Q$, we have:
\begin{eqnarray}\label{photonBoltzmannequationtemp}
	\left(\frac{\partial}{\partial\eta} + ik\mu\right)\Theta - \left(ik\mu\frac{h_{00}}{2} - \frac{1}{2}h_{ij}'\hat{p}^i\hat{p}^j\right)  = \tau'\left(\Theta - \Theta_0 - \hat p\cdot \mathbf v_{\rm b}\right)	\nonumber\\ 
	 - \frac{\tau'}{10}\sum_{m = -2}^2Y_2^m(\hat p)\int d^2\hat{p}'\left[Y_2^{m*}\Theta - \sqrt{\frac{3}{2}}\mathcal E^{m*}Q - \sqrt{\frac{3}{2}}\frac{(\mathcal B^{m*})^2}{\mathcal E^{m*}}Q\right]_{\hat{p}'}\;,
\end{eqnarray}
and:
\begin{eqnarray}\label{photonBoltzmannequationpol}
	\left(\frac{\partial}{\partial\eta} + ik\mu\right)Q  = \tau'Q \nonumber\\ 
	 + \frac{\tau'}{10}\sum_{m = -2}^2 \sqrt{\frac{3}{2}}\mathcal E^m(\hat p)\int d^2\hat p'\left[Y_2^{m*}\Theta - \sqrt{\frac{3}{2}}\mathcal E^{m*}Q - \sqrt{\frac{3}{2}}\frac{(\mathcal B^{m*})^2}{\mathcal E^{m*}}Q\right]_{\hat p'}\;. \quad
\end{eqnarray}
Note that on the left-hand side of Eq.~\eqref{photonBoltzmannequationtemp}, the dependence on the photon direction $\hat p$ enters through $\mu = \hat k\cdot \hat p$ and through $h_{ij}'\hat{p}^i\hat{p}^j$, which does not depend on $\mu$ for scalar perturbations but depends on $\mu$ and the azimuthal angle $\phi$ for tensor and vector perturbations if $\hat k = \hat z$; cf. Eqs.~\eqref{hijpipjexpansionthetaphi} and \eqref{hijVpertthetaphiBoltzeq}. Instead, on the right hand side of Eq.~\eqref{photonBoltzmannequationtemp} the dependence on the photon direction is $Y_2^m(\hat p)$.

In order to match the angular dependencies on the two sides, making the equation easier to manipulate, and also to easily express the metric contribution for tensor and vector perturbations, it is convenient to set $\hat k = \hat z$. In order to recall this, we define:
\begin{equation}
	\Theta_P \equiv Q(k\hat z)\;.
\end{equation} 
As we shall see in Chapter \ref{Chap:CMBEvo}, before performing the anti-Fourier transform to recover the physical quantities in real space, we must first apply a spatial rotation to recover a generic direction for $\hat k$. This rotation compensates if one computes straightaway the angular power spectra since they are rotationally invariant quantities. In this case, one can use the results of the equations from this section at once.

\begin{table}
\centering
\begin{tabular}{|c|c|c|}
\hline {} & {} & {}\\
		$m$ & $Y_2^m$ & ${}_2Y_2^m$\\ {} & {} & {}\\ \hline {} & {} & {}\\
		$0$ & $\frac{1}{4}\sqrt{\frac{5}{\pi}}(3\cos^2\theta - 1)$ & $\frac{3}{4}\sqrt{\frac{5}{6\pi}}\sin^2\theta$\\ {} & {} & {}\\
		$\pm 1$ & $\frac{1}{2}\sqrt{\frac{15}{2\pi}}\sin\theta\cos\theta e^{\pm i\phi}$ & $\frac{1}{4}\sqrt{\frac{5}{\pi}}\sin\theta(1 \mp \cos\theta) e^{\pm i\phi}$\\ {} & {} & {}\\
		$\pm 2$ & $\frac{1}{4}\sqrt{\frac{15}{2\pi}}\sin^2\theta e^{\pm 2i\phi}$ & $\frac{1}{8}\sqrt{\frac{5}{\pi}}(1 \mp \cos\theta)^2 e^{\pm 2i\phi}$\\ {} & {} & {}\\
		\hline
\end{tabular}
\caption{Explicit functional form of the spin-0 and spin-2 spherical harmonics. Note that the spin-$-2$ spherical harmonics can be obtained by the spin-$2$ by spatial inversion, i.e. ${}_{-2}Y_2^m(\hat p) = {}_{2}Y_2^m(-\hat p)$. In these notes, we omit the Condon-Shortley phase.}
\label{Tab:SphericalHarmonics}
\end{table}

\begin{table}
\centering
\begin{tabular}{|c|c|c|}
\hline {} & {} & {}\\
		$m$ & $\mathcal E^m$ & $\mathcal B^m$\\ {} & {} & {}\\ \hline {} & {} & {}\\		
		$0$ & $\sqrt{\frac{15}{8\pi}}\sin^2\theta$ & 0\\ {} & {} & {}\\
		$\pm 1$ & $-\frac{1}{2}\sqrt{\frac{5}{\pi}}\sin\theta\cos\theta e^{\pm i\phi}$ & $\frac{1}{2}\sqrt{\frac{5}{\pi}}\sin\theta e^{\pm i\phi}$\\ {} & {} & {}\\
		$\pm 2$ & $\frac{1}{4}\sqrt{\frac{5}{\pi}}(1 + \cos^2\theta) e^{\pm 2i\phi}$ & $-\frac{1}{2}\sqrt{\frac{5}{\pi}}\cos\theta e^{\pm 2i\phi}$\\ {} & {} & {}\\
		\hline
\end{tabular}
\caption{Explicit functional form of $\mathcal E^m$ and $\mathcal B^m$.}
\label{Tab:EandB}
\end{table}

Let us then set $\hat k = \hat z$. Comparing $Y_2^m(\mu, \phi)$ with Eqs.~\eqref{hijpipjexpansionthetaphi} and \eqref{hijVpertthetaphiBoltzeq} allows us to see that the $m = 0$ contribution of the sum on the right hand side of Eq.~\eqref{photonBoltzmannequationtemp} couples to scalar perturbations only. The $m = \pm 2$ contributions couple to tensor perturbations, and the $m = \pm 1$ contributions couple to vector perturbations.

Note that for scalar perturbations one has that $U = 0$ in the reference frame $\hat k = \hat z$, since $\mathcal B^0 = 0$. 

Unless stated otherwise, in the following formulae, the functional dependences of $\Theta$ and $\Theta_P$ are on $\eta$, $\mathbf k = k\hat z$, $\mu$, and $\phi$. The metric quantities do not depend on $\mu$ and $\phi$.

\subsection{Scalar perturbations}
\index{Photons!Perturbed Boltzmann equation!Scalar perturbations}

For scalar perturbations, we have the following:
\begin{eqnarray}
	- ik\mu\frac{h_{00}}{2} + \frac{1}{2}h_{ij}'\hat{p}^i\hat{p}^j = \Phi' + ik\mu\Psi\;,\\
	(\hat p\cdot \mathbf v_{\rm b})^{(S)} = - i\mu V_{\rm b}\;,
\end{eqnarray}
since recall that the scalar part of $\mathbf v_{\rm b}$ is $\mathbf v_{\rm b}^{(S)} = -i\hat{k}V_{\rm b}$. Note that, as in the case of neutrinos, since no azimuthal dependence appears in the differential equation, one has $\Theta^{(S)}(\eta, k\hat z, \mu)$. 

As we shall see in Chapter~\ref{Chap:CMBEvo}, the scalar contribution $\Theta^{(S)}$ is the one that dominates the CMB temperature fluctuations because it is sourced by scalar perturbations in the metric, which are the strongest since they can grow.

\hrulefill

\begin{ex}
 Perform the integration in $d\phi'$ for the $m = 0$ contribution of Eqs.~\eqref{photonBoltzmannequationtemp} and \eqref{photonBoltzmannequationpol}. Using the results of Tabs.~\ref{Tab:SphericalHarmonics} and \ref{Tab:EandB} show that:
\begin{eqnarray}
\left(\frac{\partial}{\partial\eta} + ik\mu\right)\Theta^{(S)}  + \Phi' + ik\mu\Psi = \tau'\left[\Theta^{(S)} - \Theta_0 + i\mu V_{\rm b}\right]\nonumber\\ 
	 -\frac{\tau'}{2}\mathcal P_2(\mu)\int \frac{d\mu'}{2}\left[\mathcal P_2\Theta^{(S)} - (1 - \mathcal P_2)\Theta^{(S)}_P\right]_{\mu'}\;,
\end{eqnarray} 
and 
\begin{eqnarray}
\left(\frac{\partial}{\partial\eta} + ik\mu\right)\Theta^{(S)}_P = \tau'\Theta^{(S)}_P\nonumber\\ 
	 -\frac{\tau'}{2}[1 - \mathcal{P}_2(\mu)]\int \frac{d\mu'}{2}\left[-\mathcal P_2\Theta^{(S)} + (1 - \mathcal P_2)\Theta^{(S)}_P\right]_{\mu'}\;.
\end{eqnarray}
\end{ex}

\hrulefill

Recall the partial wave expansions already used for the neutrino distribution:
\begin{equation}\label{Thetal}
	\Theta^{(S)}_{\ell} = \frac{1}{(-i)^\ell}\int_{-1}^{1}\frac{d\mu}{2}\mathcal{P}_\ell(\mu)\Theta^{(S)}(\mu)\;, \quad \Theta^{(S)}_{P\ell} = \frac{1}{(-i)^\ell}\int_{-1}^{1}\frac{d\mu}{2}\mathcal{P}_\ell(\mu)\Theta^{(S)}_P(\mu)\;,
\end{equation}
\index{Partial wave expansion}
\hrulefill

\begin{ex}
Show that the Boltzmann equations for the scalar contribution to the photon temperature and polarization are:
\begin{eqnarray}
\label{BoltzeqphotonTheta}	\boxed{\Theta^{(S)'} + ik\mu\Theta^{(S)} + \Phi' + ik\mu\Psi = -\tau'\left[\Theta_0 - \Theta^{(S)} - i\mu V_{\rm b} - \frac{1}{2}\mathcal{P}_2(\mu)\Pi\right]}\\
\label{BoltzeqphotonPol}    \boxed{\Theta^{(S)'}_P + ik\mu\Theta^{(S)}_P = -\tau'\left[- \Theta^{(S)}_P + \frac{1}{2}[1 - \mathcal{P}_2(\mu)]\Pi\right]}
\end{eqnarray}
where $\Pi$ is defined as follows:
\begin{equation}
	\boxed{\Pi \equiv \Theta^{(S)}_{2} + \Theta^{(S)}_{P2} + \Theta_{P0}}
\end{equation}
\end{ex}

\hrulefill

The anisotropic nature of Thomson scattering and polarization were neglected by \cite{Peebles:1970ag}, whereas the former was only included by \cite{Wilson:1981yi}. Polarization was considered in \cite{Bond:1984fp}.

The above two equations are usually expanded in Legendre polynomials, as we did for the neutrino equation. Using Eq.~\eqref{Thetal} applied to Eq.~\eqref{BoltzeqphotonTheta}, we obtain the following equation:
\begin{eqnarray}
	\Theta^{(S)'}_\ell + \frac{ik}{(-i)^{\ell}}\int_{-1}^{1}\frac{d\mu}{2}\mu\mathcal{P}_{\ell}\Theta^{(S)} = -\frac{\Phi'}{(-i)^\ell}\int_{-1}^{1}\frac{d\mu}{2}\mathcal{P}_{\ell} + \frac{ik\Psi}{(-i)^{\ell}}\int_{-1}^{1}\frac{d\mu}{2}\mu\mathcal{P}_{\ell}\nonumber\\
	-\tau'\left[\frac{\Theta_0}{(-i)^{\ell}}\int_{-1}^{1}\frac{d\mu}{2}\mathcal{P}_{\ell} - \Theta^{(S)}_l - \frac{iV_{\rm b}}{(-i)^{\ell}}\int_{-1}^{1}\frac{d\mu}{2}\mu\mathcal{P}_{\ell} - \frac{\Pi}{2(-i)^{\ell}}\int_{-1}^{1}\frac{d\mu}{2}\mathcal{P}_2\mathcal{P}_{\ell}\right]\;.
\end{eqnarray}
Using the orthogonality relation of the Legendre polynomials, cf. Eq.~\eqref{orthrelLegPol}, the integrals of the above equation are easily computed as follows:
\begin{eqnarray}
	\int_{-1}^{1}\frac{d\mu}{2}\mathcal{P}_{\ell} = \frac{\delta_{\ell0}}{2\ell + 1}\;, \quad \int_{-1}^{1}\frac{d\mu}{2}\mu\mathcal{P}_{\ell} = \frac{\delta_{\ell1}}{2\ell + 1}\;, \quad \int_{-1}^{1}\frac{d\mu}{2}\mathcal{P}_{2}\mathcal{P}_{\ell} = \frac{\delta_{\ell2}}{2\ell + 1}\;.
\end{eqnarray}
Therefore, we must distinguish among 4 cases: when one of the above integrals contributes or none of them does (for $\ell > 2$). We shall make use again of the recurrence relation of Eq.~\eqref{recurrencerelationPl}, which allows us to write:
\begin{equation}\label{lg2equationphoton}
	\boxed{(2\ell + 1)\Theta^{(S)'}_\ell + k\left[(\ell + 1)\Theta^{(S)}_{\ell + 1} - \ell\Theta^{(S)}_{\ell - 1}\right] = \tau'(2\ell + 1)\Theta^{(S)}_\ell\;, \qquad \ell > 2}
\end{equation}
The equation for the quadrupole $\ell = 2$:
\begin{equation}\label{le2equationphoton}
	\boxed{10\Theta^{(S)'}_2 + 2k\left(3\Theta^{(S)}_{3} - 2\Theta^{(S)}_{1}\right) = 10\tau'\Theta^{(S)}_2 - \tau'\Pi}
\end{equation}
The equation for the dipole $\ell = 1$:
\begin{equation}\label{le1equationphoton}
	\boxed{3\Theta^{(S)'}_1 + k\left(2\Theta^{(S)}_{2} - \Theta^{(S)}_{0}\right) = k\Psi + \tau'\left(3\Theta^{(S)}_1 - V_{\rm b}\right)}
\end{equation}
Finally, the equation for the monopole $\ell = 0$:
\begin{equation}\label{le0equationphoton}
	\boxed{\Theta^{'}_0 + k\Theta^{(S)}_1 = -\Phi'}
\end{equation}
For the polarization equation, the steps to be performed are the same. Therefore:
\begin{eqnarray}
	\Theta^{(S)'}_{P\ell} + \frac{k(\ell + 1)}{2\ell + 1}\Theta^{(S)}_{P(\ell + 1)} - \frac{k\ell}{2\ell + 1}\Theta^{(S)}_{P(\ell - 1)} =\nonumber\\ 
	-\tau'\left[- \Theta^{(S)}_{P\ell} + \frac{\Pi}{2(-i)^\ell}\int_{-1}^{1}\frac{d\mu}{2}\mathcal{P}_{\ell} - \frac{\Pi}{2(-i)^\ell}\int_{-1}^{1}\frac{d\mu}{2}\mathcal{P}_{\ell}\mathcal{P}_2(\mu)\right]\;.
\end{eqnarray}
So, the equation for $\ell > 2$ is as follows:
\begin{equation}\label{lg2equationpol}
	\boxed{(2\ell + 1)\Theta^{(S)'}_{P\ell} + k\left[(\ell + 1)\Theta^{(S)}_{P(\ell + 1)} - \ell\Theta^{(S)}_{P(\ell - 1)}\right] = \tau'(2\ell + 1)\Theta^{(S)}_{P\ell}\;, \qquad \ell > 2}
\end{equation}
The equation for the quadrupole $\ell = 2$:
\begin{equation}\label{le2equationpol}
	\boxed{10\Theta^{(S)'}_{P2} + 2k\left(3\Theta^{(S)}_{P3} - 2\Theta^{(S)}_{P1}\right) = 10\tau'\Theta^{(S)}_{P2} - \tau'\Pi}
\end{equation}
The equation for the dipole $\ell = 1$:
\begin{equation}\label{le1equationpol}
	\boxed{3\Theta^{(S)'}_{P1} + k\left(2\Theta^{(S)}_{P2} - \Theta^{(S)}_{P0}\right) = 3\tau'\Theta^{(S)}_{P1}}
\end{equation}
Finally, the equation for the monopole $\ell = 0$:
\begin{equation}\label{le0equationpol}
	\boxed{2\Theta^{'}_{P0} + 2k\Theta^{(S)}_{P1} = 2\tau'\Theta_{P0} - \tau'\Pi}
\end{equation}
In order to compare with \cite{Ma:1995ey}, one must make the identifications $4\Theta_0 = \delta_\gamma$, $3k\Theta^{(S)}_1 = \theta_\gamma$, $2\Theta^{(S)}_2 = \sigma_\gamma$, and $kV_{\rm b} = \theta_{\rm b}$.

Note that the monopole and the dipole are related to the density contrast and the photon fluid velocity; hence, they are gauge-dependent. In particular, the monopole can be reabsorbed into the determination of the background temperature, whereas the dipole can be reabsorbed by a suitable boost. On the other hand, the quadrupole is related to the anisotropic stress, making it gauge-invariant, as well as the higher-order multipoles.

The monopole has no superscript $(S)$ because it couples only to scalar perturbations.

\subsection{Tensor perturbations}
\index{Photons!Perturbed Boltzmann equation!Tensor perturbations}

We derive in this section the hierarchy of equations describing the evolution of fluctuations in the photon distribution caused by tensor perturbations. Consider the tensor contribution of Eq.~\eqref{photonBoltzmannequationtemp}. Using Eq.~\eqref{hijpipjexpansionthetaphi}, we have:
\begin{eqnarray}\label{tensorphotonBoltzmannequationtemp}
	\left(\frac{\partial}{\partial\eta} + ik\mu\right)\Theta^{(T)} + \frac{h'_+}{2}(1 - \mu^2)\cos 2\phi + \frac{h'_\times}{2}(1 - \mu^2)\sin 2\phi = \tau'\Theta^{(T)}\nonumber\\ 
	 -\frac{\tau'}{10}\sum_{m = -2}^2Y_2^m(\hat p)\int d^2\hat p'\left[Y_2^{m*}\Theta^{(T)} - \sqrt{\frac{3}{2}}\mathcal E^{m*} \Theta^{(T)}_P - \sqrt{\frac{3}{2}}\frac{(\mathcal B^{m*})^2}{\mathcal E^{m*}}\Theta^{(T)}_P\right]_{\hat p'}\;, \quad
\end{eqnarray}
whereas for the polarization part:
\begin{eqnarray}\label{tensorphotonBoltzmannequationpol}
	\left(\frac{\partial}{\partial\eta} + ik\mu\right)\Theta^{(T)}_P = \tau'\Theta^{(T)}_P \nonumber\\ 
	 \frac{\tau'}{10}\sum_{m = -2}^2\sqrt{\frac{3}{2}}\mathcal E^m\int d^2\hat p'\left[Y_2^{m*}\Theta^{(T)} - \sqrt{\frac{3}{2}}\mathcal E^{m*}\Theta^{(T)}_P - \sqrt{\frac{3}{2}}\frac{(\mathcal B^{m*})^2}{\mathcal E^{m*}}\Theta^{(T)}_P\right]_{\hat p'}\;. \quad
\end{eqnarray}
As we mentioned, the monopole is a scalar and therefore does not provide any tensor contribution to Eq.~\eqref{photonBoltzmannequationtemp}. Moreover, the scalar product $\mathbf v_{\rm b}\cdot \hat{p}$ also has no tensor part (only a scalar one that we have already used and a vector one that we will see later).

Mimicking the azimuthal dependence produced by tensor perturbations in the metric, we can split the tensor contribution to the temperature anisotropy as follows; see, e.g., \cite{1985SvA....29..607P} and \cite{Crittenden:1993ni}:
\begin{eqnarray}\label{tensortemperaturecontribution}
	\boxed{\Theta^{(T)}(\mu,\phi) = \Theta^{(T)}_+(\mu)(1 - \mu^2)\cos2\phi + \Theta^{(T)}_\times(\mu)(1 - \mu^2)\sin 2\phi}
	\nonumber\\ 
	\boxed{= 4\sqrt{\frac{\pi}{15}}\sum_{\lambda = \pm 2}\Theta^{(T)}_\lambda(\mu)Y^\lambda_2(\mu,\phi)}
\end{eqnarray}
where we have reproduced the sum over the helicities of Eq.~\eqref{hijpipjexpansionthetaphi}, and similarly for the polarization field:
\begin{eqnarray}
\label{ThetaPTexpThetalambda}	\boxed{\Theta_P^{(T)}(\mu,\phi) = \Theta^{(T)}_{P+}(\mu)(1 + \mu^2)\cos2\phi + \Theta^{(T)}_{P\times}(\mu)(1 + \mu^2)\sin 2\phi}	\nonumber\\ 
	\boxed{= 4\sqrt{\frac{\pi}{15}}\sqrt{\frac{3}{2}}\sum_{\lambda = \pm 2}\Theta^{(T)}_{P\lambda}(\mu)\mathcal E^\lambda(\mu,\phi)}
\end{eqnarray}
So we have to select $m = \pm 2$ in sum on the right hand side of Eq.~\eqref{photonBoltzmannequationtemp}, and we get:
\begin{eqnarray}
	\left(\frac{\partial}{\partial\eta} + ik\mu - \tau'\right)\Theta_\lambda^{(T)} + \frac{h'_\lambda}{2} = \nonumber\\ 
	 -\frac{\tau'}{10}\int d^2\hat p'\left[Y_2^{\lambda*}\Theta_\lambda^{(T)}Y_2^{\lambda} - \frac{3}{2}\mathcal E^{\lambda*} \Theta^{(T)}_{P\lambda}\mathcal E^{\lambda*} - \frac{3}{2}\frac{(\mathcal B^{\lambda*})^2}{\mathcal E^{\lambda*}}\Theta^{(T)}_{P\lambda}\mathcal E^\lambda\right]_{\hat p'}\;, \quad
\end{eqnarray}
whereas for the polarization part:
\begin{eqnarray}
	\left(\frac{\partial}{\partial\eta} + ik\mu - \tau'\right)\Theta^{(T)}_{P\lambda} = \nonumber\\ 
	 \frac{\tau'}{10}\int d^2\hat p'\left[Y_2^{\lambda*}\Theta_\lambda^{(T)}Y_2^{\lambda} - \frac{3}{2}\mathcal E^{\lambda*}\Theta^{(T)}_{P\lambda}\mathcal E^{\lambda} - \frac{3}{2}\frac{(\mathcal B^{\lambda*})^2}{\mathcal E^{\lambda*}}\Theta^{(T)}_{P\lambda}\mathcal E^{\lambda}\right]_{\hat p'}\;, \quad
\end{eqnarray}
where $\lambda = \pm 2$ (the equations are identical for the two choices). Let us work out the right hand sides. 

\hrulefill

\begin{ex}
 With the help of Tabs.~\ref{Tab:SphericalHarmonics} and \ref{Tab:EandB} show that the integral on the right hand sides becomes: 
\begin{equation}
	 -\frac{\tau'}{10}\int d\mu d\phi\frac{15}{32\pi}\left[(1 - \mu^2)^2\Theta_\lambda^{(T)} - (1 + 6\mu^2 + \mu^4)\Theta_{P\lambda}^{(T)}\right]\;.
\end{equation} 
\end{ex}

\hrulefill

Let us introduce an expansion in Legendre polynomials for $\Theta_\lambda^{(T)}(\mu)$ and $\Theta_{P,\lambda}^{(T)}(\mu)$ similar to that in Eq.~\eqref{Thetal}:
\begin{equation}\label{ThetalTensor}
	\Theta^{(T)}_{\lambda,\ell} = \frac{1}{(-i)^\ell}\int_{-1}^{1}\frac{d\mu}{2}\mathcal{P}_\ell(\mu)\Theta_\lambda^{(T)}(\mu)\;, \quad \Theta^{(T)}_{P\lambda,\ell} = \frac{1}{(-i)^\ell}\int_{-1}^{1}\frac{d\mu}{2}\mathcal{P}_\ell(\mu)\Theta_{P\lambda}^{(T)}(\mu)\;.
\end{equation}

\hrulefill

\begin{ex}
Rewrite the contributions $(1 - \mu^2)^2$ and $(1 + 6\mu^2 + \mu^4)$ in terms of Legendre polynomials, and show that:
\begin{eqnarray}
	 -\frac{3\tau'}{16}\int \frac{d\mu}{2}\left[\frac{8}{35}\mathcal P_4(\mu) - \frac{80}{105}\mathcal P_2(\mu) + \frac{8}{15}\mathcal P_0(\mu)\right]\Theta_\lambda^{(T)}\nonumber\\ + \frac{3\tau'}{16}\int \frac{d\mu}{2}\left[\frac{8}{35}\mathcal P_4(\mu) + \frac{32}{7}\mathcal P_2(\mu) + \frac{16}{5}\mathcal P_0(\mu)\right]\Theta_{P\lambda}^{(T)}\;.
\end{eqnarray} 
\end{ex}

\hrulefill

Hence, using Eq.~\eqref{ThetalTensor}, we can write the tensor Boltzmann equation for photons as follows \cite{Crittenden:1993ni}:
\begin{eqnarray}\label{tensorBoltzmannequationTheta}
	\left(\frac{\partial}{\partial\eta} + ik\mu - \tau'\right)\Theta_\lambda^{(T)} + \frac{1}{2}h'_\lambda = \nonumber\\ 
	 -\tau'\left[\frac{3}{70}\Theta^{(T)}_{\lambda,4} + \frac{1}{7}\Theta^{(T)}_{\lambda,2} + \frac{1}{10}\Theta^{(T)}_{\lambda,0} - \frac{3}{70}\Theta^{(T)}_{P\lambda,4} + \frac{6}{7}\Theta^{(T)}_{P\lambda,2} - \frac{3}{5}\Theta^{(T)}_{P\lambda,0}\right]\;,
\end{eqnarray}
and for polarization:
\begin{eqnarray}\label{tensorBoltzmannequationThetaP}
	\left(\frac{\partial}{\partial\eta} + ik\mu - \tau'\right)\Theta_{P\lambda}^{(T)} = \nonumber\\ 
	 \tau'\left[\frac{3}{70}\Theta^{(T)}_{\lambda,4} + \frac{1}{7}\Theta^{(T)}_{\lambda,2} + \frac{1}{10}\Theta^{(T)}_{\lambda,0} - \frac{3}{70}\Theta^{(T)}_{P\lambda,4} + \frac{6}{7}\Theta^{(T)}_{P\lambda,2} - \frac{3}{5}\Theta^{(T)}_{P\lambda,0}\right]\;.
\end{eqnarray}
The same equations also hold true for $\Theta^{(T)}_{+,\times}$ and $\Theta^{(T)}_{P+,\times}$. The combination of terms between square brackets in the above equations is sometimes referred to as $\Psi$ in the literature. Since $\Psi$ is already used as one of the Bardeen potentials for us, in Chapter \ref{Chap:CMBEvo} we shall use another notation.

\subsection{Vector perturbations}
\index{Photons!Perturbed Boltzmann equation!Vector perturbations}

For completeness, we present here the Boltzmann equation for photons sourced by vector perturbations in the metric, though we shall not use it. 

The scalar product $\mathbf v_{\rm b}\cdot \hat{p}$ has a vector contribution that can be written as follows:
\begin{equation}\label{vbcdothatpvectorpart}
	(\mathbf v_{\rm b}\cdot \hat{p})^{(V)} = U^1_{\rm b}\sin\theta\cos\phi + U^2_{\rm b}\sin\theta\sin\phi\;,
\end{equation}
where we have used Eq.~\eqref{hatpiinsphericalcoordinates} and the fact that $k_iU^i_{\rm b} = 0$ because of the vector nature of $U^i_{\rm b}$ and the choice of having $k^3 = k$, i.e., $\hat{k} = \hat z$. So, $\mathbf v_{\rm b}\cdot \hat{p}$ does have an azimuthal dependence.

Using Eqs.~\eqref{hijVpertthetaphiBoltzeq} and \eqref{vbcdothatpvectorpart}, we can write:
\begin{eqnarray}\label{vectorphotonBoltzmannequationTheta}
	\left(\frac{\partial}{\partial\eta} + ik\mu - \tau'\right)\Theta^{(V)} - \frac{\tau'}{\cos\theta}\sqrt{\frac{2\pi}{15}}(Y_2^1 U_{\rm b,-} + Y_2^{-1}U_{\rm b,+})\nonumber\\ - \frac{i}{2}\sqrt{\frac{2\pi}{15}}(Y_2^1 F_-' + Y_2^{-1}F_+') = \nonumber\\ 
	 -\frac{\tau'}{10}\sum_{m = -2}^2Y_2^m(\hat p)\int d^2\hat p'\left[Y_2^{m*}\Theta^{(V)} - \sqrt{\frac{3}{2}}\mathcal E^{m*}\Theta_P^{(V)} - \sqrt{\frac{3}{2}}\frac{(\mathcal B^{m*})^2}{\mathcal E^{m*}}\Theta_P^{(V)}\right]_{\hat p'}\;, \quad
\end{eqnarray}
where
\begin{equation}
	U_{\rm b,\pm} \equiv U_{\rm b1} \pm iU_{\rm b2}\;,
\end{equation}
and for the polarization:
\begin{eqnarray}\label{vectorphotonBoltzmannequationThetaP}
	\left(\frac{\partial}{\partial\eta} + ik\mu - \tau'\right)\Theta_P^{(V)} = \nonumber\\ 
	 \frac{\tau'}{10}\sum_{m = -2}^2\sqrt{\frac{3}{2}}\mathcal E^m\int d^2\hat p'\left[Y_2^{m*}\Theta^{(V)} - \sqrt{\frac{3}{2}}\mathcal E^{m*}\Theta_P^{(V)} - \sqrt{\frac{3}{2}}\frac{(\mathcal B^{m*})^2}{\mathcal E^{m*}}\Theta_P^{(V)}\right]_{\hat p'}\;. \qquad
\end{eqnarray}
Introduce the vector contribution to temperature anisotropy as follows:
\begin{equation}
\boxed{\Theta^{(V)}(\mu,\phi) = \frac{i}{\cos\theta}\sqrt{\frac{2\pi}{15}}\sum_{\lambda = \pm 1}\Theta^{(V)}_\lambda(\mu)Y^{-\lambda}_2(\mu,\phi)}
\end{equation}
where the factor $1/\cos\theta$ is due to the fact that in Eq.~\eqref{vbcdothatpvectorpart} only a $\sin\theta$ appears, and thus it is not proportional to $Y_2^{\pm 1}$. Similarly, for the polarization field:
\begin{equation}
	\boxed{\Theta_P^{(V)}(\mu,\phi) = -\sqrt{\frac{\pi}{5}}\sum_{\lambda = \pm 1}\Theta^{(V)}_{P\lambda}(\mu)\mathcal E^{-\lambda}(\mu,\phi)}
\end{equation}
The vector Boltzmann equation for the temperature can then be written as:
\begin{eqnarray}
	\left(\frac{\partial}{\partial\eta} + ik\mu - \tau'\right)\Theta_\lambda^{(V)} + i\tau'U_{\rm b\lambda} - \frac{1}{2}F_\lambda^{'}\mu = \nonumber\\ 
	 \frac{i\tau'}{10}\mu\int d^2\hat p'\left[Y_2^{-\lambda *}\Theta_\lambda^{(V)}\frac{i}{\mu'}Y_2^{-\lambda} + \frac{3}{2}\mathcal E^{-\lambda *}\Theta_{P\lambda}^{(V)}\mathcal E^{-\lambda} + \frac{3}{2}\frac{(\mathcal B^{-\lambda *})^2}{\mathcal E^{-\lambda *}}\Theta_{P\lambda}^{(V)}\mathcal E^{-\lambda}\right]\;, \qquad
\end{eqnarray}
and for polarization:
\begin{eqnarray}
	\left(\frac{\partial}{\partial\eta} + ik\mu - \tau'\right)\Theta_{P\lambda}^{(V)} = \nonumber\\ 
	 -\frac{\tau'}{10}\int d^2\hat p'\left[Y_2^{-\lambda *}\Theta_\lambda^{(V)}\frac{i}{\mu'}Y_2^{-\lambda} + \frac{3}{2}\mathcal E^{-\lambda *}\Theta_{P\lambda}^{(V)}\mathcal E^{-\lambda} + \frac{3}{2}\frac{(\mathcal B^{-\lambda *})^2}{\mathcal E^{-\lambda *}}\Theta_{P\lambda}^{(V)}\mathcal E^{-\lambda}\right]\;, \qquad
\end{eqnarray}
With the help of Tab.~\ref{Tab:SphericalHarmonics}, we then have to work out the integral:
\begin{equation}
	-\frac{\tau'}{10}\int d\mu d\phi\frac{15}{8\pi}\left[\mu(1 - \mu^2)i\Theta^{(V)}_\lambda + (1 - \mu^4)\Theta_{P\lambda}^{(V)}\right]\;,
\end{equation}
which, written in Legendre polynomial, becomes:
\begin{eqnarray}
	 -\frac{3\tau'}{4}\int \frac{d\mu}{2}\left[\frac{2}{5}\mathcal P_1(\mu) - \frac{2}{5}\mathcal P_3(\mu)\right]i\Theta_\lambda^{(V)}\nonumber\\ - \frac{3\tau'}{4}\int \frac{d\mu}{2}\left[\frac{4}{5}\mathcal P_0(\mu) - \frac{4}{7}\mathcal P_2(\mu) - \frac{8}{35}\mathcal P_4(\mu)\right]\Theta_{P\lambda}^{(V)}\;.
\end{eqnarray} 
Hence, using the usual Legendre expansion, we can write the vector Boltzmann equation for photons as follows:
\begin{eqnarray}
	\left(\frac{\partial}{\partial\eta} + ik\mu - \tau'\mu\right)\Theta_\lambda^{(V)} + i\tau'U_{\rm b,\lambda} - \frac{1}{2}F_\lambda^{'} = \nonumber\\ 
	 i\tau'\mathcal P_1(\mu)\left(\frac{3}{10}\Theta^{(V)}_{\lambda,1} + \frac{3}{10}\Theta^{(V)}_{\lambda,3} - \frac{6}{35}\Theta^{(V)}_{P\lambda,4} + \frac{3}{7}\Theta^{(V)}_{P\lambda,2} + \frac{3}{5}\Theta_{P\lambda,0}^{(V)}\right)\;.
\end{eqnarray}
and for polarization:
\begin{eqnarray}
	\left(\frac{\partial}{\partial\eta} + ik\mu - \tau'\mu\right)\Theta_{P\lambda}^{(V)} = \nonumber\\ 
	 -\tau'\left(\frac{3}{10}\Theta^{(V)}_{\lambda,1} + \frac{3}{10}\Theta^{(V)}_{\lambda,3} - \frac{6}{35}\Theta^{(V)}_{P\lambda,4} + \frac{3}{7}\Theta^{(V)}_{P\lambda,2} + \frac{3}{5}\Theta_{P\lambda,0}^{(V)}\right)\;.
\end{eqnarray}
We will conclude our treatment of vector modes here, dealing only with the scalar and tensor modes from now on.

\section{Boltzmann equation for baryons}

As baryons, we refer here generically to electrons and protons, neglecting helium nuclei. The latter can be straightforwardly included, as we are going to comment throughout the derivation. 
\index{Baryons!Boltzmann equation}
Electrons and protons interact via Coulomb scattering:
\begin{equation}
	e + p \longleftrightarrow e + p\;,
\end{equation}
and, in turn, electrons are also coupled to photons via Thomson scattering. We assume the Coulomb interaction to be so efficient that:
\begin{equation}\label{conditionsonbaryonperturbations}
	\delta_e = \delta_p = \delta_{\rm b}\;, \qquad \textbf{v}_e = \textbf{v}_p = \textbf{v}_{\rm b}\;. 
\end{equation}
Baryons are non-relativistic in the epochs of interest, and thus there is no exchange of energy via Thomson scattering with photons. For this reason, we expect that their density contrast is governed by an equation similar to Eq.~\eqref{deltaequationCDM}.  

Consider the Boltzmann equation for electrons:
\begin{equation}\label{boltzeqelec}
	\frac{d}{d\eta}\delta f_e(\eta,\textbf{x},\textbf{P}_e) = \left(\frac{\partial}{\partial \eta}\delta f_e\right)_{\rm coll-Coulomb} + \left(\frac{\partial}{\partial \eta}\delta f_e\right)_{\rm coll-Thomson}\;,
\end{equation}
where the first collisional term is relative to Coulomb scattering:
\begin{equation}
	e(\textbf{P}_e) + p(\textbf{P}_p) \leftrightarrow e(\textbf{P}_e') + p(\textbf{P}_p')\;,
\end{equation}
whereas the second is the collisional term relative to Thomson scattering:
\begin{equation}
	e(\textbf{P}_e) + \gamma(\textbf{p}) \leftrightarrow e(\textbf{P}_e') + \gamma(\textbf{p}')\;.
\end{equation}
For protons, the Boltzmann equation is similar:
\begin{equation}\label{boltzeqprot}
	\frac{d}{d\eta}\delta f_p(\eta,\textbf{x},\textbf{P}_p) = \left(\frac{\partial}{\partial \eta}\delta f_p\right)_{\rm coll-Coulomb}\;,
\end{equation}
the only difference being that we neglect their interaction with photons since the Thomson cross-section goes as $\propto 1/m^2$, and therefore it is $10^6$ less important for protons than for electrons. A Boltzmann equation for Helium nuclei would be similar to the one for protons. 

Recall that we are considering free electrons and protons here (hence the name baryonic plasma). After recombination, electrons and protons form hydrogen atoms which can be treated as a collisionless species. Therefore, their evolution is governed by equations identical to the ones we have developed earlier for CDM.

We are going to exploit the fact that electrons and protons are non-relativistic and will simply consider the first two moments of their perturbed Boltzmann equations. In particular, we use the left hand sides of Eqs.~\eqref{deltaeqgeneric} and \eqref{Vequationgeneral}, but with $w_e = w_p = 0$ and no pressure perturbations. We then have:
\begin{eqnarray}
	\delta_e' + kV_e + 3\Phi' = \frac{m_e}{\rho_e a^3}\int d^3\mathbf Q_e \left(\frac{\partial}{\partial \eta}\delta f_e\right)_{\rm coll}\;,\\
	\delta_p' + kV_p + 3\Phi' = \frac{m_p}{\rho_p a^3}\int d^3\mathbf Q_p \left(\frac{\partial}{\partial \eta}\delta f_p\right)_{\rm coll}\;.
\end{eqnarray}
Note that in considering the zero moment of the electron and proton perturbed Boltzmann equation, we have considered only the mass energy. So, the above density contrasts are also to be considered as relative fluctuations in the number densities.

For this reason, the integrals of the collisional terms on the right hand sides vanish. Indeed, the scattering processes involved here do not change the number of electrons and protons but only reshuffle their momenta.

Therefore, using the conditions stated earlier in Eq.~\eqref{conditionsonbaryonperturbations}, we obtain the following equation, similar to Eq.~\eqref{deltaequationCDM}, describing the evolution of the density contrast of baryons:
\begin{equation}\label{deltaequationbaryons}
	\boxed{\delta_{\rm b}' + kV_{\rm b} + 3\Phi' = 0}
\end{equation}
where, again, $kV_{\rm b} = \theta_{\rm b}$ in the notation of \cite{Ma:1995ey}.

As we did for Eq.~\eqref{Vequationgeneral}, we take the first moments:
\begin{eqnarray}
	V_e' + \mathcal HV_e - k\Psi = \frac{i\hat k_i}{\rho_e a^4}\int d^3\mathbf Q_e Q_e^i\left(\frac{\partial}{\partial \eta}\delta f_e\right)_{\rm coll}\;,\\
	V_p' + \mathcal HV_p - k\Psi = \frac{i\hat k_i}{\rho_p a^4}\int d^3\mathbf Q_p Q_p^i\left(\frac{\partial}{\partial \eta}\delta f_p\right)_{\rm coll}\;.
\end{eqnarray}
Now, let us again use the conditions in Eq.~\eqref{conditionsonbaryonperturbations} in summing the above two equations. We get:
\begin{eqnarray}\label{vieqbaryons}
 V_{\rm b}' + \mathcal HV_{\rm b} - k\Psi = \nonumber\\ 
 \frac{i\hat{k}_i}{\rho_{\rm b}a^4}\left[\int d^3\mathbf Q_e Q_e^i\left(\frac{\partial}{\partial \eta}\delta f_e\right)_{\rm coll} + \int d^3\mathbf Q_p Q_p^i\left(\frac{\partial}{\partial \eta}\delta f_p\right)_{\rm coll}\right]\;.
\end{eqnarray}
Now we have to calculate the terms on the right hand side. The ones related to Coulomb scattering give a vanishing result:
\begin{equation}
	\int d^3\mathbf Q_e Q_e^i\left(\frac{\partial}{\partial \eta}\delta f_e\right)_{\rm coll-Coulomb} + \int d^3\mathbf Q_p Q_p^i\left(\frac{\partial}{\partial \eta}\delta f_p\right)_{\rm coll-Coulomb} = 0\;,
\end{equation}
since $Q_e^i + Q_p^i$ is the total momentum, which is conserved. Indeed, one has:
\begin{eqnarray}
	\int d^3\mathbf Q_e Q_e^i\left(\frac{\partial}{\partial \eta}\delta f_e\right)_{\rm coll-Coulomb} = \int d^3\mathbf Q_p Q_e^i\left(\frac{\partial}{\partial \eta}\delta f_p\right)_{\rm coll-Coulomb} \nonumber\\
	= \int d^3\mathbf Q_p (Q_{\rm tot} - Q_p)^i\left(\frac{\partial}{\partial \eta}\delta f_p\right)_{\rm coll-Coulomb}\nonumber\\
	= Q_{\rm tot}^i\int d^3\mathbf Q_p\left(\frac{\partial}{\partial \eta}\delta f_p\right)_{\rm coll-Coulomb} - \int d^3\mathbf Q_p Q_p^i\left(\frac{\partial}{\partial \eta}\delta f_p\right)_{\rm coll-Coulomb}\;,
\end{eqnarray}
and, as we saw,
\begin{equation}
	\int d^3\mathbf Q_p\left(\frac{\partial}{\partial \eta}\delta f_p\right)_{\rm coll-Coulomb} = 0\;,
\end{equation}
because the particle number is conserved.

The only survivor is the collisional term for Thomson scattering, which we can manipulate as follows, exploiting the conservation of the total electron plus photon momentum:
\begin{equation}
	\int d^3\mathbf Q_e Q_e^i\left(\frac{\partial}{\partial \eta}\delta f_e\right)_{\rm coll-Thomson} = -\int d^3\mathbf q q^i\left(\frac{\partial}{\partial \eta}\delta f_\gamma\right)_{\rm coll-Thomson}\;.
\end{equation} 
We have then:
\begin{eqnarray}\label{vieqbaryonsFT}
 V_{\rm b}' + \mathcal HV_{\rm b} - k\Psi = -\frac{4\pi i}{\rho_{\rm b}a^4}\int_{-1}^1\frac{d\mu}{2}\mu\int dq\;q^2 q\left(\frac{\partial}{\partial \eta}\delta f_\gamma\right)_{\rm coll-Thomson}\;.
\end{eqnarray}
On the right hand side, we introduce the result of Eq.~\eqref{fullcolltermnopol} and the definition of $\Theta$, thus obtaining:
\begin{equation}
	V_{\rm b}' + \mathcal HV_{\rm b} - k\Psi = \frac{4i\tau'\rho_\gamma}{\rho_{\rm b}}\int_{-1}^1\frac{d\mu}{2}\mu\left[\Theta^{(S)}_0 - \Theta^{(S)} - i\mu V_{\rm b} - \frac{1}{2}\mathcal{P}_2(\mu)\Theta^{(S)}_2\right]\;.
\end{equation}
Performing the $d\mu$ integration and introducing the dipole $\Theta^{(S)}_1$, we obtain:
\begin{eqnarray}\label{Vequationbaryons}
\boxed{V_{\rm b}' + \mathcal HV_{\rm b} - k\Psi = \frac{4\tau'\rho_\gamma}{3\rho_{\rm b}}\left(V_{\rm b} - 3\Theta^{(S)}_1\right)}
\end{eqnarray}
This equation can also be obtained by exploiting momentum conservation for the two-fluid system composed of baryons and photons and using Eq.~\eqref{Vequationgeneral}. In fact, momentum is not conserved individually for baryons and photons, but, of course, it is for the photon-baryon fluid. Let us then write Eq.~\eqref{Vequationgeneral} for the latter:
\begin{eqnarray}
	\rho_{\rm b}\left(V_{\rm b}' + \mathcal HV_{\rm b} - k\Psi\right) + \frac{4}{3}\rho_\gamma V_\gamma' - \frac{k}{3}\rho_\gamma\delta_\gamma - k\hat k_l\hat k_m\pi_\gamma^{lm} - \frac{4k}{3}\rho_\gamma\Psi = 0\;,
\end{eqnarray}
where we have neglected pressure and anisotropic stress for baryons, as they are non-relativistic. Now, Eqs.~\eqref{monopoledeltanu}, \eqref{dipoleVnu}, and \eqref{quadrupoleanisotropicstressnu} hold true also for photons. So we have:
\begin{equation}
	4\Theta_0^{(S)} = \delta_\gamma\;, \qquad 4\Theta_1^{(S)} = \frac{4}{3}V_\gamma\;, \qquad 4\Theta_2^{(S)} = -\frac{3}{2\rho_\gamma}\hat k_l\hat k_m\pi_\gamma^{lm}\;.
\end{equation}

\hrulefill

\begin{ex}
	Using the above definitions and Eq.~\eqref{le1equationphoton} obtain Eq.~\eqref{Vequationbaryons}.
\end{ex}

\hrulefill

We have thus obtained all the relevant equations that describe small fluctuations in the components of the universe. In Chapter~\ref{Chap:IC} we shall tackle the issue of which initial conditions we can employ for our set of differential equations.

If we compare the above Eq.~\eqref{Vequationbaryons} with the corresponding one in \cite{Ma:1995ey}, we shall notice in this reference an extra term $k^2c^2_{\rm s}\delta_{\rm b}$, which comes from the fact that the authors did not neglect the effective speed of sound contribution for baryons (cf. Eq.~\eqref{Vequationgeneral}).

\clearpage
\chapter{Primordial modes and initial conditions}\label{Chap:IC}

{\rightskip=3truepc\leftskip=3truepc\noindent
{\it Finirai per trovarla la Via... se prima hai il coraggio di perderti\\
(You will eventually find your Way... if you have first the courage of lose yourself)}
\vskip 0.10 in
\centerline{\it ---Tiziano Terzani, Un altro giro di giostra}
\vskip 0.20 in
}

In Chapters \ref{Chap:CosmoPertTheory} and \ref{Chap:PertubedBoltzmannEquations}, we have derived the evolution equations for small fluctuations about the homogeneous and isotropic FLRW background. These are ordinary differential equations, and we need to know which initial conditions to use. This is the topic of this chapter. We mainly refer to \cite{Ma:1995ey} and \cite{Bucher:1999re}.

We discuss primordial modes for scalar perturbations only, leaving the treatment of the tensor modes for Chapter \ref{Chap:Inflation}. For this reason, we drop the superscript $(S)$.

\section{Summary of the evolution equations}

Let us recapitulate the evolution equations that we have found in Chapters \ref{Chap:CosmoPertTheory} and \ref{Chap:PertubedBoltzmannEquations}. 

\begin{itemize}
    \item Photon temperature fluctuations, cf. Eqs.~\eqref{lg2equationphoton}--\eqref{le0equationphoton}:
\begin{align}
	(2\ell + 1)\Theta_\ell' &+ k\left[(\ell + 1)\Theta_{\ell + 1} - \ell\Theta_{\ell - 1}\right] = \tau'(2\ell + 1)\Theta_\ell & (\ell > 2)\;,\\
10\Theta_2' &+ 2k\left(3\Theta_{3} - \frac{2}{3}V_\gamma\right) = 10\tau'\Theta_2 - \tau'\Pi & (\ell = 2)\;,\\
V_\gamma' &+ k\left(2\Theta_{2} - \frac{\delta_\gamma}{4}\right) = k\Psi + \tau'\left(V_\gamma - V_{\rm b}\right) & (\ell = 1)\;,\\
\label{monopoleevoeq} \delta_\gamma' &+ \frac{4}{3}kV_\gamma = -4\Phi' & (\ell = 0)\;,
\end{align}
with
\begin{equation}
	\Pi = \Theta_{2} + \Theta_{P2} + \Theta_{P0}\;,
\end{equation}
and where we have used $4\Theta_0 = \delta_\gamma$ and $3\Theta_1 = V_\gamma$. 

\item Photon polarization field, cf. Eqs.~\eqref{lg2equationpol}-\eqref{le0equationpol}:
\begin{align}
	(2\ell + 1)\Theta_{P\ell}' &+ k\left[(\ell + 1)\Theta_{P(\ell + 1)} - \ell\Theta_{P(\ell - 1)}\right] = \tau'(2\ell + 1)\Theta_{P\ell} & (\ell > 2)\;,\\
10\Theta_{P2}' &+ 2k\left(3\Theta_{P3} - 2\Theta_{P1}\right) = 10\tau'\Theta_{P2} - \tau'\Pi & (\ell = 2)\;,\\
3\Theta_{P1}' &+ k\left(2\Theta_{P2} - \Theta_{P0}\right) = 3\tau'\Theta_{P1} & (\ell = 1)\;,\\
2\Theta_{P0}' &+ 2k\Theta_{P1} = 2\tau'\Theta_{P0} - \tau'\Pi & (\ell = 0)\;.
\end{align}

\item Neutrino temperature fluctuations, cf. Eqs.~\eqref{lge2equationnu}-\eqref{le0equationnu}:
\begin{align}
	(2\ell + 1)\mathcal{N}_\ell' &+ k\left[(\ell + 1)\mathcal{N}_{\ell + 1} - \ell\mathcal{N}_{\ell - 1}\right] = 0 & (\ell > 1)\;,\\
V_\nu' &+ 2k\mathcal{N}_{2} - k\frac{\delta_\nu}{4} = k\Psi & (\ell = 1)\;,\\
\delta_\nu' &+ \frac{4}{3}kV_\nu = -4\Phi' & (\ell = 0)\;,
\end{align} 
where we have used $4\mathcal N_0 = \delta_\nu$ and $3\mathcal N_1 = V_\nu$.

\item CDM density contrast and flow velocity, cf. Eqs.~\eqref{deltaequationCDM} and \eqref{VequationCDM}:
\begin{eqnarray}
	\delta_{\rm c}' + kV_{\rm c} + 3\Phi' = 0\;, \qquad V_{\rm c}' + \mathcal HV_{\rm c} - k\Psi = 0\;.
\end{eqnarray}

\item Baryonic density contrast and flow velocity, cf. Eqs.~\eqref{deltaequationbaryons} and \eqref{Vequationbaryons}:
\begin{eqnarray}
	\delta_{\rm b}' + kV_{\rm b} + 3\Phi' = 0\;, \qquad V_{\rm b}' + \mathcal HV_{\rm b} - k\Psi = \frac{4\tau'\rho_\gamma}{3\rho_{\rm b}}\left(V_{\rm b} - V_\gamma\right)\;,
\end{eqnarray}

\item Einstein equations, cf. \eqref{relativisticPoissonequation}, \eqref{anisotropicstressscalarequation}, \eqref{EEqV} and \eqref{GiideltaPeq2}:
\begin{align}
3\mathcal{H}\left(\Phi' - \mathcal{H}\Psi\right) + k^2\Phi &= 4\pi Ga^2\left(\rho_{\rm c}\delta_{\rm c} + \rho_{\rm b}\delta_{\rm b} + \rho_\gamma\delta_\gamma + \rho_\nu\delta_\nu\right)\;,\\
k^2(\Phi + \Psi) &= -32\pi G a^2\left(\rho_\gamma\Theta_2 + \rho_\nu\mathcal{N}_2\right)\;,\\
k(-\Phi' + \mathcal H\Psi) &= 4\pi Ga^2\left(\rho_{\rm c}V_{\rm c} + \rho_{\rm b}V_{\rm b} + \frac{4}{3}\rho_\gamma V_\gamma + \frac{4}{3}\rho_\nu V_\nu\right)\;,\\
	\Phi'' + 2\mathcal{H}\Phi' - \mathcal{H}\Psi' - (2\mathcal{H}' &+ \mathcal{H}^2)\Psi + \frac{k^2}{3}(\Phi + \Psi) = - \frac{4\pi Ga^2}{3}\left(\rho_\gamma\delta_\gamma + \rho_\nu\delta_\nu\right)\;,
\end{align}
where we have expressed the anisotropic stresses of photons and neutrinos in terms of their quadrupole moments, and we used, for Eq.~\eqref{GiideltaPeq2}, that:
\begin{equation}
	\delta P = \frac{1}{3}\left(\rho_\gamma\delta_\gamma + \rho_\nu\delta_\nu\right)\;,
\end{equation}
since only photons and neutrinos have a non-negligible pressure equal to one third of their density perturbation, as they are relativistic.
\end{itemize}

\section{The \texorpdfstring{$k\eta \ll 1$}{kη ≪ 1} limit}

Now we must understand \textit{when} to set our initial conditions. These should be values for the above quantities $\Theta_\ell$, $\mathcal{N}_\ell$, $\delta_{\rm c}$, $\delta_{\rm b}$, $V_{\rm c}$, $V_{\rm b}$, $\Phi$, and $\Psi$ at a certain initial instant. This instant cannot be $\eta = 0$, because it is a singular point, and because some of the approximations that we adopted are valid well after BBN. Therefore, we need it to be some $\eta_i > 0$. But how do we establish which $\eta_i$ and, above all, the value that a fluctuation has at such $\eta_i$? 

As $\eta$ grows, any scale $k$ shall transition from a $\eta \ll 1/k$ regime called \textbf{super-horizon evolution}, to a $\eta \sim 1/k$ regime called \textbf{horizon crossing}, and finally to a $\eta \gg 1/k$ regime called \textbf{sub-horizon evolution}. The horizon mentioned above is the particle horizon. Recall from Eq.~\eqref{comdist} that:
\begin{equation}
	\chi_{\rm p}(\eta) = \int_{0}^{t}\frac{dt'}{a(t')} = \int_0^\eta d\eta' = \eta\;,
\end{equation}
i.e., the conformal time represents the comoving distance traveled by a photon since the Big Bang. 

The initial values for the perturbative quantities are set when $k\eta \ll 1$. The solutions to the equations for the perturbations in this regime are also called \textbf{primordial modes} because, for the scales of present observational interest, the conformal time guaranteeing $k\eta \ll 1$ corresponds to epochs deep in the radiation-dominated era. 

As we know, in the radiation-dominated epoch $\mathcal H = 1/\eta$, whereas in the matter-dominated epoch $\mathcal H = 2/\eta$. In both cases, demanding $k\eta \ll 1$ is equivalent to demanding $k \ll \mathcal H$.

The evolution of perturbations on super-horizon scales is somewhat peculiar. In order to be convinced of this, consider a scale $k$ today which, to be observationally interesting, must be much smaller than the particle horizon. The latter is proportional to $H_0$, thus $k \gg H_0$. At some time in the past, this condition becomes $k \ll \mathcal H$, since $\mathcal H$ diverges for $\eta \to 0$. To be more quantitative, using the Friedmann equation in the radiation-dominated epoch, one has:
\begin{equation}
	\frac{k}{\mathcal H} \approx \frac{ka}{H_0\sqrt{\Omega_{\rm r0}}} \approx \frac{10^2}{(1 + z)}\frac{k}{H_0}\;.
\end{equation}
So, even a scale $k = 100H_0$, which is today on the order of 40 Mpc (already in the non-linear regime of evolution) for redshifts larger than $10^4$, was outside the horizon.

We describe the behavior of cosmological perturbations on super-horizon scales ($k\eta \ll 1$) while assuming an expansion with respect to $k\eta$ of the perturbative variables. That is, something like this:
\begin{equation}\label{ketaexp}
	\delta X = \sum_n X_n(k\eta)^n\;,
\end{equation} 
where $\delta X$ represents a generic perturbative variable ($\delta$ or $\Phi$, for example), and $n$ starts from some value that is possibly non negative, since we do not want perturbations to diverge for $k\eta \to 0$. We say ``possibly'' because we will see that for the neutrino isocurvature velocity mode, there will be a $1/(k\eta)$ behavior of the potentials and of the monopoles.

In the next subsections, we establish a hierarchy among the various perturbative variables.

\subsection{The Einstein equations deep in the radiation-dominated epoch}

As we saw in Subsection \ref{Subsec:radplusdustuni}, the evolution of the scale factor with respect to the conformal time in the radiation-matter universe is:
\begin{equation}
	a(\eta) = \frac{(\sqrt{2} - 1)\Omega_{\rm r0}}{\Omega_{\rm m0}}\left[2\frac{\eta}{\eta_{\rm eq}} + (\sqrt{2} - 1)\frac{\eta^2}{\eta_{\rm eq}^2}\right]\;, \quad \mathcal H = \frac{2\eta_{\rm eq} + 2(\sqrt{2} - 1)\eta}{2\eta_{\rm eq}\eta + (\sqrt{2} - 1)\eta^2}\;.
\end{equation}
In Ref.~\cite{Bucher:1999re}, the authors choose $\eta_{\rm eq}$ such that $a(\eta) = \eta + \eta^2$, thereby simplifying the calculations. We assume here $\eta \ll \eta_{\rm eq}$, so that we can neglect the quadratic contribution and then employ:
\begin{equation}
	a(\eta) \simeq \sqrt{\Omega_{\rm r0}} H_0\eta\;, \qquad \mathcal H \simeq \frac{1}{\eta}\;.
\end{equation}
Equipped with these relations and using the Friedmann equation:
\begin{equation}\label{4piGa2H2}
	4\pi Ga^2 = \frac{3\mathcal H^2}{2\rho_{\rm tot}}\;,
\end{equation}
the Einstein equations can be written as:
\begin{align}
\label{Poisseqraddom} 2\left(\eta\Phi' - \Psi\right) &+ \frac{2}{3}(k\eta)^2\Phi = R_\gamma\delta_\gamma + R_\nu\delta_\nu + R_{\rm c}\delta_{\rm c} + R_{\rm b}\delta_{\rm b}\;,\\
-\frac{1}{12}(k\eta)^2(\Phi + \Psi) &= R_\gamma\Theta_2 + R_\nu\mathcal{N}_2\;,\\
\label{veleqraddom} \frac{2}{3}k\eta(-\eta\Phi' + \Psi) &= R_{\rm c}V_{\rm c} + R_{\rm b}V_{\rm b} + \frac{4}{3}R_\gamma V_\gamma + \frac{4}{3}R_\nu V_\nu\;,\\
	2\eta^2\Phi'' + 4\eta\Phi' &- 2\eta\Psi' + 2\Psi + \frac{2}{3}(k\eta)^2(\Phi + \Psi) = - R_\gamma\delta_\gamma - R_\nu\delta_\nu\;,
\end{align}
where we have defined the density fraction $R_i \equiv \rho_i/\rho_{\rm tot}$ for each component, such that: 
\begin{equation}
	R_\gamma + R_\nu + R_{\rm c} + R_{\rm b} = 1\;.
\end{equation}
Note that for $\eta \to 0$, we can neglect $R_{\rm c,b}$ with respect to $R_{\gamma,\nu}$, so $R_\gamma + R_\nu \approx 1$.

\hrulefill

\begin{ex}
	Show that $R_\nu$ is constant at early times ($\eta \to 0$) and its value is:
	\begin{equation}
		R_\nu \to \frac{N_{\rm eff}(7/8)(4/11)^{4/3}}{1 + N_{\rm eff}(7/8)(4/11)^{4/3}} = 0.4089\;,
	\end{equation}
	the last number coming from choosing $N_{\rm eff} = 3.046$.\index{Neutrinos!Fraction}
\end{ex}

\hrulefill

Note that:
\begin{equation}
	\frac{d(1/\mathcal H)}{d\eta} = -\frac{\mathcal H'}{\mathcal H^2} = \frac{3}{2}\left(R_{\rm c} + R_{\rm b} + \frac{4}{3}R_\gamma + \frac{4}{3}R_\nu\right) - 1 = \frac{1}{2}(1 + R_\gamma + R_\nu)\;.
\end{equation}
This is consistent with the approximation $R_\gamma + R_\nu \approx 1$. Moreover:
\begin{equation}
	R_{\gamma,\nu}' = -\mathcal HR_{\gamma,\nu}(R_{\rm c} + R_{\rm b})\;.
\end{equation}
By virtue of the expansion in powers of $(k\eta)$, cf. Eq. \eqref{ketaexp}, one can say that:
\begin{equation}
	\eta^2\Phi'' \simeq \eta\Phi' \simeq \Phi\;. 
\end{equation}
Therefore, inspecting Eq. \eqref{Poisseqraddom}, it is natural to establish that:
\begin{equation}
	\Phi \simeq \Psi \simeq \delta_i\;.
\end{equation}
Moreover, from Eq. \eqref{veleqraddom}, we can establish that:
\begin{equation}
	V_i \simeq (k\eta)\Psi \simeq (k\eta)\Phi \simeq (k\eta)\delta_i\;.
\end{equation}
Thus, the velocity perturbation is subdominant with respect to the gravitational potential. Finally, from the anisotropic stress equation, we have:
\begin{equation}
	\Theta_2,\mathcal N_2 \simeq (k\eta)^2\Psi \simeq (k\eta)^2\Phi\;.
\end{equation}
Therefore, the quadrupole moments are subdominant with respect to the velocity perturbations.

\subsection{The equations for the monopoles}

Given the premises of the previous subsection, the equations for the monopoles can be remarkably simplified:
\begin{equation}\label{monopolessimp}
	\delta_\gamma' = -4\Phi'\;, \quad
	\delta_\nu' = -4\Phi'\;, \quad
\delta_{\rm c}' = -3\Phi'\;, \quad
\delta_{\rm b}' = -3\Phi'\;.
\end{equation}
This approximation has an error of $O[(k\eta)V]$. The integration of the above equations is now straightforward:
\begin{align}
	\delta_\gamma &= -4\Phi + 4C_\gamma\;,\\
	\delta_\nu &= -4\Phi + C_\nu\;,\\
	\delta_{\rm c} &= -3\Phi + C_{\rm c}\;,\\
	\delta_{\rm b} &= -3\Phi + C_{\rm b}\;,
\end{align}
where we have introduced 4 integration constants. We cast the above equations as follows:
\begin{align}
	\delta_\gamma &= -4\Phi + 4C_\gamma\;,\\
\label{deltanuSnu}	\delta_\nu &= \delta_\gamma + S_\nu\;,\\
	\delta_{\rm c} &= \frac{3}{4}\delta_\gamma + S_{\rm c}\;,\\
\label{deltabSb}	\delta_{\rm b} &= \frac{3}{4}\delta_\gamma + S_{\rm b}\;,
\end{align}
for a reason that will be clearer later. We have defined:
\begin{eqnarray}
	S_\nu \equiv C_\nu - 4C_\gamma\;, \qquad S_{\rm c} \equiv C_{\rm c} - 3C_\gamma\;, \qquad S_{\rm b} \equiv C_{\rm b} - 3C_\gamma\;.
\end{eqnarray}
These are called \textbf{density isocurvature modes} , or \textbf{entropy modes}. Note that the $C$'s and thus the $S$'s are not actual constants, but functions of $\textbf{k}$. Hereafter, we inappropriately refer to time-independent quantities as ``constants''.\index{Isocurvature modes}\index{Entropy modes}   

\subsection{Multipoles in the \texorpdfstring{$k\eta \ll 1$}{kη ≪ 1} limit}

Let us first start with the neutrino hierarchy, since there is no collisional term. From the previous sections, we have established that:
\begin{equation}
	V_\nu \sim (k\eta)\delta_\nu\;, \qquad \mathcal{N}_2 \sim (k\eta)V_\nu\;.
\end{equation}
Consider the equation for $\mathcal N_\ell$ ($\ell > 2$):
\begin{equation}
	(2\ell + 1)\mathcal N_\ell' + k(\ell + 1)\mathcal N_{\ell + 1} = k\ell\mathcal N_{\ell - 1}\;.
\end{equation}
From the structure of this hierarchy, it is straightforward to establish that:
\begin{equation}
	\mathcal N_\ell \simeq (k\eta)\mathcal N_{\ell - 1}\;,
\end{equation}
so that higher order multipoles are progressively subdominant. 

In the case of photons, a similar reasoning can be applied; however, things are somewhat trickier because of the presence of the collisional term. In particular, we have that:
\begin{equation}
	-\tau' = n_e\sigma_{\rm T}a \propto a^{-2}\;.
\end{equation}
This is proportional to $1/\eta^2$ in the radiation-dominated epoch and to $1/\eta^4$ in the matter-dominated epoch (and recall that it abruptly goes to zero during recombination and decoupling).

That is, the optical depth time derivative, which is the photon-electron interaction rate, diverges for $\eta \to 0$ as $1/\eta^2$. This is expected since the universe becomes denser the further we go back in time, resulting in more interactions taking place. 

Since $\tau'$ gets larger and larger the farther back we move in time, we cannot balance the equations in the hierarchy unless the terms multiplying $\tau'$ vanish. We do assume $\tau' \to \infty$, for which the vanishing must be exact. This approximation is referred to as \textbf{tight coupling}.\index{Tight coupling}

Hence, we have that:
\begin{align}
	\Theta_\ell &= \Theta_{P\ell} = 0\;, \qquad (\ell > 2)\;,\\
\label{TemPolquadr}    9\Theta_2 &- \Theta_{P0} - \Theta_{P2} = 0\;,\\
9\Theta_{P2} &- \Theta_{P0} - \Theta_{2} = 0\;,\\
    V_\gamma &= V_{\rm b}\;, \qquad \Theta_{P1} = 0\;,\\
\label{Polmono}    \Theta_{P0} &= \Theta_{P2} + \Theta_2\;.
\end{align}

\hrulefill

\begin{ex} Combine the above equations and show that:
\begin{equation}
	\boxed{\Theta_2 = \Theta_{P2} = \Theta_{P0} = 0}
\end{equation}
\end{ex}

\hrulefill

Physically, the coupling between free electrons and photons is so efficient that it ``washes away'' the quadrupole and higher moments of the temperature fluctuations and all the polarization moments. Moreover, it glues photons and baryons into a single fluid with velocity $V_\gamma = V_{\rm b}$.

Finally, we are left with only the equation for the monopole, already discussed, and that for the velocity, for photons:
\begin{eqnarray}
\label{Photondipole} V_\gamma' - k\frac{\delta_\gamma}{4} = k\Psi\;.
\end{eqnarray}
The $k\eta \ll 1$ neutrino equations are as follows:
\begin{align}
\label{Neutrinomultipole}	(2\ell + 1)\mathcal{N}_\ell' &- k\ell\mathcal{N}_{\ell - 1} = 0\;, \qquad \ell \ge 2\;,\\
\label{Neutrinodipole} V_\nu' &- k\frac{\delta_\nu}{4} = k\Psi\;.
\end{align} 
Nothing can make the multipoles of the neutrino temperature vanish, as Thomson scattering does for photons. Therefore, we need initial conditions for all of them. 

\hrulefill

\begin{ex} Suppose that:
\begin{equation}
	\mathcal{N}_\ell = c_\ell(k\eta)^\ell\;,
\end{equation}
at the lowest order, where $c_\ell$ is some constant (in this instance a true constant number). From Eq.~\eqref{Neutrinomultipole} show that:
\begin{equation}
	c_\ell = \frac{1}{2\ell + 1}c_{\ell - 1}\;,
\end{equation}
and thus:
\begin{equation}
	\mathcal{N}_\ell = \frac{k\eta}{2\ell + 1}\mathcal{N}_{\ell - 1}\;, \qquad \ell \ge 2\;.
\end{equation}
\end{ex}

\hrulefill

Therefore, once we know the initial condition on $V_\nu$, we can determine the initial conditions for all the subsequent multipoles from the above equation. So, what about the initial condition on $V_\nu$? 

\hrulefill

\begin{ex} Subtract Eqs.~\eqref{Photondipole} and \eqref{Neutrinodipole} and, using the super-horizon equations for the monopoles, show that: 
\begin{equation}
	(V_\gamma - V_\nu)' = -\frac{k}{4}S_\nu\;.
\end{equation}
\end{ex}

\hrulefill

This gives:
\begin{equation}
	V_\gamma = V_\nu - \frac{1}{4}(k\eta)S_\nu - q_\nu\;.
\end{equation}
The constant $q_\nu$ is the \textbf{neutrino velocity isocurvature mode} or the \textbf{relative neutrino heat flux}.\index{Neutrinos!Isocurvature velocity mode}\index{Neutrinos!Relative neutrino heat flux}   

\subsection{CDM and baryon velocity equations}

Thanks to the tight-coupling approximation, the velocity equations for CDM and baryons are identical:
\begin{eqnarray}
	\eta V_{\rm c}' + V_{\rm c} = (k\eta)\Psi\;,\\
	\eta V_{\rm b}' + V_{\rm b} = (k\eta)\Psi\;,
\end{eqnarray}
and we can cast them compactly as follows:
\begin{eqnarray}
\label{vcvbeq}	(\eta V_{\rm c,b})' = (k\eta)\Psi\;.
\end{eqnarray}
This equation tells us that $V_{\rm c} = V_{\rm b}$, since taking the difference:
\begin{equation}
	(\eta V_{\rm c})' - (\eta V_{\rm b})' = 0\;,
\end{equation}
which implies that:
\begin{equation}
	V_{\rm b} = V_\gamma = V_{\rm c} + \frac{q_{\rm c}}{\eta}\;,
\end{equation}
where $q_{\rm c}/\eta$ would be a CDM velocity isocurvature mode. However, it diverges for $\eta \to 0$, so we need to put $q_{\rm c} = 0$. Therefore:
\begin{equation}
	V_{\rm b} = V_\gamma = V_{\rm c}\;,
\end{equation} 
with only neutrinos having the prerogative of a different flow velocity.

\subsection{The \texorpdfstring{$k\eta \ll 1$}{kη ≪ 1} limit of the Einstein equations}

Let us now consider the Einstein equations that we have written deep into the radiation-dominated epoch and examine their $k\eta \ll 1$ limit.

This means that we neglect $(k\eta)^2\Phi$ with respect to $\eta\Phi'$ and $\Psi$ in the Poisson equation and in the pressure equation, and that we use the solutions found for the monopoles and the velocities. Moreover, we neglect $\Theta_2$ and use $V_{\rm b} = V_{\rm \gamma} = V_{\rm c}$. 

Since the CDM and baryon isocurvature modes behave identically, it is convenient to define a single matter isocurvature mode. First, let us define:
\begin{equation}
	R_{\rm m} \equiv R_{\rm c} + R_{\rm b}\;.
\end{equation}
Then, let us define $S_{\rm m}$ such that:
\begin{equation}
	R_{\rm m} S_{\rm m} \equiv R_{\rm c} S_{\rm c} + R_{\rm b}S_{\rm b}\;.
\end{equation}
Moreover, since:
\begin{equation}
	R_{\rm m} = \frac{\rho_{\rm m}}{\rho_{\rm r}} = \frac{\Omega_{\rm m0}}{\Omega_{\rm r0}}a = \frac{\Omega_{\rm m0}}{\Omega_{\rm r0}}\sqrt{\Omega_{\rm r0}}H_0\eta\;,
\end{equation}
let us define:
\begin{equation}
	\tilde R_{\rm m} \equiv \frac{\Omega_{\rm m0}H_0}{\sqrt{\Omega_{\rm r0}}}\;,
\end{equation}
so that $R_{\rm m} = \tilde R_{\rm m}\eta$. Recall that the above calculations are valid deep into the radiation-dominated epoch, since we use $a \propto \eta$.

Doing so, we obtain:
\begin{align}
2\left(\eta\Phi' - \Psi\right)  = (-4\Phi &+ 4C_\gamma)\left(R_\gamma + R_\nu + \frac{3}{4}R_{\rm c} + \frac{3}{4}R_{\rm b}\right) + R_\nu S_\nu + \tilde R_{\rm m}S_{\rm m}\eta\;,\\
(k\eta)^2(\Phi + \Psi) &= -12R_\nu\mathcal{N}_2\;,\\
k\eta(-\eta\Phi' + \Psi) &= 2\left(R_\gamma + \frac{3}{4}R_{\rm c} + \frac{3}{4}R_{\rm b}\right)V_\gamma + 2R_\nu V_\nu\;,\\
	2\eta^2\Phi'' + 4\eta\Phi' &- 2\eta\Psi' + 2\Psi = (4\Phi - 4C_\gamma)(R_\gamma + R_\nu) - R_\nu S_\nu\;,
\end{align}
Since we are deep in the radiation-dominated epoch, we can write:
\begin{equation}
	R_\gamma + R_\nu + \frac{3}{4}R_{\rm c} + \frac{3}{4}R_{\rm b} = 1 - \frac{1}{4}R_{\rm m} \simeq 1 - \frac{1}{4}\tilde R_{\rm m}\eta\;,
\end{equation}
and we neglect the $\tilde R_{\rm m}\eta$ contribution with respect to 1. On the other hand, note that we cannot neglect the $\tilde R_{\rm m}S_{\rm m}\eta$ contribution with respect to the other entropy modes.

Using the above calculations and introducing the neutrino isocurvature velocity mode, the Einstein equations become:
\begin{align}
\label{largescaleradpoisson2} 2\left(\eta\Phi' - \Psi\right) + 4\Phi &= 4C_\gamma + R_\nu S_\nu + \tilde R_{\rm m}S_{\rm m}\eta\;,\\
\label{largescaleaniseq}(k\eta)^2(\Phi + \Psi) &= -12R_\nu\mathcal{N}_2\;,\\
\label{largescaleveleq}k\eta(-\eta\Phi' + \Psi) &= 2V_\nu - \frac{1}{2}R_\gamma(k\eta) S_\nu - 2R_\gamma q_\nu\;,\\
\label{largescalepresseq}	2\eta^2\Phi'' + 4\eta\Phi' &- 2\eta\Psi' + 2\Psi - 4\Phi = - 4C_\gamma - R_\nu S_\nu\;.
\end{align}
From the Poisson equation \eqref{largescaleradpoisson2}, we can determine $\Psi$ in favor of $\Phi$ and the constant primordial modes:
\begin{equation}\label{shPsiPhirel}
	\boxed{\Psi = \eta\Phi' + 2\Phi - 2C_\gamma - \frac{1}{2}R_\nu S_\nu - \frac{1}{2}\tilde R_{\rm m} S_{\rm m}\eta}
\end{equation}

\hrulefill

\begin{ex}
	Show that the Poisson equation \eqref{largescaleradpoisson2}, the velocity equation \eqref{largescaleveleq} and the pressure equation \eqref{largescalepresseq} are redundant. Differentiate the pressure equation \eqref{largescalepresseq} and, using the fact that $- 4C_\gamma - R_\nu S_\nu$ is time-independent, show that:
\begin{equation}
	\eta\Phi''' + 4\Phi'' - \Psi'' = 0\;.
\end{equation} 
The same equation can also be found from the Poisson equation \eqref{largescaleradpoisson2} and from the velocity equation \eqref{largescaleveleq}, by differentiating them twice. 
\end{ex}

\hrulefill

Now, let us eliminate $V_\nu$ from the velocity equation. From Eq~\eqref{Neutrinomultipole} for $\ell = 2$ we have:
\begin{equation}
	 \mathcal{N}_2' = \frac{2}{15}kV_\nu\;.
\end{equation}
Therefore, differentiating Eq.~\eqref{largescaleaniseq}, one obtains:
\begin{equation}\label{PhiPsi2eq}
	k[\eta^2(\Phi + \Psi)]' = -\frac{8}{5}R_\nu V_\nu\;.
\end{equation}
Obtaining $V_\nu$ from this equation and substituting it into the velocity equation \eqref{largescaleveleq} gives us another relation between $\Phi$ and $\Psi$:
\begin{equation}\label{shPsiPhirel2}
	\boxed{\left(1 - \frac{4}{5}R_\nu\right)\eta\Phi' + \eta\Psi' + 2\Phi + 2\left(1 + \frac{2}{5}R_\nu\right)\Psi = -\frac{2}{5}R_\gamma R_\nu\left(S_\nu + \frac{4q_\nu}{k\eta}\right)}
\end{equation}

\hrulefill

\begin{ex}
	Combine the two relations \eqref{shPsiPhirel} and \eqref{shPsiPhirel2} in order to obtain an evolution equation for $\Phi$ only:
	\begin{align}\label{eqprimordialPhi}
		\eta^2\Phi'' &+ 6\eta\Phi' +  2\left(3 + \frac{4}{5}R_\nu\right)\Phi =\nonumber\\
		4\left(1 + \frac{2}{5}R_\nu\right)C_\gamma &+ \frac{1}{5}\left(3 + 4R_\nu\right)R_\nu S_\nu + \left(\frac{3}{2} + \frac{2}{5}R_\nu\right)\tilde R_{\rm m} S_{\rm m}\eta - \frac{8}{5}R_\nu(1 - R_\nu)\frac{q_\nu}{k\eta}\;. 
	\end{align}
\end{ex}

\hrulefill

We can appreciate how all the primordial modes source the evolution of $\Phi$. 

\section{Solutions for the primordial modes}

We now look for a general solution of Eq.~\eqref{eqprimordialPhi}. As usual, one first obtains a general solution of the homogeneous equation and then adds a particular solution of the complete one.

Let us start with the homogeneous equation:
\begin{equation}
	\eta^2\Phi'' + 6\eta\Phi' + 2\left(3 + \frac{4}{5}R_\nu\right)\Phi = 0\;.
\end{equation}
The structure of this equation suggests a power law behavior of the solution, so let us employ the ansatz:
\begin{equation}
	\Phi \propto \eta^p\;.
\end{equation}
The homogeneous differential equation thus becomes an algebraic second order equation in the unknown $p$:
\begin{equation}
	p^2 + 5p + 2\left(3 + \frac{4}{5}R_\nu\right) = 0\;,
\end{equation}
whose solutions are:
\begin{equation}
	p_{1,2} = -\frac{5}{2} \pm \sqrt{1 - \frac{32}{5}R_\nu}\;.
\end{equation}
Since $R_\nu \approx 0.4$, $p_{2,3}$ are complex, we disregard these solutions.

The particular solution is immediately suggested by the structure of the source term:
\begin{equation}
	\Phi = c_1 + c_2\eta + \frac{c_3}{k\eta}\;,
\end{equation}
which, when substituted in the equation, gives us:
\begin{align}
		\frac{2c_3}{k\eta} &+ 6c_2\eta - \frac{6c_3}{k\eta} +  2\left(3 + \frac{4}{5}R_\nu\right)\left(c_1 + c_2\eta + \frac{c_3}{k\eta}\right) =\nonumber\\
		4\left(1 + \frac{2}{5}R_\nu\right)C_\gamma &+ \frac{1}{5}\left(3 + 4R_\nu\right)R_\nu S_\nu + \left(\frac{3}{2} + \frac{2}{5}R_\nu\right)\tilde R_{\rm m} S_{\rm m}\eta - \frac{8}{5}R_\nu(1 - R_\nu)\frac{q_\nu}{k\eta}\;. 
	\end{align}
From this equation, we can immediately read off the values of the integration constants $c_{1,2,3}$:
\begin{align}
	c_1 &= \frac{2\left(5 + 2R_\nu\right)}{15 + 4R_\nu}C_\gamma + \frac{\left(3 + 4R_\nu\right)R_\nu}{2\left(15 + 4R_\nu\right)}S_\nu\;,\\
	c_2 &= \frac{\left(15 + 4R_\nu\right)\tilde R_{\rm m}}{8\left(15 + 2R_\nu\right)}S_{\rm m}\;,\\
	c_3 &= \frac{4R_\nu\left(R_\nu - 1\right)}{5 + 4R_\nu}q_\nu\;.
\end{align}
Therefore, the general solution for $\Phi$ can be written as a linear combination of the primordial modes: 
\begin{equation}\label{genprimordialPhi}
	\Phi = \frac{2\left(5 + 2R_\nu\right)}{15 + 4R_\nu}C_\gamma + \frac{\left(3 + 4R_\nu\right)R_\nu}{2\left(15 + 4R_\nu\right)}S_\nu + \frac{\left(15 + 4R_\nu\right)\tilde R_{\rm m}}{8\left(15 + 2R_\nu\right)}S_{\rm m}\eta + \frac{4R_\nu\left(R_\nu - 1\right)}{5 + 4R_\nu}\frac{q_\nu}{k\eta}\;.
\end{equation}
Using the relation \eqref{shPsiPhirel}, we obtain for $\Psi$:
\begin{equation}
	\Psi = -\frac{10}{15 + 4R_\nu}C_\gamma + \frac{\left(-9 + 4R_\nu\right)R_\nu}{2\left(15 + 4R_\nu\right)}S_\nu + \frac{\left(-15 + 4R_\nu\right)\tilde R_{\rm m}}{8\left(15 + 2R_\nu\right)}S_{\rm m}\eta + \frac{4R_\nu\left(R_\nu - 1\right)}{5 + 4R_\nu}\frac{q_\nu}{k\eta}\;.
\end{equation}
The sum of the potentials is given by:
\begin{equation}
	\Phi + \Psi = \frac{4R_\nu}{15 + 4R_\nu}C_\gamma + \frac{\left(-3 + 2R_\nu\right)R_\nu}{\left(15 + 4R_\nu\right)}S_\nu + \frac{R_\nu\tilde R_{\rm m}}{\left(15 + 2R_\nu\right)}S_{\rm m}\eta + \frac{8R_\nu\left(R_\nu - 1\right)}{5 + 4R_\nu}\frac{q_\nu}{k\eta}\;.
\end{equation}
From this sum, one can straightforwardly determine the neutrino monopole from Eq.~\eqref{largescaleaniseq}:
\begin{equation}
	\mathcal{N}_2 = -\frac{(k\eta)^2(\Phi + \Psi)}{12R_\nu}\;.
\end{equation}
Knowing $\Phi$, the monopoles are given by:
\begin{equation}
	\delta_\gamma = -4\Phi + 4C_\gamma\;, \quad
	\delta_\nu = \delta_\gamma + S_\nu\;, \quad
	\delta_{\rm c} = \frac{3}{4}\delta_\gamma + S_{\rm c}\;, \quad
	\delta_{\rm b} = \frac{3}{4}\delta_\gamma + S_{\rm b}\;.
\end{equation}
The neutrino velocity is given by Eq.~\eqref{PhiPsi2eq}:
\begin{align}
	-V_\nu = \frac{5k}{8R_\nu}[\eta^2(\Phi + \Psi)]' = \nonumber\\
	\frac{5(k\eta)}{15 + 4R_\nu}C_\gamma + \frac{5\left(2R_\nu - 3\right)(k\eta)}{4\left(15 + 4R_\nu\right)}S_\nu + \frac{15\tilde R_{\rm m}(k\eta)}{8\left(15 + 2R_\nu\right)}S_{\rm m}\eta + \frac{5\left(R_\nu - 1\right)}{5 + 4R_\nu}q_\nu\;,
\end{align}
and the velocity of the other components is:
\begin{equation}
	V_\gamma = V_{\rm c} = V_{\rm b} = V_\nu - \frac{1}{4}(k\eta)S_\nu - q_\nu\;.
\end{equation}
Note that the neutrino isocurvature velocity mode is singular for the gravitational potentials and for the monopoles. Nonetheless, we have not discarded it because it can be shown \cite{Bucher:1999re} that such singularity is inherent to the choice of the Newtonian gauge and disappears in the synchronous gauge.

\section{Adiabaticity and isocurvature}

Now, let us understand the reason for the nomenclature ``adiabatic'' and ``isocurvature''.

\subsection{The adiabatic primordial mode}

The adiabatic mode is defined by $S_\nu = S_{\rm c} = S_{\rm b} = q_\nu = 0$. So, $C_\gamma$ is our only non-vanishing primordial mode. Then, we have:
\begin{eqnarray}\label{adiabaticPhip}
	\boxed{\Phi = \frac{2(5 + 2R_\nu)}{15 + 4R_\nu}C_\gamma\;, \quad \Psi = -\frac{10}{15 + 4R_\nu}C_\gamma\;, \quad \Phi = -\Psi\left(1 + \frac{2}{5}R_\nu\right)}
\end{eqnarray}
For the density contrasts:
\begin{equation}\label{deltagammarelPsiadiabatic}
	\boxed{\frac{4}{3}\delta_{\rm c} = \frac{4}{3}\delta_{\rm b} = \delta_\gamma = \delta_\nu = -2\Psi = \frac{20}{15 + 4R_\nu}C_\gamma}
\end{equation}
The primordial mode of the neutrino quadrupole moment is:
\begin{equation}
	 \boxed{\mathcal{N}_2 = -\frac{(k\eta)^2(\Phi + \Psi)}{12R_\nu} = -\frac{(k\eta)^2}{3(15 + 4R_\nu)}C_\gamma = \frac{(k\eta)^2}{30}\Psi}
\end{equation}
Since $q_\nu = 0$, we have the primordial modes for the velocities:
\begin{equation}\label{adiabaticprimordialmodevelocity}
	\boxed{V_\nu = V_\gamma = V_{\rm b} = V_{\rm c} = \frac{k\eta\Psi}{2} = -\frac{5k\eta}{15 + 4R_\nu}C_\gamma}
\end{equation}
All the above modes depend on the same constant $C_\gamma$, which is \textit{the} primordial perturbation, in the sense that its non-vanishing generates the fluctuations for all the matter components.

We will show in Chapter \ref{Chap:Inflation} how the inflationary mechanism is able to provide a value for $C_\gamma$. Note how the presence of neutrinos via $R_\nu$ prevents $\Phi = -\Psi$.

The $C_\gamma$ that we are using here corresponds to the $-2C$ of \cite{Ma:1995ey}, and in order to obtain the result of \cite{Bucher:1999re}, one has to set $C_\gamma = -1$.\footnote{Actually, doing so, $\Phi$ is not reproduced. However, there must be a typo in Eq. (28) of \cite{Bucher:1999re} because the gravitational potentials appear to be equal, and they cannot be equal because of the presence of neutrinos and the quadrupole moment of their distribution.} 

\subsubsection{Why ``adiabatic''?}

In order to understand the reason for the adjective ``adiabatic'', let us start with the familiar thermodynamic relation:
\begin{eqnarray}
	TdS = dU + pdV\;.
\end{eqnarray}
Considering:
\begin{equation}
	U = \rho_{\rm tot}V\;, \qquad V = N/n_{\rm tot}\;,
\end{equation}
we can rewrite it as follows:
\begin{equation}
	\frac{TdS}{V} = d\rho_{\rm tot} - \frac{\rho_{\rm tot} + P_{\rm tot}}{n_{\rm tot}}dn_{\rm tot}\;.
\end{equation}
Recast this equation as follows:
\begin{equation}
	\frac{TdS}{V} = \rho_{\rm c}\delta_{\rm c} + \rho_\gamma\delta_\gamma + \rho_{\rm b}\delta_{\rm b} + \rho_\nu\delta_\nu - \frac{\rho_{\rm tot} + P_{\rm tot}}{n_{\rm tot}}(\delta n_{\rm c} + \delta n_\gamma + \delta n_{\rm b} + \delta n_\nu)\;.
\end{equation}
Using:
\begin{equation}
	\delta_{\rm c} = \frac{\delta n_{\rm c}}{n_{\rm c}}\;, \qquad \delta_{\rm b} = \frac{\delta n_{\rm b}}{n_{\rm b}}\;, \qquad \delta_\gamma = \frac{4}{3}\frac{\delta n_\gamma}{n_\gamma}\;, \qquad \delta_\nu = \frac{4}{3}\frac{\delta n_\nu}{n_\nu}\;,
\end{equation}
we have:
\begin{align}\label{adiabaticpert}
	\frac{TdS}{V} &= \left(\rho_{\rm c} - n_{\rm c}\frac{\rho_{\rm tot} + P_{\rm tot}}{n_{\rm tot}}\right)\left(\delta_{\rm c} - \frac{3}{4}\delta_\gamma\right) + \left(\rho_{\rm b} - n_{\rm b}\frac{\rho_{\rm tot} + P_{\rm tot}}{n_{\rm tot}}\right)\left(\delta_{\rm b} - \frac{3}{4}\delta_\gamma\right)\nonumber\\
	&+ \left(\rho_\nu - \frac{3}{4}n_\nu\frac{\rho_{\rm tot} + P_{\rm tot}}{n_{\rm tot}}\right)\left(\delta_\nu - \delta_\gamma\right)\;. \qquad
\end{align}
Using the primordial modes of the monopoles, we obtain:
\begin{align}\label{adiabaticpert2}
	\frac{TdS}{V} &= \left(\rho_{\rm c} - n_{\rm c}\frac{\rho_{\rm tot} + P_{\rm tot}}{n_{\rm tot}}\right)S_{\rm c} + \left(\rho_{\rm b} - n_{\rm b}\frac{\rho_{\rm tot} + P_{\rm tot}}{n_{\rm tot}}\right)S_{\rm b}\nonumber\\
	&+ \left(\rho_\nu - \frac{3}{4}n_\nu\frac{\rho_{\rm tot} + P_{\rm tot}}{n_{\rm tot}}\right)S_\nu\;. \qquad
\end{align}
Therefore, when one of $S_\nu$, $S_{\rm c}$, or $S_{\rm b}$ is different from zero, we have entropy production. This is why these modes are also referred to as \textbf{entropy perturbations}. 

Note that $q_\nu$ does not contribute to the entropy and therefore one could consider it an adiabatic mode. On the other hand, as we are going to see, it is also an isocurvature mode.

\subsection{Isocurvature modes}

The terminology ``isocurvature'' refers to the fact that the gauge-invariant primordial curvature perturbations $\mathcal R$ and $\zeta$, defined in Eqs.~\eqref{Rperturb} and \eqref{zetaperturb}, are vanishing. In the Newtonian gauge, the curvature perturbations are expressed as follows:
\begin{equation}\label{Rzetaprimordial}
  \mathcal{R} = \Phi + \mathcal Hv_{\rm tot} \;, \quad \zeta = \Phi + \frac{R_\gamma\delta_\gamma + R_\nu\delta_\nu + R_{\rm c}\delta_{\rm c} + R_{\rm b}\delta_{\rm b}}{3 + R_\gamma + R_\nu}\;.
\end{equation}
In the above formula for $\mathcal R$, $v_{\rm tot}$ is not the sum of the single $v$'s. This happens because the SVT decomposition is performed not directly on $v_i$, but rather on the perturbed energy-momentum tensor $\delta T^0{}_i$. The latter is given by Eq.~\eqref{genT0imixed}:
\begin{equation}
	\delta T^0{}_i({\rm tot}) = (\rho_{\rm tot} + P_{\rm tot})v_i(\rm tot)\;.
\end{equation} 
Hence, $v_{\rm tot}$ is defined through a weighted average, with weight $\rho + P$:
\begin{equation}
	v_{\rm tot} = \frac{\sum_i (\rho_i + P_i)v_i}{\rho_{\rm tot} + P_{\rm tot}}\;.
\end{equation}
This is:
\begin{equation}
	v_{\rm tot} = \frac{4R_\gamma v_\gamma + 4R_\nu v_\nu + 3R_{\rm c}v_{\rm c} + 3R_{\rm b}v_{\rm b}}{3 + R_\gamma + R_\nu}\;.
\end{equation}
Now, using the relation $v = -V/k$, one obtains:
\begin{align}
	\mathcal{R} &= \Phi - \frac{4R_\gamma V_\gamma + 4R_\nu V_\nu + 3R_{\rm c}V_{\rm c} + 3R_{\rm b}V_{\rm b}}{(k/\mathcal H)(3 + R_\gamma + R_\nu)}\;,\\
	\zeta &= \Phi + \frac{R_\gamma\delta_\gamma + R_\nu\delta_\nu + R_{\rm c}\delta_{\rm c} + R_{\rm b}\delta_{\rm b}}{3 + R_\gamma + R_\nu}\;.
\end{align}
Using the velocity equation and the Poisson equation, one obtains:
\begin{align}
	\boxed{\mathcal{R} = \Phi + \frac{2(\Phi'/\mathcal H - \Psi)}{3 + R_\gamma + R_\nu}\;, \qquad
	\zeta = \Phi + \frac{2(\Phi'/\mathcal H - \Psi) + \frac{2}{3}(k/\mathcal H)^2\Phi}{3 + R_\gamma + R_\nu}}
\end{align}
These relations are completely general and show us that $\mathcal R = \mathcal \zeta$ on super-horizon scales, for $k \ll \mathcal H$. Note that this is true at any epoch.

Now we specify these relations for the radiation dominated epoch because we are looking for the primordial curvature perturbation. 

We can use the Eqs.~\eqref{Poisseqraddom}, \eqref{veleqraddom}, and $R_\gamma + R_\nu + R_{\rm c} + R_{\rm b} \approx R_\gamma + R_\nu \approx 1$ in order to obtain:
\begin{align}
	\mathcal{R} &= \Phi + \frac{1}{2}(\eta\Phi' - \Psi)\;,\\
	\zeta &= \Phi + \frac{1}{2}(\eta\Phi' - \Psi) + \frac{1}{6}(k\eta)^2\Phi\;.
\end{align}
Therefore, on very large scales ($k\eta \ll 1$), the two curvature potentials are equal:
\begin{equation}
	\mathcal{R} \simeq \zeta\;, \qquad (k\eta \ll 1)\;,
\end{equation}
and are equal to 4 times the left hand side of Eq.~\eqref{largescaleradpoisson2}.

Using the primordial modes for the density contrasts, we can establish that:
\begin{equation}\label{zetaprimordial2}
	\mathcal{R} = \zeta = C_\gamma + \frac{R_\nu S_\nu + \tilde R_{\rm m} S_{\rm m}\eta}{4}\;.
\end{equation}
For adiabatic perturbations, we have:
\begin{equation}\label{Rzetaprimordial2}
	\boxed{\mathcal R = \zeta = \Phi - \frac{\Psi}{2} = C_\gamma\;, \qquad \mbox{(adiabatic)}}
\end{equation}
Hence, the $C_\gamma$ is equal to the curvature perturbations. Since $C_\gamma$ is time-independent, Eq. \eqref{Rzetaprimordial2} also shows that $\mathcal{R}$ and $\zeta$ are conserved on large scales. We shall prove this important result more rigorously in Appendix \ref{App:Rconslargescales}. Actually, one can see that even if $S_\nu \neq 0$, that is, even if the primordial mode is not adiabatic, $\mathcal R = \zeta$ is conserved on large scales. This conservation is fundamental to creating a bridge between the very primordial universe (the initial conditions set by inflation) and the less primordial universe (the CMB sky).

Note that in the adiabatic case, it is not possible to have zero curvature; otherwise, all perturbations would vanish. Hence, adiabatic and isocurvature modes are independent primordial modes.

Since we have just established that $\mathcal R = \zeta$ on large scales, for the rest of the chapter we will work with $\zeta$ only.

\subsection{The neutrino density isocurvature primordial mode}
\index{Primordial modes!Neutrino density isocurvature}

Assume $S_{\rm m} = 0$ and $q_\nu = 0$. From Eq.~\eqref{zetaprimordial2} we have that: 
\begin{equation}
	\zeta = C_\gamma + \frac{R_\nu S_\nu}{4}\;.
\end{equation}
In order to make $\zeta = 0$, we need $4C_\gamma = -R_\nu S_\nu$. In this case, the potentials become:
\begin{equation}
	\Phi = \frac{-(1 - R_\nu)R_\nu}{\left(15 + 4R_\nu\right)}S_\nu\;, \qquad \Psi = \frac{-2\left(1 - R_\nu\right)R_\nu}{\left(15 + 4R_\nu\right)}S_\nu = 2\Phi\;.
\end{equation}
In order to obtain the same result as that of Eq.~(29) of \cite{Bucher:1999re}, one has to set $(1 - R_\nu)S_\nu = 1$.

\subsection{The CDM and baryons isocurvature primordial modes}
\index{Primordial modes!Baryon density isocurvature}\index{Primordial modes!Cold Dark Matter density isocurvature}

Suppose that, among the primordial modes, only $S_{\rm m} \neq 0$. Then, we have for $\zeta$:
\begin{equation}
	\zeta = \frac{\tilde R_{\rm m}S_{\rm m}\eta}{4}\;.
\end{equation}
This mode can be considered an isocurvature one since $\zeta \to 0$ as $\eta \to 0$.

The gravitational potentials are:
\begin{equation}
	\boxed{\Phi = \frac{\left(15 + 4R_\nu\right)\tilde R_{\rm m}}{8\left(15 + 2R_\nu\right)}S_{\rm m}\eta\;, \quad \Psi = \frac{\left(-15 + 4R_\nu\right)\tilde R_{\rm m}}{8\left(15 + 2R_\nu\right)}S_{\rm m}\eta\;, \quad \Phi = -\Psi\frac{15 + 4R_\nu}{15 - 4R_\nu}}
\end{equation}
Again, as we expected, the absence of neutrinos causes $\Phi = -\Psi$ since there is no other source of anisotropic stress. 

The results of \cite{Bucher:1999re} can be obtained by setting $\tilde R_{\rm m}S_{\rm m} = 4\Omega_{\rm m0}$.\footnote{Again, there seems to be a typo in Eqs.~(31) and (32) of \cite{Bucher:1999re}. Indeed, one expects to have equal (in modulus) potentials when neutrinos are absent. This does not happen in Eqs.~(31) and (32) of \cite{Bucher:1999re}.}

Note that one can have isocurvature perturbations also if $\tilde R_{\rm m}S_{\rm m} = 0$, i.e., if the entropy perturbations of CDM and baryons compensate. This mode is called ``compensated isocurvature perturbations'' and was introduced and discussed in \cite{Grin:2011tf}.

\subsection{The neutrino velocity isocurvature primordial mode}
\index{Primordial modes!Neutrino velocity isocurvature}

When we set $C_\gamma = S_\nu = S_{\rm m} = 0$, we automatically have:
\begin{equation}
	\zeta = 0\;.
\end{equation}
So, these modes are naturally isocurvature modes. The potentials are identical:
\begin{equation}
	\Phi = \Psi = -\frac{4R_\nu\left(1 - R_\nu\right)}{5 + 4R_\nu}\frac{q_\nu}{k\eta}\;.
\end{equation}
The solution of \cite{Bucher:1999re} can be found by setting $(1 - R_\nu)q_\nu = 1/3$.\footnote{Again, there is a typo in Eq.~(30) of \cite{Bucher:1999re}. Indeed, using the formulae from their Eq.~(27) applied to their Eq.~(26), one does not find Eq.~(30). The denominator of Eq.~(30) should be $3(5 + 4R_\nu)$ instead of $15 + 4R_\nu$.}

\section{Planck constraints on isocurvature modes}
\index{Primordial modes!Planck constraints}

The most up-to-date constraints on the nature of the primordial modes are given by the Planck collaboration and favor adiabaticity; see \cite{Planck:2018jri}. In fact, as we will see, isocurvature perturbations predict temperature fluctuations in the CMB that are out of phase with respect to the adiabatic case.

Moreover, if we use the general primordial mode for $\Phi$, given in Eq.~\eqref{genprimordialPhi}, in order to compute the primordial spectrum:\footnote{The reader is referred to chapter \ref{Chap:StochasticPropertiesofCP} for a definition of the power spectrum. Note that we use the notation of \cite{Planck:2018jri} in this section.}
\begin{equation}
	\langle \Phi(\mathbf k_1)\Phi^*(\mathbf k_2)\rangle = (2\pi)^3\mathcal P_{\Phi\Phi}(k)\delta^{(3)}(\mathbf k_1 - \mathbf k_2)\;,
\end{equation}
it is clear that correlations among the primordial modes (isocurvature and adiabatic) will also appear.

In \cite{Planck:2018jri}, the constraints on the primordial modes are considered at two scales:
\begin{equation}
	k_1 = k_{\rm low} = 0.002\mbox{ Mpc}^{-1}\;, \qquad k_2 = k_{\rm high} = 0.1\mbox{ Mpc}^{-1}\;,
\end{equation}
so, for the power spectra:
\begin{equation}
	\mathcal P^{(1)}_{\mathcal R\mathcal R}\;, \mathcal P^{(2)}_{\mathcal R\mathcal R}\;, \mathcal P^{(1)}_{\mathcal I\mathcal I}\;, \mathcal P^{(2)}_{\mathcal I\mathcal I}\;,
\end{equation}
where the superscript refers to the chosen scale and $\mathcal I$ to the isocurvature contribution. For the correlation spectrum, only the first scale is chosen; thus, the constraints are given for:
\begin{equation}
	\mathcal P^{(1)}_{\mathcal R\mathcal I}\;.
\end{equation}
The amount of isocurvature is then identified with the following quantities:
\begin{equation}
	\cos\Delta = \frac{\mathcal P_{\mathcal R\mathcal I}}{(\mathcal P_{\mathcal R\mathcal R}\mathcal P_{\mathcal I\mathcal I})^{1/2}}\;, \qquad \beta_{\rm iso}(k) = \frac{\mathcal P_{\mathcal I\mathcal I}(k)}{[\mathcal P_{\mathcal R\mathcal R}(k) + \mathcal P_{\mathcal I\mathcal I}(k)]}\;, 
\end{equation}
and from the fraction of non-adiabatic CMB temperature variance:
\begin{equation}
	\alpha_{\rm non-adi} = 1 - \frac{(\Delta T)_{\mathcal R\mathcal R}^2(\ell = 2,2500)}{(\Delta T)_{\rm tot}^2(\ell = 2,2500)}\;, \quad (\Delta T)_{X}^2(\ell = 2,2500) = \sum_{\ell = 2}^{2500}(2\ell + 1)C_{X,\ell}^{TT}\;.
\end{equation}
The main results are given in Tab. 14 and Fig. 38 of \cite{Planck:2018jri}. Summarizing here:
\begin{equation}
	\beta_{\rm iso}(k_{\rm low}) < 2.5\%, 7.4\%, 6.8\%\;,  
\end{equation} 
for the matter, the neutrino density and the neutrino velocity isocurvature modes, respectively. These results are given at 95\% CL. For the fraction of non-adiabatic CMB temperature variance, one obtains:
\begin{equation}
	|\alpha_{\rm non-adi}| < 1.7\%\;,
\end{equation}
for all three of the above cases, again at 95\% CL.

\clearpage
\chapter{Stochastic Properties of Cosmological Perturbations}\label{Chap:StochasticPropertiesofCP}

{\rightskip=3truepc\leftskip=3truepc\noindent
{\it The more the universe seems comprehensible, the more it also seems pointless}
\vskip 0.10 in
\centerline{\it ---Steven Weinberg, The first three minutes}
\vskip 0.20 in
}

The attentive reader must have noticed that, in fact, we have given no actual initial conditions on cosmological perturbations in Chapter~\ref{Chap:IC}. What we have done is provide the form of the primordial modes and demonstrate how, in the 5 different cases that we investigated, all scalar perturbations are sourced by a single scalar potential. For example, in the adiabatic case, we showed in Eq.~\eqref{adiabaticPhip} how $\Phi(\mathbf k)$ is related to $C_\gamma(\mathbf k)$, which we proved to be equal to $\zeta(\mathbf k)$ and $\mathcal R(\mathbf k)$.

We have already stressed the following important point: all the evolution equations that we derived for the perturbations, i.e., the evolution equations for photons, neutrinos, CDM, and baryons, plus the Einstein equations, depend only on the modulus $k$ and not on $\mathbf k$. This means that only the initial conditions can depend on the latter. We could solve the evolution equation for some initial condition in which $\mathcal R(\eta_i,\mathbf k) = 1$ is normalized to unity, and then the result will depend only on $k$: $\mathcal R(\eta, k)$. This is usually called \textbf{the transfer function}.\index{Transfer function} Multiplying it by the true initial condition $\mathcal R(\eta_i,\mathbf k)$, we obtain the correct result. Of course, this reasoning applies to any perturbation, not only to $\mathcal R$.

In this chapter, we discuss a very important point of cosmology, both from the viewpoint of theory and especially from the observational one: the stochastic nature of cosmological perturbations embedded in the stochastic character of the initial conditions. This means that $\mathcal R(\eta_i,\mathbf k)$ shall be regarded as a random variable.

This chapter mainly follows \cite{Weinberg:2008zzc} and \cite{Lyth:2009zz}.

\section{Stochastic cosmological perturbations and power spectrum} 

First of all, what do cosmological perturbations have to do with statistics? After all, we have derived differential equations describing the evolutions of the perturbative variables, and, provided that we are able to solve them, we should determine such variables exactly, with no room for randomness. 

This is true, but consider the following point: the differential equations that we have built are ordinary differential equations that describe only the time evolution of the perturbative variables, whereas their spatial (or scale $\mathbf{k}$) dependencies must be provided by the initial conditions. 

For example, focus on the CDM density contrast $\delta_{\rm c}(\eta,\textbf{x})$, for which the evolution equations Eqs.~\eqref{deltaequationCDM} and \eqref{VequationCDM} are the simplest:
\begin{eqnarray}
	\delta_{\rm c}' + kV_{\rm c} + 3\Phi' = 0\;, \qquad V_{\rm c}' + \mathcal HV_{\rm c} - k\Psi = 0\;.
\end{eqnarray}
These equations depend only on $k$; therefore, the $\mathbf k$-dependence arises from the initial condition, which we indeed know in the adiabatic case from Eq.~\eqref{deltagammarelPsiadiabatic} to be:
\begin{equation}
	\delta_{\rm c}(\mathbf k) = \frac{15}{15 + 4R_\nu}\zeta(\mathbf k)\;.
\end{equation}
Do we have such initial conditions? In other words, do we know the scale dependence of the primordial curvature perturbation? Not exactly. According to our current understanding, the universe had a quantum origin, which we are not yet able to describe in detail due to the lack of a quantum theory of gravity and because we have no experiments that penetrate the trans-Planckian scales, which allow us to see what is happening there.

The evolution equations that we found become valid well after the Planck scale, when gravity is safely classical. However, the initial conditions that we use come from a quantum phase, and as such, they carry a probabilistic feature. For example, we shall investigate the inflationary paradigm, and we shall see that it provides a prediction not on $\zeta(\textbf{k})$ itself but rather on its probability distribution. In particular, if such a distribution is Gaussian, then all the information is contained in its variance, which is called \textbf{the power spectrum}.\index{Power spectrum}

There is also another important point that stresses how useful it is to treat cosmological perturbations as random variables. Observationally, it is impossible to determine $\delta(\eta,\textbf{x})$ for any given time because we receive signals from our past light-cone only. It is impossible, therefore, to recover the initial $\zeta(\textbf{x})$ from what we observe. Even if we were able to concoct a theory predicting the initial $\zeta(\textbf{x})$, we would not be able to fully test it since we cannot determine $\delta(\eta,\textbf{x})$ completely from observation.

Moreover, determining exactly $\delta(\eta,\textbf{x})$ involves an enormous amount of information. For example, it would mean that we are able to predict the position of a certain galaxy at a specific time. Though this might be interesting to some extent, we are more concerned with averaged quantities, such as the average distance among galaxies, because these contain information about gravity and the expanding universe.

Hence, by promoting (or rather demoting) $\delta(\eta,\textbf{x})$ to a random variable, we are then able to infer information about its expectation value, variance, or other higher-order correlations from the observations limited to our past light-cone. This allows us to make contact with the \textbf{descriptive statistics} that we can apply to the distribution of structures in the sky.

\section{Random fields}\index{Random fields}

We provide a very simple and non-rigorous introduction to random fields, with an emphasis on their application in cosmology. For more details, see \cite{chung2020introduction} and references therein.

A (real) function $G(\mathbf x)$ is a random field if, for any $\mathbf{x}$, its values are realizations of a stochastic variable. Moreover, for any $n$, we can define distribution functions of the following form:
\begin{equation}\label{DistributionfunctionsFG}
	F_{1,2,\dots,n}(g_1, g_2, \dots,g_n) = F[G(\mathbf x_1) < g_1, G(\mathbf x_2) < g_2, \dots, G(\mathbf x_n) < g_n]\;,
\end{equation}
We denote by capital $G$ the random field, which has $\mathbf x$ dependence, and by $g$ a certain value that it can assume at a given $\mathbf x$ among all the possible values that form the \textbf{ensemble}. The subscripts in $g_1, g_2, \dots, g_n$ refer to the different points $\mathbf x_1, \mathbf x_2, \dots,\mathbf x_n$ in which the random field is evaluated.   

At a given point $\mathbf x_1$, a probability density function exists:
\begin{equation}
	p_1(g_1)dg_1\;,
\end{equation}
describing how probable it is for $G$ to assume a certain value $g_1$ in $\mathbf x_1$. In terms of the distribution function, $p(g_1)$ is defined as:
\begin{equation}
	p_1(g_1) = \frac{dF_1(g_1)}{dg_1}\;.
\end{equation}
Since $F$ is a cumulative probability, then $F_1(-\infty) = 0$ and $F_1(\infty) = 1$. In cosmology, $G(\mathbf x)$ is a perturbative quantity, such as the density contrast $\delta(\mathbf x)$. 

The expectation value of the random field at $\mathbf{x}_1$ is defined via an \textbf{ensemble average}\index{Ensemble average}:
\begin{equation}\label{expvalG}
	\boxed{\langle G(\mathbf x_1)\rangle \equiv \int_{\Omega_1} g_1p_1(g_1)dg_1}
\end{equation}
where $\Omega_1$ denotes the ensemble of values available at $\mathbf{x}_1$.

In general, one has:
\begin{equation}
	p_1(g_1) \neq p_2(g_2)\;,
\end{equation}
that is, the probability distribution of the values which $G$ may assume in $\mathbf x_1$ is different from that of the values which $G$ may assume in $\mathbf x_2$. Therefore, in general, the expectation value of $G$ is dependent on the position $\mathbf{x}$.

A very important simplification occurs when the probability of the realization of the random field is translationally invariant. In this case, the random field is said to be \textbf{statistically homogeneous}, and for statistically homogeneous random fields, the expectation value in Eq.~\eqref{expvalG} is independent of the position:
\begin{equation}\label{expvalGhom}\index{Statistical homogeneity}
	\boxed{\langle G\rangle \equiv \int_\Omega gp(g)dg}
\end{equation}
The value of the product of two random fields $G(\mathbf x_1)G(\mathbf x_2)$ will be described by a distribution function $p_{12}(g_1,g_2)$ of the random variables $g_1$ and $g_2$, for which, in general, one has:
\begin{equation}
	p_{12}(g_1,g_2) \neq p_1(g_1)p_2(g_2)\;,
\end{equation}
unless the realizations are independent, in which case the random process is said to be \textbf{Poissonian}. The 2-dimensional probability density allows us to define the \textbf{2-point correlation function} as follows:
\begin{equation}\label{2pointcorrelationfunction}\index{Correlation function}
	\boxed{\xi(\mathbf x_1, \mathbf x_2) \equiv \langle G(\mathbf x_1)G(\mathbf x_2)\rangle \equiv \int_{\Omega_1\times\Omega_2} g_1g_2 p_{12}(g_1,g_2)dg_1dg_2}
\end{equation}
In a similar fashion, one can define $N$-point correlation functions as:
\begin{eqnarray}
	\xi^{(N)}(\mathbf x_1, \mathbf x_2, \dots,\mathbf x_N) \equiv \langle G(\mathbf x_1)G(\mathbf x_2)\dots G(\mathbf x_N)\rangle \nonumber\\ 
	\equiv \int_{\Omega_1\times\Omega_2\times\dots\Omega_N} g_1g_2\dots g_N p_{12\dots N}(g_1,g_2,\dots,g_N)dg_1dg_2\dots dg_N\;.
\end{eqnarray}
Things are much easier if the random field is statistically homogeneous. In this case, the correlation function can depend only on the differences between the coordinates. For example, for the 2-point correlation function, one has: 
\begin{equation}
	\xi(\mathbf x_1, \mathbf x_2) = \xi(\mathbf x_1 - \mathbf x_2)\;.
\end{equation}
Moreover, consider a rotation matrix $R$ and define $\mathbf x_{R1} = R\mathbf x_1$. The random field is \textbf{statistically isotropic} if:
\begin{equation}
	p_1(g_1) = P_{R1}(g_{R1}) \qquad \forall R\;.
\end{equation}
This means that the probability of the realization is rotationally invariant.\index{Statistical isotropy}

If the random field is statistically homogeneous and statistically isotropic, then the 2-point correlation function has the following property: 
\begin{equation}
	\xi(\mathbf x_1, \mathbf x_2) = \xi(\mathbf x_1 - \mathbf x_2) = \xi(r_{12})\;,
\end{equation}
where $r_{12} =|\mathbf x_1 - \mathbf x_2|$. That is, the 2-point correlation function depends only on the distance between the two points. This is the standard case of cosmology.

The \textbf{ensemble variance}\index{Ensemble variance} of the random field is defined as:
\begin{equation}\label{variancerandomfield}
	\boxed{\sigma^2(\mathbf x_1, \mathbf x_2) \equiv \langle G(\mathbf x_1)G(\mathbf x_2)\rangle - \langle G(\mathbf x_1)\rangle\langle G(\mathbf x_2)\rangle}
\end{equation}
Therefore, if the random field is statistically homogeneous and isotropic, we obtain:
\begin{equation}
	\sigma^2(r_{12}) = \xi(r_{12}) - \langle G\rangle^2\;,
\end{equation}
where we recall that $\langle G\rangle^2$ does not depend on the position. Note that if the random process is Poissonian, i.e., $p_{12}(g_1,g_2) = p_1(g_1)p_2(g_2)$, then the variance is zero since:
\begin{equation}
	\xi(r_{12}) = \langle G\rangle^2\;.
\end{equation}
If $G$ is a variable representing the distribution of galaxies, we do not expect it to be a Poisson random variable because gravity affects the likelihood of the galaxies being closer to each other rather than farther apart. Hence, we attribute deviations from Poissonian behavior to gravity, and for this reason, it is useful to study the 2-point correlation function. 

From observation, we are able to compute only spatial averages of the field $G(\mathbf x)$:
\begin{equation}
	\bar{G} \equiv \frac{1}{V}\int_Vd^3\mathbf x\;G(\mathbf x)\;.
\end{equation}
What can we say about $X \equiv \bar{G} - \langle G\rangle$? We use this as an estimator of the error we are making by exchanging the ensemble average with the spatial average. 

\hrulefill

\begin{ex} Show that:
\begin{equation}
	\langle X\rangle = \frac{1}{V}\int_Vd^3\mathbf x\;\langle G(\mathbf x)\rangle - \langle G\rangle = 0\;,
\end{equation}
and the variance is:
\begin{equation}
	\langle X^2\rangle = \frac{1}{V^2}\int_Vd^3\mathbf x_1\int_Vd^3\mathbf x_2\langle G(\mathbf x_1)G(\mathbf x_2)\rangle - \langle G\rangle^2\;.
\end{equation}
\end{ex}

\hrulefill

Assuming statistical homogeneity, we can then obtain:
\begin{equation}
	\langle X^2\rangle = \frac{1}{V^2}\int_Vd^3\mathbf x_1\int_Vd^3\mathbf x_2\;\xi(\mathbf x_1 - \mathbf x_2) - \langle G\rangle^2\;.
\end{equation}
We see that for a Poisson process, this variance is vanishing. In fact, if the galaxies are randomly distributed, any volume is a good sample of the realization.

The above integral can be written as
\begin{equation}
	\langle X^2\rangle = \frac{1}{V}\int_Vd^3\mathbf r\;\xi(\mathbf r) - \langle G\rangle^2\;.
\end{equation}
In Sec.~\ref{Sec:ergodictheorem}, we show that this variance goes to zero if $V \to \infty$, a fact known as \textbf{the ergodic theorem}, which is based on a couple of reasonable assumptions.\index{Ergodic theorem} On the other hand, in practice, the volume is finite and thus $\langle X^2\rangle$ is, in general, different from zero. In cosmology, it is called \textbf{cosmic variance}.\index{Cosmic variance} Assuming statistical isotropy in the above integral and using a spherical volume for convenience, we can write:
\begin{equation}\label{cosmicvarianceformula}
	\langle X^2\rangle = \frac{3}{R^3}\int_0^Rdr\; r^2\xi(r) - \langle G\rangle^2\;.
\end{equation}

\section{The Fourier transform of a random field and the power spectrum}

We have found it useful to work with the Fourier transform of perturbed cosmological quantities. If we perform the Fourier transform of a random field $G(\mathbf x)$:
 \begin{equation}
	G(\mathbf x) = \int \frac{d^3\mathbf k}{(2\pi)^3}\; \tilde{G}(\mathbf k)e^{i\mathbf{k}\cdot\mathbf x}\;, \qquad \tilde{G}(\mathbf k) = \int d^3\mathbf x\; G(\mathbf x)e^{-i\mathbf{k}\cdot\mathbf x}\;,
\end{equation}
we expect $\tilde{G}(\mathbf k)$ to be a random field as well. Its expectation value is then:
\begin{align}
    \langle\tilde{G}(\mathbf k)\rangle = \int d^3\mathbf x\; \langle G(\mathbf x)\rangle e^{-i\mathbf{k}\cdot\mathbf x}\;,
\end{align}
that is, the Fourier transform of the expectation value $\langle G(\mathbf x)\rangle$. For statistical homogeneity, this function is a constant. The Fourier transform of a constant must be understood in the sense of distributions, and it is a Dirac delta function:
\begin{align}
    \langle\tilde{G}(\mathbf k)\rangle = \langle G\rangle (2\pi)^3\delta^{(3)}(\mathbf{k})\;.
\end{align}
If $G(\mathbf x)$ is a real field, then $\tilde{G}(-\textbf{k}) = \tilde{G}^*(\textbf{k})$ because of the following: 

\hrulefill

\begin{ex} Compare:
\begin{equation}
	G(\textbf{x}) = \int \frac{d^3\mathbf k}{(2\pi)^3}\;\tilde{G}(\textbf{k})e^{i\textbf{k}\cdot\textbf{x}}\;, \qquad G^*(\textbf{x}) = \int \frac{d^3\mathbf k}{(2\pi)^3}\;\tilde{G}^*(\textbf{k})e^{-i\textbf{k}\cdot\textbf{x}}\;.
\end{equation}
The two expressions are equal if $G(\mathbf x)$ is real. Therefore, show that $\tilde{G}(-\textbf{k}) = \tilde{G}^*(\textbf{k})$.
\end{ex}

\hrulefill

The relation $\tilde{G}(-\textbf{k}) = \tilde{G}^*(\textbf{k})$ is called \textbf{the reality condition}\index{Reality condition}, and we shall exploit it in the following chapters for predictions on the observed spectra.\index{Reality condition}

\subsection{Power spectrum}

The power spectrum is the Fourier transform of the 2-point correlation function:
\begin{equation}
	\langle \tilde{G}(\mathbf k)\tilde{G}(\mathbf k')\rangle = \int d^3\mathbf x\int d^3\mathbf x' \langle G(\mathbf x)G(\mathbf x')\rangle e^{-i\mathbf{k}\cdot\mathbf x}e^{-i\mathbf{k}'\cdot\mathbf x'}\;.
\end{equation}
The two Fourier modes $\mathbf{k}$ and $\mathbf{k}'$ become uncorrelated if we assume statistical homogeneity. In fact:
\begin{equation}
	\langle \tilde{G}(\mathbf k)\tilde{G}(\mathbf k')\rangle = \int d^3\mathbf x\int d^3\mathbf x' \xi_G(\mathbf x' - \mathbf x) e^{-i\mathbf{k}\cdot\mathbf x}e^{-i\mathbf{k}'\cdot\mathbf x'}\;,
\end{equation}
By changing the variable from $\mathbf x'$ to $\mathbf z = \mathbf x' - \mathbf x$, we get:
\begin{equation}
	\langle \tilde{G}(\mathbf k)\tilde{G}(\mathbf k')\rangle = \int d^3\mathbf x\;e^{-i(\mathbf{k}+\mathbf k')\cdot\mathbf x}\int d^3\mathbf z \;\xi_G(\mathbf z) e^{-i\mathbf{k}'\cdot\mathbf z}\;.
\end{equation} 
Now, using the representation of the Dirac delta, we obtain:
\begin{equation}\label{Gaussian2correlator}
	\boxed{\langle \tilde{G}(\mathbf k)\tilde{G}(\mathbf k')\rangle = (2\pi)^3\delta^{(3)}(\textbf{k} + \textbf{k}')P_G(\mathbf{k})}
\end{equation}
where:
\begin{equation}
	\boxed{P_G(\mathbf{k}) \equiv \int d^3\mathbf x \;\xi_G(\mathbf x) e^{-i\mathbf{k}\cdot\mathbf x}}
\end{equation}
is the quantity that is typically defined as the \textbf{power spectrum}. If statistical isotropy is also assumed, one has:
\begin{equation}
	P_G(k) = 2\pi\int_0^\infty dr\; r^2 \;\xi_G(r) \int_{-1}^1du\; e^{-ikru}\;,
\end{equation}
where $u$ is the cosine of the angle between $\mathbf k$ and $\mathbf x$, which we have used as the integration variable. Thanks to statistical isotropy, the power spectrum depends only on the modulus of $k$. Performing the $u$ integration, one gets:
\begin{equation}
	P_G(k) = 4\pi\int_0^\infty dr\; r^2 \;\xi_G(r) \frac{\sin(kr)}{kr}\;.
\end{equation}

\section{Gaussian random fields}\index{Random field}

Gaussian random fields are such that the $g$'s are distributed as a Gaussian:
\begin{equation}\label{defgaussmode}
	p(g) = \frac{1}{\sqrt{2\pi}\sigma_g}e^{-g^2/2\sigma^2_g}\;,
\end{equation}
where we are assuming a vanishing expectation value, and the variance has been defined in Eq.~\eqref{variancerandomfield}. If statistical homogeneity holds, the above distribution is the same at any spatial point.

In the literature, some authors define Gaussian random fields by their uncorrelated Fourier modes; that is, precisely by Eq.~\eqref{Gaussian2correlator}. However, we have seen that it is statistical homogeneity that implies Eq.~\eqref{Gaussian2correlator}. A Gaussian random field is defined independently of statistical homogeneity and isotropy via Eq. \eqref{defgaussmode}. On the other hand, the superposition of uncorrelated Fourier modes might give rise to a Gaussian random field if the hypotheses of the central limit theorem are satisfied.\footnote{For example, the most restrictive hypotheses of the theorem are that the modes must be independent, with the same expectation value and the same variance}. In this case, statistical homogeneity does imply Gaussianity.\footnote{The thread \url{https://physics.stackexchange.com/q/9106/167327} is particularly useful for understanding these points.}

If $G(\mathbf{x})$ is a Gaussian random field, then:
\begin{align}
    G(\mathbf{x}_1)G(\mathbf{x}_2)\dots G(\mathbf{x}_N)\,,
\end{align}
will draw its values from a $N$-dimensional Gaussian distribution. If we assume spatial homogeneity, the Gaussian distribution is the same for all $\mathbf{x}$; thus, the Fourier transform $\tilde G(\mathbf{k})$ is also a Gaussian random variable.

In cosmology, Gaussian random fields are characterized by vanishing correlation functions of odd order; that is:
\begin{equation}
	\langle \tilde{G}(\mathbf k)\rangle = \langle \tilde{G}(\mathbf k_1)\tilde{G}(\mathbf k_2)\tilde{G}(\mathbf k_3)\rangle = \dots = 0\;.
\end{equation}
All the correlation functions of even order can be expressed in terms of the second-order correlation function; that is, in terms of the power spectrum, which contains all the information about the random field. 

For example, the 4-point correlation function can be expressed as follows, using Wick's theorem:
\begin{eqnarray}\label{4pointcorrelator}
	\langle \tilde{G}(\mathbf k_1)\tilde{G}(\mathbf k_2)\tilde{G}(\mathbf k_3)\tilde{G}(\mathbf k_4)\rangle = \langle \tilde{G}(\mathbf k_1)\tilde{G}(\mathbf k_2)\rangle\langle \tilde{G}(\mathbf k_3)\tilde{G}(\mathbf k_4)\rangle 
	\nonumber\\ + \langle \tilde{G}(\mathbf k_1)\tilde{G}(\mathbf k_3)\rangle\langle \tilde{G}(\mathbf k_2)\tilde{G}(\mathbf k_4)\rangle 
+ \langle \tilde{G}(\mathbf k_1)\tilde{G}(\mathbf k_4)\rangle \langle \tilde{G}(\mathbf k_2)\tilde{G}(\mathbf k_3)\rangle\;.
\end{eqnarray}
As we mentioned, a typical random variable in cosmology is the density contrast $\delta$. The latter is taken to be a Gaussian random variable, although $\delta$ can take values in $[-1,\infty)$, where $-1$ represents a void, and not in $(-\infty,\infty)$, as a Gaussian random variable does. 

\subsection{Estimator of the power spectrum and cosmic variance}

Theoretically, we shall provide predictions on $P_G(k)$, whereas observationally we determine $\xi_G(r)$. Starting from Eq.~\eqref{2pointcorrelationfunction}, we have:
\begin{equation}\label{xieqaverage}
	\xi_G(r) = \langle G(\textbf{x})G(\textbf{x} + \textbf{r})\rangle = \int \frac{d^3\textbf{k}}{(2\pi)^3}\int \frac{d^3\textbf{k}'}{(2\pi)^3}\langle\tilde G(\textbf{k})\tilde G^*(\textbf{k}') e^{i\textbf{k}\cdot\textbf{x} - i\textbf{k}'\cdot(\textbf{x} + \textbf{r})}\rangle\;.
\end{equation}
We have included the Fourier modes in the average in order to investigate the difference between performing the spatial average and the ensemble average.

\paragraph{Performing the ensemble average.} Using Eq.~\eqref{Gaussian2correlator} in Eq.~\eqref{xieqaverage}, one obtains:
\begin{equation}
	\xi_G(r) = \int \frac{d^3\textbf{k}}{(2\pi)^3}\int d^3\textbf{k}'P_G(k)\delta^{(3)}(\textbf{k} - \textbf{k}')e^{i\textbf{k}\cdot\textbf{x} - i\textbf{k}'\cdot(\textbf{x} + \textbf{r})}\;.
\end{equation}
This gives:
\begin{equation}\label{xiensembleqaverage}
	\xi_G(r) = \int \frac{d^3\textbf{k}}{(2\pi)^3}P_G(k)e^{-i\textbf{k}\cdot\textbf{r}}\;.
\end{equation}
Since $\xi$ is dimensionless, $P_G(k)$ has dimensions of a volume. Working out the angular integration, we get:
\begin{equation}
	\xi_G(r) = \int \frac{dk\;k^2}{(2\pi)^2}P_G(k)\int_{-1}^1du\;e^{-ikru}\;,
\end{equation}
where again $u$ is the cosine of the angle between $\textbf{k}$ and $\textbf{r}$. Integrating in $u$, we get:
\begin{equation}\label{xiPrel}
	\boxed{\xi_G(r) = \int_0^{\infty} dk\;\frac{k^2P_G(k)}{2\pi^2}\frac{\sin(kr)}{kr}}
\end{equation}
It is customary to define a dimensionless power spectrum as follows:
\begin{equation}\label{dimensionlesspowerspectrum}
	\boxed{\Delta^2_G(k) \equiv \frac{k^3P_G(k)}{2\pi^2}}
\end{equation}\index{Power spectrum!Dimensionless}
so that Eq.~\eqref{xiPrel} becomes:
\begin{equation}\label{corrfunDelta}
	\boxed{\xi_G(r) = \int_0^{\infty} \frac{dk}{k}\;\Delta^2_G(k)\frac{\sin(kr)}{kr}}
\end{equation}

\paragraph{Performing the spatial average.} Suppose that none of the quantities in the integrand of Eq.~\eqref{xieqaverage} is a stochastic variable and that $\langle\dots\rangle$ is a spatial average. We must rewrite $\xi_G(r)$ as follows:
\begin{equation}\label{doingthespatialaverage}
	\hat\xi_G(r) = \frac{1}{V}\int_{V}d^3\mathbf x \sum_{n,m}\frac{1}{V^2}G_n G^*_me^{i\textbf{k}_n\cdot\textbf{x} - i\textbf{k}_m\cdot(\textbf{x} + \textbf{r})}\;, \quad G_n = \tilde G(\mathbf k_n)\;, \quad \mathbf k_n = \frac{2\pi}{L}\mathbf n\;,
\end{equation}
where we have used the Fourier series expansion because we are in a finite volume, which we have chosen to be a cube of side $L$, and hence $V = L^3$. This volume can be thought of as the survey volume. Note that the summation over $n$ and $m$ is intended as a sum over the three components of $\mathbf n$ and $\mathbf m$, since we cannot use statistical isotropy at this time. The spatial integration gives:
\begin{equation}
	\frac{1}{L^3}\int_V d^3\textbf{x}\;e^{i(\textbf{k}_n - \textbf{k}_m)\cdot\textbf{x}} = \delta_{nm}\;.
\end{equation}

\hrulefill

\begin{ex}
	Show in one dimension, for simplicity, that:
	\begin{equation}
		\frac{1}{L}\int_{-L/2}^{L/2} dx\;e^{i(k_n - k_m)x} = \frac{2\sin[(k_n-k_m)L/2]}{(k_n - k_m)L}\;.
	\end{equation}
\end{ex}
Since $k$ is quantized, the difference $k_n - k_m$ is always a multiple of $2\pi/L$, thereby making the sine vanish. The only exception is when the difference is zero, that is, for $n = m$.

\hrulefill

So, we have the following for the correlation function:
\begin{equation}
	\hat\xi_G(r) = \sum_{n}\frac{1}{V^2}|G_n|^2e^{-i\textbf{k}_n\cdot\textbf{r}}\;,
\end{equation}
from which we infer the power spectrum as:
\begin{equation}
	P_n \equiv P(\mathbf k_n) = \frac{|G_n|^2}{V}\;.
\end{equation}
Since the survey volume is finite, the smallest wavenumber that we can construct is $2\pi/L$, and the smallest cell in the wavenumber space thus has a volume of $(2\pi/L)^3$.

\hrulefill

\begin{ex}
	Prove that the number of independent modes between $k$ and $k + dk$ is:\footnote{This calculation is essentially identical to the one which leads to the definition of the Fermi momentum of a gas of fermions, which might be more familiar to the reader. See, for example, \cite{1987stme.book.....H}}
\begin{equation}
	N_k = 4\pi k^2\left(\frac{L}{2\pi}\right)^3dk = \frac{1}{2\pi^2}(kL)^3\frac{dk}{k}\;. 
\end{equation}

\end{ex}

\hrulefill

From this number of independent modes, we then know that the cosmic variance of the power spectrum\index{Cosmic variance!Power spectrum} is:
\begin{equation}
	\frac{\sigma_P(k)}{P(k)} \simeq \frac{1}{\sqrt{N_k}} \simeq \frac{1}{r_k^{1/2}(kL)^{3/2}}\;,
\end{equation}
where we have defined $r_k \equiv dk/k$ as the resolution of the survey. If we assume infinite resolution, then we must derive the above result again since the dimension is reduced by one.

\hrulefill

\begin{ex}
	Show that the number of independent modes on the sphere of radius $k$ in the wavenumber space are:
	\begin{equation}
		N_k = 4\pi k^2\left(\frac{L}{2\pi}\right)^2 = \frac{(kL)^2}{\pi}\;.
	\end{equation}
\end{ex}

\hrulefill

In this case, one obtains:
\begin{equation}
	\frac{\sigma_P(k)}{P(k)} \simeq \frac{1}{\sqrt{N_k}} \simeq \frac{1}{kL}\;,
\end{equation}
for the cosmic variance, which is the result obtained in \cite{Weinberg:2008zzc}.

These formulae for the cosmic variance tell us that the latter is negligible if the scale probed is much smaller than the dimension $L$ of the survey: $kL \gg 1$.\index{Cosmic variance!Power spectrum}

\subsection{Non-Gaussian perturbations}

For non-Gaussian random fields, the odd-order correlation functions are non-vanishing. For example:\index{Non-Gaussianity}
\begin{equation}
	\langle\tilde G(\textbf{k}_1)\tilde G(\textbf{k}_2)\tilde G(\textbf{k}_3)\rangle = (2\pi)^3\delta^{(3)}(\textbf{k}_1 + \textbf{k}_2 + \textbf{k}_3)B_G(k_1,k_2,k_3)\;,
\end{equation}
where $B_G(k_1,k_2,k_3)$ is called \textbf{the bispectrum}\index{Bispectrum} and can be rewritten as follows:
\begin{eqnarray}
	B_G(k_1,k_2,k_3) = \mathcal{B}_G(k_1,k_2,k_3)\left[P_G(k_1,k_2) + P_G(k_2,k_3) + P_G(k_1,k_3)\right]\;, 
\end{eqnarray}
that is, in terms of \textbf{the reduced bispectrum}, multiplied by all the possible combinations of the power spectra. This formula can be obtained using the following expansion:
\begin{equation}
	G(\mathbf x) = G_G(\mathbf x) + f_{NL}\left(G^2_G(\mathbf x) - \langle G^2_G(\mathbf x)\rangle\right) + \dots\;,
\end{equation}
where $\langle G^2_G(\mathbf x)\rangle \equiv \sigma^2_G$. This expansion is called \textbf{local type non-Gaussianity}\index{Non-Gaussianity!Local type} and is based on the fact that the square of a Gaussian random field $G_G(\mathbf x)$ is not Gaussian.\footnote{This is an example of the fact that spatial homogeneity does not necessarily imply Gaussianity. Indeed, if $G_G$ is a statistically homogeneous Gaussian random field, then $G$ is also statistically homogeneous, but not Gaussian.} 

The amount of non-Gaussianity is indicated by the free parameter $f_{NL}$, which Planck has constrained to be \cite{Planck:2019kim}:
\begin{equation}
	\boxed{f_{NL} = -0.9 \pm 5.1} \qquad \mbox{(68\% CL, statistical)}
\end{equation}\index{Non-Gaussianity!Planck constraint on $f_{NL}$}
The huge relative error shows how difficult it is to extract this kind of information and, at the same time, how Gaussianity ($f_{NL} = 0$) is fully consistent with the data. Note that we are talking here about primordial non-Gaussianity. In the structure formation process, non-Gaussianity naturally arises in the non-linear regime of evolution, since the various $\mathbf{k}$ modes mix.

\section{Matter power spectrum, transfer function, and stochastic initial conditions}

All the formulas derived up to this point can, in principle, be directly adapted to the matter density contrast field $\delta(\eta, \mathbf k)$.\footnote{We drop here the subscript of $\delta$ referring to matter, in order to maintain a concise notation. The treatment can, in principle, be applied to any $\delta$, but the most interesting case is that for matter, that is, CDM plus baryons.} However, how do we deal with the time dependence of the latter? In principle, one just needs to substitute $G(\mathbf x)$ for $\delta(\eta,\mathbf x)$, where we are leaving, on purpose, the time-dependence. We then have:
\begin{equation}
	\xi_\delta(\eta,r) = \int_0^\infty\frac{dk}{k}\Delta^2_\delta(\eta,r)\frac{\sin kr}{kr}\;, \qquad \Delta^2_\delta(\eta,r) = \frac{k^3P_\delta(\eta,k)}{2\pi^2}\;,
\end{equation}
and, from Eq.~\eqref{Gaussian2correlator}, we have:
\begin{equation}
	\langle \delta(\eta,\mathbf k)\delta^*(\eta,\mathbf k')\rangle = (2\pi)^3\delta^{(3)}(\textbf{k} - \textbf{k}')P_\delta(\eta,k)\;.
\end{equation}
In the above equation, we have assumed Gaussianity in the initial conditions. In the linear case, the modes evolve independently; hence, Gaussianity is preserved. For this reason, the above 2-point correlation function is valid even if it is considered at any time $\eta$. 

As we have anticipated, the stochastic character of $\delta(\eta, \mathbf k)$ is due to its initial value $\delta(\mathbf k)$. Hence, it is customary to write:
\begin{equation}
	\boxed{P_\delta(\eta,k) = T^2(\eta,k)P_\delta(k)}
\end{equation}
where $T(\eta,k)$ is the \textbf{transfer function} and $P_\delta(k)$ is the \textbf{primordial power spectrum}:\index{Transfer function}
\begin{equation}\label{matterpowerspectrumprimordial}
	\langle \delta(\mathbf k)\delta^*(\mathbf k')\rangle = (2\pi)^3\delta^{(3)}(\textbf{k} - \textbf{k}')P_\delta(k)\;,
\end{equation}
where $\delta(\mathbf k)$ is the initial condition of $\delta$. The primordial power spectrum is a prediction of inflation that we shall compute in Chapter~\ref{Chap:Inflation}, whereas the transfer function is the solution of the evolution equations that we have found in Chapters~\ref{Chap:CosmoPertTheory} and \ref{Chap:PertubedBoltzmannEquations}, with the primordial modes found in Chapter~\ref{Chap:IC} as initial conditions. These are characterized by constants multiplied by a function of $\mathbf k$, which becomes our stochastic initial condition.\index{Stochastic initial conditions} We call this $\alpha(\mathbf k)$ for scalar modes and $\beta(\mathbf k, \lambda)$ for tensor modes, borrowing the notation used in \cite{Weinberg:2008zzc}. Also, in the rest of these notes, we shall drop the explicit use of the transfer function $T(\eta, k)$, by adopting the renormalization:
\begin{equation}\label{scalarprimormodenorm}
	\delta(\eta, \mathbf k) \to \alpha(\mathbf k)\delta(\eta, k)\;, \qquad \Theta(\eta, \mathbf k, \hat p) \to \alpha(\mathbf k)\Theta(\eta, k, \hat p)\;,
\end{equation} 
for the density contrast and for $\Theta$, when it shall be necessary.

Let us see a practical example of how to calculate $T^2(\eta, k)$ and its normalized initial conditions. Consider adiabatic primordial modes. Then, all perturbations are sourced by $\alpha(\mathbf k) = \zeta(\mathbf k) = \mathcal R(\mathbf k) = C_\gamma(\mathbf k)$. If we normalize the initial condition $C_\gamma(\mathbf k) = 1$, then we have from Eq.~\eqref{deltagammarelPsiadiabatic} and the subsequent ones:
\begin{equation}
	\frac{1}{3}\delta_{\rm c} = \frac{1}{3}\delta_{\rm b} = \frac{1}{4}\delta_\gamma = \frac{1}{4}\delta_\nu = -\Phi + 1\;, \quad \Psi = -\frac{10}{15 + 4R_\nu}\;, \quad \Phi = -\Psi\left(1 + \frac{2}{5}R_\nu\right)\;.
\end{equation}
With this collection of initial conditions, we can compute the transfer functions for each variable, square them, and propagate the primordial power spectrum forward to any time. Hence, for matter, we can write:
\begin{equation}
	P_\delta(\eta,k) = T^2(\eta,k)P_\delta(k) = T^2(\eta,k)\left(\frac{15}{15 + 4R_\nu}\right)^2P_\zeta(k)\;.
\end{equation}
As a final comment, note that in Eq.~\eqref{matterpowerspectrumprimordial} we have the product of two $\delta$'s in $\langle \delta(\mathbf k)\delta^*(\mathbf k')\rangle$. Is this not a second order quantity and thus negligible? Locally it is, but the ensemble average can be seen as a spatial average, cf. Eq.~\eqref{doingthespatialaverage}, and thus we have a vanishing quantity integrated over an infinite volume giving a finite result. 

\section{CMB power spectra}

CMB photons do not come from a localized source, such as a galaxy, but from the whole celestial sphere and from a finite distance, the last scattering surface, which corresponds to a redshift of $z \sim 1100$. Therefore, the CMB power spectrum is analyzed differently from the matter spectrum.

The relative fluctuation in the temperature $\delta T/T$ is the same $\Theta$ as we defined in Eq.~\eqref{Thetadefinition}, provided that the perturbed distribution function $\mathcal F_\gamma$ does not depend on the photon energy $p$. Since this assumption is justified by the fact that Thomson scattering with electrons is the relevant physical process generating $\Theta$, and in Thomson scattering the energy of the photon remains unchanged, we shall use $\delta T/T = \Theta$.

Note that $\Theta = \Theta(\eta, \textbf{x},\hat{p})$; from the point of view of observation, we are just interested in temperature fluctuations here and today, that is, for $\eta = \eta_0$ and $\mathbf x = \mathbf x_0$, where the latter is the position of our laboratory, Earth. Moreover, photons travel on the light-cone (they come from our past light-cone) and from the fixed distance to the last-scattering surface. Hence:
\begin{equation}
	\mathbf x_* = - (\eta_0 - \eta_*)\hat{p} = -r_*\hat{p}\;,
\end{equation}
where $\eta_0$ is the present conformal time, $\eta_*$ the recombination time, and $r_* \equiv \eta_0 - \eta_*$ the comoving distance to recombination. Only photons satisfying this relation can be detected because their momentum is in the ``correct'' direction (towards us), and they free stream only after recombination. Therefore, of the 6 variables upon which $\Theta(\eta, \textbf{x},\hat{p})$ in principle depends, only 2 are observationally relevant since we observe 
\begin{equation}
	\Theta(\eta_0, \textbf{x}_0,\hat{n})\;,
\end{equation}
where $\hat{n} = -\hat{p}$ are indeed coordinates on the celestial sphere. The direction of the photon towards us is $\hat{p}$, so our line of sight unit vector has the opposite sign.

Now we omit the $(\eta_0, \textbf{x}_0)$ dependence of $\Theta$. It is customary to expand a function on the sphere in spherical harmonics $Y_{\ell}^m(\theta, \phi)$. Hence, for our temperature fluctuation, we have:\footnote{The expansion is usually considered for $\Delta T$, the temperature fluctuation. In this case, the $a_{T,\ell m}$'s carry dimensions of temperature. The only difference between the coefficients of the two expansions is a factor $T_0$, the temperature of the CMB.}
\begin{equation}\label{Thetaexpspherharm}\index{Cosmic Microwave Background!Temperature fluctuations expansion}
		\boxed{\Theta(\hat{n}) = \sum_{\ell=0}^\infty\sum_{m=-\ell}^\ell a_{T,\ell m}Y^m_{\ell}(\theta,\phi)} 
\end{equation}
where we have written $\hat n = (\sin\theta\cos\phi, \sin\theta\sin\phi, \cos\theta)$ (employing spherical coordinates) and $Y^m_{\ell}(\theta,\phi)$ are the spherical harmonics\index{Spherical harmonics}. These are revised in some detail in Appendix \ref{App:SphericalHarmonics}.

In these notes, we adopt a normalization that guaranties orthonormality:
\begin{equation}
	\int d^2\hat n\;Y_{\ell}^m(\hat n)Y_{\ell'}^{m'*}(\hat n) = \delta_{\ell\ell'}\delta_{mm'}\;.
\end{equation}
Note that the spherical harmonics $Y^m_\ell$ are, in general, complex; therefore, the $a_{\ell m}$ must also be complex numbers in order for $\Theta(\theta,\phi)$ to be real.\footnote{The notation $Y_{\ell m}$ usually denotes spherical harmonics in the real form, obtained by combining $Y^m_\ell$ in a suitable way to trade the imaginary exponential $\exp(im\phi)$ for a sine or cosine function of $m\phi$.} 

\hrulefill

\begin{ex}
	According to our conventions established in Appendix \ref{App:SphericalHarmonics} show that 
\begin{equation}
	Y_\ell^{m*}(\hat n) = Y_\ell^{-m}(\hat n)\;,
\end{equation}
and hence, in order for $\Theta(\hat n)$ to be real, show that:
\begin{equation}
	a_{T,\ell m}^* = a_{T,\ell,-m}\;,
\end{equation}
which is a \textbf{reality condition}\index{Reality condition} for the coefficients $a_{T,\ell m}$.
\end{ex}

\hrulefill

Note that the expansion \eqref{Thetaexpspherharm} can be done at each point of spacetime, but then the angular dependence changes because the $(\theta,\phi)$ measured from one point in space corresponds to different celestial coordinates as seen from another spot in space.

The expansion of Eq.~\eqref{Thetaexpspherharm} can be inverted as follows:
\begin{equation}\label{ThetaexpYlm}
	a_{T,\ell m} = \int d^2\hat n\; Y^{*m}_{\ell}(\hat n)\Theta(\hat n)\;,
\end{equation}
and the $a_{\ell m}$'s are promoted to stochastic variables, just as $\Theta$ is. Again, it is the initial conditions for cosmological perturbations that are actual stochastic variables for which inflation predicts a power spectrum, so we shall introduce a transfer function.

For Gaussian perturbations, the expectation value and variance of the $a_{T,\ell m}$'s are:
\begin{equation}\label{CMBPSGaussian}
	\boxed{\langle a_{T,\ell m}\rangle = 0} \qquad \boxed{\langle a_{T,\ell m}a^*_{T,\ell'm'}\rangle = \delta_{\ell\ell'}\delta_{mm'}C_{TT,\ell}}
\end{equation}
and the $C_{TT,\ell} = \langle|a_{T,\ell m}|^2\rangle$ form the CMB power spectrum\index{Cosmic Microwave Background!Power spectrum}. Introducing the Fourier transform, we have:
\begin{equation}\label{ClFourierintegrals}
	C_{TT,\ell} = \langle \int\frac{d^3\mathbf k}{(2\pi)^3}\int\frac{d^3\mathbf k'}{(2\pi)^3}e^{i(\mathbf k - \mathbf k')\cdot \mathbf x_0}\int d^2\hat n\; Y_{\ell}^{*m}(\hat{n})\Theta(\textbf{k},\hat{n})\int d^2\hat n'\; Y_{\ell}^m(\hat{n}')\Theta^*(\textbf{k}',\hat{n}')\rangle\;.
\end{equation} 
The ensemble average acts on the temperature fluctuations, which we renormalize to some primordial mode $\alpha(\mathbf k)$:
\begin{equation}\label{Primordialmodealpha}
	\langle\Theta(\textbf{k},\hat{n})\Theta^*(\textbf{k}',\hat{n}')\rangle = \langle\alpha(\textbf{k})\alpha^*(\textbf{k}')\rangle \Theta(k,\hat{n})\Theta^*(k',\hat{n}')\;.
\end{equation}
We should have perhaps written $T_\Theta(k,\hat n)$, stressing the introduction of a transfer function; however, we have decided to keep a simple notation.

For scalar perturbations, we have seen that the dependence of $\Theta$ is only on $\hat{k}\cdot \hat{p} = \mu = -\hat{k}\cdot \hat{n}$, and we have used the multipole expansion in Eq.~\eqref{Thetal}, which can be easily inverted as follows:
\begin{equation}\label{Thetapartialwaveexpansion}
	\Theta^{(S)}(k,\mu) = \sum_\ell (-i)^\ell(2\ell + 1)\mathcal P_\ell(\mu)\Theta^{(S)}_\ell(k)\;.
\end{equation}
The evolution of the multipoles $\Theta^{(S)}_\ell(\eta, k)$ until $\eta_0$ (today) is given by the hierarchy of Eqs.~\eqref{lg2equationphoton}-\eqref{le0equationphoton}, which in fact depend only on the modulus $k$.

Assuming adiabatic Gaussian perturbations $\alpha(\mathbf k) = \mathcal R(\mathbf k)$ and
\begin{equation}
	\langle\mathcal R(\mathbf k)\mathcal R^*(\mathbf k')\rangle = (2\pi)^3\delta^{(3)}(\mathbf k - \mathbf k')P_{\mathcal R}(k)\;,
\end{equation}
we have from Eq.~\eqref{ClFourierintegrals}:
\begin{eqnarray}
	C^S_{TT,\ell} = \int\frac{d^3\mathbf k}{(2\pi)^3}|\Theta^{(S)}_{\ell}(\eta_0, k)|^2P_{\mathcal R}(k)\int d^2\hat n\; Y^{m*}_{\ell}(\hat{n})\int d^2\hat n'\; Y^m_{\ell}(\hat{n}')\nonumber\\
	\sum_{\ell'}(-i)^{\ell'}(2\ell' + 1)\mathcal P_{\ell'}(\mu)\sum_{\ell''}i^{\ell''}(2\ell'' + 1)\mathcal P_{\ell''}(\mu)\;.
\end{eqnarray}
Note that a similar result can be obtained from Eq.~\eqref{ClFourierintegrals} if we perform a spatial average, i.e., an average over $\mathbf x_0$, as we saw in the case of the three-dimensional power spectrum. In this case, the square modulus $|\Theta(\textbf{k},\hat{n})|^2$ appears as our estimator of the angular power spectrum.

Recall that the Legendre polynomial is proportional to $Y_{\ell}^0$, and thus, from the orthogonality of the spherical harmonics and the addition theorem, we get:
\begin{equation}
	\int d^2\hat n\;\mathcal P_{\ell'}(\mu)Y^{m*}_{\ell}(\hat{n}) = \int d^2\hat n\;\mathcal P_{\ell'}(-\hat{k}\cdot\hat{n})Y^{m*}_{\ell}(\hat{n}) = \frac{4\pi}{2\ell + 1}Y^{m*}_{\ell}(\hat{k})\delta_{\ell\ell'}\;.
\end{equation}
Hence, we have:
\begin{equation}\label{ClfunctionThetal}\index{Cosmic Microwave Background!$C^S_{TT,\ell}$ spectrum}
	\boxed{C^S_{TT,\ell} = \frac{2}{\pi}\int dk\;k^2|\Theta^{(S)}_{\ell}(\eta_0,k)|^2 P_{\mathcal R}(k) = 4\pi\int \frac{dk}{k}|\Theta^{(S)}_{\ell}(\eta_0,k)|^2 \Delta^2_{\mathcal R}(k)}
\end{equation}
where we have used:
\begin{equation}
	\int d^2\hat k\;|Y_\ell^m(\hat{k})|^2 = 1\;.
\end{equation}
We shall discuss the tensor contribution to the CMB TT spectrum in Chapter \ref{Chap:CMBEvo}.

Not only are temperature anisotropies measurable in the CMB sky, but so are the polarization ones. Hence, we have more correlation functions and spectra than the temperature-temperature (TT) one. As we discussed in Chapter~\ref{Chap:PertubedBoltzmannEquations}, Thomson scattering provides only linear polarization, which is described by the two Stokes parameters $Q$ and $U$. It is customary to use the combinations $Q \pm iU$ since these have helicity 2, i.e., under a rotation of an angle $\theta$ in the polarization plane, they transform as:
\begin{equation}
	(Q \pm iU) \to e^{\pm 2i\theta}(Q \pm iU)\;.
\end{equation}
Similar quantities were defined for gravitational waves. Now, $Q \pm iU$ are fields of helicity 2 on the sphere, and as such, they can be expanded in spin 2-weighted spherical harmonics:
\begin{eqnarray}
	(Q + iU)(\hat{n}) = \sum_{\ell = 2}^\infty \sum_{m = -\ell}^\ell a_{P,\ell m}\;{}_{2}Y^m_\ell(\hat{n})\;,\\
	(Q - iU)(\hat{n}) = \sum_{\ell = 2}^\infty \sum_{m = -\ell}^\ell a^*_{P,\ell m}\;{}_{2}Y^{m*}_\ell(\hat{n})\;.
\end{eqnarray}
The spin-weighted spherical harmonics do not satisfy simple reality conditions as the spin-0 ones (the usual spherical harmonics) do, i.e., $Y_\ell^{m*} = Y_\ell^{-m}$, and therefore it is convenient to use the following combinations of $a_{P,\ell m}$:
\begin{equation}
	a_{E,\ell m} =  -(a_{P,\ell m} + a_{P,\ell-m}^*)/2\;, \qquad a_{B,\ell m} =  i(a_{P,\ell m} - a_{P,\ell-m}^*)/2\;.
\end{equation}
These satisfy reality conditions and have the following properties under spatial inversion: $a_{E,\ell m}$ gains a factor $(-1)^\ell$, as the $a_{\ell m}$ of the temperature-temperature correlation, whereas $a_{B,\ell m}$ gains an extra $-1$ factor.\index{Cosmic Microwave Background!Polarization spectra}

If we assume stochastic initial conditions that are invariant under spatial inversion, the only four spectra that we can build from CMB temperature and polarization measurements are as follows:
\begin{eqnarray}
\label{TTandTEspectra}	\boxed{\langle a_{T,\ell m}a^*_{T,\ell'm'}\rangle = \delta_{\ell\ell'}\delta_{mm'}C_{TT,\ell}} \quad \boxed{\langle a^*_{T,\ell m}a_{E,\ell'm'}\rangle = \delta_{\ell\ell'}\delta_{mm'}C_{TE,\ell}}\\
\label{EEandBBspectra}	\boxed{\langle a_{E,\ell m}a^*_{E,\ell'm'}\rangle = \delta_{\ell\ell'}\delta_{mm'}C_{EE,\ell}} \quad \boxed{\langle a_{B,\ell m}a^*_{B,\ell'm'}\rangle = \delta_{\ell\ell'}\delta_{mm'}C_{BB,\ell}}
\end{eqnarray}
Note that there is no dependence on $m$ in the power spectra. This is again due to statistical isotropy.

\subsection{Cosmic Variance of angular power spectra}

In order to compute the cosmic variance, let us first make an estimate. Our objective is to determine $C_\ell$ observationally, in order to compare it with our theoretical prediction. In order to do that, we probe $\langle a_{\ell m}a^*_{\ell m}\rangle$ for different values of $m$, which are $2\ell + 1$. Hence, we have $2\ell + 1$ possible samplings of $C_\ell$ for any given $\ell$ and a sampling error
\begin{equation}
	\Delta C_\ell \propto \sqrt{2\ell + 1}\;.
\end{equation}
The relative error associated with the sampling, i.e., the cosmic variance, is thus:
\begin{equation}
	\sigma_{C_\ell} = \frac{\Delta C_\ell}{2\ell + 1} \propto \frac{1}{\sqrt{2\ell + 1}}\;.
\end{equation}
Now, let us make a more precise calculation following \cite{Weinberg:2008zzc}. Consider the temperature-temperature correlation function as example:
\begin{equation}
	\langle\Theta(\hat{n})\Theta(\hat{n}')\rangle = \sum_{\ell m\ell'm'}\langle a_{\ell m}a_{\ell'm'}\rangle Y^m_{\ell}(\hat{n})Y^{m'}_{\ell'}(\hat{n}') = \sum_{\ell m}C_{\ell}Y^m_{\ell}(\hat{n})Y^{-m}_{\ell}(\hat{n}')\;,
\end{equation}
where $\cos\theta \equiv \hat{n}\cdot\hat{n}'$, and we have performed the ensemble average assuming Gaussian perturbations. The $-m$ of the second spherical harmonics arises from the reality condition by which $a^*_{\ell m} = a_{\ell,-m}$.

Using the addition theorem of spherical harmonics, we can sum over $m$ in the above formula and obtain:
\begin{equation}
	C(\theta) \equiv \langle\Theta(\hat{n})\Theta(\hat{n}')\rangle = \sum_{\ell}C_{\ell}\frac{2\ell + 1}{4\pi}\mathcal{P}_{\ell}(\hat{n}\cdot\hat{n}') = \sum_{\ell}C_{\ell}\frac{2\ell + 1}{4\pi}\mathcal{P}_{\ell}(\cos\theta)\;.
\end{equation}
Inverting this relation using the orthonormality of the Legendre polynomials, we obtain:
\begin{equation}
	C_\ell = \frac{1}{4\pi}\int d^2\hat{n}\;d^2\hat{n}'\;\mathcal{P}_{\ell}(\hat{n}\cdot\hat{n}')\langle\Theta(\hat{n})\Theta(\hat{n}')\rangle\;,
\end{equation}
which is Eq.~\eqref{ClFourierintegrals}, without introducing the Fourier Transform, which we do not need here.

The integral is, of course, over the whole sky. These are the theoretical $C_\ell$'s, and the average is the ensemble average. Observationally, the only average that we can do is the angular one, i.e.
\begin{equation}
	C_\ell^{\rm obs} = \frac{1}{4\pi}\int d^2\hat{n}\;d^2\hat{n}'\;\mathcal{P}_{\ell}(\hat{n}\cdot\hat{n}')\Theta(\hat{n})\Theta(\hat{n}')\;.
\end{equation}

\hrulefill

\begin{ex}
 Show that, substituting the spherical harmonics expansions, we have:
\begin{eqnarray}
	C_\ell^{\rm obs} = \frac{1}{4\pi}\sum_{LML'M'}\int d^2\hat{n}\;d^2\hat{n}'\;\mathcal{P}_{\ell}(\hat{n}\cdot\hat{n}')a_{LM}Y_L^{M}(\hat n)a_{L'M'}Y_{L'}^{M'}(\hat n')\nonumber\\ = \frac{1}{2\ell + 1}\sum_ma_{\ell m}a_{\ell,-m}\;.
\end{eqnarray}
\end{ex}

\hrulefill

Here it appears more clearly that for each value of $\ell$, we have $2\ell + 1$ possible realizations, and thus we expect that the counting error is $\sqrt{2\ell + 1}$. We can calculate this exactly, and the cosmic variance is the following ensemble average:
\begin{equation}
	\sigma_{C_\ell}^2 = \left\langle\left(\frac{C_\ell - C_\ell^{\rm obs}}{C_\ell}\right)^2\right\rangle = 1 - 2\frac{\langle C_\ell^{\rm obs}\rangle}{C_\ell} + \frac{1}{C_\ell^2}\langle {C_\ell^{\rm obs}}^2\rangle\;.
\end{equation}
Of course, the ensemble average of $\langle C_\ell^{\rm obs}\rangle$ is $C_\ell$, and the ensemble average of $C_\ell$ is $C_\ell$, since it is already averaged. Therefore, let us focus on
\begin{equation}
	\langle {C_\ell^{\rm obs}}^2\rangle = \frac{1}{(2\ell + 1)^2}\sum_{mm'}\langle a_{\ell m}a_{\ell,-m}a_{\ell m'}a_{\ell,-m'}\rangle\;.
\end{equation}
Since the perturbations are Gaussian, this 4-point correlation function can be split into the following sum, by Eq.~\eqref{4pointcorrelator}:
\begin{eqnarray}
	\langle a_{\ell m}a_{\ell,-m}a_{\ell m'}a_{\ell,-m'}\rangle = \langle a_{\ell m}a_{\ell,-m}\rangle\langle a_{\ell m'}a_{\ell,-m'}\rangle + \langle a_{\ell m}a_{\ell m'}\rangle\langle a_{\ell,-m}a_{\ell,-m'}\rangle\nonumber\\ + \langle a_{\ell m}a_{\ell,-m'}\rangle\langle a_{\ell m'}a_{\ell,-m}\rangle\;.
\end{eqnarray}
Using Eq.~\eqref{CMBPSGaussian}, we finally obtain:
\begin{equation}
	\boxed{\sigma_{C_\ell}^2 = \frac{2}{2\ell + 1}}
\end{equation}
as expected\index{Cosmic variance!CMB power spectrum}.

\section{Power spectrum for tensor perturbations}

When we compute the power spectrum for Gravitational Waves using the decomposition of helicities from Eq.~\eqref{hTijhelicitiesexpansion}, we obtain:
\begin{equation}
	\langle h_{ij}^T(\eta, \mathbf k)h_{lm}^{T*}(\eta, \mathbf k')\rangle = \sum_{\lambda, \lambda' = \pm 2}e_{ij}(\hat{k},\lambda)e^*_{lm}(\hat{k}',\lambda')\langle h(\eta, \mathbf k, \lambda)h^*(\eta, \mathbf k', \lambda')\rangle\;.
\end{equation}
The stochastic behavior is carried by the initial condition on $h(\eta, \mathbf k, \lambda)$, which we then renormalize as:
\begin{equation}\label{ICnormalisationh}
	\boxed{h(\eta, \mathbf k, \lambda) = \beta(\mathbf k,\lambda)h(\eta, k)}
\end{equation}
Since the evolution equation for tensor perturbations depends only on $k$ and does not depend on the helicity $\lambda$, we incorporate the latter into the stochastic initial condition.

Assuming:
\begin{equation}\label{tensorspectrum}
	\langle \beta(\mathbf k, \lambda)\beta^*(\mathbf k', \lambda')\rangle = (2\pi)^3P_h(k)\delta_{\lambda\lambda'}\delta^{(3)}(\mathbf k - \mathbf k')\;,
\end{equation}
in the ensemble average of $\langle h_{ij}^T(\eta, \mathbf k)h_{lm}^{T*}(\eta, \mathbf k')\rangle$, it appears the sum over the helicities:
\begin{eqnarray}\label{sumofthehelicities}
	\Pi_{ij,lm}(\hat{k}) \equiv \sum_{\lambda = \pm 2}e_{ij}(\hat{k},\lambda)e^*_{lm}(\hat{k},\lambda)\;.
\end{eqnarray}
We can compute this sum as follows. Consider the fact that $\Pi_{ij,lm}(\hat{k})$ is a tensor that depends on $\hat{k}$. In order to have the correct combination $ij,lm$ of indices, we can combine two $\delta_{ij}$, one $\delta_{ij}$, and two $\hat k_m$ or four $\hat k_m$. So we can put forward the following ansatz:
\begin{eqnarray}
	\Pi_{ij,lm}(\hat{k}) = A(\delta_{il}\delta_{jm} + \delta_{im}\delta_{jl}) + B\delta_{ij}\delta_{lm}\nonumber\\ + C\delta_{ij}\hat{k}_l\hat{k}_m + D\delta_{lm}\hat{k}_i\hat{k}_j + E(\delta_{il}\hat{k}_j\hat{k}_m + \delta_{jm}\hat{k}_i\hat{k}_l + \delta_{im}\hat{k}_j\hat{k}_l + \delta_{jl}\hat{k}_i\hat{k}_m)\nonumber\\
	+ F\hat{k}_i\hat{k}_j\hat{k}_l\hat{k}_m\;,
\end{eqnarray}
where we have grouped together all the terms that must be symmetrized to respect the symmetry of $\Pi_{ij,lm}(\hat{k})$, which is symmetric in $ij$ and in $lm$ separately. The coefficients introduced above are, in principle, complex.

\hrulefill

\begin{ex}
	Contract the above ansatz for $\Pi_{ij,lm}(\hat{k})$, with $\hat k^i$, with $\hat k^l$ and with $\delta^{ij}$. Since the result must be zero, show that have the following conditions on the coefficients:
	\begin{equation}
		A = - B = C = D = -E = F\;.
	\end{equation}
\end{ex}

\hrulefill

Thus, we can write:
\begin{eqnarray}
	\Pi_{ij,lm}(\hat{k}) = A(\delta_{il}\delta_{jm} + \delta_{im}\delta_{jl} - \delta_{ij}\delta_{lm}\nonumber\\ + \delta_{ij}\hat{k}_l\hat{k}_m + \delta_{lm}\hat{k}_i\hat{k}_j - \delta_{il}\hat{k}_j\hat{k}_m - \delta_{jm}\hat{k}_i\hat{k}_l - \delta_{im}\hat{k}_j\hat{k}_l - \delta_{jl}\hat{k}_i\hat{k}_m\nonumber\\
	+ \hat{k}_i\hat{k}_j\hat{k}_l\hat{k}_m)\;.
\end{eqnarray}
Now, since $\Pi_{ij,lm}^*(\hat{k}) = \Pi_{lm,ij}(\hat{k})$, i.e., complex conjugation amounts to exchanging the couple $ij$ with $lm$, it is not difficult to conclude that $A$ is a real number. In order to determine which number, consider the normalization used in Eq.~\eqref{normalizationpolarizationGW}. If we set $\hat{k} = \hat{z}$ in our improved ansatz above and choose $i = j = l = m = 1$, we get:
\begin{equation}
	\Pi_{11,11}(\hat{k}) = \sum_{\lambda = \pm 2}e_{11}(\hat{k},\lambda)e^*_{11}(\hat{k},\lambda) = 1 = A\;.
\end{equation}
Therefore, we can conclude that:
\begin{eqnarray}\label{Piijlmexpression}
	\Pi_{ij,lm}(\hat{k}) = \delta_{il}\delta_{jm} + \delta_{im}\delta_{jl} - \delta_{ij}\delta_{lm}\nonumber\\ + \delta_{ij}\hat{k}_l\hat{k}_m + \delta_{lm}\hat{k}_i\hat{k}_j - \delta_{il}\hat{k}_j\hat{k}_m - \delta_{jm}\hat{k}_i\hat{k}_l - \delta_{im}\hat{k}_j\hat{k}_l - \delta_{jl}\hat{k}_i\hat{k}_m\nonumber\\
	+ \hat{k}_i\hat{k}_j\hat{k}_l\hat{k}_m\;.
\end{eqnarray}
\index{Gravitational waves!Sum over helicities}

\section{Ergodic theorem}\label{Sec:ergodictheorem}

Following \cite{Weinberg:2008zzc}, we briefly present the ergodic theorem, which allows us to exchange ensemble and spatial averages under certain conditions. \index{Ergodic theorem}

Consider a random variable $G(\mathbf x)$ in a $D$-dimensional Euclidean space and assume statistical homogeneity:
\begin{equation}
	\langle G(\mathbf x_1)G(\mathbf x_2)\dots G(\mathbf x_n)\rangle = \langle G(\mathbf x_1 + \mathbf z)G(\mathbf x_2 + \mathbf z)\dots G(\mathbf x_n + \mathbf z)\rangle\;.
\end{equation}
That is, the correlation does not change upon the translation of the fields.

Also, let us make the following reasonable assumption. When we calculate a correlation where a certain number of points are very far from another set, the correlation breaks into two pieces.\footnote{This is also known as the Cluster Decomposition Principle; see \cite{Weinberg:1995mt}.} That is:
\begin{eqnarray}
	\langle G(\mathbf x_1 + \mathbf u)G(\mathbf x_2 + \mathbf u)\dots G(\mathbf y_1 - \mathbf u)G(\mathbf y_2 - \mathbf u)\dots\rangle\nonumber\\ \rightarrow_{|\mathbf u| \to \infty} \langle G(\mathbf x_1 + \mathbf u)G(\mathbf x_2 + \mathbf u)\dots\rangle\langle G(\mathbf y_1 - \mathbf u)G(\mathbf y_2 - \mathbf u)\dots\rangle\nonumber\\
	= \langle G(\mathbf x_1)G(\mathbf x_2)\dots\rangle\langle G(\mathbf y_1)G(\mathbf y_2)\dots\rangle\nonumber\;,
\end{eqnarray}
i.e., if we think of a simple 2-point correlation function, the points become uncorrelated over a large distance, and it is as if we have independent ensembles, which is key in order to exchange the spatial average with the ensemble average. In the last line of the above equation, we have used statistical homogeneity.

Now define what, in some sense, is closely related to cosmic variance:
\begin{eqnarray}
	\sigma_R^2(\mathbf x_1, \mathbf x_2,\dots) \equiv \nonumber\\\left\langle\left[\left(\int d^D\mathbf z N_R(\mathbf z)G(\mathbf x_1 + \mathbf z)G(\mathbf x_2 + \mathbf z)\dots\right) - \langle G(\mathbf x_1)G(\mathbf x_2)\dots\rangle\right]^2\right\rangle\;, 
\end{eqnarray}
where the integral is a spatial average about a point $\mathbf z_0$; recall that the dimension of the space is $D$, and the window function, or filter, is, for example:
\begin{equation}
	N_R(\mathbf z) \equiv \frac{1}{(\sqrt{\pi}R)^D}e^{-|\mathbf z - \mathbf z_0|^2/R^2}\;,
\end{equation}
which, as can be checked, is normalized to unity: $\int d^D\mathbf z\;N_R(\mathbf z) = 1$, the integration being over all the space. This function is a filter, and its form is not important as long as it is almost constant for $|\mathbf z - \mathbf z_0|^2 < R^2$ and decays rapidly for $|\mathbf z - \mathbf z_0|^2 > R^2$. It introduces the scale $R$ over which we perform the spatial average, We want to check how this variance depends on $R$. We can think of $R$ as the size of a survey, for example.

The ergodic theorem states that: 
\begin{equation}
	\sigma_R^2(\mathbf x_1, \mathbf x_2,\dots) \sim R^{-D} \qquad \mbox{for } R \to \infty\;.
\end{equation}
To prove it, expand the square in the variance:
\begin{eqnarray}
	\sigma_R^2 &=& \left\langle\int d^D \mathbf z N_R(\mathbf z)\int d^D \mathbf w N_R(\mathbf w)(G(\mathbf x_1 + \mathbf z)G(\mathbf x_2 + \mathbf z)\dots - \langle G(\mathbf x_1)G(\mathbf x_2)\dots\rangle)\right.\nonumber\\&\times&\left.(G(\mathbf x_1 + \mathbf w)G(\mathbf x_2 + \mathbf w)\dots - \langle G(\mathbf x_1)G(\mathbf x_2)\dots\rangle)\right\rangle\;. 
\end{eqnarray}
We can incorporate the ensemble average into the integral because of $\int d^Dz\;N_R(z) = 1$. Now apply the average and use statistical homogeneity:
\begin{eqnarray}
	\sigma_R^2 &=& \int d^D\mathbf z N_R(\mathbf z)\int d^D\mathbf w N_R(\mathbf w)(\langle G(\mathbf x_1 + \mathbf z)G(\mathbf x_2 + \mathbf z)\dots G(\mathbf x_1 + \mathbf w)G(\mathbf x_2 + \mathbf w)\dots\rangle \nonumber\\ &-& \langle G(\mathbf x_1)G(\mathbf x_2)\dots\rangle^2)\;. 
\end{eqnarray}
Changing the integration variables:
\begin{equation}
	\mathbf u = (\mathbf z - \mathbf w)/2\;, \qquad \mathbf v = (\mathbf z + \mathbf w)/2\;,
\end{equation}
and, remembering to take into account the Jacobian of the transformation, which is $2^D$, and the expression we have adopted for the window function:
\begin{eqnarray}
	\sigma_R^2 = \left(\frac{2}{\pi R^2}\right)^D\int d^D\mathbf v\int d^D\mathbf u\; e^{-|\mathbf u + \mathbf v - \mathbf z_0|^2/R^2 - |\mathbf v - \mathbf u - \mathbf z_0|^2/R^2}\qquad\nonumber\\
	(\langle G(\mathbf x_1 + \mathbf u + \mathbf v)G(\mathbf x_2 + \mathbf u + \mathbf v)\dots G(\mathbf x_1 + \mathbf v - \mathbf u)G(\mathbf x_2 + \mathbf v - \mathbf u)\dots\rangle - \nonumber\\\langle G(\mathbf x_1)G(\mathbf x_2)\dots\rangle^2)\;. 
\end{eqnarray}
Now, the $\mathbf v$ in the fields is always summed to the coordinate, so we can again use statistical homogeneity and eliminate $\mathbf v$ from the average, yielding:
\begin{eqnarray}
	\sigma_R^2 = \left(\frac{2}{\pi R^2}\right)^D\int d^D\mathbf v\int d^D\mathbf u\; e^{-2|\mathbf u|^2/R^2}e^{-|\mathbf v - \mathbf z_0|^2/R^2}\nonumber\\
	(\langle G(\mathbf x_1 + \mathbf u)G(\mathbf x_2 + \mathbf u)\dots G(\mathbf x_1 - \mathbf u)G(\mathbf x_2 - \mathbf u)\dots\rangle - \langle G(\mathbf x_1)G(\mathbf x_2)\dots\rangle^2)\;. 
\end{eqnarray}
The integration in $\mathbf v$ can be performed immediately, as it is a Gaussian integral equal to $(\pi/2)^{D/2}R^D$:
\begin{eqnarray}
	\sigma_R^2 = \left(\frac{2}{\pi R^2}\right)^{D/2}\int d^D\mathbf u\; e^{-2|\mathbf u|^2/R^2}\nonumber\\
	(\langle G(\mathbf x_1 + \mathbf u)G(\mathbf x_2 + \mathbf u)\dots G(\mathbf x_1 - \mathbf u)G(\mathbf x_2 - \mathbf u)\dots\rangle - \langle G(\mathbf x_1)G(\mathbf x_2)\dots\rangle^2)\;. \quad
\end{eqnarray}
Note that $\mathbf z_0$ is no longer present, as it should be because of statistical homogeneity. The term in parentheses approaches zero sufficiently fast for large $|\mathbf u|$, based on the hypothesis made at the beginning of our discussion, so that the $|\mathbf u|$ integral is finite. Moreover, the value of the integral does not depend on $R$ because the difference among the ensemble averages does not depend on our choice of $R$. Then, we can conclude that:
\begin{eqnarray}
	\boxed{\sigma_R = \mathcal{O}(R^{-D/2})}
\end{eqnarray}

\clearpage
\chapter{Inflation}\label{Chap:Inflation}

{\rightskip=3truepc\leftskip=3truepc\noindent
{\it And if inflation is wrong, then God missed a good trick. But, of course, we've come across a lot of other good tricks that nature has decided not to use}
\vskip 0.10 in
\centerline{\it ---Jim Peebles, interview at Princeton (1994)}
\vskip 0.20 in
}

We dedicate this chapter to inflation, a model of the primordial universe in which an almost constant $H$ provides a scale factor $a$ that grows exponentially with cosmic time. Inflation is able to solve some puzzles related to background cosmology and also provides a testable prediction of the power spectrum of primordial fluctuations. Among the first pioneering works on inflation are \cite{Starobinsky:1979ty}, \cite{Guth:1980zm}, \cite{Linde:1981mu}, and \cite{Albrecht:1982wi}.

An alternative to inflation that has garnered some interest is the so-called \textbf{bouncing cosmology}\index{Bouncing cosmology}, in which the universe is eternal and whose evolution is characterized by a contracting phase followed by an expanding one, through a bouncing phase whose physical details are governed by quantum cosmology. See \cite{Novello:2008ra} and \cite{Brandenberger:2016vhg} for recent reviews on the subject, which, though interesting, we shall not tackle here.

\section{The flatness problem}\index{Flatness problem}

We stated the flatness problem in Subsection \ref{subsec:flatnessproblem}. Let us see quantitatively how inflation can solve the flatness problem. Suppose that
\begin{equation}
	\frac{|K|}{a_i^2H_i^2} = \mathcal{O}(1)\;,
\end{equation}
at the beginning of the inflationary phase, to which we refer with a subscript $i$. That is, at the beginning of inflation, the spatial curvature might be a relevant fraction of the total energy density content of the universe.

Now, at the end of inflation, which we denote with a subscript $I$, the scale factor $a$ has increased by a factor of $e^N$. The number $N$ is called \textbf{number of e-folds}. Then we have:\index{Inflation!e-folds number}
\begin{equation}
	\frac{|K|}{a_I^2H_I^2} = \frac{|K|}{a_i^2H_i^2}e^{-2N} \approx e^{-2N}\;.
\end{equation}
Therefore, today we have the following:
\begin{equation}
	|\Omega_{ K0}| = \frac{|K|}{H_0^2} = \frac{|K|}{a_I^2H_I^2}\left(\frac{a_IH_I}{H_0}\right)^2 \approx e^{-2N}\left(\frac{a_IH_I}{H_0}\right)^2\;.
\end{equation}
In order to have $|\Omega_{ K0}| < 1$, we need
\begin{equation}\label{inflationcondition}
	\boxed{\frac{a_IH_I}{H_0} < e^N}
\end{equation}
In order to estimate the ratio on the left-hand side, let us suppose that inflation ends in the radiation-dominated epoch, so that:
\begin{equation}
	H_I^2 \approx H_0^2\Omega_{\rm r0}/a_I^4\;,
\end{equation}
and thus:
\begin{equation}
	a_I \approx \Omega_{\rm r0}^{1/4}\sqrt{\frac{H_0}{H_I}}\;.
\end{equation}
We have then:
\begin{equation}\label{energyscaleinflation}
	e^N > \Omega_{\rm r0}^{1/4}\sqrt{\frac{H_I}{H_0}} = \Omega_{\rm r0}^{1/4}\left(\frac{\rho_I}{\rho_0}\right)^{1/4} = \frac{\rho_I^{1/4}}{0.037\;h\;\mbox{eV}}\;,
\end{equation}
where $\rho_I$ is the energy scale at which inflation ends, but since $H$ is constant during the inflationary phase, then $\rho_I$ is the energy density scale of inflation \textit{tout court}.\index{Inflation!Energy scale}

Certainly, we do not want to spoil BBN; therefore, $\rho_I$ must be larger than 1 MeV$^4$. With this constraint, we get $N > 17$. Choosing the Planck scale, one obtains $N > 68$, and choosing the GUT energy scale, one obtains $N > 62$.

\section{The horizon problem}

The \textbf{horizon problem} is an issue that arises when we calculate the angular size of the particle horizon at recombination and notice that it is only a small portion of the CMB sky. Then, how is it possible that the CMB is so isotropic if no causal processes could have provided the conditions to achieve this?\index{Inflation!Horizon problem}\index{Horizon problem}

Let us again consider this issue more quantitatively. The proper particle-horizon distance is the following:
\begin{equation}
	d_{\rm H} = a(t)\int_0^t\frac{dt'}{a(t')} = a\int_0^a\frac{da'}{H(a')a^{'2}}\;.
\end{equation}
In a universe dominated by matter and radiation, the above expression becomes:
\begin{equation}\label{dHmattraduniv}
	d_{\rm H} = a\int_0^a\frac{da'}{H_0\sqrt{\Omega_{\rm m0}a' + \Omega_{\rm r0}}} = \frac{2a}{H_0\Omega_{\rm m0}}\left(\sqrt{\Omega_{\rm m0}a + \Omega_{\rm r0}} - \sqrt{\Omega_{\rm r0}}\right)\;.
\end{equation}
Now, recall that the angular diameter distance has the following form:
\begin{equation}
	d_{\rm A} = a(t)\int_t^{t_0}\frac{dt'}{a(t')} = a\int_a^1\frac{da'}{H(a')a^{'2}}\;.
\end{equation}
Again, in a universe dominated by matter and radiation, the above expression becomes:
\begin{equation}
	d_{\rm A} = a\int_a^1\frac{da'}{H_0\sqrt{\Omega_{\rm m0}a' + \Omega_{\rm r0}}} = \frac{2a}{H_0\Omega_{\rm m0}}\left(\sqrt{\Omega_{\rm m0} + \Omega_{\rm r0}} - \sqrt{\Omega_{\rm m0}a + \Omega_{\rm r0}}\right)\;.
\end{equation}
The ratio $d_{\rm H}/d_{\rm A}$ represents the angular radius of the particle horizon at a given scale factor. From the above calculations, we obtain:
\begin{equation}
	\frac{d_{\rm H}}{d_{\rm A}} = \frac{\sqrt{\Omega_{\rm m0}a + \Omega_{\rm r0}} - \sqrt{\Omega_{\rm r0}}}{\sqrt{\Omega_{\rm m0} + \Omega_{\rm r0}} - \sqrt{\Omega_{\rm m0}a + \Omega_{\rm r0}}}\;.
\end{equation}
This ratio tends to zero for $a \to 0$, and at recombination, it is equal to:
\begin{equation}
	\frac{d_{\rm H}}{d_{\rm A}}(a_{\rm rec} = 10^{-3}) = 0.018\;,
\end{equation}
which corresponds to about $1^\circ$ in the CMB sky. Therefore, we have roughly $4\pi/(0.018)^2 \approx 10^4$ causally disconnected regions in the CMB sky.

So, the particle horizon at recombination is a very small fraction of the CMB sky. If no causal process could have taken place beyond 1 degree, then what caused the high isotropy of the CMB temperature?

This issue can be solved in much the same way as we did for the flatness problem. Assume an initial inflationary phase such that:
\begin{equation}
	a(t) = a_ie^{H_I(t - t_i)} = a_Ie^{-H_I(t_I - t)}\;.
\end{equation}
Then, the proper particle-horizon distance that we have calculated in Eq.~\eqref{dHmattraduniv} acquires the following contribution for very small times:
\begin{equation}
	a\int_{t_i}^{t_I}dt\frac{e^{H_I(t_I - t)}}{a_I}\;.
\end{equation}
To solve the horizon problem, this contribution must dominate the integration in the calculation of $d_{\rm H}$. Therefore:
\begin{equation}
	d_{\rm H} \approx a\int_{t_i}^{t_I}dt\frac{e^{H_I(t_I - t)}}{a_I} = \frac{a}{a_IH_I}(e^N - 1)\;.
\end{equation}
Since $d_{\rm A} \approx a/H_0$ for small scale factors, we can conclude that:
\begin{equation}
	\frac{d_{\rm H}}{d_{\rm A}} \approx \frac{H_0}{a_IH_I}e^N\;,
\end{equation}
and so, in order to have $d_{\rm H} > d_{\rm A}$, we obtain the condition:
\begin{equation}
	\boxed{\frac{a_IH_I}{H_0} < e^N}
\end{equation}
which is the same as in Eq.~\eqref{inflationcondition}. The same solution for two different problems is a good point in favor of inflation. 

Usually, a third problem concerning standard cosmology goes along with the above two, and it is related to the abundance of unwanted relics, such as magnetic monopoles, which are produced via some symmetry breaking mechanism in the very early universe and are not observed today. See \cite{Weinberg:2008zzc} for more details. We do not treat this issue here, but it should be clear that inflation provides a mechanism for diluting any unwanted relic beyond the possibility of observation.

We will now discuss how inflation can be realized.

\section{Single scalar field slow-roll inflation}

We present here the possibility of implementing inflation via a single canonical scalar field $\varphi$, named \textbf{the inflaton}, which is subject to a potential $V(\varphi)$ with the property that the scalar field initially rolls slowly down it, attaining its minimum after the end of inflation. This kind of behavior is called \textbf{slow-roll}, and a single slowly rolling canonical scalar field seems to be the favored model of inflation \cite{Martin:2013nzq, Planck:2018jri}.\index{Inflation!Inflaton}\index{Inflation!Slow-roll}

The Lagrangian of a canonical scalar field is:
\begin{equation}
	\mathcal L = \frac{1}{2}g^{\mu\nu}\partial_\mu\varphi\partial_\nu\varphi + V(\varphi)\;,
\end{equation}
where the plus sign might be deceiving (we are used to a difference between the kinetic term and the potential term in Lagrangian mechanics) if we do not recall that the signature in use is $(-,+,+,+)$.

The energy-momentum tensor of a canonical scalar field has the following form:
\begin{equation}\label{Tmunucanphi}
	T^\alpha{}_\beta = g^{\alpha\nu}\partial_\nu\varphi\partial_\beta\varphi - \delta^\alpha{}_\beta\left[\frac{1}{2}g^{\mu\nu}\partial_\mu\varphi\partial_\nu\varphi + V(\varphi)\right]\;,
\end{equation}
where $V(\varphi)$ is some potential. In the background FLRW metric $ds^2 = -dt^2 + a(t)^2\gamma_{ij}dx^idx^j$, the scalar field must depend only on $t$, and thus the energy density and pressure are:
\begin{equation}\label{rhophiPphi}
	\rho_\varphi = -T^0{}_0 = \frac{1}{2}\dot\varphi^2 + V(\varphi)\;, \qquad P_\varphi = \frac{1}{3}\delta^i{}_jT^j{}_i = \frac{1}{2}\dot\varphi^2 - V(\varphi)\;.
\end{equation}

\hrulefill

\begin{ex} Derive the Klein-Gordon equation for $\varphi$ using the continuity equation. Show that:
\begin{equation}\label{KGeq}\index{Scalar field!Klein-Gordon equation}
	\ddot\varphi + 3H\dot\varphi + V_{,\varphi} = 0\;,
\end{equation}
where $V_{,\varphi} \equiv dV/d\varphi$. Show that in the conformal time
\begin{equation}\label{KGeqconftime}
	\varphi'' + 2\mathcal H\varphi' + a^2V_{,\varphi} = 0\;.
\end{equation}
\end{ex}

\hrulefill

During the inflationary phase, the scalar field density is supposed to dominate over the densities of the other components of the universe, if they are present. Therefore, the Friedmann equations are as follows:
\begin{equation}
	H^2 = \frac{8\pi G}{3}\left[\frac{1}{2}\dot\varphi^2 + V(\varphi)\right] - \frac{K}{a^2}\;, \qquad \frac{\ddot a}{a} = -\frac{8\pi G}{3}\left[\dot\varphi^2 - V(\varphi)\right]\;.
\end{equation}
We now need to understand under which conditions $H$ is almost constant. Clearly, the first one is that:
\begin{align}
    \frac{8\pi G}{3}\rho_\varphi \gg \frac{K}{a^2}\,.
\end{align}
Otherwise, there would be no inflation, and introducing a scalar field would be useless. So, the density of the scalar field \textit{must dominate} the energy density associated with the spatial curvature. This effectively amounts to taking $K = 0$. Note that this does not mean that we start with zero spatial curvature. In fact, at the end of inflation, the scalar field would disappear, possibly decaying into various particle species, and the spatial curvature will have, in principle, to be taken into account again. On the other hand, due to the fact that $H$ has been constant for a sufficiently long time, $\Omega_K$ will be sufficiently small for $\Omega_{K0}$ to be compatible with the observational results, as we have seen in Subsection \ref{subsec:flatnessproblem}.

With the above caveat, let us take $K = 0$ from now on and tackle the following problem: in order to have an accelerated phase of expansion induced by the scalar field, we need $H$ to slowly vary, $\dot H \sim 0$. How do we better quantify this condition?

\hrulefill

\begin{ex}
	Show that the acceleration equation can be written in the following form:
	\begin{equation}\label{Hdotphidotrel}
		\dot H = \frac{\ddot a}{a} - H^2 = -4\pi G\dot\varphi^2\;.
	\end{equation}
\end{ex}

\hrulefill

Since the only time scale available in the problem is $1/H$ itself, we ask that the following condition holds true:
\begin{equation}\label{slowrollcondH}
	-\frac{\dot H}{H^2} = \frac{d}{dt}\left(\frac{1}{H}\right) \ll 1\;,
\end{equation}
because this is the only dimensionless combination possible with $\dot H$. It means that, during an expansion time $1/H$, the relative variation of $H$ is much less than unity. Using the Friedmann equation and the expression \eqref{Hdotphidotrel} for $\dot H$, we can write the condition \eqref{slowrollcondH} as follows:
\begin{align}
    \frac{3\dot\varphi^2}{\dot\varphi^2 + 2V(\varphi)} \ll 1\,,
\end{align}
which amounts to:
\begin{equation}\label{slowrollcond}\index{Slow-roll!Conditions}
	\boxed{\dot\varphi^2 \ll V(\varphi)}
\end{equation}
This is the first \textbf{slow-roll condition}: the kinetic term of the scalar field is negligible with respect to the potential. From Eq.~\eqref{rhophiPphi}, this implies that: 
\begin{equation}
	P_\varphi \approx -\rho_\varphi \approx -V(\varphi) \approx \mbox{constant}\;.
\end{equation}
That is, not unexpectedly, the potential of the scalar field behaves like a cosmological constant. 

Note that condition \eqref{slowrollcond} does not necessarily require the potential $V(\varphi)$ to be flat, as, for example, the Coleman-Weinberg (Erick, not Steven) potential \cite{Coleman:1973jx} used in the ``new'' inflationary scenario \cite{Albrecht:1982wi}. The kinetic term can be much smaller than the potential, even for $V(\varphi)\propto \varphi^2$ or $\varphi^4$. We shall see this in more detail later.

The condition given in Eq.~\eqref{slowrollcondH} is usually reformulated in terms of a parameter $\epsilon$, called \textbf{slow-roll parameter}, and defined as follows:
\begin{equation}\label{epsilonparam}
	\boxed{\epsilon \equiv -\frac{\dot H}{H^2} = \frac{d}{dt}\left(\frac{1}{H}\right)}
\end{equation}
For an exponential expansion, $H$ is constant (de Sitter space), and thus $\epsilon = 0$. The slow-roll condition is then $\epsilon \ll 1$. During the radiation-dominated epoch, $a \propto \sqrt{t}$ and thus $1/H = 2t$ and $\epsilon = 2$.\index{Slow-roll!Parameters} 

\hrulefill

\begin{ex} The minimal requirement for inflation is that it must produce an acceleration, i.e. $\ddot a > 0$. For the limiting case in which we have $\ddot a = 0$, show that $\epsilon = 1$.	
\end{ex}

\hrulefill

\begin{figure}[ht]
\center
\includegraphics[width=\columnwidth]{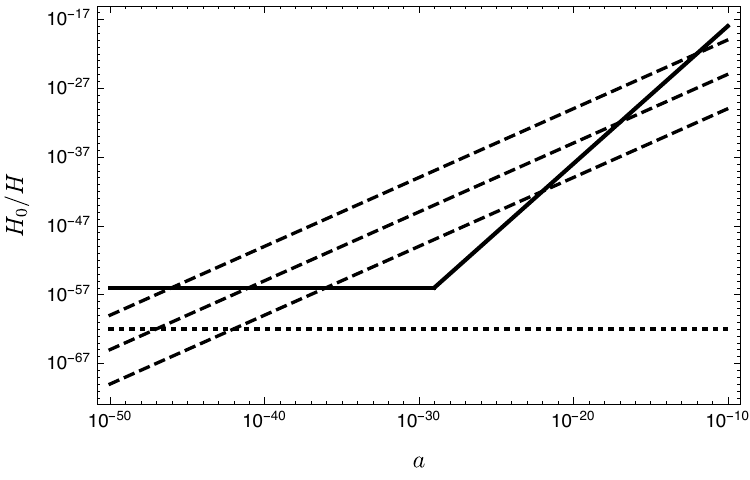}
\caption{The solid line represents the evolution of the Hubble radius $1/H$, normalized to the present one $1/H_0$ for the $\Lambda$CDM model. The dashed lines represents different scales, which grow as $a$, exiting the Hubble radius and then re-entering again during the radiation-dominated epoch. The dotted line represents the Planck scale. We have chosen inflation to end at $a = 10^{-29}$, corresponding to the GUT scale $10^{16}$ GeV.}
\label{Fig:HubbleRadiusInflation}
\end{figure}

In Fig.~\ref{Fig:HubbleRadiusInflation}, the dashed lines represent generic physical scales $\lambda^{\rm phys} \propto a/k$ that grow proportionally to the scale factor. Since $1/H$ remains almost constant (it is actually constant in the figure) during inflation, some physical scales cross the Hubble radius, becoming super-horizon scales. After the end of inflation, these scales cross the Hubble radius again, becoming sub-horizon scales once more. This happens because, during radiation-domination $1/H \propto a^2$ and during matter-domination $1/H \propto a^{3/2}$; both growths are faster than $a$. At late times, when the cosmological constant dominates, $1/H$ tends again to a constant ($\propto 1/\sqrt{\Lambda}$); therefore, in principle, scales would exit the Hubble radius again. On the other hand, it must be understood that the scales we are discussing here are the Fourier modes of small fluctuations; therefore, the above description of exiting and entering the Hubble radius makes sense only as long as we are dealing with small fluctuations. We shall see later that when a perturbative scale crosses the Hubble horizon during inflation, it attains a specific value that remains constant during its super-horizon evolution and then serves as the initial condition when re-entering the horizon during the radiation-dominated epoch.

According to the scheme outlined above, the last scale to exit is also the first one to re-enter. Therefore, we can set a lower bound on the scale factor of exiting and, thus, an upper bound on the number of e-folds to which we have, in principle, observational access. If a scale enters the horizon today, then $\lambda_0^{\rm phys} = 1/H_0$. When it exited the horizon during inflation, $\lambda_I^{\rm phys} = 1/H_I$. Therefore, the scale factor at the time of exit is:
\begin{equation}
	a_{\rm exit} = \frac{H_0}{H_I}\;.
\end{equation} 
Using Eq.~\eqref{energyscaleinflation}, we can thus write down the maximum number of e-folds to which we could have access observationally as:
\begin{equation}
	N_{\rm max} = \ln\left(\frac{a_I}{a_{\rm exit}}\right) = \ln\left(\frac{a_IH_I}{H_0}\right) = \ln\left(\frac{\rho_I^{1/4}}{0.037\;h\;\mbox{eV}}\right)\;.
\end{equation}
For the energy scale of GUT, $10^{16}$ GeV, one obtains $N_{\rm max} \approx 61$. Note that this is not the duration of inflation. It can last much longer (for example, $N \approx 145$ as computed in \cite{Bolliet:2017czc}), but we can observationally probe only the last $61$ e-folds or so, depending on the energy scale of inflation.\index{Inflation!Maximum number of e-folds}

In Fig.~\ref{Fig:HubbleRadiusInflation}, the horizontal dotted line represents the Planck scale. In the same figure, three scales are displayed, growing from below the Planck scale to beyond the Hubble radius and then reentering after the end of inflation during the radiation-dominated epoch. In this latter phase, such scales are observable, for example, as fluctuations in the CMB. On the other hand, they originated in a trans-Planckian regime! This is generally considered a problem known as \textbf{the Trans-Planckian problem} because we are treating gravitational fluctuations as if they were classical when, perhaps, they should be treated according to quantum gravity.\footnote{On the other hand, such a problem is not unexpected if inflation is ``dangerously'' placed close to the Planck scale.} See \cite{Brandenberger:2012aj} for a recent review on the subject.\index{Inflation!Trans-Planckian problem}

\subsection{More slow-roll parameters}

It is not only important that the inflaton slow-rolls, but also that it does so for a sufficiently long time to provide at least $N = 60$ e-folds. In order to make this claim more quantitative, let us investigate how $\epsilon$ varies by computing its time-derivative. Using the definition \eqref{epsilonparam}, we get:
\begin{equation}
	\dot\epsilon = \frac{2\dot H^2}{H^3} - \frac{\ddot H}{H^2}\;,
\end{equation}
and from Eq.~\eqref{Hdotphidotrel} we get:
\begin{equation}
	\dot\epsilon = 2H\epsilon^2 + \frac{8\pi G\dot\varphi}{H^2}\ddot\varphi = 2H\epsilon^2 - 2\frac{\dot H\ddot\varphi}{H^2\dot\varphi} = 2H\epsilon^2 + 2\epsilon\frac{\ddot\varphi}{\dot\varphi}\;.
\end{equation}
Let us define the \textbf{second slow-roll parameter} as follows:
\begin{equation}\label{etadefinition}
	\boxed{\eta \equiv -\frac{1}{H}\frac{\ddot\varphi}{\dot\varphi}}
\end{equation}
We follow the standard notation in which, unfortunately, the second slow-roll parameter is indicated by the same Greek letter as the conformal time. When confusion might arise, we will indicate the conformal time with $\tau$. The derivative of $\epsilon$ can thus be written as follows:
\begin{equation}\label{dotepsiloneq}
	\dot\epsilon = 2H\epsilon(\epsilon - \eta)\;.
\end{equation}
The smallness of $\epsilon$ and $\eta$ then makes $\epsilon$ vary slowly. The requirement $\eta \ll 1$ provides, from the KG equation \eqref{KGeq}, the following \textbf{second slow-roll condition}: 
\begin{equation}\label{secondslowrollcond}
	\boxed{3H\dot\varphi \approx -V_{,\varphi}}
\end{equation}
Considering $\epsilon,\eta \ll 1$ as first-order quantities, the time-derivative $\dot\epsilon$ is thus a second-order quantity, cf. Eq.~\eqref{dotepsiloneq}.

Nothing prevents us from considering now the time-derivative of $\eta$ and then defining a third slow-roll parameter $\alpha$, of which we could again consider the time-derivative, defining a fourth slow-roll parameter $\beta$, and so on, constructing a hierarchy of slow-roll parameters. This was done in \cite{Liddle:1994dx}, but we limit ourselves here to $\epsilon$ and $\eta$, upon which the predictions for the scalar and tensor spectral indices depend, as we shall see. It must be stressed, on the other hand, that future observations might be able to constrain with great precision the \textit{running} of the spectral indices, which depend on the higher-order slow-roll parameters, namely $\alpha$ and $\beta$ mentioned above; see \cite{Munoz:2016owz}.

Since it is the scalar field that triggers the inflationary phase, it is useful to express $\epsilon$ and $\eta$ in terms of quantities related to the scalar field itself, namely the potential $V(\varphi)$ and its derivatives. This can be done as follows. Combining the definition of $\epsilon$ in Eq.~\eqref{epsilonparam} with the Friedmann equation and Eq.~\eqref{Hdotphidotrel}, one gets:
\begin{equation}\label{epsilonslowroll1}
	\epsilon = \frac{3\dot\varphi^2}{2V + \dot\varphi^2} = \frac{3\dot\varphi^2}{2V} + \mathcal{O}\left[\left(\frac{\dot\varphi^2}{2V}\right)^2\right]\;,
\end{equation}
where we are implementing the slow-roll condition \eqref{slowrollcond}. Using now Eq.~\eqref{secondslowrollcond}, we find:
\begin{equation}\label{epsilonVdefinition}
	\boxed{\epsilon \approx \frac{1}{16\pi G}\left(\frac{V_{,\varphi}}{V}\right)^2 \equiv \epsilon_V}
\end{equation}
at the lowest order in $\dot\varphi^2/2V$. In this equation, we have defined $\epsilon_V$ as a quantity describing the steepness of the inflaton potential. 

Since $\eta$ depends on the second time derivative of the scalar field, we expect it to depend on the second derivative of the potential (the curvature $V_{,\varphi\varphi}$) in the slow-roll limit. Using Eq.~\eqref{secondslowrollcond} and the above definition \eqref{etadefinition}, we can write:
\begin{equation}
	\eta \approx \frac{V_{,\varphi\varphi} + 3\dot H}{3H^2}\;.
\end{equation}
Now, recalling the definition of $\epsilon$ and that, at the lowest order in the slow-roll approximation, $3H^2 \approx 8\pi G V$, we can conclude that:
\begin{eqnarray}\label{etaVdefinition}
	\boxed{\eta + \epsilon \approx \frac{1}{8\pi G}\frac{V_{,\varphi\varphi}}{V} \equiv \eta_V}
\end{eqnarray}
Thus, there exists a hierarchy of slow-roll parameters based on the derivatives of the Hubble factor and another one based on those of the potential. They can, of course, be related, as we did above for the first two slow-roll parameters.

So, in order to have $\epsilon_V,\eta_V \ll 1$ and thus trigger inflation, $V$ does not necessarily have to be constant; rather, its first and second derivatives must be much smaller than the value of $V$ itself.

\subsection{Relation between the number of e-folds and the slow-roll parameters}

We can relate the number of e-folds to the slow-roll parameter $\epsilon$ as follows. Recalling that the number of e-folds is $N = \ln a$, we can immediately write that:
\begin{equation}
	\dot N = H \qquad \Rightarrow \qquad \Delta N_{12} = \int_{t_1}^{t_2}Hdt\;,
\end{equation}
where $t_1 < t_2$ are two generic instants during the inflationary phase. By changing the variable in favor of the inflaton field, we can write:
\begin{equation}
	\Delta N_{12} = \int_{\varphi_1}^{\varphi_2}\frac{H}{\dot\varphi}d\varphi\;,
\end{equation}
where $\varphi_1 \equiv \varphi(t_1)$ and $\varphi_2 \equiv \varphi(t_2)$. Using the slow-roll conditions presented earlier, one has:
\begin{equation}\label{numberofefoldsbetween1and2}
	\Delta N_{12} = 8\pi G\int_{\varphi_2}^{\varphi_1}\frac{V}{V_{,\varphi}}d\varphi\;. 
\end{equation}
Since $V/V_{,\varphi}$ is almost constant and very large during inflation (its square is proportional to $1/\epsilon$), we can pull it out of the integral and approximate the above equation as:
\begin{equation}
	\Delta N_{12} \approx \frac{8\pi G V}{V_{,\varphi}}(\varphi_1 - \varphi_2) = \frac{V}{V_{,\varphi}}\frac{\varphi_1 - \varphi_2}{M^2_{\rm Pl}}\;,
\end{equation}
where we have introduced the Planck mass $M_{\rm Pl}$ instead of $1/\sqrt{8\pi G}$. In order to produce a large $N$, it might be possible that $|\varphi_1 - \varphi_2| > M_{\rm Pl}$, but this does not necessarily lead to a trans-Planckian problem because the energy scale of inflation is $V(\varphi)$, not the field itself, and so it is the potential that has to be smaller than the Planck scale in order for a classical treatment to be valid. This can be achieved, for example, if there is a sufficiently small coupling constant.

Using Eq.~\eqref{epsilonVdefinition}, we can generally write that:
\begin{equation}
	\Delta N = \frac{1}{\sqrt{2\epsilon}}\frac{\Delta\varphi}{M_{\rm Pl}}\;. 
\end{equation}
Suppose that $\varphi_1$ and $\varphi_2$ are the values of the inflaton field for which the wavenumbers corresponding to the CMB multipoles $\ell = 1$ and $\ell = 100$ exit the horizon. Since, as we are going to see in Chapter~\ref{Chap:CMBEvo}, $\ell \propto k$, then:
\begin{equation}
	\Delta N = \ln 100 \approx 4.6\;.
\end{equation}
The above formula, then, gives: 
\begin{equation}
	\frac{\Delta\varphi}{M_{\rm Pl}} \approx 6.5\sqrt{\epsilon}\;. 
\end{equation}
We will see later in this chapter that $\varepsilon$ also regulates the ratio between the amounts of gravitational waves and scalar fluctuations produced during inflation. Because of cosmic variance, a sufficiently large $\epsilon$ is needed for us to have the possibility of detecting primordial gravitational waves. This establishes a lower bound on the total variation of the inflaton field, known as \textbf{the Lyth bound} \cite{Lyth:1996im}. See also \cite{DiMarco:2017ihz}. According to \cite{Lyth:1996im}, one needs $\epsilon > 4.4\times 10^{-3}$, and therefore $\Delta\varphi\gtrsim 0.5M_{\rm Pl}$. Inflationary models in which the excursion of the inflaton field is near or super-Planckian are called \textbf{Large-field inflation}.\index{Lyth bound}\index{Large-field Inflation}

\subsection{Reheating}

When $\epsilon_V$ and $\eta_V$ cease to be very small and attain values of order unity, inflation ends. Suppose that the slow-rolling phase ends in proximity of a local minimum $\varphi_0$ of the potential $V(\varphi)$. Close to $\varphi_0$, the inflaton potential can be expanded in the following way:
\begin{equation}
	V(\varphi) = V_0 + \frac{1}{2}\left.V_{,\varphi\varphi}\right|_{\varphi = \varphi_0}(\varphi - \varphi_0)^2 + \dots \approx \frac{1}{2}m_\varphi^2(\varphi - \varphi_0)^2\;,
\end{equation}
where $V_0 \equiv V(\varphi_0)$ is the minimum of the potential, which we assume to be very small; in fact, negligible, in order not to generate an important vacuum energy contribution (which dominates only at late times). We have also introduced the inflaton mass as the second derivative of the potential evaluated at $\varphi_0$.

The above approximated potential is that of a harmonic oscillator with proper frequency $m_\varphi$, and so we expect the inflaton to perform oscillations about the minimum, damped by the Hubble flow $H$. Moreover, since the radiation-dominated epoch must start after the end of inflation, we need to couple the inflaton field to other matter fields for the inflaton to lose energy in favor of the latter, thereby heating up, or reheating, the universe.\footnote{The use of the prefix ``re'' in this context seems to be more or less obsolete, as in the case of recombination.}

Treating the inflaton field and the matter content in the fluid approximation, we can write the following coupled continuity equations:
\begin{eqnarray}
	\dot\rho_\varphi + 3H\rho_\varphi(1 + w_\varphi) = -\Gamma\rho_\varphi\;,\\
	\dot\rho_{\rm M} + 3H\rho_{\rm M}(1 + w_{\rm M}) = \Gamma\rho_\varphi\;,
\end{eqnarray}
where $\Gamma$ is some scattering rate governing the decay of the inflaton, and with $M$ we refer to matter in general, which has to be relativistic in order to give rise to a radiation-dominated epoch, and thus $w_{\rm M} \approx 1/3$.

Therefore, the inflaton oscillations are also damped by the presence of $\Gamma$. This final phase of inflation, transitioning to the radiation-dominated epoch, is called \textbf{reheating}.\index{Reheating}

\hrulefill

\begin{ex}
	Assuming $w_\varphi$ constant, show that:
	\begin{equation}
		\rho_\varphi(t) = \rho_\varphi(t_I)\left[\frac{a(t_I)}{a(t)}\right]^{3(1 + w_\varphi)}\exp\left(-\int_{t_I}^tdt'\Gamma(t')\right)\;,
	\end{equation}
\end{ex}
where $t_I$ is the time at which inflation ends.

\hrulefill

With this formal solution, we can find another formal one for the material part.

\hrulefill

\begin{ex}
	Assuming $w_{\rm M} = 1/3$, show that:
	\begin{equation}
		\rho_M(t) = \frac{\rho_\varphi(t_I)a(t_I)^{3(1 + w_\varphi)}}{a(t)^4}\int_{t_I}^tdt'\Gamma(t')a(t')^{1 - 3w_\varphi}\exp\left(-\int_{t_I}^{t'}dt''\Gamma(t'')\right)\;.
	\end{equation}
\end{ex}

\hrulefill

From the above equation, one sees that the energy density of the matter fields is zero at the end of inflation; then it rises more or less abruptly (depending on $\Gamma$) and finally decreases again as $1/a^4$ after the inflation has given up all its energy. Now, assume $\Gamma$ to be constant and $w_\varphi = 0$, for simplicity. We get for the matter density:
\begin{equation}
	\rho_M(t) = \frac{\rho_\varphi(t_I)\Gamma a(t_I)^{3}}{a(t)^4}\int_{t_I}^tdt'a(t')\exp\left[-\Gamma(t' - t_I)\right]\;.
\end{equation}
Integrating once by parts and considering the limit $\Gamma(t - t_I) \gg 1$, i.e., a very large decay rate, we obtain:
\begin{equation}
	\rho_M(t) = \frac{\rho_\varphi(t_I) a(t_I)^{4}}{a(t)^4}\;.
\end{equation}
This means that the decay takes place so rapidly that all the energy of the inflaton is passed to the matter.

If, on the other hand, $\Gamma(t - t_I) \ll 1$, i.e., a very small decay rate, we can approximate the matter density as:
\begin{equation}
	\rho_M(t) \approx \frac{\rho_\varphi(t_I)\Gamma a(t_I)^{3}}{a(t)^4}\int_{t_I}^tdt'a(t')\;,
\end{equation}
and since $\Gamma$ is so small, the inflaton is still dominating the dynamics, thus giving for the scale factor
\begin{equation}
	a(t) = a(t_I)(t/t_I)^{2/3}\;,
\end{equation}
because we have chosen $w_\varphi = 0$. Integrating, we have thus:
\begin{equation}
	\rho_M(t) \approx \frac{3}{5}\frac{\rho_\varphi(t_I)\Gamma t_I}{(t/t_I)^{8/3}}\left[(t/t_I)^{5/3} - 1\right]\;,
\end{equation}
and the maximum is attained at $t_{\rm max}/t_I = (8/3)^{3/5}$, and its value is:
\begin{equation}
	\rho_{M,\rm max}(t) \approx 0.139\frac{\Gamma}{H(t_I)}\rho_\varphi(t_I)\;.
\end{equation}
Hence, in this case, the energy density of matter is much smaller than that of the inflaton. Most of the latter is still spent driving the expansion of the universe instead of generating matter because of the small $\Gamma$.

We have seen how to produce an inflationary phase and how to quantify it through the slow-roll parameters. Now we are going to see what inflation has to say about quantum fluctuations. We hypothesize that before inflation, the universe was quantum and that quantum fluctuations were turned into classical ones by inflation itself; however, we do not address the details of the quantum-to-classical transition of the primordial fluctuations. About the latter topic, see e.g. \cite{Kiefer:2008ku, Sudarsky:2009za}.\index{Quantum-to-classical transition}

\section{Production of gravitational waves during inflation}\label{Sec:ProductionofGWduringinfl} 

Consider the equation that we found for the evolution of gravitational waves, i.e., from Eq.~\eqref{Gijtens}:
\begin{equation}\label{GWequationwithnabla}
	h^{T''}_{ij} + 2\mathcal Hh^{T'}_{ij}  - \nabla^2 h^T_{ij} = 0\;,
\end{equation} 
where we recall that $h^T_{ij}$ is traceless and transverse, i.e.
\begin{equation}
	h^T_{ii} = 0\;, \qquad \partial^j h^T_{ij} = 0\;,
\end{equation}
and no source term $\pi_{ij}^T$ appears in the GW equation since this vanishes for a scalar field. This can be understood intuitively, since the inflaton $\varphi$ is a scalar and thus unable to produce a tensor perturbation. Mathematically, in Eq.~\eqref{deltaTijphi}, no anisotropic stresses appear in the $\delta T^i{}_j$ of a scalar field, since this is diagonal. 

The above Eq.~\eqref{GWequationwithnabla} has no source term, is gauge-invariant and paves the way for a similar treatment that we shall conduct for scalar perturbations. For this reason, we tackle the tensor ones first.

The Fourier transform of the above equation is Eq.~\eqref{GWeqconftime}, which we write here in the following compact form:
\begin{equation}\label{heqGW}
	h'' + 2\mathcal Hh' + k^2h = 0\;,
\end{equation}
where $h(\eta,k)$ represents either $h_{+,\times}$ or $h(\lambda = \pm 2)$. Since the above equation only contains the modulus $k$, only the initial condition on $h$ shall have a dependence on $\mathbf k$, cf. Eq.~\eqref{ICnormalisationh}. 

It shall be useful to cast Eq.~\eqref{heqGW} in the form of a harmonic oscillator with no damping term by defining:
\begin{equation}\label{htildetransf}
	\boxed{g \equiv \frac{ah}{\sqrt{32\pi G}} = \frac{M_{\rm Pl}}{2}ah}
\end{equation}
The normalization comes in order to give $g$ dimensions of mass. Indeed, $h$ is dimensionless and $\sqrt{8\pi G} = 1/M_{\rm Pl}$. This is necessary to provide the correct dimensionality (a cube length or inverse cube mass) to the power spectrum.

A very direct way to see that $\sqrt{32\pi G}$ is the correct normalization is, e.g., to look at the calculations in \cite{Mukhanov:1990me}. Here it is found the action that yields Eq.~\eqref{GWequationwithnabla} by perturbing up to the second order a $f(R)$ action about a generic FLRW background. In the Einstein-Hilbert case $f(R) = R$ and for a spatially flat FLRW background, one has:
\begin{equation}
	\delta_2S = \frac{1}{64\pi G}\int a^2\left(h^{Ti'}{}_kh^{Tk'}{}_i - \partial_l h^{Ti}{}_k\partial^lh^{Tk}{}_i\right)d^4x\;.
\end{equation}
In \cite{Mukhanov:1990me}, the corresponding second order action for the scalar field is also found, and the factor outside the integral is $1/2$. Hence, in order to properly compare tensor and scalar fluctuations, we must rescale the former as:
\begin{equation}
	h^{Ti}{}_j \to \frac{h^{Ti}{}_j}{\sqrt{32\pi G}}\;.
\end{equation}

\hrulefill

\begin{ex} Use Eq.~\eqref{htildetransf} into Eq.~\eqref{heqGW} in order to find:
\begin{equation}\label{geqGW}
	\boxed{g'' + \left(k^2 - \frac{a''}{a}\right)g = 0}
\end{equation}
\end{ex}

\hrulefill

The gravitational wave can be expressed as:
\begin{eqnarray}\label{GWexpansionplanewaves}
	h^T_{ij}(\eta,\mathbf x) = \sum_{\lambda=\pm 2}\int\frac{d^3\mathbf k}{(2\pi)^3}\left[h(\eta,k)e^{i\mathbf k\cdot\mathbf x}a(\mathbf k,\lambda)e_{ij}(\hat{k},\lambda)\right. \nonumber\\
	\left. h^*(\eta,k)e^{-i\mathbf k\cdot\mathbf x}a^*(\mathbf k,\lambda)e^*_{ij}(\hat{k},\lambda)\right]\;,
\end{eqnarray}
where the sum is over the helicities and $e_{ij}(\hat{k},\lambda)$ is the polarization tensor defined in Sec.~\ref{Sec:EEtenspert}. We have put the initial dependence on $\mathbf k$ in $a(\mathbf k,\lambda)$ and its complex conjugate, which is introduced for $h^T_{ij}(\eta,\mathbf x)$ to be real. 

Now, we assume the initial state of the $h$ field to be a quantum one on very small scales $k \gg aH$. Thus, we promote $h^T_{ij}$ and $a$ to operators and impose the canonical commutation relations:
\begin{equation}\label{commutationrelationsGW}
	\left[a(\mathbf k,\lambda), a(\mathbf k',\lambda')\right] = 0\;, \qquad \left[a(\mathbf k,\lambda), a^\dagger(\mathbf k',\lambda')\right] = (2\pi)^3\delta^{(3)}(\mathbf k - \mathbf k')\delta_{\lambda\lambda'}\;,
\end{equation}
which tell us that $a^\dagger(\mathbf k,\lambda)$ creates a graviton of momentum-energy $k$ and helicity $\lambda$, whereas $a(\mathbf k,\lambda)$ destroys it. 

The quantum state of the universe during inflation is typically assumed to be the vacuum $|0\rangle$, which, by definition, is annihilated by $a(\mathbf k,\lambda)$. The expectation value on the vacuum:
\begin{equation}\label{VEVGW}
	\langle 0|h^T_{ij}(\eta,\mathbf x)h^T_{lm}(\eta,\mathbf x')|0\rangle\;,
\end{equation}
is what shall define our primordial spectrum. Two comments are in order here. First of all, we could choose another quantum state for the universe that is not necessarily the vacuum state. Second, the above \eqref{VEVGW} is a vacuum expectation value, which we expect to become an ensemble average over classical random fields well outside the horizon.

\hrulefill

\begin{ex}Compute the vacuum expectation value in Eq.~\eqref{VEVGW} using the plane-wave expansion \eqref{GWexpansionplanewaves} and the commutation relations \eqref{commutationrelationsGW}. Show that:
\begin{eqnarray}
	\langle 0|h^T_{ij}(\eta,\mathbf x)h^T_{lm}(\eta,\mathbf x')|0\rangle = \int\frac{d^3\mathbf k}{(2\pi)^3}|h(\eta,k)|^2e^{i\mathbf k\cdot(\mathbf x - \mathbf x')}\Pi_{ij,lm}(\hat{k})\;,
\end{eqnarray}
where
\begin{equation}
	\Pi_{ij,lm}(\hat k) \equiv \sum_{\lambda = \pm 2}e_{ij}(\hat{k},\lambda)e^*_{lm}(\hat{k},\lambda)\;,
\end{equation}
is the sum over the helicities, defined in Eq.~\eqref{sumofthehelicities}.
\end{ex}

\hrulefill

Comparing the result found in the exercise with Eq.~\eqref{xiensembleqaverage}, it is quite straightforward to see that if the tensor quantum perturbations are Gaussian, we completely characterize them by their power spectrum:
\begin{equation}
	P_h(\eta,k) \propto |h(\eta,k)|^2\;.
\end{equation}
Hence, we need to solve Eq.~\eqref{heqGW}, or rather Eq.~\eqref{geqGW}, in order to determine $|h(\eta,k)|^2$.

There are two regimes of interest. The first is for short wavelengths $k^2 \gg a''/a$, for which the equation becomes:
\begin{equation}\label{geqveryshortwavelengths}
	g'' + k^2g = 0\;, \qquad (k^2 \gg a''/a)\;,
\end{equation}
which is the usual harmonic oscillator equation. This means that the details of the inflationary evolution, encoded in $a(\eta)$, are not important on very small scales since the time scale of the variation of $g$ is much smaller than the expansion rate of the universe. Therefore, we can treat the quantization of $g$ as that of a free field in Minkowski space. In this case, one has independent modes:
\begin{equation}\label{initialconditionsgminkowski}
	g(\eta,k) = \frac{1}{\sqrt{2k}}e^{-ik\eta}\;, \quad g^*(\eta,k) = \frac{1}{\sqrt{2k}}e^{ik\eta}\;, \qquad (k^2 \gg a''/a)\;,
\end{equation}
where the normalization $1/\sqrt{2k}$ comes from the quantization procedure in Minkowski space \cite{Weinberg:1995mt}. We can take this solution as the ``initial condition'' in solving Eq.~\eqref{geqGW}, therefore addressing only those modes that satisfy the condition $k^2 \gg a''/a$ during the inflationary epoch.

The second regime of evolution is for long wavelengths $k^2 \ll |a''/a|$, for which Eq.~\eqref{geqGW} becomes:
\begin{equation}\label{geqsverylargewavelenghts}
	g'' - \frac{a''}{a}g = 0\;, \qquad (k^2 \ll |a''/a|)\;.
\end{equation}
This equation has the following formal solution:
\begin{equation}\label{smallksol}
	g(\eta,k) = C_1(k)a + C_2(k)a\int^\eta\frac{d\bar\eta}{a(\bar\eta)^2}\;, \qquad (k^2 \ll |a''/a|)\;,
\end{equation}
characterized by two $k$-dependent functions that are to be determined by a match with the short wavelength solution. Equation \eqref{smallksol} contains the solution:
\begin{equation}
	\frac{g}{a} = \frac{h}{\sqrt{32\pi G}} = \mbox{constant}\;,
\end{equation}
representing the constant value that $h$ reaches outside the horizon and that will eventually become the initial condition upon re-entry. The other independent mode:
\begin{align}
    a\int^\eta\frac{d\bar\eta}{a(\bar\eta)^2}\,,
\end{align}
is, in general, decaying because $a \propto 1/|\eta|$ during the inflationary phase, with $\eta$ negative and decreasing in modulus, as we will see briefly. So, the non-constant independent mode scales as $1/a^2$ and exponentially vanishes.

The existence of the above constant solution for $h$ is not a fortuitous event. It is a particular case of a much more general result due to Weinberg \cite{Weinberg:2003sw, Weinberg:2003ur, Weinberg:2008zzc}, by virtue of which, \textit{regardless of the content of the universe, for $k \ll \mathcal{H}$ there is always one constant solution for $h$}.\index{Weinberg's theorem}

Recall that $k$ is the comoving wavenumber and is a constant. Therefore, how is it possible that, due to inflation, a given $k$ starts satisfying the condition $k^2 \gg |a''/a|$ and afterwards satisfies the condition $k^2 \ll |a''/a|$? This is due to the behavior of $|a''/a|$, of course. From a dimensional argument, $|a''/a| \propto 1/\eta^2$ and, again, $\eta$ negative and decreasing in modulus. Therefore, $|a''/a|$ grows until $k^2 \gg |a''/a|$ is broken and $k^2 \ll |a''/a|$ is satisfied.

Let us try to be more quantitative with a first simple calculation. 

\hrulefill

\begin{ex} Assume that the inflationary phase is described by a constant $H = a'/a^2$, say $H_\Lambda$. Show that: 
\begin{equation}\label{aofetadesitterspace}
	a(\eta) = -\frac{1}{H_\Lambda\eta}\;.
\end{equation}
Since the scale factor and $H_\Lambda$ are positive, then $\eta$ must be negative. Moreover, since $a$ grows, then $\eta$ must decrease in modulus. 
\end{ex}

\hrulefill

Since:
\begin{equation}
	\frac{a''}{a} =  2H_\Lambda^2a^2 = \frac{2}{\eta^2}\;,
\end{equation}
Equation~\eqref{geqGW} then becomes:
\begin{equation}\label{geqGWdeSitter}
	g'' + \left(k^2 - \frac{2}{\eta^2}\right)g = 0\;.
\end{equation}
In this case, the equation can be solved exactly, so we do not need to consider the two separate cases $k^2 \gg |a''/a|$ and $k^2 \ll |a''/a|$.

\hrulefill

\begin{ex} Show that a solution of Eq.~\eqref{geqGWdeSitter} can be written as:
\begin{equation}
	g(\eta,k) = C(k)\left(1 - \frac{i}{k\eta}\right)e^{-i\eta k}\;.
\end{equation}
\end{ex}

\hrulefill

When $k^2 \gg |a''/a| = 2/\eta^2$, that is $k|\eta| \gg 2$, the matching with Eq.~\eqref{initialconditionsgminkowski} yields:
\begin{equation}
	C(k) = \frac{1}{\sqrt{2k}}\;,
\end{equation}
and so we can write:
\begin{equation}
	g(\eta,k) = \frac{1}{\sqrt{2k}}\left(1 - \frac{i}{k\eta}\right)e^{-i\eta k}\;.
\end{equation}
Using now Eq.~\eqref{htildetransf}, we obtain the power spectrum:
\begin{equation}
	P_h(\eta,k) = \frac{32\pi G}{a(\eta)^2}P_g(\eta,k) = \frac{16\pi G}{a(\eta)^2k}\left(1 + \frac{1}{k^2\eta^2}\right) = \frac{16\pi G}{k}H_\Lambda^2\eta^2\left(1 + \frac{1}{k^2\eta^2}\right)\;,
\end{equation}
and the dimensionless one, according to the definition of Eq.~\eqref{dimensionlesspowerspectrum}, is:
\begin{equation}
	\Delta^2_h(\eta,k) = \frac{k^3P_h(\eta,k)}{2\pi^2} = \frac{8\pi G}{\pi^2}H_\Lambda^2\left[1 + (k^2\eta^2)\right] = \frac{H_\Lambda^2}{\pi^2M_{\rm Pl}^2}\left[1 + (k^2\eta^2)\right]\;.
\end{equation}
This spectrum depends on time, so which $\eta$ should we choose? We have seen that when a scale $k$ exits the horizon, its corresponding $h(\eta,k)$ becomes constant. Its value will be the initial condition during the radiation-dominated epoch, when the scale re-enters the horizon. The condition $k/a \ll H$ translates to $k|\eta| \ll 1$. Therefore, for $k\eta \to 0$, we have:
\begin{equation}\label{powerspectrumdesittercase}
	\Delta^2_h(\eta,k) = \frac{H_\Lambda^2}{\pi^2M_{\rm Pl}^2}\;,
\end{equation}
which no longer depends on the scale $k$: it is a \textbf{scale-invariant} power spectrum. 

Note that for a constant $H_\Lambda$, the slow-roll parameter $\epsilon$ is vanishing. In order to describe the inflationary phase more realistically, we should consider a small but non-vanishing $\epsilon$. In this case, $H$ varies, and a $k$-dependence is gained by the power spectrum.

Let us write the definition of $\epsilon$ in Eq.~\eqref{epsilonparam}, but using the conformal time:
\begin{equation}
	\epsilon = \left(\frac{a}{\mathcal H}\right)'\frac{1}{a} = 1 - \frac{\mathcal H'}{\mathcal H^2}\;.
\end{equation}

\hrulefill

\begin{ex}
	Solve the above differential equation for $\mathcal H$ assuming a constant $\epsilon$. Show that:
	\begin{equation}\label{mathcalHforetaconstant}
		\mathcal H = -\frac{1}{(1 - \epsilon)\eta}\;.
	\end{equation}
\end{ex}

\hrulefill

From this solution, it is not difficult to show that:
\begin{equation}
	\frac{a''}{a} = \mathcal H' + \mathcal H^2 = \frac{1}{(1 - \epsilon)\eta^2} + \frac{1}{(1 - \epsilon)^2\eta^2} \approx \frac{2 + 3\epsilon}{\eta^2}\;,
\end{equation}
where in the last approximation we have considered $\epsilon \ll 1$, as it should be during inflation, and kept $\epsilon$ to first order.

Substituting into Eq.~\eqref{geqGW}, we then get:
\begin{equation}\label{geq2plus3epsilon}
	g'' + \left(k^2 - \frac{2 + 3\epsilon}{\eta^2}\right)g = 0\;,
\end{equation}
which can be solved exactly, providing a general solution as a combination of Hankel functions:
\begin{equation}
	g(\eta,k) = C_1(k)\sqrt{-\eta}H^{(1)}_\nu(-k\eta) + C_2(k)\sqrt{-\eta}H^{(2)}_\nu(-k\eta)\;, \quad \nu = \frac{\sqrt{3}}{2}\sqrt{3 + 4\epsilon}\;,
\end{equation}
where the minus sign must be introduced to account for the fact that $\eta < 0$. We have chosen to express the solution in terms of the Hankel functions since it is easier to see which one of them matches the initial condition of Eq.~\eqref{initialconditionsgminkowski}. Indeed, consider the asymptotic expansion, cf. \cite{Abramowitz1972}, of $H_\nu^{(1)}(-k\eta)$:
\begin{equation}
	H_\nu^{(1)}(-k\eta) \sim \sqrt{\frac{2}{-\pi k\eta}}e^{-ik\eta - i\nu\pi/2 -i\pi/4}\;, \qquad (k|\eta| \gg 1)\;.
\end{equation}
Hence, this is the correct behavior at very small scales for $g$, and the integration constant must be:
\begin{equation}
	C_1(k) = \frac{\sqrt{\pi}}{2}e^{i\nu\pi/2 + i\pi/4}\;.
\end{equation}
The solution we are looking for is as follows:
\begin{equation}
	g(\eta,k) = \frac{\sqrt{\pi}}{2}e^{i\nu\pi/2 + i\pi/4}\sqrt{-\eta}H^{(1)}_\nu(-k\eta)\;, \quad \nu = \frac{\sqrt{3}}{2}\sqrt{3 + 4\epsilon}\;,
\end{equation}
and the power spectrum is thus:
\begin{equation}
	P_h(\eta,k) = \frac{4}{M_{\rm Pl}^2a(\eta)^2}|g(\eta,k)|^2 = \frac{\pi|\eta|}{M_{\rm Pl}^2a(\eta)^2}|H^{(1)}_\nu(-k\eta)|^2\;.
\end{equation}
For $k|\eta| \to 0$, one has that
\begin{equation}
	H^{(1)}_\nu(-k\eta) \sim -i\frac{\Gamma(\nu)}{\pi}\left(\frac{k|\eta|}{2}\right)^{-\nu}\;, \qquad (k|\eta| \to 0)\;,
\end{equation}
and hence the dimensionless power spectrum becomes:
\begin{equation}
	\Delta^2_h(\eta, k) = \frac{k^3|\eta|\Gamma(\nu)^2}{2\pi^3 M_{\rm Pl}^2a^2}\left(\frac{k|\eta|}{2}\right)^{-2\nu}\;.
\end{equation}
Considering $\epsilon$ at first order in the exponent of $k|\eta|$ and negligible elsewhere, we can write:
\begin{equation}\label{Delta2h-2epsilon}
	\Delta^2_h(\eta, k) = \frac{k^3|\eta|\Gamma(\nu)^2}{2\pi^3M_{\rm Pl}^2 a^2}\left(\frac{k|\eta|}{2}\right)^{-3 - 2\epsilon} = \frac{1}{\pi^2 M_{\rm Pl}^2a^2|\eta|^2}\left(k|\eta|\right)^{- 2\epsilon}\;.
\end{equation}
Now, the power spectrum is no longer scale-invariant but has gained a small $k$-dependence: a power law with an exponent $-2\epsilon$. The latter is known as \textbf{the tensor spectral index} and is denoted as $n_T$. We shall see a little more detail about it later.\index{Spectral index!Tensor}

On large scales, $h$ is time-independent, and so is its power spectrum. Hence, we can choose any convenient value of $\eta$ when evaluating $\Delta_h^2$. It is customary to use the time $\eta_k$ at which a given scale $k$ crosses the horizon:
\begin{equation}
	k = \mathcal H(\eta_k)\;.
\end{equation}
Using Eq.~\eqref{mathcalHforetaconstant}, the horizon crossing condition gives the following relation between $|\eta|$ and $k$:
\begin{equation}
	k = \mathcal H(\eta_k) = \frac{1}{(1 - \epsilon)|\eta_k|}\;,
\end{equation}
which at first order in $\epsilon$ gives:
\begin{equation}
	k|\eta_k| = 1 + \epsilon\;.
\end{equation}
Using again Eq.~\eqref{mathcalHforetaconstant}, we can write the power spectrum evaluated at horizon crossing as:
\begin{equation}
	\Delta^2_h(k) = \frac{1}{\pi^2 M_{\rm Pl}^2}\frac{\mathcal H(\eta_k)^2}{a^2(\eta_k)}\;,
\end{equation}
which is usually formulated as follows:
\begin{equation}\label{Deltah2}
	\Delta^2_h(k) =  \left.\frac{H^2}{\pi^2M_{\rm Pl}^2}\right|_{k = aH}\;.
\end{equation}
It is very interesting to note that if we could directly measure the gravitational wave background, we would be able to determine the energy scale of inflation.

\section{Production of scalar perturbations during inflation}

Consider the perturbed FLRW metric with scalar perturbations written in the conformal Newtonian gauge:
\begin{equation}
	ds^2 = a(\eta)^2\left[-\left(1 + 2\Psi\right)d\eta^2 + \left(1 + 2\Phi\right)\delta_{ij}dx^idx^j\right]\;.
\end{equation}
We have again set $K = 0$, but it should be understood as $\mathcal{H}^2 \gg K$, as explained in the previous section.

Since the metric is perturbed, we must also consider the perturbations of the inflaton field:
\begin{equation}\label{scalarfieldexp}
	\boxed{\varphi = \bar{\varphi}(\eta) + \delta\varphi(\eta,\textbf{x})}
\end{equation}

\hrulefill

\begin{ex} Using Eqs.~\eqref{Tmunucanphi} and \eqref{scalarfieldexp}, write down the perturbed energy-momentum tensor. Show that:
\begin{eqnarray}
\label{deltaT00phi}	T^0{}_0 = -\frac{1}{2a^2}\bar{\varphi}^{'2} - V(\varphi) - \frac{1}{a^2}\bar{\varphi}'\delta\varphi' + \frac{1}{a^2}\Psi\bar{\varphi}^{'2} - V_{,\varphi}\delta\varphi\;,\\
\label{deltaT0iphi}	T^0{}_i = -\frac{1}{a^2}\bar{\varphi}'\partial_i\delta\varphi\;,\\
\label{deltaTijphi}	T^i{}_j = \delta^i{}_j\left[\frac{1}{2a^2}\bar{\varphi}^{'2} - V(\varphi)\right] + \delta^i{}_j\left(\frac{1}{a^2}\bar{\varphi}'\delta\varphi' - \frac{1}{a^2}\Psi\bar{\varphi}^{'2} - V_{,\varphi}\delta\varphi\right)\;.
\end{eqnarray}
\end{ex}

\hrulefill

The perturbed contributions in the above expressions are grouped to the right. Note the following important fact: $\delta T^i{}_j$ is diagonal, and thus no anisotropic stresses can be sourced by a single scalar field. From Eq.~\eqref{anisotropicstressscalarequation}, this implies $\Psi = -\Phi$.

\hrulefill

\begin{ex} Obtain the perturbed Klein-Gordon equation. Start from:
\begin{equation}
	\nabla_\mu T^\mu{}_\nu = 0\;,
\end{equation}
and then put $\nu = 0$. Show that:
\begin{equation}\label{pertKGeq}
	\boxed{\delta\varphi'' + 2\mathcal H\delta\varphi' + \left(k^2 + V_{,\varphi\varphi}a^2\right)\delta\varphi = 2\Phi V_{,\varphi}a^2 - 4\Phi'\bar\varphi'}
\end{equation}
where we have used the result $\Psi = -\Phi$. Compare this equation with the corresponding one found in \cite{Mukhanov:1990me}. It is also useful to consider that found in \cite{Weinberg:2008zzc} and recover the above one by transforming from the cosmic time to the conformal one.
\end{ex}

\hrulefill

Unfortunately, the contributions $V_{,\varphi\varphi}a^2$, $2\Phi V_{,\varphi}a^2$, and $4\Phi'\bar\varphi'$ make life more difficult. Indeed, if they were absent, we would get the following equation:
\begin{equation}\label{niceformpertKGeq}
	\delta\varphi'' + 2\mathcal H\delta\varphi' + k^2\delta\varphi = 0\;,
\end{equation}
which is formally identical to Eq.~\eqref{heqGW} and, therefore, all the calculations performed for the tensor case would follow in the same fashion. Even if we could neglect $V_{,\varphi\varphi}$ in Eq.~\eqref{pertKGeq} during the slow-roll phase, there would be no \textit{a priori} reason to neglect the whole right hand side of Eq.~\eqref{pertKGeq}. 

Should we proceed with computing the power spectrum for $\delta\varphi$? If yes, how do we treat it at the end of inflation and during reheating? And how do we relate it to the primordial modes of Chapter \ref{Chap:IC} for the standard components of the universe (photons, neutrinos, CDM, and baryons)?

These problems can be bypassed if we switch our analysis from $\delta\varphi$ to $\mathcal R$ because, as we saw in Eq.~\eqref{Rzetaprimordial2}, for adiabatic perturbations, $\mathcal R$ sources all the other perturbations. Moreover, $\mathcal R$ is conserved on large scales, see Appendix~\ref{App:Rconslargescales}. This means that we just need to determine $\mathcal{R}$ at the horizon exit during inflation, and we automatically have our primordial adiabatic mode, ready to source the perturbations in photons, neutrinos, CDM, and baryons.

But what about the isocurvature modes? How do we treat these? Weinberg's theorem \cite{Weinberg:2003sw, Weinberg:2003ur, Weinberg:2008zzc} also applies to scalar modes: \textit{regardless of the content of the universe, for $k \ll \mathcal{H}$ there are always two adiabatic constant modes, one of which is $\mathcal{R}$}.\index{Weinberg's theorem} This means that in the case of single scalar field inflation, \textit{there are no isocurvature modes}. In fact, in this class of models, there are only two independent modes for scalar perturbations. By virtue of Weinberg's theorem, these must be adiabatic. To allow for the production of isocurvature modes, one needs to consider more complicated models of inflation, such as multi-field inflation \cite{Wands:2007bd}.

\hrulefill

\begin{ex} Express $v$ in Eq.~\eqref{Rperturb} in terms of the inflaton field velocity potential. From Eq.~\eqref{genT0imixed} we know that:
\begin{equation}
	T^{0(\varphi)}{}_{i}= \left(\bar\rho_\varphi + \bar{P}_\varphi\right)v^{(\varphi)}_{i} = \frac{\bar\varphi'^2}{a^2}v^{(\varphi)}_{i}\;.
\end{equation}	
Using Eq.~\eqref{deltaT0iphi}, show that:
\begin{equation}
	v^{(\varphi)}_{i} = -\frac{\partial_i\delta\varphi}{\bar\varphi'}\;.
\end{equation}
Thus, writing 
\begin{equation}\label{scalarfieldvelocitydecomposition}
	v^{(\varphi)}_{i} = \partial_i v^{(\varphi)}\;,
\end{equation}
we can identify the velocity potential of the inflation field (we drop the superscript $\varphi$ now) as:
\begin{equation}\label{vscalarfield}
	v = -\frac{\delta\varphi}{\bar\varphi'}\;.
\end{equation}
\end{ex}

\hrulefill

Using the results of the exercise, we can write:
\begin{equation}
	\mathcal R = \Phi -\frac{\mathcal H}{\bar\varphi'}\delta\varphi\;.
\end{equation}
We need to find an evolution equation for $\mathcal R$. In principle, by combining the above expression with the Klein-Gordon equation \eqref{pertKGeq}, we can eliminate $\delta\varphi$. Then, $\Phi$ can be addressed using the Einstein equations, which are:
\begin{eqnarray}
	3\mathcal H(\Phi' + \mathcal H\Phi) + k^2\Phi = 4\pi G\left(\bar\varphi'\delta\varphi' + \Phi\bar\varphi^{'2} + V_{,\varphi}a^2\delta\varphi\right)\;,\\
	\Phi' + \mathcal H \Phi = -4\pi G\bar\varphi'\delta\varphi\;,\\
	\Phi'' + 3\mathcal H\Phi' + (2\mathcal H' + \mathcal H^2)\Phi = -4\pi G\left(\bar\varphi'\delta\varphi' + \Phi\bar\varphi^{'2} - V_{,\varphi}a^2\delta\varphi\right)\;.
\end{eqnarray}
These equations are obtained by combining Eq.~\eqref{relativisticPoissonequation} with Eq.~\eqref{deltaT00phi}, Eq.~\eqref{0iEinsteineq} with Eqs.~\eqref{scalarfieldvelocitydecomposition} and Eq.~\eqref{vscalarfield}, Eq.~\eqref{GiideltaPeq2} with Eq.~\eqref{deltaTijphi}, and using $\Phi = -\Psi$.

Though in principle possible, it involves a mountain of calculations that we can at least limit by choosing a different gauge. Following \cite{Weinberg:2008zzc}, let us consider the following gauge:
\begin{equation}\label{newgaugedeltaphi0}
	ds^2 = -a^2(1 + E)d\eta^2 + 2a^2F_{,i}d\eta dx^i + a^2(1 + A)\delta_{ij}dx^idx^j\;, \qquad \delta\hat\varphi = 0\;.
\end{equation}
This is a gauge in which the perturbed scalar field (which we denote with a hat in the new gauge) is zero. In this gauge one has:
\begin{equation}
	\mathcal R = \frac{A}{2}\;.
\end{equation}
So, our objective is to find a closed equation for $A$. The price to pay in order to set the perturbed scalar field equal to zero is to introduce one more scalar perturbation in the metric. Now, let us calculate the energy-momentum tensor in the new gauge.

\hrulefill

\begin{ex} Using Eqs.~\eqref{Tmunucanphi} and \eqref{newgaugedeltaphi0}, write down the perturbed energy-momentum tensor. Show that:
\begin{eqnarray}
\label{deltaT00phinewgauge}	\hat{T}^0{}_0 = -\frac{1}{2a^2}(1 - E)\bar{\varphi}^{'2} - V(\varphi)\;,\\
\label{deltaT0iphinewgauge}	\hat{T}^0{}_i = 0\;, \qquad \hat{T}^i{}_0 = \frac{1}{a^2}F_{,i}\bar\varphi^{'2}\;,\\
\label{deltaTijphinewgauge}	\hat{T}^i{}_j = \delta^i{}_j\left[\frac{1}{2a^2}(1 - E)\bar{\varphi}^{'2} - V(\varphi)\right]\;.
\end{eqnarray}
\end{ex}

\hrulefill

The above is a quite simple energy-momentum tensor, and indeed we are now going to exploit its simplicity. First of all, since $\delta \hat{T}^0{}_i = 0$, we have from Eq.~\eqref{deltaG0igen} that:
\begin{equation}\label{eq1formathcalR}
	-\mathcal HE + A' = 0\;,
\end{equation}
that is a simple algebraic relation between two of the three scalar potentials.

The second relation that we are going to use is the continuity equation.

\hrulefill

\begin{ex}
	Compute the continuity equation $\nabla_\nu T^\nu{}_0 = 0$ using the energy-momentum tensor given in Eqs.~\eqref{deltaT00phinewgauge}-\eqref{deltaTijphinewgauge}. Compute from the metric in Eq.~\eqref{newgaugedeltaphi0} the only necessary Christoffel symbol:
	\begin{equation}
		\Gamma^i_{0j} = \mathcal H\delta^i{}_j + \frac{A'}{2}\delta^i{}_j\;. 
	\end{equation}
	Then show that:
	\begin{equation}\label{eq2formathcalR}
		\frac{a^2}{2}\left[\frac{E}{a^2}(\mathcal H^2 - \mathcal H')\right]' + (\mathcal H^2 - \mathcal H')\left(\nabla^2F + 3\mathcal HE - \frac{3}{2}A'\right) = 0\;, 
	\end{equation}
	where we have used the relation $\mathcal H^2 - \mathcal H' = 4\pi G\bar\varphi^{'2}$.
\end{ex}

\hrulefill

As a final relation, we shall use $\delta R_{ij}$ computed from the metric in Eq.~\eqref{newgaugedeltaphi0}. So, using Eq.~\eqref{deltaRijgeneralhmunu}, we have:
\begin{eqnarray}
	\delta R_{ij} = -\frac{1}{2}E_{,ij} - \frac{\mathcal H}{2}E'\delta_{ij} - (2\mathcal H^2 + \mathcal H')E\delta_{ij} - \frac{1}{2}(\nabla^2 A\delta_{ij} + A_{,ij})\nonumber\\ + \frac{1}{2}A''\delta_{ij} + \frac{5}{2}\mathcal HA'\delta_{ij} + (2\mathcal H^2 + \mathcal H')A\delta_{ij}\nonumber\\ - \mathcal H\nabla^2F\delta_{ij} -F'_{ij} - 2\mathcal HF_{,ij}\;.
\end{eqnarray}
Through the Einstein equations, we know that:
\begin{equation}
	\delta R_{ij} = 8\pi G\left(\delta T_{ij} - a^2A\delta_{ij}\frac{\bar T}{2} - a^2\delta_{ij}\frac{\delta T}{2}\right)\;,
\end{equation}
where $\bar T$ and $\delta T$ are the background and perturbed traces, respectively, of the energy-momentum tensor.

\hrulefill

\begin{ex}
	Calculate the right hand side of the above equation. Show that:
	\begin{equation}
		\delta R_{ij} = 8\pi G a^2A\delta_{ij}V = (2\mathcal H^2 + \mathcal H')A\delta_{ij}\;.
	\end{equation}
\end{ex}

\hrulefill

Extracting only the part of $\delta R_{ij}$ that is proportional to $\delta_{ij}$, we finally obtain:
\begin{eqnarray}\label{eq3formathcalR}
	- \frac{\mathcal H}{2}E' - (2\mathcal H^2 + \mathcal H')E - \frac{1}{2}\nabla^2 A + \frac{1}{2}A'' + \frac{5}{2}\mathcal HA' - \mathcal H\nabla^2F = 0\;.
\end{eqnarray}

\hrulefill

\begin{ex}
	Combine Eqs.~\eqref{eq1formathcalR}, \eqref{eq2formathcalR} and \eqref{eq3formathcalR}, eliminating $E$ and $F$ and thus finding the following equation for $\mathcal R$:
	\begin{equation}
		\mathcal R'' + \frac{2}{\mathcal H}\mathcal R'\left[\mathcal H^2 - \mathcal H' + \frac{\mathcal H}{2}\frac{(\mathcal H^2 - \mathcal H')'}{\mathcal H^2 - \mathcal H'}\right] + k^2\mathcal R = 0\;.
	\end{equation}
	Show that this equation can be written in a more compact form as:
	\begin{equation}\label{MukhanovSasakiequation}
		\boxed{\mathcal R'' + 2\frac{z'}{z}\mathcal R' + k^2\mathcal R = 0}
	\end{equation}
	with
	\begin{equation}
		z \equiv \frac{a\bar\varphi'}{\mathcal H}\;,
	\end{equation}
    not to be confused with the redshift.
\end{ex}\index{Mukhanov-Sasaki equation}

\hrulefill

Equation~\eqref{MukhanovSasakiequation} is called \textbf{the Mukhanov-Sasaki equation}, cf. \cite{Mukhanov:1985rz} and \cite{Sasaki:1986hm}. Notice that this equation has the very same structure as Eq.~\eqref{heqGW} if one makes the change $a \to z$. Therefore, a similar analysis applies, and a constant solution for $\mathcal R$ is allowed when:
\begin{equation}
	k^2 \ll |z''/z|\;,
\end{equation}
which is a similar but not identical condition to $k^2 \ll |a''/a|$. In fact, note that:
\begin{equation}
	z^2 = a^2\left(\frac{\bar\varphi'}{\mathcal H}\right)^2 = a^2\frac{3\bar\varphi'^2}{8\pi G(\bar\varphi'^2/2 + Va^2)} = a^2\frac{\epsilon}{4\pi G}\;,
\end{equation}
where we have used Eq.~\eqref{epsilonslowroll1}, written in the conformal time. The above relation between $z$ and $a$ is exact, and it invokes the slow-roll parameter $\epsilon$.

Now, the procedure that we need in order to obtain the power spectrum for $\mathcal R$ is the same as the one used for $h$. We promote $\mathcal R$ to a quantum field and compute its vacuum expectation value, which will become the power spectrum itself. The vacuum expectation value is computed first by solving the Mukhanov-Sasaki equation, choosing the correct Minkowski behavior at large $k$.

Finally, we can use the same result from Eq.~\eqref{Deltah2}, remembering to divide by the $32\pi G$ factor that we have introduced in order to give dimensionality to $h$, and multiplying by $a^2/z^2 = 4\pi G/\epsilon$ in order to recover the $2z'/z$ factor in front of $\mathcal R'$, instead of $2\mathcal H$.

Thus, we can finally write down the scalar power spectrum as:
\begin{equation}\label{PRandDeltaR}
	\boxed{\left.P_\mathcal{R} = \frac{H^2}{4M_{\rm Pl}^2\epsilon k^3}\right|_{k=aH}} \qquad \boxed{\left.\Delta^2_\mathcal{R} = \frac{H^2}{8\pi^2M_{\rm Pl}^2\epsilon}\right|_{k=aH}}
\end{equation}
This is the most important result of this chapter since it provides a prediction that can be (and indeed is) tested observationally. Using Eq.~\eqref{Rzetaprimordial2}, one can relate this power spectrum to those of the other quantities that become relevant at horizon re-entry during the radiation-dominated epoch. For example, using Eq.~\eqref{adiabaticPhip}, we can write that:
\begin{equation}
	\Delta_\Phi^2 = \frac{4(5 + 2R_\nu)^2}{(15 + 4R_\nu)^2}\Delta^2_{\mathcal R}\;.
\end{equation}
In general, $\Phi$ is not constant on large scales during inflation; however, it is indeed constant on large scales during radiation domination, as we saw in Chapter~\ref{Chap:IC}. 

Now, let us perform a calculation similar to that in the previous section, culminating in Eq.~\eqref{Deltah2}.

\hrulefill

\begin{ex}
	Employing Eq.~\eqref{dotepsiloneq} written in the conformal time, show that:
	\begin{equation}
		\frac{z'}{z} = \mathcal H(1 + \epsilon - \eta)\;.
	\end{equation}
Afterwards, assume $\epsilon$ and $\eta$ to be constant. Then, use Eq.~\eqref{mathcalHforetaconstant} and recast the Mukhanov-Sasaki equation as follows:
	\begin{equation}
		\mathcal R'' - \frac{2(1 + 2\epsilon - \eta)}{\tau}\mathcal R' + k^2\mathcal R = 0\;,
	\end{equation}
	where we are now employing $\tau$ as conformal time, in order to avoid confusion with the second slow-roll parameter $\eta$. Finally, show that the above equation can be written as:
	\begin{equation}
		(z^2\mathcal R)'' + \left(k^2 - \frac{2 + 6\epsilon - 3\eta}{\tau^2}\right)(z^2\mathcal R) = 0\;.
	\end{equation}
\end{ex}

\hrulefill

This equation has the same form as Eq.~\eqref{geq2plus3epsilon}, and hence a similar analysis applies. In particular, its solution is of the form:
\begin{equation}
	z^2(\tau)\mathcal R(\tau, k) = C_1(k)\sqrt{-\tau}H^{(1)}_\nu(-k\tau) + C_2(k)\sqrt{-\tau}H^{(2)}_\nu(-k\tau)\;,
\end{equation}
with order
\begin{equation}
	\nu = \frac{\sqrt{3}}{2}\sqrt{3 + 8\epsilon - 4\eta}\;.
\end{equation}

\hrulefill

\begin{ex}
	Following the same steps that we saw in the tensor case, show that:
	\begin{equation}\label{Delta2R-4epsilon+2eta}
		\Delta^2_{\mathcal R}(\tau, k) = \frac{1}{8\pi^2M_{\rm Pl}^2\epsilon}\frac{1}{a^2\tau^2}(k|\tau|)^{-4\epsilon + 2\eta}\;.
	\end{equation}
\end{ex}

\hrulefill

For the scalar case, a different $k$-dependence of the spectrum appears, involving the second slow-roll parameter. The combination $-4\epsilon + 2\eta$ is the first order expression of the \textbf{scalar spectral index}. Evaluating the above spectrum at horizon crossing, we get the result already shown in Eq.~\eqref{PRandDeltaR}.\index{Spectral index!Scalar}

\section{Spectral indices}

It is customary to express the dimensionless scalar and tensor power spectra as follows:
\begin{eqnarray}
\label{Delta2S}	\Delta_S^2 \equiv \Delta_\mathcal{R}^2 \equiv \frac{k^3 P_\mathcal{R}(k)}{2\pi^2} = \left.\frac{H^2}{8\pi^2M_{\rm Pl}^2\epsilon}\right|_{k=aH} \equiv A_S \left(\frac{k}{k_*}\right)^{n_S(k) - 1}\;,\\
\label{Delta2T}	\Delta_T^2 \equiv 2\Delta_h^2 \equiv \frac{k^3 P_h(k)}{\pi^2} = \left.\frac{2H^2}{\pi^2M_{\rm Pl}^2}\right|_{k = aH} \equiv A_T\left(\frac{k}{k_*}\right)^{n_T(k)}\;,
\end{eqnarray}
where the general $k$-dependence (given by the specific model of inflation) is embedded in $n_S(k)$ and $n_T(k)$, the \textbf{scalar spectral index} and \textbf{tensor spectral index}.

We have introduced the pivot scale $k_*$, which is taken to be 0.05 Mpc$^{-1}$ in \cite{Planck:2018jri}, and the factor of 2 in $\Delta_T^2 \equiv 2\Delta_h^2$ accounts for the two polarizations of the tensor modes. The \textbf{spectral amplitudes} $A_S$ and $A_T$, of which only the first is constrained by observations since we do not yet detect the primordial gravitational wave background, are related to the energy scale of inflation.\index{Power spectrum!Amplitude}

Since the spectral indices are not constant, the spectra can be written in the following more general form:
\begin{eqnarray}\label{lnkexpansionspectra}
	\ln\frac{\Delta_S^2}{A_S} = \left[n_S - 1 + \frac{1}{2}\frac{dn_S}{d\ln k}\ln\frac{k}{k_*} + \frac{1}{6}\frac{d^2n_S}{d(\ln k)^2}\left(\ln\frac{k}{k_*}\right)^2 + \cdots\right]\ln\frac{k}{k_*}\;,\\
	\ln\frac{\Delta_T^2}{A_T} = \left[n_T + \frac{1}{2}\frac{dn_T}{d\ln k}\ln\frac{k}{k_*} + \cdots\right]\ln\frac{k}{k_*}\;.
\end{eqnarray}
The derivative of the spectral index with respect to $\ln k$ is called \textbf{the running of the spectral index}. We do not consider running in detail here for simplicity, and Planck has not constrained it very tightly, as we shall see. However, they will probably become of great interest in future CMB experiments \cite{Munoz:2016owz}.\index{Spectral index!Running} 

From the above Eq.~\eqref{lnkexpansionspectra} at first order, it is straightforward to write:
\begin{equation}
	n_S - 1 = \frac{d\ln \Delta_S^2}{d\ln k}\;, \qquad n_T = \frac{d\ln \Delta_T^2}{d\ln k}\;.
\end{equation}
For the tensor case:
\begin{equation}
	n_T = 2\frac{k}{H}\left.\frac{dH}{dk}\right|_{aH= k}\;.
\end{equation}
Recall that $H$ does not depend on the scale in general, as it being a background quantity; however, it is the evaluation at $aH = k$ that makes $H$ depend on $k$. This is because different scales cross the horizon at different times during inflation. In particular, recall from Eq.~\eqref{aofetadesitterspace} that:
\begin{equation}
	\eta = \int_0^a\frac{da}{Ha^2} \approx \frac{1}{H}\int_0^a\frac{da}{a^2} = - \frac{1}{Ha}\;,
\end{equation}
since $H$ is almost constant during inflation. Now, when we evaluate $aH = k$, we find that $\eta = -1/k$. Therefore, we obtain the following for the tensor spectral index:
\begin{equation}
	n_T = 2\frac{k}{H}\frac{dH}{d\eta}\left.\frac{d\eta}{dk}\right|_{aH= k} = 2\frac{1}{kH}\left.\frac{dH}{d\eta}\right|_{aH= k}\;.
\end{equation}
By the definition of $\epsilon$, cf. Eq.~\eqref{epsilonparam}, we know that:
\begin{equation}
	\frac{dH}{d\eta} = -aH^2\epsilon\;,
\end{equation}
so that we can finally determine:
\begin{equation}
	\boxed{n_T = -2\epsilon}
\end{equation}
which is the result already derived in Eq.~\eqref{Delta2h-2epsilon}.

The calculation for the scalar spectral index is essentially the same. We start from:
\begin{equation}
	n_S - 1 = \left.\frac{d\ln (H^2/\epsilon)}{d\ln k}\right|_{aH= k} = -2\epsilon - \left.\frac{d\ln\epsilon}{d\ln k}\right|_{aH= k}\;.
\end{equation}
We can deal with the last term as follows:
\begin{equation}
	\left.\frac{d\ln\epsilon}{d\ln k}\right|_{aH= k} = \frac{k}{\epsilon}\epsilon'\left.\frac{d\eta}{dk}\right|_{aH= k} = \left.\frac{1}{k\epsilon}\epsilon'\right|_{aH= k}\;.
\end{equation}
Using Eq.~\eqref{dotepsiloneq} one has:
\begin{equation}
	\epsilon' = 2aH\epsilon(\epsilon - \eta)\;.
\end{equation}
The scalar spectral index is thus written as:
\begin{equation}\label{scalarespectralindexformula}
	\boxed{n_S - 1 = -4\epsilon + 2\eta = -6\epsilon_V + 2\eta_V}
\end{equation}
Again, in agreement with the result found in Eq.~\eqref{Delta2R-4epsilon+2eta}.

Finally, it is customary to define the \textbf{tensor-to-scalar ratio}:
\begin{equation}\label{tensortoscalarratioformula}\index{Tensor-to-scalar ratio}
	\boxed{r_{*} \equiv \frac{\Delta^2_T(k_*)}{\Delta^2_S(k_*)} = \frac{A_T}{A_S} = 16\epsilon = - 8n_T} 
\end{equation}
and to use $r$ to express the energy scale of Inflation at the time when the pivot scale exits the Hubble radius. From Eq.~\eqref{Delta2T} we have:
\begin{equation}
	H_*^2 = \frac{\pi^2M_{\rm Pl}^2}{2}\Delta^2_T(k_*)\;.
\end{equation}
Using the definition of $r_*$ and Eq.~\eqref{Delta2S}, one obtains:
\begin{equation}
	H_*^2 = \frac{\pi^2M_{\rm Pl}^2}{2}r_*\Delta^2_S(k_*) = \frac{\pi^2M_{\rm Pl}^2}{2}r_*A_S\;,
\end{equation}
from which:
\begin{equation}
	\boxed{V_* = \frac{3\pi^2M_{\rm Pl}^4}{2}r_*A_S}
\end{equation}

\section{Observational results}

In \cite{Planck:2018jri}, one can find recent constraints on the inflationary parameters. They slightly change with respect to the different datasets considered. We report here the spectral index and its running, as well as the running of the running at 68\% CL:
\begin{eqnarray}
	n_S = 0.9587 \pm 0.0056\;,\\
	\frac{dn_S}{d\ln k} = 0.013 \pm 0.012\;,\\
	\frac{d^2n_S}{d(\ln k)^2} = 0.022 \pm 0.012\;,
\end{eqnarray}
using the pivot scale $k_* = 0.05$ Mpc$^{-1}$. For the scalar amplitude at 68\% CL:
\begin{equation}
	\ln(10^{10}A_S) = 3.044 \pm 0.014\;.
\end{equation}
For the tensor-to-scalar ratio:
\begin{equation}\label{rconstraintPlanck}
	r_{0.002} < 0.058\;,
\end{equation}
at a 95\% confidence level. From these numbers, we can express the energy scale of Inflation at the time when the pivot scale exits the Hubble radius as follows:
\begin{equation}
	V_* = \left(1.88 \times 10^{16}\;\mbox{GeV}\right)^4\frac{r}{0.10}\;,
\end{equation}
and therefore obtain an upper bound that corresponds to the GUT energy scale.



\section{Examples of models of inflation}

We conclude this chapter by presenting some important models of inflation, limiting ourselves to the class of single field models. There is a huge number of inflationary models; in \cite{Martin:2013nzq}, 193 of them are analyzed, and it seems that the data favor the simplest category of single field slow-roll inflation that we have presented in this chapter. See also \cite{Tsujikawa:2014rta} for a review of many inflationary models.

\subsection{General power law potential}\index{Inflation!Power-law}

A very simple model of inflation consists of an inflaton field described by the potential:
\begin{equation}\label{powerlawpotentialinflation}
	V(\varphi) = \lambda_n\frac{\varphi^n}{n}\;,
\end{equation}
from which one can easily determine the slow-roll parameters from Eqs.~\eqref{epsilonVdefinition} and \eqref{etaVdefinition}:
\begin{equation}
	\epsilon_V = \frac{n^2}{16\pi G\varphi^2} = \frac{n^2M_{\rm Pl}^2}{16\pi\varphi^2}\;, \qquad \eta_V = \frac{n(n - 1)}{8\pi G\varphi^2} = \frac{n(n - 1)M_{\rm Pl}^2}{8\pi\varphi^2}\;.
\end{equation}
In order for these parameters to be very small and thus trigger inflation, one has to have $\varphi \gg M_{\rm Pl}$; note that this condition does not depend on the coupling. We can put the predictions on the spectral indices as functions of the number of e-folds to the end of inflation by defining the value of the field at which inflation ends as $\epsilon_V(\varphi_f) = 1$.

For the general power law potential, this condition becomes:
\begin{equation}
	\frac{n^2}{16\pi G\varphi_f^2} = 1 \qquad \Rightarrow \qquad \varphi_f = \frac{n}{\sqrt{16\pi G}}\;.
\end{equation}
The number of e-folds to the end of inflation can be obtained exactly from Eq.~\eqref{numberofefoldsbetween1and2}:
\begin{equation}
	N = \frac{8\pi G}{n}\int_{\varphi_f}^{\varphi_i}\varphi\; d\varphi = \frac{4\pi G}{n}\left(\varphi_i^2 - \frac{n^2}{16\pi G}\right)\;,
\end{equation}
from which the initial value of the field is:
\begin{equation}\label{initialvaluefieldpowerlawinflation}
	\varphi_i^2 = \frac{n}{4\pi G}\left(N + \frac{n}{4}\right)\;.
\end{equation}
From this, the slow-roll parameters can be written as:
\begin{equation}
	\epsilon_V = \frac{n}{4N + n}\;, \qquad \eta_V = \frac{2(n - 1)}{4N + n}\;.
\end{equation}
The scalar spectral index and the tensor-to-scalar ratio are as follows:
\begin{equation}
	n_S - 1 = -6\epsilon_V + 2\eta_V = -\frac{2(n + 2)}{4N + n}\;, \quad r = 16\epsilon_V = \frac{16n}{4N + n} = \frac{8n}{n + 2}(1 - n_S)\;.
\end{equation}
If we substitute Eq.~\eqref{initialvaluefieldpowerlawinflation} into Eq.~\eqref{powerlawpotentialinflation}, we get:
\begin{equation}
	V(\varphi) \approx \lambda_n n^{n/2 - 1}N^{n/2}M_{\rm Pl}^n\;.
\end{equation}
In order for the classical treatment to be valid, we need $V(\varphi) \ll M_{\rm Pl}^4$, and hence:
\begin{equation}
	\lambda_n \ll \frac{M_{\rm Pl}^{4 - n}}{n^{n/2 - 1}N^{n/2}}\;.
\end{equation}
For $n = 4$ and $N = 60$, we have $\lambda_4 \ll 10^{-5}$.

\subsection{The Starobinsky model}\index{Inflation!Starobinsky model}

The Starobinsky model \cite{Starobinsky:1979ty} is based on quadratic gravity; that is, a correction to the vacuum action of General Relativity that includes terms proportional to the square of the Riemann tensor $R_{\mu\nu\rho\sigma}R^{\mu\nu\rho\sigma}$, the square of the Ricci tensor $R_{\mu\nu}R^{\mu\nu}$, and the square of the Ricci scalar $R^2$, induced by the quantum back-reaction of the matter fields.

The Starobinsky model is usually framed in the class of modifications of General Relativity known as $f(R)$ gravity. These theories of gravity are described by the following action:
\begin{equation}
	S = \frac{M_{\rm Pl}^2}{2}\int d^4x\sqrt{-g}f(R)\;,
\end{equation}
where $f(R)$ is a generic function of the Ricci scalar. When $f(R) = R$ we recover General Relativity. The above action can be recast as follows:
\begin{equation}\label{JordanFrameaction}
	S = \int d^4x\sqrt{-g}\left[\frac{M_{\rm Pl}^2}{2}\varphi R - V(\varphi)\right]\;,
\end{equation}
i.e., as a \textbf{non-minimally coupled scalar-tensor theory}, where the scalar field $\varphi$ and its potential are defined as:
\begin{equation}
	\varphi \equiv M_{\rm Pl}\frac{df}{dR}\;, \qquad V(\varphi) \equiv \frac{\rm M_{Pl}^2}{2}\left(R\frac{df}{dR} - f\right)\;.
\end{equation}
This kind of theory has a non-minimal coupling because of the term $\varphi R$ (the minimal coupling to geometry occurs only with $\sqrt{-g}$). The action written as in Eq.~\eqref{JordanFrameaction} is also said to be in the \textbf{Jordan frame} and corresponds to the Brans-Dicke theory \cite{Brans:1961sx} with a special choice for its free parameter ($\omega = 0$).\index{Jordan frame} 

\hrulefill

\begin{ex}
Performing the \textbf{conformal transformation} 
\begin{equation}
	\hat{g}_{\mu\nu} = \frac{df}{dR} g_{\mu\nu} = \frac{\varphi}{M_{\rm Pl}} g_{\mu\nu}\;,
\end{equation}\index{Conformal transformation}
show that:
\begin{equation}\label{hatSaction}
	\hat{S} = \int d^4x\sqrt{-g}\left[\frac{M_{\rm Pl}^2}{2}\hat{R} - \frac{1}{2}\hat{g}^{\mu\nu}\partial_\mu\chi\partial_\nu\chi - U(\chi)\right]\;,
\end{equation}
where $\hat R$ is the Ricci scalar computed from $\hat g_{\mu\nu}$ and:
\begin{equation}\label{Upotchidef}
	U \equiv \frac{VM_{\rm Pl}^2}{\varphi^2} = \frac{M_{\rm Pl}^2}{2(df/dR)^2}\left(R\frac{df}{dR} - f\right)\;, \qquad \chi \equiv \sqrt{\frac{3}{2}}M_{\rm Pl}\ln\frac{\varphi}{M_{\rm Pl}}\;.
\end{equation}
Note that, in order for the above definition of $\chi$ to make sense, $df/dR > 0$.
\end{ex}

\hrulefill

The action $\hat{S}$ is said to be in the \textbf{Einstein frame}. Thus, any $f(R)$ theory can be reformulated as a scalar-tensor theory, the scalar field being $\varphi \equiv M_{\rm Pl}df/dR$. The $f(R)$ theories have been intensively studied in the past decades; see the comprehensive reviews \cite{Sotiriou:2008rp} and \cite{DeFelice:2010aj}.\index{Einstein frame} 

Since any $f(R)$ can be conformally mapped in General Relativity plus a canonical scalar field, it seems natural to investigate how this kind of modified gravity accounts for inflation. Indeed, probably the most successful application of a $f(R)$ theory is in the inflationary paradigm, as we are going to show in a moment.\index{$f(R)$ gravity} 

Now it is $\chi$ in action~\eqref{hatSaction} that plays the role of the inflaton. The Starobinsky model \cite{Starobinsky:1979ty} is given by a quadratic $R^2$ correction to the usual Einstein-Hilbert term in the action for gravity:
\begin{equation}
	f(R) = R + \frac{R^2}{6M^2}\;.
\end{equation}
Then, one can easily calculate from Eq.~\eqref{Upotchidef}:
\begin{equation}
	U = \frac{3M_{\rm Pl}^2M^2R^2}{4(R + 3M^2)^2}\;, \qquad \chi = \sqrt{\frac{3}{2}}M_{\rm Pl}\ln\frac{R + 3M^2}{3M^2}\;.
\end{equation}

\hrulefill

\begin{ex}
Invert the above relation for $\chi(R)$, obtaining thus a $R(\chi)$, and substitute it into the expression for $U(\chi)$, obtaining thus the expression for the inflaton potential in the Starobinsky model:
\begin{equation}\label{StarobisnkyPotential}
	\boxed{U(\chi) = \frac{3}{4}M_{\rm Pl}^2M^2\left(1 - e^{-\sqrt{2/3}\chi/M_{\rm Pl}}\right)^2}
\end{equation}
\end{ex}

\hrulefill

\begin{figure}[ht]
\center
\includegraphics[width=\columnwidth]{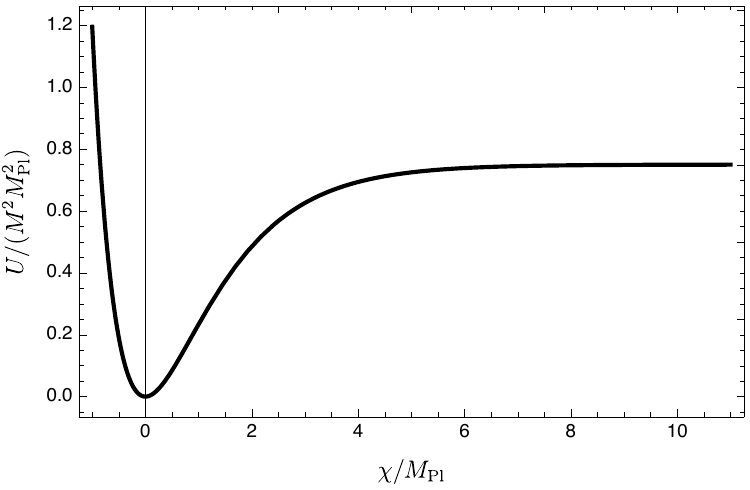}
\caption{Plot of the Starobinsky potential, in Eq.~\eqref{StarobisnkyPotential}.}
\label{Fig:StarobinskyPotential}
\end{figure}

In Fig.~\ref{Fig:StarobinskyPotential} we plot the Starobinsky potential $U(\chi)$ as a function of the scalar field $\chi$. The most important feature is the plateau at large values of the field, which is the ideal feature for a slow-roll evolution of the inflaton field.

The slow-roll parameters are easily calculated from Eq.~\eqref{StarobisnkyPotential} and Eqs.~\eqref{epsilonVdefinition} and \eqref{etaVdefinition}:
\begin{equation}
	\epsilon_U = \frac{4}{3}\frac{e^{-2\sqrt{2/3}\chi/M_{\rm Pl}}}{\left(1 - e^{-\sqrt{2/3}\chi/M_{\rm Pl}}\right)^2}\;, \qquad \eta_U = \frac{4}{3}\frac{2e^{-2\sqrt{2/3}\chi/M_{\rm Pl}} - e^{-\sqrt{2/3}\chi/M_{\rm Pl}}}{\left(1 - e^{-\sqrt{2/3}\chi/M_{\rm Pl}}\right)^2}\;.
\end{equation}
The number of e-folds until the end of inflation is obtained by solving the following integral:
\begin{equation}
	N = \sqrt{\frac{3}{8}}\int_{\chi_f}^{\chi}\frac{d\chi}{M_{\rm Pl}}\left(e^{\sqrt{2/3}\chi/M_{\rm Pl}} - 1\right) \approx \frac{3}{4}e^{\sqrt{2/3}\chi/M_{\rm Pl}}\;,
\end{equation}
so that we can write the slow-roll parameters as:
\begin{equation}
	\epsilon_U = \frac{12}{(4N - 3)^2} \approx \frac{3}{4N^2}\;, \qquad \eta_U = 4\frac{9 - 4N}{(4N - 3)^2} \approx -\frac{1}{N}\;,
\end{equation}
where we have kept the dominant contributions for large $N$. The scalar spectral index and the tensor-to-scalar ratio are thus written, using Eqs.~\eqref{scalarespectralindexformula} and \eqref{tensortoscalarratioformula}, as follows:
\begin{equation}
	n_S = 1 - \frac{2}{N}\;, \qquad r = \frac{12}{N^2}\;.
\end{equation}
Substituting $N = 50$ and $60$, the predictions obtained are in excellent agreement with the Planck constraints, thus making the Starobinsky model successful.\footnote{Many consider the Starobinsky model the correct model for the primordial universe because it partially takes into account the quantum nature of the world. In fact, as mentioned at the beginning of the subsection, quadratic corrections to the Einstein-Hilbert action naturally occur when coupling quantum fields to a classical geometry. Such quadratic corrections, in turn, have inflation as a natural solution.}

\clearpage
\chapter{Evolution of perturbations}\label{Chap:Evopert}

{\rightskip=3truepc\leftskip=3truepc\noindent
{\it On peut braver les lois humaines, mais non r\'esister aux lois naturelles\\ (One can challenge human laws, but not the natural ones)
}
\vskip 0.10 in
\centerline{\it ---Jules Verne, Vingt mille lieues sous le mers}
\vskip 0.20 in
}

In this chapter, we solve some of the equations that we found in Chapter~\ref{Chap:CosmoPertTheory}. In particular, we distinguish 4 cases of evolution:
\begin{enumerate}
	\item On super-horizon scales,
	\item In the matter-dominated epoch,
	\item In the radiation-dominated epoch,
	\item Deep inside the horizon,
\end{enumerate}
for which it is possible to perform analytic calculations and thus gain a clearer physical insight.

Our scope is to understand the shape of the matter power spectrum. Since we already know the form of the primordial power spectrum, cf. Eq.~\eqref{Delta2S}, the above task amounts to determining the matter (CDM plus baryons) transfer function. 

It must be noted that data are derived from the observation of the distribution of galaxies in the sky and hence provide information on the galaxy density contrast $\delta_g$, which is a \textbf{biased tracer}\index{Bias} of the underlying distribution of matter in the sense that the galaxy correlation function is not equal to the total matter correlation function \cite{Kaiser:1984sw}. At low redshift, this bias is usually considered to be constant:
\begin{equation}\label{biasrelationlowz}
	\delta_g = b\delta_{\rm m}\;,
\end{equation} 
with $\delta_{\rm m}$ being the matter density contrast (baryonic plus CDM). However, for larger redshifts, it might be a function of redshift and of the wavenumber, i.e. $b = b(k,z)$. 

Beyond the cosmic variance affecting any large-scale observation in a relevant way, the determination of the power spectrum is also afflicted by another noise, known as \textbf{shot noise}, which is due to the fact that $\delta_g$ is given by a discrete distribution (that of galaxies) tracing a continuous one (that of the underlying matter).\index{Shot noise}

Finally, power spectra are, in general, 3-dimensional in the sense that they are computed from the spatial distribution of galaxies. While determining the angular positions on the celestial sphere is not complicated, the only direct measure of distance that we have is the redshift. It is possible, of course, to transform the redshift into an actual distance (a proper distance, for example) through the cosmological model that we want to test, but determining redshift is time-consuming, especially if it is done via spectroscopy, and it introduces extra errors due to peculiar motions and photometry (if $z$ is determined photometrically). Hence, it is perhaps more convenient to work with a 2-dimensional power spectrum, the angular one $w(\theta)$, since angular positions on the celestial sphere are easily, precisely, and rapidly determined.  


\section{Evolution on super-horizon scales}

A given comoving wavenumber $k$ is super-horizon at a certain conformal time $\eta$ if
\begin{equation}
	\boxed{k\eta \ll 1}
\end{equation}
Since $\mathcal H \propto 1/\eta$, the above condition amounts to:
\begin{equation}
	\boxed{k \ll \mathcal H}
\end{equation}
which can be rewritten for the physical scale as follows:
\begin{equation}
	\boxed{\frac{k}{a} \ll H}
\end{equation}
The super-horizon regime is the same one that we used in Chapter~\ref{Chap:IC} when we investigated the primordial modes. The main difference in what we are going to see here is that we do not limit ourselves to the radiation-dominated epoch but investigate what happens through radiation-matter equality and also through matter-dark energy equality.

Thus, the above conditions can be written as follows in the epochs of interest:
\begin{equation}
	k \ll \mathcal H = \frac{1}{\eta} \quad \Rightarrow \quad k\eta \ll 1\;,
\end{equation}
during the radiation-dominated epoch (for which $a \propto \eta$),
\begin{equation}
 k \ll \mathcal H = \frac{2}{\eta} \quad \Rightarrow \quad k\eta \ll 2\;,
\end{equation}
during the matter-dominated epoch (for which $a \propto \eta^2$) and
\begin{equation}\label{LambdadominationmathcalH}
	k \ll \mathcal H = \frac{H_\Lambda}{1 + H_\Lambda(\eta_0 - \eta)}\;,
\end{equation}
for the $\Lambda$-dominated era. The solution in Eq. \eqref{LambdadominationmathcalH} is similar to that used for the inflationary phase in Chapter \ref{Chap:Inflation}. The difference is that here $\eta > 0$.

We use Eqs.~\eqref{monopolessimp}:
\begin{equation}
	\delta_\gamma' = -4\Phi'\;, \quad \delta_\nu' = -4\Phi'\;, \quad \delta_{\rm c}' = -3\Phi'\;, \quad \delta_{\rm b}' = -3\Phi'\;,
\end{equation}
the Poisson equation is written using Eq.~\eqref{4piGa2H2} as follows:
\begin{eqnarray}\label{PoisseqSHevo}
	\frac{3}{\mathcal{H}}\left(\Phi' - \mathcal{H}\Psi\right) + \frac{k^2}{\mathcal H^2}\Phi = \frac{3}{2\rho_{\rm tot}}\left(\rho_{\rm c}\delta_{\rm c} + \rho_{\rm b}\delta_{\rm b} + \rho_\gamma\delta_\gamma + \rho_\nu\delta_\nu\right)\;,
\end{eqnarray}
and the anisotropic stress equation \eqref{anisotropicstressscalarequation}:
\begin{equation}
	k^2(\Phi + \Psi) = -32\pi G a^2\rho_\nu\mathcal{N}_2\;.
\end{equation}
We could use Eq.~\eqref{4piGa2H2} again in order to highlight the $k^2/\mathcal H^2$ term on the left, but it is not necessary because the neutrino quadrupole is of the same order. Thus, we need to perform differentiations as we did in Chapter~\ref{Chap:IC} to obtain a closed equation for the potentials.

Note that we have neglected the photon quadrupole contribution. It is an approximation motivated by the fact that, before recombination, the tight coupling with electrons washes out $\Theta_2$, whereas, after radiation-matter equality, $R_\gamma$ becomes rapidly ($\propto 1/a \propto 1/\eta^2$) negligible. 

When matter dominates, $R_\nu$ is also negligible, so we expect the potentials to become equal. Since our objective here is to perform an analytic calculation, we already assume $\Phi = -\Psi$. This is, strictly speaking, incorrect when considering the radiation-matter domination transition, but it is acceptable when considering the matter-dark energy one. Therefore, let us rewrite Eq.~~\eqref{PoisseqSHevo} as follows:
\begin{equation}\label{PoisseqSHevo1a}
	3\mathcal{H}\left(\Phi' + \mathcal{H}\Phi\right) = \frac{3\mathcal H^2}{2\rho_{\rm tot}}\left(\rho_{\rm c}\delta_{\rm c} + \rho_{\rm b}\delta_{\rm b} + \rho_\gamma\delta_\gamma + \rho_\nu\delta_\nu\right)\;.
\end{equation}
This equation holds true also in the presence of dark energy, provided that the latter does not cluster (as is the case for $\Lambda$).

\subsection{Evolution through radiation-matter equality}

In the presence of radiation and matter only, the Friedmann equation can be written as follows:
\begin{equation}
	\mathcal H^2 = \frac{8\pi Ga^2}{3}(\rho_{\rm m} + \rho_{\rm r}) = \frac{8\pi G}{3}\rho_{\rm m}\left(1 + \frac{1}{y}\right)\;,
\end{equation}
where we have grouped together the species that evolve in the same way and indicated them with a subscript r, i.e., radiation (photons and neutrinos), and with a subscript m, i.e., matter (CDM and baryons). We have also employed the definition:
\begin{equation}
	y \equiv \frac{\rho_{\rm m}}{\rho_{\rm r}} = \frac{a}{a_{\rm eq}}\;,
\end{equation}
where $a_{\rm eq}$ is the equivalence scale factor. 

\hrulefill

\begin{ex}
Assume adiabaticity, i.e.
\begin{equation}
	\delta_{\rm c} = \delta_{\rm b} = \frac{3}{4}\delta_\gamma = \frac{3}{4}\delta_\nu \equiv \delta_{\rm m}\;.
\end{equation}
Rewrite Eq.~\eqref{PoisseqSHevo1a} using $y$ as independent variable. Show that:
\begin{equation}\label{PoisseqSHevo2}
	y\frac{d\Phi}{dy} + \Phi = \frac{4 + 3y}{6(y + 1)}\delta_{\rm m}\;. 
\end{equation}
\end{ex}

\hrulefill

\begin{ex} We are looking now for a closed equation for $\Phi$. Therefore, differentiate Eq.~\eqref{PoisseqSHevo2} with respect to $y$ and use $\delta_{\rm m}' = -3\Phi'$ in order to find:
\begin{equation}\label{PoisseqSHevo3}
	\boxed{\frac{d^2\Phi}{dy^2} + \frac{(7y + 8)(3y + 4) + 2y}{2y(y + 1)(3y + 4)}\frac{d\Phi}{dy} + \frac{1}{y(y + 1)(3y + 4)}\Phi = 0}
\end{equation}
\end{ex}

\hrulefill

\begin{ex} Quite unexpectedly, the above equation \eqref{PoisseqSHevo3} can be solve exactly. Indeed, use the following transformation \cite{Kodama:1985bj}:
\begin{equation}
	u \equiv \frac{y^3\Phi}{\sqrt{1 + y}}\;,
\end{equation}
and show that
\begin{equation}\label{PoisseqSHevo4}
	\frac{d^2u}{dy^2} + \left[-\frac{2}{y} + \frac{3}{2(y + 1)} - \frac{3}{3y + 4}\right]\frac{du}{dy} = 0\;.
\end{equation}
\end{ex}

\hrulefill

\begin{ex} Integrate once Eq.~\eqref{PoisseqSHevo4} and show that:
\begin{equation}
	\ln\frac{du}{dy} = C_1 + 2\ln y - \frac{3}{2}\ln(y + 1) + \ln(3y + 4)\;,
\end{equation}
where $C_1$ is an integration constant. Upon exponentiation and integrating again show that:
\begin{equation}
	\frac{y^3\Phi}{\sqrt{1 + y}} = A\int_0^ydy\frac{y^2(3y + 4)}{(1 + y)^{3/2}}\;,
\end{equation}
where $A$ is an integration constant related to $C_1$ and assume that $y^3\Phi \to 0$ for $y \to 0$.
\end{ex}

\hrulefill

\begin{ex} Solve the above integration and show that:
\begin{equation}
	\Phi(y) = \frac{\Phi_{\rm P}}{10y^3}\left(16\sqrt{1 + y} + 9y^3 + 2y^2 - 8y - 16\right)\;,
\end{equation}
where $\Phi_{\rm P}$ is the primordial gravitational potential, which we introduced in Eq.~\eqref{adiabaticPhip} in place of $C_\gamma = \mathcal R = \zeta$. Show that $\Phi(y\to 0) = \Phi_{\rm P}$.\index{Evolution of perturbation!Super-horizon}\index{Kodama-Sasaki equation}
\end{ex}

\hrulefill

\begin{figure}[htbp]
\center
\includegraphics[width=\columnwidth]{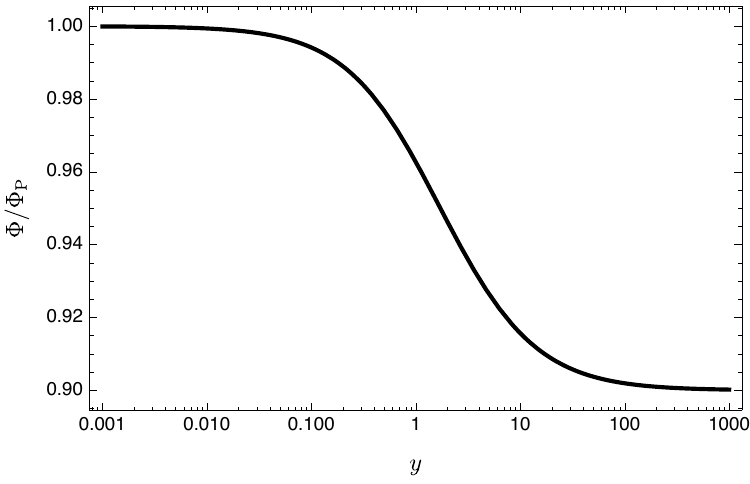}
\caption{Evolution of the gravitational potential $\Phi$ on super-horizon scales ($k = 0$) through radiation-matter equality. From Eq.~\eqref{Phi910res}.}
\label{Fig:PhiSuperHplot}
\end{figure}

From the above solution, we see that for $y \to \infty$ (which here means deep into the matter-dominated epoch) the gravitational potential drops by 10\%, i.e.
\begin{equation}\label{Phi910res}
	\Phi \to \frac{9}{10}\Phi_{\rm P}\;, \qquad \mbox{for } y \to \infty\;.
\end{equation}
In Fig.~\ref{Fig:PhiSuperHplot} we display the evolution of $\Phi/\Phi_{\rm P}$ as a function of $y$, as given by Eq.~\eqref{Phi910res}.

It can be shown that $\Phi$ is constant on super-horizon scales for any background evolution with $w \neq -1$ constant and for adiabatic perturbations \cite{Mukhanov:1990me}.

\hrulefill

\begin{ex}
	Setting $\Phi = -\Psi$, combine Eq.~\eqref{relativisticPoissonequation} with Eq.~\eqref{GiideltaPeq2} and use Eq.~\eqref{entropypert}. Show that:
	\begin{equation}\label{MukhanoveqPhi}
		\Phi'' + 3\mathcal H\left(1 + c_{\rm ad}^2\right)\Phi' + \left[2\mathcal H' + \mathcal H^2\left(1 + 3c_{\rm ad}^2\right)\right]\Phi + k^2c_{\rm ad}^2\Phi = -4\pi Ga^2\Gamma\;.
	\end{equation} 
	where we have defined $c_{\rm ad}^2 \equiv P'/\rho'$ as the \textbf{adiabatic speed of sound}. Show that this equation can be cast as follows:\index{Adiabatic speed of sound}
	\begin{equation}\label{Mukhanovequ}
		\boxed{u'' + \left(k^2c_{\rm a}^2 - \frac{\theta''}{\theta}\right)u = -a^2(\rho + P)^{-1/2}\Gamma}
	\end{equation}
	where
	\begin{equation}\label{uandthetadefinitions}
		u \equiv \frac{\Phi}{4\pi G(\rho + P)^{1/2}}\;, \qquad \theta \equiv \frac{1}{a}\left(\frac{\rho}{\rho + P}\right)^{1/2}\;.
	\end{equation}
\end{ex}

\hrulefill

It is clear that the above transformation cannot treat the cosmological constant, for which $P = -\rho$. Nevertheless, it is interesting to see that Eq.~\eqref{Mukhanovequ} has a form similar to Eq.~\eqref{geqGW}, and hence, for $k = 0$ and $\Gamma = 0$ (adiabatic perturbations), its general solution is:
\begin{equation}\label{usolk0Gamma0}
	u = C_1\theta + C_2\theta\int \frac{d\eta}{\theta^2}\;.
\end{equation}

\hrulefill

\begin{ex}
	Consider a single fluid model $P = w\rho$, with $w$ constant and different from $-1$. Show, from solving the Friedmann equation, that:
	\begin{equation}
		a = (\eta/\eta_0)^{2/(1 + 3w)}\;,
	\end{equation}
where $w \neq -1/3$ (in this case the solution grows exponentially with the conformal time). Show then that:
\begin{equation}
	\theta\int \frac{d\eta}{\theta^2} \propto \frac{a}{\mathcal H}\;,
\end{equation}
and thus show, using Eq.~\eqref{uandthetadefinitions}, that $\Phi$ is constant.
\end{ex}

\hrulefill

Hence, the $9/10$ drop of $\Phi$ between the radiation-dominated era and the matter-dominated one can be extended to any kind of adiabatic fluid with $w \neq - 1$ constant, using the constancy of $\mathcal{R}$ on large scales. See e.g. \cite{Mukhanov:2005sc}. Indeed, from \eqref{RPhirhoPrel} we have
\begin{equation}
	\mathcal{R} = \Phi + \mathcal H\frac{\Phi' - \mathcal H\Psi}{4\pi Ga^2(\rho + P)} = \Phi + \frac{2}{3}\frac{\mathcal{H}^{-1}\Phi' + \Phi}{1 + w}\;.
\end{equation}
Now, assume that $w$ changes from a constant value $w_i$ to another constant value $w_f$. For each of the two cases, $\Phi$ is a constant: $\Phi_i$ and $\Phi_f$, respectively. Then, taking advantage of the constancy of $\mathcal{R}$, we can say that:
\begin{equation}
\Phi_i + \frac{2}{3}\frac{\Phi_i}{1 + w_i} = \Phi_f + \frac{2}{3}\frac{\Phi_f}{1 + w_f}\;,
\end{equation}
i.e
\begin{equation}
\Phi_f = \Phi_i\frac{5 + 3w_i}{5 + 3w_f}\frac{1 + w_f}{1 + w_i}\;,
\end{equation}
with which we can easily check the result of Eq.~\eqref{Phi910res}. 

\subsection{Evolution in the \texorpdfstring{$\Lambda$}{Lambda}-dominated epoch}

As already mentioned, we cannot use the above formula for the case of greatest interest, which is for $w = -1$, i.e., the cosmological constant. In this case, we have to start directly from Eq.~\eqref{MukhanoveqPhi}. Since $P = -\rho$ is constant, then $P' = 0$, and thus $c_{\rm ad} = 0$.

\hrulefill

\begin{ex}
	Using Eq.~\eqref{LambdadominationmathcalH} into Eq.~\eqref{MukhanoveqPhi} with $c_{\rm ad} = 0$ and $\Gamma = 0$, show that:
	\begin{equation}\label{PhiequationLambdadomination}
		\Phi'' + \frac{3H_\Lambda}{1 + H_\Lambda(\eta_0 - \eta)}\Phi' + \frac{3H_\Lambda^2}{[1 + H_\Lambda(\eta_0 - \eta)]^2}\Phi = 0\;.
	\end{equation}
\end{ex} 

\hrulefill

Note that $\eta$ does not go to infinity but to a maximum value $\eta_\infty = \eta_0 - 1/H_\Lambda$, for which the scale factor diverges. Moreover, note that this equation and its solution are also valid for small scales because for $c_{\rm ad} = 0$ the $k$-dependence is suppressed.

\hrulefill

\begin{ex}
	Find the relation between the cosmic time $t$ and the conformal time using Eq.~\eqref{LambdadominationmathcalH} and show that $\eta_\infty = \eta_0 - 1/H_\Lambda$ corresponds to an infinite $t$.
\end{ex}

\hrulefill

\begin{ex}
	Change variable to the scale factor in Eq.~\eqref{PhiequationLambdadomination}, and show that:
\begin{equation}
	\frac{d^2\Phi}{da^2} + \frac{5}{a}\frac{d\Phi}{da} + \frac{3}{a^2}\Phi = 0\;.
\end{equation}
Solve this equation, using a power-law ansatz, and show that:
\begin{equation}
	\Phi = C_1a^{-1} + C_2a^{-3}\;. 
\end{equation}
\end{ex}

\hrulefill

Hence, when $\Lambda$ dominates, $\Phi$ is not constant on super-horizon scales, but vanishes rapidly, as $\Phi \propto 1/a$ or $\Phi \propto 1 + H_\Lambda(\eta_0 - \eta)$.\index{Evolution of perturbations!$\Lambda$-dominated epoch}

\subsection{Evolution through matter-DE equality}

In order to study the transition between the matter-dominated epoch and the DE-dominated one, we consider Eq.~\eqref{PoisseqSHevo1a} neglecting radiation:
\begin{equation}\label{PoisseqSHevo1anoradiation}
	3\mathcal{H}\left(\Phi' + \mathcal{H}\Phi\right) = \frac{3\mathcal H^2\rho_{\rm m}}{2\rho_{\rm tot}}\delta_{\rm m}\;,
\end{equation}
where we have already considered adiabatic perturbations. Now, let us introduce the following variable:
\begin{equation}
	x \equiv \frac{\rho_x}{\rho_{\rm m}} = \frac{\tilde\rho(a/a_x)^{-3(1 + w_x)}}{\tilde\rho (a/a_x)^{-3}} = \left(\frac{a}{a_x}\right)^{-3w_x}\;,
\end{equation}
where $\rho_x$ is a DE component with an equation of state $w_x$, which we assume to be constant, and $a_x$ is the scale factor at matter-DE equivalence, for which both densities are equal to $\tilde\rho$. Note that $w_x < -1/3$, in order to have a useful DE (it has to produce an accelerated expansion), and recall that, for Eq.~\eqref{PoisseqSHevo1anoradiation} to be valid, DE must not cluster.

\hrulefill

\begin{ex}
	Following the same steps which brought us to Eq.~\eqref{PoisseqSHevo3}, find a closed equation for $\Phi$, using $x$ as independent variable. Show that:
	\begin{equation}
		\frac{d^2\Phi}{dx^2} + \frac{6w_x(x + 1) - 2x - 5}{6w_xx(x + 1)}\frac{d\Phi}{dx} - \frac{1}{3w_xx(x + 1)}\Phi = 0\;.
	\end{equation}
\end{ex}

\hrulefill

This equation can be cast as a hypergeometric equation and thus solved exactly \cite{Piattella:2014lba}. Only one of the two independent solutions is well-behaved for $x \to 0$, i.e., is a constant:
\begin{equation}\label{PhiSHmattertoDE}
	\Phi \propto -2w_x + \frac{4w_x(1 + x)}{5}{}_2F_1\left(1,1 - \frac{1}{3w_x},1 - \frac{5}{6w_x},-x\right) \to_{x \to 0} -\frac{6w_x}{5}\;. 
\end{equation}
The integration constant must be selected in order to match $9\Phi_{\rm P}/10$.

\begin{figure}[htbp]
\center
\includegraphics[width=\columnwidth]{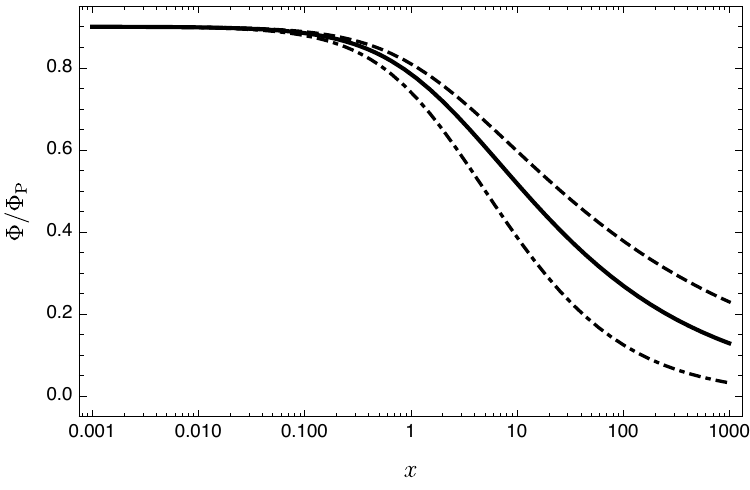}
\caption{Evolution of the gravitational potential $\Phi$ on super-horizon scales ($k = 0$) through matter-DE equality, from Eq.~\eqref{PhiSHmattertoDE}. The solid line represents the cosmological constant case $w_x = -1$, the dashed line $w_x = -1.5$ and the dash-dotted one $w_x = -0.5$.}
\label{Fig:PhiSHmattertoDE}
\end{figure}

In Fig.~\ref{Fig:PhiSHmattertoDE}, we display the evolution of $\Phi$, computed for $w_x = -1.5,-1,-0.5$ from Eq.~\eqref{PhiSHmattertoDE}. 

\begin{figure}[htbp]
\center
\includegraphics[width=\columnwidth]{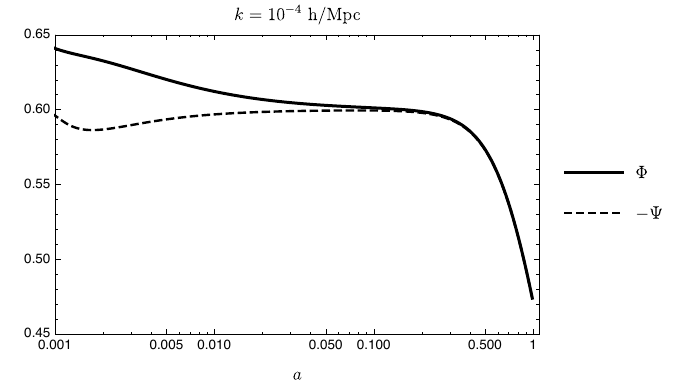}
\caption{Evolution of the potentials $\Phi$ (solid line) and $-\Psi$ (dashed-line) for the $\Lambda$CDM model with adiabatic initial condition using CLASS. The wavenumber chosen here is $k = 10^{-4}$ Mpc$^{-1}$.}
\label{Fig:PhiPsievok1em4CLASS}
\end{figure}

In Fig.~\ref{Fig:PhiPsievok1em4CLASS} we display the evolution of the potentials $\Phi$ (solid line) and $-\Psi$ (dashed-line) for the $\Lambda$CDM model with adiabatic initial conditions using CLASS. The wavenumber chosen here is $k = 10^{-4}$ Mpc$^{-1}$, which corresponds to a scale larger than the present time horizon that has spent the entire cosmic evolution outside the horizon. Note how the two potentials display a difference at early times due to the presence of neutrinos.

The investigation of the evolution of scales larger than the horizon today is not very useful because they are not observable. However, the scales that we do observe today were outside the horizon in the past. We shall see in the next section that in the matter-dominated regime, the gravitational potential $\Phi$ is constant at all scales, even during horizon-crossing; thus, it is interesting to know the behavior of super-horizon modes (the same behavior is not shared by the density contrast). The same does not occur when radiation dominates; instead, the gravitational potential decays and oscillates rapidly for those scales that enter the horizon.

Since the behavior in the radiation-dominated and matter-dominated epochs is so dramatically different, it is useful to introduce the so-called \textbf{equivalence wavenumber}, i.e., the wavenumber corresponding to a scale that enters the horizon at the equivalence epoch and is thus defined as:\index{Equivalence wavenumber}
\begin{equation}
	k_{\rm eq} \equiv \mathcal H_{\rm eq}\;.
\end{equation}
Neglecting dark energy, the Friedmann equation in the presence of radiation and matter is written as follows:
\begin{equation}\label{Friedmannequationmatterandradiation}
 \mathcal H^2 = \frac{8\pi G}{3}(\rho_{\rm m} + \rho_{\rm r})a^2 = H_0^2(\Omega_{\rm m0}a^{-1} + \Omega_{\rm r0}a^{-2})\;.
\end{equation}
From this, we can establish that, at equivalence, the conformal Hubble parameter has the following expression:
\begin{equation}\label{conformalHubblefactoratequality}
 \mathcal H_{\rm eq}^2 = \frac{16\pi G}{3}\frac{\rho_{\rm R0}}{a_{\rm eq}^2} = 2H_0^2\Omega_{\rm r0}(1 + z_{\rm eq})^2 = k_{\rm eq}^2\;,
\end{equation}
or, using the matter density parameter:
\begin{equation}\label{keqdefinitionOm}
 \mathcal H_{\rm eq}^2 = \frac{16\pi G}{3}\frac{\rho_{\rm M0}}{a_{\rm eq}} = 2H_0^2\Omega_{\rm m0}(1 + z_{\rm eq}) = k_{\rm eq}^2\;,
\end{equation}
from which it is clear that:
\begin{equation}
	1 + z_{\rm eq} = \frac{\Omega_{\rm m0}}{\Omega_{\rm r0}}\;.
\end{equation}
Using the observed values for the density parameters, we have that:
\begin{equation}\label{keqdefinitionOmOr}
	\boxed{k_{\rm eq} = \frac{\sqrt{2}H_0\Omega_{\rm m0}}{\sqrt{\Omega_{\rm r0}}} \approx 0.014\; h\; \mbox{Mpc}^{-1}}
\end{equation}
The behavior of super-horizon modes is especially important in CMB physics. Indeed, in Chapter~\ref{Chap:CMBEvo} we shall see that for a given multipole $\ell$ the CMB temperature correlation spectrum $C_\ell$ is determined mostly by those wavenumbers that satisfy:
\begin{equation}
	\ell \approx k(\eta_0 - \eta_*) = kr_* \approx k\eta_0\;,
\end{equation}
where $\eta_0$ is the present conformal time, $\eta_*$ is the one corresponding to recombination, and $r_* \equiv \eta_0 - \eta_*$ is the comoving distance to recombination. The last approximation is motivated by the fact that $\eta_* \approx 3\times 10^2$ Mpc, whereas $\eta_0 \approx 10^4$ Mpc.

The above expression can be manipulated as follows:
\begin{equation}
	\ell \approx k\eta\frac{\eta_0}{\eta} \approx k\eta\frac{1}{\sqrt{a}}\;,
\end{equation}
where $\eta$ is a generic past conformal time in the matter-dominated epoch; for this reason, we have used $a \propto \eta^2$. So, e.g., at radiation-matter equality, i.e., $a \approx 10^{-4}$, the super-horizon scales $k\eta < 1$ contribute to the multipoles $\ell < 100$, and at recombination $a_* \approx 10^{-3}$, $\ell < 30$.

\section{The matter-dominated epoch}

Let us now completely disregard radiation. Since there are no photon and neutrino quadrupoles, then $\Phi = -\Psi$ and, since matter dominates, $\delta P_{\rm tot} = 0$.

Equation~\eqref{GiideltaPeq2} then provides us with a closed equation for $\Phi$:
\begin{equation}\label{PhieqdeltaP0}
	\Phi'' + 3\mathcal{H}\Phi' + 2\mathcal{H}'\Phi + \mathcal{H}^2\Phi = -4\pi Ga^2\delta P_{\rm tot} = 0\;,
\end{equation}
which can be solved exactly since, being $a \propto \eta^2$, we have:
\begin{equation}
	\Phi'' + \frac{6}{\eta}\Phi' = 0\;.
\end{equation}
The general solution is:
\begin{equation}\label{constantpotentialMD}
	\boxed{\Phi(k,\eta) = A(k) + B(k)(k\eta)^{-5} = A(k) + \hat{B}(k)a^{-5/2}}
\end{equation}
where $A(k)$, $B(k)$, and $\hat{B}(k)$ are functions of $k$, the latter being equal to $B$ times the proportionality factor between the conformal time and the scale factor, which are related by $a \propto \eta^2$. We have kept a $k\eta$ dependence above because it is dimensionless, as $\Phi$, $A$, and $B$ are.  

Now, let us neglect the decaying mode $(k\eta)^{-5}$ since it disappears very quickly. The important result here is that the gravitational potential is constant at all scales, through horizon crossing, during matter domination. We shall see a very different behavior when radiation dominates. 

Hence, provided that we are deep into the matter-dominated epoch, at large scales, when $k\eta < 1$, we can match the results for $\Phi$ of this section and the previous one and see that:
\begin{equation}\label{APhiprelationMD}
	A(k < 1/\eta) = \frac{9}{10}\Phi_{\rm P}(k)\;, \qquad (\eta > \eta_{\rm eq})\;.
\end{equation} 
So, the gravitational potential $\Phi$ for those scales that enter the horizon during the matter dominated epoch is a constant with a value:
\begin{equation}
	\boxed{\Phi(k) = \frac{9}{10}\Phi_{\rm P}(k)\;, \qquad (k\eta_{\rm eq} < 1)}
\end{equation}
The condition $k\eta_{\rm eq} < 1$, which corresponds to $k < k_{\rm eq}$ guarantees that the mode was outside the horizon at equivalence and hence that it entered during matter domination. 

Considering the generalized Poisson equation with $\Phi$ constant, we have:
\begin{eqnarray}
	3\mathcal H^2\Phi + k^2\Phi = \frac{3\mathcal H^2}{2}\delta_{\rm m}\;,
\end{eqnarray}
where $\delta_{\rm m}$ is the density contrast of matter, which we define here as:
\begin{equation}\label{newdeltaMdefinition}
	\delta_{\rm m} \equiv (1 - \Omega_{\rm b0})\delta_{\rm c} + \Omega_{\rm b0}\delta_{\rm b}\;,
\end{equation}
which comes from the fact that, being in the matter-dominated epoch, $\rho_{\rm tot} \propto a^{-3}$ and $\Omega_{\rm c0} + \Omega_{\rm b0} = 1$ (of course, we continue to consider a spatially flat universe). 

We must be careful that even in the matter-dominated epoch, but before recombination, for those modes inside the horizon we can have that $\delta_{\rm c} \gg \delta_{\rm b}$ because baryons are tightly coupled to photons and thus $\delta_{\rm b}$ cannot grow, whereas CDM fluctuations can. For these modes, soon after recombination, $\delta_{\rm b}$ becomes equal to $\delta_{\rm c}$ or, in other words, \textit{baryons fall into the potential wells of CDM}. We shall see this in more detail later. For large scales, those which enter the horizon after recombination, we have $\delta_{\rm c} = \delta_{\rm b} = \delta_{\rm m}$ (assuming, as usual, adiabaticity).  

\begin{figure}[htbp]
\center
\includegraphics[width=\columnwidth]{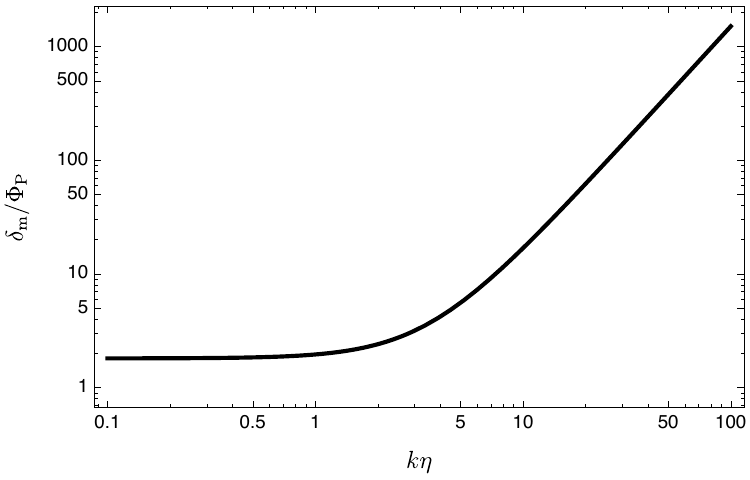}
\caption{Evolution of the matter density contrast $\delta_{\rm m}$ normalized to the primordial potential, in the matter-dominated era. From Eq.~\eqref{deltamattevo}.}
\label{Fig:PhideltaMattplot}
\end{figure}

For the moment, let us focus on the evolution for $\delta_{\rm m}$. It follows immediately that:
\begin{equation}\label{deltamattevo}
	\delta_{\rm m}(k,\eta) = 2A(k)\left(1 + \frac{k^2}{3\mathcal H^2}\right) = \frac{9\Phi_{\rm P}(k)}{5}\left(1 + \frac{k^2\eta^2}{12}\right)\;, \qquad (k < k_{\rm eq})\;. 
\end{equation}
Therefore, $\delta_{\rm m}$ is constant on super-horizon scales, but when a scale crosses the horizon, it starts to grow as $\delta \propto \eta^2 \propto a$, as the plot in Fig.~\eqref{Fig:PhideltaMattplot} shows. Note that the first equality holds true at any scale, but the second one only for $k < k_{\rm eq}$, because only in this regime are we allowed to use Eq.~\eqref{APhiprelationMD}.

The power spectrum at any time during the matter dominated epoch can thus be immediately related to the primordial one:
\begin{equation}
	P_\delta(k,\eta) = \frac{81}{25}\left(1 + \frac{k^2\eta^2}{12}\right)^2P_\Phi(k)\;, \qquad (k < k_{\rm eq})\;.
\end{equation}
For $n_S = 1$, the primordial power spectrum goes as $P_\Phi \propto 1/k^3$, cf. Eq.~\eqref{Delta2S}, and thus the above matter grows linearly with $k$ (when $k\eta > 1$). From the above equation, we can read off the matter transfer function:
\begin{equation}\label{mattertransferfunctionmatterdom}
	T_\delta(k,\eta) = \frac{9}{5}\frac{k^2\eta^2}{12}\;, \qquad (1/\eta < k < k_{\rm eq})\;.
\end{equation}
From this solution, we can infer that the larger $k$ is, i.e., the smaller the scale under consideration is, the more it grows. This scenario is called \textbf{bottom-up} because smaller scales become non-linear before the larger ones. In other words, small structures form first, and then these merge in order to form larger structures. The bottom-up scenario is the opposite mechanism of the \textbf{top-down} scenario \cite{1984ASPRv...3....1Z}, according to which the largest structures form first and then fragment to create the smaller ones.\index{Structure formation!Top-down scenario}\index{Structure formation!Bottom-up scenario}

In Eq.~\eqref{mattertransferfunctionmatterdom}, we can appreciate that the $k$-dependence and the $\eta$ dependence are separate. The latter is proportional to $\eta^2 \propto a$ and is usually called \textbf{growth factor}. The separation of the $k$ and $\eta$ dependencies occurs because matter has a vanishing adiabatic speed of sound, and the $k$-dependence of the equation governing $\delta_{\rm m}$ is multiplied by the gravitational potential, which is a constant (during matter domination).\index{Growth factor}

We can see this in detail by combining Eqs.~\eqref{deltaeqgeneric} and \eqref{Vequationgeneral}.

\hrulefill

\begin{ex}
Combine Eqs.~\eqref{deltaeqgeneric} and \eqref{Vequationgeneral} after putting to zero $w$, $\delta P$ and the anisotropic stresses, which are indeed vanishing for CDM and also for baryons, after recombination. Show that:
\begin{equation}\label{deltaMeq}
	\boxed{\delta_{\rm m}'' + \mathcal H\delta'_{\rm m} = -k^2\Psi - 3\Phi'' - 3\mathcal H\Phi'}
\end{equation}
\end{ex}

\hrulefill

In the above equation, the $k$-dependence enters only through $k^2\Psi$ and then, in the matter-dominated epoch we have that:
\begin{equation}\label{deltaMeqMD}
	\delta_{\rm m}'' + \mathcal H\delta'_{\rm m} = k^2\Phi\;,
\end{equation}
with $\Phi$ constant (if we neglect the decaying mode already).

\hrulefill

\begin{ex}
	Check that the solution of the above equation is the same as Eq.~\eqref{deltamattevo}.
\end{ex}

\hrulefill

The transfer function that we have determined in this section is valid for small values of $k$, i.e., $k < k_{\rm eq} \approx 0.014$ $h$ Mpc$^{-1}$, which correspond to very large scales that we do not actually observe or for which the errors and cosmic variance are too large. It is necessary, therefore, to understand how matter fluctuations behave during radiation-domination.\index{Evolution of perturbations!Matter-dominated epoch}

\subsection{Baryons falling into the CDM potential wells}

We offer a simple calculation here that should convey the idea of how important CDM is for structure formation. This is often stated as the fact that, after recombination, \textit{baryons fall into the gravitational potential wells of CDM}.

As we anticipated earlier, before recombination, baryons were tightly coupled to photons. When they decouple and their over-densities are free to grow, in general, we have $\delta_{\rm b} \ll \delta_{\rm c}$ for those modes that were well inside the horizon during recombination.

Let us see this more quantitatively. In the same fashion by which we obtained Eq.~\eqref{deltaMeqMD}, we can write the following coupled equations for CDM and baryons:
\begin{eqnarray}
\label{deltaceqMD}	\delta_{\rm c}'' + \mathcal H\delta_{\rm c}' = k^2\Phi\;,\\
\label{deltabeqMD}	\delta_{\rm b}'' + \mathcal H\delta_{\rm b}' = k^2\Phi\;,
\end{eqnarray}
of which the baryon one is valid only after recombination. These equations are coupled since $\Phi$ is determined by both components. Indeed, from the Poisson equation, we have that:
\begin{eqnarray}
	(3\mathcal H^2 + k^2)\Phi = \frac{3\mathcal H^2}{2\rho_{\rm tot}}(\rho_{\rm c}\delta_{\rm c} + \rho_{\rm b}\delta_{\rm b}) \equiv \frac{3\mathcal H^2}{2}\delta_{\rm m}\;,
\end{eqnarray}
Now, the two Eqs.~\eqref{deltaceqMD} and \eqref{deltabeqMD} have solutions (neglecting the decaying mode):
\begin{equation}
	\delta_{\rm c}(k,\eta) = C_1(k) + \frac{k^2\eta^2}{6}A(k)\;, \qquad \delta_{\rm b}(k,\eta) = C_2(k) + \frac{k^2\eta^2}{6}A(k)\;,
\end{equation}
where we have used the constant potential solution in Eq.~\eqref{constantpotentialMD} (also neglecting the decaying mode here). Using this solution in Eq.~\eqref{newdeltaMdefinition} and comparing it with Eq.~\eqref{deltamattevo}, we can conclude that:
\begin{equation}
	C_1(k) + \Omega_{\rm b0}[C_2(k) - C_1(k)] = 2A(k)\;.
\end{equation}
For large scales $k\eta \ll 1$ we already know that $\delta_{\rm c} = \delta_{\rm b} = \delta_{\rm m}$, because of adiabaticity, and hence $C_1 = C_2 = 2A$.

For small scales, at recombination $\delta_{\rm c} \gg \delta_{\rm b}$ because baryons were tightly coupled to photons and thus $\delta_{\rm b}$ could not grow. This, combined with the crucial fact that $\Omega_{\rm b0} = 0.04$ is small, makes $\delta_{\rm m}$ (and the gravitational potential) dominated by $\delta_{\rm c}$. However, we have again $\delta_{\rm c} = \delta_{\rm b}$ soon after recombination, with the detailed transient coming from the decaying modes that we have neglected. 

In other words, baryons fall into the potential wells already created by CDM. Without CDM, $\delta_{\rm b}$ would grow proportionally to $a$, by a factor $10^3$ by today, being of order $10^{-2}$. This is too small to account for the structures that we observe. This fact is one of the most compelling arguments for the necessity of CDM, as we saw in Chapter \ref{Chap:Cosmology}.

\begin{figure}[htbp]
\center
\includegraphics[width=\columnwidth]{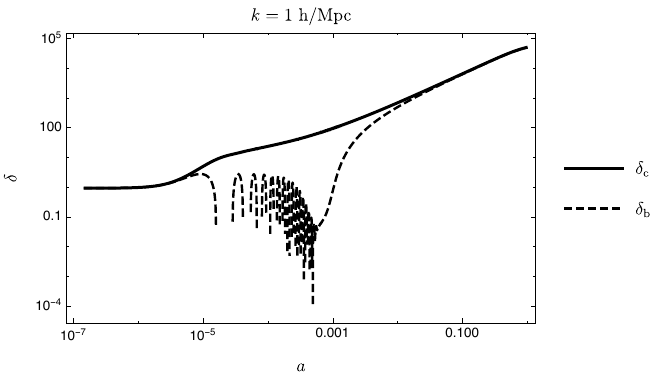}
\caption{Evolution of $\delta_{\rm c}$ (solid line) and $\delta_{\rm b}$ (dashed line) for $k = 1$ Mpc$^{-1}$ computed with CLASS.}
\label{Fig:deltaevok1CLASS}
\end{figure}

In Fig.~\ref{Fig:deltaevok1CLASS}, we plot the evolution of $\delta_{\rm c}$ (solid line) and $\delta_{\rm b}$ (dashed line) for $k = 1$ Mpc$^{-1}$ computed with CLASS. Note the oscillations in $\delta_{\rm b}$, which are related to the \textbf{baryon acoustic oscillations} (BAO) of the matter power spectrum. The oscillations in the plot of Fig.~\ref{Fig:deltaevok1CLASS} are a function of the time (scale factor) evolution, whereas those in the matter power spectrum are a function of the wavenumber $k$.\index{Baryon Acoustic Oscillations}

The oscillations of Fig.~\ref{Fig:deltaevok1CLASS} are caused by the tight coupling of baryons with photons before recombination, when structure formation is impossible. On the other hand, CDM can grow even when radiation dominates; thus, for the chosen scale $k = 1$ Mpc$^{-1}$, the ratio $\delta_{\rm c}/\delta_{\rm b}$ is about 3 orders of magnitude.

\begin{figure}[htbp]
\center
\includegraphics[width=\columnwidth]{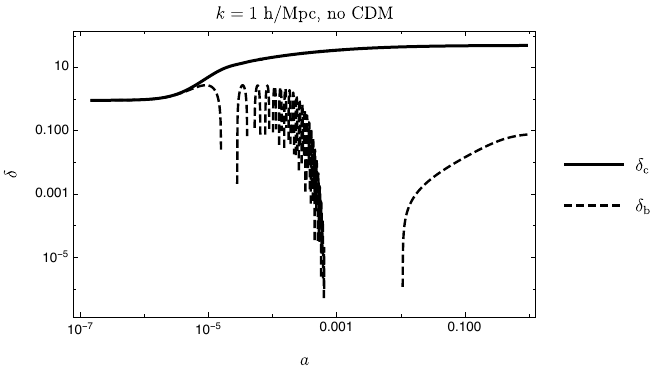}
\caption{Evolution of $\delta_{\rm c}$ (solid line) and $\delta_{\rm b}$ (dashed line) for $k = 1$ Mpc$^{-1}$ computed with CLASS with a negligible amount of CDM ($\Omega_{\rm c0}h^2 = 10^{-6}$).}
\label{Fig:deltaevok1NODMCLASS}
\end{figure}

In Fig.~\ref{Fig:deltaevok1NODMCLASS}, we again plot the evolution of $\delta_{\rm c}$ (solid line) and $\delta_{\rm b}$ (dashed line) for $k = 1$ Mpc$^{-1}$ computed with CLASS, but this time with a negligible amount of CDM ($\Omega_{\rm c0}h^2 = 10^{-6}$). Note how $\delta_{\rm b}$ grows six orders of magnitude less today than in the standard case.

\section{The radiation-dominated epoch}

Consider now the case of full radiation dominance and neglect $\delta_{\rm c}$ and $\delta_{\rm b}$ as sources of the gravitational potentials. Moreover, assume adiabaticity, so that $\delta_\gamma = \delta_\nu = \delta_{\rm r}$ and neglect the neutrino anisotropic stress, so that $\Phi = -\Psi$. 

Since we are deep into the radiation-dominated epoch, $w = c_{\rm ad}^2 = 1/3$, and thus from Eq.~\eqref{MukhanoveqPhi}, we can immediately write:
\begin{equation}\label{PhieqRaddom}
 \Phi'' + \frac{4}{\eta}\Phi' + \frac{k^2}{3}\Phi = 0\;,
\end{equation}
where we have used $a \propto \eta$. Defining
\begin{equation}
u \equiv \Phi\eta\;,
\end{equation}
the above equation becomes
\begin{equation}
u'' + \frac{2}{\eta}u' + \left(\frac{k^2}{3} - \frac{2}{\eta^2}\right)u = 0\;.
\end{equation}
This is a Bessel equation with solutions $j_1(k\eta/\sqrt{3})$ and $n_1(k\eta/\sqrt{3})$. Since $n_1$ diverges for $k\eta \to 0$, we discard it as nonphysical.

Recovering the gravitational potential $\Phi$ and using the fact that \cite{Abramowitz1972}
\begin{equation}
j_1(x) = \frac{\sin x}{x^2} - \frac{\cos x}{x}\;,
\end{equation}
and
\begin{equation}
\lim_{x\to 0}\frac{\sin x - x\cos x}{x^3} = \frac{1}{3}\;,
\end{equation}
we can write
\begin{equation}\label{Phiradevolution}
\Phi = 3\Phi_{\rm P}\frac{\sin(k\eta/\sqrt{3}) - (k\eta/\sqrt{3})\cos(k\eta/\sqrt{3})}{(k\eta/\sqrt{3})^3}\;.
\end{equation}
This solution shows that as soon as a mode $k$ of the gravitational potential enters the horizon, it rapidly decays as $1/\eta^3$ or $1/a^3$ while oscillating.

\begin{figure}[htbp]
\center
\includegraphics[width=\columnwidth]{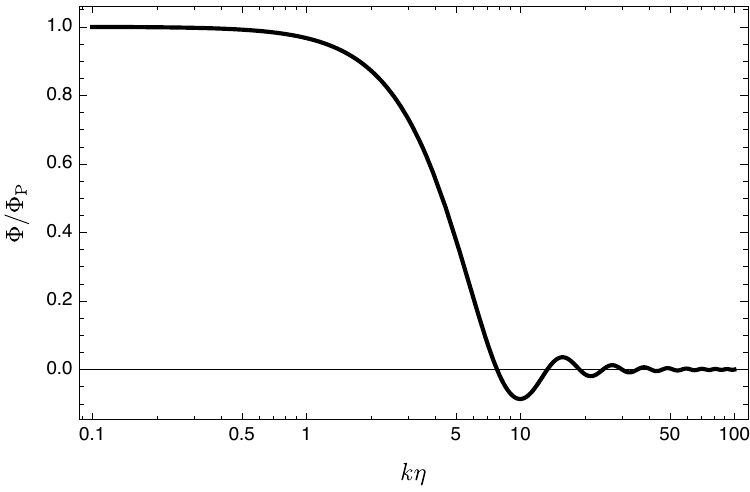}
\caption{Evolution of the gravitational potential $\Phi$ deep into the radiation-dominated era. From Eq.~\eqref{Phiradevolution}.}
\label{Fig:Phiradplot}
\end{figure}

In Fig.~\ref{Phiradevolution} we plot the evolution of the gravitational potential according to Eq.~\eqref{Phiradevolution}. Note how $\Phi$ starts to decay right after $k\eta > 1$, i.e., after horizon crossing.

\begin{figure}[htbp]
\center
\includegraphics[width=\columnwidth]{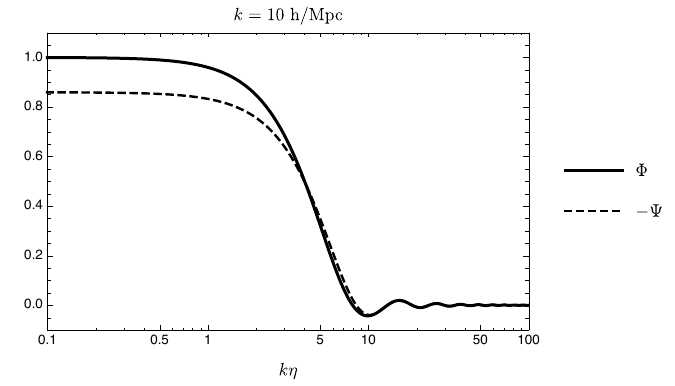}
\caption{Evolution of the gravitational potentials $\Phi$ (solid line) and $-\Psi$ (dashed line) deep into the radiation-dominated era computed with CLASS for the $\Lambda$CDM model and $k = 10$ Mpc$^{-1}$. Adiabatic perturbations have been used and the initial values have been normalised to that of $\Phi$.}
\label{Fig:PhiPsievok10CLASS}
\end{figure}

The goodness of the solution \eqref{Phiradevolution} plotted in Fig.~\ref{Fig:Phiradplot} can be appreciated in Fig.~\ref{Fig:PhiPsievok10CLASS}, where both $\Phi$ (solid line) and $-\Psi$ (dashed line) are plotted. Outside the horizon, the two potentials are constant, with a difference due to the neutrino fraction $R_\nu$. As soon as they enter the horizon, they rapidly decay to zero.

Now, recall Eq.~\eqref{deltaMeq} that we derived for the matter density contrast. It can be used in the radiation-dominated epoch, but only for CDM:
\begin{equation}
	\delta_{\rm c}'' + \frac{1}{\eta}\delta_{\rm c}' = k^2\Phi - 3\Phi'' - \frac{3}{\eta}\Phi'\;.
\end{equation}
Using Eq.~\eqref{PhieqRaddom}, we can write
\begin{equation}
	\delta_{\rm c}'' + \frac{1}{\eta}\delta_{\rm c}' = 2k^2\Phi + \frac{9}{\eta}\Phi' \equiv S(k,\eta)\;.
\end{equation}
As stated at the beginning of the section, we are so deep into the radiation-dominated epoch that the matter density contrast does not contribute to the gravitational potential, but only feels it.

Using Eq.~\eqref{Phiradevolution} with $x \equiv k\eta$ as the new independent variable, the function $S(x)$ has the following form:
\begin{equation}
  \frac{S(x)}{k^2\Phi_{\rm P}} = \frac{9\left[\left(27x - 2x^3\right)\cos\left(x/\sqrt{3}\right) + \sqrt{3}\left(5x^2 - 27\right)\sin\left(x/\sqrt{3}\right)\right]}{x^5}\;,
\end{equation}
and the equation for $\delta_{\rm c}$ becomes:
\begin{equation}
	\frac{d}{dx}\left(x\frac{d\delta_{\rm }}{dx}\right) = 9\Phi_{\rm P}\frac{\left[\left(27x - 2x^3\right)\cos\left(x/\sqrt{3}\right) + \sqrt{3}\left(5x^2 - 27\right)\sin\left(x/\sqrt{3}\right)\right]}{x^4}\;.
\end{equation}
The homogeneous part of this equation has a simple solution:
\begin{equation}
\delta_{\rm hom} = C_1 + C_2\ln x\;.
\end{equation}
A particular solution is obtained by integrating the right-hand side twice, thus obtaining:
\begin{equation}
\delta_{\rm part} = \frac{9\Phi_{\rm P}\left[-x^3\text{Ci}\left(x/\sqrt{3}\right) + \sqrt{3}\left(x^2 - 3\right)\sin \left(x/\sqrt{3}\right) + 3x\cos\left(x/\sqrt{3}\right)\right]}{x^3}\;,
\end{equation}
where $\text{Ci}(z)$ is defined as the cosine integral function:
\begin{equation}
\text{Ci}(z) \equiv -\int_z^\infty dt\frac{\cos t}{t}\;.
\end{equation}
The above is a particular solution; therefore, the integration constants that stem from the indefinite integration can be incorporated in $C_1$. The general solution for $\delta_{\rm c}$ is then:
\begin{equation}\label{deltacevoRD}
\delta_{\rm c} = C_1 + C_2\ln x + \frac{9\Phi_{\rm P}\left[-x^3\text{Ci}\left(x/\sqrt{3}\right) + \sqrt{3}\left(x^2 - 3\right)\sin \left(x/\sqrt{3}\right) + 3x\cos\left(x/\sqrt{3}\right)\right]}{x^3}\;.
\end{equation}
For $x \to 0$, we can expand the above solution as follows:
\begin{equation}
\delta(x \to 0) = C_1 + C_2\ln(x) + \Phi_{\rm P}\left[- 9\ln(x) - 9\gamma + 6 + \frac{9\ln(3)}{2}\right] + \mathcal{O}\left(x^2\right)\;,
\end{equation}
where $\gamma$ is the Euler constant. \index{Evolution of perturbations!Radiation-dominated epoch}

\begin{figure}[htbp]
\center
\includegraphics[width=\columnwidth]{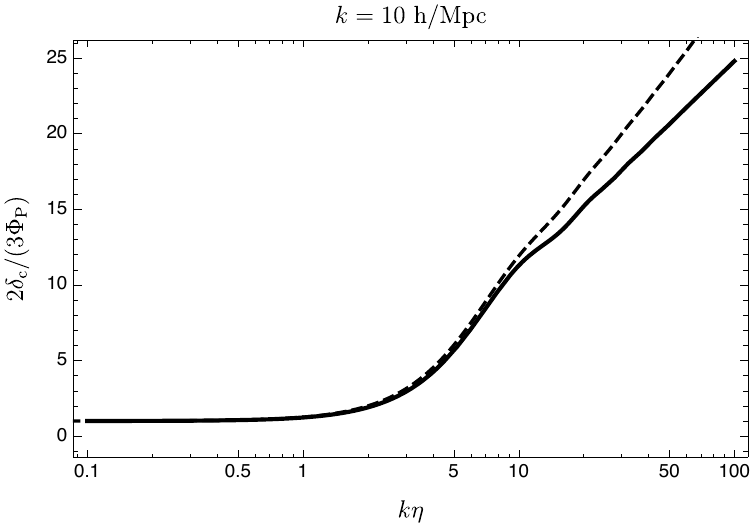}
\caption{Evolution of $\delta_{\rm c}$ deep into the radiation-dominated era computed from Eqs.~\eqref{deltacevoRD}, \eqref{C2constdeltacRD} and \eqref{C1constdeltacRD} (solid line) compared with the numerical calculation performed with CLASS for $k = 10$ Mpc$^{-1}$ (dashed line). Note the semi-logarithmic scale employed.}
\label{Fig:deltaradplot}
\end{figure}

Since $\ln x$ is divergent for $x \to 0$ and we do not want $\delta_{\rm c}$ to diverge, we have to ask:
\begin{equation}\label{C2constdeltacRD}
C_2 = 9\Phi_{\rm P}\;.
\end{equation}
Moreover, we know that $\delta_{\rm c}(x \to 0) = 3\Phi_{\rm P}/2$ when we choose adiabatic initial conditions (and neglect neutrinos), cf. Eq.~\eqref{deltagammarelPsiadiabatic}, thus:
\begin{equation}\label{C1constdeltacRD}
C_1 = -\frac{9}{2}\Phi_{\rm P}\left[-2\gamma + 1 + \ln(3)\right]\;.
\end{equation}
We plot the evolution of $\delta_{\rm c}$ through horizon-crossing in Fig.~\ref{Fig:deltaradplot}.

For $x \gg 1$, deep inside the horizon, we can neglect the contribution $\delta_{\rm part}$ to the solution for $\delta_{\rm c}$ since it decays rapidly. Thus, we can write the density contrast as follows:
\begin{equation}\label{deltasolradAlnB}
\delta_{\rm c} = -\frac{9}{2}\Phi_{\rm P}\left[-2\gamma + 1 + \ln(3)\right] + 9\Phi_{\rm P}\ln x = A\Phi_{\rm P}\ln(Bk\eta)\;,
\end{equation}
where
\begin{equation}
A = 9\;, \qquad B = \exp\left[\gamma - \frac{1}{2} - \frac{\ln(3)}{2}\right] \approx 0.62\;.
\end{equation}

\section{Deep inside the horizon}

The last domain in which it is possible to analytically solve the equations for the perturbations is when $k \gg \mathcal H$, i.e., deep inside the horizon. We shall neglect baryons in this calculation (this is imprecise, and we shall see why numerically) and assume $\Phi = -\Psi$ (which is a good approximation, based on the results of the previous section). 

The relevant equations are thus the following:
\begin{eqnarray}
	\delta_{\rm c}' + kV_{\rm c} = -3\Phi'\;,\\
	V_{\rm c}' + \mathcal HV_{\rm c} = -k\Phi\;,\\
\label{smallscalesPoissoneqDM}	k^2\Phi = 4\pi Ga^2\rho_{\rm c}\delta_{\rm c}\;.
\end{eqnarray}
In the latter equation, we have neglected all the potential terms except for the one accompanied by $k^2$, and we have also neglected radiation perturbations. It is not evident why we should neglect $\rho_{\rm r}\delta_{\rm r}$ with respect to $\rho_{\rm c}\delta_{\rm c}$, even when $\rho_{\rm r} \gg \rho_{\rm c}$ deep into the radiation-dominated epoch. Neglecting $\delta_{\rm b}$, at least, is justified by the fact that before recombination it behaves as the fluctuation in radiation and afterwards as that in CDM, while being $\Omega_{\rm b}$ always subdominant.

An explanation of why we can neglect $\rho_{\rm r}\delta_{\rm r}$ is offered by Weinberg, who shows that new modes appear (dubbed \textit{fast}), which rapidly decay and oscillate \cite{Weinberg:2002kg}. He also takes into account baryons at first order in $\Omega_{\rm b0}$.

\hrulefill

\begin{ex} Use again the variable $y \equiv a/a_{\rm eq}$ and manipulate the three equations above in order to obtain a single second-order equation for $\delta_{\rm c}$:
\begin{equation}
\frac{d^2\delta_{\rm c}}{dy^2} + \frac{2 + 3y}{2y(y + 1)}\frac{d\delta_{\rm c}}{dy} - \frac{3}{2y(y + 1)}\delta_{\rm c} = 0\;.
\end{equation}
\end{ex}

\hrulefill

This equation is known as \textbf{M\'esz\'aros equation} \cite{Meszaros:1974tb}. A solution can be found at once by multiplying by $2y(y + 1)$:
\begin{equation}
2y(y + 1)\frac{d^2\delta_{\rm c}}{dy^2} + 2\frac{d\delta_{\rm c}}{dy} + 3y\frac{d\delta_{\rm c}}{dy} - 3\delta_{\rm c} = 0\;.
\end{equation}
A linear ansatz $\delta_{\rm c} \propto y$ kills the second derivative and the last two terms on the left hand side. Therefore, the simple solution we looked for is:
\begin{equation}
\boxed{D_1(y) \equiv y + \frac{2}{3}}
\end{equation}
This is also the growing mode. For $y \ll 1$, i.e., before matter-radiation equality, $\delta$ is practically constant, whereas for $y \gg 1$ we have the known growth linear with respect to the scale factor.\index{Evolution of perturbations!M\'esz\'aros equation}

In order to find the other independent solution, say $D_2$, we can use the Wronskian:
\begin{equation}\label{WronskianD1D2}
W(y) = D_1\frac{dD_2}{dy} - \frac{dD_1}{dy}D_2\;.
\end{equation}

\hrulefill

\begin{ex} Show that the Wronskian satisfies the simple first-order differential equation:
\begin{equation}
\frac{dW}{dy} = - \frac{2 + 3y}{2y(y + 1)}W\;,
\end{equation}
from which one gets
\begin{equation}
W = \frac{1}{y\sqrt{1 + y}}\;.
\end{equation}
\end{ex}

\hrulefill

\begin{ex} From the very definition of the Wronskian in Eq.~\eqref{WronskianD1D2}, write a first-order equation also for $D_2$, which is the following:
\begin{equation}
\left(y + 2/3\right)^2\frac{d}{dy}\left(\frac{D_2}{y + 2/3}\right) = \frac{1}{y\sqrt{1 + y}}\;.
\end{equation}
Integrate it and show that the result is:
\begin{equation}
\boxed{D_2 = \frac{9}{2}\sqrt{1 + y} - \frac{9}{4}\left(y + 2/3\right)\ln\left(\frac{\sqrt{y + 1} + 1}{\sqrt{y + 1} - 1}\right)}
\end{equation}
\end{ex}

\hrulefill

This mode grows logarithmically when $y \ll 1$, recovering the logarithmic solution of the previous section. It decays as $1/y^{3/2}$ for $y \gg 1$.

The complete solution for $\delta_{\rm c}$ on small scales $k \gg \mathcal H$ and through radiation-matter equality is as follows:
\begin{equation}
\delta_{\rm c}(k,a) = C_1(k)D_1(a) + C_2(k)D_2(a)\;.
\end{equation}
The dependence of $C_1$ and $C_2$ on $k$ can be established by matching this solution with that of the previous section in Eq.~\eqref{deltasolradAlnB}.

\section{Matching and CDM transfer function}

As we discussed earlier, the observed scales in the matter power spectrum today are those for which $k > k_{\rm eq}$, i.e., those that entered the horizon before matter equality.

In the last two sections, we have obtained the exact solution for $\delta_{\rm c}$ deep into the radiation-dominated epoch for all scales and through radiation-matter equality at very small scales. There is then the possibility of matching the two solutions on very small scales and thus obtaining, in this regime, the CDM transfer function until today.

Consider the following two solutions found in the previous sections:
\begin{eqnarray}
\delta_{\rm c}(k,\eta) = A\Phi_{\rm P}(k)\ln(Bk\eta)\;,\\
\delta_{\rm c}(k,a) = C_1(k)D_1(a) + C_2(k)D_2(a)\;,
\end{eqnarray}
which are valid on very small scales, $k\eta \gg 1$. The purpose is to find the functional forms of $C_1(k)$ and $C_2(k)$.

Using Eqs.~\eqref{Friedmannequationmatterandradiation} and \eqref{conformalHubblefactoratequality}, one can approximate the Hubble parameter deep into the radiation era as follows:
\begin{equation}
\mathcal H \approx \frac{\mathcal H_{\rm eq}a_{\rm eq}}{\sqrt{2}a}\;,
\end{equation}
and solving using the conformal time, one has:
\begin{equation}
a = \frac{\mathcal H_{\rm eq}a_{\rm eq}}{\sqrt{2}}\eta\;.
\end{equation}
The proportionality constant is the correct one that gives $\mathcal H = 1/\eta$ when substituted in the approximated formula for $\mathcal H$.

We can thus write the logarithmic solution for $\delta_{\rm c}$, deep into the radiation-dominated era, as follows:
\begin{eqnarray}
\delta_{\rm c}(k,a) = A\Phi_{\rm P}(k)\ln\left(Bk\frac{\sqrt{2}a}{\mathcal H_{\rm eq}a_{\rm eq}}\right)\;.
\end{eqnarray}
Introducing $y \equiv a/a_{\rm eq}$, the equivalence wavenumber $k_{\rm eq} = \mathcal H_{\rm eq}$, and the rescaled wavenumber:
\begin{equation}\label{kappadefinition}
	\boxed{\kappa \equiv \frac{\sqrt{2}k}{k_{\rm eq}} = \frac{k\sqrt{\Omega_{\rm r0}}}{H_0\Omega_{\rm m0}} = \frac{k}{0.052\;\Omega_{\rm m0}\;h^2\mbox{ Mpc}^{-1}}}
\end{equation}
one has:
\begin{equation}
\delta_{\rm c}(k,a) = A\Phi_{\rm P}(k)\ln\left(B\kappa y\right)\;.
\end{equation}
Recall that this solution is valid deep into the radiation-dominated epoch, thus $y \ll 1$. At the same time, it holds true only on very small scales, $k\eta \gg 1$. Using the above formulae, this means:
\begin{equation}\label{ketaequaltokappay}
	k\eta = k\frac{\sqrt{2}a}{\mathcal H_{\rm eq}a_{\rm eq}} = k\frac{\sqrt{2}a}{k_{\rm eq}a_{\rm eq}} = \kappa y \gg 1\;.
\end{equation}
Therefore, if we want to match the radiation-dominated solution with the solution of the M\'esz\'aros equation, we need to choose a suitable $y_m$ at which to perform the junction of the two solutions such that $1/\kappa \ll y_m \ll 1$.

Asking for the equality of the two solutions and their derivatives at $y_m$ implies solving the following system:
\begin{eqnarray}
A\Phi_{\rm P}\ln(B\kappa y_m) = C_1\left(y_m + 2/3\right) + C_2D_2(y_m)\;,\\
\frac{A\Phi_{\rm P}}{y_m} = C_1 + C_2\left.\frac{dD_2}{dy}\right|_{y=y_m}\;.
\end{eqnarray}

\hrulefill

\begin{ex} Solve the above system and take the dominant contribution for $y_m \to 0$ (because the junction condition has to be imposed for $y_m \ll 1$). Show that:
\begin{eqnarray}
C_1 = \frac{3}{2}A\Phi_{\rm P}\ln\left(4B\kappa e^{-3}\right)\;, \qquad C_2 = \frac{2}{3}A\Phi_{\rm P}\;.
\end{eqnarray}
\end{ex}

\hrulefill

Therefore, the solution for $\delta_{\rm c}$ valid at all times and at small scales is the following:
\begin{equation}
\delta_{\rm c}(k,y) = \frac{3}{2}A\Phi_{\rm P}\ln\left(4\sqrt{2}B\kappa e^{-3}\right)(y + 2/3) + \frac{2}{3}A\Phi_{\rm P}D_2(y)\;, \quad (\kappa \gg 1)\;.
\end{equation}
We can project this solution at late times, when matter dominates, and thus neglect the decaying mode $D_2$ and write:
\begin{equation}\label{deltalatetimes}
\boxed{\delta_{\rm c}(k,a) = \frac{3}{2}A\Phi_{\rm P}\ln\left(\frac{4\sqrt{2}Be^{-3}k}{k_{\rm eq}}\right)\frac{a}{a_{\rm eq}}\;, \quad (a \gg a_{\rm eq}, k \gg k_{\rm eq})}
\end{equation}
Note again the separated dependence from $k$ and from the scale factor, since the growth function is $D(a) = a$. This allows us to write the transfer function for CDM as follows, using Eq.~\eqref{keqdefinitionOm} to eliminate $a_{\rm eq}$:
\begin{equation}
	\boxed{T_\delta(k) = \frac{Ak_{\rm eq}^2}{2H_0^2\Omega_{\rm m0}}\ln\left(\frac{4\sqrt{2}Be^{-3}k}{k_{\rm eq}}\right)D(a)\;, \quad (a \gg a_{\rm eq},k \gg k_{\rm eq})}
\end{equation}
Recall that the factor $3\Phi_{\rm P}/2$ is the adiabatic initial condition on $\delta_{\rm c}$, and thus enters the primordial power spectrum. The above transfer function can be generalized to include dark energy, if the latter does not cluster. If it is the case, as for the cosmological constant, its effect enters only the growth factor and in $k_{\rm eq}$, since, being another component and having the fixed total $\Omega_{\rm r0} + \Omega_{\rm m0} + \Omega_{\rm \Lambda} = 1$, the relative amount of radiation and matter has to change (usually, the amount of radiation is fixed) and thus $k_{\rm eq}$ changes.

We can also determine the transfer function for the gravitational potential $\Phi$ at late times. From Eq.~\eqref{smallscalesPoissoneqDM} we have that:
\begin{equation}\label{smallscalesPoissoneqDM2}
k^2\Phi = 4\pi Ga^2\rho_{\rm m}\delta_{\rm m} = \frac{3H_0^2\Omega_{\rm m0}}{2a}\delta_{\rm m}\;,
\end{equation}
where we have included baryons since we are considering late times and we have seen that after recombination $\delta_{\rm b} = \delta_{\rm c}$. We are neglecting any dark energy contribution to the expansion of the universe and considering a radiation plus matter model. Hence, $\Omega_{\rm 0} \approx 1$.

We thus have for the gravitational potential, combining Eqs.~\eqref{deltalatetimes} and \eqref{smallscalesPoissoneqDM2}:
\begin{equation}
\boxed{\Phi(k) = \frac{9Ak_{\rm eq}^2}{8k^2}\Phi_{\rm P}(k)\ln\left(\frac{4\sqrt{2}Be^{-3}k}{k_{\rm eq}}\right)\;, \qquad k \gg k_{\rm eq}}
\end{equation}
The transfer function for the gravitational potential is usually normalized to $9\Phi_{\rm P}/10$ and therefore:
\begin{equation}
\boxed{T_\Phi(k) = \frac{5Ak_{\rm eq}^2}{4k^2}\ln\left(\frac{4\sqrt{2}Be^{-3}k}{k_{\rm eq}}\right)\;, \qquad k \gg k_{\rm eq}}
\end{equation}

\hrulefill

\begin{ex} Using $\Omega_{\rm r0}h^2 = 4.15\times 10^{-5}$ in Eq.~\eqref{keqdefinitionOmOr} and defining 
\begin{equation}
	q \equiv k\times\frac{\mbox{Mpc}}{\Omega_{\rm m0}h^2}\;,
\end{equation}
show that we can cast the above transfer function in the following form:
\begin{equation}\label{TPhiapproximated}
\boxed{T_\Phi(k) = \frac{\ln\left(2.40q\right)}{(4.07q)^2}}
\end{equation}
See also \cite[pag. 310]{Weinberg:2008zzc}.\index{Transfer function!Matter}
\end{ex}

\hrulefill

We have written the transfer function as in Eq.~\eqref{TPhiapproximated} because it is simpler to compare it with the numerical fit of Bardeen, Bond, Kaiser, and Szalay (BBKS) of the exact transfer function \cite{Bardeen:1985tr}, which is as follows:
\begin{equation}\label{BBKStransferfunction}
T_{\rm BBKS}(k) = \frac{\ln\left(1 + 2.34q\right)}{2.34q}\left[1 + 3.89q + (16.2q)^2 + (5.47q)^3 + (6.71q)^4\right]^{-1/4}\;,
\end{equation}
and see that for large $q$ it goes as $\ln(2.34q)/(3.96q)^2$, which is in good agreement with our analytic estimate \eqref{TPhiapproximated}.\index{Transfer function!BBKS}

Given the transfer function $T_{\Phi}$, $\delta_{\rm m}$ can be written, from Eq.~\eqref{smallscalesPoissoneqDM2}, as:
\begin{equation}
\delta_{\rm m}(k,a) = \frac{2a}{3H_0^2\Omega_{\rm m0}}k^2\Phi = \frac{3k^2}{5H_0^2\Omega_{\rm m0}}\Phi_{\rm P}(k)T_\Phi(k)a = \frac{3k^2}{5H_0^2\Omega_{\rm m0}}\Phi_{\rm P}(k)T_\Phi(k)D(a)\;.
\end{equation}
In the last equality, we have recovered the growth factor $D(a)$ in order to provide a more general formula. From this solution, we can obtain the power spectrum for $\delta_{\rm m}$ starting from the primordial one for $\Phi$, i.e.
\begin{equation}
P_\delta(k,a) = \frac{9k^4}{25H_0^4\Omega_{\rm m0}^2}P_\Phi(k)T_\Phi^2(k)D^2(a)\;.
\end{equation}
Using Eq.~\eqref{Rzetaprimordial2}, with $\Phi = -\Psi$ since we are neglecting neutrinos, $\Phi_{\rm P} = 2\mathcal R/3$ and thus the primordial power spectrum of $\Phi$ can be substituted with the one from $\mathcal{R}$, and we get:
\begin{equation}
P_\delta(k,a) = \frac{4k^4}{25H_0^4\Omega_{\rm m0}^2}P_\mathcal{R}(k)T_\Phi^2(k)D^2(a)\;.
\end{equation}
From Eq.~\eqref{Delta2S}, we can write the explicit dependence of the primordial power spectrum of $k$ as follows:
\begin{equation}\label{matterP}
\boxed{P_\delta(k,a) = \frac{8\pi^2k}{25H_0^4\Omega_{\rm m0}^2}A_S \left(\frac{k}{k_*}\right)^{n_S - 1}T_\Phi^2(k)D^2(a)}
\end{equation}
The power spectrum $P_\delta(k,a)$ can be determined through the observation of the distribution of galaxies in the sky. Therefore, through the above formula, we can probe many quantities of great interest, such as the primordial tilt of the power spectrum.

\begin{figure}[htbp]
\center
\includegraphics[width=\columnwidth]{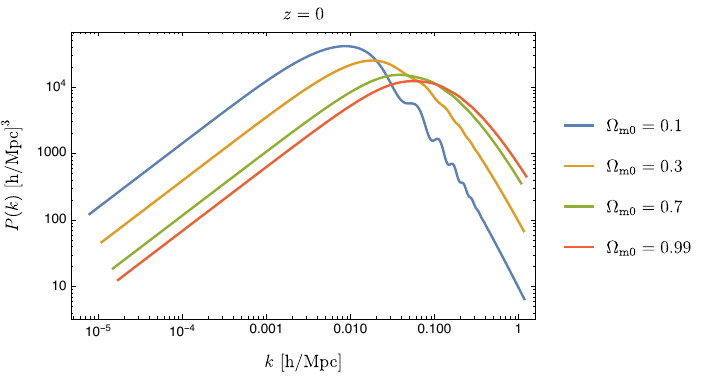}
\caption{Matter power spectra for (left to right, in case no colours are available) $\Omega_{\rm m0} = 0.1$ (blue), $\Omega_{\rm m0} = 0.3$ (yellow), $\Omega_{\rm m0} = 0.7$ (green) and $\Omega_{\rm m0} = 0.99$ (red).}
\label{Fig:Pkplot}
\end{figure}

Now, let us make a plot of the power spectrum today ($z = 0$) using CLASS and fixing all the parameters to the $\Lambda$CDM best fit values except for $\Omega_{\rm m0}$, which we let free. We show in Fig.~\ref{Fig:Pkplot} the shape of the matter power spectrum for $\Omega_{\rm m0} = 0.1$ (blue), $\Omega_{\rm m0} = 0.3$ (yellow), $\Omega_{\rm m0} = 0.7$ (green), and $\Omega_{\rm m0} = 0.99$ (red). 

The first interesting feature of the power spectrum is that it has a maximum. This maximum occurs roughly at $k_{\rm eq}$ for the following reason: entering the horizon at equivalence is the best time for a matter fluctuation to grow. In fact, as we have seen, scales that entered the horizon earlier ($k > k_{\rm eq}$) are suppressed because radiation is dominating and $\delta$ grows logarithmically during this epoch, cf. Eq.~\eqref{deltacevoRD}. These are the scales of great observational interest.

On the other hand, the scales that entered after the equivalence grow proportionally to $a$, cf.~Eq.~\eqref{deltamattevo}. Evidently, the scale that entered at equivalence has had more time to grow than all the others; hence, the maximum, or \textbf{turnover} in the power spectrum. 

The second interesting feature of the power spectrum is that its maximum is shifted to the left when we reduce $\Omega_{\rm m0}$. This means that the equivalence wavenumber $k_{\rm eq}$ is smaller when $\Omega_{\rm m0}$ is smaller, and this can be immediately seen from Eq.~\eqref{keqdefinitionOmOr} if we fix $\Omega_{\rm r0}$.

Remarkably, observation favors the line for $\Omega_{\rm m0} = 0.3$ of Fig.~\ref{Fig:Pkplot}. Since the radiation content is very well established from CMB observation (and our knowledge of neutrinos) as well as the spatial flatness of the universe (from the CMB, we shall see this in Chapter~\ref{Chap:CMBEvo}), and its age (at least a lower bound) and $H_0$ are also well-determined, the only way to make the total is to add a further component, which clearly is dark energy. The important point here is that the necessity for dark energy can already be seen by analyzing the large scale structure of the universe (together with CMB) and this was realized well before type Ia supernovae started to be used as standard candles \cite{Maddox:1990hb}, \cite{Efstathiou:1990xe}.

\begin{figure}[htbp]
\center
\includegraphics[width=\columnwidth]{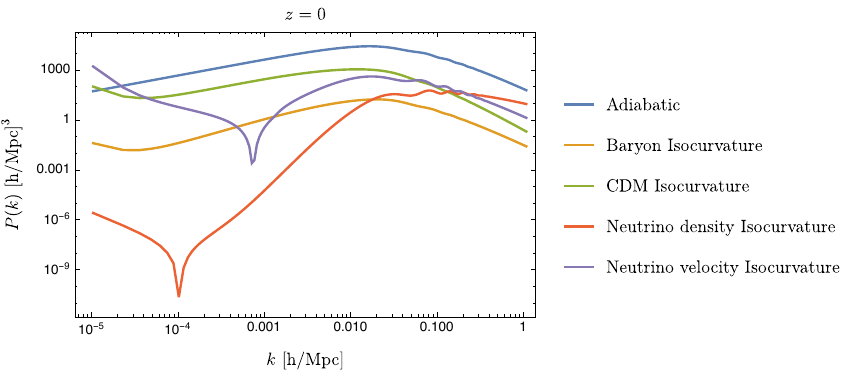}
\caption{Matter power spectra for different initial conditions: Adiabatic (blue), Baryon Isocurvature (yellow), CDM Isocurvature (green), Neutrino density Isocurvature (red), and Neutrino velocity Isocurvature (purple).}
\label{Fig:PlotPkvarIC}
\end{figure}

For completeness, in Fig.~\ref{Fig:PlotPkvarIC} we plot with CLASS the matter power spectrum at $z = 0$ for the $\Lambda$CDM model, with different choices of the initial conditions.

The transfer function thus tells us how the shape of the primordial power spectrum is changed through cosmological evolution, and through the analytic estimates that we have done in this chapter, we understand that the $k$-dependence is set up during radiation-domination. 

However, we have made two important assumptions in our calculations: we have neglected baryons and neutrino anisotropic stress (we have taken into account neutrinos from the point of view of the background expansion). If we aim for precise predictions, and we have to in order to keep pace with the increasing sophistication of the observational techniques, we must take them into account. A more precise fitting formula (to numerical calculations performed with CMBFAST) taking into account baryons and neutrino anisotropic stress is given by \cite{Eisenstein:1997ik}.

\begin{figure}[htbp]
\center
\includegraphics[width=0.5\columnwidth]{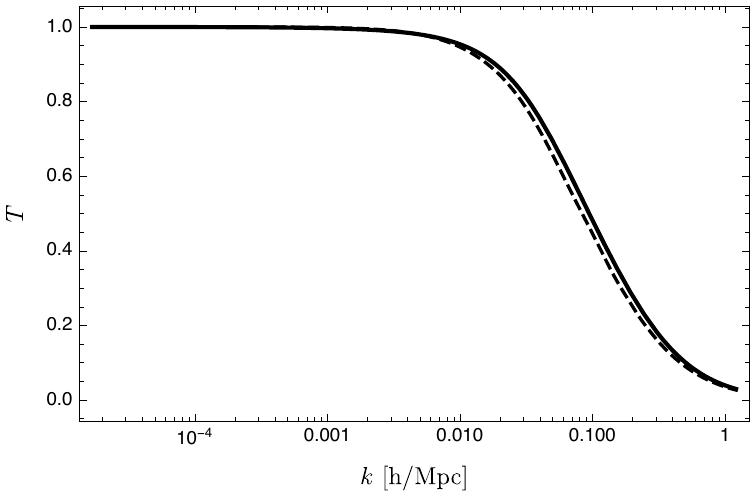}\includegraphics[width=0.5\columnwidth]{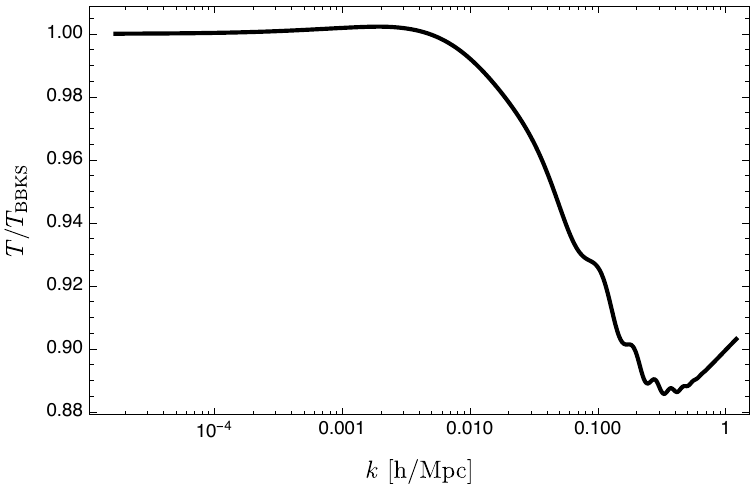}
\caption{\textit{Left Panel.} Evolution of BBKS transfer function of Eq.~\eqref{BBKStransferfunction} (solid line) compared with the numerical computation of CLASS, using $\Omega_{\rm m0} = 0.95$ (which means negligible DE). \textit{Right Panel.} Ratio between the numerical result and the BBKS transfer function.}
\label{Fig:CompTBBKSTClassnoDEPlotandTkPhinoDEnormBBKSPlot}
\end{figure}

In Fig.~\ref{Fig:CompTBBKSTClassnoDEPlotandTkPhinoDEnormBBKSPlot}, we compare the BBKS transfer function with the numerical calculation of CLASS, adopting $\Omega_{\rm m0} = 0.95$, while leaving all the remaining cosmological parameters as in the $\Lambda$CDM model, except for $\Omega_\Lambda$, which is adjusted to the value $\Omega_\Lambda = 1.632908\times 10^{-3}$ in order to match the correct budget of energy density. So, in practice, we are neglecting dark energy.

As can be appreciated from the plots, the BBKS transfer function overestimates the correct transfer function by about 5\% for scales $k \gtrsim 0.1$ h Mpc$^{-1}$; see also \cite[pag. 208]{Dodelson:2003ft}. The reason is that baryons behave like radiation before decoupling because of their tight coupling due to Thomson scattering. Therefore, they contribute further to thwart scales entering the horizon before equivalence.

\begin{figure}[htbp]
\center
\includegraphics[width=0.5\columnwidth]{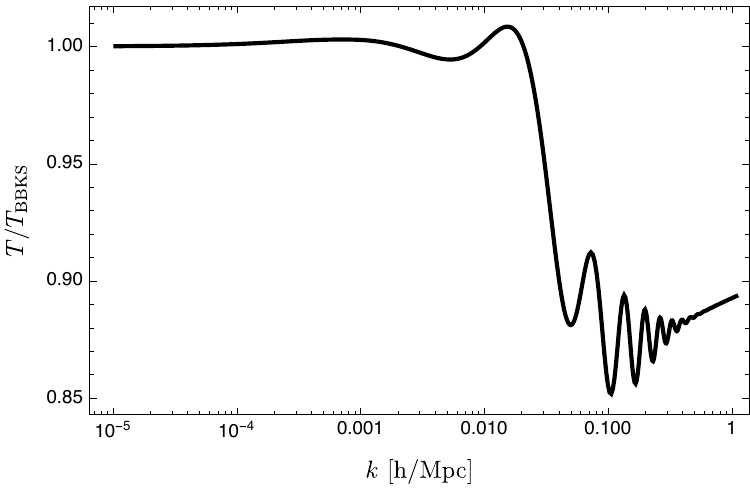}\includegraphics[width=0.5\columnwidth]{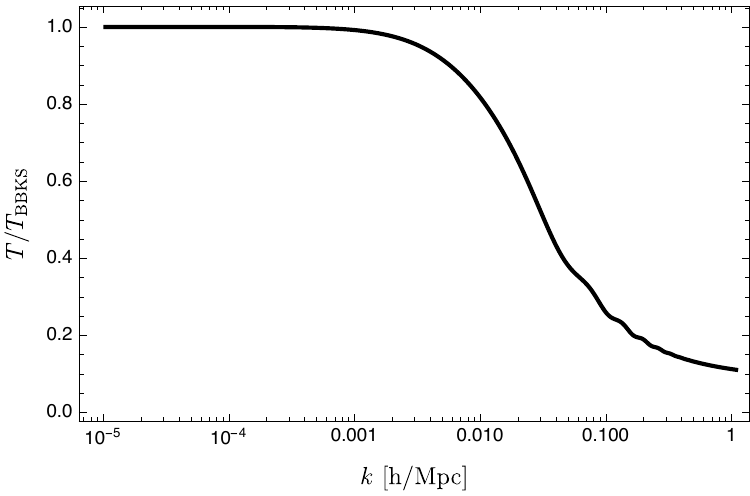}
\caption{\textit{Left Panel.} Ratio between the numerical result computed with CLASS assuming the $\Lambda$CDM model, and the BBKS transfer function with $\Omega_{\rm m0}h^2 = 0.12038$. \textit{Right Panel.} Ratio between the numerical result computed with CLASS assuming the $\Lambda$CDM model, and the BBKS transfer function with $\Omega_{\rm m0} = 1$.}
\label{Fig:TkPhiLCDMnormBBKSPlot1and2}
\end{figure}

In Fig.~\ref{Fig:TkPhiLCDMnormBBKSPlot1and2}, we display the ratio between the numerical transfer function computed with CLASS for the $\Lambda$CDM model and the BBKS transfer function in two cases. In the left panel, the same matter density parameter of the $\Lambda$CDM model is used for both transfer functions. In the right panel, we used the BBKS transfer function with $\Omega_{\rm m0} = 1$.

The latter choice was made in order to reproduce the plot of \cite[pag. 208]{Dodelson:2003ft} and thus to show the very large correction due to the cosmological constant which, if not taken into account, leads to an 80\% error at the scale $k = 0.1$ $h$ Mpc$^{-1}$. 

When $\Lambda$ is taken into account properly, the BBKS transfer function still overestimates the correct transfer function by at least 10\% on small scales, which implies an imprecision of 1\% in the power spectrum (since this depends on the squared transfer function). Moreover, since it does not include baryons, the BBKS cannot describe the BAO, which appear as oscillations in the transfer function at about $k = 0.1$ $h$ Mpc$^{-1}$ in Fig.~\ref{Fig:TkPhiLCDMnormBBKSPlot1and2}.

\section{The transfer function for tensor perturbations}

Treating the evolution of tensor perturbations, which we have done in the context of inflation in Chapter~\ref{Chap:Inflation}, is much easier than for scalar modes for two reasons. The first is that CDM and baryons are very non-relativistic and so do not possess anisotropic stress, which would source gravitational waves. The second reason is that, though photons and neutrinos do have anisotropic stresses, these are always very small, so it is a good approximation to neglect them.

Therefore, what we have to do is to recover the calculations of Sec.~\ref{Sec:ProductionofGWduringinfl} and solve Eq.~\eqref{heqGW} with a $\mathcal H$ given in the radiation- and matter-dominated epoch. As we already know, on very large scales, meaning $k^2 \ll a''/a$, $h$ is a constant (we shall call it $h_{\rm P}$), and this holds true for whatever background expansion we choose. We used this fact in order to determine the primordial power spectrum, and in the present section we use it again to determine the initial value for solving Eq.~\eqref{heqGW}.

Using Eqs.~\eqref{Friedmannequationmatterandradiation} and \eqref{conformalHubblefactoratequality}, we can write the conformal Hubble parameter as follows:
\begin{equation}
	\mathcal H = \frac{\mathcal H_{\rm eq}\sqrt{1 + y}}{\sqrt{2}y}\;,
\end{equation}
where we have introduced again $y \equiv \rho_{\rm m}/\rho_{\rm r}$ as new independent variable.

\hrulefill

\begin{ex}
	Cast Eq.~\eqref{heqGW} using $y$ as independent variable and the above form of the conformal Hubble factor. Show that:
	\begin{equation}\label{GWequationy}
		\boxed{(1 + y)\frac{d^2h}{dy^2} + \left[\frac{1}{2} + \frac{2(1 + y)}{y}\right]\frac{dh}{dy} + \kappa^2h = 0}
	\end{equation}
	using also Eq.~\eqref{kappadefinition}.
\end{ex}

\hrulefill

The above equation cannot be solved analytically, but we can find an exact solution in the same four instances that we worked out for scalar perturbations. On super-horizon scales, we already know that $h = h_{\rm P}$ irrespective of the background evolution (this is relevant only for the decaying mode), so we skip this case.

\subsection{Radiation-dominated epoch}

In this case, we consider $y \ll 1$ in Eq.~\eqref{GWequationy}:
\begin{equation}
		\frac{d^2h}{dy^2} + \frac{2}{y}\frac{dh}{dy} + \kappa^2h = 0\;,
\end{equation}
which can be rewritten as:
\begin{equation}
		\frac{d^2(\kappa hy)}{d(\kappa y)^2} + \kappa hy = 0\;,
\end{equation}
and hence is a harmonic oscillator equation for the quantity $\kappa hy$ with respect to the variable $\kappa y$, with unitary frequency. Hence, the solution for $h$ is:
\begin{equation}
	h(k,y) = C_1(k)\frac{\sin(\kappa y)}{\kappa y} + C_2(k)\frac{\cos(\kappa y)}{\kappa y}\;,
\end{equation}
where, as usual, $C_1$ and $C_2$ are generic $k$-dependent functions. For $y \to 0$, or equivalently on very large scales, $\kappa y \to 0$ only the sine tends to a finite result, and hence we must put $C_2$ to zero:
\begin{equation}
	h(k,y) = h_{\rm P}(k)\frac{\sin(\kappa y)}{\kappa y} = h_{\rm P}(k)j_0(\kappa y)\;.
\end{equation}
Equivalently, putting $\mathcal H = 1/\eta$ into the original equation \eqref{heqGW}, one finds:
\begin{equation}
	h'' + \frac{2}{\eta}h' + k^2h = 0\;,
\end{equation}
and so:
\begin{equation}\label{hGWdeepraddom}
	h(k,\eta) = h_{\rm P}(k)\frac{\sin(k\eta)}{k\eta} = h_{\rm P}(k)j_0(k\eta)\;,
\end{equation}
since, indeed, $\kappa y = k\eta$ when radiation dominates, cf. Eq.~\eqref{ketaequaltokappay}.

\subsection{Matter-dominated epoch}

In this case, we consider $y \gg 1$ in Eq.~\eqref{GWequationy}:
\begin{equation}
		y\frac{d^2h}{dy^2} + \frac{5}{2}\frac{dh}{dy} + \kappa^2h = 0\;,
\end{equation}
or, using Eq.~\eqref{heqGW} with $\mathcal H = 2/\eta$:
\begin{equation}
	h'' + \frac{4}{\eta}h' + k^2h = 0\;.
\end{equation}
This equation can be cast in the form of a Bessel equation. 

\hrulefill

\begin{ex}
	Introduce the new function:
	\begin{equation}
		h \equiv g(k\eta)^{\alpha}\;,
	\end{equation}
	where $\alpha$ is a number to be determined. Then, show that:
	\begin{equation}
		\eta^2 g'' + (2\alpha + 4)\eta g' + g[k^2\eta^2 + \alpha^2 + 3\alpha] = 0\;.
	\end{equation}
\end{ex}

\hrulefill

Then, we recover the form of the Bessel function for $\alpha = -3/2$, for which the order is $3/2$, and hence:
\begin{equation}
	g \propto J_{3/2}(k\eta) = \sqrt{\frac{2k\eta}{\pi}}j_1(k\eta)\;,
\end{equation}
where we have introduced the spherical Bessel function and neglected the one of the second kind, since it diverges for $k\eta \to 0$. The gravitational wave amplitude thus evolves as:
\begin{equation}
	h(k,\eta) = C_1(k)\frac{
	j_1(k\eta)}{k\eta} = C_1(k)\left[\frac{
	\sin(k\eta)}{(k\eta)^3} - \frac{\cos(k\eta)}{(k\eta)^2}\right]\;.
\end{equation}
In the limit $k\eta \to 0$:
\begin{equation}
	\frac{
	\sin(k\eta)}{(k\eta)^3} - \frac{\cos(k\eta)}{(k\eta)^2} \to \frac{1}{3}\;,
\end{equation}
hence, we can write:
\begin{equation}
	h(k,\eta) = 3h_{\rm P}(k)\left[\frac{
	\sin(k\eta)}{(k\eta)^3} - \frac{\cos(k\eta)}{(k\eta)^2}\right]\;,
\end{equation}
but this is valid only for those modes that enter the horizon well deep into the matter-dominated epoch. Using the definition of $\kappa$ in Eq.~\eqref{kappadefinition} and the fact that, deep in the matter-dominated epoch, we have that:
\begin{equation}
	\mathcal H = \frac{2}{\eta} = \frac{\mathcal H_{\rm eq}}{\sqrt{2y}}\;,
\end{equation}
we can conclude that:
\begin{equation}
	k\eta = 2\kappa\sqrt{y}\;.
\end{equation}
Thus, the solution for $h(k,y)$ is:
\begin{equation}
	h(k,y) = \frac{3h_{\rm P}(k)}{4}\left[\frac{
	\sin(2\kappa\sqrt{y})}{2(\kappa\sqrt{y})^3} - \frac{\cos(2\kappa\sqrt{y})}{(\kappa\sqrt{y})^2}\right]\;,
\end{equation}
which is again valid only for those modes $\kappa\sqrt{y} \ll 1$, and hence, since $y \gg 1$, then $\kappa \ll 1$, i.e., modes that entered the horizon well deep into the matter-dominated epoch.

\subsection{Deep inside the horizon}

In the case $k \gg \mathcal H$, we work directly on Eq.~\eqref{heqGW} and, following \cite{Weinberg:2008zzc}, we introduce a new variable
\begin{equation}
	x \equiv \int\frac{d\eta}{a^2}\;,
\end{equation}
with the purpose of eliminating the first-order derivative, we can thus obtain the following equation:
\begin{equation}\label{hxequationsmallscales}
	\frac{d^2h}{dx^2} + k^2a^4h = 0\;.
\end{equation}
The latter is not analytically solvable for a generic $a(x)$, but it is possible to find an approximated WKB solution. Indeed, if $a(x)$ is approximately constant, the above equation becomes that of a harmonic oscillator, and the solution is simply:
\begin{equation}
	h(k,x) \propto e^{\pm ika^2}\;.
\end{equation}
Taking into account the $x$ dependence of $a$, we can use the following ansatz:
\begin{equation}
	h(k,x) = A(x)\exp\left(\pm ik\int a^2dx\right)\;,
\end{equation}
and substitute it back into Eq.~\eqref{hxequationsmallscales}, obtaining:
\begin{equation}
	\frac{d^2A}{dx^2} + 2\frac{dA}{dx}(\pm ika^2) + A\left(\pm 2ika\frac{da}{dx}\right) + k^2a^4A = 0\;.
\end{equation}
Equating the imaginary part to zero, we then have:
\begin{equation}
	\frac{dA}{dx}a^2 + Aa\frac{da}{dx} = 0\;,
\end{equation}
for which the solution is:
\begin{equation}
	A(x) \propto 1/a(x)\;.
\end{equation}
Hence, the approximated WKB solution is:
\begin{equation}
	h(k,\eta) \propto \frac{1}{a}\exp\left(\pm ik\int a^2dx\right) = \frac{1}{a}\exp\left(\pm ik\eta\right)\;.
\end{equation}
Note that this solution is valid for any background expansion since we have made no assumptions about it. Moreover, note the oscillatory behavior $\exp\left(\pm ik\eta\right)$ with amplitude damped by a factor $1/a$. Recall from \cite{Weinberg:1972} that the energy-momentum tensor of a gravitational wave is:
\begin{equation}
	\langle t_{\mu\nu}\rangle = \frac{p_\mu p_\nu}{16\pi G}\left(|h_+|^2 + |h_\times|^2\right)\;,
\end{equation}  
where $t_{\mu\nu}$ is the gravitational pseudo-tensor, the average over it is taken over a sufficiently large region of spacetime such that the oscillatory terms give a constant result, and $p_\mu$ is the physical momentum $p_\mu = dx_\mu/d\lambda$ of the gravitational wave. We know that in a FLRW metric for a massless particle $p_0 = p \propto 1/a$, and hence $\langle t_{00}\rangle \propto 1/a^4$, as we expected for the energy density of a relativistic species (the graviton, in this case).

Now, we can match the above WKB solution (for $y \ll 1$) with the one we found in Eq.~\eqref{hGWdeepraddom} (for large $\kappa y$), where we use $y$ instead of $\eta$ because it is straightforward to project the matching at late times. It is the same procedure we employed in order to find the transfer function for CDM. So, we have:
\begin{equation}
	\frac{C_1(k)}{ya_{\rm eq}}\sin(\kappa y) = h_{\rm P}(k)\frac{\sin(\kappa y)}{\kappa y}\;, 
\end{equation}
and thus
\begin{equation}
	C_1(k) = h_{\rm P}(k)\frac{a_{\rm eq}}{\kappa} = h_{\rm P}(k)\frac{a_{\rm eq}H_0\Omega_{\rm m0}}{k\sqrt{\Omega_{\rm r0}}} = h_{\rm P}(k)\frac{H_0\sqrt{\Omega_{\rm r0}}}{k}\;.
\end{equation}
So, the solution for small scales is:
\begin{equation}
	\boxed{h(k,\eta) = h_{\rm P}(k)\frac{H_0\sqrt{\Omega_{\rm r0}}}{ka(\eta)}\sin(k\eta)}
\end{equation}
Note that this solution is valid for any content of the universe since there is no longer a $y$ dependence appearing (any content after radiation-domination, since we have matched the solutions there). Close to the present time $\eta_0$, we can set as usual $a(\eta_0) = 1$, and then the gravitational wave solution for small scales is:
\begin{equation}
	\boxed{h(k,t) = h_{\rm P}(k)\frac{H_0\sqrt{\Omega_{\rm r0}}}{k}\sin\left[k\eta_0 + k(t - t_0)\right]}
\end{equation}
for $t$ close to $t_0$ and where we have used $d\eta = dt/a(t)$ close to the present time. This profile is very different from those detected by LIGO for the merging of black holes, the first in \cite{Abbott:2016blz}, and from the spectacular event GW170817 of the merging of two neutron stars detected both by LIGO and VIRGO \cite{TheLIGOScientific:2017qsa}, whose electromagnetic counterpart was also seen as a short gamma-ray burst (GRB) \cite{GBM:2017lvd}. The main characteristics of these profiles are their growth in frequency and amplitude for times close to the merging time. The cosmological gravitational wave profile is a simple sine function (for very small scales), with an amplitude containing cosmological information of great relevance such as the primordial amplitude, which we have seen to be related to the energy scale of inflation.\index{Transfer function!Gravitational waves}

\clearpage
\chapter{Anisotropies in the Cosmic Microwave Background}\label{Chap:CMBEvo}

{\rightskip=3truepc\leftskip=3truepc\noindent
{\it Long the realm of armchair philosophers, the study of the origins and evolution of the universe became a physical science with falsifiable theories}
\vskip 0.10 in
\centerline{\it ---Wayne Hu, PhD Thesis}
\vskip 0.20 in
}

In this chapter, we address the hierarchy of Boltzmann equations that we have found for photons and present an approximate, semi-analytic solution that will allow us to understand the temperature correlation in the CMB sky and its relation to the cosmological parameters. Our scope is to understand the features of the CMB power spectra.
In these spectra, the definition
\begin{equation}
	\mathcal D^{TT}_\ell \equiv \frac{\ell(\ell + 1)C_{TT,\ell}}{2\pi}\;,
\end{equation}
is used. We shall see the reason for the $\ell(\ell + 1)$ normalization, whereas the $C_{TT,\ell}$'s are given in Eq.~\eqref{ClfunctionThetal} as functions of the multipole moments of the temperature distribution and the primordial power spectrum for scalar perturbations.\index{Cosmic Microwave Background!Planck TT spectrum}

Data points are worked out by Planck up to $\ell \approx 2500$. What can we say from this number about the angular sensitivity of Planck? It can be roughly computed as follows. For a given $\ell_{\rm max}$, how many realizations of $a_{\ell m}$ do we have? 

\hrulefill

\begin{ex}
For each $\ell$ we have $2\ell + 1$ possible values of $m$, thus show that:
\begin{equation}
	N_{\ell_{\rm max}} = \sum_{\ell = 0}^{\ell_{\rm max}}(2\ell + 1) = (\ell_{\rm max} + 1)^2\;.
\end{equation}
\end{ex}

\hrulefill

The full sky has:
\begin{equation}
	4\pi \mbox{ rad}^2 = \frac{4}{\pi}(180 \mbox{ deg})^2 \approx 41000\mbox{ deg}^2\;.
\end{equation}
If an experiment has a sensitivity of $7$ deg, then we can have at most
\begin{equation}
	\frac{4}{\pi}(180/7)^2 \approx 842\;,
\end{equation}
pieces of independent information, and therefore we can determine as many $a_{\ell m}$. This gives $\ell_{\rm max} \approx 28$, and it was the sensitivity of \textit{CoBE}. For \textit{Planck}, the angular sensitivity was $5$ arcmin, which corresponds to 
\begin{equation}
	\frac{4}{\pi}(180\times 60/5)^2 \approx 10^6\;,
\end{equation}
pieces of independent information and then to $\ell_{\rm max} = 2436$. 

In this chapter, we omit the superscript $S$ referring to the scalar perturbations contribution to $\Theta$, since we focus almost exclusively on this type of perturbation. We shall use the $T$ superscript to distinguish the tensor contribution.

\section{Free-streaming}

It is convenient to start neglecting the collisional term in the Boltzmann equation and thus consider the phase of \textbf{free-streaming}\index{Photons!Free-streaming}. The following discussion is similar to the one in Sec.~\ref{Sec:masslessnuBoltzeq}. Consider Eq.~\eqref{pertBoltzeqgeneralscalarpertFT} with no collisional term. Using the definition of $\Theta$ in Eq.~\eqref{Thetadefinition}, we can write for scalar perturbations:
\begin{equation}
	\left(\frac{\partial}{\partial\eta} + \frac{dx^i}{d\eta}\frac{\partial}{\partial x^i}\right)(\Theta + \Psi) = \Psi' - \Phi'\;.
\end{equation}
As we know from the Boltzmann equation, the differential operator on the left hand side is a convective derivative, i.e., a derivative along the photon path:
\begin{equation}\label{Freestreamingphotons}
	\frac{d}{d\eta}(\Theta + \Psi) = \Psi' - \Phi'\;,
\end{equation}
whose inversion is the basis of the \textbf{line-of-sight integration} approach to CMB anisotropies \cite{Seljak:1996is}, which is an alternative to attacking the hierarchy of coupled Boltzmann equations (which still must be attacked but can be truncated at much lower $\ell$'s) as it was done in \cite{Ma:1995ey}. We shall see this technique in some detail in Sec.~\ref{Sec:lineofsightintegration}.

For time-independent potentials, as they are in the matter-dominated epoch, the collisionless Boltzmann equation for photons tells us that $\Theta + \Psi$ is constant along the photon paths, i.e., along our past light-cone. 

Recall that the scalar-perturbed metric that we are using is given in Eq.~\eqref{confnewtmetric}:
\begin{equation}
	ds^2 = -a^2(\eta)(1 + 2\Psi)d\eta^2 + a^2(\eta)(1 + 2\Phi)\delta_{ij}dx^idx^j\;.
\end{equation}
Inside a potential well, $\Psi$ is negative. In order to be convinced of this, one has just to think about the Newtonian limit and realize that $2\Psi$ is the Newtonian gravitational potential; hence, it is negative. So, since $\Theta + \Psi$ stays constant, we have that:
\begin{equation}
	\Theta(\eta_*,\mathbf x_*,\hat{p}) + \Psi(\eta_*,\mathbf x_*) = \Theta(\eta_0,\mathbf x_0,\hat{p}) + \Psi(\eta_0,\mathbf x_0)\;.
\end{equation}
On the left-hand side, we have evaluated the quantities at recombination, whereas on the right-hand side, we have chosen the present time. Note that $\mathbf x_0$ is where our laboratory (the CMB experiment) is located, and as such, it is fixed. Therefore, since we can only detect photons on our past light-cone, and those from the CMB come from a fixed comoving distance $r_* = \eta_0 - \eta_*$, we have that:
\begin{equation}
	\mathbf x_* = \mathbf x_0 - r_*\hat{p}\;.
\end{equation}
Note that $\hat{p}$ is the photon direction, and so it is opposite to the direction of the line of sight $\hat n = -\hat p$. Therefore, the only independent variables are 2: the components of $\hat{p}$. As we discussed in Chapter \ref{Chap:PertubedBoltzmannEquations}, they become just a single one ($\mu$).

The potential $\Psi(\eta_0,\mathbf x_0)$ is usually neglected or incorporated into the potential at recombination since it is not detectable.\footnote{Only potential differences are detectable, as we know from classical Physics. In General Relativity this is true only in the linear approximation with which we are working here.} The above equation then tells us that:
\begin{equation}
	\Theta(\eta_*,\mathbf x_0 - r_*\hat{ p},\hat{ p}) + \Psi(\eta_*,\mathbf x_0 - r_*\hat{ p}) = \Theta(\eta_0,\mathbf x_0,\hat{ p})\;,
\end{equation}
i.e., the observed temperature fluctuation (on the right hand side) accounts for the energy loss due to climbing out of the potential well or falling down a potential hill. This is the so-called \textbf{Sachs-Wolfe effect}\index{Sachs-Wolfe effect} \cite{Sachs:1967er}. Writing the above equation in Fourier modes, we have:
\begin{equation}
	\int\frac{d^3\mathbf k}{(2\pi)^3}\Theta(\eta_0,\mathbf k,\hat{ p})e^{i\mathbf k\cdot\mathbf x_0} = \int\frac{d^3\mathbf k}{(2\pi)^3}\left[\Theta(\eta_*,\mathbf k,\hat{ p}) + \Psi(\eta_*,\mathbf k)\right]e^{i\mathbf k\cdot(\mathbf x_0 - r_*\hat{ p})}\;.
\end{equation}
We can set now $\mathbf x_0 = 0$, without loss of generality, and change the functional dependencies to the modulus $k$ and $\mu$, normalizing to the scalar primordial mode $\alpha(\mathbf k)$, cf. Eq.~\eqref{scalarprimormodenorm}. Hence, we have for the Fourier modes:
\begin{equation}\label{freestreamingphotonsFouriermodes}
	\Theta(\eta_0, k,\mu) = \left[\Theta(\eta_*, k,\mu) + \Psi(\eta_*, k)\right]e^{-i k\mu r_*}\;.
\end{equation}
Using the partial wave expansion, we get:
\begin{equation}
	\Theta_\ell(\eta_0, k) = \frac{1}{(-i)^\ell}\int_{-1}^1\frac{d\mu}{2}\mathcal P_\ell(\mu)\left[\Theta(\eta_*, k,\mu) + \Psi(\eta_*, k)\right]e^{-i k\mu r_*}\;,
\end{equation}
and using the relation:
\begin{equation}\label{intdmuPleikmur}
	\int_{-1}^{1}\frac{d\mu}{2}\;\mathcal{P}_\ell(\mu)e^{-ik\mu r_*} = (-i)^\ell j_\ell\left(kr_*\right)\;,
\end{equation}
which can be obtained by inverting the expansion of Eq.~\eqref{Partialwaveexpansionplanewave}, we can write:
\begin{equation}
	\Theta_\ell(\eta_0, k) = \Psi(\eta_*, k)j_\ell(kr_*) + \frac{1}{(-i)^\ell}\int_{-1}^1\frac{d\mu}{2}\mathcal P_\ell(\mu)\Theta(\eta_*, k,\mu)e^{-i k\mu r_*}\;.
\end{equation}
Using the partial wave expansion again, we can write the above formula as:
\begin{eqnarray}
	\Theta_\ell(\eta_0, k) = \Psi(\eta_*, k)j_\ell(kr_*)\nonumber\\ + \frac{1}{(-i)^\ell}\sum_{\ell'}(-i)^{\ell'}(2\ell' + 1)\Theta_{\ell'}(\eta_*, k)\int_{-1}^1\frac{d\mu}{2}\mathcal P_\ell(\mu)\mathcal P_{\ell'}(\mu)e^{-i k\mu r_*}\;.
\end{eqnarray}
We shall see later that, because of tight-coupling, the monopole and the dipole contribute the most at recombination. Hence, we can write, truncating the summation at $\ell' = 1$:
\begin{equation}
	\Theta_\ell(\eta_0, k) = \left(\Theta_0 + \Psi\right)(\eta_*, k)j_\ell(kr_*) + \frac{3\Theta_{1}(\eta_*, k)}{(-i)^{\ell - 1}}\int_{-1}^1\frac{d\mu}{2}\mathcal P_\ell(\mu)\mu e^{-i k\mu r_*}\;.
\end{equation}
The integral can be performed as follows:
\begin{equation}
	\int_{-1}^1\frac{d\mu}{2}\mathcal P_\ell(\mu)\mu e^{-i k\mu r_*} = i\frac{d}{d(kr_*)}\int_{-1}^1\frac{d\mu}{2}\mathcal P_\ell(\mu)e^{-i k\mu r_*} = \frac{1}{i^{\ell - 1}}\frac{d}{d(kr_*)}j_\ell\left(kr_*\right)\;.
\end{equation}
The same technique can be used, in principle, to calculate the integral for any $\ell'$: for each power of $\mu$ one gains a derivative of the spherical Bessel function. Recalling the formula \cite{Abramowitz1972}:\footnote{The website \url{http://functions.wolfram.com} is also very useful.}
\begin{equation}\label{derivativejl}
	\frac{dj_\ell(x)}{dx} = j_{\ell - 1}(x) - \frac{\ell + 1}{x}j_\ell(x)\;,
\end{equation}
we can write:
\begin{eqnarray}\index{Free-streaming!Solution}
	\Theta_\ell(\eta_0, k) = \left(\Theta_0 + \Psi\right)(\eta_*, k)j_\ell(kr_*)\nonumber\\ + 3\Theta_{1}(\eta_*, k)\left[j_{\ell - 1}(kr_*) - \frac{\ell + 1}{kr_*}j_\ell(kr_*)\right]\;.
\end{eqnarray}
So, the spherical Bessel functions that we have mentioned in Chapter~\ref{Chap:Evopert} start to appear. We have obtained the above free-streaming solution by neglecting the derivatives of the gravitational potentials in Eq.~\eqref{Freestreamingphotons}. Taking them into account is not difficult, since an additional piece containing the integration of such derivatives would appear in Eq.~\eqref{freestreamingphotonsFouriermodes}:
\begin{equation}
	\Theta(\eta_0, k,\mu) = \left[\Theta(\eta_*, k,\mu) + \Psi(\eta_*, k)\right]e^{-i k\mu r_*} + \int_{\eta_*}^{\eta_0}d\eta\;(\Psi' - \Phi')(\eta, k)e^{-i k\mu (\eta_0 - \eta)}\;.
\end{equation}
The exponential factor in the integral comes from the Fourier transform of the potentials and from considering:
\begin{equation}
	\mathbf x = \mathbf x_0 - (\eta_0 - \eta)\hat{ p}\;,
\end{equation}
at any given time $\eta$ along the photon trajectory (this is the ``line of sight'', in practice). Performing again the expansion in partial waves, we get:
\begin{eqnarray}\label{freestreamingformulaphotons}
	\Theta_\ell(\eta_0, k) = \left(\Theta_0 + \Psi\right)(\eta_*, k)j_\ell(kr_*)\nonumber\\ + 3\Theta_{1}(\eta_*, k)\left[j_{\ell - 1}(kr_*) - \frac{\ell + 1}{kr_*}j_\ell(kr_*)\right] \nonumber\\
	+ \int_{\eta_*}^{\eta_0}d\eta\;[\Psi'(\eta, k) - \Phi'(\eta, k)]j_\ell(kr)\;,
\end{eqnarray}
where 
\begin{equation}
	r \equiv \eta_0 - \eta\;.
\end{equation}
As we are going to see, the first two terms of the above formula contain the \textbf{primary anisotropies}\index{Cosmic Microwave Background!Primary anisotropies} of the CMB, which are the \textbf{acoustic oscillations}\index{Cosmic Microwave Background!Acoustic oscillations} and the \textbf{Doppler effect}.\index{Cosmic Microwave Background!Doppler effect} The $\Psi(\eta_*, k)$ contribution in the first term describes, as we have already anticipated, the \textbf{Sachs-Wolfe effect}. The last term is the \textbf{Integrated Sachs-Wolfe}\index{Cosmic Microwave Background!Integrated Sachs-Wolfe effect} (ISW) effect \cite{Sachs:1967er} and contributes only when the gravitational potentials are time-varying. This happens, as we have seen in Chapter \ref{Chap:Evopert}, when radiation and dark energy cannot be neglected in the energy budget. For this reason, the ISW effect is usually separated into the \textbf{early-times ISW}, arising due to the effect of radiation at decoupling, and the \textbf{late-times ISW}, whose cause is the time variation of the gravitational potentials due to dark energy, cf. Chapter \ref{Chap:Evopert}.

Once we know all the contributions to the above formula, we can use Eq.~\eqref{ClfunctionThetal} and provide the prediction on the $C_{TT,\ell}^S$ spectrum. In principle, Eq.~\eqref{freestreamingformulaphotons} has the very same form for neutrinos, but with an initial conformal time $\eta_i$ that is well before $\eta_*$, since neutrinos do not interact and therefore they only free-stream (at least for temperatures of the primordial plasma below 1 MeV).

\begin{figure}[htbp]
\center
\includegraphics[width=\columnwidth]{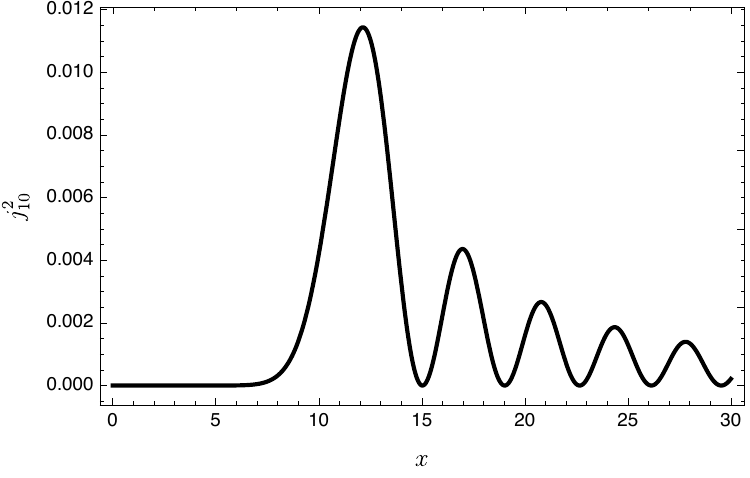}
\caption{Evolution of the spherical Bessel function $j_{10}^2(x)$. We have chosen to plot the squared spherical Bessel function because it is the relevant window function when computing the $C_\ell$'s.}
\label{Fig:j102Plot}
\end{figure}

The spherical Bessel function $j_\ell(x)$ attains a global maximum roughly when $x \approx \ell$ and rapidly vanishes when moving away from it. See Fig.~\ref{Fig:j102Plot}. Therefore, for a given multipole number $\ell$, the scale that contributes most to the observed anisotropy is:
\begin{equation}
	k \approx \frac{\ell}{\eta_0 - \eta_*}\;.
\end{equation}
We have anticipated this already in Chapter~\ref{Chap:Evopert}.

Let us now focus on the monopole and the dipole contributions at recombination, in Eq.~\eqref{freestreamingformulaphotons}. We shall commence in the next section by discussing very large scales. 

\section{Anisotropies on large scales}\index{Cosmic Microwave Background!Large scale anisotropies}

On large scales $k\eta \ll 1$, the relevant equations are those for the monopoles of Chapter~\ref{Chap:IC}, which we report here:
\begin{eqnarray}
	\delta_\gamma' = -4\Phi'\;, \quad \delta_\nu' = -4\Phi'\;, \quad \delta_{\rm c}' = -3\Phi'\;, \quad \delta_{\rm b}' = -3\Phi'\;.
\end{eqnarray}
Since we want to describe the CMB, let us focus on the photon density contrast, which can be written as:
\begin{equation}
	\delta_\gamma(k,\eta) = 4\Theta_0(k,\eta)\;,
\end{equation}
introducing the monopole of the temperature fluctuation. The equation $\Theta_0' = -\Phi'$ can be immediately integrated, obtaining:
\begin{equation}
	\Theta_0(k,\eta) = -\Phi(k,\eta) + C_\gamma(k)\;.
\end{equation}
For the adiabatic primordial mode, the only one that we are going to consider in detail, we know from Eq.~\eqref{Rzetaprimordial2} that $C_\gamma(k) = \Phi_{\rm P}(k) - \Psi_{\rm P}(k)/2$ and thus:
\begin{equation}
	\Theta_0(k,\eta) = -\Phi(k,\eta) + \Phi_{\rm P}(k) - \frac{1}{2}\Psi_{\rm P}(k)\;.
\end{equation}
As we know from Chapter~\ref{Chap:Evopert}, we can consider the gravitational potentials to be equal in modulus, and on large scales $\Phi(k,\eta)$ is independent of time. Since recombination $\eta_* \gg \eta_{\rm eq}$ takes place well after radiation-matter equality, we know that $\Phi(k,\eta_*) = 9\Phi_{\rm P}(k)/10$. Therefore:
\begin{equation}\label{Theta0Phirelrec}
	\Theta_0(k,\eta_*) = \frac{3}{5}\Phi_{\rm P}(k) = \frac{2}{3}\Phi(k,\eta_*) = -\frac{2}{3}\Psi(k,\eta_*)\;.
\end{equation}
As we saw earlier in Eq.~\eqref{freestreamingformulaphotons}, the observed anisotropy is not $\Theta_0(k, \eta_*)$ but $\Theta_0(k, \eta_*) + \Psi(k, \eta_*)$ because of the gravitational redshift. Again, this is the Sachs-Wolfe effect, amounting to a shift in the photons frequency when they decouple from the baryon plasma depending on whether they are in a well or hill of the gravitational potential. So, we have from Eq.~\eqref{Theta0Phirelrec} that:
\begin{equation}
	(\Theta_0 + \Psi)(k,\eta_*) = \frac{1}{3}\Psi(k,\eta_*)\;.
\end{equation}
On the other hand, for $\delta_{\rm c}$ we know that
\begin{equation}
	\delta_{\rm c}(k, \eta) = -3\Phi(k, \eta) + \frac{9\Phi_{\rm P}(k)}{2}\;,
\end{equation}
again assuming adiabatic primordial modes. Using again $\Phi(k,\eta_*) = 9\Phi_{\rm P}(k)/10$, we get:
\begin{equation}
	\delta_{\rm c}(k, \eta_*) = 2\Phi(k,\eta_*) = -2\Psi(k,\eta_*)\;.
\end{equation}
The fluctuations in CDM contribute more to generating the potential wells than photons, by a factor of 2 against a factor $-2/3$. Combining the two equations:
\begin{equation}
	\boxed{(\Theta_0 + \Psi)(k,\eta_*) = -\frac{\delta_{\rm c}(k,\eta_*)}{6}}
\end{equation}
This result tells us that, on large scales, colder spots represent larger overdensities; a counter-intuitive result. One expects hotter photons the deeper the well is, and, in fact, this is the case with just $\Theta_0(k,\eta_*)$, since we have:
\begin{equation}
	\Theta_0(k, \eta_*) = -\frac{2}{3}\Psi(k, \eta_*) = \frac{\delta_{\rm c}(k, \eta_*)}{3}\;,
\end{equation}
i.e., the larger the CDM overdensity is, the larger the potential well and $\Theta_0(k, \eta_*)$ are. However, photons' response to the gravitational potential is only a factor $-2/3$, whereas the gravitational redshift adds a $\Psi$ contribution, thus changing the sign of the observed anisotropy. In the limit of $\delta_{\rm c} \to -1$, one gets $(\Theta_0 + \Psi)(k,\eta_*) \to 1/6$, so cosmic voids correspond to hot spots! 

The results found here are valid only on large scales, for $k\eta_* \ll 1$; these are scales much larger than the horizon at recombination, which has an angular size of approximately 1 degree. Moreover, they also depend on the choice of initial conditions. We have opted for the adiabatic ones, as usual.

\hrulefill

\begin{ex}
	Reproduce the above argument for the other primordial modes.
\end{ex}

\hrulefill

Let us use the theoretical prediction on the $C_{TT,\ell}^S$ given in Eq.~\eqref{ClfunctionThetal} together with the first contribution only from Eq.~\eqref{freestreamingformulaphotons}. The latter approximation is justified by the fact that we are considering large scales; hence, the dipole contribution is negligible, and the ISW effect is vanishing because the potentials are constant. Since:
\begin{equation}\label{Theta0Psireclargescales}
	(\Theta_0 + \Psi)(k,\eta_*) = \frac{1}{3}\Psi(k,\eta_*) = -\frac{1}{3}\Phi(k,\eta_*) = -\frac{3}{10}\Phi_{\rm P}(k) = -\frac{1}{5}\mathcal R(k)\;,
\end{equation} 
the transfer function is just the constant $-1/5$ (recall that we are neglecting the neutrino fraction $R_\nu$), and thus the angular power spectrum is:
\begin{equation}
	C^{S}_{TT,\ell}({\rm SW}) = \frac{4\pi}{25}\int_0^\infty \frac{dk}{k}\Delta^2_{\mathcal R}(k)j^2_\ell(k\eta_0)\;,
\end{equation}
since $\eta_* \ll \eta_0$. Note that $k\eta_* \ll k\eta_0$, and we have seen in Fig.~\ref{Fig:j102Plot} that the spherical Bessel function contributes the most about $k\eta_0 \approx \ell$. Thus, for small $\ell$, i.e., large angular scales, $k\eta_0$ is small, and $k\eta_*$ is very small, where in fact $|\left(\Theta_0 + \Psi\right)(k,\eta_*)|^2$ is constant. In other words, the above approximation is valid for small $\ell$, typically $\ell \lesssim 30$.

In the above integral, we can look at $j^2_\ell(k\eta_0)$ as a very peaked window function and approximate it as:
\begin{equation}
	C^{S}_{TT,\ell}({\rm SW}) \approx \frac{4\pi}{25}\Delta^2_{\mathcal R}(\ell/\eta_0)\int_0^\infty \frac{dk}{k}j^2_\ell(k\eta_0)\;.
\end{equation}
Using the result:
\begin{equation}
	\int_0^\infty \frac{dx}{x}j^2_\ell(x) = \frac{1}{2\ell(\ell + 1)}\;,
\end{equation}
we have then:
\begin{equation}
	\frac{\ell(\ell + 1)C^{S}_{TT,\ell}({\rm SW})}{2\pi} \approx \frac{1}{25}\Delta^2_{\mathcal R}(\ell/\eta_0)\;.
\end{equation}
Hence, for a scale-invariant spectrum $n_S = 1$, the combination $\ell(\ell + 1)C^{S}_{TT,\ell}({\rm SW})$ is constant, and it is called \textbf{the Sachs-Wolfe plateau}.\index{Sachs-Wolfe plateau} This also explains why CMB power spectra are usually normalized with $\ell(\ell + 1)$. 

If $n_S \neq 1$, then $\ell(\ell + 1)C^{S}_{TT,\ell}({\rm SW})$ is proportional to $\ell^{n_S - 1}$; the \textbf{primordial tilt} in the power spectrum leaves its mark on a tilted plateau for small $\ell$.\index{Primordial tilt}

\section{Tight-coupling and acoustic oscillations}

We have seen that to predict the $C_{TT,\ell}$, we need to know what happened at recombination. We devote this section to that purpose, explaining why the monopole and the dipole contribute the most.

Let us recover here the hierarchy of Boltzmann equations for the $\Theta_\ell$'s (not taking into account polarization) that we derived in Chapter~\ref{Chap:PertubedBoltzmannEquations}: 
\begin{eqnarray}
	(2\ell + 1)\Theta_\ell' + k[(\ell + 1)\Theta_{\ell + 1} - \ell\Theta_{\ell - 1}] = (2\ell + 1)\tau'\Theta_\ell\;, \qquad (\ell > 2)\;,\\
	10\Theta_2' + 2k\left(3\Theta_3 - 2\Theta_1\right) = 10\tau'\Theta_2 - \tau'\Pi\;,\\
3\Theta_1' + k(2\Theta_2 - \Theta_0) = k\Psi + \tau'\left(3\Theta_1 - V_{\rm b}\right)\;,\\
\Theta_0' + k\Theta_1 = -\Phi'\;,
\end{eqnarray}
where we recall that $\delta_\gamma = 4\Theta_0$ and $3\Theta_1 = V_\gamma$. The best way to deal with these equations is to solve them numerically by using Boltzmann codes such as CAMB or CLASS; however, in this way, the physics behind $C_{TT,\ell}$ remains hidden or unclear. For this reason, we tackle these equations in an approximate fashion, but analytically.

We take the limit $-\tau' \gg \mathcal H$, which is called \textbf{tight-coupling} (TC) approximation.\index{Tight coupling} This limit physically means that the Thomson scattering rate between photons and electrons is much larger than the Hubble rate until recombination and then drops abruptly since the free electron fraction $X_e$ goes to zero very rapidly, as we have seen when studying thermal history in Chapter~\ref{Chap:ThermalHistory}. We shall first consider the case of \textbf{sudden recombination}, where all the photons last scatter at the same time. It is a fair approximation, though unrealistic.\index{Sudden recombination}

\hrulefill

\begin{ex}
From the definition of the optical depth:
	\begin{equation}
\tau \equiv \int_\eta^{\eta_0}d\eta'\;n_e\sigma_{\rm T}a\;,
\end{equation}
show that $\tau \propto 1/\eta^3$ when matter dominates and $\tau \propto 1/\eta$ when radiation dominates.
\end{ex}

\hrulefill

We can be more quantitative and write:
\begin{equation}
	-\tau' = n_e\sigma_{\rm T}a = n_{\rm b}\sigma_{\rm T}a = \frac{\rho_{\rm b}}{m_{\rm b}}\sigma_{\rm T}a\;,
\end{equation}
where we have used the definition of $\tau$ and assumed it to be in an epoch before recombination, so that we can approximate $n_e$ with $n_{\rm b}$, since all the electrons are free.

\hrulefill

\begin{ex} Introducing the baryon density parameter and using $m_{\rm b} = $ 1 GeV, the mass of the proton, show that: 
\begin{equation}\label{dottaueq}
 -\tau' \approx 1.46\times 10^{-19}\frac{\Omega_{\rm b0}h^2}{a^2} \mbox{ s}^{-1}\;.
\end{equation}
\end{ex}

\hrulefill

Now we need to compare this scattering rate with the Hubble rate in order to check the goodness of the TC approximation. Assuming matter-domination and using the conformal time Friedmann equation (this is because $\tau'$ is derived with respect to the conformal time), we have:
\begin{equation}
	\mathcal H = H_0\sqrt{\Omega_{\rm m0}}a^{-1/2} \approx 3.33\times 10^{-18} h\sqrt{\Omega_{\rm m0}}\;a^{-1/2}\mbox{ s}^{-1}\;.
\end{equation}
Therefore, the ratio:
\begin{equation}
	\frac{-\tau'}{\mathcal H} = 0.044\;\frac{\Omega_{\rm b0}h^2}{\sqrt{\Omega_{\rm m0}h^2}}a^{-3/2}\;,
\end{equation}
diverges for $a \to 0$ as expected (though the formula should be generalized to the case of radiation-domination), so if it is sufficiently big at recombination, then the TC approximation would be reliable. Substituting the \textit{Planck} values $\Omega_{\rm b0}h^2 = 0.022$ and $\Omega_{\rm m0}h^2 = 0.12$, one gets at recombination, i.e., for $a = 10^{-3}$:
\begin{equation}
	\frac{-\tau'}{\mathcal H} \approx 10^2\;.
\end{equation}
This means that the scattering rate is much larger than the Hubble rate even at recombination, as long as there are free electrons around. Thus, we are going to use the tight-coupling approximation with reliability.

Let us see in detail how the TC limit works. Let us compare in the hierarchy for $\ell \ge 2$ the terms $\Theta_\ell'$, $k\Theta_\ell$, and $\tau'\Theta_\ell$, which all have the same dimensions of inverse time. There are two physical time scales in our problem; one is given by the expansion rate, and the other by the scattering rate. Hence:
\begin{equation}
	\Theta_\ell' \propto \mathcal H\Theta_\ell, \tau'\Theta_\ell\;,
\end{equation}  
from a dimensional analysis. However, the mode for which $\Theta_\ell' \propto \tau'\Theta_\ell$ implies that $\Theta_\ell \propto \exp\tau$ diverges at early times, which is unacceptable for a small fluctuation. We then dismiss this mode as nonphysical and take into account just that for which $\Theta_\ell' \propto \mathcal H\Theta_\ell$, which is small compared to $\tau'\Theta_\ell$.

Now, let us inspect the ratio
\begin{equation}
	\frac{-\tau'}{k}\;.
\end{equation}
This is the number of collisions that occur on a scale $1/k$. Hence, this number is very large, provided that we consider sufficiently large scales, i.e., small $k$. If the scale is too small, i.e., large $k$, then the TC approximation does not work well, and we must take into account the multipole moments for $\ell \ge 2$. We will see this when investigating the \textbf{diffusion damping} or \textbf{Silk damping} effect.\index{Diffusion damping}\index{Silk damping}

From the above analysis, for sufficiently large scales, we can conclude that $\Theta_\ell \approx 0$ for $\ell \ge 2$. Sufficiently large means much larger than the mean free path $-1/\tau'$, which is approximately of the order of 10 Mpc at recombination. This number can be computed from Eq.~\eqref{dottaueq} and is a comoving scale; the physical one is divided by a factor of a thousand, so it is 10 kpc.

Eliminating all the multipoles $\ell \ge 2$, the relevant equations are the following two:
\begin{eqnarray}
\label{Theta0eqTC}	\Theta_0' + k\Theta_1 = -\Phi'\;,\\
\label{Theta1eqTC}	3\Theta_1' - k\Theta_0 = k\Psi + \tau'\left(3\Theta_1 - V_{\rm b}\right)\;.
\end{eqnarray}
The TC approximation allows us to treat photons as a fluid until recombination. Note the coupling to baryons via the baryon velocity $V_{\rm b}$. Thus, we also need the equations for baryons:
\begin{eqnarray}
\label{deltabeqTC}	\delta_{\rm b}' + kV_{\rm b} = -3\Phi'\;,\\
\label{VbeqTC}	V_{\rm b}' + \mathcal H V_{\rm b} = k\Psi + \frac{\tau'}{R}(V_{\rm b} - 3\Theta_1)\;,
\end{eqnarray}
where we have introduced $R \equiv 3\rho_{\rm b}/4\rho_\gamma$: the baryon density to photon density ratio. This number can be expressed as:
\begin{equation}
	R = \frac{3\Omega_{\rm b0}}{4\Omega_{\gamma 0}}a \approx 600a\;, 
\end{equation}
using the known values for the density parameters. Therefore, $R$ grows from zero at early times to $R_* \approx 0.6$ at recombination. Let us rewrite the velocity equation for baryons in the following way:
\begin{equation}\label{vbsuccapproxeq}
	V_{\rm b} = 3\Theta_1 + \frac{R}{\tau'}\left(V_{\rm b}' + \mathcal H V_{\rm b} - k\Psi\right)\;.
\end{equation}
We can solve this equation via successive approximation, exploiting the fact that $R < 1$ before recombination. That is, assume the expansion:
\begin{equation}
	V_{\rm b} = V_{\rm b}^{(0)} + RV_{\rm b}^{(1)} + R^2V_{\rm b}^{(2)} + \dots\;.
\end{equation}
The solution for $R = 0$ simply gives $V_{\rm b}^{(0)} = 3\Theta_1$, which we have used in Chapter~\ref{Chap:IC} in order to investigate the primordial modes. This solution is reliable well before recombination, say at $a = 10^{-7}$, for example, because $R \approx 6\times 10^{-5}$ there; however, it is not satisfactory at recombination. We need to take into account at least the first order in $R$ in the above expansion.

\subsection{The acoustic peaks for \texorpdfstring{$R = 0$}{R = 0}}

Let us start with the simple case of $R = 0$, which amounts to neglecting baryons.

\hrulefill

\begin{ex} Combine the photon Eqs.~\eqref{Theta0eqTC}-\eqref{Theta1eqTC} with the zeroth-order TC condition $V_{\rm b} = 3\Theta_1$ and find the following second-order equation for $\Theta_0$: 
\begin{equation}
 \Theta_0'' + \frac{k^2}{3}\Theta_0 = -\frac{k^2\Psi}{3} - \Phi''\;.
\end{equation}
\end{ex}

\hrulefill

We already have here the first fundamental piece of physics of the CMB. This is the equation of motion of a driven harmonic oscillator where, instead of the position, we have the monopole of the temperature fluctuation, and the driving force is given by the gravitational potential. This equation describes the \textbf{acoustic oscillations}\index{Acoustic oscillations} of the baryon-photon fluid until recombination. After recombination, we expect to observe these fluctuations in the $C_{TT,\ell}$'s, using the free-streaming formula \eqref{freestreamingformulaphotons}, and in fact, we do. 

Note that these oscillations are in the baryon-photon fluid and therefore also affect baryons. We therefore expect to see oscillations in the baryon distribution after recombination, called \textbf{baryon acoustic oscillations} (BAO),\index{Baryon Acoustic Oscillations} and detected in 2005 \cite{Eisenstein:2005su}. The BAO are the manifestation of a special length, the sound horizon at recombination, in the correlation function of galaxies, which appears as a bump, i.e., an excess probability. In the Fourier space, i.e., for the power spectrum, a given scale is represented with various oscillations. We have already encountered BAO in Chapter~\ref{Chap:Evopert}. BAO and weak gravitational lensing are among the main observables on which current and future observational experiments are based.

\hrulefill

\begin{ex} Combine Eqs.~\eqref{deltabeqTC} and \eqref{VbeqTC} and the TC condition $V_{\rm b} = 3\Theta_1$ and find the following equation for $\delta_{\rm b}$: 
\begin{equation}
 \delta_{\rm b}' = 3\Theta_0'\;.
\end{equation}
Hence, the same oscillatory solution of $\Theta_0$ holds true for $\delta_{\rm b}$.
\end{ex}

\hrulefill

Now, consider the fact that close to recombination, CDM is already dominating the energy density content of the universe; thus, the potentials are equal and constant at all scales. We get:
\begin{eqnarray}
	\Theta_0'' + \frac{k^2}{3}\Theta_0 = -\frac{k^2\Psi}{3}\;.
\end{eqnarray}
This equation can be put in the following form:
\begin{equation}\label{Theta0R0eq}
	(\Theta_0 + \Psi)'' + \frac{k^2}{3}(\Theta_0 + \Psi) = 0\;,
\end{equation}
where we have used the constancy of $\Psi$. Note how the observed temperature fluctuation, used in Eq.~\eqref{freestreamingformulaphotons}, has appeared. The solution is:
\begin{equation}
	(\Theta_0 + \Psi)(\eta,k) = A(k)\sin\left(\frac{k\eta}{\sqrt{3}}\right) + B(k)\cos\left(\frac{k\eta}{\sqrt{3}}\right)\;, 
\end{equation}
with the driving potential providing just an offset for the oscillations. At recombination, we have
\begin{equation}\label{Theta0solutionR0}
	(\Theta_0 + \Psi)(\eta_*,k) = A(k)\sin\left(\frac{k\eta_*}{\sqrt{3}}\right) + B(k)\cos\left(\frac{k\eta_*}{\sqrt{3}}\right)\;, 
\end{equation}
with peaks and troughs in the temperature fluctuations given by this combination of sine and cosine, therefore dependent on the functions $A(k)$ and $B(k)$. Inserting these formulas into Eq.~\eqref{freestreamingformulaphotons} in order to compute the $\Theta_\ell(\eta_0, k)$ (the anisotropies today) and then into Eq.~\eqref{ClfunctionThetal} in order to compute the $C_{TT,\ell}$'s, we are able to explain the \textbf{acoustic oscillations} feature of the CMB TT spectrum.

The functions $A(k)$ and $B(k)$ are determined by the initial condition, i.e., for $k\eta_* \ll 1$:
\begin{equation}
	(\Theta_0 + \Psi)(k\eta_* \ll 1) \sim A(k)\frac{k\eta_*}{\sqrt{3}} + B(k)\;. 
\end{equation}
Hence, if we choose adiabatic modes, we must put $A(k) = 0$. So, considering different initial conditions changes the position of the acoustic peaks, and observation allows us to test the choice made. As we saw in Chapter~\ref{Chap:IC}, Planck limits the presence of isocurvature modes to a few percent. With $A(k) = 0$, i.e., for adiabatic perturbations, using the large-scale solution that we found in Eq.~\eqref{Theta0Psireclargescales}, we have:
\begin{equation}\label{Theta0solutionR0adiabatic}
	(\Theta_0 + \Psi)(\eta_*,k) = -\frac{1}{5}\mathcal R(k)T(k)\cos\left(\frac{k\eta_*}{\sqrt{3}}\right)\;, 
\end{equation} 
where $T(k)$ is the transfer function of $\Theta_0 + \Psi$. We did not calculate it in Chapter~\ref{Chap:Evopert}, but it can be shown that it is limited to a range $0.4$-$2$, approximately. See \cite{Mukhanov:2005sc}.

The extrema of the effective temperature fluctuations are thus given by:
\begin{equation}
	\frac{k\eta_*}{\sqrt{3}} = n\pi\;, \qquad (n = 1, 2, \dots)\;,
\end{equation}
where the odd values provide peaks, corresponding to the highest temperature fluctuations and thus to scales at which photons are maximally compressed and hot, whereas the even values provide throats, corresponding to the lowest temperature fluctuations and thus to scales at which photons are maximally rarefied and cold. In the spectrum, only peaks appear because of the quadratic nature of the $C_{TT,\ell}$'s as functions of the $\Theta_\ell$'s, but it should be clear that the first and the third peaks are compressional.

From Eqs.~\eqref{Theta0eqTC} and \eqref{Theta0solutionR0adiabatic}, we can easily determine the dipole contribution:
\begin{equation}
	\Theta_1(\eta_*, k) = -\frac{\Theta_0'(\eta_*, k)}{k} = -\frac{1}{5\sqrt{3}}\mathcal R(k)T(k)\sin\left(\frac{k\eta_*}{\sqrt{3}}\right)\;,
\end{equation}
where we are still continuing to keep the potentials constant. Substituting this equation and Eq.~\eqref{Theta0solutionR0adiabatic} into Eq.~\eqref{freestreamingformulaphotons} and then into Eq.~\eqref{ClfunctionThetal} in order to compute the angular power spectrum, we get:
\begin{equation}\label{Clnobaryonseq1}
	C_{TT,\ell} = \frac{4\pi}{25}\int_0^\infty \frac{dk}{k}\Delta^2_{\mathcal R}(k)\left[\cos\left(\frac{k\eta_*}{\sqrt{3}}\right)j_\ell(k\eta_0) + \sqrt{3}\sin\left(\frac{k\eta_*}{\sqrt{3}}\right)\frac{dj_\ell(k\eta_0)}{d(k\eta_0)}\right]^2\;,
\end{equation}
where the derivative of the spherical Bessel function is given in Eq.~\eqref{derivativejl}. We have put $T(k) = 1$ here for simplicity. 

We can manipulate this integral analytically, following the technique used in \cite{Mukhanov:2003xr} and \cite{Mukhanov:2005sc}. In these references, baryon loading and diffusion damping are taken into account, but here we just tackle a simpler case.

\begin{figure}[htbp]
\center
	\includegraphics[width=\columnwidth]{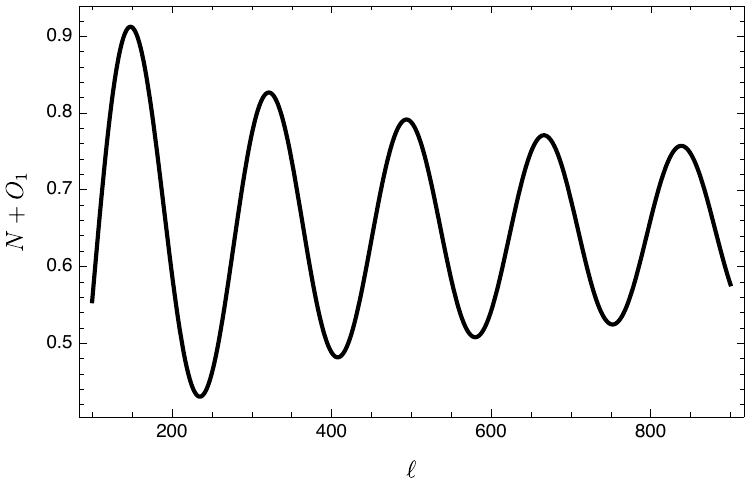}
\caption{Sum of the $N$ and $O_1$ contributions.}
\label{Fig:ApproxFormNobarPlot}
\end{figure}

The idea is to avoid the oscillatory nature of the Bessel function and the trigonometric ones (which are also problematic from a numerical perspective) by approximating $j_\ell(x)$ as follows, for large $\ell$:
\begin{equation}
j_\ell(x) \approx 
	\begin{cases}
		0\;, \qquad (x < \ell)\;,\\
		\frac{1}{\sqrt{x}(x^2 - \ell^2)^{1/4}}\cos\left[\sqrt{x^2 - \ell^2} - \ell\arccos(\ell/x) - \pi/4\right]\;, \qquad (x > \ell)\;.
	\end{cases}
\end{equation}
This approximation is identical for both $j_\ell(x)$ and $j_{\ell - 1}(x)$, since we are assuming $\ell$ to be large. Hence, when we deal with the derivative of the spherical Bessel function in Eq.~\eqref{Clnobaryonseq1}, we can factor a $j^2_\ell(x)$ and approximate the squared cosine from the above approximation with its average, i.e., a factor $1/2$. We thus have the following integration:
\begin{equation}\label{Clnobaryonseq2}
	C_{TT,\ell} = \frac{2\pi\Delta^2_{\mathcal R}}{25}\int_{\ell/\eta_0}^\infty \frac{dk}{k^2\eta_0\sqrt{(k\eta_0)^2 - \ell^2}}\left[\cos\left(\frac{k\eta_*}{\sqrt{3}}\right) + \sqrt{3}\left(1 - \frac{\ell}{k\eta_0}\right)\sin\left(\frac{k\eta_*}{\sqrt{3}}\right)\right]^2\;,
\end{equation}
where we have already assumed a scale-invariant spectrum for simplicity. Using now the variable
\begin{equation}
	x \equiv \frac{k\eta_0}{\ell}\;,
\end{equation}
we can write:
\begin{equation}\label{Clnobaryonseq3}
	\ell^2C_{TT,\ell} = \frac{2\pi\Delta^2_{\mathcal R}}{25}\int_{1}^\infty \frac{dx}{x^2\sqrt{x^2 - 1}}\left[\cos\left(\ell\varrho x\right) + \sqrt{3}\frac{x - 1}{x}\sin\left(\ell\varrho x\right)\right]^2\;,
\end{equation}
where note the appearance of the factor $\ell^2$ on the left hand side and we have defined the quantity:
\begin{equation}
	\varrho \equiv \frac{\eta_*}{\sqrt{3}\eta_0}\;.
\end{equation}
Now, developing the square and using the trigonometric formulae:
\begin{equation}
	\cos^2\alpha = \frac{1 + \cos 2\alpha}{2}\;, \qquad \sin^2\alpha = \frac{1 - \cos 2\alpha}{2}\;, \qquad 2\sin\alpha\cos\alpha = \sin2\alpha\;,
\end{equation}
we can write:
\begin{eqnarray}\label{Clnobaryonseq4}
	\ell^2C_{TT,\ell} = \frac{2\pi\Delta^2_{\mathcal R}(k)}{25}\int_{1}^\infty \frac{dx}{x^2\sqrt{x^2 - 1}}\nonumber\\
	\left[\frac{x^2 + 3(x - 1)^2}{2x^2} + \frac{x^2 - 3(x - 1)^2}{2x^2}\cos\left(2\ell\varrho x\right) + \frac{\sqrt{3}(x - 1)}{x}\sin\left(2\ell\varrho x\right)\right]\;.
\end{eqnarray}
Now, let us treat the three integrands separately. The first non-oscillatory one is the simplest one:
\begin{equation}
	N \equiv \int_{1}^\infty \frac{dx}{x^2\sqrt{x^2 - 1}}\frac{x^2 + 3(x - 1)^2}{2x^2} = 3\left(1 - \frac{\pi}{4}\right)\;,
\end{equation}
but also the less interesting. The oscillatory ones can be dealt with as follows. Define:
\begin{equation}
	O_1 \equiv \int_{1}^\infty \frac{dx}{\sqrt{x - 1}}\frac{x^2 - 3(x - 1)^2}{2x^4\sqrt{x + 1}}\cos\left(2\ell\varrho x\right)\;.
\end{equation}
Then, solving the problem in \cite[page 383]{Mukhanov:2005sc}, we can use the formula:
\begin{equation}
	\int_1^\infty\frac{dx}{\sqrt{x - 1}}f(x)\cos(bx) \approx f(1)\sqrt{\frac{\pi}{b}}\cos(b +\pi/4)\;,
\end{equation}
for large values of $b$ and a slowly varying $f(x)$. A similar result also holds true for the sine function. Using this formula, we then have:
\begin{equation}
	O_1 = \frac{1}{2\sqrt{2}}\sqrt{\frac{\pi}{2\ell\varrho}}\cos(2\ell\varrho +\pi/4)\;,
\end{equation}
whereas for the integral containing the sine:
\begin{equation}
	O_2 \equiv \int_{1}^\infty \frac{dx}{\sqrt{x - 1}}\frac{\sqrt{3}(x - 1)}{x^3\sqrt{x + 1}}\sin\left(2\ell\varrho x\right) \approx 0\;,
\end{equation}
since $f(1) = 0$ here. The contribution $O_2$ comes from the cross product between the monopole and dipole terms, and it is usually neglected in the calculations. We have explicitly shown why here. Gathering the $N$ and $O_1$ contributions, we plot the sum $N + O_1$ in Fig.~\ref{Fig:ApproxFormNobarPlot}. 

In order to make this plot, we have used $a \propto \eta^2$, since we are in the matter-dominated epoch; thus, we have evaluated $\varrho$ as follows:
\begin{equation}
	\varrho = \frac{\eta_*}{\sqrt{3}\eta_0} = \frac{1}{\sqrt{3(1 + z_*)}} = \frac{1}{\sqrt{3000}} \approx 0.0183\;.
\end{equation}
The agreement between the spectrum in Fig.~\ref{Fig:ApproxFormNobarPlot} and the one observed is poor, but at least we have understood how the acoustic oscillations free-stream until today and are seen in the CMB TT power spectrum. There are several features missing in Fig.~\ref{Fig:ApproxFormNobarPlot}: there are too many peaks, their relative height diminishes too slowly, and the overall trend does not decay. The reason is that we have neglected baryons and diffusion damping, which we are going to tackle in the next sections.

\subsection{Baryon loading}

The oscillations in Eq.~\eqref{Theta0R0eq} take place with frequency $k/\sqrt{3}$, i.e., as if the speed of sound were $1/\sqrt{3}$, i.e., the speed of sound of a pure photon fluid. We have been too radical in assuming $V_{\rm b} = 3\Theta_1$ in the equation for baryons. In fact, we saw that this assumption is equivalent to saying that $R = 0$, i.e., the baryon density is negligible with respect to the photon density. That is why photons do not feel baryons at all, and baryon fluctuations oscillate in the same way as photons do. 

We now take into account $R$ up to first-order. If we consider $V^{(0)}_{\rm b} = 3\Theta_1$ substituted in Eq.~\eqref{vbsuccapproxeq}, we get up to order $R$:
\begin{equation}\label{VbfirstorderR}
	V_{\rm b} = 3\Theta_1 + \frac{R}{\tau'}\left(3\Theta_1' + 3\mathcal H\Theta_1 - k\Psi\right)\;.
\end{equation}

\hrulefill

\begin{ex} Combine the above equation and Eqs.~\eqref{Theta0eqTC}-\eqref{Theta1eqTC} in order to find the following second-order equation for $\Theta_0$: 
\begin{equation}\label{Theta0eqwithR}
 \Theta_0'' + \mathcal H\frac{R}{1 + R}\Theta_0' + \frac{k^2}{3(1 + R)}\Theta_0 = -\frac{k^2\Psi}{3} - \Phi'' - \mathcal H\frac{R}{1 + R}\Phi'\;.
\end{equation}
\end{ex}

\hrulefill

Now the speed of sound, i.e., the quantity multiplying $k^2$, has been reduced:
\begin{equation}\label{speedofsoundbaryonphotonplasma}
	c_s^2 = \frac{1}{3(1 + R)}\;.
\end{equation}
The extrema of the temperature fluctuations at recombination are now expected to be slightly changed, since:
\begin{equation}
	\frac{k\eta_*}{\sqrt{3(1 + R)}} = n\pi\;, \qquad (n = 1, 2, \dots)\;,
\end{equation}
is now the condition defining them. Moreover, baryons are also responsible for the damping term $\mathcal HR\Theta_0'/(1 + R)$; hence, we also expect the extrema to have increasingly smaller amplitudes. These features translate, once free-streamed, into a relative suppression of the second peak with respect to the first one.

This effect is due to the \textbf{baryon loading}\index{Baryon loading} and it is also called \textbf{baryon drag}. Physically, baryons are heavy and prevent the oscillations in $\Theta_0$ from being symmetric, favoring compression over rarefaction. Since $R \propto a$, we have that:
\begin{equation}
	R' = \mathcal H R\;.
\end{equation}
Let us write Eq.~\eqref{Theta0eqwithR} in the following form:
\begin{equation}\label{EqTheta0Phi}
 \left(\frac{d^2}{d\eta^2} + \frac{R'}{1 + R}\frac{d}{d\eta} + k^2 c_s^2\right)(\Theta_0 + \Phi) = \frac{k^2}{3}\left(\frac{\Phi}{1 + R} - \Psi\right)\;.
\end{equation}
The above equation cannot be solved analytically, but we can use a semi-analytic approximation provided by \cite{Hu:1995en}. Let us employ the WKB method and use the following ansatz:
\begin{equation}
	(\Theta_0 + \Phi)(\eta, k) = A(\eta)e^{iB(\eta, k)}\;,
\end{equation}
where $A(\eta)$ and $B(\eta, k)$ are functions to be determined via Eq.~\eqref{EqTheta0Phi}. 

\hrulefill

\begin{ex} Substitute this ansatz into the homogeneous part of Eq.~\eqref{EqTheta0Phi} and find the following couple of equations, by separately equating the real and imaginary parts to zero:
\begin{eqnarray}
	-A(B')^2 + A'' + \frac{R'}{1 + R}A' + k^2c_s^2A = 0\;,\\
	2B'A' + AB'' + \frac{R'}{1 + R}AB' = 0\;.
\end{eqnarray}
\end{ex}

\hrulefill

In the first equation, let us neglect the second and third terms with respect to the first one. That is, the oscillations provide almost at any time (except at the extrema) a much larger derivative than that of the amplitude or $R$. Then, the first equation is readily solved as:
\begin{equation}
	\boxed{B(\eta, k) = k\int_0^\eta c_s(\eta')d\eta' \equiv kr_s(\eta)}
\end{equation}
where in the last step we have defined the \textbf{sound horizon}, i.e., the conformal distance traveled by a sound wave propagating in the baryon photon fluid. When evaluated at recombination, $r_s(\eta_*) = 150$ Mpc, and this scale is fundamental for BAO, making them \textbf{standard rulers}.\index{Standard rulers}

\hrulefill

\begin{ex} Determine now $A(\eta)$. Show that the above equations, together with the found solution for $B(\eta, k)$, can be cast as:
\begin{equation}
	\frac{A'}{A} = -\frac{1}{4}\frac{R'}{1 + R}\;,
\end{equation}
which gives:
\begin{equation}
	A(\eta) = (1 + R)^{-1/4}\;.
\end{equation}
\end{ex}

\hrulefill

The general, approximate solution of the homogeneous equation is:
\begin{equation}\label{HuSugiyamasolutionhom}
	\boxed{(\Theta_0 + \Phi)(\eta, k) = \frac{1}{(1 + R)^{1/4}}\left[C(k)\sin(kr_s) + D(k)\cos(kr_s)\right]}
\end{equation}
The condition $|A'|, R' \ll |B'|$, which was employed to find the above solution, can be checked as follows:
\begin{equation}
	\frac{R'}{4(1 + R)^{5/4}}, R' \ll \frac{k}{\sqrt{3(1 + R)}}\;,
\end{equation}
which essentially amounts to saying that:
\begin{equation}
	k \gg R'\;,
\end{equation}
i.e., the solution found is good on sufficiently small scales. Since $R' = \mathcal HR \sim R/\eta$, we must have that $k\eta \gg R$. Since $R$ is at most $R_* \approx 0.6$ at recombination, this condition means any scale at early times, but sub-horizon scales at recombination. 

Equation \eqref{HuSugiyamasolutionhom} gives us the general solution of the homogeneous part of Eq.~\eqref{EqTheta0Phi}. In order to find the general solution of the full equation, we need to find a particular solution of Eq.~\eqref{EqTheta0Phi}. This can be obtained via the Green's functions method. Let us define, in order to keep a more compact notation, the independent solutions of the homogeneous equation that we have just found in Eq.~\eqref{HuSugiyamasolutionhom} as follows:
\begin{equation}
	S_1(\eta, k) \equiv \frac{1}{(1 + R)^{1/4}}\sin(kr_s)\;, \qquad S_2(\eta, k) \equiv \frac{1}{(1 + R)^{1/4}}\cos(kr_s)\;.
\end{equation}
Taking into account the non-homogeneous term, the general solution of Eq.~\eqref{EqTheta0Phi} is:
\begin{equation}
	(\Theta_0 + \Phi)(\eta, k) = C(k)S_1 + D(k)S_2 + \frac{k^2}{3}\int_0^\eta d\eta'\;\left[\frac{\Phi(\eta')}{1 + R} - \Psi(\eta')\right]G(\eta, \eta')\;,
\end{equation}
where $G(\eta, \eta')$ is the Green's function.

\hrulefill

\begin{ex} Determine the Green's function:
\begin{equation}
	G(\eta, \eta') = \frac{S_1(\eta')S_2(\eta) - S_1(\eta)S_2(\eta')}{W(\eta')}\;,
\end{equation}
using the homogeneous solution. Show that:
\begin{equation}
	G(\eta, \eta') = \frac{1}{\sqrt{1 + R)}}\frac{\sin[kr_s(\eta') - kr_s(\eta)]}{W(\eta')}\;,
\end{equation}
and
\begin{equation}
	W(\eta') = -\frac{1}{\sqrt{3}(1 + R)}\;.
\end{equation}
We are omitting the $k$-dependence for simplicity.
\end{ex}

\hrulefill

With the above results, we can write:
\begin{eqnarray}
	(\Theta_0 + \Phi)(\eta, k) = C(k)S_1 + D(k)S_2\nonumber\\ + \frac{k}{\sqrt{3}}\int_0^\eta d\eta'\;\left[\frac{\Phi(\eta')}{1 + R(\eta')} - \Psi(\eta')\right]\sqrt{1 + R(\eta')}\sin[kr_s(\eta) - kr_s(\eta')]\;.
\end{eqnarray}
For the primordial modes, in the limit $k\eta \to 0$, one obtains the dominant order:
\begin{equation}
	(\Theta_0 + \Phi)(0, k) = D(k)\;.
\end{equation}
Hence, it is the adiabatic mode that multiplies the cosine. As we have already commented, since sine and cosine have a $\pi/2$ phase difference, the effect of different initial conditions is to change the scales for which the effective temperature fluctuations are maximum or minimum, and hence the positions of the peaks in the $C_{TT,\ell}$'s.

In the adiabatic case, we have:
\begin{eqnarray}\label{HuSugiyamasolution}
	\boxed{(\Theta_0 + \Phi)(\eta, k) = (\Theta_0 + \Phi)(0, k)\frac{\cos[kr_s(\eta)]}{(1 + R)^{1/4}}}\nonumber\\ \boxed{+ \frac{k}{\sqrt{3}}\int_0^\eta d\eta'\;\left[\frac{\Phi(\eta')}{1 + R} - \Psi(\eta')\right]\sqrt{1 + R}\sin[kr_s(\eta) - kr_s(\eta')]}
\end{eqnarray}
This is the semi-analytic (semi because the integral has to be performed numerically) formula of \cite{Hu:1995en}.

The above solution \eqref{HuSugiyamasolution} can also be used for baryons. Indeed, combining Eq.~\eqref{deltabeqTC} with Eq.~\eqref{VbfirstorderR} and then with Eq.~\eqref{Theta0eqTC}, we get:
\begin{equation}
	\delta_{\rm b}' = 3\Theta_0' + \frac{3R}{\tau'}\left[\Theta_0'' + \Phi'' + \mathcal H(\Theta_0' + \Phi') + \frac{k^2\Psi}{3}\right]\;.
\end{equation}
Eliminating the second derivative by means of the differential equation \eqref{EqTheta0Phi}, we have:
\begin{equation}
	\delta_{\rm b}' = 3\Theta_0' + \frac{R}{\tau'(1 + R)}\left[-k^2\Theta_0 + 3\mathcal H(\Theta_0' + \Phi')\right]\;.
\end{equation}
Just to make a rough estimate, let us neglect the second contribution (which is divided by $\tau'$ anyway, which is much larger than $\mathcal H$ and also than $k$, for suitable scales) and use the homogeneous part of Eq.~\eqref{HuSugiyamasolution}. It is straightforward then to integrate $\delta_{\rm b}'$ and obtain at recombination:
\begin{equation}
	\delta_{\rm b}(\eta_*, k) \propto \cos[kr_s(\eta_*)] = \cos\left[2\pi \frac{r_s(\eta_*)}{\lambda}\right]\;.
\end{equation}
So the scale $r_s(\eta_*) \approx 150$ Mpc is relevant for baryons, too. Indeed, at about this scale, the matter power spectrum displays the BAO feature, as we saw in Chapter~\ref{Chap:Evopert}.

\section{Diffusion damping}\index{Diffusion damping}

In order to understand what happens to the $C_{TT,\ell}$ when $\ell$ grows larger and larger, we need to take into account smaller and smaller scales because of the relation $\ell \approx k\eta_0$. As discussed earlier, for larger and larger $k$, the ratio $-\tau'/k$ becomes smaller and smaller, and so the TC approximation must be relaxed. 

In this section, we investigate what happens to the temperature fluctuations when the quadrupole moment $\Theta_2$ is taken into account. Since this analysis accounts for the behavior of very small scales which entered the horizon deep into the radiation-dominated epoch, we can neglect the gravitational potentials since these, as we saw in Chapter~\ref{Chap:Evopert}, rapidly decay.

Moreover, being deep into the radiation-dominated epoch, we can also neglect $R$ and thus take $3\Theta_1 = V_{\rm b}$. Neglecting also polarization, we have the following set of three equations for $\Theta_0$, $\Theta_1$, and $\Theta_2$:
\begin{eqnarray}
	\Theta_0' + k\Theta_1 = 0\;,\\
	3\Theta_1' + 2k\Theta_2 - k\Theta_0 = 0\;,\\
	10\Theta_2' - 4k\Theta_1 = 9\tau'\Theta_2\;.
\end{eqnarray}
In the last equation we can neglect $\Theta'_2$ with respect $\tau'\Theta_2$, as we already did earlier, and then find:
\begin{equation}
	\Theta_2 = -\frac{4k}{9\tau'}\Theta_1\;.
\end{equation}
The minus sign might raise some alarm, but recall that $\tau'$ is negative by definition.

\hrulefill

\begin{ex} Combine the above condition with the remaining equations in order to find a closed equation for $\Theta_0$:
\begin{equation}
	\boxed{\Theta_0'' + \left(-\frac{8k^2}{27\tau'}\right)\Theta_0' + \frac{k^2}{3}\Theta_0 = 0}
\end{equation}
\end{ex}

\hrulefill

This is the equation for a harmonic oscillator that we have already found earlier in Eq.~\eqref{Theta0R0eq}, only that now there appears a damping term which is relevant on small scales, i.e., when $k \sim -\tau'$. Baryons also provide a damping term, cf. Eq.~\eqref{Theta0eqwithR}, but they are irrelevant in the present case since we set $R = 0$.

This damping term here depends on $\Theta_2$ and is time-dependent. Let us consider it constant and assume a solution of the type $\Theta_0 \propto \exp(i\omega\eta)$. Substituting this ansatz into the equation, we find:
\begin{equation}
	-\omega^2 + \left(-\frac{8k^2}{27\tau'}\right)i\omega + \frac{k^2}{3} = 0\;.
\end{equation}
The frequency must have an imaginary part, which accounts for the damping; thus, let us stipulate:
\begin{equation}
	\omega = \omega_R + i\omega_I\;,
\end{equation}

\hrulefill

\begin{ex} Substitute this ansatz in the equation and find:
\begin{equation}
	\omega_R = \frac{k}{\sqrt{3}}\;, \qquad \omega_I = -\frac{4k^2}{27\tau'}\;.
\end{equation}
\end{ex}

\hrulefill

Hence, we can write the general solution for $\Theta_0$ as:
\begin{equation}
	\Theta_0 \propto e^{ik\eta/\sqrt{3}}e^{-k^2/k_{\rm Silk}^2}\;,
\end{equation}
where we have introduced the comoving diffusion length, the \textbf{Silk length},\index{Silk length} as
\begin{equation}
	\lambda_{\rm Silk}^2 = \frac{1}{k_{\rm Silk}^2} \equiv -\frac{4\eta}{27\tau'}\;.
\end{equation}
What does the diffusion length physically represent? It is the comoving distance traveled by a photon in a time $\eta$, while taking into account the collisions it suffers, i.e., its diffusion. Let us see this in some more detail.

Since $-\tau'$ is the scattering rate, i.e., how many collisions take place per unit conformal time, $-1/\tau'$ is the average conformal time between 2 consecutive collisions, which, for a photon, is also the average comoving distance between two collisions, i.e., the mean free path. 

Now, we have:
\begin{equation}
	\lambda_{\rm Silk}^2 \propto -\frac{\eta}{\tau'} \propto  \lambda_{\rm MFP}\eta\;,
\end{equation}
where we have used the comoving mean free path, $\lambda_{\rm MFP}$. Now, multiply and divide by $\lambda_{\rm MFP}$ and take the square root:
\begin{equation}
	\lambda_{\rm Silk} \propto \lambda_{\rm MFP}\sqrt{\frac{\eta}{\lambda_{\rm MFP}}}\;,
\end{equation}
Under the square root, we have the comoving distance $\eta$ divided by the photon comoving mean free path. This gives us the average number of collisions $N$ that the photons experience up to the time $\eta$, and hence:
\begin{equation}
	\lambda_{\rm Silk} \propto \sqrt{N}\lambda_{\rm MFP}\;,
\end{equation}
which is the typical relation for diffusion. Below this scale $\lambda_{\rm Silk}$ all fluctuations are suppressed because photons cannot agglomerate since they escape away. This effect is known as \textbf{Silk damping} \cite{silk1967fluctuations}. Therefore, the behavior of the $C_l$'s for large $l$'s is also decaying, though not exactly as in the above solution since this has to be free-streamed first.

We can do a more detailed calculation of the damping scale as follows. Let us neglect the gravitational potential and the $\ell \ge 3$ multipoles as before, but let us deal with more care for baryons and take into account polarization. From Eq.~\eqref{vbsuccapproxeq} we have:
\begin{equation}
	V_{\rm b} = 3\Theta_1 + \frac{R}{\tau'}\left(V_{\rm b}' + \mathcal H V_{\rm b}\right)\;,
\end{equation}
and the six equations for the monopole, dipole, and quadrupole of the temperature fluctuations and polarization:
\begin{eqnarray}
	\Theta_0' + k\Theta_1 = 0\;,\\
	3\Theta_1' + 2k\Theta_2 - k\Theta_0 = \tau'(3\Theta_1 - V_{\rm b})\;,\\
	10\Theta_2' - 4k\Theta_1 = 9\tau'\Theta_2 - \tau'\Theta_{P0} - \tau'\Theta_{P2}\;,\\
	2\Theta_{P0}' + 2k\Theta_{P1} = \tau'\Theta_{P0} - \tau'\Theta_{P2} - \tau'\Theta_2\;,\\
	3\Theta_{P1}' + 2k\Theta_{P2} - k\Theta_{P0} = 3\tau'\Theta_{P1}\;,\\
	10\Theta_{P2}' - 4k\Theta_{P1} = 9\tau'\Theta_{P2} - \tau'\Theta_{P0} - \tau'\Theta_{2}\;.
\end{eqnarray}
Now, assuming a solution of the type $\exp(i\int\omega d\eta)$ for all the above 7 variables and also assuming that $\omega \gg \mathcal H$, we have:
\begin{eqnarray}
	V_{\rm b} = \frac{3\Theta_1}{1 + Ri\omega\eta_c}\;,
\end{eqnarray}
where we have defined $\eta_c \equiv -1/\tau'$ as the average conformal time between 2 consecutive collisions. We thus have a closed system for $\Theta_0$, $\Theta_1$, $\Theta_2$, $\Theta_{P0}$, $\Theta_{P1}$, and $\Theta_{P2}$:
\begin{eqnarray}
	i\omega\Theta_0 + k\Theta_1 = 0\;,\\
	- k\Theta_0 + 3i\omega\Theta_1\left(1 + \frac{R}{1 + Ri\omega\eta_c}\right) + 2k\Theta_2 = 0\;,\\
	- 4k\eta_c\Theta_1 + (10i\omega\eta_c + 9)\Theta_2 - \Theta_{P0} - \Theta_{P2}= 0\;,\\
	- \Theta_2 + (2i\omega\eta_c + 1)\Theta_{P0} + 2k\eta_c\Theta_{P1} - \Theta_{P2} = 0\;,\\
	- k\Theta_{P0} + 3(i\omega\eta_c + 1)\Theta_{P1} + 2k\eta_c\Theta_{P2} = 0\;,\\
	-\Theta_{2} -\Theta_{P0} - 4k\eta_c\Theta_{P1} + (10i\omega\eta_c + 9)\Theta_{P2}  = 0\;.
\end{eqnarray}
We have already arranged the variables in order for the system matrix to appear clearly. The determinant of this matrix, in order to have a non trivial solution, must be zero. Considering the limit $\omega\eta_c \ll 1$, and keeping only the first-order in $\omega\eta_c$, we get:
\begin{equation}
	\frac{k^2}{3} - \omega^2(1 + R) + \frac{2i}{30}\omega\eta_c\left[37k^2 - 285(1 + R)\omega^2 + 15\omega^2R^2\right] = 0\;.
\end{equation}
In order to solve for $\omega$, let us again employ the smallness of $\omega\eta_c$ and stipulate that:
\begin{equation}
	\omega = \omega_0 + \delta\omega\;,
\end{equation}
where $\delta\omega$ is a small correction. From the above equation, it is then straightforward to obtain:
\begin{eqnarray}
	\frac{k^2}{3} - \omega_0^2(1 + R) = 0\;,\\
	 -2\omega_0\delta\omega(1 + R) + \frac{2i}{30}\omega_0\eta_c\left[37k^2 - 285(1 + R)\omega_0^2 + 15\omega_0^2R^2\right] = 0\;.
\end{eqnarray}
The first equation gives the result that we have already encountered:
\begin{equation}
	\boxed{\omega_0^2 = \frac{k^2}{3(1 + R)} = k^2c_s^2}
\end{equation}
When substituted into the second equation, it gives us:
\begin{equation}
	\boxed{\delta\omega = \frac{i\eta_ck^2}{6(1 + R)}\left[\frac{16}{15} + \frac{R^2}{1 + R}\right]}
\end{equation}
This result was obtained for the first time by \cite{1983MNRAS.202.1169K}. See also the derivation of \cite{Weinberg:2008zzc}.

Therefore, the evolution of the multipoles is proportional to the following factor:
\begin{equation}
	\exp\left(i\int\omega d\eta\right) = e^{ikr_s(\eta)}e^{-k^2/k_{\rm Silk}^2}\;,
\end{equation}
where
\begin{equation}\label{Silkscale}
	\boxed{\frac{1}{k_{\rm Silk}^2} \equiv -\int_0^\eta d\eta'\frac{1}{6\tau'(1 + R)}\left(\frac{16}{15} + \frac{R^2}{1 + R}\right)}
\end{equation}
From the best fit values of the parameters of the $\Lambda$CDM model, we have:
\begin{equation}
	\boxed{d_{\rm Silk} = 0.0066\mbox{ Mpc}}
\end{equation}

\section{Line-of-sight integration}\label{Sec:lineofsightintegration}

The approximate solutions found earlier are based on the TC limit, which allows us to take into account just the monopole and the dipole until recombination and then to better understand the physics behind the CMB anisotropies. On the other hand, observation demands more precise calculations to be compared with; therefore, in the end, numerical computations and codes such as CLASS are needed. Even so, there is a more efficient way of computing predictions on the CMB anisotropies than dealing directly with the hierarchy of the Boltzmann equation, and that is to formally integrate along the photon past light-cone according to a semi-analytic technique called \textbf{line-of-sight integration},\index{Line-of-sight integration} due to Seljak and Zaldarriaga \cite{Seljak:1996is}, and which was the basis for the CMBFAST code.\footnote{\url{https://lambda.gsfc.nasa.gov/toolbox/tb_cmbfast_ov.cfm}}

Recall the photon Boltzmann equations \eqref{BoltzeqphotonTheta} and \eqref{BoltzeqphotonPol}:
\begin{eqnarray}
	\Theta' + ik\mu\Theta = -\Phi' - ik\mu\Psi - \tau'\left[\Theta_0 - \Theta - i\mu V_{\rm b} - \frac{1}{2}\mathcal{P}_2(\mu)\Pi\right]\;,\\
	\Theta_P' + ik\mu\Theta_P = -\tau'\left[- \Theta_P + \frac{1}{2}[1 - \mathcal{P}_2(\mu)]\Pi\right]\;,
\end{eqnarray}
where $\Pi = \Theta_{2} + \Theta_{P2} + \Theta_{P0}$. Let us rewrite them as follows:
\begin{eqnarray}
	\Theta' + (ik\mu - \tau')\Theta = -\Phi' - ik\mu\Psi - \tau'\left[\Theta_0 - i\mu V_{\rm b} - \frac{1}{2}\mathcal{P}_2(\mu)\Pi\right] \equiv \mathcal S(\eta,k,\mu)\;,\qquad\\
	\Theta_P' + (ik\mu - \tau')\Theta_P = -\frac{\tau'}{2}[1 - \mathcal{P}_2(\mu)]\Pi \equiv \mathcal S_P(\eta,k,\mu)\;,\qquad
\end{eqnarray}
where we have introduced two source functions on the right hand sides. Note that the dependence is on $k$ and not on $\mathbf k = k\hat z$ because we are considering the equations for the transfer functions. Afterwards, before performing the anti-Fourier transform, we must rotate back $\hat k$ in a generic direction.

Let us write the left-hand sides as follows:
\begin{equation}
	\Theta' + (ik\mu - \tau')\Theta = e^{-ik\mu\eta + \tau}\frac{d}{d\eta}\left(\Theta\;e^{ik\mu\eta - \tau}\right)\;,
\end{equation}
with a similar expression for $\Theta_P$. Substituting these into the Boltzmann equations and integrating formally from a certain initial $\eta_i \to 0$ to $\eta_0$, we get:
\begin{eqnarray}
	\Theta(\eta_0)e^{-\tau(\eta_0)} = \Theta(\eta_i)e^{ik\mu(\eta_i - \eta_0) - \tau(\eta_i)} + \int_{\eta_i}^{\eta_0}d\eta\;e^{ik\mu(\eta - \eta_0) - \tau(\eta)}\mathcal S(\eta, k ,\mu)\;,\qquad\\
	\Theta_P(\eta_0)e^{-\tau(\eta_0)} = \Theta_P(\eta_i)e^{ik\mu(\eta_i - \eta_0) - \tau(\eta_i)} + \int_{\eta_i}^{\eta_0}d\eta\;e^{ik\mu(\eta - \eta_0) - \tau(\eta)}\mathcal S_P(\eta, k ,\mu)\;,\qquad
\end{eqnarray}
Now recall the definition of the optical depth:
\begin{equation}
	\tau \equiv \int_{\eta}^{\eta_0}d\eta'\;n_e\sigma_Ta\;.
\end{equation}
It is clear then that $\tau(\eta_0) = 0$, and since $\eta_i \to 0$ is deep into the radiation-dominated epoch, $\tau \propto 1/\eta$ is very large, and we can neglect $\exp[-\tau(\eta_i)]$. Therefore, we are left with 
\begin{eqnarray}
	\Theta(\eta_0, k, \mu) = \int_{0}^{\eta_0}d\eta\;e^{ik\mu(\eta - \eta_0) - \tau(\eta)}\mathcal S(\eta,k,\mu)\;,\\
	\Theta_P(\eta_0, k, \mu) = \int_{0}^{\eta_0}d\eta\;e^{ik\mu(\eta - \eta_0) - \tau(\eta)}\mathcal S_P(\eta,k,\mu)\;.
\end{eqnarray}
where we have already implemented the limit $\eta_i \to 0$. Now we calculate the $\Theta_\ell$'s by inverting the Legendre expansion as done in Eq.~\eqref{Thetal} and obtain:
\begin{eqnarray}
	\Theta_\ell(\eta_0, k) = \frac{1}{(-i)^\ell}\int_{-1}^{1}\frac{d\mu}{2}\;\mathcal{P}_\ell(\mu)\int_{0}^{\eta_0}d\eta\;e^{ik\mu(\eta - \eta_0) - \tau(\eta)}\mathcal S(k,\eta,\mu)\;,\\
	\Theta_{P\ell}(\eta_0, k) = \frac{1}{(-i)^\ell}\int_{-1}^{1}\frac{d\mu}{2}\;\mathcal{P}_\ell(\mu)\int_{0}^{\eta_0}d\eta\;e^{ik\mu(\eta - \eta_0) - \tau(\eta)}\mathcal S_P(k,\eta,\mu)\;,
\end{eqnarray}
The source terms have $\mu$-dependent contributions (up to $\mu^2$) that we can handle integrating by parts. Take, for example, the $-ik\mu\Psi$ contribution of $\mathcal S(k,\eta,\mu)$. Let $I_\Psi$ be its integral, which can be rewritten as follows:
\begin{equation}
	I_\Psi \equiv -\int_{0}^{\eta_0}d\eta\;ik\mu\Psi e^{ik\mu(\eta - \eta_0) - \tau(\eta)} = - \int_{0}^{\eta_0}d\eta\;\Psi e^{-\tau(\eta)}\frac{d}{d\eta}\left[e^{ik\mu(\eta - \eta_0)}\right]\;, 
\end{equation}
and now it is easy to integrate by parts and obtain:
\begin{equation}
	I_\Psi = -\left.\Psi e^{-\tau(\eta)}e^{ik\mu(\eta - \eta_0)}\right|_0^{\eta_0} + \int_{0}^{\eta_0}d\eta\; e^{ik\mu(\eta - \eta_0)}\frac{d}{d\eta}\left[\Psi e^{-\tau(\eta)}\right]\;. 
\end{equation}
The first contribution gives $-\Psi(\eta_0)$, i.e., the gravitational potential evaluated at present time. This is just an undetectable offset that we incorporate into the definition of $\Theta_\ell(\eta_0, k)$, as the observed anisotropy, like we did at the beginning of this chapter when dealing with the free-streaming solution.

\hrulefill

\begin{ex} Take care of the term containing $\mu^2$, in $\mathcal P_2(\mu)$. Show that:
\begin{equation}
	\int_0^{\eta_0}d\eta\;\tau'\mu^2\Pi e^{ik\mu(\eta - \eta_0) - \tau(\eta)} = -\frac{1}{k^2}\int_0^{\eta_0}d\eta\;e^{ik\mu(\eta - \eta_0)}\frac{d^2}{d\eta^2}\left[\tau'\Pi e^{-\tau(\eta)}\right]\;.
\end{equation}
\end{ex}

\hrulefill

Combining all the terms treated with integration by parts, we get:
\begin{eqnarray}
	\Theta_\ell(k,\eta_0) = \frac{1}{(-i)^\ell}\int_{-1}^{1}\frac{d\mu}{2}\;\mathcal{P}_\ell(\mu)\int_{0}^{\eta_0}d\eta\;e^{ik\mu(\eta - \eta_0)}\nonumber\\
	\left[-\left(\Phi' + \tau'\Theta_0 + \frac{\tau'\Pi}{4}\right)e^{-\tau} + \left(\Psi e^{-\tau} - \frac{\tau'V_{\rm b}e^{-\tau}}{k}\right)' - \frac{3}{4k^2}\left(\tau'\Pi e^{-\tau}\right)''\right]\;,\qquad\\
	\Theta_{P\ell}(k,\eta_0) = -\frac{3}{4(-i)^\ell}\int_{-1}^{1}\frac{d\mu}{2}\;\mathcal{P}_\ell(\mu)\int_{0}^{\eta_0}d\eta\;e^{ik\mu(\eta - \eta_0)}\left[\tau'\Pi e^{-\tau} + \frac{1}{k^2}\left(\tau'\Pi e^{-\tau}\right)''\right]\;.\qquad
\end{eqnarray}
Using now the relation of Eq.~\eqref{intdmuPleikmur}, we can cast the above equations as:
\begin{eqnarray}
	\Theta_\ell(\eta_0, k) = \int_{0}^{\eta_0}d\eta\;S(\eta, k)j_\ell\left[k(\eta_0 - \eta)\right]\;,\\
	\Theta_{P\ell}(\eta_0, k) = \int_{0}^{\eta_0}d\eta\;S_P(\eta, k)j_\ell\left[k(\eta_0 - \eta)\right]\;.
\end{eqnarray}
with
\begin{eqnarray}
	\label{sourcetermTheta} S(\eta, k) \equiv (\Psi' - \Phi')e^{-\tau} + g\left(\Theta_0 + \frac{\Pi}{4} + \Psi\right) + \frac{1}{k}(gV_{\rm b})' + \frac{3}{4k^2}\left(g\Pi\right)''\;,\\
	\label{sourcetermThetaP} S_P(\eta, k) \equiv \frac{3}{4}g\Pi + \frac{3}{4k^2}\left(g\Pi\right)''\;,
\end{eqnarray}
where we have introduced the \textbf{visibility function}:\index{Visibility function}
\begin{equation}\label{visbilityfunction}
	\boxed{g(\eta) \equiv -\tau'e^{-\tau}}
\end{equation}

\hrulefill

\begin{ex} Show that the visibility function is normalized to unity:
\begin{equation}
	\int_0^{\eta_0}d\eta\;g(\eta) = 1\;.
\end{equation}
\end{ex}

\hrulefill

The visibility function represents the Poisson probability that a photon was last scattered at a time $\eta$. It is peaked at a time that we define as the moment of recombination, i.e., at $\eta = \eta_*$, because for $\eta > \eta_*$ it is basically zero, since $\tau' = 0$. Before recombination, in the radiation-dominated epoch, we saw that $-\tau' \propto 1/\eta^2$, and thus $\tau \propto 1/\eta$ and $g \propto \exp(-1/\eta)/\eta^2$, i.e., it goes to zero exponentially fast.

\begin{figure}[htbp]
\center
	\includegraphics[width=\columnwidth]{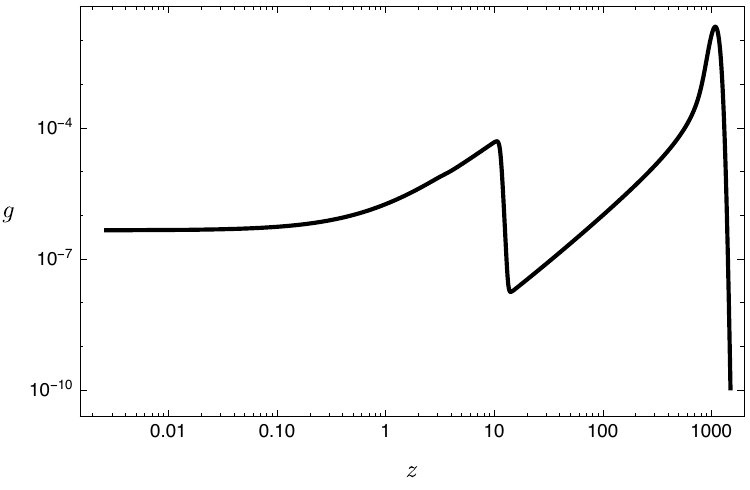}
\caption{Visibility function $g$ as function of the redshift from the numerical calculation performed with CLASS for the standard model.}
\label{Fig:VisibilityfunPlot}
\end{figure}

In Fig.~\ref{Fig:VisibilityfunPlot}, we plot the numerical calculation of the visibility function performed with CLASS for the standard model. Note the peak at about $z = 1000$, which has always been our reference for the recombination redshift. Note also another peak at about $z = 10$, representing the epoch of \textbf{reionization}.\index{Reionization} Until now, we have treated the visibility function as if it were a Dirac delta $\delta(\eta - \eta_*)$: this is known as the \textbf{sudden recombination approximation}.\index{Sudden recombination} From Fig.~\ref{Fig:VisibilityfunPlot}, we can appreciate that it is a good approximation (mind the logarithmic scale there). As usual, in cosmology but not only, the calculations get more and more complicated and impossible to do analytically the more precision we demand.

Inserting the source terms \eqref{sourcetermTheta} and \eqref{sourcetermThetaP} into the expressions for $\Theta_\ell(\eta_0, k)$ and $\Theta_{P\ell}(\eta_0, k)$ in the expression for $\Theta_\ell$ and integrating by parts, we get:
\begin{eqnarray}\label{lineofsightintegralThetal}
	\Theta_\ell(k,\eta_0) = \int_{0}^{\eta_0}d\eta\;g\left(\Theta_0 + \Psi + \frac{\Pi}{4}\right)j_\ell\left[k(\eta_0 - \eta)\right]\nonumber\\ - \int_{0}^{\eta_0}d\eta\frac{gV_{\rm b}}{k}\frac{d}{d\eta}j_\ell\left[k(\eta_0 - \eta)\right] + \int_{0}^{\eta_0}d\eta\frac{3g\Pi}{4k^2}\frac{d^2}{d\eta^2}j_\ell\left[k(\eta_0 - \eta)\right]\nonumber\\
	 + \int_{0}^{\eta_0}d\eta\;e^{-\tau}(\Psi' - \Phi')j_\ell\left[k(\eta_0 - \eta)\right]\;,\\
\label{lineofsightintegralThetaPl}	 \Theta_{P\ell}(k,\eta_0) = \int_{0}^{\eta_0}d\eta\;\frac{3g\Pi}{4}j_\ell\left[k(\eta_0 - \eta)\right] + \int_{0}^{\eta_0}d\eta\frac{3g\Pi}{4k^2}\frac{d^2}{d\eta^2}j_\ell\left[k(\eta_0 - \eta)\right]\;.
\end{eqnarray}
Assuming the visibility function to be a Dirac delta $\delta(\eta - \eta_*)$, and neglecting $\Pi$, we recover formula \eqref{freestreamingformulaphotons}. Note that neglecting $\Pi$ results in no polarization being present. Indeed, from the above equation, we see that a non-zero quadrupole moment of the photon distribution at recombination is essential in order to have polarization.

The above equations still need the Boltzmann hierarchy in order to be integrated, but just up to $\ell = 4$ (because $\Theta_2$ and $\Theta_4$ moments are contained in the equation for $\Theta_3'$) and hence are much more convenient from a computational point of view. 

The partial wave expansion of $\Theta$ given in Eq.~\eqref{Thetapartialwaveexpansion}:
\begin{equation}
	\Theta(k,\mu) = \sum_\ell (-i)^\ell(2\ell + 1)\mathcal P_\ell(\mu)\Theta_\ell(k)\;,
\end{equation}
and that we have used in the above calculations is valid as long as $\hat k = \hat z$. Now we have to rotate it in a general direction before performing the Fourier anti-transform. The task is simple because the temperature fluctuation is a scalar. Therefore:
\begin{equation}
	\Theta(k,\hat k\cdot\hat p) = \sum_\ell (-i)^\ell(2\ell + 1)\mathcal P_\ell(\hat k\cdot\hat p)\Theta_\ell(k)\;.
\end{equation}
The same is not true for $\Theta_P$, since the Stokes parameters are not scalars.

Using the definition of $a_{T,\ell m}$ given in Eq.~\eqref{ThetaexpYlm}, we can then write:
\begin{equation}
	a_{T,\ell m}^S = \int d^2\hat n\;Y^{m*}_{\ell}(\hat n)\sum_l(-i)^\ell(2\ell + 1)\int\frac{d^3\mathbf k}{(2\pi)^3}\mathcal P_\ell(\hat k\cdot\hat p)\alpha(\mathbf k)\Theta_\ell(k)\;.
\end{equation}
The integration is over $d^2\hat n$, hence we must change $\hat p \to \hat n = -\hat p$ in the Legendre polynomial. This gives an extra $(-1)^\ell$ factor due to the parity of the Legendre polynomials, and then, using the addition theorem, we obtain:
\begin{eqnarray}
	a_{T,\ell m}^S = \int d^2\hat n\; Y^{m*}_{\ell}(\hat n)\sum_li^\ell(2\ell + 1)\int\frac{d^3\mathbf k}{(2\pi)^3}\alpha(\mathbf k)\frac{4\pi}{2\ell' + 1}\nonumber\\ \sum_{m' = -\ell'}^{\ell'}Y^{*m'}_{\ell'}(\hat k)Y^{m'}_{\ell'}(\hat n)\Theta_\ell(k)\;.
\end{eqnarray}
Now the integration over the whole solid angle can be performed, and the orthonormality of the spherical harmonics can be employed, thus obtaining:
\begin{equation}\label{aTlmscalar}
	\boxed{a_{T,\ell m}^S = 4\pi i^\ell\int\frac{d^3\mathbf k}{(2\pi)^3}Y^{m*}_{\ell}(\hat k)\alpha(\mathbf k)\Theta_\ell(k)}
\end{equation}
This formula, together with Eq.~\eqref{lineofsightintegralThetal}, allows us to explicitly calculate the scalar contribution to the $a_{T,\ell m}$'s. Earlier, we have focused on the $C_{TT,\ell}$'s only, for which the calculations are simpler because there is no need to perform a spatial rotation. However, we need to know the explicit form of the $a_{T,\ell m}$'s in order to compute the TE correlation spectrum of Eq.~\eqref{TTandTEspectra}.

\section{Finite thickness effect and reionization}

In this section, we discuss two more effects that influence the CMB spectrum, namely the finite thickness effect and reionization. The first one is related to the fact that the visibility function $g$ in Fig.~\ref{Fig:VisibilityfunPlot} is peaked, but it is not a Dirac delta. In other words, CMB photons do not last scatter all at once at $\eta_*$, but during a finite amount of time, say $\Delta\eta_*$. This is the \textbf{finite thickness effect}\index{Finite thickness effect}. Physically, on scales smaller than the thickness $\Delta\eta_*$, we expect fluctuations to be washed out because they are averaged over a finite amount of time. This is similar to what is called \textbf{Landau damping}\index{Landau damping} (although the latter arises from a spread in frequency and not in time). It may seem that Landau damping is a small effect, but it is actually of the same order as Silk damping, and therefore, it must be taken into account.

Let us take advantage of this investigation and derive the form of the visibility function here. The question is: what is the probability that a photon last scatters during some sufficiently small interval between the instants $\eta$ and $\eta + \Delta\eta$? The time interval $\Delta\eta$ is sufficiently small so that only one collision can take place in it. The attentive reader has noticed that this is the same requirement we make when we derive the Poisson distribution, cf. Appendix \ref{App:Poisson}, which in fact rules the statistics of e.g., the scattering process.

So, we divide the time interval $\eta_0 - \eta$ into many intervals, i.e., $N \equiv (\eta_0 - \eta)/\Delta\eta$, and write the probability as:
\begin{equation}
	\Delta P = \frac{\Delta\eta}{\eta_c(\eta)}\left[1 - \frac{\Delta\eta}{\eta_c(\eta_1)}\right]\left[1 - \frac{\Delta\eta}{\eta_c(\eta_2)}\right]\dots\left[1 - \frac{\Delta\eta}{\eta_c(\eta_N)}\right]\;,
\end{equation}
where we recall that $\eta_c \equiv -1/\tau'$ is the average time between two consecutive collisions and it is time-dependent. We have chosen the time interval $\Delta\eta$ small enough for $\Delta\eta/\eta_c$ to be the probability of having one scattering during its duration and hence $1 - \Delta\eta/\eta_c$ being that of having no scattering. Now, in the limit $\Delta\eta \to 0$ we can write:
\begin{equation}
	dP = \frac{d\eta}{\eta_c}\exp\left(-\int_{\eta}^{\eta_0}\frac{d\eta}{\eta_c}\right) = -\tau'\exp\left[-\tau(\eta)\right]d\eta = g(\eta)d\eta\;,
\end{equation}
i.e., the visibility function defined in Eq.~\eqref{visbilityfunction} appears. As we have anticipated, the maximum of the visibility function occurs in a time that we dub $\eta_*$, i.e., the recombination time if we make the assumption of sudden recombination. From the condition for the extrema of a function:
\begin{equation}
	g'(\eta) = -\tau''e^{-\tau} + (\tau')^2e^{-\tau} = 0\;,
\end{equation}
we get:
\begin{equation}
	-\tau'' = (\tau')^2\;,
\end{equation}
as the condition which defines $\eta_*$. Therefore, employing the definition of the optical depth, we get:
\begin{equation}
	(n_e\sigma_{\rm T} a)'_* = -(n_e\sigma_{\rm T}a)^2_*\;,
\end{equation}
where the derivative is evaluated at $\eta_*$, as well as the function on the right hand side. Now, let us write the free-electron number density as $n_e = X_en_{\rm b}$, i.e., introducing the free-electron fraction and the baryon number density. Recall from the Boltzmann equation, cf. Chapter \ref{Chap:ThermalHistory}, that:
\begin{equation}
	X'_e = - \frac{1.44\times 10^4}{z}\mathcal HX_e\;.
\end{equation}
Evaluating the above equation at $\eta_*$ and combining it with the extremum condition for the visibility function, we get:
\begin{equation}
	X_e(\eta_*) \approx \frac{1.44\times 10^4}{z_*(n_{\rm b}\sigma_{\rm T} a)_*}\mathcal H(\eta_*) \equiv \frac{K}{(n_{\rm b}\sigma_{\rm T} a)_*}\mathcal H(\eta_*)\;,
\end{equation}
and from this, we get the recombination redshift $z_* \approx 1050$.

Let us approximate the visibility function by expanding it about its maximum:
\begin{equation}
	g(\eta) = \exp[\ln(-\tau') - \tau] \approx \exp\left[-\frac{1}{2}[\tau - \ln(-\tau')]''_*(\eta - \eta_*)^2\right]\;,
\end{equation}
i.e., we have a Gaussian function. Now we determine the second derivative and hence the variance of the distribution by using the Boltzmann equation and employing the following approximation: we consider only the first derivative of $X_e$ to be different from zero. All the other quantities are approximately constant indeed since recombination takes place quite rapidly. So, we can write:
\begin{equation}
	\left(\tau' - \frac{\tau''}{\tau'}\right)'_* = \left(-n_{\rm b}X_e\sigma_{\rm T}a - \frac{n_{\rm b}X_e'\sigma_{\rm T}a}{n_{\rm b}X_e\sigma_{\rm T}a}\right)'_*\;. 
\end{equation}
Now, within our approximation, $X_e'/X_e$ is constant, and thus:
\begin{equation}
	\left(\tau' - \frac{\tau''}{\tau'}\right)'_* = -(n_{\rm b}X_e'\sigma_{\rm T}a)_* = (n_{\rm b}X_e\sigma_{\rm T}a)^2_* = K^2 \mathcal H(\eta_*)^2\;. 
\end{equation}
Hence, the visibility function can be approximated as a Gaussian function:
\begin{equation}
	g(\eta) \approx \frac{K\mathcal H_*}{\sqrt{2\pi}}\left[-\frac{1}{2}(K\mathcal H_*)^2(\eta - \eta_*)^2\right]\;,
\end{equation}
with variance $1/(K\mathcal H_*)$.

When we substitute this approximation of the visibility function into the line-of-sight integrals of Eqs.~\eqref{lineofsightintegralThetal} and \eqref{lineofsightintegralThetaPl}, we can extract the spherical Bessel functions as $j_l[k(\eta_0 - \eta_*)]$, since $\eta_0$ is much larger than any conformal time about recombination, and the other integrals are oscillating functions of $kr_s(\eta)$, as we saw in Eq.~\eqref{HuSugiyamasolution}. We can approximate this about $\eta_*$ as follows:
\begin{equation}
	kr_s(\eta) = k\int_0^\eta d\eta'\frac{1}{\sqrt{3(1 + R)}} \approx kr_s(\eta_*) + \frac{k}{\sqrt{3(1 + R_*)}}(\eta - \eta_*)\;.
\end{equation}
Hence, we finally have a conformal time Gaussian integral of the following type:
\begin{equation}
	\int_{-\infty}^\infty d\eta\exp\left[-\frac{1}{2}(K\mathcal H_*)^2(\eta - \eta_*)^2 + \frac{ik}{\sqrt{3(1 + R_*)}}(\eta - \eta_*)\right]\;.
\end{equation}
Now, using the formula for the Gaussian integral:
\begin{equation}
	\int_{-\infty}^\infty e^{-ax^2 + bx}dx = \sqrt{\frac{\pi}{a}}e^{b^2/4a}\;,
\end{equation}
we can conclude that the $\Theta_{\ell}$'s get an additional damping factor of the following form:
\begin{equation}
	\exp\left[-\frac{k^2}{6(1 + R_*)(K\mathcal H_*)^2}\right]\;,
\end{equation}
so a new damping scale appears:
\begin{equation}
	\boxed{d_{\rm Landau}^2 = \frac{1}{k_{\rm Landau}^2} \equiv \frac{1}{6(1 + R_*)(K\mathcal H_*)^2}}
\end{equation}
which, following \cite{Weinberg:2008zzc}, we call \textbf{the Landau damping scale}. For the $\Lambda$CDM model best fit parameters, we have:
\begin{equation}
	\boxed{d_{\rm Landau} \approx 0.0048\mbox{ Mpc}}
\end{equation}
Now let us turn to reionization. At a redshift of about 10, hydrogen gets ionized again by the ultraviolet radiation of the first structures. Hence, the free-electron fraction grows again, increasing the probability of a CMB photon being scattered again; cf. the extra bump in the visibility function in Fig.~\ref{Fig:VisibilityfunPlot}. Following the calculation done above to obtain the visibility function, we know that the probability for a photon not to be scattered from reionization until today is:
\begin{equation}
	\exp[-\tau(\eta_{\rm reion})]\;,
\end{equation}
and of course, the one for being scattered is 1 minus the above quantity, which we compute now:
\begin{equation}
	\tau(\eta_{\rm reion}) = \int_{\eta_{\rm reion}}^{\eta_0}d\eta\;n_e\sigma_{\rm T} a = \sigma_{\rm T}\int_0^{z_{\rm reion}}\frac{dz}{(1 + z)^2\mathcal H}n_e\;.
\end{equation}
For the free-electron number density, we can write:
\begin{equation}
	n_e = 0.88 n_{\rm b}X_e = 0.88\frac{3H_0^2}{8\pi m_{\rm b}}\Omega_{\rm b0}(1 + z)^3\;,
\end{equation}
where the factor $0.88$ is due to the fact that not all baryons are electrons or protons; there are also neutrons in Helium nuclei. Assuming matter-domination and instantaneous reionization, i.e.
\begin{equation}
	\mathcal H^2 = H_0^2\Omega_{\rm m0}(1 + z)^3\;,
\end{equation}
and $X_e = 1$ for $z < z_{\rm reion}$, we get:
\begin{equation}
	\tau(z_{\rm reion}) \approx 0.04\frac{\Omega_{\rm b0}h^2}{\sqrt{\Omega_{\rm m0}h^2}}z_{\rm reion}^{3/2}\;.
\end{equation}
Hence, the probability of a CMB photon not being scattered during reionization taking place at redshift $z_{\rm reion} = 10$ is about $0.99$. Those photons that are scattered are mixed up; hence, the correlation in their temperature is destroyed. So, the effect of reionization on the $C_{TT,\ell}$'s is simply to weigh them by a factor $\exp(-2\tau_{\rm reion})$, the factor of 2 appearing because the spectrum is a quadratic function of the temperature fluctuations.

\section{Cosmological parameters determination}

In this section, we discuss how the CMB TT spectrum is sensitive to the cosmological parameters. We have learned in this chapter about many quantities that are relevant in forming the shape of the spectrum, but we have not actually derived an analytic, approximated formula in order to see this explicitly. These can be found in \cite{Mukhanov:2005sc} and \cite{Weinberg:2008zzc}. Here instead, we plot with CLASS various spectra for varying parameters and discuss the physics behind the changes.

Note that, for the standard $\Lambda$ model, 6 of the overall parameters are usually left free and constrained by observation:
\begin{enumerate}
	\item The amplitude of the primordial power spectrum: $A_S$;
	\item The primordial tilt: $n_S$;
	\item The baryon abundance: $\Omega_{\rm b0}h^2$;
	\item The CDM abundance: $\Omega_{\rm c0}h^2$;
	\item The reionization epoch: $z_{\rm reion}$;
	\item The sound horizon at recombination: $r_s(\eta_*)$, which is related to the Hubble constant value $H_0$.
\end{enumerate}
The other cosmological parameters can be derived from these ones. In particular, the amount of radiation is already well known by measuring the CMB temperature, and the amount of $\Lambda$ and curvature are determined via the positions of the peaks, which depend on $r_s(\eta_*)$, which in turn depends on the baryon content.

\begin{figure}[htbp]
\center
	\includegraphics[width=\columnwidth]{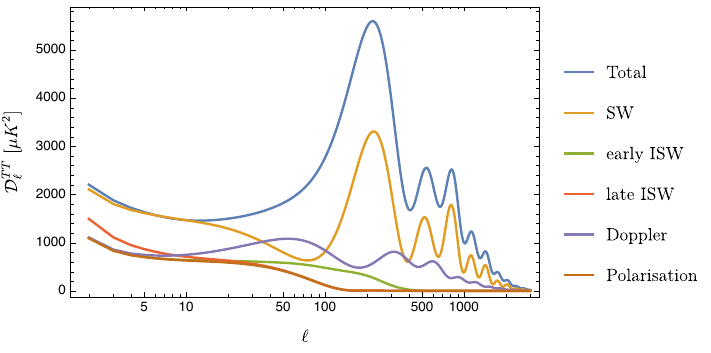}
\caption{Total CMB TT power spectrum (blue line) computed with CLASS and decomposed in the physically different contributions: Sachs-Wolfe effect (yellow line), early-times ISW effect (green line), late-times ISW effect (red line), Doppler effect (purple line), and polarization (brown line).}
\label{Fig:CldecompositionPlot}
\end{figure}

\begin{figure}[htbp]
\center
	\includegraphics[width=\columnwidth]{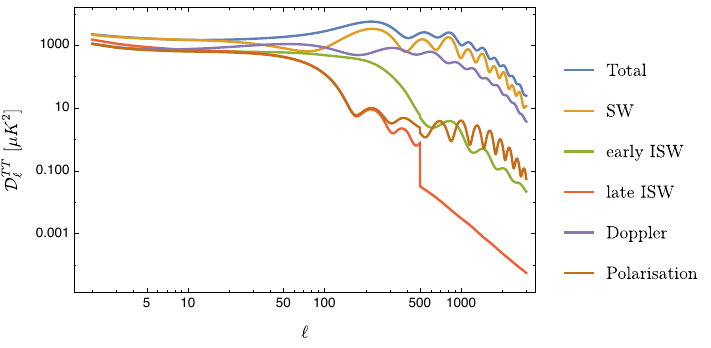}
\caption{Same as Fig.~\ref{Fig:CldecompositionPlot} but in logarithmic scale, in order to better distinguish the weakest contributions.}
\label{Fig:CldecompositionPlotLogLog}
\end{figure}

In Figs.~\ref{Fig:CldecompositionPlot} and \ref{Fig:CldecompositionPlotLogLog}, we start to show the numerical calculation of the CMB TT power spectrum decomposed into the contributions discussed in this Chapter. See also \cite{Wands:2015fua}. We consider the $\Lambda$CDM as a fiducial model.

\begin{figure}[htbp]
\center
	\includegraphics[width=\columnwidth]{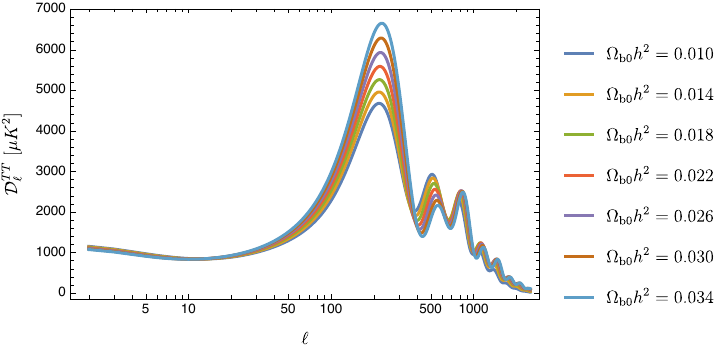}
\caption{CMB TT power spectrum computed with CLASS and varying $\Omega_{\rm b0}h^2$. From the lowest first peak to the highest: $\Omega_{\rm b0}h^2 = 0.010, 0.014, 0.018, 0.022, 0.026, 0.030, 0.034$.}
\label{Fig:VaryingObPlot}
\end{figure}

In Fig.~\ref{Fig:VaryingObPlot} we show what happens to the CMB TT spectrum for $\Omega_{\rm b0}h^2 = 0.010, 0.014, 0.018, 0.022, 0.026, 0.030, 0.034$. Taking the first peak height as a reference, the larger the value of $\Omega_{\rm b0}h^2$ is, the higher the peak becomes. When we vary one of the density parameters, because of the closure relation, something else must vary. We have chosen to vary $\Omega_\Lambda$.

Why so? We have seen that baryon loading makes compression favored over rarefaction, and hence the first and third peaks are higher for higher values of $\Omega_{\rm b0}h^2$, but the second one is lower. In other words, the peaks relative height is very sensitive to the baryon content. The position of the first peak does not change much because it is most sensitive to the spatial curvature, and this has been fixed to zero. Finally, the curves for larger $\Omega_{\rm b0}h^2$, as we commented, have less $\Omega_\Lambda$ and therefore less ISW effect. For this reason, they are slightly lower for small $\ell$.

\begin{figure}[htbp]
\center
	\includegraphics[width=\columnwidth]{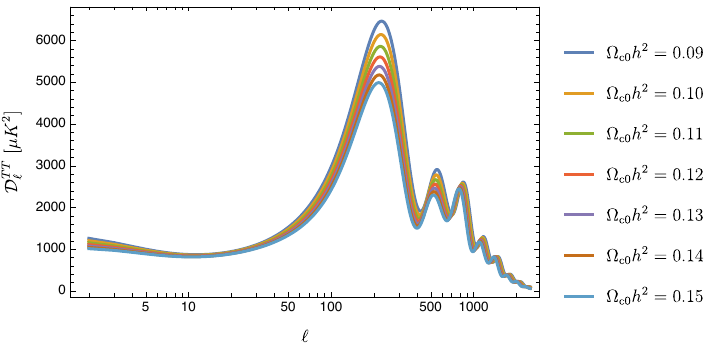}
\caption{CMB TT power spectrum computed with CLASS and varying $\Omega_{\rm c0}h^2$. From the highest first peak to the lowest: $\Omega_{\rm c0}h^2 = 0.09, 0.10, 0.11, 0.12, 0.13, 0.14, 0.15$.}
\label{Fig:VaryingOcPlot}
\end{figure}

In Fig.~\ref{Fig:VaryingOcPlot}, we show what happens to the CMB TT spectrum for $\Omega_{\rm c0}h^2 = 0.09, 0.10, 0.11, 0.12, 0.13, 0.14, 0.15$. Taking the first peak height as a reference, the larger the value of $\Omega_{\rm c0}h^2$ is, the lower the peak becomes. This behavior is the opposite of what we found by varying $\Omega_{\rm b0}h^2$. Mostly, CDM intervenes through the SW effect since it dominates the gravitational potential $\Psi$ at recombination. The first peak is affected more because it corresponds to large scales, basically the horizon at recombination, and there the transfer function is approximately one, meaning that $-\Psi$ is as large as possible. The subsequent peaks correspond to scales that entered the horizon much earlier, and therefore, the CDM influence there is weak.

In this case, we have also chosen to vary $\Omega_\Lambda$ in order to keep the total density budget constant. Indeed, the more abundant CDM is, the less abundant $\Lambda$ becomes, and the late ISW effect is less pronounced, as expected.

\begin{figure}[htbp]
\center
	\includegraphics[width=\columnwidth]{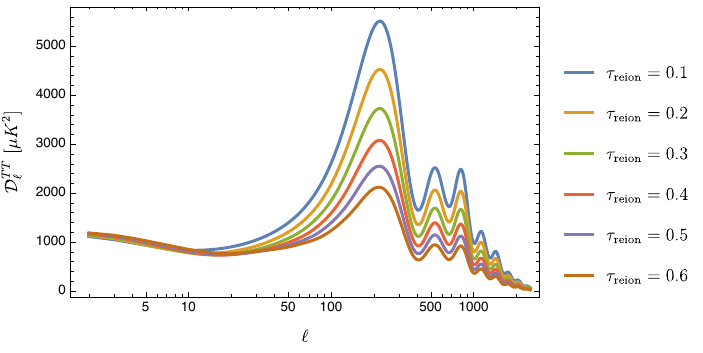}
\caption{CMB TT power spectrum computed with CLASS and varying $\tau_{\rm reion}$. From the highest first peak to the lowest: $\tau_{\rm reion} = 0.1, 0.2, 0.3, 0.4, 0.5, 0.6, 0.7$.}
\label{Fig:VaryingtauPlot}
\end{figure}

In Fig.~\ref{Fig:VaryingtauPlot}, we show what happens to the CMB TT spectrum for $\tau_{\rm reion} = 0.1, 0.2, 0.3, 0.4, 0.5, 0.6, 0.7$. As we have commented on in the previous section, the overall effect of reionization is simple: a damping of the order $\exp(-2\tau_{\rm reion})$ for multipoles larger than a certain $\ell_{\rm reion}$, which we infer to be about 10 from the plots in Fig.~\ref{Fig:VaryingtauPlot}.

\begin{figure}[htbp]
\center
	\includegraphics[width=\columnwidth]{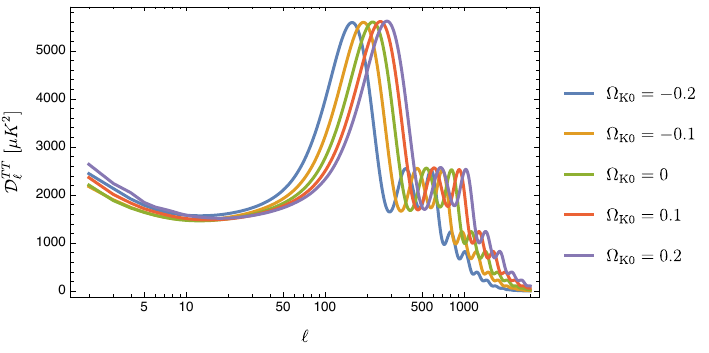}
\caption{CMB TT power spectrum computed with CLASS and varying $\Omega_{\rm K0}$. From the left to the right: $\Omega_{\rm K0} = -0.2, -0.1, 0, 0.1, 0.2$.}
\label{Fig:VaryingomegakPlot}
\end{figure}

From Fig.~\ref{Fig:VaryingomegakPlot}, we can appreciate how the CMB TT power spectrum is affected by the spatial geometry of the universe. From the leftmost spectrum to the rightmost one $\Omega_{\rm K0} = -0.2, -0.1, 0, 0.1, 0.2$. Hence, the position of the first peak is of great importance in determining whether our universe is closed or open.

As we saw in Eq.~\eqref{HuSugiyamasolution}, the length scale associated with the acoustic peaks is the sound horizon at recombination:
\begin{equation}
	r_s(\eta_*) = \int_0^{\eta_*}c_sd\eta\;,
\end{equation}
where the speed of sound of the baryon-photon plasma is given by Eq.~\eqref{speedofsoundbaryonphotonplasma}:
\begin{equation}
	c_s^2 = \frac{1}{3(1 + R)} = \frac{4\Omega_{\gamma 0}}{3(4\Omega_{\gamma 0} + 3\Omega_{\rm b0}a)}\;.
\end{equation}
The physical sound horizon is given by:
\begin{equation}
	r^{\rm phys}_s(z_*) = \int_0^{t_*}c_s(t)dt = \int_{z_*}^{\infty}dz\frac{c_s(z)}{H(z)(1 + z)}\;,
\end{equation}
i.e., integrating the lookback time. We need the physical quantity in order to relate it to the angular-diameter distance to recombination:
\begin{equation}
	d_{\rm A}(z_*) = \frac{1}{(1 + z_*)}\int_0^{z_*}\frac{dz}{H(z)}\;,
\end{equation}
and thus estimate the multipole corresponding to the first peak:
\begin{equation}
	\ell_{\rm 1st} \approx \frac{1}{\theta_{\rm 1st}} = \frac{d_{\rm A}(z_*)}{r^{\rm phys}_s(z_*)}\;.
\end{equation}
Let us approximate the physical sound horizon by assuming a constant $c_s$ and a matter-dominated universe. We have thus:
\begin{equation}
	r^{\rm phys}_s(z_*) \approx \frac{c_s}{H_0\sqrt{\Omega_{\rm m0}}}\int_{z_*}^{\infty}\frac{dz}{(1 + z)^{5/2}} = \frac{2c_s}{3H_0\sqrt{\Omega_{\rm m0}}}\frac{1}{(1 + z_*)^{3/2}}\;,
\end{equation}
and for the angular-diameter distance, we also assume a matter plus $\Lambda$ universe:
\begin{equation}
	d_{\rm A}(z_*) = \frac{1}{H_0(1 + z_*)}\int_0^{z_*}\frac{dz}{\sqrt{\Omega_{\rm m0}(1 + z)^3 + (1 - \Omega_{\rm m0})}}\;.
\end{equation}

\hrulefill

\begin{ex}
Show that $d_{\rm A}(z_*)$ can be approximated as:
\begin{equation}
	d_{\rm A}(z_*) \approx \frac{2}{7H_0(1 + z_*)\sqrt{\Omega_{\rm m0}}}(9 - 2\Omega_{\rm m0}^3)\;.
\end{equation}
\end{ex}

\hrulefill

Hence, we have:
\begin{equation}
	\ell_{\rm 1st} \approx 0.74\sqrt{1 + z_*}(9 - 2\Omega_{\rm m0}^3) \approx 220\;,
\end{equation}
which clearly shows how the position of the first peak changes as a function of the total matter content.

\begin{figure}[htbp]
\center
	\includegraphics[width=\columnwidth]{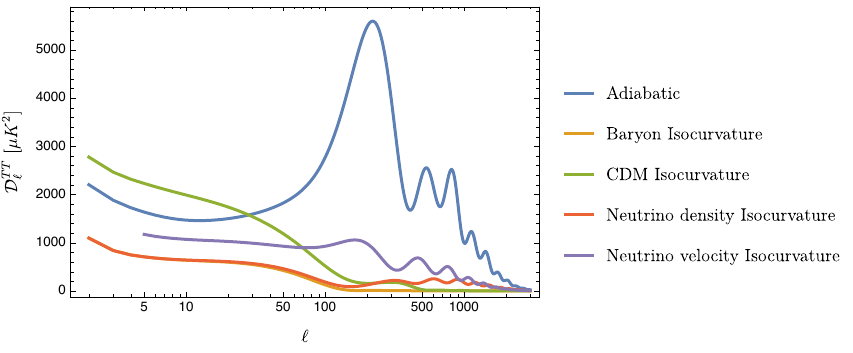}
\caption{CMB TT power spectrum computed with CLASS and varying initial conditions: adiabatic (blue line), baryon isocurvature (yellow line), CDM isocurvature (green line), neutrino density isocurvature (red line), neutrino velocity isocurvature (purple line).}
\label{Fig:VaryingICPlot}
\end{figure}

In Fig.~\ref{Fig:VaryingICPlot}, we show how the initial conditions dramatically affect the CMB TT power spectrum and how the adiabatic ones are favored by observation.

\section{Tensor contribution to the CMB TT correlation}

Tensor perturbations also contribute to generating temperature anisotropies, as we see in Eq.~\eqref{tensorBoltzmannequationTheta}, which we report here after renormalizing to the primordial mode $\beta(\mathbf k,\lambda)$, cf. Eq.~\eqref{ICnormalisationh}:
\begin{eqnarray}
	\left(\frac{\partial}{\partial\eta} + ik\mu - \tau'\right)\Theta^{(T)}(\eta, k, \mu) + \frac{h^{'T}}{2} = \nonumber\\ - \tau'\left[\frac{3}{70}\Theta^{(T)}_{4} + \frac{1}{7}\Theta^{(T)}_{2} + \frac{1}{10}\Theta^{(T)}_{0} - \frac{3}{70}\Theta_{P4}^{(T)} + \frac{6}{7}\Theta_{P2}^{(T)} - \frac{3}{5}\Theta_{P0}^{(T)}\right]\nonumber\\ \equiv - \tau'\mathcal S^T(\eta, k)\;,\qquad\\
	\left(\frac{\partial}{\partial\eta} + ik\mu - \tau'\right)\Theta_{P}^{(T)}(\eta, k, \mu) = \tau'\mathcal S^T(\eta, k)\;,\qquad
\end{eqnarray}
The label $\lambda$ representing the two possible states of helicity is absent because of the renormalization with $\beta(\mathbf k,\lambda)$. It represents the fact that the evolution of the two helicities is the same.
 
The line-of-sight solutions of the above equations are as follows:
\begin{eqnarray}\label{lineofsightintegralThetatens}
	\Theta^{(T)}(\eta_0, k, \mu) = \int_{0}^{\eta_0}d\eta\;e^{ik\mu(\eta - \eta_0) -\tau}\left[-h'^T/2 -\tau'\mathcal S^T(\eta, k)\right]\;,\qquad\\
\label{lineofsightintegralThetaPtens}	\Theta_{P}^{(T)}(\eta_0, k, \mu) = \int_{0}^{\eta_0}d\eta\;e^{ik\mu(\eta - \eta_0) -\tau}\tau'\mathcal S^T(\eta, k)\;. \qquad
\end{eqnarray}
We now focus on $\Theta^{(T)}(\eta_0, k, \mu)$. Defining: 
\begin{equation}
	S^T(\eta, k) \equiv e^{-\tau}\left[-h'^T/2 -\tau'\mathcal S^T(\eta, k)\right]\;,
\end{equation}
and using Eq.~\eqref{tensortemperaturecontribution} and Eq.~\eqref{hijpipjexpansionthetaphi}, the tensor contribution to the temperature fluctuation is made up of the sum of the following two contributions:
\begin{equation}\label{ThetaTexpansionkparz}
	f_\lambda(k\hat z, \hat{p}) \equiv 4\sqrt{\frac{\pi}{15}}Y^\lambda_2(\hat p)\int_0^{\eta_0}d\eta\;S^T(\eta, k)e^{-i\mu kr(\eta)}\;,
\end{equation}
where $r(\eta) \equiv \eta_0 - \eta$ and where we stress that the result holds true for $\hat k = \hat z$, since this was the condition under which we derived the Boltzmann equation for photons. We cannot yet sum over $\lambda$ because we have to include $\beta(\mathbf k,\lambda)$ first. For this reason, we shall work on $f_\lambda(k\hat z, \hat{p})$.

In order to investigate temperature fluctuations in the sky, we need to anti-transform $\Theta^{(T)}(\mathbf k, \hat{p})$ to employ the usual expansion:
\begin{equation}\label{ThetaTYexpans}
	\Theta^{(T)}(\hat{ n}) = \sum_{\ell m}a_{T,\ell m}^TY_\ell^m(\hat{ n})\;, \quad a_{T,\ell m}^T = \int d^2\hat n\;Y_\ell^{m*}(\hat{ n})\Theta^{(T)}(\hat{ n})\;. 
\end{equation}
So, let us proceed as follows. We trade $\hat p$ for the line-of-sight $\hat n = -\hat p$ and use the expansion of a plane wave in spherical harmonics:
\begin{equation}
	e^{i\hat k\cdot\hat{n}kr} = 4\pi\sum_{LM}i^LY_L^{M*}(\hat{k})Y_L^{M}(\hat{n})j_L(kr)\;,
\end{equation}
in Eq.~\eqref{ThetaTexpansionkparz}.

\hrulefill

\begin{ex}
	First of all, since $\hat k = \hat z$ then show that:
\begin{equation}
	Y_L^{M*}(\hat{k}) = Y_L^{M*}(\hat{z}) = \delta_{M0}\sqrt{\frac{2L + 1}{4\pi}}\;,
\end{equation}
i.e. for $\theta = 0$ (representing the $\hat z$ direction) the spherical harmonics are non-vanishing only if $M = 0$.
\end{ex}

\hrulefill

Therefore, we can write:
\begin{equation}\label{ThetaTexpansionkparz2}
	f_\lambda(k\hat z, \hat{n}) = \frac{8\pi}{\sqrt{15}}Y^\lambda_2(\hat n)\sum_L i^L\sqrt{2L + 1}Y_L^0(\hat n)\int_0^{\eta_0}d\eta\;S^T(\eta, k)j_L(kr)\;.
\end{equation}
The idea is now to perform a rotation in order to put $\hat k$ in a generic direction. But then $\hat n$ also rotates, and therefore we need to know how spherical harmonics behave under rotations. In order to deal with just one spherical harmonic, we take advantage of the following decomposition:
\begin{eqnarray}\label{productYexpansion}
	Y^{\pm 2}_2(\hat n)Y_L^0(\hat n) = \sqrt{\frac{5(2L + 1)}{4\pi}}\sum_{L'}\sqrt{2L' + 1}\nonumber\\
	\left(\begin{array}{ccc}
		L & 2 & L'\\ 0 & \pm 2 & \mp 2 
	\end{array}\right)\left(\begin{array}{ccc}
		L & 2 & L'\\ 0 & 0 & 0
	\end{array}\right)Y_{L'}^{\pm 2}(\hat n)\;,
\end{eqnarray}
where we have employed the \textbf{Wigner 3$j$-symbols},\index{Wigner 3$j$-symbols} which are coefficients appearing in the quantum theory of angular momentum, when we combine two angular momenta and we want to write the state of total angular momentum as a linear combination on the basis of the tensor product of the two combined angular momenta. They are an alternative to the (perhaps more commonly used) Clebsch-Gordan coefficients See \cite{Landau:1991wop} and \cite{weinberg2015lectures}.

This expansion allows us to deal with just one spherical harmonic. Now we take advantage of the properties of the spherical harmonics under spatial rotation:
\begin{equation}
	\boxed{Y_\ell^m(R\hat n) = \sum_{m' = -\ell}^\ell D_{m'm}^{(\ell)}(R^{-1})Y_\ell^{m'}(\hat n)}
\end{equation}
where the $D_{m'm}^{(\ell)}$ are the elements of the \textbf{Wigner D-matrix}.\index{Wigner D-matrix} See \cite{Landau:1991wop} for more detail. The above $R$ is a generic rotation. Of course, we are interested in a $R(\hat k)$ rotation that brings $\hat k$ in a generic direction. Hence, we can write:
\begin{eqnarray}\label{ThetaTexpansionkrot}
	f_\lambda(\mathbf k, \hat{n}) = 4\pi\sum_L i^L\frac{2L + 1}{\sqrt{3\pi}}\sum_{L'}\sqrt{2L' + 1}
	\left(\begin{array}{ccc}
		L & 2 & L'\\ 0 & \lambda & -\lambda 
	\end{array}\right)\nonumber\\
	\left(\begin{array}{ccc}
		L & 2 & L'\\ 0 & 0 & 0
	\end{array}\right)
	\sum_{m'}D^{(L')}_{m'\lambda}[R(\hat k)]Y_{L'}^{m'}(\hat n)\int_0^{\eta_0}d\eta\;S^T(\eta, k)j_L(kr)\;.
\end{eqnarray}
Here we have dubbed $R\hat n$ the original line of sight and $\hat n$ the resulting one after the rotation.

Now we can perform the Fourier anti-transform. Let us multiply $f_\lambda(\mathbf k, \hat{n})$ by $\beta(\mathbf k, \lambda)$ and $Y_\ell^{m*}(\hat n)$ and integrate over $d^2\hat n$ in order to obtain the $a_{T,\ell m}^T$'s. We obtain:
\begin{eqnarray}\label{almTformula1}
	a_{\ell m,\pm 2}^T = 4\pi\sum_L i^L\frac{2L + 1}{\sqrt{3\pi}}\sqrt{2\ell + 1}
	\left(\begin{array}{ccc}
		L & 2 & \ell\\ 0 & \pm 2 & \mp 2 
	\end{array}\right)\nonumber\\
	\left(\begin{array}{ccc}
		L & 2 & \ell\\ 0 & 0 & 0
	\end{array}\right)\int\frac{d^3\mathbf k}{(2\pi)^3}D^{(\ell)}_{m\pm 2}[R(\hat k)]\beta(\mathbf k, \pm 2)\int_0^{\eta_0}d\eta\;S^T(\eta, k)j_L(kr)\;.
\end{eqnarray}
We have used the orthonormality relation of the spherical harmonics here and distinguished the contributions from different helicities. Of course $a_{\ell m} = a_{\ell m,+2} + a_{\ell m, -2}$.

It is now time to compute the $3j$ symbols and to perform the summation over $L$. A general formula for those was obtained in \cite{racah1942theory}, but we can read their expression from \cite{Landau:1991wop}. We then have the only following non-vanishing occurrences:
\begin{eqnarray}
	\left(\begin{array}{ccc}
		\ell & 2 & \ell\\ 0 & 0 & 0
	\end{array}\right) = (-1)^{\ell + 1}\sqrt{\frac{\ell(\ell + 1)}{(2\ell - 1)(2\ell + 1)(2\ell + 3)}}\;,\\
	\left(\begin{array}{ccc}
		\ell + 2 & 2 & \ell\\ 0 & 0 & 0
	\end{array}\right) = (-1)^{\ell}\sqrt{\frac{3(\ell + 1)(\ell + 2)}{2(2\ell + 1)(2\ell + 3)(2\ell + 5)}}\;,\\
	\left(\begin{array}{ccc}
		\ell - 2 & 2 & \ell\\ 0 & 0 & 0
	\end{array}\right) = (-1)^{\ell}\sqrt{\frac{3\ell(\ell - 1)}{2(2\ell - 3)(2\ell - 1)(2\ell + 1)}}\;.
\end{eqnarray}
In particular, there is no contribution coming from $L = \ell \pm 1$. The other three relevant (i.e., not considering those for $L = \ell \pm 1$, which are non-vanishing in this case) symbols are:
\begin{eqnarray}
	\left(\begin{array}{ccc}
		\ell & 2 & \ell\\ 0 & \pm 2 & \mp 2 
	\end{array}\right) = (-1)^{\ell}\sqrt{\frac{3(\ell - 1)(\ell + 2)}{2(2\ell - 1)(2\ell + 1)(2\ell + 3)}}\;,\\
	\left(\begin{array}{ccc}
		\ell + 2 & 2 & \ell\\ 0 & \pm 2 & \mp 2 
	\end{array}\right) = (-1)^{\ell}\frac{1}{2}\sqrt{\frac{(\ell - 1)\ell}{(2\ell + 1)(2\ell + 3)(2\ell + 5)}}\;,\\
	\left(\begin{array}{ccc}
		\ell - 2 & 2 & \ell\\ 0 & \pm 2 & \mp 2 
	\end{array}\right) = (-1)^{\ell}\frac{1}{2}\sqrt{\frac{(\ell + 1)(\ell + 2)}{(2\ell - 3)(2\ell - 1)(2\ell + 1)}}\;,
\end{eqnarray}

\hrulefill

\begin{ex}
Derive the above expressions for the relevant Wigner 3$j$ symbols given in \cite{Landau:1991wop} and put them in Eq.~\eqref{almTformula1}. Show that:
\begin{eqnarray}
	a_{T,\ell m,\pm 2}^T = -i^\ell\sqrt{\frac{2\pi(2\ell + 1)(\ell + 2)!}{(\ell - 2)!}}\int\frac{d^3\mathbf k}{(2\pi)^3}D^{(\ell)}_{m\pm 2}[R(\hat k)]\beta(\mathbf k,\pm 2)\int_0^{\eta_0}d\eta\;S^T(\eta, k)\nonumber\\
	\left[\frac{j_{\ell-2}(kr)}{(2\ell - 1)(2\ell + 1)} + \frac{2j_{\ell}(kr)}{(2\ell - 1)(2\ell + 3)} + \frac{j_{\ell+2}(kr)}{(2\ell + 1)(2\ell + 3)}\right]\;.\qquad
\end{eqnarray}
Recall that $r = r(\eta) \equiv \eta_0 - \eta$.
\end{ex}

\hrulefill

\begin{ex}
	Show that, using the recurrence relation \cite{Abramowitz1972}:
	\begin{equation}\label{recurrencereljl}
		\frac{j_\ell(x)}{x} = \frac{j_{\ell - 1}(x) + j_{\ell + 1}(x)}{2\ell + 1}\;,
	\end{equation}
	we can write: 
\begin{eqnarray}
		a_{T,\ell m}^T = -i^\ell\sqrt{\frac{2\pi(2\ell + 1)(\ell + 2)!}{(\ell - 2)!}}\sum_{\lambda = \pm 2}\int\frac{d^3\mathbf k}{(2\pi)^3}D_{m,\lambda}^{(\ell)}[R(\hat k)]\beta(\mathbf k,\lambda)\nonumber\\\int_0^{\eta_0}d\eta\;S^T(\eta, k)\frac{j_{\ell}(kr)}{(kr)^2}\;.
\end{eqnarray}
\end{ex}

\hrulefill

The Wigner $D$-matrix can be related to the spin-weighted spherical harmonics as follows:
\begin{equation}
	D_{m,\pm 2}^{(\ell)}(\hat k) = \sqrt{\frac{4\pi}{2\ell + 1}}{}_{\pm 2}Y_\ell^{-m}(\hat k) = \sqrt{\frac{4\pi}{2\ell + 1}}{}_{\mp 2}Y_\ell^{m*}(\hat k)\;,
\end{equation}
so we have:
\begin{equation}\label{aTTlmTfinalformula}
		\boxed{a_{T,\ell m}^T = -4\pi i^\ell\sqrt{\frac{(\ell + 2)!}{2(\ell - 2)!}}\sum_{\lambda = \pm 2}\int\frac{d^3\mathbf k}{(2\pi)^3}{}_{\lambda}Y_{\ell}^{m*}(\hat k)\beta(\mathbf k,\lambda)\int_0^{\eta_0}d\eta\;S^T(\eta, k)\frac{j_{\ell}(kr)}{(kr)^2}}
\end{equation}

This is the main result of this section. 

In order to compute the tensor contribution to the $C_{TT,\ell}$'s, we perform the ensemble average:
\begin{equation}
	\langle a_{T,\ell m}^Ta_{T,\ell'm'}^{T*} \rangle = C_{TT,\ell}^T\delta_{\ell\ell'}\delta_{mm'}\;.
\end{equation}

\hrulefill

\begin{ex}
	Assuming Gaussian perturbations, using Eq.~\eqref{tensorspectrum} and the orthogonality property of the Wigner D-matrices or the spin-weighted spherical harmonics:
	\begin{equation}
		\int d^2\hat k\;D_{m,\pm 2}^{(\ell)}[R(\hat k)]D_{m',\pm 2}^{(\ell')*}[R(\hat k)] = \frac{4\pi}{2\ell + 1}\delta_{\ell\ell'}\delta_{mm'}\;,
	\end{equation}
	show that:
	\begin{equation}\label{ClTformula}
		\boxed{C_{TT,\ell}^{T} = \frac{4\pi(\ell + 2)!}{(\ell - 2)!}\int_0^\infty\frac{dk}{k}\Delta_h^2(k)\left|\int_0^{\eta_0}d\eta\;S^T(\eta, k)\frac{j_{\ell}(kr)}{(kr)^2}\right|^2}
	\end{equation}
Note that a factor $2$ arises because of the two helicities states.	
	
The above result was originally obtained in \cite{Abbott:1984fp} (though not exactly in the same way and final form).\index{Cosmic Microwave Background!$C_{TT,\ell}^{T}$ spectrum}
\end{ex}

\hrulefill

The main difficulty we faced in computing the $a_{T,\ell m}^T$ was the spatial rotation, which brought $\hat k$ in a generic direction. This can be avoided if we calculate straightaway $C_{TT,\ell}^T$ because it is rotationally invariant. 
Note that no correlation exists between scalar and tensor modes. In fact, if we compute:
\begin{equation}
	\langle a_{T,\ell m}^Ta_{T,\ell'm'}^{S*} \rangle\;,
\end{equation}
we would get zero, mathematically, because of the integral:
\begin{equation}
		\int d^2\hat k\;{}_2Y_\ell^m(\hat k)Y_{\ell'}^{m'}(\hat k) = 0\;,
\end{equation}
between a spin-2 spherical harmonic and a spin-0 one. Physically, we know that at the linear order, scalar and tensor perturbations do not couple.

\begin{figure}[htbp]
\center
	\includegraphics[width=\columnwidth]{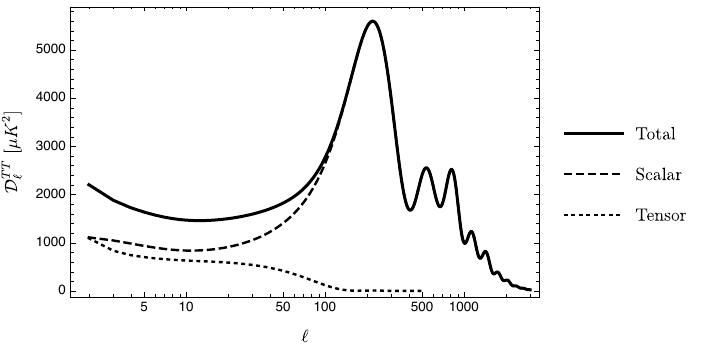}
\caption{Numerical calculations done with CLASS of the total (solid line), scalar (dashed line) and tensor (dotted line) angular power spectra}
\label{Fig:ScalarTensorClPlot}
\end{figure}

\begin{figure}[htbp]
\center
	\includegraphics[width=\columnwidth]{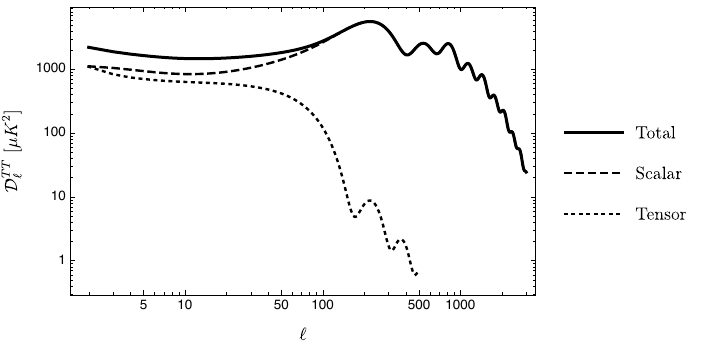}
\caption{Same as Fig.~\ref{Fig:ScalarTensorClPlot} but using a logarithmic scale.}
\label{Fig:ScalarTensorClPlotLogLog}
\end{figure}

We can again approximate this angular power spectrum for large values of $\ell$ as follows. First, $S^T(\eta, k)$ contains the derivative of $h$, which is maximum when a mode enters the horizon for $k\eta \approx 1$, being almost zero elsewhere. Therefore, assuming instantaneous recombination, we can write:
\begin{equation}
	C_{TT,\ell}^T = 4\pi(\ell - 1)\ell(\ell + 1)(\ell + 2)\int_0^\infty\frac{dk}{k}\Delta^2_h(k)\frac{j^2_\ell(k\eta_0)}{(k\eta_0)^4}\;.
\end{equation}
Defining the new variable $x \equiv k\eta_0$ and introducing the primordial tensor power spectrum, we get:
\begin{equation}
	C_{TT,\ell}^T \propto 4\pi(\ell - 1)\ell(\ell + 1)(\ell + 2)\int_0^\infty dx\;x^{n_T - 5}j^2_\ell(x)\;.
\end{equation}
The integral can be performed exactly:
\begin{equation}
	\int_0^\infty dx\;x^{n_T - 5}j^2_\ell(x) = \frac{\sqrt{\pi}}{2}\frac{\Gamma[1 - (n_T - 4)/2]\Gamma[(n_T - 4)/2 + \ell]}{(4 - n_T)\Gamma[1/2 - (n_T - 4)/2]\Gamma[\ell + 2 - (n_T - 4)/2]}\;,
\end{equation}
but in the case of $n_T = 0$, a scale-invariant primordial tensor spectrum, we get:
\begin{equation}
	\frac{\ell(\ell + 1)C_{TT,\ell}^T}{2\pi} \propto \frac{\ell(\ell + 1)}{(\ell - 2)(\ell + 3)}\;.
\end{equation}
The behavior of the tensor contribution to the TT power spectrum is thus very different from the one coming from scalar perturbations. In Fig.~\ref{Fig:ScalarTensorClPlot} and \ref{Fig:ScalarTensorClPlotLogLog} we display the numerical calculations done with CLASS of the total (solid line), scalar (dashed line), and tensor (dotted line) angular power spectra.

The tensor contribution is practically irrelevant on very small angular scales (large $\ell$), and on large angular scales, it can be as large as 10\% of the total. Typically, one can then give upper limits on $C_{TT,\ell}^T/C_{TT,\ell}^S$ for small multipoles ($\ell = 2$ or $\ell =10$), and this ratio is proportional to $A_T/A_S$ and therefore to the parameter $r$, the tensor-to-scalar ratio. From Eq.~\eqref{rconstraintPlanck}, we saw that $r < 0.1$. In order to determine this constraint, one must also use polarization data since with this we are able to disentangle the $A_S\exp(-2\tau_{\rm reion})$ dependence coming from the scalar contribution only to the temperature power spectrum.

\section{Polarization}\label{Sec:CMBpolarization}

In this section, we address CMB polarization. Recall that before recombination, polarization is also erased because of tight-coupling. Polarization is generated thanks to the fact that recombination does not take place instantaneously, so the finite-thickness effect is indeed important. Moreover, since Thomson scattering is axially-symmetric, circular polarization is not produced.

In Appendix~\ref{App:polarization} we recall the main terminology regarding polarization and, in particular, the Stokes parameters. 

\subsection{Scalar perturbations contribution to polarization}

Now, let us focus on scalar perturbations only and write down from Eq.~\eqref{BoltzeqphotonPol} the line of sight solution for the combination $Q + iU$. Since we have chosen a reference frame in which $\hat k = \hat z$, there is no $U$ polarization. This can also be seen from the fact that $\mathcal B^0 = 0$. Hence, we shall again perform a rotation in order to compute the $a_{P,\ell m}$.

We have called $\Theta_P$ the Stokes parameter $Q$ in the $\hat k = \hat z$ frame. So, let us work on its line-of-sight solution.

\hrulefill

\begin{ex}
	Show that:
	\begin{eqnarray}
	\Theta_P(k\hat z, \hat n) = \frac{3}{2}\sqrt{\frac{8\pi}{15}}{}_2Y_2^0({\hat n})\int_0^{\eta_0}d\eta\;e^{-i\mu kr}S_P^S(\eta, k)\;,
\end{eqnarray}
where we have defined a new source term:
\begin{equation}
	S_P^S(\eta, k) \equiv g(\eta)\Pi(\eta, k)\;.
\end{equation}
\end{ex}

\hrulefill

We could have written ${}_{-2}Y_2^0({\hat n})$ instead of ${}_2Y_2^0({\hat n})$, since they are equal. However, we are going to deal with $Q + iU$ first. The above equation can be written as:
\begin{eqnarray}
	\Theta_P(k\hat z, \hat n) = 6\pi\sqrt{\frac{2}{15}}{}_2Y_2^0(\hat p)\sum_Li^L\sqrt{2L + 1}Y_L^0(\hat n)\int_0^{\eta_0}d\eta\;S_P^S(\eta, k)j_L(kr)\;, \quad
\end{eqnarray}
where again $r \equiv \eta_0 - \eta$ and we have used the expansion of a plane wave into spherical harmonics plus the fact that $\hat k = \hat z$.

Now, as in Eq.~\eqref{productYexpansion}, we can write the product of spherical harmonics as follows:
\begin{eqnarray}\label{productYexpansionPol}
	{}_2Y^{0}_2(\hat n)Y_L^0(\hat n) = \sqrt{\frac{5(2L + 1)}{4\pi}}\sum_{L'}\sqrt{2L' + 1}\nonumber\\
	\left(\begin{array}{ccc}
		L & 2 & L'\\ 0 & - 2 & + 2 
	\end{array}\right)\left(\begin{array}{ccc}
		L & 2 & L'\\ 0 & 0 & 0
	\end{array}\right){}_2Y_{L'}^{0}(\hat n)\;,
\end{eqnarray}
and thus obtain:
\begin{eqnarray}\label{QplusiUscalexpansionkrot}
	(Q + iU)^S(\hat{n}) = \sqrt{6\pi}\sum_L i^L(2L + 1)\sum_{L'}\sqrt{2L' + 1}
	\left(\begin{array}{ccc}
		L & 2 & L'\\ 0 & -2 & 2 
	\end{array}\right)\nonumber\\
	\left(\begin{array}{ccc}
		L & 2 & L'\\ 0 & 0 & 0
	\end{array}\right)
	\sum_{m'}{}_2Y_{L'}^{m'}(\hat n)\int\frac{d^3\mathbf k}{(2\pi)^3}D^{(L')}_{m'0}(\hat k)\alpha(\mathbf k)\int_0^{\eta_0}d\eta\;S_P^S(\eta, k)j_L(kr)\;,\qquad
\end{eqnarray}
where we have already considered the rotation which brings $\hat k$ in a generic direction.

Now, from the expansion:
\begin{equation}
	(Q + iU)^S(\hat{n}) = \sum_{\ell m}a^S_{P,\ell m}\;{}_2Y_\ell^m(\hat n)\;,
\end{equation}
we are able to calculate the coefficients $a_{P,\ell m}^S$ by taking advantage of the orthonormality of the spin-2 spherical harmonics. We can therefore write:
\begin{eqnarray}
	a_{P,\ell m}^S = \sqrt{6\pi}\sum_L i^L(2L + 1)\sqrt{2\ell + 1}
	\left(\begin{array}{ccc}
		L & 2 & \ell\\ 0 & -2 & 2 
	\end{array}\right)\nonumber\\
	\left(\begin{array}{ccc}
		L & 2 & \ell\\ 0 & 0 & 0
	\end{array}\right)\int\frac{d^3\mathbf k}{(2\pi)^3}D^{(\ell)}_{m0}(\hat k)\alpha(\mathbf k)\int_0^{\eta_0}d\eta\;S_P^S(\eta, k)j_L(kr)\;.
\end{eqnarray}
Remarkably, the sum over $L$ can be performed in the very same way we did for the $a_{T,\ell m}^T$, since the $3j$ symbols are the same. Therefore, we have:
\begin{eqnarray}
	a_{P,\ell m}^S = -\frac{3i^\ell}{2}\sqrt{\frac{\pi(2\ell + 1)(\ell + 2)!}{(\ell - 2)!}}\int\frac{d^3\mathbf k}{(2\pi)^3}D^{(\ell)}_{m0}(\hat k)\alpha(\mathbf k)\nonumber\\
	\int_0^{\eta_0}d\eta\;S_P^S(\eta, k)\frac{j_\ell(kr)}{(kr)^2}\;,\qquad
\end{eqnarray}
and using
\begin{equation}
	D_{m0}^{(\ell)}(\hat k) = \sqrt{\frac{4\pi}{2\ell + 1}}Y_\ell^{-m}(\hat k) = \sqrt{\frac{4\pi}{2\ell + 1}}Y_\ell^{m*}(\hat k)\;,
\end{equation}
we can write:
\begin{eqnarray}\label{apLMscalar}
	\boxed{a_{P,\ell m}^S = -3\pi i^\ell\sqrt{\frac{(\ell + 2)!}{(\ell - 2)!}}\int\frac{d^3\mathbf k}{(2\pi)^3}Y_\ell^{m*}(\hat k)\alpha(\mathbf k)\int_0^{\eta_0}d\eta\;S_P^S(\eta, k)\frac{j_\ell(kr)}{(kr)^2}}
\end{eqnarray}
The expansion for $(Q - iU)^S(\hat n)$ can be obtained by complex conjugation:
\begin{eqnarray}
	(Q - iU)(\hat{n}) = \sum_{\ell m}a_{P,\ell m}^{*}\;{}_2Y_\ell^{m*}(\hat n) = \sum_{\ell m}a_{P,\ell m}^{*}\;{}_{-2}Y_\ell^{-m}(\hat n)\nonumber\\
	= \sum_{\ell m}a_{P,\ell,-m}^{*}\;{}_{-2}Y_\ell^{m}(\hat n)\;.
\end{eqnarray}
There is no reality condition here holding true for the $a_{P,\ell m}$ as the one holding true for the $a_{T,\ell m}$, because $Q + iU$ is not real and is not a scalar. It is thus convenient to define the following combinations:
\begin{equation}
	a_{E,\ell m} \equiv -(a_{P,\ell m} + a^*_{P,\ell, -m})/2\;, \quad a_{B,\ell m} \equiv i(a_{P,\ell m} - a^*_{P,\ell, -m})/2\;,
\end{equation}
because the first has parity $(-1)^\ell$, whereas the second $(-1)^{\ell + 1}$. Thus $Q\pm iU$ can be expanded as:
\begin{eqnarray}
	(Q \pm iU)(\hat{n}) = \sum_{\ell m}(-a_{E,\ell m} \mp ia_{B,\ell m})\;{}_2Y_\ell^{m}(\hat n)\;.
\end{eqnarray}
Now, if we compute $a_{P,\ell m}^{S*}$, we obtain:
\begin{eqnarray}
	a_{P,\ell m}^{S*} = -3\pi(-i)^\ell\sqrt{\frac{(\ell + 2)!}{(\ell - 2)!}}\int\frac{d^3\mathbf k}{(2\pi)^3}Y_\ell^{-m*}(\hat k)\alpha(-\mathbf k)\nonumber\\
	\int_0^{\eta_0}d\eta\;S_P^S(\eta, k)\frac{j_\ell(kr)}{(kr)^2}\;,
\end{eqnarray}
since $\alpha(\mathbf k)^* = \alpha(-\mathbf k)$ because of the reality condition of the power spectrum. Changing the integration variable to $\mathbf k$ and using the parity of the spherical harmonic:
\begin{equation}
	Y_\ell^{-m*}(-\hat k) = (-1)^\ell Y_\ell^{-m*}(\hat k)\;,
\end{equation}
we can finally conclude that:
\begin{equation}\label{aPlmScomplconjcond}
	a_{P,\ell m}^{S} = a_{P,\ell, -m}^{S*}\;,
\end{equation}
and therefore scalar perturbations only affect the $E$-mode, i.e.
\begin{equation}
	a^S_{E,\ell m} = -a^S_{P,\ell m}\;, \qquad a^S_{B,\ell m} = 0\;.
\end{equation}
This means that, if the $B$-mode was detected, it would be a clear indication of the existence of primordial gravitational waves.

From Eq.~\eqref{apLMscalar}, we can then obtain the scalar contribution to the EE spectrum. Assuming adiabatic Gaussian perturbations:\index{Cosmic Microwave Background!$C_{TE,\ell}^{S}$ spectrum}\index{Cosmic Microwave Background!$C_{EE,\ell}^{S}$ spectrum}
\begin{eqnarray}\label{ClEEscalar}
	\boxed{C_{EE,\ell}^S = \frac{9\pi}{4}\frac{(\ell + 2)!}{(\ell - 2)!}\int\frac{dk}{k}\Delta^2_{\mathcal R}\left|\int_0^{\eta_0}d\eta\;S_P^S(\eta, k)\frac{j_\ell(kr)}{(kr)^2}\right|^2}
\end{eqnarray}
Using Eq.~\eqref{aTlmscalar} instead, we can compute the cross-correlation TE multipole coefficients:
\begin{eqnarray}\label{ClTEscalar}
	\boxed{C_{TE,\ell}^S = -3\pi\sqrt{\frac{(\ell + 2)!}{(\ell - 2)!}}\int\frac{dk}{k}\Delta^2_{\mathcal R}\Theta_\ell(k)\int_0^{\eta_0}d\eta\;S_P^S(\eta, k)\frac{j_\ell(kr)}{(kr)^2}}
\end{eqnarray}

\subsection{Tensor perturbations contribution to polarization}

Let us now calculate the contribution to CMB polarization coming from tensor perturbations. From Eq.~\eqref{lineofsightintegralThetaPtens} we have
\begin{equation}
	\Theta_{P}^{(T)}(\eta_0, k\hat z, \mu) = \int_{0}^{\eta_0}d\eta\;e^{ik\mu(\eta - \eta_0)}S_{P}^{T}(\eta, k)\;,
\end{equation}
with
\begin{eqnarray}
	S_{P}^{T}(\eta, k) \equiv g(\eta)\mathcal S^{T}(\eta, k)\;.
\end{eqnarray}
and then use Eq.~\eqref{ThetaPTexpThetalambda} to write part of the tensor contribution to polarization:
\begin{equation}\label{ThetaPTexpansionkparz}
	\mathcal Q_\lambda^{(T)}(k\hat z, \hat{p}) \equiv \sqrt{\frac{8\pi}{5}}\mathcal E^\lambda(\hat p)\int_0^{\eta_0}d\eta\;e^{-i\mu kr(\eta)}S_P^{T}(\eta, k)\;,
\end{equation}
where $r(\eta) \equiv \eta_0 - \eta$ and where we stress that the result holds true for $\hat k = \hat z$, since this was the condition under which we derived the Boltzmann equation for photons.

In the scalar case, no $U$ contribution to polarization is produced, so the above expression already furnishes the quantity $Q + iU$. However, the same is not true for tensor perturbation. We thus have to add the $iU$ contribution. As we saw in chapter~\ref{Chap:PertubedBoltzmannEquations}, this is equal to $\mathcal B^mQ/\mathcal E^m$, and for this reason, we have just one polarization hierarchy. Hence, summing up, we get:
\begin{equation}\label{QpiUtenskz}
	(\mathcal Q_\lambda + i\mathcal U_\lambda)^{(T)}(k\hat z, \hat{n}) = \sqrt{\frac{32\pi}{5}}{}_2Y_2^\lambda(\hat n)\int_0^{\eta_0}d\eta\;e^{i\hat k\cdot\hat n kr(\eta)}S_P^{T}(\eta, k)\;.
\end{equation}
Using the usual plane-wave expansion and recalling that $\hat k = \hat z$, we get:
\begin{eqnarray}\label{QpiUtenskz2}
	(\mathcal Q_\lambda + i\mathcal U_\lambda)^{(T)}(k\hat z, \hat{n}) = 4\pi\sqrt{\frac{8}{5}}{}_2Y^\lambda_2(\hat n)\sum_L i^L\sqrt{2L + 1}Y_L^0(\hat n)\nonumber\\
	\int_0^{\eta_0}d\eta\;S_P^{T}(\eta, k)j_L(kr)\;.
\end{eqnarray}
The product of the two spherical harmonics can be written via the Wigner $3j$-symbols as follows:
\begin{eqnarray}
	{}_2Y^{\pm 2}_2(\hat n)Y_L^0(\hat n) = \sqrt{\frac{5(2L + 1)}{4\pi}}\sum_{L'}\sqrt{2L' + 1}\nonumber\\
	\left(\begin{array}{ccc}
		L & 2 & L'\\ 0 & \pm 2 & \mp 2 
	\end{array}\right)\left(\begin{array}{ccc}
		L & 2 & L'\\ 0 & -2 & 2
	\end{array}\right){}_2Y_{L'}^{\pm 2}(\hat n)\;,
\end{eqnarray}
and we rotate in a generic $\hat k$ direction the only spherical harmonic left, i.e. 
\begin{equation}
	{}_2Y_{L'}^{\pm 2}(R\hat n) = \sum_{m' = -L'}^{L'} D_{m',\pm 2}^{(L')}[R^{-1}(\hat k)]{}_2Y_{L'}^{m'}(\hat n)\;.
\end{equation}
Hence, we can write:
\begin{eqnarray}\label{QpiUtenskgeneric}
	(\mathcal Q_\lambda + i\mathcal U_\lambda)^{(T)}(\mathbf k, \hat{n}) = 4\sqrt{2\pi}\sum_L i^L(2L + 1)\sum_{L'}\sqrt{2L' + 1}
	\left(\begin{array}{ccc}
		L & 2 & L'\\ 0 & \lambda & -\lambda 
	\end{array}\right)\nonumber\\
	\left(\begin{array}{ccc}
		L & 2 & L'\\ 0 & -2 & 2
	\end{array}\right)
	\sum_{m'}D^{(L')}_{m'\lambda}[R(\hat k)]Y_{L'}^{m'}(\hat n)\int_0^{\eta_0}d\eta\;S_P^{T}(\eta, k)j_L(kr)\;. \qquad
\end{eqnarray}
Now we can perform the Fourier anti-transform. Multiply by $\beta(\mathbf k,\lambda)$ and ${}_2Y_\ell^{m*}(\hat n)$ and integrate over $d^2\hat n$ in order to obtain the $a_{P,\ell m}^{T}$'s. We obtain:
\begin{eqnarray}\label{aPlmTformula1}
	a_{P,\ell m,\pm 2}^{T} = 4\sqrt{2\pi}\sum_L i^L(2L + 1)\sqrt{2\ell + 1}
	\left(\begin{array}{ccc}
		L & 2 & \ell\\ 0 & \pm 2 & \mp 2 
	\end{array}\right)\nonumber\\
	\left(\begin{array}{ccc}
		L & 2 & \ell\\ 0 & -2 & 2
	\end{array}\right)\int\frac{d^3\mathbf k}{(2\pi)^3}D^{(\ell)}_{m\pm 2}[R(\hat k)]\beta(\mathbf k,\lambda)\int_0^{\eta_0}d\eta\;S_P^{T}(\eta, k)j_L(kr)\;.
\end{eqnarray}
We have used the orthonormality relation of the spin-$2$ spherical harmonics and distinguished the contributions of different helicities. Of course $a_{P,\ell m}^T = a_{P,\ell m,+2}^T + a_{P,\ell m, -2}^T$.

We have already computed some of the $3j$ symbols earlier for the tensor case, but now two more enter: those for $L = \ell \pm 1$. We shall see that these contributions will characterize the $B$-mode of polarization. They are: 
\begin{eqnarray}
	\left(\begin{array}{ccc}
		\ell + 1 & \ell & 2\\ 0 & - 2 & + 2 
	\end{array}\right) = (-1)^{\ell + 1}\sqrt{\frac{(\ell - 1)}{2(2\ell + 1)(2\ell + 3)}}\;,\\
	\left(\begin{array}{ccc}
		\ell & \ell - 1 & 2\\ -2 & 0 & + 2 
	\end{array}\right) = (-1)^{\ell}\sqrt{\frac{(\ell + 2)}{2(2\ell - 1)(2\ell + 1)}}\;.
\end{eqnarray}
Extra care has to be taken when manipulating these terms. The reason is that the 3$j$ symbols gain an overall phase factor 
\begin{equation}
	(-1)^{j_1 + j_2 + j_3}\;,
\end{equation}
where $j_{1,2,3}$ are the momenta that are being combined; each time we swap two columns or change simultaneously all the signs of the bottom row.\footnote{These signs can be changed only simultaneously since the sums $m_1 + m_2 + m_3 = 0$ are always the same. This is a selection rule.} Therefore, as long as $j_1 + j_2 + j_3$ is even, no matter how many times we perform the above operations. This is the case for $L = \ell \pm 2$ or $L = \ell$. However, for $L = \ell \pm 1$, we have that
\begin{equation}
	L + \ell + 2 = 2(\ell + 1) \pm 1\;,
\end{equation}
which is odd, and thus we have to keep track of the correct sign.

\hrulefill

\begin{ex}
Using the formulas for the $3j$ symbols, show that:
\begin{eqnarray}
	a_{P,\ell m,\pm 2}^{T} = \frac{-4\pi i^\ell\sqrt{2\ell + 1}}{\sqrt{8\pi}}\int\frac{d^3\mathbf k}{(2\pi)^3}D^{(\ell)}_{m\pm 2}[R(\hat k)]\beta(\mathbf k,\lambda)\int_0^{\eta_0}d\eta\;S_P^{T}(\eta, k)\nonumber\\
	\left[\frac{(\ell + 1)(\ell + 2)}{(2\ell - 1)(2\ell + 1)}j_{\ell-2}(kr) - \frac{6(\ell - 1)(\ell + 2)}{(2\ell - 1)(2\ell + 3)}j_{\ell}(kr) + \frac{(\ell - 1)\ell}{(2\ell + 1)(2\ell + 3)}j_{\ell+2}(kr)\right.\nonumber\\
	\left.\pm 2i\frac{\ell - 1}{2\ell + 1}j_{\ell + 1}(kr) \mp 2i\frac{\ell + 2}{2\ell + 1}j_{\ell - 1}(kr)\right]\;. \qquad
\end{eqnarray}
Recall that $r = r(\eta) \equiv \eta_0 - \eta$.
\end{ex}

\hrulefill

\begin{ex}
	Show that, using the recurrence relation of Eq.~\eqref{recurrencereljl} and the following ones for the derivatives:
	\begin{equation}
		j_\ell'(x) = j_{\ell - 1}(x) - \frac{\ell + 1}{x}j_{\ell}(x)\;, \quad j_\ell'(x) = \frac{\ell}{x}j_{\ell}(x) - j_{\ell + 1}(x)\;,
	\end{equation}
	we can write: 
\begin{eqnarray}
		a_{P,\ell m}^{T} = \frac{-4\pi i^\ell\sqrt{2\ell + 1}}{\sqrt{8\pi}}\sum_{\lambda = \pm 2}\int\frac{d^3\mathbf k}{(2\pi)^3}D_{m,\lambda}^{(\ell)}[R(\hat k)]\beta(\mathbf k,\lambda)\int_0^{\eta_0}d\eta\;S_P^{T}(\eta, k)\nonumber\\
		\left[\frac{2}{kr}j_\ell' - 2j_\ell + \frac{2 + \ell(\ell + 1)}{(kr)^2}j_\ell - i\lambda\left(j_\ell' + \frac{2}{kr}j_\ell\right)\right]\;.
\end{eqnarray}
Show that the term between square brackets is equal to the corresponding one in \cite[pag. 389]{Weinberg:2008zzc}. In order to make the second derivative of the spherical Bessel function to appear one must use Bessel differential equation:
\begin{equation}
	j_\ell'' + \frac{2}{x}j_\ell + \frac{1 - \ell(\ell + 1)}{x^2}j_\ell = 0\;.
\end{equation} 
\end{ex}

\hrulefill

Now we are ready to investigate the reality property of $a_{P,\ell m}^{T}$ and discern from it the $E$-mode and $B$-mode contributions. Recall that the Wigner $D$-matrix can be related to the spin-weighted spherical harmonics as follows:
\begin{equation}
	D_{m,\pm 2}^{(\ell)}(\hat k) = \sqrt{\frac{4\pi}{2\ell + 1}}{}_{\pm 2}Y_\ell^{-m}(\hat k) = \sqrt{\frac{4\pi}{2\ell + 1}}{}_{\mp 2}Y_\ell^{m*}(\hat k)\;.
\end{equation}
So, taking the complex conjugate, we find:
\begin{eqnarray}
		a_{P,\ell m}^{T*} = -2\sqrt{2}\pi(-i)^\ell\sum_{\lambda = \pm 2}\int\frac{d^3\mathbf k}{(2\pi)^3}{}_{\mp 2}Y_\ell^{m}(\hat k)\beta(-\mathbf k,\lambda)\int_0^{\eta_0}d\eta\;S_P^T(\eta, k)\nonumber\\
		\left[\frac{2}{kr}j_\ell' - 2j_\ell + \frac{2 + \ell(\ell + 1)}{(kr)^2}j_\ell + i\lambda\left(j_\ell' + \frac{2}{kr}j_\ell\right)\right]\;.
\end{eqnarray}
Note that $\beta(\mathbf k, \lambda)^* = \beta(-\mathbf k, \lambda)$ because of the reality condition and beware that the sign of the imaginary unit inside the square brackets has changed. Now, changing integration variable
\begin{equation}
	\mathbf k \to -\mathbf k\;,
\end{equation}
and taking advantage of the parity property and the complex conjugation property:
\begin{equation}
	{}_{\mp 2}Y_\ell^{m}(-\hat k) = (-1)^\ell{}_{\pm 2}Y_\ell^{m}(\hat k) = (-1)^\ell{}_{\mp 2}Y_\ell^{-m*}(\hat k)\;,
\end{equation}
we get:
\begin{eqnarray}
		a_{P,\ell m}^{T*} = -2\sqrt{2}\pi i^\ell\sum_{\lambda = \pm 2}\int\frac{d^3\mathbf k}{(2\pi)^3}{}_{\mp 2}Y_\ell^{-m*}(\hat k)\beta(\mathbf k,\lambda)\int_0^{\eta_0}d\eta\;S_P^T(\eta, k)\nonumber\\
		\left[\frac{2}{kr}j_\ell' - 2j_\ell + \frac{2 + \ell(\ell + 1)}{(kr)^2}j_\ell + i\lambda\left(j_\ell' + \frac{2}{kr}j_\ell\right)\right]\;.
\end{eqnarray}
This time we do not have the same situation as in Eq.~\eqref{aPlmScomplconjcond} because of the $i\lambda$ contribution. Hence, we can compute the $E$-mode:
\begin{eqnarray}
		a_{E,\ell m}^{T} = 2\sqrt{2}\pi i^\ell\sum_{\lambda = \pm 2}\int\frac{d^3\mathbf k}{(2\pi)^3}{}_{\mp 2}Y_\ell^{m*}(\hat k)\beta(\mathbf k,\lambda)\int_0^{\eta_0}d\eta\;S_P^T(\eta, k)\nonumber\\
		\left[\frac{2}{kr}j_\ell' - 2j_\ell + \frac{2 + \ell(\ell + 1)}{(kr)^2}j_\ell\right]\;.
\end{eqnarray}
and the $B$-mode is also present:
\begin{eqnarray}\label{aBlmTfinalformula}
		\boxed{a_{B,\ell m}^{T} = -2\sqrt{2}\pi i^\ell\sum_{\lambda = \pm 2}\lambda\int\frac{d^3\mathbf k}{(2\pi)^3}{}_{\mp 2}Y_\ell^{m*}(\hat k)\beta(\mathbf k,\lambda)\int_0^{\eta_0}d\eta\;S_P^T(\eta, k)\left(j_\ell' + \frac{2}{kr}j_\ell\right)}\qquad
\end{eqnarray}
Now we are in a position to give the formulas for the angular power spectra. Assuming Gaussian perturbations, we have:\index{Cosmic Microwave Background!$C_{EE,\ell}^{T}$ spectrum}
\begin{eqnarray}
		C_{EE,\ell}^{T} = 4\pi\int\frac{dk}{k}\Delta^2_h(k)\left|\int_0^{\eta_0}d\eta\;S_P^{T}(\eta, k)\left[\frac{2}{kr}j_\ell' - 2j_\ell + \frac{2 + \ell(\ell + 1)}{(kr)^2}j_\ell\right]\right|^2\;,\qquad
\end{eqnarray}
and\index{Cosmic Microwave Background!$C_{BB,\ell}^{T}$ spectrum}
\begin{eqnarray}
		C_{BB,\ell}^{T} = 4\pi\int\frac{dk}{k}\Delta_h^2\left|\int_0^{\eta_0}d\eta\;S_P^T(\eta, k)\left(2j_\ell' + \frac{4}{kr}j_\ell\right)\right|^2\;.\quad
\end{eqnarray}
The cross-correlation $C^T_{TE,\ell}$, using Eq.~\eqref{aTTlmTfinalformula} gives:\index{Cosmic Microwave Background!$C_{TE,\ell}^{T}$ spectrum}
\begin{eqnarray}
		C_{TE,\ell}^{T} = -2\pi\sqrt{\frac{(\ell + 2)!}{(\ell - 2)!}}\int\frac{dk}{k}\Delta^2_h(k)\int_0^{\eta_0}d\eta\;S^{T}(\eta, k)\frac{j_\ell}{(kr)^2}\nonumber\\
		\int_0^{\eta_0}d\eta'S_P^{T}(\eta', k)\left[\frac{2}{kr}j_\ell' - 2j_\ell + \frac{2 + \ell(\ell + 1)}{(kr)^2}j_\ell\right]\;.\qquad
\end{eqnarray}
If we try to compute the cross correlations $C_{TB,\ell}^T$ and $C_{EB,\ell}^T$, we obtain a vanishing result, as expected, because of the term $\lambda$ in the sum of Eq.~\eqref{aBlmTfinalformula}. In fact, we get, considering for example $C_{EB,\ell}^T$:
\begin{eqnarray}
		C_{EB,\ell}^{T} = -2\pi\sum_{\lambda = \pm 2}\lambda\int\frac{dk}{k}\Delta_h^2\int_0^{\eta_0}d\eta\;S_P^T(\eta, k)\left[\frac{2}{kr}j_\ell' - 2j_\ell + \frac{2 + \ell(\ell + 1)}{(kr)^2}j_\ell\right]\nonumber\\
		\int_0^{\eta_0}d\bar\eta\;S_P^T(\bar\eta, k)\left(2j_\ell' + \frac{4}{kr}j_\ell\right)\;.\qquad
\end{eqnarray}
Now, the sum over $\lambda$ is equivalent to a difference, and since nothing else depends on $\lambda$, the result is zero. The same happens with the correlation $C_{TB,\ell}^T$.

\begin{figure}[htbp]
\center
	\includegraphics[width=0.5\columnwidth]{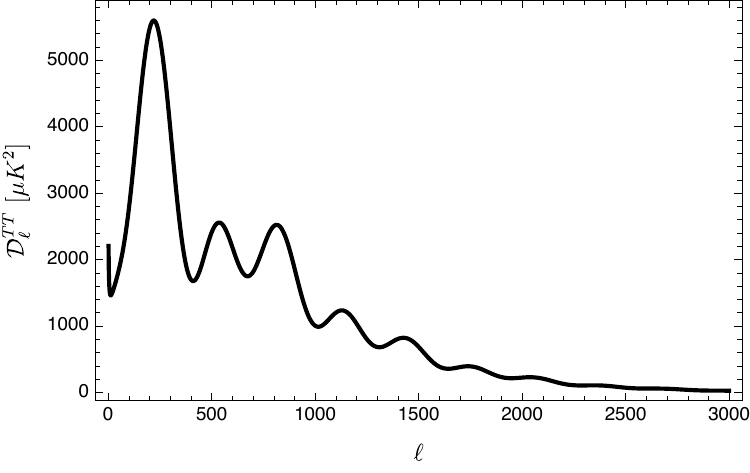}\includegraphics[width=0.5\columnwidth]{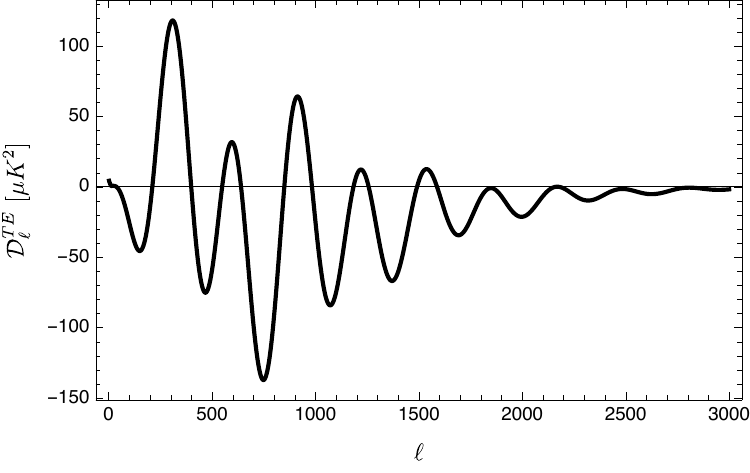}\\
	\includegraphics[width=0.5\columnwidth]{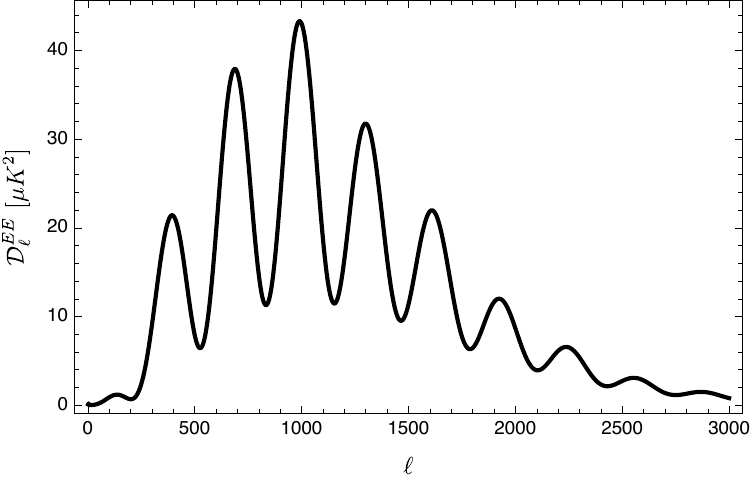}\includegraphics[width=0.5\columnwidth]{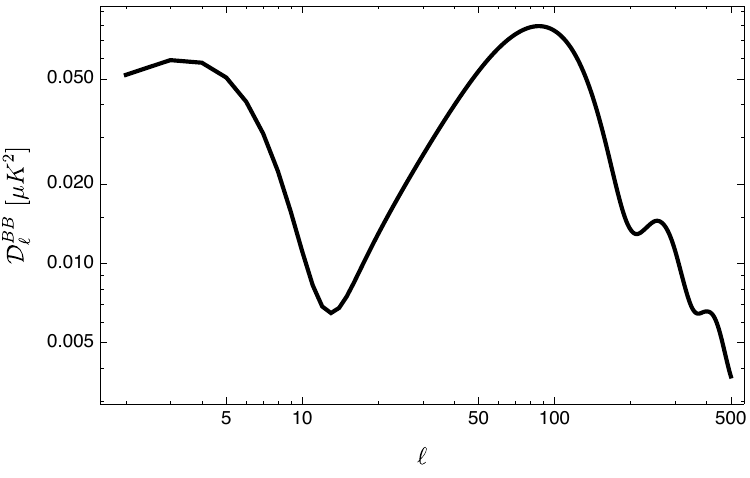}
\caption{The four angular power spectra characterising the CMB, computed with CLASS for the standard model. Top left: TT. Top right: TE. Bottom left: EE. Bottom right: BB. As for the TT spectrum, $\mathcal D \equiv \ell(\ell + 1)C_\ell/(2\pi)$.}
\label{Fig:CMBangularpowerspectra}
\end{figure}

In Fig.~\ref{Fig:CMBangularpowerspectra} we display the 4 angular CMB power spectra that constitute a wealth of cosmological information. The only possible cross-correlation is the one between temperature and the E-mode of polarization. The largest signal is the TT one, followed by the TE, EE, and BB ones in order of decreasing power. The latter is 5 orders of magnitude smaller than the TT one, and it has not yet been detected. The bump it displays for small $\ell$'s is due to reionization.

From Fig.~\ref{Fig:CMBangularpowerspectra} we can appreciate how small the polarization spectra are with respect to the TT one. This is due to the fact that a quadrupole moment in the distribution of photons is needed in order to have production of polarization. Before recombination, the Thomson scattering rate is so high that photons are in nearly perfect thermal equilibrium, and any moment from the quadrupole up is washed out. After recombination, photons free stream and thus have no more chance of being polarized by Thomson scattering.

\clearpage
\chapter{Miscellanea}\label{Chap:Miscellanea}

{\rightskip=3truepc\leftskip=3truepc\noindent
{\it Like all people who try to exhaust a subject, he exhausted his listeners}
\vskip 0.10 in
\centerline{\it ---Oscar Wilde, The picture of Dorian Gray}
\vskip 0.20 in
}


In this chapter, we collect some extra material that extends the topics treated so far.

\section{Magnitudes}
 
The \textbf{apparent magnitude}\index{Magnitude! Apparent} $m$ is a measure of how much energy we receive per unit of time from a certain source. It depends on the intrinsic luminosity of the source, its distance, and how much light is lost, for example, scattered or absorbed along the line of sight (\textbf{extinction}). 

Suppose that we measure a certain flux $F(X)$ in a passband $X$. The apparent magnitude is defined as:
\begin{equation}
	m(X) = -2.5\log_{10}\left[\frac{F(X)}{F_0(X)}\right]\,,
\end{equation}
where $F_0(X)$ is a reference flux (for example, that of the star Vega).

The \textbf{absolute magnitude}\index{Magnitude! Absolute} $M$ is a measure of the intrinsic luminosity of a source, defined as the apparent magnitude of the source if it were placed at 10 pc from the observer without extinction. Using the \textbf{inverse-square law}, the luminosity $L$ of a source can be related to its flux as follows:
\begin{equation}
	F = \frac{L}{4\pi d^2}\,,
\end{equation}
where $d$ is the distance to the source. So, forgetting the passband dependence, one has:
\begin{equation}
	M = -2.5\log_{10}\left[\frac{F}{F_0}\left(\frac{d}{10\mbox{ pc}}\right)^2\right] = m - 5\log_{10}\left(\frac{d}{10\mbox{ pc}}\right)\,.
\end{equation}
The \textbf{distance modulus}\index{Distance modulus} is a way of expressing $d$, as follows:
\begin{equation}\label{distancemodulus}
	\mu \equiv m - M = 5\log_{10}\left(\frac{d}{10\mbox{ pc}}\right)\,.
\end{equation}
If the flux is integrated over all wavelengths, the magnitude is called \textbf{bolometric}.\index{Magnitude! Bolometric}

Relative motions would invalidate the comparison of the magnitudes of different sources in a given passband and the determination of the distance modulus of a given source.\footnote{ This is because the intrinsic luminosity is calculated in the rest frame of the source.} This is compensated for by a term called the \textbf{$K$ correction} \cite{Hogg:2002yh}:\index{Magnitude! $K$ correction}
\begin{equation}
	m = M + \mu + K\,.
\end{equation}

\section{The cosmic distance ladder}\label{Sec:cosmicdistladder}

The cosmic distance ladder\index{Distance ladder} is a sequence of measuring techniques that allows astronomers to determine distances up to cosmological scales. It is a ``ladder'' because different techniques work only in some ranges of distances, and these overlap, allowing calibration. In this way, step by step, we can reach large distances. 
\\\\
The first step is the \textbf{parallax},\index{Parallax} which is a direct method to determine distances by trigonometry (it is the same as the \textit{triangulation} used on Earth). Stars that are not too far from us will be seen to move with respect to the background of ``fixed'' stars. This motion is apparent because it is due to the revolution of the Earth around the Sun (technically, this motion is called the \textit{annual apparent parallax motion}). 


Imagine a triangle whose vertices are the Earth, the Sun, and the star whose distance we want to determine. Take the Earth-Sun distance as the basis for this triangle, and let $\theta$ be the opposite angle. If $\theta \ll 1$, the distance to the star is:
\begin{equation}
	d_{\rm parallax} = \frac{1\mbox{ AU}}{\theta}\,,
\end{equation}
where 1 AU $= 1.5\times 10^{11}$ m is the \textbf{astronomical unit},\index{Astronomical unit} the Earth-Sun distance. If $\theta = 1$ arcsecond, equivalent in radians to: 
\begin{equation}
	\theta = \frac{\pi}{6.48\times 10^5}\,,
\end{equation}
then the corresponding parallax distance is:
\begin{equation}
	d_{\rm parallax} = \frac{6.48\times 10^5\mbox{ AU}}{\pi} \approx 3.09\times 10^{16}\mbox{ m} \equiv 1 \mbox{ pc}\,,
\end{equation}
which defines the unit of length called \textbf{parsec}.\index{Parsec} The \textit{Gaia} spacecraft of the \textit{European Space Agency} has been able to measure the distances of stars up to 100 pc.

The parallax method is very direct and reliable, but it works only for nearby stars. Our Milky Way is about 30 kpc in diameter, so via parallax, we cannot even measure the distances to most of the stars within our own galaxy.

In order to measure larger distances, we need to rely on indirect methods in which the distance is inferred by measuring something else. These methods are based on \textbf{standard candles},\index{Standard candles} which are sources whose luminosity is known and whose distance can therefore be obtained by measuring their flux. Indeed, by the inverse square law, one has:
\begin{equation}
	F = \frac{L}{4\pi d^2}\,,
\end{equation}
with $F$ flux, $L$ luminosity, and $d$ distance. Therefore:
\begin{equation}\label{distancefluxluminosity}
	d = \sqrt{\frac{L}{4\pi F}}\,.
\end{equation}
The difficult part is identifying the sources that are standard candles. Two of them are typically used in building the distance ladder: \textbf{Cepheid variables and type Ia supernovae}.\index{Cepheid variables}\index{Supernovae, type Ia} 
\\\\
Cepheid variables are pulsating stars whose periods of pulsation are related to their luminosity. This period-luminosity relationship can be calibrated if we determine the distance of a Cepheid variable through parallax. Then, we can use this relationship to determine the luminosity of Cepheids that are too far for their distances to be determined via parallax: Once we know their luminosity, we can establish their distance by measuring their flux and using the above formula \eqref{distancefluxluminosity}. This is how the cosmic distance ladder works.


A type Ia supernova is a kind of supernova explosion that takes place in a binary star system in which one of the two stars is a white dwarf accreting material from the companion until it exceeds the Chandrasekhar limit. What makes type Ia supernovae standard candles is their peculiar light curves. In fact, the faster the light curve decays from its peak emission, the fainter the absolute magnitude at the peak. This behavior can be quantified in a relationship called \textbf{Phillips's relation}\index{Phillips's relation} \cite{Phillips1993ApJ}:
\begin{equation}\label{Phillipsrel}
	\boxed{M_{\rm max} = a + b\,\Delta m_{15}(B)}
\end{equation}
where $M_{\rm max}$ is the absolute magnitude at the peak, $a$ and $b$ are parameters to be fitted (they depend on the passband used but are the same for all supernovae; this being the crucial point), and $\Delta m_{15}(B)$ is the variation of the apparent magnitude in the $B$ band 15 days after the peak. Phillips's relation must be calibrated first; that is, we must know beforehand the values of $M_{\rm max}$ and $\Delta m_{15}(B)$ for many type Ia supernovae in order to find $a$ and $b$. This means that we must already know their distances in order to determine $M_{\rm max}$. In Ref.~\cite{Phillips1993ApJ}, Tully-Fisher methods are used, but one can also use the very same Cepheid variables \cite{Riess:2021jrx} that we have discussed earlier (one needs to be sure that the Cepheid variable and the type Ia supernova are in the same galaxy, thus at the same distance). Thus, we take another step on the cosmic distance ladder, and since type Ia supernovae are very bright, we are able to reach very far (hundreds of Mpc, in fact) in the realm of cosmology.


The last step of the ladder is the Hubble-Lemaître law, which we discuss in detail in Chapter \ref{Chap:Cosmology}.

\section{The redshift}

At some point, astronomers look so deep into the cosmos that they do not bother anymore to express distances in Mpc but rather use Hubble's law and the cosmological redshift.

The redshift\index{Redshift} is a fundamental observable in cosmology. Its definition is the usual one, from spectroscopy:
\begin{equation}\label{redshiftdef}
	\boxed{z = \frac{\lambda_{\rm obs}}{\lambda_{\rm em}} -1}
\end{equation}
The remarkable fact is that, in cosmology, it is always positive; i.e., observed radiation is always redder than the emitted one. The reason for this is \textbf{the expansion of the universe}. 

Note that the expansion of the universe is not evidence that we can find on distance scales that are too small, such as within our galaxy or in our Solar System. Even within our local group of galaxies, Andromeda displays a negative redshift (that is, a blueshift). The reason is that local gravitational fields are sufficiently strong to mask the effect of the recession of the sources (also known as \textbf{the Hubble flow}).\index{Hubble flow} 

For the moment, we can think of the redshift as the Doppler effect due to the relative motion of the sources. In Chapter \ref{Chap:ExpandingUniverse} we describe the expansion of the universe in GR as a geometric effect.  

The redshift is measured in two ways: spectroscopically or photometrically. For the former, one needs to detect known emission or absorption lines from a source and compare their wavelengths with the ones measured in a laboratory on Earth. In this case, one uses directly Eq.~\eqref{redshiftdef} and thus calculates $z$.\index{Redshift! Spectroscopic}

Photometric redshifts are calculated by assuming certain spectral features of the sources and measuring their relative brightness in certain wavebands using filters.\index{Redshift! Photometric} 

A simple example is the following. The spectrum of the Sun is approximately a blackbody one with a temperature of about 6000 K, and by using Wien's displacement law, it has a peak emission at a wavelength of 500 nm. Therefore, if a star similar to the Sun had a peak emission of, say, 600 nm, then using Eq.~\eqref{redshiftdef} one would calculate $z = 0.2$.

The reason for using photometry instead of spectroscopy is that it is less time-consuming and allows for obtaining redshifts of very distant sources, for which it is difficult to do spectroscopy. However, photometric redshifts are less precise.

\section{Bayesian analysis using type Ia supernovae data}\label{Sec:BayesiananalysisSNIa}

In this section, it is shown how we can use type Ia supernovae to infer cosmological information. We do not address important issues such as the calibration of the light curves, how the distance moduli are calculated, and the treatment of the systematic errors. Some knowledge of statistics and Bayesian analysis is required. The latter can be helped by, e.g., \cite{Trotta:2017wnx}.

Let us rewrite Eq. \eqref{distancemodulus} as:
\begin{equation}
	\mu = 5\log_{10}\left(\frac{d_{\rm L}}{1\mbox{ Mpc}}\right) + 25\;,
\end{equation}
where we have introduced the Megaparsec (Mpc) as a more appropriate distance scale for cosmology.

For the sake of simplicity, we use the 31 binned data of \cite{Betoule:2014frx}, which we report in Tab.~\ref{Tab:JLAbinneddata}, and the relative binned 31$\times$31 covariant matrix $\textbf{C}_{b}$, which we do not report here.

\begin{table}
\centering
\begin{tabular}{|c c|c c|c c|}
\hline
		$z_b$ & $\mu_b$ & $z_b$ & $\mu_b$ & $z_b$ & $\mu_b$\\ \hline
		0.010 & 32.9538 & 0.051 & 36.6511 & 0.257 & 40.5649\\
		0.012 & 33.8790 & 0.060 & 37.1580 & 0.302 & 40.9052\\
		0.014 & 33.8421 & 0.070 & 37.4301 & 0.355 & 41.4214\\
		0.016 & 34.1185 & 0.082 & 37.9566 & 0.418 & 41.7909\\
		0.019 & 34.5934 & 0.097 & 38.2532 & 0.491 & 42.2314\\
		0.023 & 34.9390 & 0.114 & 38.6128 & 0.578 & 42.6170\\
		0.026 & 35.2520 & 0.134 & 39.0678 & 0.679 & 43.0527\\
		0.031 & 35.7485 & 0.158 & 39.3414 & 0.799 & 43.5041\\
		0.037 & 36.0697 & 0.186 & 39.7921 & 0.940 & 43.9725\\
		0.043 & 36.4345 & 0.218 & 40.1565 & 1.105 & 44.5140\\
		      &         &       &         & 1.300 & 44.8218\\
		\hline
\end{tabular}
\caption{Binned type Ia supernovae data. From \cite{Betoule:2014frx}.}
\label{Tab:JLAbinneddata}
\end{table}

Combining Eq.~\eqref{distancemodulus} with Eq.~\eqref{lumdistfunzredshift} allows us to find the theoretical prediction on $\mu(z)$, which we can adjust to the observed binned values of Tab.~\ref{Tab:JLAbinneddata} and thus find the best-fit values for $H_0$ and $q_0$.

Note that there are binned redshifts larger than 1 in Tab.~\ref{Tab:JLAbinneddata}, and for these, the expansion of Eq.~\eqref{lumdistfunzredshift} is certainly inaccurate. However, let us not worry about this now but focus only on how the analysis is performed.

The $\chi^2$ is calculated as follows:
\begin{equation}\label{chi2hq0}
	\chi^2(h,q_0) = \textbf{r}^T\cdot \textbf{C}_{\rm b}^{-1} \cdot \textbf{r}\;,
\end{equation}
where $\textbf{r}$ is the vector of the differences between the observed $\mu_b$ and the predicted values, i.e.
\begin{equation}
	r_i \equiv \mu_{bi} - 5\log_{10}\left[\frac{3000}{h}\left(z_{bi} + \frac{1}{2}(1 - q_0)z_{bi}^2\right)\right] - 25\;,
\end{equation}
for $i = 1,\cdots,31$. The $\mu_{bi}$ and $z_{bi}$ are those of Tab.~\ref{Tab:JLAbinneddata}. Note that we have expressed $H_0$ as 100 $h$ km s$^{-1}$ Mpc$^{-1}$ and $c = 3\times 10^5$ km s$^{-1}$.

Through the $\chi^2$ we define the \textbf{likelihood}\index{Likelihood}:
\begin{equation}\label{likelihoodhq0}
	\mathcal{L}(h,q_0) = Ne^{-\chi^2(h,q_0)/2}\;,
\end{equation} 
where $N$ is a normalization constant. The likelihood represents the probability of having a dataset given a cosmological model. We are interested in the opposite: the probability of having a certain cosmological model given a dataset. This is called \textbf{posterior probability}. If there is no \textit{a priori} reason for which some values of the parameters are preferred with respect to others, then the likelihood and the posterior probability are equal.

Let us see what happens in our case. What we do is the following: we set up a 100$\times$100 grid in the parameter space $(h,q_0)$ with $0.65 < h < 0.75$ and $-0.7 < q_0 < 0.2$. We choose these values because we already know the results; however, in general, one must explore the parameter space. Also, the 100$\times$100 grid is purely arbitrary, as one may choose a finer one to improve precision.

For each point of the grid, we compute the $\chi^2$ of Eq.~\eqref{chi2hq0} and thus construct the likelihood as a function of the parameter space. Its maximum corresponds to the minimum $\chi^2$, which, in turn, corresponds to the best fit values for the parameters. In our case, we find:
\begin{equation}\label{bestfitvalueshq0}
	\chi^2_{\rm min} = 37.94\;, \quad h^{(\rm bf)} = 0.69\;, \quad q_0^{(\rm bf)} = -0.18\;,
\end{equation} 
where the superscript \textit{bf} means ``best fit'', of course. Note the negative best fit value of $q_0$, which corresponds to an accelerated expansion.

\begin{figure}[ht]
\centering
	\includegraphics[width=\columnwidth]{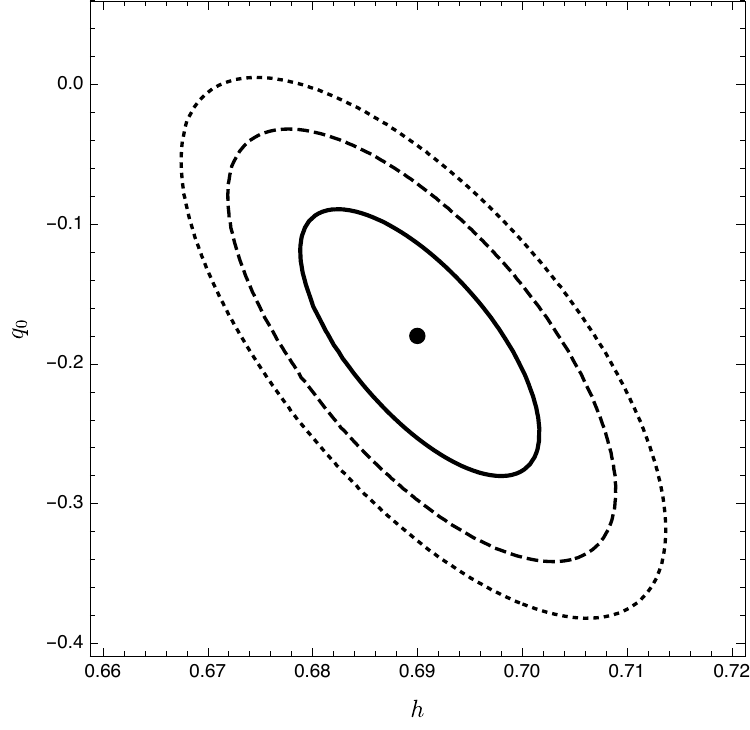}
	\caption{Contour plots at 68\%, 95\% and 99\% confidence level (from the inner to the outer contour). The big dot represent the best fit values of Eq.~\eqref{bestfitvalueshq0}.}
	\label{fig:CPlot}
\end{figure}

We could, in principle, plot the likelihood as a function of the parameters in a 3-dimensional plot, but it is more convenient and visually clearer to plot contours, i.e., horizontal slices of the likelihood, where horizontal means parallel to the $(h,q_0)$ plane. This is what we have in Fig.~\ref{fig:CPlot}.

In Fig.~\ref{fig:CPlot}, there are three contour plots, which correspond to the 68\%, 95\%, and 99\% \textbf{confidence levels}\index{Confidence level}. What are these? Starting from the maximum value of the likelihood (represented by the big dot in Fig.~\ref{fig:CPlot}), we slice its graph with a horizontal plane such that the enclosed volume is 68\%, 95\%, and 99\%. If the contours are very close to one another, it means that the likelihood is very peaked; thus, the measure has been very precise.

Therefore, from Fig.~\ref{fig:CPlot}, we can say that $q_0$ is negative with almost 99\% confidence, and this is remarkable.

\begin{figure}[ht]
\centering
	\includegraphics[width=0.5\columnwidth]{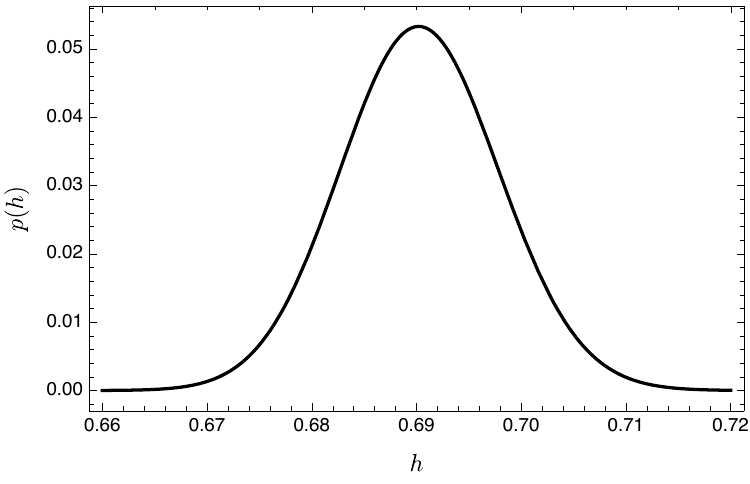}\includegraphics[width=0.5\columnwidth]{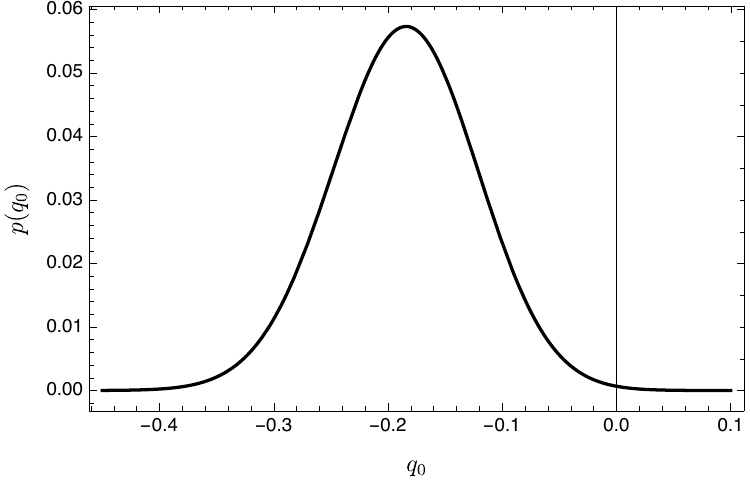}
	\caption{Normalized probability distribution functions for $h$ and $q_0$.}
	\label{fig:hq0pdf}
\end{figure}

What happens if there are more than two parameters? In this case, the likelihood cannot be plotted, nor its 2-dimensional contours. Then, one performs \textbf{marginalization}, i.e., one integrates the likelihood with respect to all the parameters except for two. In our case, we did not need to do this because we had two parameters from the beginning, but we can marginalize over one and obtain the probability distribution function (PDF) for the other. This is done in Fig.~\ref{fig:hq0pdf}.\index{Marginalization}

Now we can perform a calculation similar to that of the contour plots based on the confidence levels. For each of the PDF in Fig.~\ref{fig:hq0pdf}, starting from the maximum and going down, we calculate the values of the parameters for which the area encompassed is 68\%, 95\%, and 99\%. These values represent the uncertainty regarding the best fit values of our parameters. In our case, we have:
\begin{equation}
	h = 0.69^{+0.015}_{-0.015}\;, \qquad q_0 = -0.18^{+0.12}_{-0.13}\;,
\end{equation}
at the 95\% confidence level.

Both the PDF's for $h$ and $q_0$ are beautifully symmetric. This happens because of the special way $h$ and $q_0$ enter into the $\chi^2$; that is, the latter is quadratic in $h$ and $q_0$, and therefore the PDF for each of these parameters is a Gaussian function. This can also be seen in the contour plots of Fig.~\ref{fig:CPlot}, which are ellipses. On the other hand, if a parameter enters in a more complicated way into the $\chi^2$, then the contour plots might be very asymmetric (typically, they are ``banana'' shaped).

\hrulefill

\begin{ex}
	Reproduce the analysis of this section developing a suitable numerical code.
\end{ex}

\hrulefill

\section{Doing statistics in the sky}

In Chapter \ref{Chap:StochasticPropertiesofCP}, we have tackled the stochastic character of cosmological perturbations from a theoretical perspective. Here, we offer a simplistic approach to how statistical methods are applied observationally. More detailed and comprehensive treatments can be found in \cite{1980large} and \cite{bonometto2008cosmologia}.

As we learn in the very first year of our Physics course, determining the value of some physical quantity is no trivial task. The true value that we hope to find is only an abstraction, being our measurements imperfect, i.e., characterized by an error that we try to harness with some mathematical tools. Statistics is one of them and tells us, for example, that the more we repeat our measurements, the closer to a certain value the average goes. This certain value might be the true value, or not, if systematic errors are present. In other words, we might have a very precise but inaccurate experiment. The approach of repeating experiments and measurements in order to extract information about an underlying pattern is called \textbf{frequentist}. The question is: how do we apply that machinery to the universe, being this the only one?  

We want to learn something about gravity by observing how structures (galaxies) are distributed in the universe. We need statistics in order for our results to be meaningful, and, thankfully, there are many galaxies. Imagine a big volume $V$ which contains $N$ galaxies. The galaxy number density is $n = N/V$. This, for example, represents the volume of a certain survey. Getting information on the volume $V$ in itself is not meaningful because we have just a single realization of it. Therefore, let us consider small spheres of radius $R$ and volume
\begin{equation}
	V_R = \frac{4\pi}{3}R^3\;,
\end{equation}
with, of course, $V_R \ll V$. Now, inside these spheres, we would expect a number $\bar{N}_R \equiv nV_R$ of galaxies if these were randomly distributed, i.e., distributed according to a Poisson distribution. But actually, this is not the case because of gravity. The latter is attractive, and therefore it is more probable to find a galaxy closer to a big cluster rather than to a smaller one. Therefore, studying the large scale structure of the universe amounts to studying the properties of gravity on large scales. 

Now, let us count the galaxies in each one of the spheres that we have constructed. We have different numbers depending on the sphere considered, say $N_R(\textbf{x})$, because the sphere has its center in $\textbf{x}$. The number density thus becomes
\begin{equation}
	n_R(\textbf{x}) = \frac{N_R(\textbf{x})}{V_R(\textbf{x})}\;.
\end{equation}
It is not really a function of a continuous variable because, in practice, $\textbf{x}$ takes only a countable number of values, as many as the centers of the chosen spheres. Formally, we have:
\begin{equation}
	\langle n_R(\textbf{x})\rangle = \lim_{\nu \to \infty}\frac{1}{\nu}\sum_{i = 1}^{\nu}n_R(\textbf{x}_i) = n\;.
\end{equation}
For $\textbf{x}$ being a continuous variable, we would have an integral:
\begin{equation}
	\langle n_R(\textbf{x})\rangle = \frac{1}{V}\int_Vd^3\textbf{x}\;n_R(\textbf{x}) = n\;.
\end{equation}
So, we have divided the initial big volume $V$ into small spherical regions $V_R$ in order to gain statistics, i.e., in order to define averages. In particular, the \textbf{mass variance}
\begin{equation}\label{massvariance}
	\boxed{\sigma_R^2 = \frac{\langle (N_R(\textbf{x}) - \bar{N}_R)^2\rangle}{\bar{N}_R^2}}
\end{equation}
is a very important indicator, as we shall see in a moment.\index{Mass variance}

\hrulefill

\begin{ex} Assume that all galaxies have the same mass $m_g$. Show that from Eq.~\eqref{massvariance} we get
\begin{equation}
	\sigma_R^2 = \frac{\langle (\rho_R(\textbf{x}) - \rho)^2\rangle}{\rho^2} = \left\langle \left(\frac{\delta\rho_R(\textbf{x})}{\rho}\right)^2\right\rangle = \langle\delta_R(\textbf{x})^2\rangle\;,
\end{equation}
where $\rho_R(\textbf{x}) \equiv n_R(\textbf{x})m_g$ is the mass-density inside a sphere of radius $R$, centered in $\textbf{x}$ and $\rho = nm_g$, i.e. it is the background density, averaged over all $V$.
\end{ex}

\hrulefill

The important point shown by the above exercise is that the mass variance is related to the averaged smoothed squared density contrast. From the theory developed in these lecture notes, we know how to get $\delta(\textbf{x})$. The point now is to learn how to smooth it, and for this purpose, we need to use filters. 

\subsection{Top Hat and Gaussian filters}

The density contrast field smoothed over a sphere of radius $R$ can be formally written as:
\begin{equation}
	\delta_R(\textbf{x}) = \int_{V_R(\textbf{x})}d^3\textbf{u}\;\delta(\textbf{u})\;.
\end{equation} 
Note that the $\mathbf x$ dependence of $\delta_R(\mathbf x)$ comes from where we have centered the sphere. We can transform the above integral into an equivalent one over the whole space, which we need in order to use the Fourier transform:
\begin{equation}\label{smootheddelta}
	\delta_R(\textbf{x}) = \int_{V_R(\textbf{x})}d^3\textbf{u}\;\delta(\textbf{u}) = \int d^3\textbf{u}\;W_R(|\textbf{x} - \textbf{u}|)\delta(\textbf{u})\;,
\end{equation} 
where
\begin{equation}
	W_R(y) = \begin{cases}
		1/V_R &\mbox{ for } y < R\;,\\
		0  &\mbox{ for } y > R\;,
	\end{cases}
\end{equation}
is the \textbf{Top Hat filter},\index{Top Hat filter} or window function (hence the symbol $W$ for denoting it). It is a simple trick to exclude all the information content beyond a certain scale $R$. There is no reason for this to depend on the direction; therefore, the argument of $W_R$ is the modulus $|\textbf{x} - \textbf{u}|$. A filter is a function that must not add or subtract any information; therefore, it must be normalized to unity:
\begin{equation}
	\int d^3\textbf{y}\;W_R(y) = 1\;.
\end{equation}
In general, the functional form of a filter is:
\begin{equation}
	W_R(y) = \frac{1}{V_R}w(y/R)\;,
\end{equation}
where $w(y/R)$ is a generic function that goes to zero rapidly as $y/R$ grows. In the case of the Top Hat filter, for example, $w$ is a Heaviside function. One can also have a \textbf{Gaussian filter}:\index{Gaussian filter}
\begin{equation}
	W_R(y) = (2\pi R^2)^{-3/2}e^{-y^2/(2R^2)}\;.
\end{equation}
Equation \eqref{smootheddelta} is a convolution between the density contrast and the filter; hence, its Fourier transform is the product of the two Fourier transforms:
\begin{equation}\label{deltaRFT}
	\delta_R(\textbf{k}) = \delta(\textbf{k})\tilde{W}(kR)\;,
\end{equation}
where we have already guessed the dependence $kR$ for the Fourier transformed filter since $W_R$ is a function of $y/R$.

\hrulefill

\begin{ex} Calculate the Fourier transform of the Top Hat filter. Show that:
\begin{equation}\label{TopFilterFT}
	\tilde{W}(kR) = \frac{3}{(kR)^3}\left[\sin(kR) - (kR)\cos(kR)\right]\;.
\end{equation}
\end{ex}

\hrulefill

Of course, the Fourier transform of the Gaussian filter is still a Gaussian. The Top Hat filter is very effective in the configuration space since it cuts out all the scales above a given one, i.e. $W_R = 0$ if $y/R > 1$ or $kR < 2\pi$. However, its Fourier transform scales as $1/(kR)^3$, so it is not cut out drastically. This gives rise to spurious effects and therefore, sometimes it might be safer to use the Gaussian filter, for which one has the same cutoff also for the Fourier transform.

So, in Eq.~\eqref{deltaRFT} the left hand side is the smoothed density contrast, which on should compare with observation. On the right hand side we have the full density contrast and the window function. Actually, there is another observational limitation. We have integrated over the whole space, together with a window function, in order to smooth out but often a survey does not cover the full sky. Certainly not in depth (we cannot observe up to infinite redshift) and also not the full celestial sphere. There is a \textbf{mask} say $f(\textbf{x})$ by which the effective density contrast that we are going to use is
\begin{equation}\label{effectivedelta}
	\delta_{\rm eff}(\textbf{x}) = f(\textbf{x})\delta(\textbf{x})\;.
\end{equation}
Hence, its Fourier transform is the convolution of the two Fourier transforms:
\begin{equation}
	\delta_{\rm eff}(\textbf{k}) = (\tilde{f}*\delta)(\textbf{k}) = \int d^3\textbf{k}'\tilde{f}(\textbf{k}')\delta(\textbf{k} - \textbf{k}')\;.
\end{equation}
This convolution has important effects on the final predictions of a model. In fact, suppose that we have a theory providing a density contrast that oscillates; then we expect, of course, an oscillating power spectrum. This would be ruled out if compared directly with the data, but since the result of the convolution smooths out oscillations, it might happen that the model still remains viable. See e.g. \cite{2011arXiv1112.4386M} for a discussion on this point.

\subsection{Sampling and shot noise}

We can formally define the following quantity:
\begin{equation}
	n(\textbf{x}) \equiv \lim_{R \to 0}n_R(\textbf{x})\;,
\end{equation}
as the number density field. On the other hand, we do not observe such a field, but only a finite number of galaxies. Hence, observationally $n(\textbf{x})$ is realized by
\begin{equation}\label{obsnofx}
	n(\textbf{x}) = \sum_{i = 1}^N\delta^{(3)}(\textbf{x} - \textbf{x}_i)\;,
\end{equation}
i.e., abusing (since it is a distribution) the Dirac delta of the positions of the galaxies $\textbf{x}_i$. Avoiding any abuse, we could write:
\begin{equation}
	n(\textbf{x}) = \sum_{i = 1}^N\frac{1}{a^3\pi^{3/2}}e^{-|\textbf{x} - \textbf{x}_i|^2/a^2}\;,
\end{equation}
for $a \to 0$, i.e., using a representation of the Dirac delta. 

Technically, when integrating Eq.~\eqref{obsnofx} over a certain volume $V$, we should obtain the number of galaxies contained therein. This is formally achieved by using the Top Hat filter:
\begin{equation}
	N = \int_VdV\;n(\textbf{x}) = \sum_{i = 1}^N\int dV\;w(\textbf{x})\delta^{(3)}(\textbf{x} - \textbf{x}_i) = \sum_{i = 1}^Nw(\textbf{x}_i)\;,
\end{equation}
where $w(\textbf{x}) = V W(\textbf{x})$ and $W(\textbf{x})$ are equal to $1/V$ when $\textbf{x}$ is inside the volume $V$ and zero otherwise (this ensures normalization to unity). In the last sum above, $w(\textbf{x}_i) = 1$ only exists when $\textbf{x}_i$ is inside the volume $V$, and hence the equation holds true.

Now let us keep our sampling volume $V$, imagining that it is the volume probed by a certain survey. As we commented earlier, the effective density contrast field is then given by Eq.~\eqref{effectivedelta}. The difference now is that we are constructing the observed density contrast field, whereas in that equation it was our theoretical prediction. The density field is:
\begin{equation}
	\rho(\textbf{x}) = \sum_{i = 1}^Nm_i\delta^{(3)}(\textbf{x} - \textbf{x}_i)\;,
\end{equation}
where $m_i$ is the mass of the $i$-th galaxy. The effective, or sampled, density contrast is
\begin{equation}
	\delta_{\rm s}(\textbf{x}) = \left(\frac{\rho(\textbf{x})}{\rho} - 1\right)w(\textbf{x}) = \left(\frac{\sum_{i = 1}^Nm_i\delta^{(3)}(\textbf{x} - \textbf{x}_i)}{\sum_{i = 1}^Nm_i/V} - 1\right)w(\textbf{x})\;.
\end{equation}
where $\rho$ is the background density, which is equal to $\sum_{i = 1}^Nm_i/V$ in the sampled volume. Assuming the same mass for each galaxy, one obtains:
\begin{equation}
	\delta_{\rm s}(\textbf{x}) = \left(\frac{V}{N}\sum_{i = 1}^N\delta^{(3)}(\textbf{x} - \textbf{x}_i) - 1\right)w(\textbf{x})\;.
\end{equation}

\hrulefill

\begin{ex} Show that the Fourier transform of the density contrast is then:
\begin{equation}
	\delta_{\rm s}(\textbf{k}) = \frac{1}{N}\sum_{i = 1}^N w(\textbf{x}_i)e^{-i\textbf{k}\cdot\textbf{x}_i} - \tilde{W}(\textbf{k})\;,
\end{equation}
where $\tilde{W}(\textbf{k})$ is the Fourier transform of the Top Hat filter, computed in Eq.~\eqref{TopFilterFT}.
\end{ex}

\hrulefill

When we compute the power spectrum from the above realization and assume a Gaussian distribution, we get for the mode $\textbf{k}$:
\begin{eqnarray}
	\langle\delta_{\rm s}(\textbf{k})\delta_{\rm s}^*(\textbf{k})\rangle = \frac{1}{N^2}\sum_{i,j = 1}^N \langle w(\textbf{x}_i)w(\textbf{x}_j)\rangle e^{-i\textbf{k}\cdot(\textbf{x}_i - \textbf{x}_j)} - \frac{\tilde{W}(\textbf{k})}{N}\sum_{i = 1}^N \langle w(\textbf{x}_i)\rangle e^{-i\textbf{k}\cdot\textbf{x}_i}\nonumber\\ - \frac{\tilde{W}(\textbf{k})}{N}\sum_{i = 1}^N \langle w(\textbf{x}_i)\rangle e^{i\textbf{k}\cdot\textbf{x}_i} + \tilde{W}(\textbf{k})^2\;,
\end{eqnarray}
where the average is an ensemble average. Since, by definition:
\begin{equation}
	\langle\delta_{\rm s}(\textbf{k})\rangle = 0\;,
\end{equation}
we have that:
\begin{equation}
	\frac{1}{N}\sum_{i = 1}^N \langle w(\textbf{x}_i)\rangle e^{-i\textbf{k}\cdot\textbf{x}_i} = \tilde{W}(\textbf{k})\;.
\end{equation}
Therefore:
\begin{eqnarray}
	\langle\delta_{\rm s}(\textbf{k})\delta_{\rm s}^*(\textbf{k})\rangle = \frac{1}{N^2}\sum_{i,j = 1}^N \langle w(\textbf{x}_i)w(\textbf{x}_j)\rangle e^{-i\textbf{k}\cdot(\textbf{x}_i - \textbf{x}_j)} - \tilde{W}(\textbf{k})^2\;.
\end{eqnarray}
When $i = j$ there is a contribution of the above form:
\begin{equation}
	\frac{1}{N^2}\sum_{i = 1}^N \langle w(\textbf{x}_i)^2\rangle = \frac{1}{N}\;.
\end{equation}
This is an error (a variance) on the Fourier mode $\textbf{k}$ that does not depend on the wavenumber and is called \textbf{shot noise}\index{Shot noise}. It comes from the fact that we are mapping a continuous density field with a discrete distribution or sampling. It is the Poisson part of the distribution of galaxies, which must be subtracted to obtain the true spectrum given by the correlations $\langle w(\textbf{x}_i)w(\textbf{x}_j)\rangle$.

Multiplying by $V$ for dimensional reasons, we thus have the true power spectrum:
\begin{equation}
	P(\textbf{k}) = \frac{V}{N^2}\sum_{i \neq j}^N \langle w(\textbf{x}_i)w(\textbf{x}_j)\rangle e^{-i\textbf{k}\cdot(\textbf{x}_i - \textbf{x}_j)} - V\tilde{W}(\textbf{k})^2\;,
\end{equation}
and the noise spectrum:
\begin{equation}
	P_n = \frac{V}{N}\;.
\end{equation}

\subsection{Correlation function}

How do we measure $P(k)$ and put the data on a plot? All we observe are galaxies in the sky. Consider again the very large volume $V$ upon which we have worked until now, containing $N$ galaxies. So, the mean number density is $n = N/V$. Assuming the same mass $m_g$ for every galaxy, $\rho = Nm_g/V$ is the mean density. Consider two small volumes $\delta V_1$ and $\delta V_2$, centered in $\textbf{x}_1$ and $\textbf{x}_2$ respectively and much smaller than $V$. If the distribution of galaxies is random, i.e., Poissonian, one would expect $\delta N_1 = n\delta V_1$. Deviations from this randomness are due to gravity and are encoded in the \textbf{correlation function} $\xi(\textbf{x}_1,\textbf{x}_2)$:\index{Correlation function}
\begin{equation}
	\langle\delta N_1(\textbf{x}_1)\delta N_2(\textbf{x}_2)\rangle = n^2\delta V_1\delta V_2\left[1 + \xi(\textbf{x}_1,\textbf{x}_2)\right]\;.
\end{equation}
We have already met the correlation function in Chapter~\ref{Chap:StochasticPropertiesofCP}, in Eq.~\eqref{2pointcorrelationfunction}. There it was defined through the ensemble average, whereas here the average is made over all the couples of volumes, and clearly these must be chosen large enough in order to contain many galaxies, but sufficiently small in order to identify many of them in $V$. Usually, one assumes that 
\begin{equation}
	\xi(\textbf{x}_1,\textbf{x}_2) = \xi(r)\;,
\end{equation}
where $r = |\textbf{x}_1 - \textbf{x}_2|$, i.e., the correlation function depends only on the distance between the volumes, not on their positions or the direction in which they are aligned. This is again statistical homogeneity and isotropy. Dividing the above equation by $\delta V_1\delta V_2$, one gets
\begin{equation}
	\langle\delta n_1(\textbf{x})\delta n_2(\textbf{x} + \textbf{r})\rangle = n^2\left[1 + \xi(r)\right]\;,
\end{equation}
Where now, the average can be interpreted as the integration over $\textbf{x}$.

\hrulefill

\begin{ex} Show that the above equation can be written as:
\begin{equation}
	\langle\delta(\textbf{x})\delta(\textbf{x} + \textbf{r})\rangle = \xi(r)\;,
\end{equation}
and this gives a direct relation between the density contrast and the correlation function. Compare with Eq.~\eqref{xieqaverage}.
\end{ex}

\hrulefill

The correlation function measures galaxy clustering. From observation, we have the following empirical formula:
\begin{equation}
	\xi(r) = \left(\frac{r_0}{r}\right)^\gamma\;,
\end{equation}
where $r_0 = 5.5$ $h^{-1}$ Mpc and $\gamma \approx 1.77$ \cite{1980large}. The mass variance can be related to the correlation function as follows:
\begin{equation}
	\sigma_R^2 = G(\gamma)\xi(R)\;,
\end{equation}
where $G(\gamma) \approx 2$. We can argue that the non-linear regime of cosmology becomes dominant when $\sigma_R^2 = 1$, and from the above formulae, one can find that this happens at $R = 8$ $h^{-1}$ Mpc. Hence, this is why $\sigma_8$ is so often used in cosmology.\index{$\sigma_8$}

The mass variance can be expressed as:
\begin{equation}
	\sigma_R^2 = \langle\delta^2_R(\textbf{x})\rangle = \int\frac{d^3\textbf{k}}{(2\pi)^3}\int\frac{d^3\textbf{k}'}{(2\pi)^3}\;\langle\delta(\textbf{k})\tilde{W}(kR)\delta(\textbf{k}')\tilde{W}(k'R)\rangle\;,
\end{equation}
where we have used the Fourier transform of the smoothed density contrast, which is the product of the density contrast times the window function (or filter). Using the ensemble average, we get:
\begin{equation}
	\boxed{\sigma_R^2 = \int\frac{d^3\textbf{k}}{(2\pi)^3}\;P_\delta(k)\tilde{W}(kR)^2 = \int_0^{\infty} \frac{dk}{k}\;\Delta_\delta^2(k)\tilde{W}(kR)^2}
\end{equation}
The above expression for the mass variance and that for the correlation function in Eq.~\eqref{corrfunDelta} are very similar, with the difference being in the function that weighs the dimensionless power spectrum $\Delta^2$. For the correlation function, it is $\sin(kr)/(kr)$, whereas for the mass variance, it is the squared Fourier transform of the filter. If we take the Top Hat filter, then from Eq.~\eqref{TopFilterFT} we see that $\tilde{W}(kR)^2$ decays as $(kR)^6$.

\subsection{Bias}

The bias\index{Bias} is the deviation of the clustering behavior of ordinary matter from that of CDM. In general, we might suppose that structures on the scale $R$ are formed in those sites where
\begin{equation}
	\delta_R(\textbf{x}) > \nu\sigma_R\;,
\end{equation} 
where $\nu$ is some parameter. In this way, we are stating that galaxies or clusters do not have exactly the same fluctuation pattern as CDM. Define the biased density field:
\begin{equation}
	\rho_{R,\nu} = \theta[\delta_R(\textbf{x}) - \nu\sigma_R]\;,
\end{equation} 
where $\theta$ is the step function. For a Gaussian process, it can be shown that \cite{bonometto2008cosmologia}
\begin{equation}
	\langle \rho_{R,\nu}\rangle \approx \frac{1}{\sqrt{2\pi}}\frac{1}{\nu}e^{-\nu^2/2}\;,
\end{equation}
which gives
\begin{equation}
	\langle n_{R,\nu}\rangle \approx \frac{3}{(2\pi)^{3/2}R^3\nu}e^{-\nu^2/2}\;.
\end{equation}
The correlation function can then be written as
\begin{equation}
	\xi^{(\nu,R)}(r) = e^{\nu^2\xi^R(r)/\sigma_R^2} - 1\;,
\end{equation}
which, for a small exponential, can be cast as
\begin{equation}
	\xi^{(\nu,R)}(r) = \frac{\nu^2}{\sigma_R^2}\xi^R(r) \equiv b^2\xi^R(r)\;,
\end{equation}
In general, the relationship between galactic and CDM density contrast is expressed as:
\begin{equation}
	\delta_g = b\delta\;,
\end{equation}
i.e. the same Eq.~\eqref{biasrelationlowz} that we presented in Chapter~\ref{Chap:Evopert}. As we mentioned there, typically $b$ is treated as a parameter, though it may depend on the scale and on time.

\clearpage
\appendix
\chapter{Thermal distributions}\label{App:Thermaldistr}

In this Appendix, we derive the functional forms of the Maxwell-Boltzmann, Fermi-Dirac, and Bose-Einstein thermal distributions in the grand-canonical ensemble. See, for example, \cite{1987stme.book.....H} for a textbook reference.\index{Thermal distributions}

\section{Derivation of the Maxwell-Boltzmann distribution}

Suppose that we have a system of $N$ classical particles distributed in $K$ states of different energies. For each state $i$, there are $n_i$ particles with energy $e_i$, and the total energy of the system is fixed and equal to $E$. Therefore, we have two constraints:
\begin{equation}\label{MBdistrconstr}
	N = \sum_{i = 1}^Kn_i\;, \qquad E = \sum_{i = 1}^Kn_ie_i\;.
\end{equation}
The number $\Omega$ of configurations that correspond to the same macroscopic state is as follows:
\begin{equation}
	\Omega\left(\{n_i\}\right) = \frac{N!}{n_1!n_2!\cdots n_K!}\;.
\end{equation}
The numerator is $N!$ because classical particles are \textit{distinguishable}, so each of their permutations counts as a different configuration. However, permutations done within the same state $i$ do not count; hence, we eliminate these possibilities by dividing by $n_1!n_2!\cdots n_K!$.

Now take the logarithm of $\Omega$ and use Stirling's approximation (since $N$ is very large):
\begin{equation}
	\log\Omega = \log (N!) - \sum_{i = 1}^K\log (n_i!) \approx N\log N - N - \sum_{i = 1}^K(n_i\log n_i - n_i)\;.
\end{equation}
We have to find the maximum of this expression (this is required by the condition of equilibrium or maximum entropy), but taking into account the constraints \eqref{MBdistrconstr}. We then introduce Lagrange multipliers $\alpha$ and $\beta$ and calculate the differential:
\begin{equation}
	d\left[N\log N - N - \sum_{i = 1}^K(n_i\log n_i - n_i) - \alpha\sum_{i = 1}^Kn_i - \beta\sum_{i = 1}^Kn_ie_i\right] = 0\;. 
\end{equation}
The variables are the $n_i$'s, thus:
\begin{equation}
	\sum_{i = 1}^K\left(-\log n_i - \alpha - \beta e_i\right)dn_i = 0\;.
\end{equation}
From this equation, we obtain:
\begin{equation}
	n_i = \exp(-\alpha - \beta e_i) \equiv \exp\left(-\frac{e_i - \mu}{k_{\rm B}T}\right)\;,
\end{equation}
where in the last equality we have interpreted the Lagrange multiplier $\alpha$ as the chemical potential (divided by $k_{\rm B}T$) and $1/\beta$ as the thermal energy of the bath.\index{Maxwell-Boltzmann distribution}

Now the claim: ``the chemical potential is zero because the number of particles is not conserved'' is hopefully more transparent. In fact, when the particle number is not conserved, we do not need to introduce the Lagrange multiplier $\alpha$.

\section{Derivation of the Fermi-Dirac distribution}

When calculating the Fermi-Dirac distribution\index{Fermi-Dirac distribution}, two main differences with respect to the Maxwell-Boltzmann case appear: the first is that we now have quantum particles, which are \textit{indistinguishable}, and the second is that fermions obey Pauli's exclusion principle, so there could be at most one fermion per quantum state. 

Let us assume again $N$ quantum particles and $K$ energy states. In general, for each energy state $i$ there are $g_i$ sub-states; i.e., the energy states are degenerate (e.g., there are quantum numbers other than the one referred to as the energy that characterize the state). Let us focus on the energy state $i$. Here we must place $n_i$ fermions. Because of Pauli's exclusion principle, $g_i \ge n_i$, otherwise we would have more than one fermion in each state. 

\hrulefill

\begin{ex}
	In how many ways can we fit $n_i$ indistinguishable fermions in $g_i$ ``slots''? Prove that the answer is:
\begin{equation}\label{OmegaiFD}
	\Omega_i = \frac{g_i!}{n_i!(g_i - n_i)!}\;.
\end{equation}
Since combination math might be irksome (and, after all, these are lecture notes) here is the solution. However, try not jumping to it right away and to work out the solution yourself.

\paragraph{Solution.} We can choose $g_i$ slots for the first fermion, $g_i - 1$ for the second and so on until finally we can choose $g_i - n_i + 1$ slots for the $n_i$-th fermion. This gives $g_i!/(g_i - n_i)!$ and we only used Pauli's exclusion principle. Until here, then, the order of the chosen particles matters: having say fermion 1 in the first slot is different from having fermion 2 in the first slot. But fermions are indistinguishable, therefore we must divide by $n_i!$.
\end{ex}

\hrulefill

The total number of micro-states is then:
\begin{equation}
	\Omega\left(\{n_i\}\right) = \prod_{i=1}^K\Omega_i = \prod_{i=1}^K\frac{g_i!}{n_i!(g_i - n_i)!}\;.
\end{equation}
Now we proceed as before, taking the logarithm of $\Omega$:
\begin{equation}
	\log\Omega = \sum_{i = 1}^K\left[\log(g_i!) - \log(n_i!) - \log((g_i - n_i)!)\right]\;,
\end{equation}
using Stirling's approximation and calculating the constrained maximum:
\begin{equation}
	\sum_{i = 1}^K\left[-\log n_i + \log(g_i - n_i) - \alpha - \beta e_i\right]dn_i = 0\;, 
\end{equation}
one obtains:
\begin{equation}
	\log\frac{g_i - n_i}{n_i} = \alpha + \beta e_i\;,
\end{equation}
and finally:
\begin{equation}
	n_i = g_i\frac{1}{1 + \exp(\alpha + \beta e_i)} \equiv g_i\frac{1}{1 + \exp\left(\frac{e_i - \mu}{k_{\rm B}T}\right)}\;,
\end{equation}
with the same physical meanings for $\alpha$ and $\beta$ as those stated for the Maxwell-Boltzmann distribution.\index{Fermi-Dirac distribution} 

\section{Derivation of the Bose-Einstein distribution}

The setup for deriving the Bose-Einstein distribution\index{Bose-Einstein distribution} is the same as the one used in the previous subsection for the Fermi-Dirac distribution, except for the fact that now Pauli's exclusion principle does not apply, and so the constraint $n_i \le g_i$ does not hold true anymore. This changes the way in which we calculate $\Omega_i$.

\hrulefill

\begin{ex}
	Show that:
	\begin{equation}\label{OmegaiBE}
	\Omega_i = \binom{n_i + g_i - 1}{g_i - 1} = \frac{(n_i + g_i - 1)!}{n_i!(g_i - 1)!}\;.
\end{equation}
This is the same calculation of the number of $n_i$-th partial derivatives of a function of $g_i$ variables. This is related to the fact that the wave-function of an ensemble of bosons is symmetric, just as partial derivatives are. Again, here is the proof.

\paragraph{Solution.}Imagine $n_i$ particles and $g_i$ slots where to fit them. These slots are separated by $g_i - 1$ walls. So, compute all the permutations among these objects, which are $(n_i + g_i - 1)!$, but do not consider the permutations among the walls $(g_i - 1)!$ and the particles, $n_i!$, because they are indistinguishable. So, we find Eq.~\eqref{OmegaiBE}.
\end{ex}

Proceeding as we did in the previous two subsections, we find:
\begin{equation}
	\sum_{i = 1}^K\left[\log(n_i + g_i - 1) - \log n_i - \alpha - \beta e_i\right]dn_i = 0\;, 
\end{equation}
and finally:
\begin{equation}
	n_i = g_i\frac{1}{\exp(\alpha + \beta e_i) - 1} \equiv g_i\frac{1}{\exp\left(\frac{e_i - \mu}{k_{\rm B}T}\right) - 1}\;.
\end{equation}\index{Bose-Einstein distribution}

\chapter{Derivation of the Poisson distribution}\label{App:Poisson}

The Poisson distribution describes stochastic \textit{independent} events happening randomly over time but with a certain average rate, say $\lambda$. It is important for the description of particle scattering or decaying processes, and we have used it in Chapter~\ref{Chap:ThermalHistory} to take into account the neutron decay during BBN, as well as in Chapter~\ref{Chap:CMBEvo} when we discussed the visibility function.

Let $P(n;\lambda,t)$ be the probability that $n$ events occur in a time interval $t$, given the rate $\lambda$. Consider a sufficiently small time interval $\delta t$, for which:
\begin{equation}
	P(1;\lambda,\delta t) = \lambda\delta t\;, \qquad P(0;\lambda,\delta t) = 1 - \lambda\delta t\;.
\end{equation}
Indeed, we can regard the first formula as the \textit{definition} of $\lambda$. Therefore, calculating $P(0;\lambda,t + \delta t)$, one gets:
\begin{equation}\label{P0lambdatdt}
	P(0;\lambda,t + \delta t) = P(0;\lambda,t)(1 - \lambda\delta t)\;,
\end{equation}
where we have used the independence of the randomly occurring events.

\hrulefill

\begin{ex} From Eq.~\eqref{P0lambdatdt} show that:
\begin{equation}\label{P0lambdat}
	P(0;\lambda,t) = e^{-\lambda t}\;.
\end{equation}
\end{ex}

\hrulefill

Now, exploiting the independence of the events, we can find a recurrence relation for $P(n;\lambda,t)$. Consider the following:
\begin{equation}
	P(n;\lambda,t + \delta t) = P(n;\lambda,t)(1 - \lambda\delta t) + P(n - 1;\lambda,t)\lambda\delta t\;, 
\end{equation}
i.e. $n$ events in a time interval $t + \delta t$ can either occur as $n$ in the time interval $t$ and none in the subsequent $\delta t$, or $n - 1$ in the time interval $t$ and just one in the subsequent $\delta t$. This is because we have chosen $\delta t$ sufficiently small to accommodate at most one event.

From the above equation, we then have:
\begin{equation}\label{recurrrelpoisson}
	\frac{dP(n;\lambda,t)}{dt} + \lambda P(n;\lambda,t) = \lambda P(n - 1;\lambda,t)\;. 
\end{equation}

\hrulefill

\begin{ex} From Eq.~\eqref{recurrrelpoisson} show that:
\begin{equation}\label{P1lambdat}
	P(1;\lambda,t) = \lambda te^{-\lambda t}\;.
\end{equation}
By induction show that:
\begin{equation}\label{Pnlambdat}
	P(n;\lambda,t) = \frac{(\lambda t)^n}{n!}e^{-\lambda t}\;.
\end{equation}
\end{ex}

\hrulefill

This is the Poisson distribution. Another way to find it is from the binomial distribution:
\begin{equation}\label{binomialdistr}
	P(N;n,p) = \binom{N}{n} p^n (1 - p)^{N -n}\;,
\end{equation}
where $N$ is the number of trials, $n$ is the number of successful ones, and $p$ is the probability of success for a single trial.

\hrulefill

\begin{ex} Assume $N \to \infty$ and $p \to 0$ such that $Np \equiv \mu$ stays constant. Performing these limits in Eq.~\eqref{binomialdistr} and using Stirling's approximation show that:
\begin{equation}
	P(n,Np \equiv \mu) = \frac{\mu^n}{n!}e^{-\mu}\;.
\end{equation}
Defining $Np \equiv \lambda t$, we recover again Poisson distribution \eqref{Pnlambdat}.\index{Poisson distribution}

\paragraph{Solution.} We have:
\begin{equation}
	\binom{N}{n} p^n (1 - p)^{N -n} = \binom{N}{n} \left(\frac{\mu/N}{1 - \mu/N}\right)^n (1 - \mu/N)^{N} = *\;.
\end{equation}
When $N\to \infty$ the last factor tends to $e^{-\mu}$. Using Stirling's approximation we have:
\begin{align}
	* \sim \frac{\sqrt{2\pi N}N^N}{n!\sqrt{2\pi(N-n)}(N-n)^{N-n}} e^{-n}\left(\frac{\mu/N}{1 - \mu/N}\right)^n e^{-\mu} \sim \nonumber\\
	\sim \frac{1}{n!}\frac{N^Ne^{-n}}{N^N(1-n/N)^N}\left(\frac{\mu - \mu n/N}{1 - \mu/N}\right)^n e^{-\mu}\;.
\end{align}
The second factor tends to 1, while the third tends to $\mu^n$. So, we are done.
\end{ex}

\hrulefill

Note that in the Poisson distribution, there is no need to introduce a time variable, as we did in its derivation. Indeed, the stochastic variable is $n$ and $\mu$ is a parameter. One could also use the distance instead of the time if the physical situation requires it.

\chapter{Liouville theorem}\label{App:LiouvilleTh}

In order to complete this brief introduction to the Boltzmann equation, we prove the Liouville theorem.\index{Liouville theorem!Proof} First of all, let us expand the time derivative of $\rho$:
\begin{equation}
	\frac{d\rho(t,\textbf{x}_i,\textbf{p}_i)}{dt} = \frac{\partial\rho}{\partial t} + \sum_{i=1}^N\left(\frac{d\textbf{x}_i}{dt}\cdot\nabla_{\textbf{x}_i}\rho + \frac{d\textbf{p}_i}{dt}\cdot\nabla_{\textbf{p}_i}\rho\right)\;,
\end{equation}
where $\nabla_{\textbf{x}_i}$ is the gradient computed with respect to $\textbf{x}_i$, and $\nabla_{\textbf{p}_i}$ is the one computed with respect to $\textbf{p}_i$.

Let $H$ be the Hamiltonian of the system. The Hamilton equations tell us that:
\begin{equation}
	\frac{d\textbf{x}_i}{dt} = \nabla_{\textbf{p}_i}H\;, \qquad \frac{d\textbf{p}_i}{dt} = -\nabla_{\textbf{x}_i}H\;,
\end{equation}
so we can also write the total derivative of $\rho$ as follows:
\begin{equation}
	\frac{d\rho(t,\textbf{x}_i,\textbf{p}_i)}{dt} = \frac{\partial\rho}{\partial t} + \left\{\rho,H\right\}\;,
\end{equation}
i.e., with the Poisson brackets. Through a simple manipulation, we can write:
\begin{equation}\label{Liouvilltheoproof}
	\frac{d\rho(t,\textbf{x}_i,\textbf{p}_i)}{dt} = \frac{\partial\rho}{\partial t} + \sum_{i=1}^N\left[\nabla_{\textbf{x}_i}\cdot(\rho\dot{\textbf{x}}_i) + \nabla_{\textbf{p}_i}\cdot(\rho\dot{\textbf{p}}_i)\right]\;,
\end{equation}
since the extra term
\begin{equation}
	\nabla_{\textbf{x}_i}\cdot\dot{\textbf{x}}_i + \nabla_{\textbf{p}_i}\cdot\dot{\textbf{p}}_i = \nabla_{\textbf{x}_i}\cdot\nabla_{\textbf{p}_i}H - \nabla_{\textbf{p}_i}\cdot\nabla_{\textbf{x}_i}H = 0\;, 
\end{equation}
is vanishing (we have used the Hamilton equations of motion here). Equation~\eqref{Liouvilltheoproof} can be cast as follows:
\begin{equation}\label{Liouvilltheoproof2}
	\frac{d\rho(t,\textbf{x}_i,\textbf{p}_i)}{dt} = \frac{\partial\rho}{\partial t} + \nabla_{\textbf{y}}\cdot (\rho\dot{\textbf{y}})\;,
\end{equation} 
where we have indicated $\textbf{y}$ as the generic variable of the phase space, i.e., $\textbf{y} = \{\textbf{x}_i,\textbf{p}_i\}$ for $i = 1, \cdots, N$. Equation~\eqref{Liouvilltheoproof2} is a continuity equation. Therefore, if the particle number $N$ is conserved, then Eq.~\eqref{Liouvilletheorem} must hold.

\chapter{Helmholtz theorem}\label{App:Helmholtztheorem}

We follow here Appendix B of \cite{2017inel.book.....G}. Let $\textbf{F}(\textbf{r})$ be a real vector field on $\mathbb R^3$. Let its divergence and curl be the following:
\begin{equation}\label{DandEdef}
	\nabla\cdot \textbf{F}(\textbf{r}) \equiv D(\textbf{r})\;, \qquad \nabla\times \textbf{F}(\textbf{r}) \equiv \textbf{C}(\textbf{r})\;.
\end{equation}
By construction, the divergence of the curl is vanishing:
\begin{equation}
	\nabla\cdot\left[\nabla\times \textbf{F}(\textbf{r})\right] = \nabla\cdot\textbf{C}(\textbf{r}) = 0\;,
\end{equation} 
Hence, $\textbf{C}(\textbf{r})$ is a solenoidal, i.e., divergenceless vector field.

Helmholtz theorem\index{Helmholtz theorem} states that $\textbf{F}(\textbf{r})$ can be decomposed as:
\begin{equation}\label{Helmholtzdec}
	\textbf{F}(\textbf{r}) = -\nabla U(\textbf{r}) + \nabla\times\textbf{W}(\textbf{r})\;,
\end{equation}
where:
\begin{equation}\label{UandWdefs}
	U(\textbf{r}) \equiv \frac{1}{4\pi}\int_Vd^3\textbf{r}'\frac{D(\textbf{r}')}{|\textbf{r} - \textbf{r}'|}\;, \qquad \textbf{W}(\textbf{r}) = \frac{1}{4\pi}\int_Vd^3\textbf{r}'\frac{\textbf{C}(\textbf{r}')}{|\textbf{r} - \textbf{r}'|}\;.
\end{equation}
Since the integration is over the whole space, in order for it to be well-defined, $D(\textbf{r})$ and $\textbf{C}(\textbf{r})$ must tend to zero more rapidly than $1/r^2$ for $r \to \infty$. This can be seen from the fact that:
\begin{equation}
	\frac{d^3\textbf{r}'}{|\textbf{r} - \textbf{r}'|} \sim dr'\;r'\;, \qquad \mbox{for } r' \to \infty\;,
\end{equation} 
and then, if the two functions $D$ and $\mathbf C$ scaled as $1/r^{'2}$, we would have a logarithmic divergence.

By taking the divergence of Eq.~\eqref{Helmholtzdec}, remembering that the divergence of a curl is vanishing, and using Eq.~\eqref{DandEdef}, we can easily check that:
\begin{equation}
	D(\textbf{r}) = \nabla\cdot \textbf{F}(\textbf{r}) = -\nabla^2U(\textbf{r})\;,
\end{equation}
which is a Poisson equation, and its solution is the first equation of Eq.~\eqref{UandWdefs}.

On the other hand, by taking the curl of \eqref{Helmholtzdec}, remembering that the curl of the gradient is vanishing, we can check that:
\begin{equation}
	\textbf{C}(\textbf{r}) = \nabla\times\left[\nabla\times\textbf{W}(\textbf{r})\right] = \nabla\left[\nabla\cdot\textbf{W}(\textbf{r})\right] - \nabla^2\textbf{W}\;.
\end{equation}
The Laplacian term only gives the second equation of Eq.~\eqref{UandWdefs}, but what about the $\nabla\left[\nabla\cdot\textbf{W}(\textbf{r})\right]$ term? Does it vanish? Let us check that this is the case:
\begin{eqnarray}
	\nabla\cdot\textbf{W}(\textbf{r}) = \frac{1}{4\pi}\int_Vd^3\textbf{r}'\textbf{C}(\textbf{r}')\cdot\nabla\left(\frac{1}{|\textbf{r} - \textbf{r}'|}\right)\nonumber\\ = -\frac{1}{4\pi}\int_Vd^3\textbf{r}'\textbf{C}(\textbf{r}')\cdot\nabla'\left(\frac{1}{|\textbf{r} - \textbf{r}'|}\right) =\nonumber\\ -\frac{1}{4\pi}\int_{\partial V}d\textbf{S}'\cdot\textbf{C}(\textbf{r}')\frac{1}{|\textbf{r} - \textbf{r}'|} + \frac{1}{4\pi}\int_Vd^3\textbf{r}'\nabla'\cdot\textbf{C}(\textbf{r}')\frac{1}{|\textbf{r} - \textbf{r}'|}\;.
\end{eqnarray}
In the last line, both contributions vanish. The first is a surface integral at infinity; in the second, the divergence of $\textbf{C}(\textbf{r})$ is zero by construction.

In principle, the decomposition of Eq.~\eqref{Helmholtzdec} might not be unique because we could add to $\textbf{F}(\textbf{r})$ a vector field, say $\textbf{G}(\textbf{r})$, which has both a vanishing divergence and curl. On the other hand, it can be proved that there is no such $\textbf{G}(\textbf{r})$ with both a vanishing divergence and curl that goes to zero at infinity. Therefore, if $\textbf{F}(\textbf{r})$ goes to zero sufficiently fast at infinity, then the decomposition in Eq.~\eqref{Helmholtzdec} is indeed unique.

\chapter{Spherical harmonics}\label{App:SphericalHarmonics}

The spherical harmonics are the eigenfunctions of the Laplacian operator on the sphere, or equivalently, the eigenfunctions of the square of the angular momentum operator. They form a complete orthonormal system for the $L^2$ space defined on the sphere; therefore, square-integrable functions on the sphere can be expanded in a linear combination of spherical harmonics. That is why they are extensively used in cosmology to analyze CMB anisotropies in temperature and polarization: the latter are fields (of spin zero and spin 2, respectively) on the celestial sphere. In this section, we shall follow principally \cite{butkovmathematical}. However, many references exist that deal with the theory of angular momentum and spherical harmonics, so the reader is encouraged to find the treatment that best suits them.  

\hrulefill

\begin{ex}
	Find the expression of the Laplacian operator on the sphere. The line element on the unit sphere is:
	\begin{equation}
		ds_S^2 = g_{ab}dx^adx^b = d\theta^2 + \sin^2\theta d\phi^2\;,
	\end{equation}
with $a,b =\theta,\phi$, the standard angular coordinates. Using the formula:
	\begin{equation}
		\nabla^2 = \frac{1}{\sqrt{g}}\frac{\partial}{\partial x^a}\left(g^{ab}\sqrt{g}\frac{\partial}{\partial x^b}\right)\;,
	\end{equation}
where $\sqrt{g}$ is the determinant of the metric, one finds:
	\begin{equation}
		\nabla^2 = \frac{1}{\sin\theta}\frac{\partial}{\partial\theta}\left(\sin\theta\frac{\partial}{\partial\theta}\right) + \frac{1}{\sin^2\theta}\frac{\partial^2}{\partial\phi^2}\;.
	\end{equation}
\end{ex}

\hrulefill

Let us then set up the eigenvalue equation:
\begin{equation}
	\frac{1}{\sin\theta}\frac{\partial}{\partial\theta}\left[\sin\theta\frac{\partial Y(\theta,\phi)}{\partial\theta}\right] + \frac{1}{\sin^2\theta}\frac{\partial^2 Y(\theta,\phi)}{\partial\phi^2} = \lambda Y(\theta,\phi)\;.
\end{equation}
Assume that the $\theta$ and $\phi$ dependencies can be factorized:
\begin{equation}
	Y(\theta,\phi) = \Theta(\theta)\Phi(\phi)\;,
\end{equation}
so that:
\begin{equation}
	\frac{\sin\theta}{\Theta}\frac{d}{d\theta}\left[\sin\theta\frac{d\Theta}{d\theta}\right] - \lambda\sin^2\theta = - \frac{1}{\Phi}\frac{d^2\Phi}{d\phi^2}\;.
\end{equation}
Now, the left hand side depends only on $\theta$, whereas the right hand side depends only on $\phi$. Therefore, the only possible way for them to be equal is for both to equal the same constant, which we call $m^2$. In this way, we now have two equations:
\begin{equation}
	\frac{d^2\Phi}{d\phi} + m^2\Phi = 0\;, \qquad \sin\theta\frac{d}{d\theta}\left[\sin\theta\frac{d\Theta}{d\theta}\right] - \lambda\sin^2\theta\Theta = m^2\Theta\;.
\end{equation}
The first one is very simple to solve and gives us:
\begin{equation}
	\Phi = e^{\pm im\phi}\;,
\end{equation}
with some normalization, which we will determine afterwards for the full $Y(\theta,\phi)$. On the other hand, $\Phi$ must be periodic since $\phi$ and $\phi + 2\pi$ denote the same angular position. Hence:
\begin{equation}
	e^{im\phi} = e^{im(\phi + 2\pi)}\;,
\end{equation}
which implies:
\begin{equation}
	e^{i2m\pi} = 1\;,
\end{equation}
and therefore $m$ must be an integer, positive or negative, or zero. The equation for $\Theta$ can be treated as follows. 

\hrulefill

\begin{ex}
	Consider the new variable:
\begin{equation}
	x \equiv \cos\theta\;.
\end{equation}
Show that the derivative with respect to $\theta$ satisfies:
\begin{equation}
	\sin\theta\frac{d}{d\theta} = -(1 - x^2)\frac{d}{dx}\;,
\end{equation}
and hence the equation for $\Theta(x)$ can be written as follows:
\begin{equation}\label{Thetaeqx}
	\frac{d}{dx}\left[(1 - x^2)\frac{d\Theta(x)}{dx}\right] - \lambda\Theta(x) -\frac{m^2}{1 - x^2}\Theta(x) = 0\;.
\end{equation}
\end{ex}

\hrulefill

Now, we need $\Theta(x)$ to be regular at $x = \pm 1$, i.e., for $\theta = 0$ or $\theta = \pi$. A way to investigate this is to change the variable:
\begin{equation}
	z = 1 \mp x\;,
\end{equation}
in order to trade the neighborhood of $x = \pm 1$ for that of $z = 0$.

\hrulefill

\begin{ex}
	Obtain the new equation for $\Theta(z)$ and $z = 1 - x$:
	\begin{equation}\label{Thetaeqz}
		z(2 - z)\frac{d^2\Theta}{dz^2} + 2(1 - z)\frac{d\Theta}{dz} - \lambda\Theta -\frac{m^2}{z(2 - z)}\Theta = 0\;.
	\end{equation}
\end{ex}

\hrulefill

Now, using the Frobenius method, we look for a solution of the form:
\begin{equation}
	\Theta(z) = z^s\sum_{n = 0}^\infty a_nz^n\;.
\end{equation}
where $s \ge 0$ in order for $\Theta(z)$ to be regular for $z \to 0$. 

\hrulefill

\begin{ex}
	Substitute the above ansatz in Eq.~\eqref{Thetaeqz} and equate to zero power by power. Show that we get from the lowest power (which is $z^s$) that:
\begin{equation}
	s = \pm \frac{m}{2}\;.
\end{equation}

\end{ex}

\hrulefill

Let us stipulate that $m \ge 0$. Then we must choose the positive sign, i.e. $s = m/2$, since $z^{-m/2} \to \infty$ for $z \to 0$. 

\hrulefill

\begin{ex}
	Carry on a similar analysis for $x = -1$, by transforming variable to $z = 1 + x$. Show that again $s = m/2$.
\end{ex}

\hrulefill

Combining the results from the above exercises, we can conclude that $\Theta(x)$ must have the following form:
\begin{equation}\label{Thetaexprf}
	\Theta(x) = (1 - x^2)^{m/2}f(x)\;,
\end{equation}
where $f(x)$ is an analytic function that is non-vanishing for $x = \pm 1$. We have also learned that $\Theta(x)$ does vanish for $x \pm 1$, except when $m = 0$. Now, let us find an equation for $f$. 

\hrulefill

\begin{ex}
	Substituting Eq.~\eqref{Thetaexprf} into Eq.~\eqref{Thetaeqx} show that:
\begin{equation}\label{fequation}
	(1 - x^2)\frac{d^2f}{dx^2} - 2x(m + 1)\frac{df}{dx} - (\lambda + m + m^2)f = 0\;.
\end{equation}
\end{ex}

\hrulefill

We now prove that $\lambda = -\ell(\ell + 1)$, with $\ell$ being an integer such that $\ell \ge m$. This result will stem from the requirement of regularity of $f$. Adopting the Frobenius method once again, let us stipulate that:
\begin{equation}
	f(x) = x^s\sum_{n = 0}^\infty a_nx^n\;.
\end{equation}

\hrulefill

\begin{ex}
	Substitute the above ansatz into Eq.~\eqref{fequation} and find the following relation:
	\begin{eqnarray}
		\sum_{n = 0}^\infty a_n(n + s)(n + s - 1)x^{n + s - 2} = \nonumber\\
		\sum_{n = 0}^\infty a_n\left[m(m + 1) + \lambda + 2(m + 1)(n + s) + (n + s)(n + s -1)\right]x^{n + s}\;.
	\end{eqnarray}
\end{ex}

\hrulefill

The two series can be combined starting from $n = 2$.

\hrulefill

\begin{ex}
	Show that:
	\begin{eqnarray}
		a_0s(s - 1)x^{s - 2} + a_1s(s + 1)x^{s - 1} + \sum_{n = 2}^\infty a_n(n + s)(n + s - 1)x^{n + s - 2} = \nonumber\\
		\sum_{n = 2}^\infty a_{n - 2}\left[m(m + 1) + \lambda + 2(m + 1)(n + s - 2) + (n + s - 2)(n + s - 3)\right]x^{n + s - 2}\;.\nonumber\\
	\end{eqnarray}
\end{ex}

\hrulefill

Now the two series can be merged into a single one, and equating each power to zero, we have either $s = 0$ or $s = 1$. These conditions lead to the following recursion relations for the series coefficients:
\begin{eqnarray}
	a_{n + 2} = \frac{n(n - 1) + \lambda + m(m + 1) + 2n(m + 1)}{(n + 2)(n + 1)}a_n\;, \qquad (s = 0)\;,\\
	a_{n + 2} = \frac{n(n + 1) + \lambda + m(m + 1) + 2(n + 1)(m + 1)}{(n + 3)(n + 2)}a_n\;, \qquad (s = 1)\;.
\end{eqnarray}
The integral test of convergence fails because $a_{n + 2}/a_n \to 1$ for $n \to \infty$; hence, the series solution would diverge unless:
\begin{eqnarray}
	n(n - 1) + \lambda + m(m + 1) + 2n(m + 1) = 0\;, \quad (s = 0)\;,\\
	n(n + 1) + \lambda + m(m + 1) + 2(n + 1)(m + 1) = 0\;, \quad (s = 1)\;.
\end{eqnarray}
These are constraints on $\lambda$, which then has to assume the following form:
\begin{eqnarray}
	\lambda = -(m + n)(m + n + 1)\;, \quad (s = 0)\;,\\
	\lambda = -(m + n + 1)(m + n + 2)\;, \quad (s = 1)\;,
\end{eqnarray}
or, in general:
\begin{equation}
	\boxed{\lambda \equiv -\ell(\ell + 1)}
\end{equation}
with $\ell$ integer and $\ell \ge m$, which is what we wanted to prove. This result can also be obtained in quantum mechanics by exploiting the commutation relations of the components of the angular momentum operator; see e.g. \cite{weinberg2015lectures}. We have adopted here an approach based on calculus rather than on operator algebra.

The above condition on $\lambda$ implies that the series solution for $f$ terminates after the $(\ell - m)$-th term, and thus $\Theta(x)$ is a polynomial. The polynomials thus obtained for all the possible choices of $(\ell,m)$ are known as \textbf{associated Legendre polynomials}\index{Associated Legendre polynomials} and denoted as $P_\ell^m(x)$. Equation \eqref{Thetaeqx} can thus be written as:
\begin{equation}\label{generalLegendreequation}
	\boxed{\frac{d}{dx}\left[(1 - x^2)\frac{dP_\ell^m(x)}{dx}\right] + \left[\ell(\ell + 1) -\frac{m^2}{1 - x^2}\right]P_\ell^m(x) = 0}
\end{equation}
It is known as the \textbf{general Legendre equation}. By some manipulation, it is possible to obtain it from the \textbf{Legendre equation}:
\begin{equation}\label{Legendreequation}
	\boxed{\frac{d}{dx}\left[(1 - x^2)\frac{d\mathcal P_\ell(x)}{dx}\right] + \ell(\ell + 1)\mathcal P_\ell(x) = 0}
\end{equation}
and therefore, write the associated Legendre polynomials as:
\begin{equation}
	P^m_\ell(x) = (1 - x^2)^{m/2}\frac{d^{m}}{dx^{m}}[\mathcal P_\ell(x)]\;.
\end{equation}
Using Rodrigues' formula\index{Rodrigues' formula}:
\begin{equation}
	\mathcal{P}_\ell(x) = \frac{1}{2^\ell\ell!}\frac{d^{\ell}}{dx^{\ell}}[(x^2 -1)^\ell]\;,
\end{equation}
we have then 
\begin{equation}\label{rodformulaassPlm}
	P^m_\ell(x) = \frac{(1 - x^2)^{m/2}}{2^\ell\ell!}\frac{d^{\ell + m}}{dx^{\ell + m}}[(x^2 -1)^\ell]\;.
\end{equation}
This formula allows us to extend the definition of $P_\ell^m(x)$ to negative values of $m$, such that $\ell \ge |m|$ or $-\ell \le m \le \ell$, as we are accustomed to from the theory of angular momentum in quantum mechanics. In these notes, we have adopted the following convention:
\begin{equation}
	P_\ell^{-m} = \frac{(\ell - m)!}{(\ell + m)!}P_\ell^m\;.
\end{equation}
Another, perhaps more standard, convention includes a $(-1)^m$ factor, but we choose to omit it in order to have the following property for the spherical harmonics under complex conjugation:\index{Spherical harmonics!Complex conjugation}
\begin{equation}
	\boxed{Y_\ell^{m*}(\theta,\phi) = Y_{\ell}^{-m}(\theta,\phi)}
\end{equation}
We can now finally write down the explicit formula for the spherical harmonics:\index{Spherical harmonics!Normalisation}
\begin{eqnarray}
	\boxed{Y_{\ell}^m(\theta,\phi) = \sqrt{\frac{(2\ell + 1)(\ell - m)!}{4\pi(\ell + m)!}}P^m_\ell(\cos\theta)e^{im\phi}}
\end{eqnarray}
where the normalization has been chosen in order to have\index{Spherical harmonics!Orthonormality}
\begin{equation}
	\boxed{\int_0^\pi d\theta\;\sin\theta\int_0^{2\pi} d\phi\;Y_{\ell}^m(\theta,\phi)Y_{\ell'}^{m'*}(\theta,\phi) = \delta_{\ell\ell'}\delta_{mm'}}
\end{equation}
This is probably the single most important relation regarding spherical harmonics because it tells us that they form an \textbf{orthonormal system}. The spherical harmonics satisfy the following \textbf{completeness relation}:\index{Spherical harmonics!Completeness relation}
\begin{equation}
	\sum_{\ell = 0}^\infty\sum_{m = -\ell}^\ell Y_{\ell}^m(\theta,\phi)Y_{\ell}^{m*}(\theta', \phi') = \frac{1}{\sin\theta}\delta(\theta - \theta')\delta(\phi - \phi')\;,
\end{equation} 
Therefore, any square-integrable function $f(\theta,\phi)$ can be expanded as:
\begin{equation}
	f(\theta,\phi) = \sum_{\ell = 0}^\infty\sum_{m = -\ell}^\ell a_{\ell m}Y_{\ell}^m(\theta,\phi)\;,
\end{equation}
in a unique way (meaning that the $a_{\ell m}$'s are unique for that function). The $\delta_{mm'}$ in the orthonormality can be easily understood from the $d\phi$ integration, since:
\begin{equation}
	\int_0^{2\pi} d\phi\;e^{i(m - m')\phi} = \left.\frac{e^{i(m - m')\phi}}{i(m - m')}\right|_0^{2\pi} = 0\;,
\end{equation}
for $m \neq m'$, leaving only the $m = m'$ possibility and the $2\pi$ value of the integral. The $\delta_{\ell\ell'}$ part of the orthonormality relation depends, of course, on the properties of the associated Legendre polynomials, but we do not provide further details here.

It is important to know the parity of the spherical harmonics, i.e., how $Y_\ell^m(\hat n)$ changes under $\hat n \to - \hat n$. In terms of the angles:
\begin{equation}
	\hat n \to -\hat n \quad \Rightarrow \quad (\theta,\phi) \to (\pi - \theta,\pi + \phi)\;.
\end{equation}
Since $\cos(\pi - \theta) = -\cos\theta$, spatial inversion amounts to $x \to -x$ for the associated Legendre polynomial and therefore a $(-1)^{\ell + m}$ factor comes from the derivative in the formula of Eq.~\eqref{rodformulaassPlm}. The exponential $\exp(im\phi)$ instead becomes:
\begin{equation}
	e^{im\phi} \to e^{im(\pi + \phi)} = e^{im\pi}e^{im\phi} = (-1)^m e^{im\phi}\;.
\end{equation} 
Therefore, we finally have:\index{Spherical harmonics!Parity}
\begin{equation}
	\boxed{Y_\ell^m(\pi - \theta,\pi + \phi) = (-1)^\ell Y_\ell^m(\theta,\phi)}
\end{equation} 
A formula that we have intensively employed in these notes is the expansion of a plane wave into spherical harmonics:\index{Spherical harmonics!Plane wave expansion}
\begin{equation}\label{planewaveYexpansion}
	\boxed{e^{i\mathbf k\cdot\mathbf r} = 4\pi\sum_{\ell = 0}^\infty\sum_{m = -\ell}^{\ell}i^{\ell}Y_{\ell}^{m*}(\hat k)Y_{\ell}^{m}(\hat r)j_{\ell}(kr)}
\end{equation}
where the complex conjugation can be switched from one spherical harmonic to the other, since $\mathbf k\cdot\mathbf r = \mathbf r\cdot\mathbf k$. Also, we have made great use of the \textbf{addition theorem}:\index{Spherical harmonics!Addition theorem}
\begin{equation}\label{additiontheorem}
	\boxed{\mathcal P_\ell(\mathbf x\cdot\mathbf y) = \frac{4\pi}{2\ell + 1}\sum_{m = -\ell}^{\ell}Y_{\ell}^{m*}(\mathbf y)Y_{\ell}^{m}(\mathbf x)}
\end{equation}
We can then combine the plane wave expansion with the addition theorem and obtain:
\begin{equation}\label{planewaveYexpansion2}
	e^{i\mathbf k\cdot\mathbf r} = \sum_{\ell = 0}^\infty i^{\ell}(2\ell + 1)\mathcal P_\ell(\hat k\cdot\hat r)j_{\ell}(kr)\;.
\end{equation}
We now turn to the spin-weighted spherical harmonics.

\section{Spin-weighted spherical harmonics}

The usual spherical harmonics allow for the expansion of a scalar quantity such as the temperature on the sphere. Consider a rotation $R$ such that: 
\begin{equation}
	\hat n \to \hat n' = R\hat n\;.
\end{equation}
Since the relative temperature fluctuation is a scalar, we have the following:
\begin{equation}
	\Theta'(\hat{ n}') = \Theta(\hat{ n})\;.
\end{equation}
Using the expansion in spherical harmonics, we can write:
\begin{equation}
	\sum_{\ell m}a_{\ell m}'Y_\ell^m(\hat{ n}') = \sum_{\ell m}a_{\ell m}Y_\ell^m(\hat{ n})\;. 
\end{equation}
Now we take advantage of the properties of the spherical harmonics under spatial rotation:
\begin{equation}
	\boxed{Y_\ell^m(R\hat n) = \sum_{m' = -\ell}^\ell D_{m'm}^{(\ell)}(R^{-1})Y_\ell^{m'}(\hat n)}
\end{equation}
where the $D_{m'm}^{(\ell)}$ are the elements of the \textbf{Wigner D-matrix}.\index{Wigner D-matrix} See \cite{Landau:1991wop} for more details. Hence, we can write:
\begin{equation}
	\sum_{\ell m}a_{\ell m}'Y_\ell^m(\hat{ n}') = \sum_{\ell mm'}a_{\ell m}D_{m'm}^{(\ell)}(R)Y_\ell^{m'}(\hat{ n}')\;. 
\end{equation}
Readjusting the indices, we finally have:
\begin{equation}\label{almtransrotation}
	\boxed{a_{\ell m}' = \sum_{m'}D_{m'm}^{(\ell)}(R)a_{\ell m'}}
\end{equation}
since the spherical harmonics form an orthonormal basis. This is an important relation because it tells us that the angular power spectrum is rotationally invariant:
\begin{eqnarray}
	C_\ell' = \langle a_{\ell m}'a^{'*}_{\ell m}\rangle = \sum_{MM'} D_{mM}^{(\ell)}D_{mM'}^{(\ell)*}\langle a_{\ell M}a^{*'}_{\ell M}\rangle = C_\ell \sum_{M} D_{mM}^{(\ell)}D_{mM}^{(\ell)*} = C_\ell\;.
\end{eqnarray}
The product of Wigner D-matrices being unity since it represents a product of one rotation with its inverse. Hence, we have found that the power spectrum is rotationally invariant. This is a consequence of the fact that $\Theta$ is a scalar; but what about another function that is not? In Sec.~\ref{App:polarization}, we shall see that indeed the Stokes parameters are not scalars: upon a rotation $\hat n \to \hat n' = R\hat n$, we have, in general, that:
\begin{equation}
	Q'(\hat n') \neq Q(\hat n)\;, \qquad U'(\hat n') \neq U(\hat n)\;.
\end{equation}
Therefore, if we expand them on a spherical harmonics basis, we would obtain power spectra depending on the orientation of our coordinate frame.

In order to avoid this, we must employ the \textbf{spin-weighted spherical harmonics}.\index{Spherical harmonics!Spin-weighed} These were introduced in \cite{Newman:1966ub} in the context of gravitational waves, which also have polarization. Using their notation, $\eta$ is a spin-$s$ quantity if it transforms as:
\begin{equation}
	\eta(\hat n) \to \eta(\hat n)' = e^{si\psi}\eta(\hat n)\;,
\end{equation} 
under a rotation of an angle $\psi$ about the line of sight $\hat n$. Then, one defines the differential operators $\eth$ and $\bar\eth$ as:
\begin{eqnarray}\label{ethdefinition}
	\eth\eta = -(\sin\theta)^s\left[\frac{\partial}{\partial\theta} + \frac{i}{\sin\theta}\frac{\partial}{\partial\phi}\right]\left[(\sin\theta)^{-s}\eta\right]\;,\\
	\bar\eth\eta = -(\sin\theta)^{-s}\left[\frac{\partial}{\partial\theta} + \frac{i}{\sin\theta}\frac{\partial}{\partial\phi}\right]\left[(\sin\theta)^{s}\eta\right]\;,
\end{eqnarray}
in spherical coordinates, it is proven that $\eth\eta$ is a spin-$(s + 1)$ quantity, whereas $\bar\eth\eta$ is a spin-$(s - 1)$ quantity. These operators are essentially the covariant derivative on the sphere, as we shall see in Sec.~\ref{App:polarization}. The spin-$s$ spherical harmonics are thus defined as:
\begin{eqnarray}\label{spinweightedsphericalharmonicsdef}
	{}_sY_\ell^m = \sqrt{\frac{(\ell - s)!}{(\ell + s)!}}\eth^sY_\ell^m \qquad (0 \le s \le \ell)\;,\\
	{}_sY_\ell^m = (-1)^s\sqrt{\frac{(\ell + s)!}{(\ell - s)!}}\bar\eth^{-s}Y_\ell^m \qquad (-\ell \le s \le 0)\;.
\end{eqnarray}
The spin-weighted spherical harmonics have fundamental properties similar to those characterizing the usual ones. First of all:
\begin{equation}
	{}_0Y_{\ell}^m(\theta,\phi) = Y_{\ell}^m(\theta,\phi)\;,
\end{equation}
i.e. the usual spherical harmonics can be regarded as spin-0. Then, the ${}_sY_\ell^m$ form a complete orthonormal system; they are orthonormal\index{Spherical harmonics!Orthonormality}
\begin{equation}
	\boxed{\int_0^\pi d\theta\;\sin\theta\int_0^{2\pi} d\phi\;{}_sY_{\ell}^m(\theta,\phi){}_sY_{\ell'}^{m'*}(\theta,\phi) = \delta_{\ell\ell'}\delta_{mm'}}
\end{equation}
and complete:\index{Spherical harmonics!Completeness relation}
\begin{equation}
	\boxed{\sum_{\ell = 0}^\infty\sum_{m = -\ell}^\ell {}_sY_{\ell}^m(\theta,\phi){}_sY_{\ell}^{m*}(\theta', \phi') = \frac{1}{\sin\theta}\delta(\theta - \theta')\delta(\phi - \phi')}
\end{equation} 
and therefore any spin-$s$ quantity $\eta$ can be expanded as:
\begin{equation}
	\boxed{\eta(\theta,\phi) = \sum_{\ell = 0}^\infty\sum_{m = -\ell}^\ell \eta_{\ell m}\;{}_sY_{\ell}^m(\theta,\phi)}
\end{equation}
Under complex conjugation and spatial inversion, the spin-weighted spherical harmonics transform as:
\begin{equation}
	{}_sY_\ell^{m*} = (-1)^s{}_{-s}Y_\ell^{-m}\;, \qquad {}_sY_\ell^{m}(-\hat n) = (-1)^\ell{}_{-s}Y_\ell^{m}(\hat n)\;,
\end{equation}
and they can be related to the Wigner D-matrix as follows:
\begin{equation}
	D_{ms}^{(\ell)}(\phi,\theta,\psi) = \sqrt{\frac{4\pi}{2\ell + 1}}{}_{s}Y_\ell^{-m}(\theta,\phi)e^{is\psi}\;,
\end{equation}
where $(\phi,\theta,\psi)$ are the Euler angles.

\chapter{Polarization}\label{App:polarization}

In this appendix, we review the polarization of an electromagnetic wave. The standard reference is \cite{1998clel.book.....J}. See also \cite{2017inel.book.....G}.

\section{Electromagnetic waves in vacuum}

Let us start with Maxwell equations in vacuum and in Minkowski space:\index{Maxwell's equations}
\begin{equation}\label{Maxwellseqs}
	\nabla\cdot\mathbf E = 0\;,\quad \nabla\times\mathbf E = -\frac{\partial\mathbf B}{\partial t}\;,\quad  \nabla\cdot\mathbf B = 0\;,\quad\nabla\times\mathbf B = \frac{1}{c^2}\frac{\partial\mathbf E}{\partial t}\;.
\end{equation}

\hrulefill

\begin{ex}
	Combine Maxwell equations in order to find the wave equations:
	\begin{equation}\label{waveeqEandB}
		\square\mathbf E = \frac{1}{c^2}\frac{\partial^2\mathbf E}{\partial t^2} - \nabla^2\mathbf E = 0\;, \quad \square\mathbf B = \frac{1}{c^2}\frac{\partial^2\mathbf B}{\partial t^2} - \nabla^2\mathbf B = 0\;.
	\end{equation}
\end{ex}

\hrulefill

The general solutions of the two wave equations are as follows:
\begin{equation}
	\mathbf E(t, \mathbf x) = \mathbf E_0f(\hat{ p}\cdot\mathbf x - ct)\;, \quad \mathbf B(t, \mathbf x) = \mathbf B_0g(\hat{ p}'\cdot\mathbf x - ct)\;,
\end{equation}
where $\hat{ p}$ and $\hat{ p}'$ are, in principle, distinct directions of propagation; $f$ and $g$ are two distinct functions; and $\mathbf E_0$ and $\mathbf B_0$ are the amplitudes of the wave describing the electric field and the magnetic field, respectively.

Now we impose some constraints using Maxwell's equations. We have:
\begin{eqnarray}
	\label{diveE=0eq}\nabla\cdot\mathbf E = 0 \quad\Rightarrow\quad \mathbf E_0\cdot\hat{ p}f' = 0 \quad\Rightarrow\quad \boxed{\mathbf E_0\cdot\hat{ p} = 0}\;,\\
	\label{rotE=0eq}\nabla\times\mathbf E = -\frac{\partial\mathbf B}{\partial t} \quad\Rightarrow\quad \hat{ p}\times\mathbf E_0 f' = -\frac{\partial\mathbf B}{\partial t} \quad\Rightarrow\quad \hat{ p}\times\mathbf E_0 f' = \mathbf B_0 g'c\;, 
\end{eqnarray}
where $f'$ and $g'$ are the derivatives of $f$ and $g$ with respect to their arguments. We assume $f'$ and $g'$ not to be identically vanishing; otherwise, $f$ and $g$ would be constant, and there would be no electromagnetic wave. 

From Eq. \eqref{diveE=0eq}, we learn that the electric field oscillates perpendicularly to the direction $\hat{p}$ of its propagation. 

\hrulefill

\begin{ex}
Show from the other couple of Maxwell equations that:
\begin{eqnarray}
	\label{divB=0eq}\nabla\cdot\mathbf B = 0 \quad\Rightarrow\quad \boxed{\mathbf B_0\cdot\hat{ p}' = 0}\;,\\
	\label{rotB=0eq}\nabla\times\mathbf B = \frac{1}{c^2}\frac{\partial\mathbf E}{\partial t} \quad\Rightarrow\quad \hat{ p}'\times\mathbf B_0 g' = -\frac{1}{c}\mathbf E_0f'\;. 
\end{eqnarray}
Here we learn that also the magnetic field oscillates perpendicularly to the direction $\hat{p}'$ of its propagation.
\end{ex}

\hrulefill

Since $\hat{ p}\cdot(\hat{ p}\times\mathbf E_0) = 0$, we can conclude from Eq. \eqref{rotE=0eq} that $\hat{ p}\times\mathbf B_0 = 0$; therefore, the magnetic field oscillates perpendicularly to both directions of propagation, $\hat{p}$ and $\hat{p}'$. On the other hand:
\begin{equation}
	\hat{ p}'\times(\hat{ p}\times\mathbf E_0)f' = \hat{ p}'\times\mathbf B_0 g'c = -\mathbf E_0f'\;.
\end{equation}
Hence, being $f'$ not identically vanishing, we can conclude that:
\begin{equation}
	\hat{ p}'\times(\hat{p}\times\mathbf E_0) = -\mathbf E_0\;.
\end{equation}

\hrulefill

\begin{ex}
	Use the rule for the triple cross product and show that:
	\begin{equation}
		\hat{p}(\hat{p}\cdot\mathbf E_0) - \mathbf E_0(\hat{p}\cdot\hat{p}') = -\mathbf E_0\;.
	\end{equation}
\end{ex}

\hrulefill

Therefore, being $\hat{p}\cdot\mathbf E_0 = 0$ and $\mathbf E_0$ non-vanishing, we must conclude that:
\begin{equation}
	\hat{p}\cdot\hat{p}' = 1 \quad\Rightarrow\quad \boxed{\hat{p} = \hat{p}'}
\end{equation}
Thus, we have proven that electric and magnetic fields propagate in the same direction and, from Eqs. \eqref{rotE=0eq} or \eqref{rotB=0eq}, that they are perpendicular.

Now we prove that we can take $f = g$. Consider the modulus of Eqs. \eqref{rotE=0eq} or \eqref{rotB=0eq}:
\begin{equation}
	\frac{|\mathbf E_0|}{|\mathbf B_0|c} = \frac{g'}{f'}\;,
\end{equation}
where we have considered $f'$ and $g'$ to be positive, without loss of generality, because a negative sign could be incorporated in $\mathbf E_0$ or $\mathbf B_0$. We have that:
\begin{equation}
	g = \frac{|\mathbf E_0|}{|\mathbf B_0|c}f + K\;,
\end{equation}
where $K$ is an integration constant. Hence, we can write:
\begin{equation}
	\mathbf E = \mathbf E_0f = |\mathbf E_0|\hat{E}_0f\;, \qquad \mathbf B = \mathbf B_0\left(\frac{|\mathbf E_0|}{|\mathbf B_0|c}f + K\right) = \frac{|\mathbf E_0|}{c}\hat{B}_0f + K\mathbf B_0\;.
\end{equation}
Maxwell's equations \eqref{Maxwellseqs} are invariant if we add constant vectors to $\mathbf{E}$ and $\mathbf{B}$. Therefore, we redefine:
\begin{align}
    \mathbf{B} - K\mathbf B_0 \longrightarrow \mathbf{B}\,.
\end{align}
In this way:
\begin{equation}
	\boxed{\mathbf B = \frac{1}{c}\hat{p}\times\mathbf E}
\end{equation}
For this reason, the polarization properties of an electromagnetic wave can be analyzed with respect to the electric field only.

\section{Polarization ellipse and Stokes parameters}

Now we focus on a monochromatic electromagnetic plane wave propagating in the direction $\hat{p} = \hat{ z}$. From Fourier analysis, we know that any wave profile can be realized as a superposition of plane waves.

In electrodynamics, it is often useful to promote the electric field to a complex vector, written as follows:
\begin{equation}
	\mathbf E(t, \mathbf x) = \mathbf E_0 e^{i(z - ct)} = \left(\begin{array}{c}
		E_{x} \\ E_{y} \\ 0
	\end{array}\right)e^{i(z - ct)}\;.
\end{equation}
The two-dimensional vector is called the \textbf{Jones vector}. The real part is the actual electric field:
\begin{equation}
	\mathbf E(t, \mathbf x) = \left[\begin{array}{c}
		E_{x}(t, \mathbf x) \\ E_{y}(t, \mathbf x) \\ 0
	\end{array}\right] = \left[\begin{array}{c}
		A_{x}\cos(z - ct + \phi_x) \\ A_{y}\cos(z - ct + \phi_y) \\ 0
	\end{array}\right]\;.
\end{equation} 
The wave equation \eqref{waveeqEandB} that we solved in the previous section for the electric field is actually three wave equations, one for each component. However, the electromagnetic wave is transverse; therefore, one of the components can always be set to zero, as we have done above by choosing $\hat{p} = \hat{ z}$. We then have two second-order differential equations; thus, we expect 4 integration constants, which we have introduced above as the amplitudes $A_x$ and $A_y$ and the phases $\phi_x$ and $\phi_y$. The polarization state of the wave depends on these four quantities. 

We can always redefine our clock such that:
\begin{equation}
	\mathbf E = \left[\begin{array}{c}
		E_{x}(t, \mathbf x) \\ E_{y}(t, \mathbf x) \\ 0
	\end{array}\right] = \left[\begin{array}{c}
		A_{x}\cos(z - ct) \\ A_{y}\cos(z - ct + \beta) \\ 0
	\end{array}\right]\;,
\end{equation} 
where $\beta \equiv \phi_y - \phi_x$ is the relative phase, which is indeed the relevant physical quantity. So we actually have 3 independent parameters that describe polarization. When $\beta = 0$, the polarization is purely linear, whereas when $\beta = \pi/2$, it is purely circular. This can generally be seen by studying the time-evolution of the field in the $E_y-E_x$ plane. In fact, we have:
\begin{equation}
	\boxed{\frac{E_x^2}{A_x^2} + \frac{E_y^2}{A_y^2} - \frac{2E_xE_y}{A_xA_y}\cos\beta = \sin^2\beta}
\end{equation}
This is the equation of an ellipse, the \textbf{polarization ellipse}, rotated by an angle $\alpha$ in the $E_y-E_x$ plane.\index{Polarization ellipse} 

\hrulefill

\begin{ex}
	Defining:
	\begin{equation}
		A_x = A\cos\theta\;, \qquad A_y = A\sin\theta\;,
	\end{equation}
	show that:
	\begin{equation}
		\tan2\alpha = \tan2\theta\cos\beta\;, 
	\end{equation}
	and that the semi-major and semi-minor axis of the polarisation ellipse are given by:
	\begin{eqnarray}
		a^2 = \frac{A^2}{2}\left(1 + \sqrt{1 - \sin^22\theta\sin^2\beta}\right)\;,\\
		b^2 = \frac{A^2}{2}\left(1 - \sqrt{1 - \sin^22\theta\sin^2\beta}\right)\;.
	\end{eqnarray}
\end{ex}

\hrulefill

When $\beta = 0$ we have $b = 0$: the ellipse degenerates into a straight line, tilted at an angle $\alpha$ with respect to the $E_x$ axis. In this case, the wave is purely linearly polarized. On the other hand, when $\beta = \pi/2$ the ellipse degenerates into a circle: we have a purely circularly polarized wave.

The \textbf{Stokes parameters} are defined as follows in terms of the polarization ellipse parameters:\index{Stokes parameters}
\begin{eqnarray}
	I \equiv a^2 + b^2 = A^2\;,\\
	Q \equiv (a^2 - b^2)\cos2\alpha = A^2\sqrt{1 - \sin^22\theta\sin^2\beta}\cos2\alpha = A^2\cos2\theta\;,\\
	U \equiv (a^2 - b^2)\sin2\alpha = A^2\sqrt{1 - \sin^22\theta\sin^2\beta}\sin2\alpha = A^2\sin2\theta\cos\beta\;,\\
	V \equiv 2ab h = (A^2\sin2\theta\sin\beta)h\;,
\end{eqnarray}
where $h = \pm 1$ simply establishes the direction of rotation: clockwise ($h = -1$) or counterclockwise ($h = 1$). The Stokes parameter $I$ is the intensity of the wave.

\paragraph{Pure $Q$ polarization.} If $U = V = 0$, then $\beta = 0$, and therefore we have pure linear polarization. Since $\sin2\alpha = 0$, either $\alpha = 0$ or $\alpha = \pi/2$. In the former case, the electric field oscillates along the $x$-axis and $Q = A^2 = I > 0$. If $\alpha = \pi/2$, then the electric field oscillates along the $y$-axis and $Q = -A^2 = -I < 0$.

\paragraph{Pure $U$ polarization.} If $Q = V = 0$, then $\beta = 0$, and therefore we have again pure linear polarization. This time $\cos2\alpha = 0$, then either $\alpha = \pi/4$ or $\alpha = 3\pi/4$. In the former case, the electric field oscillates along the $y = x$ line and $Q = A^2 = I > 0$. If  $\alpha = 3\pi/4$, then the electric field oscillates along the $y = -x$ line and $Q = -A^2 = -I < 0$.

\paragraph{Pure $V$ polarization.} If $Q = U = 0$, then $\theta = \pi/4$ and $\beta = \pi/2$; therefore, we have pure circular polarization. We have $V = A^2h = \pm A^2 = \pm I$, with the sign determined by the direction of rotation.

As we have seen, the polarization is fully determined by 3 parameters. This means that the 4 Stokes parameters are not independent.

\hrulefill

\begin{ex}
	Show that the Stokes parameters satisfy the following constraint:
	\begin{equation}
		\boxed{Q^2 + U^2 + V^2 = I^2}
	\end{equation}
\end{ex}

\hrulefill

Other definitions of the Stokes parameters are as follows. Using the complex electric field:
\begin{equation}
	\mathbf E(t, \mathbf x) = \left(\begin{array}{c}
		E_{x} \\ E_{y} \\ 0
	\end{array}\right)e^{i(z - ct)}\;,
\end{equation}
they are defined as:
\begin{eqnarray}
	I \equiv |E_x|^2 + |E_y|^2\;,\quad Q \equiv |E_x|^2 - |E_y|^2\;,\\
	U \equiv 2\mbox{Re}(E_xE_y^*)\;,\quad V \equiv -2\mbox{Im}(E_xE_y^*)\;.
\end{eqnarray}
We have considered so far a monochromatic wave, possibly made up of many waves, but all coherent. On the other hand, if we consider the superposition of incoherent waves, then we can write a generic time-dependence for the electric field:
\begin{equation}
	\mathbf E(t,\mathbf x) = \left[\begin{array}{c}
		E_{x}(t, \mathbf x) \\ E_{y}(t, \mathbf x) \\ 0
	\end{array}\right]\;.
\end{equation} 
In this case, the Stokes parameters are defined as time averages:
\begin{eqnarray}
	I \equiv \langle E_x^2\rangle + \langle E_y^2\rangle\;,\quad Q \equiv \langle E_x^2\rangle - \langle E_y^2\rangle\;,\\
	U \equiv 2\langle E_xE_y\rangle\cos\beta\;,\quad V \equiv 2\langle E_xE_y\rangle\sin\beta\;,
\end{eqnarray}
where: 
\begin{equation}
	\langle X\rangle = \frac{1}{T}\int_0^Tdt\;X(t)\;.
\end{equation}
Moreover, one has the following:
\begin{equation}
		\boxed{Q^2 + U^2 + V^2 \le I^2}
	\end{equation}
since each Stokes parameter is defined as the sum of the corresponding Stokes parameters of the individual components of the wave. See, for example, \cite{1960ratr.book.....C}.

\hrulefill

\begin{ex}
	Show that the three definitions of Stokes parameters given above are in fact equivalent for a single monochromatic sinusoidal wave.
\end{ex}

\hrulefill

Now, suppose we make successive counterclockwise rotations of $\pi/4$ in the polarization plane. What happens is the following:
\begin{equation}
	Q \to U \to -Q \to -U \to Q\;.
\end{equation}
That is, after a rotation of $\pi$ about the propagation direction, we recover the initial polarization state as if we had applied an identity operator. This suggests that we are dealing with a spin-2 field. Indeed, let us introduce the intensity matrix for an electromagnetic wave propagating along $-\hat z$:
\begin{equation}
	J_{ij}(-\hat z) = \frac{1}{2}\left(\begin{array}{ccc}
		I(\hat z) + Q(\hat z) & U(\hat z) - iV(\hat z) & 0\\
		U(\hat z) + iV(\hat z) & I(\hat z) - Q(\hat z) & 0\\
		0 & 0 & 0
	\end{array}\right)\;,
\end{equation}
where the factor $1/2$ serves to reproduce the correct result that the trace $J_{ii} = I$. This intensity matrix reminds us of $h_{ij}^T$, the tensor perturbation of the metric, defined in Eq.~\eqref{tensorperthphx}. 

\hrulefill

\begin{ex}
	Show that applying the a rotation of an angle $\theta$ about the axis $\hat z$ we have that:
	\begin{eqnarray}
	I \to I\;,\\
\label{QpmiUtransfrot}		Q \pm iU \to e^{\pm 2i\theta}(Q \pm iU)\;,\\
		V \to V\;.
	\end{eqnarray}
\end{ex}

\hrulefill

The different sign with respect to the GW case is due to the fact that the rotation is made about the propagation direction $-\hat z$. Hence, a rotation by $\theta$ about $-\hat z$ corresponds to a rotation by $-\theta$ about $\hat z$, which is the line-of-sight direction.

The matrix defined on the two-dimensional subspace orthogonal to the propagation direction is:
\begin{equation}
	\rho_{ij}(-\hat z) = \frac{1}{2}\left(\begin{array}{cc}
		I(\hat z) + Q(\hat z) & U(\hat z) - iV(\hat z)\\
		U(\hat z) + iV(\hat z) & I(\hat z) - Q(\hat z)
	\end{array}\right)\;.
\end{equation}
It is called \textbf{the photon density matrix}\index{Photon density matrix} and can be expressed in the following form:
\begin{equation}
	\rho = \frac{1}{2}(I\mathbf{1} + Q\sigma_3 + U\sigma_1 + V\sigma_2)\;,
\end{equation}
where $\mathbf 1$ is the $2\times 2$ identity matrix and the $\sigma_{1,2,3}$ are the Pauli matrices. It might seem strange that these could be associated with a photon since they are usually employed for the description of spin-$1/2$ particles, whereas the photon is a spin-1 boson. However, since the photon is massless, it is characterized by only two spin states $\pm 1$ (or helicities), just as any spin-$1/2$ particle, whether massive or massless, is characterized by two spin states $\pm 1/2$.

The polarization vectors introduced in Eq.~\eqref{polarizationvectors} can be defined equivalently here. Along the $\hat z$ axis, they are simply:\index{Polarization vectors}
\begin{equation}
	e_{\pm}(\hat z) = (1,\pm i, 0)/\sqrt{2}\;,
\end{equation}
and the Stokes parameters are defined as:
\begin{eqnarray}
	Q(\hat z) \pm iU(\hat z) = 2e_{\pm,i}(\hat z)e_{\pm,j}(\hat z)J_{ij}(-\hat z)\;.
\end{eqnarray}
For a generic direction of propagation, say $-\hat n$, we define:
\begin{eqnarray}
	Q(\hat n) \pm iU(\hat n) = 2e_{\pm,i}(\hat n)e_{\pm,j}(\hat n)J_{ij}(-\hat n)\;,
\end{eqnarray}
where in spherical coordinates:
\begin{eqnarray}
	\hat n = (\sin\theta\cos\phi,\sin\theta\sin\phi,\cos\theta)\;.
\end{eqnarray}
The basis vectors that form the polarization vectors can be chosen as follows:
\begin{eqnarray}
	\hat \theta = (\cos\theta\cos\phi,\cos\theta\sin\phi,-\sin\theta)\;, \qquad \hat \phi = (-\sin\phi,\cos\phi,0)\;.
\end{eqnarray}
Hence:
\begin{eqnarray}
	e_{\pm}(\hat n) = \frac{\hat\theta \pm i\hat\phi}{\sqrt{2}}\;.
\end{eqnarray}
As we have already discussed, $Q \pm iU$ are spin-$2$ fields and thus are expanded as:
\begin{equation}
	(Q + iU)(\hat n) = \sum_{\ell m}a_{P,\ell m}\;{}_2Y_\ell^m(\hat n)\;.
\end{equation}
Through the polarization vectors, we can characterize the operator $\eth$ as 
\begin{equation}\label{ethtildenablarelation}
	\boxed{\eth^s = (-1)^s2^{s/2}e_{+,i_1}(\hat n)\cdots e_{+,i_s}(\hat n)\tilde\nabla_{i_1}\cdots\tilde\nabla_{i_s}}
\end{equation}
where:
\begin{equation}
	\tilde\nabla = \hat\theta\frac{\partial}{\partial\theta} + \frac{\hat\phi}{\sin\theta}\frac{\partial}{\partial\phi}\;,
\end{equation}
is the covariant derivative on the sphere.

\hrulefill

\begin{ex}
	Show that the definition of Eq.~\eqref{ethdefinition} and formula \eqref{ethtildenablarelation} are identical for $s = 1$ and $s = 2$.
\end{ex}

\hrulefill

\chapter{Thomson scattering}\label{App:Thomsonscattering}

In this Appendix, we work out the collisional term of Eq.~\eqref{photonBoltzmannequationfull}. We follow the calculations of \cite{1960ratr.book.....C}, but we take advantage of the properties of spherical harmonics, as done in \cite{Hu:1997hp}. See also \cite{Kosowsky:1994cy}.

Consider an incoming wave propagating in the direction $\hat p'$ and scattered in the direction $\hat p$. The plane in which $\hat p'$ and $\hat p$ lie is called \textbf{the scattering plane}. 

\begin{figure}[ht]
\centering
	\begin{tikzpicture}
	\draw[-] (-0.25,0) -- (7,0);
	\draw[->] (0,-0.25) -- (0,0.75);
	\draw (0,0.75) node[above] {$\hat x'$};
	\fill (0,0) circle (2pt) node[below left] {$\hat y'$};
	\draw[->] (-0.25,0) -- (0.75,0);
	\draw (1.25,0) node[above] {$\hat p' = \hat z'$};
	\draw[->] (3.5,0) -- (5.5, 2);
	\fill (3.5,0) circle (2pt) node[below] {$e^-$};
	\draw (5.5,2) node[above] {$\hat p$};
	\draw (3,0) arc (180:45:0.5);
	\draw (3,0.4) node[above] {$\beta$};
	\fill (5,1.5) circle (2pt) node[below] {$\hat y$};
	\draw[->] (5,1.5) -- (4.5, 2);
	\draw (4.5,2) node[left] {$\hat x$};
\end{tikzpicture}
\caption{Scattering plane. The unit vectors $\hat y'$ and $\hat y$ go out from the page orthogonally, towards the reader.}
\label{Fig:scatteringplane}
\end{figure}

Choosing the simple geometry of Fig.~\ref{Fig:scatteringplane}, the incoming electric field can be decomposed as follows:
\begin{equation}
	\mathbf E' = E'_x\hat x' + E'_y\hat y'\;.
\end{equation}
When the electric field interacts with the electron, it begins to oscillate and emit electromagnetic waves in almost all directions. The ``almost'' is made quantitative by \textbf{the Larmor formula},\index{Larmor formula} which establishes that the irradiated power per solid angle is:
\begin{equation}
	\frac{dP}{d\Omega} = \frac{e^2}{4\pi c^3}\left|\hat p\times(\hat p \times \mathbf a)\right|^2\;,
\end{equation}
where $\mathbf a$ is the electron acceleration. See, for example, \cite{1998clel.book.....J} and note that here we are using Gaussian units for which $[F] = [e^2/r^2]$, i.e., the force can be dimensionally expressed as squared Coulomb divided by a squared length. For a particular outgoing polarization, say $\hat\epsilon$, we make the scalar product with $\hat\epsilon$ in the square modulus of the above formula and obtain:
\begin{equation}
	\frac{dP}{d\Omega} = \frac{e^2}{4\pi c^3}\left|\hat\epsilon\cdot\mathbf a\right|^2\;.
\end{equation}
The acceleration of the electron is produced by the incident electric field $\mathbf E'$:
\begin{equation}
	\mathbf a = \frac{e}{m}\mathbf E'\;,
\end{equation}
so that:
\begin{equation}
	\frac{dP}{d\Omega} = \frac{e^4}{4\pi m^2c^3}\left|\hat\epsilon\cdot\mathbf E'\right|^2\;.
\end{equation}
Writing now $\mathbf E' = \hat\epsilon'|\mathbf E'|$, we obtain:
\begin{equation}
	\frac{dP}{d\Omega} = \frac{c}{4\pi}|\mathbf E'|^2\left(\frac{e^2}{mc^2}\right)^2\left|\hat\epsilon\cdot\hat\epsilon'\right|^2\;.
\end{equation}
The contribution $c|\mathbf E'|^2/(4\pi)$ is the Poynting vector and, hence, the flux of the incoming wave, leaving:
\begin{equation}
	\boxed{\frac{d\sigma}{d\Omega} = \left(\frac{e^2}{mc^2}\right)^2\left|\hat\epsilon\cdot\hat\epsilon'\right|^2 \equiv \frac{3\sigma_{\rm T}}{8\pi}\left|\hat\epsilon\cdot\hat\epsilon'\right|^2}
\end{equation}
as the differential cross section, where of course $\sigma_{\rm T}$ is the \textbf{Thomson cross section}.\index{Thomson scattering!Cross section} The above derivation is based on the Larmor formula and is thus purely classical, valid only for a non-relativistic motion of the electron and for photon energies much smaller than the electron mass. If these conditions are not met, one should employ the Klein-Nishina formula; see, for example, \cite{Weinberg:1995mt}.

Let us assume $\sigma_{\rm T} = 1$ for simplicity. We shall recover the correct dimensions only at the end of our derivation via the scattering rate $-\tau' = n_e\sigma_{\rm T}a$. The electric field scattered in the direction $\hat p$ is of the form:
\begin{equation}
	\mathbf E = A\left[\hat p\times(\hat p \times \mathbf E')\right]\;,
\end{equation}
where $A^2 = 3/(8\pi)$. Let us now use the geometry in the scattering plane of Fig.~\ref{Fig:scatteringplane} and calculate the double cross product.

\hrulefill

\begin{ex}
	Show that:
	\begin{equation}
		\mathbf E = -A\left(\hat x'E_x'\cos^2\beta + \hat y'E_y' + \hat z'E_x'\sin\beta\cos\beta\right)\;.
	\end{equation}
\end{ex}

\hrulefill

Since $\hat y = \hat y'$, we can already conclude that:
\begin{equation}
	E_y = -AE_y'\;,
\end{equation}
i.e., the electric field contribution perpendicular to the scattering plane gains no angular dependence.

In order to calculate $E_x$, which is the contribution of the electric field parallel to the scattering plane, we need to know $\hat x$. One has:
\begin{equation}
	\hat x = \hat y \times \hat p\;,
\end{equation}
and
\begin{equation}
	\hat p = (\sin\beta, 0,-\cos\beta)\;.
\end{equation}
Hence:
\begin{equation}
	\hat x = -\hat x'\cos\beta - \hat z'\sin\beta\;.
\end{equation}
Therefore, we have:
\begin{equation}
	E_x = \mathbf E\cdot\hat x = AE_x'\cos\beta\;.
\end{equation}
We are now in the position of relating the Stokes parameters of the outgoing wave to those of the incoming one:
\begin{eqnarray}
	I = \langle E_x^2\rangle + \langle E_y^2\rangle = A^2\langle E_x^{'2}\rangle\cos^2\beta + A^2\langle E_{y'}^2\rangle \equiv A^2I_x'\cos^2\beta + A^2I_y'\;,\\ 
	Q = \langle E_x^2\rangle - \langle E_y^2\rangle = A^2I_x'\cos^2\beta - A^2I_y'\;,\\
	U = 2\langle E_xE_y\rangle\cos\beta = A^2U'\cos\beta\;,\\
	V = 2\langle E_xE_y\rangle\sin\beta = A^2V'\cos\beta\;. 
\end{eqnarray}
Using the definitions:
\begin{eqnarray}
	I' = \langle E_x^{'2}\rangle + \langle E_{y'}^2\rangle\;,\\
	Q' = \langle E_x^{'2}\rangle - \langle E_{y'}^2\rangle\;,
\end{eqnarray}
we have the complete transformation:
\begin{equation}
	\left(\begin{array}{c}
		I\\ Q\\ U\\ V 
	\end{array}\right) = \frac{3}{8\pi}\left(\begin{array}{cccc}
		\frac{1 + \cos^2\beta}{2} & -\frac{\sin^2\beta}{2} & 0 & 0\\
		-\frac{\sin^2\beta}{2} & \frac{1 + \cos^2\beta}{2} & 0 & 0\\
		0 & 0 & \cos\beta & 0\\
		0 & 0 & 0 & \cos\beta
	\end{array}\right)\left(\begin{array}{c}
		I'\\ Q'\\ U'\\ V' 
	\end{array}\right)\;.
\end{equation}
These Stokes parameters have the dimension of intensity. We now introduce $\Theta$, the temperature fluctuation instead of $I$, and suitably redefine all the other Stokes parameters. Moreover, we introduce the combination $Q \pm iU$, since we know how it transforms under a rotation in the polarization plane, cf. Eq.~\eqref{QpmiUtransfrot}. We shall need this rotation in a moment. We have then:
\begin{equation}
	\left(\begin{array}{c}
		\Theta\\ Q + iU\\ Q - iU 
	\end{array}\right) = \frac{3}{16\pi}\left(\begin{array}{ccc}
		1 + \cos^2\beta & -\frac{\sin^2\beta}{2} & -\frac{\sin^2\beta}{2}\\
		-\sin^2\beta & \frac{(1 + \cos\beta)^2}{2} & \frac{(1 - \cos\beta)^2}{2}\\
		-\sin^2\beta & \frac{(1 - \cos\beta)^2}{2} & \frac{(1 + \cos\beta)^2}{2} 
	\end{array}\right)\left(\begin{array}{c}
		\Theta'\\ Q' + iU'\\ Q' - iU' 
	\end{array}\right)\;,
\end{equation}
and $V$ does not mix with the other Stokes parameters, so we consider it separately:
\begin{equation}\label{Vscatteredeqcosbeta}
	V = \frac{3\cos\beta}{16\pi}V'\;.
\end{equation}
Recall that the above transformations hold true in the scattering plane, and the primed quantities depend on the incoming direction $\hat p'$, whereas the unprimed quantities depend on $\hat p$. 

\begin{figure}[ht]
\centering
\tdplotsetmaincoords{70}{100}
\begin{tikzpicture}[tdplot_main_coords]
	\coordinate (O) at (0,0,0);
	\draw[->] (0,0,0) -- (0,0,5) node[right] {$z$};
	\draw[->] (0,0,0) -- (0,5,0) node[right] {$y$};
	\draw[->] (0,0,0) -- (5,0,0) node[left] {$x$};
	\tdplotdrawarc{(O)}{4.5}{0}{90}{}{}
	\tdplotdrawarc{(0,0,3.897)}{2.25}{0}{90}{}{}
	\tdplotdrawarc{(0,0,2.25)}{3.897}{0}{90}{}{}
	\tdplotsetthetaplanecoords{0}
	\tdplotdrawarc[tdplot_rotated_coords]{(O)}{4.5}{0}{90}{}{}
	\tdplotsetthetaplanecoords{22.5}
	\tdplotdrawarc[tdplot_rotated_coords]{(O)}{4.5}{0}{90}{}{}
	\tdplotsetthetaplanecoords{45}
	\tdplotdrawarc[tdplot_rotated_coords]{(O)}{4.5}{0}{90}{}{}
	\tdplotsetthetaplanecoords{67.5}
	\tdplotdrawarc[tdplot_rotated_coords]{(O)}{4.5}{0}{90}{}{}
	\tdplotsetthetaplanecoords{90}
	\tdplotdrawarc[tdplot_rotated_coords]{(O)}{4.5}{0}{90}{}{}
	\tdplotsetcoord{P}{4.5}{30}{22.5}
	\draw[-,thick] (O) -- (P) node[above] {$(\theta,\phi)$};
	\tdplotsetcoord{Q}{4.5}{60}{67.5}
	\draw[-,thick] (O) -- (Q) node[right] {$(\theta',\phi')$};
	\tdplotdefinepoints(0,0,0)(2.0787,0.861,3.897)(1.491,3.6,2.25)
	\tdplotdrawpolytopearc[thick]{4.5}{}{}
	\tdplotdrawpolytopearc[thick]{0.5}{above right}{$\beta$}
	\tdplotsetrotatedcoords{30}{22.5}{0}
    \tdplotsetrotatedcoordsorigin{(P)}
    \tdplotdrawarc[tdplot_rotated_coords,color=blue,thick]{(0,0,0)}{0.5}{-10}{50}{anchor=north west,color=blue}{$\gamma$}
    \tdplotsetrotatedcoordsorigin{(Q)}
    \tdplotdrawarc[tdplot_rotated_coords,color=red,thick]{(0,0,0)}{0.6}{180}{200}{anchor=south east,color=red}{$\alpha$}
\end{tikzpicture}
\caption{Scattering geometry.}
\label{Fig:scatteringgeometry}
\end{figure}
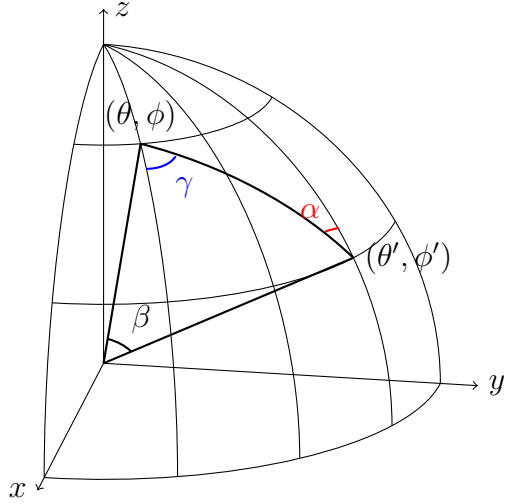

We now have to transform to a generic reference frame in which the incident wave comes from a direction $(\theta',\phi')$ and the outgoing wave goes along a direction $(\theta,\phi)$. This calculation is performed in \cite{1960ratr.book.....C}, but we follow \cite{Hu:1997hp} because it is faster and takes advantage of the properties of the spherical harmonics that we have seen earlier. In particular, we refer to Fig.~\ref{Fig:scatteringgeometry}.

We need to perform a rotation $R(-\alpha)$ in order to pass from the laboratory frame to the scattering frame. Recall that the rotation is performed in the polarization plane, i.e., about $-\hat p'$. Hence, it is a clockwise rotation, and for this reason, we have $-\alpha$. Then, we apply the above matrix, say $S(\beta)$, in order to obtain the outgoing Stokes parameters. Finally, we apply another rotation $R(-\gamma)$ in order to obtain the Stokes parameters in the laboratory frame. This rotation seems to be anti-clockwise because it is in the direction opposite to $R(-\alpha)$. However, notice that $\hat p$ is now outgoing and thus the polarization plane is reflected, giving $R(-\gamma)$ instead of $R(\gamma)$.

\hrulefill

\begin{ex}
	Using the transformation of Eq.~\eqref{QpmiUtransfrot} and the fact that $\Theta$ is invariant under rotation in the polarisation plane, show that:
\begin{equation}
	R(-\gamma)S(\beta)R(-\alpha) = \frac{3}{16\pi}\left(\begin{array}{ccc}
		1 + \cos^2\beta & -\frac{\sin^2\beta}{2}e^{-2i\alpha} & -\frac{\sin^2\beta}{2}e^{2i\alpha}\\
		-e^{-2i\gamma}\sin^2\beta & e^{-2i\gamma}\frac{(1 + \cos\beta)^2}{2}e^{2i\alpha} & e^{-2i\gamma}\frac{(1 - \cos\beta)^2}{2}e^{2i\alpha}\\
		-e^{2i\gamma}\sin^2\beta & e^{2i\gamma}\frac{(1 - \cos\beta)^2}{2}e^{-2i\alpha} & e^{2i\gamma}\frac{(1 + \cos\beta)^2}{2}e^{2i\alpha}
	\end{array}\right)\;,
\end{equation}	
\end{ex}

\hrulefill

Since $V$ is also invariant under rotation in the polarization plane, Eq.~\eqref{Vscatteredeqcosbeta} is valid even in the laboratory frame.

\hrulefill

\begin{ex}
	Using the definitions of spherical harmonics given in Tab.~\ref{Tab:SphericalHarmonics}, write the above matrix as:
\begin{equation}
	R(\gamma)S(\beta)R(-\alpha) = \frac{1}{8\pi}\sqrt{\frac{4\pi}{5}}\left(\begin{array}{ccc}
		Y_2^0 + 2\sqrt{5}Y_0^0 & -\sqrt{3/2}Y_2^{-2} & -\sqrt{3/2}Y_2^2\\
		-\sqrt{6}e^{-2i\gamma}{}_2Y_2^0 & 3e^{-2i\gamma}{}_2Y_2^{-2} & 3e^{-2i\gamma}{}_2Y_2^{2}\\
		-\sqrt{6}e^{2i\gamma}{}_{-2}Y_2^0 & 3e^{2i\gamma}{}_{-2}Y_2^{-2} & 3e^{2i\gamma}{}_{-2}Y_2^{2}
	\end{array}\right)\;.
\end{equation}
The spherical harmonics inside this matrix are function of $(\beta,\alpha)$. For $V$ show that:
\begin{equation}\label{Vscatteredeqcosbeta2}
	V = \frac{1}{4}\sqrt{\frac{3}{4\pi}}Y_1^0(\beta,\alpha)V'\;.
\end{equation}
\end{ex}

\hrulefill

Now, using the addition theorem:
\begin{equation}
	\sum_{m = -\ell}^\ell{}_{s_1}Y_\ell^{m*}(\theta',\phi'){}_{s_2}Y_\ell^{m}(\theta,\phi) = \frac{2\ell + 1}{4\pi}{}_{s_2}Y_\ell^{-s_1}e^{-is_2\gamma}\;,
\end{equation}
we can introduce the matrix:
\begin{equation}
	P^{(m)}(\hat p,\hat p') = \left(\begin{array}{ccc}
		Y_2^{m*}Y_2^m & -\sqrt{3/2}{}_2Y_2^{m*}Y_2^m & -\sqrt{3/2}{}_{-2}Y_2^{m*}Y_2^m\\
		-\sqrt{6}Y_2^{m*}{}_2Y_2^m & 3{}_2Y_2^{m*}{}_2Y_2^m & 3{}_{-2}Y_2^{m*}{}_2Y_2^m\\
		-\sqrt{6}Y_2^{m*}{}_{-2}Y_2^m & 3{}_2Y_2^{m*}{}_{-2}Y_2^m & 3{}_{-2}Y_2^{m*}{}_{-2}Y_2^m
	\end{array}\right)\;,
\end{equation}
where the complex conjugate spherical harmonics depend on $\hat p'$, whereas the others depend on $\hat p$. Hence, the scattered Stokes parameters can then be written as:
\begin{equation}
	\left(\begin{array}{c}
		\Theta\\ Q + iU\\ Q - iU 
	\end{array}\right) = \frac{1}{4\pi}\left(\begin{array}{c}
		\Theta'\\ 0\\ 0 
	\end{array}\right) + \frac{1}{10}\sum_{m = -2}^2P^{(m)}(\hat p,\hat p')\left(\begin{array}{c}
		\Theta'\\ Q' + iU'\\ Q' - iU' 
	\end{array}\right)\;,
\end{equation}
whereas for $V$ we can write:
\begin{equation}
	V(\hat p) = \frac{1}{4}\sum_{m = -1}^1Y_1^m(\hat p)Y_1^{m*}(\hat p')V(\hat p')\;.
\end{equation}
Integrating over all the incoming directions and multiplying it by the Thomson scattering rate:
\begin{equation}
	-\tau' = n_e\sigma_{\rm T}a\;,
\end{equation}
we obtain part of the collisional terms for the Boltzmann equation \eqref{photonBoltzmannequationfull}. This is the scattering rate into photons with momentum in the direction $\hat p$, which we have taken as the reference direction. There is also a contribution that takes into account the scattering of photons into directions different from $\hat p$. This generates the term:
\begin{equation}
	\tau'\left(\begin{array}{c}
		\Theta\\ Q + iU\\ Q - iU 
	\end{array}\right)\;.
\end{equation}
Finally, all the calculations performed here assume the electron fluid to be at rest. Performing a boost, we get the Doppler shift term $-\tau'\hat p\cdot\mathbf v_{\rm b}$. Indeed, the photon energy in the new frame is given by:
\begin{equation}
	\tilde p = p(1 - \mathbf v_{\rm b}\cdot\hat p)\;,
\end{equation}
upon a boost with velocity $\mathbf v_{\rm b}$. The Lorentz factor on the right hand side is unity since $|\mathbf v_{\rm b}|^2$ is negligible, being a first-order quantity. The perturbed distribution $\mathcal F_\gamma$, being a scalar, transforms as:
\begin{equation}
	\mathcal F_\gamma(p) = \tilde{\mathcal F}_\gamma[p(1 - \mathbf v_{\rm b}\cdot\hat p)]\;,
\end{equation}
Thus, by developing the distribution function up to first-order, we obtain:
\begin{equation}
	\mathcal F_\gamma(p) = \tilde{\mathcal F}_\gamma(p) - \frac{\partial\bar f_\gamma}{\partial p}p\mathbf v_{\rm b}\cdot\hat p\;,
\end{equation}
and using Eq.~\eqref{Thetadefinition} we have that:
\begin{equation}
	\Theta = \tilde\Theta + \mathbf v_{\rm b}\cdot\hat p\;.
\end{equation}
We have thus fully recovered the collisional term of Eq.~\eqref{photonBoltzmannequationfull}.

Finally, consider initially unpolarized photons. Then, upon scattering:
\begin{eqnarray}
	\left(\begin{array}{c}
		\Theta\\ Q + iU\\ Q - iU 
	\end{array}\right)(\hat p) = \left(\begin{array}{c}
		\int\frac{d^2\hat p'}{4\pi}\Theta(\hat p')\\ 0\\ 0 
	\end{array}\right)\nonumber\\ + \frac{1}{10}\sum_{m = -2}^2\left(\begin{array}{c}
		Y_2^m\\
		-\sqrt{6}{}_2Y_2^m\\
		-\sqrt{6}{}_{-2}Y_2^m
	\end{array}\right)(\hat p)\int d^2\hat p'Y_2^{m*}(\hat p')\Theta(\hat p')\;.
\end{eqnarray}
Expanding the relative temperature fluctuations in spherical harmonics:
\begin{equation}
	\Theta(\hat p') = \sum_{\ell m}a_{\ell m}Y_\ell^m(\hat p')\;,
\end{equation}
due to the orthogonality properties of the latter, polarization will be produced only if the incident $\Theta(\hat p')$ has a quadrupole moment.

\chapter{Conservation of \texorpdfstring{$\mathcal{R}$}{R} on large scales and for adiabatic perturbations}\label{App:Rconslargescales}

In order to show the constancy of $\mathcal{R}$ on large scales and for adiabatic perturbations, let us start from Eq.~\eqref{0iEinsteineq} and combine its scalar part with the definition of $\mathcal R$ given in Eq.~\eqref{Rperturb}:
\begin{equation}\label{RPhirhoPrel}
	\mathcal{R} = \Phi + \mathcal H\frac{\Phi' - \mathcal H\Psi}{4\pi Ga^2(\rho + P)}\;.
\end{equation}

\hrulefill

\begin{ex}
Differentiate Eq.~\eqref{Rperturb} with respect to the conformal time and show that:
\begin{eqnarray}\label{RPhirelder}
	4\pi Ga^2(\rho + P)\mathcal{R}' = 4\pi Ga^2(\rho + P)\Phi' + \mathcal{H}'\left(\Phi' - \mathcal{H}\Psi\right)\nonumber\\ + \mathcal{H}\left(\Phi'' - \mathcal{H}\Psi' - \mathcal{H}'\Psi\right)	+ \mathcal{H}^2\left(\Phi' - \mathcal{H}\Psi\right) + 3\mathcal{H}^2\frac{P'}{\rho'}\left(\Phi' - \mathcal{H}\Psi\right)\;,
\end{eqnarray}
where $\rho' = -3\mathcal H(\rho + P)$ has been used.
\end{ex}

\hrulefill

\begin{ex}
Use the generalised Poisson equation \eqref{relativisticPoissonequation}, written as
\begin{equation}
	3\mathcal{H}\left(\Phi' - \mathcal{H}\Psi\right) + k^2\Phi = 4\pi Ga^2\delta\rho\;,
\end{equation}
and the background relation:
\begin{equation}
	4\pi Ga^2(\rho + P) = \mathcal{H}^2 - \mathcal{H}'\;,
\end{equation}
in order to cast Eq.~\eqref{RPhirelder} as follows:
\begin{eqnarray}\label{RPhirelder2}
	4\pi Ga^2(\rho + P)\mathcal{R}' = \mathcal{H}\left(\Phi'' + 2\mathcal{H}\Phi' - \mathcal H\Psi' - 2\mathcal{H}'\Psi - \mathcal{H}^2\Psi\right)\nonumber\\
	 + \mathcal{H}\frac{P'}{\rho'}\left(4\pi Ga^2\delta\rho - k^2\Phi\right)\;.
\end{eqnarray}
\end{ex}

\hrulefill

The first term on the right hand side can be simplified using the Einstein equation \eqref{GiideltaPeq2}, so that we have:
\begin{equation}
	4\pi Ga^2(\rho + P)\mathcal{R}' = -4\pi Ga^2\mathcal{H}\delta P - \frac{k^2\mathcal H}{3}(\Phi + \Psi) + \mathcal{H}\frac{P'}{\rho'}\left(4\pi Ga^2\delta\rho - k^2\Phi\right)\;.
\end{equation}
Recalling the gauge-invariant entropy perturbation that we introduced in Eq.~\eqref{entropypert}:
\begin{equation}
	\Gamma \equiv \delta P - \frac{P'}{\rho'}\delta\rho\;,
\end{equation}
we can finally write:
\begin{equation}
	\boxed{\mathcal{R}' = -\mathcal{H}\frac{\Gamma}{\rho + P}
	 - \mathcal{H}\frac{P'}{\rho'}\frac{k^2\Phi}{\mathcal H^2 - \mathcal H'} - \mathcal H\frac{k^2(\Phi + \Psi)}{3(\mathcal H^2 - \mathcal H')}}
\end{equation}
The first term on the right-hand side vanishes if the perturbations are adiabatic, whereas the second and third terms on the right-hand side vanish on large scales, leaving $\mathcal{R}$ constant.

The gauge-invariant variables $\mathcal R$ and $\zeta$ can be related in the following way. First, using their definitions \eqref{Rperturb} and \eqref{zetaperturb} in the Newtonian gauge we can write:
\begin{equation}\label{zetaRrelation}
	\zeta = \mathcal R - \mathcal Hv + \frac{\delta\rho}{3(\rho + P)}\;.
\end{equation}
Second, write down the relativistic Poisson equation \eqref{relativisticPoissonequation} and the velocity equation \eqref{0iEinsteineq} in the following form:
\begin{eqnarray}
	3\mathcal H\Phi' - 3\mathcal H^2\Psi + k^2\Phi = 4\pi Ga^2\delta\rho\;,\\
	\Phi' - \mathcal H\Psi = 4\pi Ga^2\left(\rho + P\right)v\;.
\end{eqnarray}
Combine them to provide the constraint:
\begin{equation}
	12\pi G\mathcal Ha^2(\rho + P)v + k^2\Phi = 4\pi G\delta\rho\;.
\end{equation}
Using this constraint, the relation in Eq.~\eqref{zetaRrelation} becomes:
\begin{equation}\label{zetaRrelationfinal}
	\boxed{\zeta = \mathcal R + \frac{k^2\Phi}{12\pi Ga^2(\rho + P)}}
\end{equation}
Since $(\rho + P)a^2$ is of order $\mathcal H^2$, then $\zeta - \mathcal R$ is of order $k^2/\mathcal H^2$, which vanishes for large scales. This means that if $\mathcal R$ is conserved, then $\zeta$ is as well.

Weinberg proved a stronger result in \cite{Weinberg:2003sw, Weinberg:2003ur}. See also \cite{Weinberg:2008zzc}. That is, independently of the content of the universe, for $k \ll \mathcal{H}$ there are always two adiabatic scalar modes, for which $\mathcal{R}$ is constant, and one tensor mode for which $h$ is constant.

\clearpage
\bibliographystyle{apalike}
\addcontentsline{toc}{chapter}{Bibliography}
\bibliography{Cosmology-LN}
\printindex
\end{document}